\documentclass[11pt,letterpaper,oneside,openany,extrafontsizes]{memoir}

\usepackage{fontspec}
\IfFontExistsTF{EBGaramond-Regular.otf}{%
  \setmainfont{EBGaramond}[
    Extension     = .otf,
    UprightFont   = *-Regular,
    BoldFont      = *-Bold,
    ItalicFont    = *-Italic,
    BoldItalicFont= *-BoldItalic,
    Numbers       = OldStyle,
    Ligatures     = TeX,
  ]
}{%
  \IfFontExistsTF{texgyrepagella-regular.otf}{%
    \setmainfont{texgyrepagella}[
      Extension     = .otf,
      UprightFont   = *-regular,
      BoldFont      = *-bold,
      ItalicFont    = *-italic,
      BoldItalicFont= *-bolditalic,
      Ligatures     = TeX,
    ]
  }{%
    \setmainfont{Latin Modern Roman}
  }
}
\newcommand{\chapterfont}{\normalfont}
\usepackage{microtype}
\microtypesetup{
  protrusion = true,
  tracking   = smallcaps,
}

\usepackage[
  letterpaper,
  top=1.15in, bottom=1.25in,
  inner=1.25in, outer=1.0in,
  headsep=0.35in,
]{geometry}

\usepackage{setspace}
\usepackage{xcolor}
\definecolor{darkblue}{RGB}{0,38,84}
\definecolor{darkgreen}{RGB}{0,84,38}
\definecolor{chaptergrey}{RGB}{88,88,88}
\definecolor{rulegold}{RGB}{180,155,80}
\definecolor{lightgrey}{RGB}{200,200,200}
\definecolor{alertred}{RGB}{180,40,40}

\usepackage[
  colorlinks  = true,
  linkcolor   = darkblue,
  citecolor   = darkgreen,
  urlcolor    = darkblue,
  pdfauthor   = {Peter Woodbridge and John J. O'Hare},
  pdftitle    = {Dream Machine: The New Creative Economy},
  pdfsubject  = {Creative economy and artificial intelligence},
  bookmarksnumbered = true,
  breaklinks  = true,
]{hyperref}
\usepackage{bookmark}
\hypersetup{pdfkeywords={AI, creative economy, generative AI, digital labour}}
\usepackage{url}

\usepackage[
  backend  = biber,
  style    = numeric-comp,
  sorting  = none,
  maxbibnames = 99,
]{biblatex}
\usepackage{booktabs}
\usepackage{longtable}
\usepackage{tabularx}

\usepackage{graphicx}
\graphicspath{{figures/}}

\usepackage{amsmath}

\usepackage{csquotes}
\usepackage{enumitem}

\usepackage{lettrine}
\usepackage{tikz}
\usetikzlibrary{arrows.meta,calc,positioning,shapes.geometric,decorations.pathmorphing}

\usepackage{cleveref}

\usepackage{epigraph}
\makepagestyle{dream}
\makeevenhead{dream}{\small\textit{Dream Machine}}{}{}
\makeoddhead {dream}{}{}{\small\textit{\leftmark}}
\makeevenfoot{dream}{}{\thepage}{}
\makeoddfoot {dream}{}{\thepage}{}
\makeheadrule{dream}{\textwidth}{0.4pt}
\aliaspagestyle{chapter}{empty}

\makechapterstyle{dreammachine}{%
  \setlength{\beforechapskip}{40pt}
  \setlength{\afterchapskip}{30pt}
  \setlength{\midchapskip}{20pt}

}
\chapterstyle{dreammachine}

\footmarkstyle{{\footnotesize\textsuperscript{#1}\enspace}}
\newcommand{\orcid}[1]{%
  \href{https://orcid.org/#1}{\textcolor{darkgreen}{ORCID: #1}}}

\settocdepth{section}
\setsecnumdepth{section}

\abnormalparskip{0pt}

\begin{document}

\thispagestyle{empty}
\begin{center}
  \vspace*{\fill}
  \includegraphics[width=0.85\textwidth]{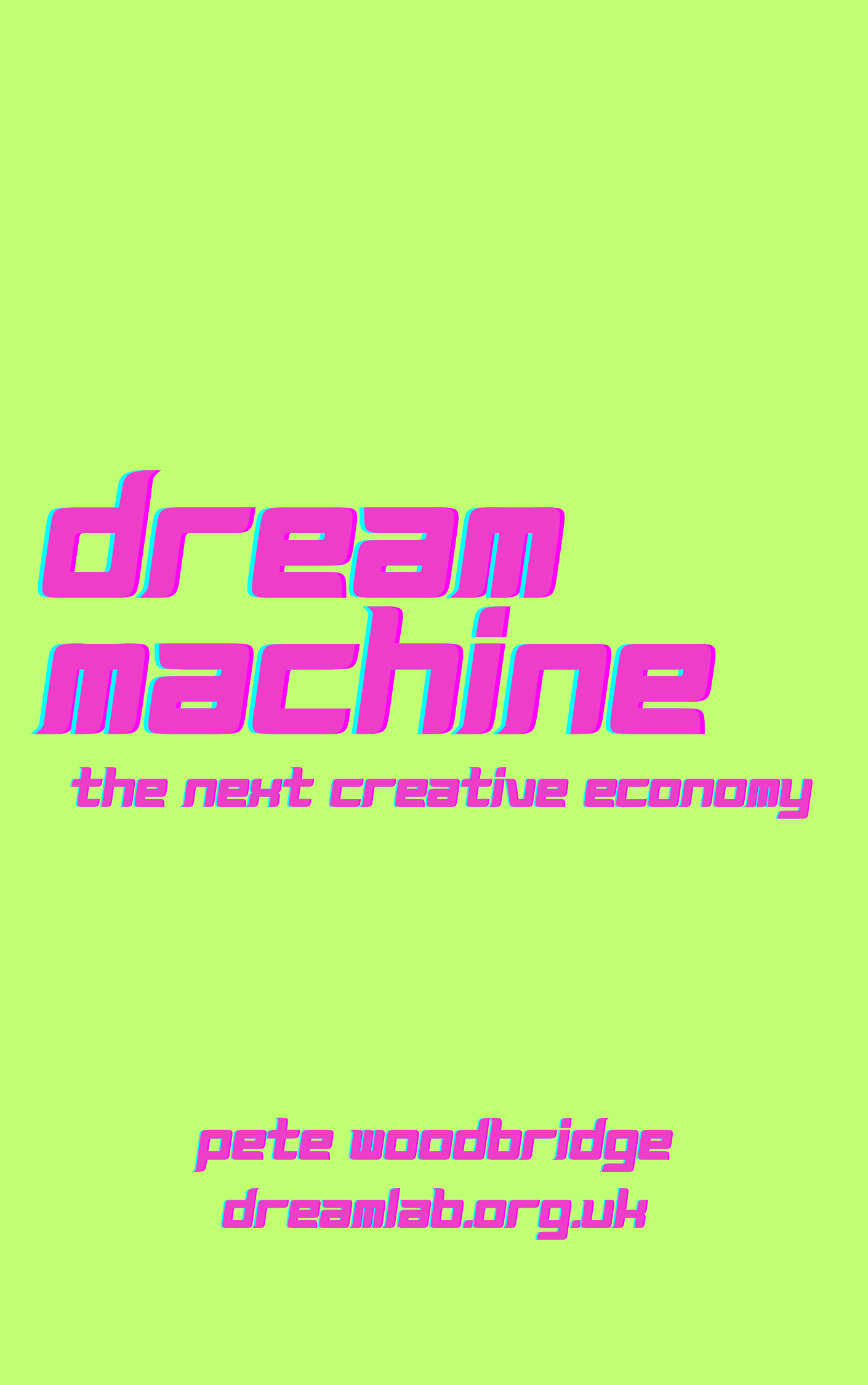}
  \vspace*{\fill}
\end{center}
\cleardoublepage

\thispagestyle{empty}
\vspace*{\stretch{1}}
\begin{center}
  {\chapterfont\fontsize{28}{34}\selectfont\textcolor{darkblue}{Dream Machine}}
\end{center}
\vspace*{\stretch{3}}
\cleardoublepage

\frontmatter
\begin{titlingpage}

\calccentering{\unitlength}
\begin{adjustwidth*}{\unitlength}{-\unitlength}

\vspace*{2cm}

\begin{center}
  {\fontsize{36}{42}\selectfont\textbf{Dream Machine}}\\[0.6em]
  {\fontsize{18}{24}\selectfont\textit{The New Creative Economy}}\\[3.5cm]

  {\large Peter Woodbridge}\\[0.3em]
  {\normalsize\href{https://dreamlab.org.uk}{DreamLab AI Collective},
    North West, UK}\\[0.4em]
  {\small\orcid{0000-0002-2858-3426}}\\[1.5em]

  {\large John J.\ O'Hare}\\[0.3em]
  {\normalsize\href{https://dreamlab.org.uk}{DreamLab AI Collective},
    North West, UK}\\[0.4em]
  {\small\orcid{0000-0001-5209-7754}}\\[0.4em]
  {\small\href{https://scholar.google.com/citations?user=Etx-Au4AAAAJ&hl=en}{%
    \textcolor{darkblue}{Google Scholar profile}}}\\[3.5cm]

  {\normalsize 21 May 2026}\\[1em]
  {\normalsize\textit{\href{https://dreamlab.org.uk}{DreamLab AI Collective}
    -- a non-profit collective of academics and practitioners}}
\end{center}

\vfill

\end{adjustwidth*}
\end{titlingpage}

\thispagestyle{empty}
\vspace*{\fill}

\begin{flushleft}
\small

\textit{Dream Machine: The New Creative Economy}\\[0.5em]

\textcopyright{} 2026 Peter Woodbridge. All rights reserved.\\[0.5em]

No part of this publication may be reproduced, distributed, or transmitted
in any form or by any means, including photocopying, recording, or other
electronic or mechanical methods, without the prior written permission of the
publisher, except in the case of brief quotations embodied in critical reviews
and certain other non-commercial uses permitted by copyright law.\\[0.5em]

Edition of 21 May 2026\\[0.5em]

Written at DreamLab in the North West of England.\\[0.5em]

Published by the DreamLab AI Collective, a non-profit collective of
academics and practitioners in the North West of England.\\
\href{https://dreamlab.org.uk}{dreamlab.org.uk}\\[0.5em]

Typeset in \LaTeX{} using the \textit{memoir} document class
with EB Garamond typeface.

\end{flushleft}

\vspace*{\fill}
\clearpage

\chapter*{About This Book}
\addcontentsline{toc}{chapter}{About This Book}
\thispagestyle{empty}

\noindent
In September 2025, a synthetic actress walked onto a Zurich film festival stage,
OpenAI shipped Sora~2, and Pete Woodbridge sat down to write a newsletter about
it. Thirty weekly issues later, what began as a one-month experiment had become
\textit{Dream Machine} -- the most-read working-creative record of the AI
transition.

\medskip
\noindent
This is the book that record produced. From the slop ceiling that the audience
is already enforcing, to the 88\% of UK creators demanding licensing in all
cases, to the chess-grandmaster strategy of deliberately playing the move the
machine wouldn't make -- Woodbridge holds the whole picture together in a way
no journalist, academic or platform-company keynote has managed.

\medskip
\noindent
Equal parts history, manifesto and operating manual, \textit{Dream Machine} is
the field guide to the most consequential year in creative work since cinema
learned to talk.

\bigskip
\begin{center}
\textit{The age of the \textup{How} is ending. Welcome the \textup{Why}.}
\end{center}

\clearpage

  \tableofcontents
  \listoffigures
  \listoftables
\chapter*{Foreword}
\label{ch:foreword}
\addcontentsline{toc}{chapter}{Foreword}
\markboth{Foreword}{Welcome to the Dream Machine}

\section*{Welcome to the Dream Machine}

\lettrine[lines=3,lhang=0.15,findent=0.1em]{T}{here was a week,} at the very end of
September 2025, when two things happened on opposite sides of the world that I knew,
while I was watching them, would change the rest of my year -- and probably most of yours.

In Zurich, an AI company called Particle6 walked an entirely synthetic actress called
Tilly Norwood onto the Zurich Film Festival stage and announced that talent agencies were
already in conversation about representing her. She had a face, a personality, a showreel
and, by the founder's own framing, a future. Within forty-eight hours, the U.S.\ actors'
union SAG-AFTRA had issued a statement that she was ``not an actor'' but ``a character
generated by a computer program that was trained on the work of countless professional
performers.''\footnote{%
  \href{https://variety.com/2025/film/news/sag-aftra-tilly-norwood-ai-actress-1236534779/}{Variety},
  ``SAG-AFTRA Condemns Tilly Norwood: AI Actress Is Not an Actor,'' 30 September 2025.
  See also \href{https://www.nbcnews.com/pop-culture/pop-culture-news/tilly-norwood-fully-ai-actor-blasted-actors-union-sag-aftra-devaluing-rcna234685}{NBC News},
  ``Tilly Norwood, fully AI `actor,' blasted by actors union SAG-AFTRA for `devaluing human artistry'.''
  Discussed in \emph{Dream Machine} Issue~1 (6 October 2025).}
The U.K.\ union Equity followed.\footnote{%
  \href{https://www.hollywoodreporter.com/movies/movie-news/tilly-norwood-ai-actress-uk-union-equity-sag-aftra-debate-1236391739/}{\textit{The Hollywood Reporter}},
  ``U.K.\ Union Equity Condemns Tilly Norwood: `AI Tool, Not a Performer'.''
  See also \href{https://variety.com/2025/film/global/tilly-norwood-slammed-equity-ai-tool-concerned-origin-1236537042/}{Variety},
  ``Tilly Norwood Slammed by Equity as AI Tool, Concerned About Origin.''
  \emph{Dream Machine} Issue~1.}
By the time the weekend was over, Whoopi Goldberg, Emily Blunt, Melissa Barrera and a
long list of other working performers had told the world, in their own words, what they
thought of all this.\footnote{%
  \href{https://www.cnn.com/2025/09/30/tech/hollywood-ai-actor-backlash}{CNN},
  ``Tilly Norwood: Hollywood is fuming over a new `AI actress','' 30 September 2025.}

The other thing was that OpenAI released Sora~2.\footnote{%
  \href{https://openai.com/index/sora-2/}{OpenAI}, ``Sora~2 is here,'' announcement,
  30 September 2025. The model launched alongside an invite-only iOS app of the same name
  in the U.S.\ and Canada. \emph{Dream Machine} Issue~1 carried the launch alongside
  contemporaneous coverage from NBC News and \textit{The Guardian} on the model's first
  copyright and safety incidents.}

I sat down on a Monday morning in early October, opened a blank LinkedIn article, and
wrote my first edition of a newsletter called \textit{Dream Machine}. I didn't have a
plan. I had a sentence:

\begin{quote}
\textit{``It's time to take AI seriously.''}
\end{quote}

I sent that issue out to a few hundred people in my network. I assumed I'd do it for a
month, maybe two -- that the wave would pass, or that I'd run out of material, or that
someone with a bigger platform would do it better. Eight months later, the newsletter has
3,800-odd subscribers, twenty-nine published editions, several thousand curated links, and
a small community of people in the North West of England -- the \textbf{DreamLab
Collective} -- who help me read, sift, argue and build around it every week.\footnote{%
  \textit{Dream Machine | Creative AI}, LinkedIn newsletter, archive of Issues 1–29,
  October 2025–May 2026.
  \href{https://www.linkedin.com/newsletters/dream-machine-creative-ai-7379776527871381505/}{linkedin.com/newsletters/dream-machine-creative-ai}.}

This book is the thing that happens when you keep a careful, public record of an industry
coming apart and putting itself back together inside the same eight months.

\bigskip\noindent\rule{\textwidth}{0.4pt}\bigskip

I should say who I am, because the rest of this book is in the first person and you should
know what kind of ``I'' is talking to you.

I am a Creative Technologist. I have spent twenty years working in and around what used to
be called ``new media'' -- virtual production, immersive, experiential, R\&D -- and is now
mostly called whatever the platform companies want to call it that quarter. I run a studio
called \textbf{DreamLab} in the North West of the UK. We are about fifty people: artists,
technologists, directors, games developers and storytellers, some of whom have won Emmys
and BAFTAs, some of whom finished their PhDs last year, all of whom are trying to figure
out, alongside everyone else, what it means to make creative work right now.\footnote{%
  \href{https://dreamlab-ai.com/team}{DreamLab AI Collective team page}.
  Referenced from \emph{Dream Machine} Issue~16 onward.}

I am not an AI evangelist. I am also not an AI sceptic. I am the kind of practitioner who
has had Marble running on a beta key for months and who has also sat in a room with a
games studio CEO who used the phrase ``AI was an expensive mistake'' without breaking eye
contact.\footnote{%
  Charles Cecil (Revolution Software, \textit{Broken Sword}) quoted in
  \href{https://www.gamesindustry.biz/ai-was-an-expensive-mistake-charles-cecil-on-innovation-insolvency-and-broken-sword}{\textit{gamesindustry.biz}},
  ```AI was an expensive mistake': Charles Cecil on innovation, insolvency, and Broken
  Sword.'' \emph{Dream Machine} Issue~3.}
I have built things with these tools and I have watched them break. I am writing this book
from inside the work.

Through the eight months the book covers, I have been doing three things in parallel:

\begin{itemize}
  \item \textbf{Running DreamLab}, the fifty-person studio above, which uses every major
    AI platform discussed in this book in live production. We have been a closed-beta
    partner for World Labs' Marble since October 2025.

  \item \textbf{Writing the \textit{Dream Machine} newsletter} -- twenty-nine weekly
    issues, several thousand curated source links across film, music, games, advertising
    and broadcast, a subscriber base of roughly 3,800 working creatives at the time of
    publication.

  \item \textbf{Building out a small set of analytical frameworks} -- the
    \textit{Human–AI Agency Continuum} (Chapter~\ref{ch:3}), the \textit{Slop Ceiling}
    (Chapter~\ref{ch:5}), the \textit{Four Positions} map of studio strategy
    (Chapter~\ref{ch:7}), the \textit{Year of the Orchestrator} (Chapter~\ref{ch:11}),
    \textit{Coordination Collapse} and the \textit{Consumption Gap} (Chapter~\ref{ch:13}),
    the \textit{AI Literacy Premium} and the \textit{Apprenticeship Gap}
    (Chapter~\ref{ch:14}), the \textit{Four Principles} of a humane creative economy
    (Chapter~\ref{ch:15}), and the \textit{Seven-Layer Toolchain} model
    (Chapter~\ref{ch:16}) -- that have begun, over the period, to circulate in the wider
    industry conversation.
\end{itemize}

The combination of the three is unusual. The trade-press journalists write the coverage
but do not, on the whole, run studios. The studio operators run studios but do not, on the
whole, publish a public weekly record. The academics produce research, sometimes excellent,
but at the cadence of academia rather than the cadence of the transition. The platform
companies produce material at the cadence of their own product cycles. Working from across
the three positions at once, week by week, has produced a particular kind of vantage point
-- neither outside the work nor confined to one slice of it -- that I have not seen anyone
else holding consistently through this period.

The book is what that vantage point produces. There are good histories of cinema written
by people who never ran a film studio, and good histories of the music industry written by
people who never produced a record. There are also, sometimes, books written from inside
the work -- by people who were making the decisions the book describes, in real time, under
the same conditions. The second kind tends, when honest, to tell you something the
outsider accounts cannot: what the moment \textit{felt like} to be inside, what the
practitioners thought they were doing, and what -- looking back -- the moment was actually
for.

This is that kind of book.

I am writing it because, in early October 2025, I realised that nobody I respected was
doing what I needed somebody to do: hold the whole picture in one place. Not the boosters.
Not the doomers. Not the niche tool-reviewers. The whole picture, week by week, including
the contradictions. Including the bits where Adobe was telling us that 86\% of creators
already use generative AI in their
workflow,\footnote{%
  Adobe, ``Inaugural Adobe Creators' Toolkit Report: 86 Percent of Global Creators Use
  Creative Generative AI.''
  \href{https://news.adobe.com/news/2025/10/adobe-max-2025-creators-survey}{Adobe press release}.
  Survey of 16,000 creators across the U.S., U.K., France, Germany, South Korea, Japan,
  India and Australia, released at Adobe MAX 2025.
  \emph{Dream Machine} Issue~6.}
\textit{and} where 88\% of creators who replied to the UK government's copyright
consultation said AI companies should have to license their work in every
case.\footnote{%
  UK Department for Science, Innovation and Technology (DSIT),
  \href{https://www.gov.uk/government/publications/copyright-and-artificial-intelligence-progress-report/copyright-and-artificial-intelligence-statement-of-progress-under-section-137-data-use-and-access-act}{\textit{Statement of Progress on Copyright and AI}},
  December 2025. See also
  \href{https://ipwatchdog.com/2025/12/16/respondents-uk-ai-consultation-overwhelmingly-want-ai-companies-license-copyrighted-works-all-cases/}{IPWatchdog},
  ``Respondents to UK AI Consultation Overwhelmingly Want AI Companies to License
  Copyrighted Works in All Cases.'' \emph{Dream Machine} Issue~12.}

The whole point -- to me -- is that those two numbers are both true.

\bigskip\noindent\rule{\textwidth}{0.4pt}\bigskip

The arc of these eight months turned out to be tighter than I expected.

In \textbf{October 2025}, the question on the table was whether AI in the creative
industries was real. By the time I finished writing \emph{Dream Machine} Issue~5, it was
no longer a question.\footnote{%
  \emph{Dream Machine} Issue~5, ``Adobe's Latest AI Announcements -- Is every tool going
  AI?'', 31 October 2025.
  \href{https://www.linkedin.com/pulse/dream-machine-creative-ai-news-insight-oct-25-issue-5-woodbridge-f7jnc/}{linkedin.com}.}

In \textbf{November}, the question shifted to whose tool stack it would run on. Adobe
announced -- and I am quoting them, this isn't editorial flourish -- ``AI in everything,
everywhere, all at once''.\footnote{%
  Adobe MAX 2025 keynote messaging, October 2025. Coverage:
  \href{https://www.creativeboom.com/news/adobe-is-putting-ai-in-everything-everywhere-all-at-once/}{Creative Boom},
  ``Adobe is putting AI in everything everywhere all at once.''
  \emph{Dream Machine} Issue~5.}
World Labs released Marble for public use a couple of weeks later, and the entire shape of
what ``a creative asset'' can be quietly changed.\footnote{%
  World Labs, \textit{Marble} -- first commercial spatial-AI world model, public launch
  November 2025. \href{https://marble.worldlabs.ai/}{marble.worldlabs.ai}. Technical
  context: \href{https://techcrunch.com/2025/11/12/fei-fei-lis-world-labs-speeds-up-the-world-model-race-with-marble-its-first-commercial-product/}{TechCrunch},
  ``Fei-Fei Li's World Labs speeds up the world model race with Marble, its first
  commercial product.'' DreamLab participated in the closed beta during October–November
  2025. \emph{Dream Machine} Issue~7.}

In \textbf{December}, the question was about consent and money. The UK government's
copyright consultation closed with eleven and a half thousand responses -- one of the
largest copyright consultations the country has ever run -- and a number that has stayed
with me: 88\%.\footnote{%
  11,514 responses across the Citizen Space portal and email, of which 10,112 came through
  Citizen Space; 88\% of those supported licensing as a default rule, against 3\% who
  supported the government's preferred opt-out model. UK DSIT,
  \href{https://www.gov.uk/government/publications/copyright-and-artificial-intelligence-progress-report/copyright-and-artificial-intelligence-statement-of-progress-under-section-137-data-use-and-access-act}{\textit{Statement of Progress}},
  December 2025; analysis in \emph{Dream Machine} Issue~12 (18 December 2025). Final
  report and economic impact assessment to be laid before Parliament by 18 March 2026.}

In \textbf{January 2026}, the question was who decides the rules. Sundance launched an AI
Literacy Initiative for filmmakers. Bandcamp banned AI music outright. Steam clarified,
then re-clarified, what counts as AI in a video game. Almost 800 creators signed an open
declaration with the line, \textit{``Stealing our work is not
innovation.''}\footnote{%
  \href{https://www.digitalmusicnews.com/2026/01/22/stealing-isnt-innovation/}{\textit{Digital Music News}},
  ``Nearly 800 Creatives, Including Jason Aldean and One Republic, Sign Responsible AI
  Declaration -- `Stealing Our Work Is Not Innovation'.''
  \emph{Dream Machine} Issue~16.}

In \textbf{February}, \textbf{March}, \textbf{April} and \textbf{May}, the question
started to feel like a different question altogether. It wasn't really \textit{should we
use these tools.} It was: \textit{now that the tools are inside the production pipeline at
every studio, every label, every newsroom, every agency, what kind of creative economy do
we actually want on the other side?}

That last question is the one this book is about.

\subsection*{What this book is and is not}

This is \textbf{not a tools guide}. Chapter~\ref{ch:16} lists every significant tool that
surfaced in the \textit{Dream Machine} archive in the period the book covers, but the rest
of the chapters are organised around the \textit{transition} -- the economics, the
audience, the labour, the unions, the law, the institutions, the rails -- rather than
around the apps. The tools change weekly. The transition is what will still be true in
2030.

This is \textbf{not a manifesto}. I do not believe the cleanest five-point plans for the
future of creative work, and I have refused to write one. What this book argues for, in
Chapter~\ref{ch:15}, is a \textit{test} you can apply to any policy, any contract, any
platform decision: \textit{Agency, Attribution, Access, Audience.} The test is not a
programme. It is a way of staying oriented in a fast-moving environment.

This is \textbf{not a chronicle}. The newsletter is the chronicle. Every issue is online,
every link is preserved, and the comprehensive thematic source index at the back of this
book (Appendix~\ref{app:a8}) catalogues the entire archive by topic for the reader who
wants to follow specific threads.

What this book \textit{is} is an argument, in sixteen chapters and eight appendices, that
creative work is being re-platformed in a twelve-month window -- that this is not the
internet of 1995 or the mobile phone of 2007, this is a faster, deeper, more thorough
re-platforming of the economic and cultural rails on which creative work travels -- and
that the choices being made \textit{right now}, by studios, by unions, by governments, by
toolmakers, by individual creatives at their kitchen tables, will set the terms for the
next decade. The book is here to help you make those choices on better information than
you would otherwise have.

That re-platforming has a name. It's the title of the book. \textit{The New Creative
Economy}. I don't think it's a metaphor and I don't think we have very long to decide what
we want it to look like.

\subsection*{What I believe, stated plainly}

The book lays out my position chapter by chapter, but if you want the headline conviction
up front, it is this:

\begin{quote}
\textbf{AI is best understood as an assistive instrument that amplifies human creativity.
Not a replacement for it. Not a substitute for it. An amplifier of it.}
\end{quote}

The working creative economy that emerges from this transition will be -- has to be -- the
one that does not lose sight of which side of that relationship is the master and which is
the servant. The human creativity is the master. The AI is the servant.
Chapter~\ref{ch:15} is the long-form version of that argument; everything else in the book
sits inside it.

I want to flag a second framing the book leans on, because it is the one I use most often
in the talks I have been giving. \textbf{We are leaving the age of the \textit{How} and
entering the age of the \textit{Why}.} The \textit{How} -- the technical labour of
executing a creative thought, the ability to draw, light, mix, model, edit, render -- has
been the bottleneck of professional creative work for a century. The \textit{How} is, in
2026, becoming a utility. A teenager with a midrange GPU can now produce work whose
surface quality sits on a continuum with what a full studio could produce in 2020. When
the \textit{How} becomes a utility, the \textit{Why} -- taste, intent, authenticity, the
willingness to take a risk on the move the data does not yet endorse -- is the only thing
left with commercial leverage. Chapter~\ref{ch:15} anchors this argument in a story I
borrow from elite chess; the rest of the book lives inside the strategic implication.

If you read no further than this front matter, you have the heart of the book.

\subsection*{What this book does not do well}

A critic-friendly note, because it makes for a more honest read.

This book is more confident about the \textit{creative-industries} layer of the AI
transition than it is about the layers above and below. The environmental and energy
footprint of the systems the book describes is something I touch on in Chapter~\ref{ch:15}
and otherwise under-treat; a fuller account is the subject of a different book, by a
different writer, that I hope is being written now. The labour conditions of the global
data-supply chain -- the labellers, evaluators and content moderators that the platform
companies depend on -- sit underneath every chapter of this book without being centred in
any of them; the same caveat applies. The geopolitics of AI, the macro-economic question
of the platform-company stock-market valuations, the wider policy questions about national
AI strategy, the philosophical questions about machine consciousness -- none of these is
the book's subject, and the book is shorter and more useful for not pretending otherwise.

The book is also, by design, anglophone-skewed and Global-North-skewed in its primary
sourcing. The 88\% in Chapter~\ref{ch:6} is a UK number. The Sundance turn in
Chapters~\ref{ch:8} and~\ref{ch:9} is a US story. The platform-company analysis in
Chapter~\ref{ch:9} is, in the main, an analysis of US and European companies, with
significant Chinese coverage but less than the Chinese open-source ecosystem deserves. The
Indian, African, Latin American and Southeast Asian stories I cover in
Chapters~\ref{ch:7}, \ref{ch:11} and~\ref{ch:12} are real but I cover them, in places,
from the wrong side of a translation gap. The next edition, if there is one, should fix
this. The deep-dive appendices begin the work but do not finish it.

Finally: the book is written \textit{while the transition is happening}. Some of the
specific claims, particularly in the tools chapter and in the predictions, will age in
ways I am not yet able to predict. The frameworks should outlast their evidence. The
evidence should be checked, when you read the book, against whatever the state of play is
by then.

\subsection*{Who this book is for}

I have written this book for \textbf{working creatives} -- the writers, directors,
songwriters, games designers, photographers, illustrators, editors, producers, agency
creatives, indie filmmakers, YouTubers, freelance designers, students and senior
practitioners who are, right now, trying to figure out what creative life looks like in
2026 and beyond.

It is also, secondarily, written for the \textbf{studio, agency and label leadership}
trying to make organisational decisions about AI integration in a year in which the cost
of getting it wrong is, by my read, the next decade of cultural authority.

And it is, thirdly, written for the \textbf{policy, union, institute and platform} people
who are deciding the rails the next decade of creative work will run on. The 88\% turned
up to the UK consultation. The Cannes Disclosure Standard, the Academy rule, the
SAG-AFTRA contract -- these are the institutional decisions that shape the field. The
people making them are part of the audience for this book.

Whoever you are: read with a pen. The chapters do not need to be read in order -- there is
a Reader Paths guide (page~\pageref{ch:reader-paths}) for different routes through the
material. The Source Index at the back lets you follow any thread back to its primary
sources.

\subsection*{A practical note, and an ask}

I've written the book in the same voice I write the newsletter -- talkative, opinionated,
North-West English, occasionally too fond of a bracket. But I have tried, in every
chapter, to put my opinions on top of evidence rather than the other way around. Every
claim that matters is footnoted. Every footnote points either to a primary source (a
research report, a court filing, an official announcement) or to the \textit{Dream
Machine} edition where I first wrote about it, where the original link is preserved. There
are several thousand citations. If you only ever read this book once, you can ignore them
entirely; if you ever want to know whether I made something up, follow the trail.

I have one ask of you before we start.

The temptation, when reading a book about AI in the creative industries in 2026, is to
take a side before chapter one. To decide, on page one, whether this is going to be about
how the machines are coming for us or about how the machines are setting us free. Please
don't. The most honest thing I can tell you about what I have learned over these eight
months is that the truth is almost always both at once, and that the most interesting
people in this story -- the directors, the songwriters, the games developers, the union
reps, the platform engineers, the indie filmmakers in bomb shelters and the policy
officers in Whitehall -- are the ones who can hold both sides at the same time without
flinching.

That's the kind of book I want this to be.

Welcome to the Dream Machine.

\bigskip
\noindent--~\textit{Pete Woodbridge}\\[0.2em]
\textit{DreamLab AI Collective}\\[0.2em]
\textit{The North West of England}\\[0.2em]
\textit{May 2026}

\bigskip\noindent\rule{\textwidth}{0.4pt}\bigskip

\noindent\textit{The complete book in reading order:}

\begin{itemize}
  \item \textbf{Foreword: Welcome to the Dream Machine} -- this piece
  \item \textbf{Reader Paths} -- six routes through the material
  \item Chapter~\ref{ch:1}: \textbf{The Day Sora Landed} -- the watershed week
  \item Chapter~\ref{ch:2}: \textbf{A History of Resistance} -- the 1839–2025 pattern,
    and where AI sits inside it
  \item Chapter~\ref{ch:3}: \textbf{The Human–AI Agency Continuum} -- the foundational
    frame, and the case for opening the black box
  \item Chapter~\ref{ch:4}: \textbf{Dead Internet, Living Web} -- the web infrastructure
    question, and the provenance stack
  \item Chapter~\ref{ch:5}: \textbf{The Slop Ceiling} -- the audience's verdict, and the
    Authenticity Premium
  \item Chapter~\ref{ch:6}: \textbf{The 88\%} -- the political watershed, and the
    Petrillo template applied to AI
  \item Chapter~\ref{ch:7}: \textbf{The Studios Decide} -- four strategic positions, and
    the trap legacy built for itself
  \item Chapter~\ref{ch:8}: \textbf{Worlds, Not Pictures} -- the spatial turn
  \item Chapter~\ref{ch:9}: \textbf{AI in Everything} -- the platform layer, and its
    underlying economics
  \item Chapter~\ref{ch:10}: \textbf{What is Newly Possible} -- six categories of
    newly-possible work, and the finite-attention ceiling
  \item Chapter~\ref{ch:11}: \textbf{The Orchestrator} -- the working creative's new role
  \item Chapter~\ref{ch:12}: \textbf{Authenticity as the New Scarcity} -- the provenance
    economy
  \item Chapter~\ref{ch:13}: \textbf{Coordination Collapse} -- shadow AI, the consumption
    gap, organisational re-shaping
  \item Chapter~\ref{ch:14}: \textbf{The New Jobs} -- the labour-market restructuring, the
    AI literacy premium, the Apprenticeship Gap
  \item Chapter~\ref{ch:15}: \textbf{Choosing the Future} -- the four principles, the
    assistive-amplifier conviction, the Age of the Why
  \item Chapter~\ref{ch:16}: \textbf{The Tools} -- the complete categorised inventory
  \item Chapter~\ref{ch:17}: \textbf{Five Years Inside the Dream Machine} -- a speculative
    five-year future-cast, argued from the book
  \item Chapter~\ref{ch:epilogue}: \textbf{Epilogue} -- a letter to 2030
  \item Appendix~\ref{app:a1}: \textbf{Quantitative Anatomy} -- the numbers behind the book
  \item Appendix~\ref{app:a2}: \textbf{Glossary} -- terms of art
  \item Appendix~\ref{app:a3}: \textbf{Bibliography by Topic} -- curated reading
  \item Appendix~\ref{app:a4}: \textbf{The Shadow AI Paradox} -- covert adoption,
    displacement, hypocrisy
  \item Appendix~\ref{app:a5}: \textbf{Dynamics of Generative AI Adoption} -- the
    consumption-gap evidence
  \item Appendix~\ref{app:a6}: \textbf{AI, Stigma, Privilege, Democratisation} -- the class
    question
  \item Appendix~\ref{app:a7}: \textbf{The Age of Intent} -- the philosophical spine
  \item Appendix~\ref{app:a8}: \textbf{The Dream Machine Source Index} -- thematic
    catalogue of every significant source across the 29 issues
\end{itemize}

\begin{figure}[htbp]
  \centering
  \includegraphics[width=0.95\textwidth]{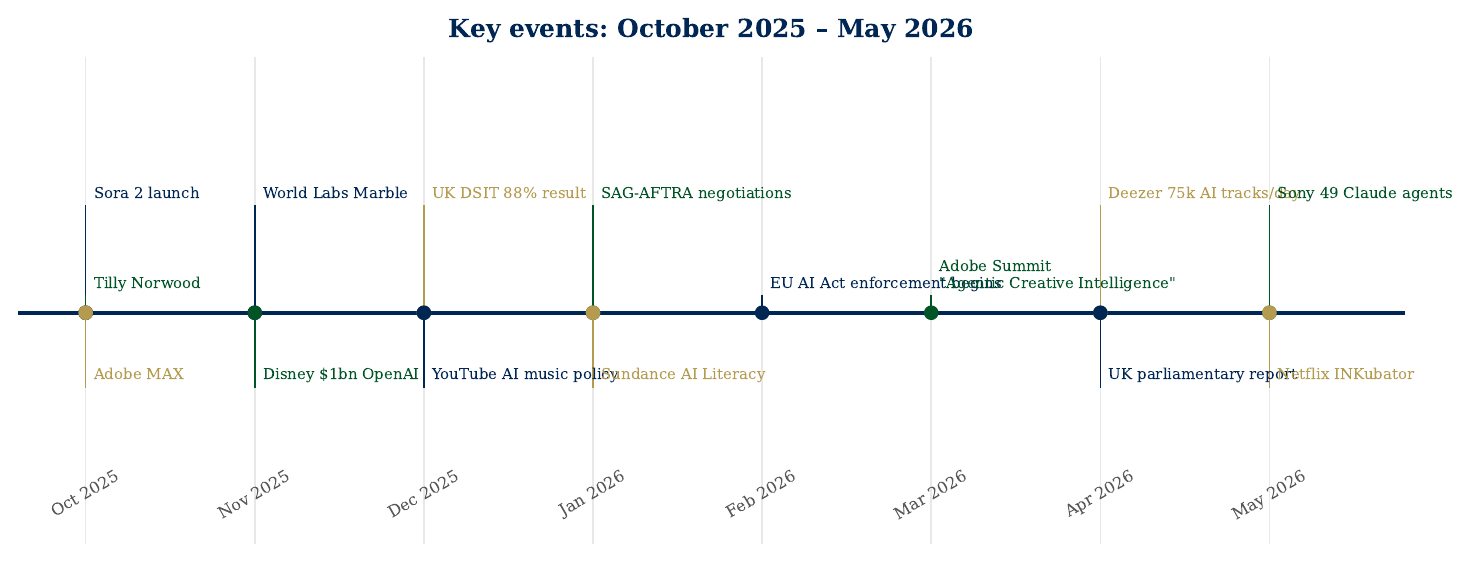}
  \caption{Key events in AI and the creative industries, October 2025 -- May 2026.}
  \label{fig:timeline}
\end{figure}

\chapter*{Reader Paths}
\label{ch:reader-paths}
\addcontentsline{toc}{chapter}{Reader Paths}
\markboth{Reader Paths}{Reader Paths}

\textit{This is a book about creative AI in the six months between October 2025 and
May 2026. It is not a tools guide. It is an argument, with the evidence underneath,
about what kind of creative economy is being built right now and what we should do
about it.}

You can read this book straight through. It is built to reward that. The book has
eighteen sections -- a combined Foreword, seventeen chapters, an Epilogue -- plus eight
appendices, sequenced so that each one earns the next.

If you don't have time for that, here are six ways into the book that will save you from
reading something that doesn't serve you yet. Pick one. Come back to the rest later.

\section*{If you are a working creative trying to figure out what to do next}

Read the \textbf{Foreword}, then Chapter~\ref{ch:2} (\emph{A History of Resistance}) for
the historical pattern you are inside, then Chapter~\ref{ch:3} (\emph{The Human–AI Agency
Continuum}), then Chapter~\ref{ch:10} (\emph{What is Newly Possible}) for the new
categories of work, then Chapter~\ref{ch:11} (\emph{The Orchestrator}), then
Chapter~\ref{ch:14} (\emph{The New Jobs}) for the labour-market evidence, then
Chapter~\ref{ch:15} (\emph{Choosing the Future}) -- particularly the section \emph{What
working creatives should do on Monday morning.} That is the spine of the
\textit{practitioner's} argument. Everything else in the book is evidence supporting it.

\section*{If you run a studio, agency, label or production company}

Read the \textbf{Foreword}, then Chapter~\ref{ch:7} (\emph{The Studios Decide}) --
particularly the section \emph{The trap the legacy industries built for themselves} --
then Chapter~\ref{ch:13} (\emph{Coordination Collapse}) -- particularly the section
\emph{What organisations should do.} Then read Chapter~\ref{ch:9} (\emph{AI in
Everything, Everywhere, All at Once}) for the platform-economics frame,
Chapter~\ref{ch:10} (\emph{What is Newly Possible}) for the new business categories, and
Chapter~\ref{ch:14} (\emph{The New Jobs}) for the labour-market restructuring evidence.
Chapter~\ref{ch:15} is the closing argument. Chapter~\ref{ch:16} (\emph{The Tools}) is
the practical inventory.

\section*{If you are working in policy or law}

Read the \textbf{Foreword}, then Chapter~\ref{ch:6} (\emph{The 88\%}) and
Chapter~\ref{ch:12} (\emph{Authenticity as the New Scarcity}) as a pair.
Chapter~\ref{ch:14} (\emph{The New Jobs}) for the labour-market policy framework.
Chapter~\ref{ch:15} for the four principles. Appendix~\ref{app:a1} for the data the policy
arguments rest on; Appendix~\ref{app:a6} for the class-and-democratisation analysis;
Appendix~\ref{app:a8} for the comprehensive source archive.

\section*{If you are working in (or covering) the music industry}

Read Chapter~\ref{ch:5} (\emph{The Slop Ceiling}) first -- it's where most of the
music-specific analysis lives. Then Chapter~\ref{ch:6} (\emph{The 88\%}) for the
rights-and-licensing argument. Then Chapter~\ref{ch:12} (\emph{Authenticity as the New
Scarcity}) for where it's heading. Chapter~\ref{ch:16} (\emph{The Tools}) for the
comprehensive music-AI tool inventory.

\section*{If you are working in (or covering) film, TV or games}

Read Chapter~\ref{ch:1} (\emph{The Day Sora Landed}) for the watershed scene-setter,
Chapter~\ref{ch:7} (\emph{The Studios Decide}) for the strategic map of how the industry
has positioned itself, Chapter~\ref{ch:8} (\emph{Worlds, Not Pictures}) for what is
coming next, Chapter~\ref{ch:11} (\emph{The Orchestrator}) for what it means for working
roles, and Chapter~\ref{ch:14} (\emph{The New Jobs}) for the labour-market data.

\section*{If you are reading this in a class, a book club, or as part of training}

Read the \textbf{Foreword} and Chapter~\ref{ch:1} to get oriented, then
Chapter~\ref{ch:4} (\emph{Dead Internet, Living Web}) to understand the structural
argument, then Chapter~\ref{ch:14} (\emph{The New Jobs}) and Chapter~\ref{ch:15}
(\emph{Choosing the Future}) to see the labour-market story and the four principles the
book argues for. The other chapters are evidence and elaboration. The \textbf{Glossary}
(Appendix~\ref{app:a2}), Citation Index, and Source Index (Appendix~\ref{app:a8}) are
designed to be the back-pocket reference set for the rest of the year.

\bigskip\noindent\rule{\textwidth}{0.4pt}\bigskip

A note about reading order: the chapters are designed to be load-bearing on each other.
If you skip a chapter and a later one references it without re-explanation, the Glossary
should fill the gap. If the Glossary doesn't fill the gap, that is my failure and not
yours -- please write to me through the newsletter and I will improve the next edition.

\medskip\noindent--~\textit{Pete}

\mainmatter
  \chapter{The Day Sora Landed}\label{ch:1}

\lettrine[lines=3,lhang=0.15,findent=0.1em]{T}{he first thing} to understand about the week of 30 September 2025 is that nothing in it was supposed to be a watershed. Sora had existed, in some form, for a year and a half. Runway, Pika, Kling, Luma, Veo and a dozen others had been releasing video models on a near-monthly cadence since the back end of 2023.\footnote{For a contemporaneous overview of the AI video model release cadence through 2024 and 2025, see \emph{Dream Machine} Issues 1--8 (October--November~2025), which logged near-weekly releases from Runway, Luma, Pika, Kling, Veo, Wan, Higgsfield, Hunyuan and a long tail of smaller labs.} AI-assisted post-production had been quietly integrated into nearly every studio pipeline in Hollywood for months. The major U.K. and U.S. actors' unions had been negotiating over digital replicas since the 2023 SAG-AFTRA strike. The Tilly Norwood character had been on Instagram, posting selfies and pretending to drink coffee in caf\'{e}s, since the previous July.

And yet that week~-- the seven days I would later mark as the start of this book, the start of the newsletter, and the start of a year that nobody in the creative industries has fully recovered from~-- three things lined up in a sequence so neat that I almost didn't believe it at the time.

On the Friday and Saturday, at the Zurich Film Festival, the founder of an AI studio called Particle6 stood on a panel and announced that several talent agencies were interested in representing her company's flagship product: a fully AI-generated actress called Tilly Norwood.\footnote{\emph{The Hollywood Reporter}, ``AI Performer Tilly Norwood Sparks Hollywood Backlash.'' \url{https://www.hollywoodreporter.com/movies/movie-news/tilly-norwood-ai-actress-uk-union-equity-sag-aftra-debate-1236391739/}. \emph{Dream Machine} Issue~1.}

On the Tuesday, the U.S. actors' union SAG-AFTRA issued a public condemnation calling her ``not an actor'' but ``a character generated by a computer program that was trained on the work of countless professional performers''~-- adding, in a line I have thought about more than any other this year, that ``audiences aren't interested in watching computer-generated content untethered from the human experience.''\footnote{SAG-AFTRA statement, 30 September 2025, reported in \emph{Variety}, ``SAG-AFTRA Condemns Tilly Norwood: AI Actress Is Not an Actor.'' \url{https://variety.com/2025/film/news/sag-aftra-tilly-norwood-ai-actress-1236534779/}.}

In between those two moments, on the Tuesday of the same week, OpenAI released Sora~2.\footnote{OpenAI, ``Sora~2 is here,'' 30 September 2025. \url{https://openai.com/index/sora-2/}. \emph{Dream Machine} Issue~1.}

I want to take each of these in turn, because the order matters, and then I want to argue that the actual watershed was something else entirely~-- something that happened underneath all three, that almost nobody noticed at the time, and that I think we will be living with for the rest of our working lives.

\section*{The actress}

Tilly Norwood was the creation of Eline Van der Velden~-- a comedian, writer and producer who had spent the better part of a year building her under the banner of a U.K.--Netherlands company called Particle6.\footnote{Particle6 background and Van der Velden interview: \emph{The Hollywood Reporter}, ``Meet the Creator of the AI Actress Hollywood Loves to Hate: `You're Gonna See a Lot of Tilly Norwood Next Year'.'' \url{https://www.hollywoodreporter.com/movies/movie-features/tilly-norwood-creator-particle6-eline-van-der-velden-talks-1236428824/}. \emph{Dream Machine} Issue~8.} Tilly had a face that looked like it had been assembled by committee from the most marketable features of the late 2010s. She had a voice. She had a small social-media following. She had, by the time of the Zurich announcement, ``another forty AI actors in the pipeline,'' according to Van der Velden's later interview in \emph{Deadline}.\footnote{\emph{Deadline}, ``Tilly Norwood Creator Eline Van Der Velden Talks Backlash, Reveals Another 40 AI Actors Are In The Pipeline.'' \url{https://deadline.com/2025/11/tilly-norwood-creator-interview-backlash-more-ai-actors-coming-1236601334/}.}

The Zurich announcement wasn't subtle. Van der Velden told the audience that several talent agencies were ``looking'' at signing Tilly. She framed her as an industrial product: a character who could be cast in feature films and television, who came with all the upsides of a real performer (a marketing footprint, an emotional connection with audiences, a brand) and none of the downsides (no salary, no per diems, no licensing complications, no aging, no scandals).\footnote{Northeastern Global News, ``Why AI `Actress' Tilly Norwood Has Hollywood Angry.'' \url{https://news.northeastern.edu/2025/10/02/ai-actress-tilly-norwood-hollywood-backlash/}.}

It is hard, looking back, to remember how loaded the word \emph{agency} was that weekend. In Hollywood and the wider acting industry, the moment a talent agency takes on a new client is the moment that client moves from aspiration to product. The agencies are the gatekeepers of the working economy. Van der Velden's announcement, in essence, was that the gate had cracked.

The response was almost immediate, and it came from the only side of the gate that had anything to lose.

SAG-AFTRA's statement, issued on the Tuesday after the festival closed, called Tilly Norwood ``not an actor.'' The full quote went further than the headlines tended to carry, and the next sentence was the one that actually mattered for the industry that read it: \emph{``Signatory producers should be aware that they may not use synthetic performers without complying with our contractual obligations, which require notice and bargaining whenever a synthetic performer is going to be used.''}\footnote{SAG-AFTRA, official statement reproduced in \emph{Variety}, \emph{op.~cit.}; also NBC News, ``Tilly Norwood, fully AI `actor,' blasted by actors union SAG-AFTRA for `devaluing human artistry'.'' \url{https://www.nbcnews.com/pop-culture/pop-culture-news/tilly-norwood-fully-ai-actor-blasted-actors-union-sag-aftra-devaluing-rcna234685}.} That sentence~-- terse, contractual, almost dull~-- was the union reminding every signatory studio in Hollywood that the 2023 strike had already settled this question, and that the rules already on paper applied here too.

The U.K.'s actors' union Equity issued its own condemnation within hours, focusing less on the philosophical question of what an actor is and more on the practical one: where had the training data come from? Their general secretary put it more sharply than any of the other initial responses: \emph{``We're at the stage in AI where so much data has been used that the original source becomes more and more unclear. And that's something that should worry every viewer, every working person, because that's not really the way our data should be used.''}\footnote{Equity (U.K.), statement of 2 October 2025: \emph{Variety}, ``Tilly Norwood Slammed by Equity as AI Tool, Concerned About Origin.'' \url{https://variety.com/2025/film/global/tilly-norwood-slammed-equity-ai-tool-concerned-origin-1236537042/}.} If you build a model on the labour of working actors without consent, payment or attribution, you have, in their view, not created a new performer~-- you have repackaged a stolen one.

Within seventy-two hours, the public response from working actors followed. Emily Blunt called it ``really, really scary.'' Whoopi Goldberg, on \emph{The View}, dismissed the entire premise. Melissa Barrera, Kiersey Clemons, Lukas Gage~-- a roll-call of working performers, mostly the ones with enough security in their careers to be willing to say anything on the record at all~-- each took their position.\footnote{CNN, ``Tilly Norwood: Hollywood is fuming over a new `AI actress'.'' \url{https://www.cnn.com/2025/09/30/tech/hollywood-ai-actor-backlash}.}

Eline Van der Velden, Particle6's founder, replied. \emph{``To those who have expressed anger over the creation of our AI character Tilly Norwood: she is not a replacement for a human being, but a creative work~-- a piece of art.''} The defence carried, in miniature, the entire shape of the argument the working actors were about to spend two years pushing back against. \emph{Art.} \emph{Creation.} \emph{Not a replacement.} The whole rhetoric of the AI-native studios, in one sentence, on the Monday morning of October 2025.

The line that got repeated most often, in the threads and the green rooms and the union briefings I read that week, was a variant of: \emph{we already gave you eighteen months of strike to settle this question.} The 2023 SAG-AFTRA strike had not been about AI in the abstract. It had been, in significant part, about digital replicas and the right of performers to control the use of their likeness in synthetic content. The deal that ended that strike had~-- supposedly~-- set the rules for the next era. Tilly Norwood, dropped onto a festival stage eighteen months later, was a test of whether those rules meant anything.

The answer was: they were going to have to be re-tested in public, on every individual case, for years.

\section*{The model}

What turned the Tilly Norwood weekend from a single-news-cycle controversy into the moment everyone in the creative industries started paying serious attention was that, on the Tuesday she was being condemned, OpenAI released Sora~2.

Sora~2 was not just an upgrade. It was, by OpenAI's own framing, a step-change in three things at once: physical realism (a ball that bounces correctly off a backboard), audio integration (sound effects and synchronised dialogue baked in, not added in post) and what the company called ``world state''~-- the ability to follow instructions across multiple shots while keeping the scene logically consistent.\footnote{OpenAI, ``Sora~2 is here,'' \url{https://openai.com/index/sora-2/}. Technical capabilities summary including physics modelling, multi-shot world-state persistence and synchronised audio.} If Sora~1, a year and a half earlier, had been the model that made people sit up and notice that AI video was a thing, Sora~2 was the model that made people sit up and notice that AI video might be a \emph{medium}.

The launch came with a second thing that, looking back, was the actually significant part: an invite-only iOS app, also called Sora, that worked like TikTok. You scrolled. You remixed. You did ``cameos''~-- the in-app feature that let you drop a generated likeness of yourself, or a friend, or anyone you had a clip of, into the model's output.\footnote{\emph{Dream Machine} Issue~1, ``Editor's Pick''; further launch context in NBC News, ``OpenAI's Sora~2: a major leap in AI video and audio.'' \url{https://www.nbcnews.com/tech/tech-news/openai-sora-2-app-video-chatgpt-creation-rcna234973}.}

Within five days, the Sora app had hit a million downloads.\footnote{LinkedIn News aggregation: ``Sora Tops 1 Million Downloads in 5 Days.'' \url{https://www.linkedin.com/news/story/sora-tops-1m-downloads-in-5-days-6684988/}. \emph{Dream Machine} Issue~3.}

The thing I want you to hold in your head about this is the timeline. The model was announced on the Tuesday. The app launched the same week. By the Friday, it was on the front page of the App Store. By the following Monday, the first wave of celebrity deepfakes~-- Robin Williams's daughter calling them ``gross''; Michael Jackson generated into music videos he never made; dead historical figures being put through new dialogue by users who didn't realise they were doing anything illegal because the app's design hadn't told them so\footnote{\emph{The Guardian}, ``OpenAI Sora~2 violence racism.'' \url{https://www.theguardian.com/us-news/2025/oct/04/openai-sora-violence-racism}. \emph{Dream Machine} Issue~1.}~-- was being written up in \emph{The Guardian} and the \emph{NBC News} tech vertical.\footnote{NBC News, \emph{op.~cit.}; \emph{The Guardian}, \emph{op.~cit.}}

OpenAI's own likeness-protection rules, \emph{The Hollywood Reporter} noted that week, had a specific carve-out: dead celebrities and ``historical figures'' weren't covered.\footnote{\emph{Digital Music News}, ``OpenAI's Sora~2 includes likeness protections for celebrities who don't opt in, but that doesn't apply to `historical figures' and dead celebrities.'' \url{https://www.digitalmusicnews.com/2025/10/08/openais-likeness-protections-dont-apply-to-dead-celebrities/}. \emph{Dream Machine} Issue~2.} The carve-out was treated, in the first days of the app, as a feature rather than a problem.

The clearest line of the entire launch week came not from OpenAI's blog post but from \emph{The Guardian}'s technology reporter, in the lead of their coverage of the Sora app's first violent and racist outputs. \emph{``In 2022, [the tech companies] would have made a big deal about how they were hiring content moderators \ldots{} In 2025, this is the year that tech companies have decided they don't give a shit.''}\footnote{Quoted in \emph{The Guardian}, ``OpenAI launch of video app Sora plagued by violent and racist images: `The guardrails are not real'.'' \url{https://www.theguardian.com/us-news/2025/oct/04/openai-sora-violence-racism}. \emph{Dream Machine} Issue~1.} I think that sentence will end up being remembered as one of the best one-line summaries of the moment. It was, if anything, optimistic about 2022. It was wholly accurate about 2025.

The reason this matters is not that Sora~2 was the first AI video model. It wasn't even the best one, by some metrics, that month~-- Veo~3.1 from Google would land in mid-October with arguably more sophisticated cinematic controls.\footnote{Google DeepMind, Veo~3.1 launch, mid-October 2025. \emph{Dream Machine} Issue~3, ``Editor's Pick: Veo~3.1 and the Rise of AI Filmmaking.'' Coverage: \url{https://www.cometapi.com/veo-3-1-is-comingand-whats-rumor/}.} The reason it matters is that the \emph{combination} of a major model release and a consumer iOS app, in a single week, collapsed a distinction that had~-- until that week~-- kept the conversation about AI in the creative industries safely in the hands of the people who made the creative industries.

Until Sora's iOS app, AI video was something that happened on a desktop, with a subscription, with a prompt window. After Sora's iOS app, AI video was something that happened on the phone of every teenager with an invite code, on a swipeable, remixable, social feed.

The line between ``an AI tool a working filmmaker might use'' and ``a default app on a default phone'' had been the line that held the cultural debate in shape for two years. That week, it didn't.

\section*{The phrase}

The line that has stayed with me from those first seven days came not from a celebrity, or from OpenAI's launch page, or from a union release. It came from a small Florida public-radio station's website. The headline read: \emph{Kiss reality goodbye: AI-generated social media has arrived.}\footnote{WUFT, ``Kiss reality goodbye: AI-generated social media has arrived,'' 3 October 2025. \url{https://www.wuft.org/2025-10-03/kiss-reality-goodbye-ai-generated-social-media-has-arrived}. \emph{Dream Machine} Issue~1.}

I have read that headline a hundred times since the week it was written. It is a perfect sentence. It is also~-- and I think this is what made the early days of Sora~2 so vertiginous~-- \emph{prematurely} true. Reality, as a category, did not end the week that Sora~2 launched. People still had real lives, real friends, real coffees in the morning, real bills at the end of the month. What ended, that week, was the easy assumption that what you saw on your phone had been made by somebody~-- a person, a team~-- with a recognisable connection to a recognisable place.

What's striking about the WUFT piece, and the few hundred similar pieces that followed it, is that they were not, in the main, written by people in the AI industry. They were written by reporters at local stations, by columnists at regional papers, by the kind of journalist who covers schools and county budgets and the planning department. The collapse of the line between ``made by a person'' and ``made by a machine'' was being noticed first not by the experts but by the audience.

That, more than anything else that happened that week, was the thing I wrote in the first issue of the newsletter. \emph{It's time to take AI seriously.} The line, six months on, embarrasses me a bit~-- it sounds like the kind of thing you say when you don't have anything else to say~-- but it was, on the morning of 6 October 2025, the only sentence I could write that felt like it was about the actual situation.

\section*{The director}

There was one more piece of the early reaction that I want to flag, because it set the template for almost every ``high-end'' creative response that followed.

In late September, the website \emph{No Film School} ran an interview with James Cameron in which he said~-- bluntly, on the record, in a quote that travelled~-- that AI was ``never going to take the place'' of humans in filmmaking. ``Filmmaking is subconscious,'' he said, ``and can't be quantified.''\footnote{\emph{No Film School}, ``James Cameron Says AI Is `Never Going to Take the Place' of Humans.'' \url{https://nofilmschool.com/james-cameron-ai}. \emph{Dream Machine} Issue~1.}

Two months later, in promotion for \emph{Avatar: Fire and Ash}, Cameron expanded on the point in a CBS \emph{Sunday Morning} interview that became one of the most-shared creative-industry stories of the entire winter. Asked about generative AI's ability to ``make up a performance from scratch with a text prompt,'' Cameron said: ``It's like, no. That's horrifying to me.'' And then, in a line I have used in talks and in arguments and in this book: \emph{``The act of performance, the act of actually seeing an artist creating in real time, will become sacred.''}\footnote{\emph{The Guardian}, ``James Cameron says AI actors are `horrifying to me','' 1 December 2025. \url{https://www.theguardian.com/film/2025/dec/01/james-cameron-says-ai-actors-are-horrifying-to-me}. Original quote from CBS \emph{Sunday Morning}. \emph{Dream Machine} Issue~10.}

He went on, in a passage that almost never travelled with the headline quote: \emph{``It also causes us to have to set our bar to a very disciplined level, and to continue to be out-of-the-box imaginative.''} The whole interview, taken together, is not the doom-laden refusal the press cycle reduced it to. It is a working filmmaker articulating an argument about \emph{craft} in an era of automated production~-- that the existence of cheap synthetic performance does not retire the human performer, it \emph{raises the discipline required of the human performer.} The audience's bar moves up. The work that wins it has to move up too.

The companion line Cameron gave \emph{Variety} the same month was even more telling: \emph{``For years, there was this sense that, `Oh, they're doing something strange with computers and they're replacing actors,' when in fact, once you really drill down and you see what we're doing, it's a celebration of the actor-director moment.''}\footnote{\emph{Variety}, ``James Cameron Says It's `Horrifying' that AI Can `Make Up an Actor'.'' \url{https://variety.com/2025/film/news/james-cameron-horrifying-ai-replace-actors-1236595864/}.} Cameron has spent forty years building digital filmmaking. He is, as a working artist, perhaps the most sophisticated user of computer-aided performance in the history of cinema. His argument is not against AI; it is against AI \emph{that displaces the moment in the room where an artist creates.} That distinction~-- \emph{between AI that augments the actor-director relationship and AI that substitutes for it}~-- is the one most working creatives are trying to draw, and the one most press coverage of the AI debate still flattens.

What almost every report of those quotes glossed over~-- and what made them more interesting, not less~-- was that Cameron was, and still is, a board member of Stability AI.\footnote{Stability AI, board composition, 2024--2026. Reported across multiple outlets including \emph{Deadline}, ``James Cameron Calls AI Replacing Actors `Horrifying'; Art `Sacred'.'' \url{https://deadline.com/2025/11/james-cameron-gen-ai-horrifying-human-art-sacred-avatar-1236631387/}.} He is not an opponent of AI in filmmaking. He is, by any reasonable definition, an investor in it. His position was not ``no AI.'' His position was ``no AI that replaces the actor in the room.'' That distinction is the one most working creatives, in my experience, are trying to make. The early coverage, looking for the clean villain-or-saviour story, mostly missed it.

I think Cameron's \emph{Sunday Morning} line is the most important thing said by a working filmmaker in the entire six months I have been writing the newsletter. Not because it is right about everything~-- I think the ``sacred'' framing makes a smaller and more brittle claim than it sounds~-- but because it is the first time, in the AI era, that one of the people who \emph{built} the apparatus of digital cinema in the 1990s and 2000s drew a line that he himself was prepared to defend.

The line, again, is not ``no machines.'' Cameron has been a machine-builder for forty years. The line is: a human creates \emph{in real time}, and that creation is the work. Everything else is just delivery.

\section*{The state of the field, the week before}

I want to spend a section on what was already in motion when the Tilly Norwood announcement happened, because the historians of this period are, I suspect, going to under-tell the \emph{cumulative} story underneath the \emph{catalysing} one. The week of 30 September 2025 was not the moment AI arrived in the creative industries. It was the moment the \emph{audience} arrived in the AI debate. The infrastructure underneath had been forming, in a slow, uneven, partly-public, partly-private way, for at least two years before it.

Let me sketch the field.

By the time Sora~2 launched, the major AI-video model release cadence was running at roughly one significant model per fortnight. Runway had shipped Gen-4 in late 2024 and was deep into the public roll-out of Gen-4 Image-to-Video and the Workflows product by September 2025. Luma had released its Dream Machine consumer app; the Genie~3 demo from Google DeepMind in late summer 2025 had been described by \emph{Time} as one of the year's best inventions. Pika~2.0 was shipping. Higgsfield, which would close \$80M on a \$1.3B valuation by January 2026, was already on its third major product cycle. Hunyuan Video and Wan~2.2~-- the open-source Chinese-built models from Tencent and Alibaba~-- had been freely available, on commodity GPUs, for months. Kling, the Kuaishou model that would, by mid-2026, be the model many professional filmmakers actually used for production-grade clips, had been in continuous public release. Sora~2 was the headline of the week. It was not, in any meaningful sense, the entire field. The field was already crowded.

The same is true of the studio-side adoption. Lionsgate had publicly partnered with Runway on a deal to produce AI-augmented studio films a year before \emph{Futurism}'s headline-grabbing ``crumbled into disaster'' piece. Netflix's quiet integration of AI tools into background-plate generation, animated short development and de-aging post-production work had been documented across 2024 and the first half of 2025. Disney's \emph{House of David}~-- the show whose creator would, in November 2025, defend the use of more than 350 AI-generated visual-effects shots in its second season~-- had been in production with that pipeline in place months before the Sora~2 launch. The major broadcast and streaming companies had been integrating AI under the hood at a pace that the public conversation had not yet caught up with. The Tilly Norwood week was, in part, the moment that pace became impossible to keep quiet.

The unions had been working on this even longer. The 2023 SAG-AFTRA strike had~-- through the Writers Guild and through the actors' bargaining~-- produced contract language on ``digital replicas'' that was, by 2025, already two years old. Equity in the UK had been running consultations and ballots through 2024 and into 2025. The Authors Guild's class-action lawsuits against OpenAI had been filed in mid-2023 and were grinding through discovery. The European AI Act had been finalised in 2024 and was beginning to bite on copyright disclosures by the time of the Zurich announcement. The Music Performance Trust Fund's emerging conversations about an AI-era levy mechanism~-- the Petrillo-template applied to neural-network outputs that I argue for in Chapter~\ref{ch:2} and Chapter~\ref{ch:15}~-- were already on the agenda at AFM Local 802 and at the UK Musicians' Union long before Tilly walked on stage. The institutional response had pre-existed the cultural rupture. The cultural rupture is what made the institutional response politically possible.

The adoption telemetry on the platform side, in the months before the Sora~2 week, was already at the level that would later be revealed in Adobe's MAX 2025 Creators' Toolkit Report and the Stanford AI Index 2025. Firefly was already on track for its 22-billion-asset milestone. ChatGPT, by the time Sora~2 launched, was already at roughly 700--800 million weekly active users on its way to 900 million. The 86\% of global creators who reported using generative AI in their workflow in the Adobe survey~-- published a few weeks after the Tilly Norwood week, but capturing data collected before it~-- was a number the platform-companies had been quietly watching for months. The Adobe survey did not produce the adoption. It documented the adoption that was already complete by the time the public was paying attention.

On the \emph{consumer} side, the Sora app was, again, the catalyst rather than the inventor of the dynamic. The TikTok-style consumer surface for AI generation had been visible for at least a year. ByteDance's Dreamina and CapCut tools had been integrating Seedance, Seedream and the wider ByteDance generative stack into consumer-facing video editing through 2024 and into 2025. The Sora app's million-downloads-in-five-days number~-- the headline that defined the week's consumer dynamics~-- landed inside a market that had been \emph{prepared for it} by Dreamina, CapCut, the Krea consumer app and the Suno and Udio consumer-facing music platforms. What was new about the Sora app was not the \emph{concept} of swipeable AI generation. What was new was that an American flagship-AI company had chosen to ship it as a \emph{primary} consumer surface alongside the model.

Particle6, finally, had been building Tilly Norwood for the better part of a year before the Zurich panel. Van der Velden's Instagram and TikTok rollouts had been running since the previous summer. The character had a follower count, a posting cadence, a slowly-built visual identity, a small but real fanbase that engaged with her as if she were a person. The Particle6 strategy~-- the deliberate cultivation of a \emph{parasocial} relationship between an AI character and a human audience, in the months before the studio-system pitch~-- was a play borrowed from the playbook of the Korean virtual-idol economy, the Japanese vocaloid scene, and the long history of cultivated personas in the influencer industry. What was new about Tilly was not the \emph{idea} of a synthetic personality with a fanbase. What was new was the framing of that synthetic personality as a \emph{casting option} for legacy film and television. The Zurich announcement was the moment the parasocial-character economy and the working-actor economy were proposed, on a festival stage, as overlapping markets. The reaction was, in retrospect, the audience and the union refusing to let the markets overlap.

What all of this means, when you stack it together, is that the Tilly Norwood / Sora~2 week did not invent the AI moment in the creative industries. It \emph{named} it. It made it impossible to keep treating AI as a technical category that working creatives could opt out of. The model release was overdue. The Particle6 announcement was overdue. The union response was already drafted in some form. The audience reaction was already, on the slop-ceiling logic I lay out in Chapter~\ref{ch:5}, structurally inevitable. What the week did was \emph{force everything to happen in public, at the same time, in front of an audience that had not previously been part of the conversation.}

I think that is the deeper reason the week mattered. The conditions were ripe. The catalyst was small. The reaction was big. The change in the \emph{visibility} of the AI transition~-- from a back-room toolchain conversation to a front-page audience question~-- was, by my reading, the actual watershed. Everything else in this book is downstream of that visibility shift.

\section*{The watershed under the watershed}

I said earlier that the actual watershed of that week was not the Tilly Norwood announcement, and not the Sora~2 launch, and not the union responses, and not the Cameron quote. The actual watershed was something underneath all four.

For two years, the conversation about AI in the creative industries had been a conversation among insiders. Toolmakers talked to creators. Creators talked to studios. Studios talked to unions. Unions talked to governments. The general public~-- the audience for the things being made~-- had largely been a backdrop. They had been the people \emph{for whom} this argument was happening, not the people in the argument.

In the week of Sora~2, that ended.

The Sora app put a generative video tool on the phone of anyone with an iOS device and an invite code, and the invite codes were not hard to come by. The Tilly Norwood announcement put the abstract concept of ``a synthetic actress'' into the \emph{Daily Mail}, \emph{The Guardian}, \emph{The View}, the breakfast television circuit, and three different morning radio shows I happened to be listening to that week. The Cameron quote, when it came, ran on every wire service that covers the entertainment business.

The audience joined the argument. Not as a side, but as a participant.

And once the audience is in the argument, the argument changes. It is no longer about what the unions can negotiate, what the studios will adopt, what the toolmakers will ship. It is about what the people who watch films and listen to music and play games and scroll feeds will accept, demand, refuse and forgive.

The rest of this book~-- the eleven chapters that follow this one~-- is, in one way or another, an account of what the audience has been doing with the argument since it became theirs. The artists' boycotts and the streaming platforms' counter-moves; the eleven and a half thousand UK citizens who turned up to a government consultation; the 50,000 AI-generated tracks uploaded to Deezer every day and the 1 to 3 per cent of streams those tracks actually got;\footnote{Deezer, ``AI-generated tracks now represent 44\% of all new uploaded music,'' April 2026 newsroom release. \url{https://newsroom-deezer.com/2026/04/ai-generated-tracks-represent-44-of-new-uploaded-music/}. Companion analysis: \emph{Music Business Worldwide}, ``75,000 AI-generated tracks now flood Deezer daily, representing 44\% of all new music uploaded to the platform.'' \url{https://www.musicbusinessworldwide.com/75000-ai-generated-tracks-now-flood-deezer-daily-representing-44-of-all-new-music-uploaded-to-the-platform-says-streamer/}. Daily AI uploads to Deezer rose from approximately 50,000 per day in November~2025 (\emph{Dream Machine} Issue~7, citing Deezer / \emph{Musically}) to 75,000 per day by April 2026, with consumer streams of fully-AI tracks holding between 1\% and 3\% of total platform plays~-- and up to 85\% of those streams identified as fraudulent in 2025. \emph{Dream Machine} Issues 7, 26, 27, 28.} the death threats sent to Tilly Norwood's creator in early 2026; the moment in the middle of January 2026 when the U.S. actors' union and SAG-AFTRA went back to the negotiating table because the audience, having looked at the new landscape, had decided what it wanted.

The week of 30 September 2025 was the last week before any of that.

I started writing the \emph{Dream Machine} newsletter the Monday after. By the time I sent the first edition out, the conversation had already moved on.

  \chapter{A History of Resistance}\label{ch:2}

\lettrine[lines=3,lhang=0.15,findent=0.1em]{I}{} want to start this chapter with a piece of writing.

\begin{quote}
\emph{``Sweeping across the country with the speed of a transient fashion in slang or Panama hats, political war cries or popular novels, comes now the mechanical device to sing for us a song or play for us a piano, in substitute for human skill, intelligence, and soul\ldots{} Let us not hamper it with a machine that tells the story day by day, without variation, without soul, barren of the joy, the passion, the ardor that is the inheritance of man alone.}

\emph{Singing will no longer be a fine accomplishment; vocal exercises, so important a factor in the curriculum of physical culture, will be out of vogue! Then what of the national throat? Will it not weaken? What of the national chest? Will it not shrink?}

\emph{When a mother can turn on the phonograph with the same ease that she applies to the electric light, will she croon her baby to slumber with sweet lullabys, or will the infant be put to sleep by machinery? Children are naturally imitative, and if, in their infancy, they hear only phonographs, will they not sing, if they sing at all, in imitation and finally become simply human phonographs~-- without soul or expression?''}
\end{quote}

Read that paragraph once more, and try, before I tell you when it was published, to date it.

If you guessed \emph{autumn 2024 trade-press editorial on Suno}, you would not be alone. The language sits comfortably alongside the take-pieces about AI music slop that filled the music press in the months I started writing the newsletter. \emph{Machinery in substitute for human skill, intelligence and soul.} \emph{The story told day by day without variation.} \emph{Children growing up as imitative human phonographs, without soul or expression.} Every clause has a 2024--25 equivalent.

The essay is not from 2024. It was published in \textbf{September 1906}, in \emph{Appleton's Magazine}, by \textbf{John Philip Sousa}~-- the most popular bandleader in the United States at the time~-- and it is titled \emph{``The Menace of Mechanical Music.''}\footnote{John Philip Sousa, ``The Menace of Mechanical Music,'' \emph{Appleton's Magazine}, Vol.~8, September 1906, pp.~278--284. Full text via ExplorePAHistory: \url{https://explorepahistory.com/odocument.php?docId=1-4-1A1}. Academic context: Patrick Warfield, ``John Philip Sousa and `The Menace of Mechanical Music','' \emph{Journal of the Society for American Music}, Cambridge University Press: \url{https://www.cambridge.org/core/journals/journal-of-the-society-for-american-music/article/abs/john-philip-sousa-and-the-menace-of-mechanical-music/A9E621587BE7580ABD73AEF64D4B2DC8}. The 1906 essay was, in part, lobbying for what would become the 1909 Copyright Act.} The machine Sousa was warning his readers against was the phonograph. The ``soul-barren'' recording technology Sousa feared was Edison's flat disc, spinning at 78~rpm, playing back music captured by a brass horn.

The same essay was~-- almost word for word, with a few changes in the names of the machines~-- written about the player piano in 1900, the microphone in 1932, the synthesiser in 1980, the drum machine in 1991, Auto-Tune in 1998, non-linear video editing in the early 1990s, the digital camera in the 2000s, the smartphone-as-camera in the 2010s, and generative AI in 2023, 2024 and 2025.

That is the subject of this chapter. The structured, recurring, almost ritual \emph{pattern} by which every major creative-technology introduction in the modern era has been received by its working practitioners~-- and what that pattern, eighteen iterations later, tells us about how to think about the AI moment the rest of this book describes.

I want to be careful about how I do this. The historical-analogy move is, in tech writing, a famously cheap one. \emph{``Every disruptive technology has been resisted; therefore your resistance to this technology is wrong''} is the rhetorical operating system of two decades of platform-company keynotes, and it has been wielded so dishonestly that working creatives in 2026 are right to be suspicious of any version of it. I am not, in this chapter, making that argument. I am making a \emph{more specific} one. The historical pattern has~-- across twenty distinct technologies in the period from 1839 to 2022~-- a recognisable five-act shape, and the \emph{five-act shape} is what the pattern tells you. Not that resistance is wrong. Not that the new tool will be fine. Something more useful than either: which institutional moves work, which fail, what gets preserved, what gets lost, where the \emph{new} creative forms come from, and what the working practitioner's actual leverage in the period is.

The history is, in other words, \emph{operationally} informative. That is the use I want to make of it.

\section*{The five-act curve}

Every resistance I look at in this chapter~-- and there are many more I do not have room for~-- moves through roughly the same five stages, in something like the same order.

\textbf{Act One: Ridicule.} The new tool is dismissed as a toy. It cannot compete. It sounds awful, looks crude, has the wrong specifications. The serious practitioners are unconcerned because the work it produces is not, on inspection, work. The Roland TR-808 drum machine, released in 1980 and a commercial failure, was reviewed as ``toy robot drums.'' The Canon 5D Mark~II, the DSLR that started the death of dedicated cinema cameras, was~-- in 2008~-- dismissed as a stills camera with a video gimmick. Auto-Tune, between 1997 and roughly 2003, was used as an \emph{undisclosed} studio tool because no working singer wanted to admit that their pitch was being machine-corrected.

\textbf{Act Two: Moral panic.} The new tool is reframed as a threat to public morals, aesthetic standards, or the integrity of the form. It is degenerate. It is theft. It is not real \emph{singing}, not real \emph{photography}, not real \emph{art}. The 1932 sermon by \textbf{Cardinal O'Connell} of Boston that crystallised this stage for the microphone~-- that crooning was ``a degenerate form of singing,'' that ``no true American man would practice this base art,'' and that crooners were ``whiners and bleaters defiling the air''~-- is unimprovable as a template.\footnote{William Henry Cardinal O'Connell, Archbishop of Boston, sermon to the Holy Name Society, Boston, 10 January 1932. Reported widely in the contemporaneous press, including the \emph{Daily Courier} (Connellsville, PA), 12 January 1932 (\url{https://www.newspapers.com/newspage/38168082/}). Cultural context: KUOW/NPR, ``\,`Imbecile Slush': Surprising Early Reactions to Crooning,'' \url{https://www.kuow.org/stories/imbecile-slush-surprising-early-reactions-crooning}. JSTOR Daily, ``The Gender Politics of the First Boy Bands,'' \url{https://daily.jstor.org/the-gender-politics-of-the-first-boy-bands/}.} \emph{``Imbecile slush,''} O'Connell called it, in language that anyone who has read a 2024 anti-AI op-ed will recognise. The 1991 federal court ruling in \emph{Grand Upright Music v.\ Warner Bros.}~-- the case in which Biz Markie was sued for sampling Gilbert O'Sullivan~-- opened with the words \emph{``Thou shalt not steal,''} quoting the Seventh Commandment, in a US copyright opinion.\footnote{\emph{Grand Upright Music, Ltd.\ v.\ Warner Bros.\ Records Inc.}, 780 F.~Supp.~182 (S.D.N.Y.~1991). Full text: \url{https://law.justia.com/cases/federal/district-courts/FSupp/780/182/1445286/}. The ``Thou shalt not steal'' opening is the most-quoted line from a US copyright opinion of the late twentieth century.} In 2025, the moral-panic stage of AI is mostly behind us; the analogous \emph{Thou shalt not steal} language is in the UK's 88\% copyright consultation response, the \emph{Stealing Our Work Is Not Innovation} declaration, and the union statements quoted throughout this book.

\textbf{Act Three: Existential professional alarm.} The displaced practitioners realise the tool is not a toy and not just morally suspect~-- it is structural. It is going to take their work, change their craft, and reshape the institutions that support them. The classic statement is \textbf{Phil Tippett's,} the stop-motion master who saw ILM's first digital test of a \emph{Jurassic Park} T.~rex in 1992 and said: \emph{``I think I'm extinct.''}\footnote{Tippett's account of the \emph{Jurassic Park} digital test is documented across multiple ASC and contemporaneous press accounts. American Society of Cinematographers, ``Jurassic Park: Effects Team Brings Dinosaurs Back from Extinction,'' \url{https://theasc.com/articles/jurassic-park-effects-team-brings-dinosaurs-back}. Wikipedia, ``Phil Tippett,'' \url{https://en.wikipedia.org/wiki/Phil_Tippett}. The dialogue paraphrase Spielberg incorporated into the film is Goldblum/Malcolm's response to Grant's ``I think we're out of a job'': ``Don't you mean \emph{extinct}?''} Spielberg liked the line enough to put a paraphrase of it in the film. Tippett was right about himself in some local sense~-- his go-motion craft did not survive \emph{Jurassic Park}'s release. He was wrong about himself in a wider one~-- his studio went on to produce digital animation work for \emph{Starship Troopers} and is still operating in 2026. Both readings can be true at the same time. Most of the existential-alarm moments work like this.

\textbf{Act Four: Institutional and legal counter-attack.} Unions strike. Lawmakers legislate. Lawsuits get filed. The 1942 and 1948 \textbf{Petrillo strikes}~-- when the American Federation of Musicians, led by James Caesar Petrillo, refused to record for the major labels~-- are the canonical version of this stage, and I will spend longer on them in a moment. The UK Musicians' Union's \textbf{1980 ``Massacre of the Musicians'' BBC strike} and its \textbf{1982 motion to ban synthesisers} are the British version. The 1991 \emph{Grand Upright} ruling that sampling was theft is the legal version. The 2007 \emph{Viacom v.\ YouTube} \$1bn lawsuit is the platform-distribution version. The 2023 SAG-AFTRA and WGA strikes~-- which gave us the contract language that frames Chapter~\ref{ch:1}~-- are the most recent before the period this book covers.

\textbf{Act Five: Settlement.} The dust settles into one of three forms. The displaced craft \emph{dies}: miniature painting after 1840, hand-drawn feature animation after \emph{Toy Story}, the photo-lab business after digital. The new tool is \emph{taxed and the revenue redistributed}: the Petrillo settlement created the \textbf{Music Performance Trust Fund}, still distributing payments to live musicians in 2026; the DMCA Section~512 plus YouTube's Content ID created a parallel pool of platform-paid royalties for the music industry; needletime in the UK forced the BBC to pay for live sessions until 1988. Or~-- most often~-- the two creative categories \emph{coexist}: photography did not kill painting, it forced painting toward what photography could not do (Impressionism, abstraction); the microphone did not kill singing, it \emph{redefined} what counts as singing; sampling did not kill composition, it redefined what counts as composition.

That is the shape. It is, by my count, the shape of every single technological transition in the creative industries in the period 1839 to 2022. It is happening, around AI, right now~-- and where we are on the curve is part of what this chapter is for.

Let me walk briefly through a handful of the cases, because the \emph{texture} of the historical record is, I think, more useful than the abstract pattern alone.

\section*{The photograph (1839)}

The daguerreotype was unveiled in Paris on 7 January 1839 and publicly described to the Acad\'{e}mie des Sciences on 19 August. By 1849, roughly 100,000 daguerreotypes had been produced in Paris alone; by 1861, about 33,000 people in Paris were making their living from photography and photographic supplies. The professional class most immediately wiped out was the \textbf{portrait miniaturist}~-- painters of small ivory-based likenesses, the working photographers of the pre-photographic age. They could not compete on speed, price or fidelity. Within a single working generation, the craft was effectively gone.

The most articulate resistance was not from the miniaturists themselves but from the literary intelligentsia. \textbf{Charles Baudelaire's} 1859 essay \emph{``The Modern Public and Photography,''} published in the \emph{Revue Fran\c{c}aise}, made the case in language a 2024 AI-sceptic would recognise: \emph{``this industry, by invading the territories of art, has become art's most mortal enemy.''} And, harder: \emph{``The photographic industry was the refuge of all failed painters, too ill-equipped or too lazy to complete their studies.''}\footnote{Charles Baudelaire, ``Le Public Moderne et la Photographie,'' \emph{Revue Fran\c{c}aise}, 1859 (part of the \emph{Salon de 1859} essays). English translation widely available; the original French in PDF form: \url{https://gallowayexploringart.wordpress.com/wp-content/uploads/2014/08/baudelaire_the-modern-public-photography.pdf}. Smithsonian Archives institutional overview: ``Photography Murdered Painting, Right?'', \url{https://siarchives.si.edu/blog/photography-murdered-painting-right}.}

The famous \textbf{Paul Delaroche} line~-- \emph{``From today, painting is dead''}~-- is, the historians who have looked carefully tell us, almost certainly apocryphal. The earliest sourced version of it appears in an 1873 survey, thirty-four years after Delaroche supposedly said it; the painter's own contemporary writing on the daguerreotype called it \emph{``an immense service to the arts,''} and he continued painting until his death in 1856.\footnote{The Delaroche apocrypha is documented in Quote Investigator: \url{https://quoteinvestigator.com/2022/10/16/photo-mortal/}. The earliest sourced version is in an 1873 survey, 34 years after Delaroche reportedly said it. Delaroche's own contemporary writing on the daguerreotype, in Gernsheim's standard 1959 monograph, characterised the new technology as ``an immense service to the arts.''} The story has outlived the saying. (This is itself a pattern. I will come back to it.)

The settlement took seventy years. Alfred Stieglitz founded the Photo-Secession on 17 February 1902. \emph{Camera Work} ran from 1903 to 1917. The ``291'' gallery opened in 1905. MoMA established the first photography department at a major museum on 31 December 1940. From invention to \emph{full institutional acceptance of the new form as art}: about a century. The compensating gain: the entire history of modern photography as a fine-art tradition, an industrial portrait business, a documentary-journalism profession, and~-- eventually~-- the cultural substrate on which the smartphone-camera moment of the 2010s rests.

What painting did, meanwhile, was redefine itself. Impressionism (light, atmosphere, the subjective moment), Post-Impressionism (the interior state), Cubism (multiple viewpoints), and ultimately abstraction~-- every one of these moves makes more sense if you read them as painting's response to the daguerreotype's having taken representation. The standard art-history reading, which I think is correct, is that \emph{photography liberated painting from the burden of representation}. The fear-namers were locally right and structurally wrong. Painting did not die. It became something else.

\section*{The phonograph (1877--1920s)}

I have already quoted Sousa. The whole 1906 \emph{Appleton's} essay is worth reading; almost every paragraph of it could be republished, with names changed, in 2025.

The institutional response to the phonograph~-- and this is the part of the story that working creatives in 2026 need to know~-- came in the form of the \textbf{1909 Copyright Act}, the first statutory acknowledgement in US law that \emph{machine reproduction of human creative work required a legal regime.} The 1909 Act created the \textbf{compulsory mechanical licence} for recorded music, the structural ancestor of every machine-licensing argument we are now having about AI training. Sousa, in part because his lobbying helped pass it, prospered in the recording era; his compositions still generate royalties under that licensing structure today.

The phonograph also created entirely new occupations that the parlour-music economy could not have anticipated. The recording engineer. The A\&R executive. The producer. The mastering engineer. The sleeve designer. The pressing-plant operator. The retail-store buyer. The pop \emph{single} as a commercial form. Jazz on record. The LP. The concept album. The bedroom-studio that, in the 2000s, would take all of those occupations apart again. Sousa got the \emph{parlour} prediction right~-- recorded music did dent amateur home music-making~-- and the \emph{industry} prediction completely wrong. The recording industry was the largest expansion of working-musician employment in the history of music, and it was made possible by the technology Sousa thought would destroy it.

\section*{The Petrillo bans (1942, 1948)}

The \textbf{American Federation of Musicians} had, by the 1940s, watched the phonograph, the talking picture (which alone wiped out roughly 22,000 cinema-orchestra jobs in the US in 1927), and commercial radio progressively displace live performance. \textbf{James Caesar Petrillo}, AFM president from 1940, did the thing every union threatened by a creative technology has tried, with varying success, to do since: he turned off the recording machine.

\textbf{On 1 August 1942}, AFM members stopped recording. The strike lasted \textbf{27 months}. Decca settled in 1943; RCA Victor and Columbia in November 1944. The settlement: a per-record royalty paid into an AFM fund for unemployed musicians.

\textbf{On 1 January 1948} Petrillo did it again, triggered by the Taft-Hartley Act outlawing the first royalty arrangement. The 1948 settlement created the \textbf{Music Performance Trust Fund} under Section~302 of Taft-Hartley~-- a jointly-administered labour-management fund paid into by the labels and broadcasters, used to subsidise free live performances by working musicians. The MPTF still exists. It still pays out, in 2026, several million dollars a year for live music.\footnote{The 1942--44 Petrillo strike: Wikipedia, ``1942--44 musicians' strike,'' \url{https://en.wikipedia.org/wiki/1942\%E2\%80\%931944_musicians\%27_strike}; Mainspring Press, ``The Man Who Crippled the American Recording Industry: James Caesar Petrillo and the American Federation of Musicians Recording Bans,'' \url{https://mainspringpress.org/2024/11/23/the-man-who-crippled-the-recording-industry-james-caesar-petrillo-and-the-american-federation-of-musicians-recording-bans/}; DownBeat, ``The Petrillo Ban of 1942--`44: Past \& Future at War,'' \url{https://downbeat.com/news/detail/the-petrillo-ban-of-194244-past-future-at-war}; Local 802 AFM, ``The Silence Was Deafening,'' \url{https://www.local802afm.org/allegro/articles/the-silence-was-deafening/}. The Music Performance Trust Fund's institutional history: \url{https://musicpf.org/establishment-of-mptf-led-to-the-formation-of-afms-pension-and-residual-funds/}.}

I want to dwell on this for a moment, because the Petrillo settlement is~-- by some distance~-- the \textbf{most operationally important precedent} for how the AI debate could land, and almost nobody in the current creative-AI conversation talks about it.

The Petrillo template has four parts. \emph{One,} the displacing technology is not banned. It is allowed to displace. \emph{Two,} the platform owner pays an ongoing per-unit tribute to the displaced labour pool. \emph{Three,} the tribute is collected centrally, by a joint labour--management body, not negotiated individual-by-individual. \emph{Four,} the tribute is paid out to subsidise the \emph{displaced creative practice itself}~-- live music, in this case~-- keeping it alive as a category even as the market for it shrinks.

The SAG-AFTRA \emph{Tilly Tax} provisions in the 2026 contract, the UK 88\% licensing-by-default proposal, the \emph{Creative Weight Attribution} musical-AI infrastructure I described in Chapter~\ref{ch:5}, the C2PA / SynthID provenance stack from Chapter~\ref{ch:12}~-- these are all, on inspection, \emph{attempts to reconstruct the Petrillo template for AI}. Per-unit tribute. Joint collection. Redistribution to the displaced practice. The mechanism is the same. The political question is whether the platforms will accept it.

The mechanism worked once. It can work again. The pattern of the resistance that fails~-- the 1982 UK Musicians' Union motion to \emph{ban synthesisers outright}, the European \emph{Right to be Forgotten} style absolutism on training data~-- is the pattern of resistance that tries to legislate against the machine rather than to tax it. Levy beats ban, every time. Royalty pool beats injunction. \emph{Mechanism, not prohibition.}

\section*{The microphone (1932)}

Cardinal O'Connell's January 1932 sermon to 3,000 men of the Holy Name Society of Boston is the single funniest moment in the resistance literature. \emph{Crooning}~-- the conversational, intimate, microphone-enabled vocal style that \textbf{Bing Crosby}, Rudy Vall\'{e}e and others had built into a mass commercial form by the late 1920s~-- was, for O'Connell, ``a degenerate form of singing. No true American man would practice this base art\ldots{} If you will listen closely [to crooners' songs] you will discern the basest appeal to sex emotion in the young.'' Crooners were ``whiners and bleaters defiling the air.'' Their work was ``imbecile slush.''\footnote{William Henry Cardinal O'Connell, \emph{op.~cit.}}

The cultural fight was effectively over by the end of the decade. Bing Crosby was the biggest male voice in America. By the late 1930s no popular vocalist \emph{not} trained in microphone technique could make a competitive career.

What is interesting about the microphone case, for our purposes, is \emph{what it did to the underlying definition of the craft}. Before the microphone, \emph{good singing} meant volume, projection, throat technique~-- the operatic, theatrical, music-hall tradition that O'Connell had grown up inside. After the microphone, \emph{good singing} meant timbre, intimacy, breath control at low dynamics, conversational diction~-- a fundamentally different skill set. Almost every popular vocalist since 1935 has been a ``crooner'' in the technical sense, including those who would not call themselves that. \emph{The microphone redefined what counted as singing.} The vocalists who refused the microphone are mostly remembered as period figures. The ones who absorbed it defined the rest of the century of popular music.

This is the deeper pattern I will come back to. Resistance, in the named-fear form, almost always defends \emph{the existing definition} of the craft. The settlement, almost always, \emph{redefines} the craft. The fear is real and the language is sincere; what's actually happening is bigger than what the fear is naming.

\section*{The synthesiser (1980--82)}

The UK part of this story is the cleanest because it generated public archives. The \textbf{Yamaha DX7}, released in 1983, put credible electric piano, brass, strings, marimba and dozens of other sounds into a single keyboard at a price point~-- about \pounds{}1,500~-- accessible to working session players. The DX7 was used on roughly \textbf{40\% of US Billboard Hot 100 \#1 singles in 1986}. Session keyboard players and orchestral string sections previously hired for adverts, TV scoring, library music and pop sessions were directly displaced.

The \textbf{Musicians' Union} of the UK responded in two stages.

\textbf{1980, the Massacre of the Musicians.} The BBC announced in March 1980 that it would cut 172 staff orchestra posts and disband five of its eleven in-house orchestras. 83\% of MU BBC members voted to strike. The strike began 16 May 1980. \textbf{The First Night of the Proms was cancelled for the first time in its history.} The strike ran until 1 August 1980, ending with a compromise: the BBC Northern Ireland Orchestra and BBC Midland Radio Orchestra were disbanded as planned; the others survived.\footnote{Musicians' Union History, ``The Strike That Made History~-- Massacre of the Musicians 1980,'' \url{https://www.muhistory.com/the-massacre-of-the-musicians-1980/}. Academic context on the broader MU--BBC dispute landscape: ``Negotiating Needletime'' (Tandfonline), \url{https://www.tandfonline.com/doi/full/10.1080/03071022.2016.1215098}.}

\textbf{1982, the synthesiser ban motion.} On 23 May 1982~-- by coincidence Bob Moog's birthday~-- the MU's Central London Branch passed a motion to ban synthesiser, drum-machine and electronic-device use by union members entirely.\footnote{MusicRadar, ``The Day the Loony Musicians Union Tried to Kill the Synthesizer (Which Also Happened to be Bob Moog's Birthday),'' \url{https://www.musicradar.com/news/the-union-passed-a-motion-to-ban-the-use-of-synths-drum-machines-and-any-electronic-devices-the-day-the-loony-musicians-union-tried-to-kill-the-synthesizer-which-also-happened-to-be-bob-moogs-birthday}. Far Out Magazine, ``Why did the Musicians Union outlaw synthesisers in 1982?'', \url{https://faroutmagazine.co.uk/musicians-union-outlaw-synthesisers/}.} The trigger was that Barry Manilow's UK tour had replaced its string section with synth players. The motion was never adopted as full union policy. The MU's Executive Committee passed a more measured resolution in November 1982. A breakaway ``Union of Sound Synthesists'' was formed. \emph{Top of the Pops}, for a period, required bands to record their backing tracks the afternoon before the show ``to prove they could actually play it.'' None of it held. The DX7 sold over 200,000 units. By 1990 the cultural debate was over and the synthesiser was, simply, another instrument.

What the \emph{MU} got wrong, in retrospect, was the same thing every resistance-by-prohibition gets wrong: they tried to ban the tool rather than to tax it. There was no Petrillo-style fund. There was no per-output levy on synthesiser use. There was no royalty pool subsidising live orchestral work. The MU's defensive posture preserved nothing structurally and lost the cultural argument decisively. The session-musician economy contracted. Some of the displaced players retrained as programmers and prospered. Others did not. \textbf{Trevor Horn}~-- the Buggle who, three years before the dispute, had sung \emph{Video Killed the Radio Star}~-- became the defining pop producer of the 1980s.

The lesson, for the working creative in 2026, is not that the MU was wrong to resist. The MU was \emph{right} to read the displacement signal three years ahead of the rest of the industry. The MU was wrong about \emph{what kind} of resistance to mount. Prohibition was always going to lose. A levy-and-pool argument~-- what an \emph{AI-era Petrillo} could look like~-- might have held.

\section*{Drum machines, sampling and the lawsuits (1980s--2000s)}

The Roland TR-808 (1980) and TR-909 (1983) both \emph{failed} commercially on release. The 808 was criticised as ``toy robot drums''; the 909 was reviewed as ``still sounds like a drum machine, instead of a machine playing drums.'' Both became cult instruments only after being dumped at secondhand prices to young hip-hop and dance producers in the mid-1980s. The 808's hand-clap, snare and signature deep kick are now the defining percussion sounds of contemporary popular music. The same dismissed-then-canonised arc is, in 2026, beginning to play out for Suno and Udio at the consumer-music end of the spectrum.

The legal resistance to sampling produced two rulings every working creative should know about, because they are the cleanest available templates for how the courts may treat AI-training disputes.

\textbf{\emph{Grand Upright Music v.\ Warner Bros.}} (S.D.N.Y.~1991). Biz Markie sampled three bars of Gilbert O'Sullivan's \emph{Alone Again (Naturally)} without clearing it. Judge Kevin Thomas Duffy's opinion opened with the words \emph{``Thou shalt not steal,''} quoting the Seventh Commandment, in a US federal court ruling. He referred the case to the US Attorney for potential \emph{criminal} investigation. The album was pulled. The sample-dense Bomb Squad / Public Enemy style of production it had been built on became commercially impossible.\footnote{\emph{Grand Upright Music, Ltd.\ v.\ Warner Bros.\ Records Inc.}, 780 F.~Supp.~182 (S.D.N.Y.~1991), \emph{op.~cit.}}

\textbf{\emph{Bridgeport Music v.\ Dimension Films}} (6th Cir.~2005). NWA's \emph{100 Miles and Runnin'} sampled a two-second guitar chord from Funkadelic's \emph{Get Off Your Ass and Jam}, looped and pitched down. The Sixth Circuit eliminated the \emph{de minimis} defence for sampling sound recordings and issued the rule that has, for twenty years, defined how clearance works: \textbf{``Get a license or do not sample.''}\footnote{\emph{Bridgeport Music, Inc.\ v.\ Dimension Films}, 410 F.3d~792 (6th Cir.~2005). Full text: \url{https://law.justia.com/cases/federal/appellate-courts/F3/410/792/574458/}. The ``Get a licence or do not sample'' rule is the most-cited line in the opinion.}

Hip-hop, of course, did not die. It became the dominant global popular music form. What changed was that the \emph{aesthetic} of dense, layered, sample-heavy production gave way to a more clearance-friendly style. The Bomb Squad lineage continued, but on different terms.

The AI-training analogy here is direct, and I would commend the rulings to anyone trying to think clearly about \emph{UMG v.\ Anthropic} and the cases that will follow it. \emph{Thou shalt not steal} and \emph{Get a license or do not sample} are not, despite their archaic phrasing, particularly anti-technology rulings. They are \emph{pro-licensing} rulings. They say: the tool can be used, but the inputs have to be paid for. That is what working creatives are asking for in the UK 88\% and the \emph{Stealing Our Work Is Not Innovation} declaration. The line of legal reasoning is already in the books.

\section*{Auto-Tune (1998--2010)}

I want to spend less time on Auto-Tune than the dossier supports, because the case is so clean it almost makes itself. Andy Hildebrand, a former Exxon seismic-data engineer, released Auto-Tune in 1997. \textbf{Cher's \emph{Believe}} (1998), produced by Mark Taylor and Brian Rawling with the retune speed maxed out, produced the now-iconic ``Cher effect''~-- the audible warble. Taylor and Rawling claimed it was a vocoder for several years to protect the trick.

\textbf{T-Pain} made the effect his signature from 2005 onwards and was rewarded with \textbf{Jay-Z's} \emph{``D.O.A.\ (Death of Auto-Tune)''}~-- released June 2009 on \emph{The Blueprint 3}~-- a direct moral-panic attack on what Auto-Tune was doing to vocal authenticity. \textbf{\emph{TIME} magazine's \emph{50 Worst Inventions} list} in 2010 ranked Auto-Tune at \#15: software that ``can make bad singers sound good, and really bad singers sound like robots.''\footnote{\emph{TIME}, ``50 Worst Inventions,'' 2010, Auto-Tune at \#15: \url{https://content.time.com/time/specials/packages/article/0,28804,1991915_1991909_1991903,00.html}. Wikipedia, ``Auto-Tune,'' \url{https://en.wikipedia.org/wiki/Auto-Tune}. NPR, ``25 Years of Believe,'' \url{https://www.npr.org/2023/10/19/1207028349/25-years-ago-cher-released-a-song-that-would-change-the-sound-of-pop-music}. Wikipedia, ``D.O.A.\ (Death of Auto-Tune),'' \url{https://en.wikipedia.org/wiki/D.O.A._(Death_of_Auto-Tune)}.}

Sixteen years later, Auto-Tune is on every pop vocal you hear, used as both correction and effect. Bon Iver's \emph{22, A Million} (2016) deployed it as a self-conscious aesthetic instrument. Billie Eilish has used it across her career, transparently. The cultural rehabilitation is complete. The moral-panic stage, in retrospect, looks parochial.

But notice what \emph{did} happen. The microphone-era settlement~-- \emph{you have to be able to sing}~-- was, in the Auto-Tune era, definitively broken. The new settlement is \emph{you have to be able to perform the post-corrected vocal as a self-conscious artistic choice}. The redefinition is real. Cardinal O'Connell, transported eighty years forward, would have hated Auto-Tune for exactly the same reasons he hated crooning, and would have been just as wrong about it.

\section*{Non-linear editing and the death of the splice}

Avid Media Composer launched in 1989. Through the early 1990s it displaced the Moviolas, Steenbecks and KEM flatbeds that working film editors had used for sixty years. The American Cinema Editors did not strike. The Motion Picture Editors Guild did not stop the transition. The settlement was generational: editors trained on film retired; editors trained on Avid became standard.

The witness I want to quote here is \textbf{Walter Murch}, ACE~-- possibly the most respected film editor of the last fifty years. Murch edited \emph{The Conversation} and \emph{Apocalypse Now} on physical film. He edited \emph{The English Patient} on Avid, winning Oscars for both picture and sound. He then edited Anthony Minghella's \emph{Cold Mountain} (2003) on \textbf{Apple Final Cut Pro running on commodity Power Mac G4 hardware}~-- for a \$79m feature. The story is in Charles Koppelman's book \emph{Behind the Seen} (Peachpit, 2004).\footnote{Walter Murch, \emph{In the Blink of an Eye: A Perspective on Film Editing}, Silman-James Press, 1995 (2nd edition 2001). PDF: \url{https://www.craftfilmschool.com/userfiles/files/Walter\%20Murch\%20-\%20In\%20the\%20Blink\%20of\%20an\%20Eye\%20Revised\%202nd\%20Edition\%20(2001,\%20Silman-James\%20Pr).pdf}. Charles Koppelman, \emph{Behind the Seen: How Walter Murch Edited Cold Mountain Using Apple's Final Cut Pro and What This Means for Cinema}, Peachpit Press, 2004: \url{https://www.peachpit.com/store/behind-the-seen-how-walter-murch-edited-cold-mountain-9780735714267}.} The decision saved about \$1m versus an equivalent Avid rental, and Murch chose it on practical grounds. The editor who literally wrote the textbook on film editing~-- \emph{In the Blink of an Eye}, the canonical philosophical text on the cut~-- was, by 2003, working on the tool the editing-room conservatives were warning the industry against.

The compensating gain, from the non-linear editing transition, was that the toolkit shipped on a laptop. Indie cinema benefited enormously. The grammar of the cut accelerated~-- the average shot length in Hollywood drama dropped from roughly ten seconds in the 1960s to roughly four seconds by the 2000s, a change made trivial by NLE that would have been physically punishing to execute on a Moviola. The contemporary visual grammar that ranges from \emph{The Bourne Identity}'s hyper-cuts to Wong Kar-wai's asynchronous editing aesthetic to TikTok's stitched, layered, fast-moving native form is, in operational terms, \emph{what non-linear editing made possible.} The form the new tool enabled was bigger than the form it replaced.

\section*{Kodak (1975--2012)}

Steven Sasson, an engineer at Kodak, built \textbf{the first digital camera prototype in December 1975}. 0.01 megapixel, black and white, the size of a toaster, with a 23-second save time to magnetic tape. Sasson's own description, in multiple interviews, is that the executive response to the demo was \emph{curious but concerned about the implications for film}. Kodak suppressed the project to protect its film business.\footnote{Sasson's account documented at the National Inventors Hall of Fame: \url{https://www.invent.org/blog/inventors/Legacy-Steve-Sasson}. Snopes verification of the ``Kodak suppressed the digital camera'' claim: \url{https://www.snopes.com/fact-check/kodak-digital-camera-invention/}. Knowledge@Wharton on the Kodak collapse: \url{https://knowledge.wharton.upenn.edu/podcast/knowledge-at-wharton-podcast/whats-wrong-with-this-picture-kodaks-30-year-slide-into-bankruptcy/}. Bankruptcy filing: 19 January 2012, S.D.N.Y., \$5.1bn assets / \$6.8bn liabilities.}

Kodak's peak headcount was about 145,000 in the early 1980s. At the company's Chapter~11 bankruptcy filing on \textbf{19 January 2012}, the workforce was about \textbf{19,000.} A nearly \$7bn liability stack. A company that had invented the technology that would destroy it, and then suppressed that technology, and then been destroyed by it anyway when the rest of the market~-- Nikon, Canon, Sony~-- built around the patents Kodak had not commercialised.

The Kodak story is the standard cautionary tale for incumbents about to be disrupted by the technology they themselves built. It is being told, with increasing force, about the legacy entertainment industries in 2026. \emph{The companies that already have the IP, the audience and the distribution are, by structural inheritance, in the position Kodak was in in 1990}. What we will find out, in the rest of this decade, is whether they have learned the Kodak lesson~-- that the new technology is going to displace the old whether or not you commercialise it, so you may as well be the company that does~-- or whether the next few years will produce the entertainment-industry equivalent of the 2012 bankruptcy filing.

\section*{Photoshop, social media, smartphones~-- a faster montage}

Three more cases, briefly, because the pattern is by this point clear and the texture matters more than the recital.

\textbf{Photoshop and photojournalism (1990--2015).} The famous \textbf{Brian Walski} firing, 1 April 2003~-- LA Times staff photographer dismissed by satellite phone from Iraq for compositing two combat photographs into one stronger image~-- is the single canonical example.\footnote{Wikipedia, ``Brian Walski,'' \url{https://en.wikipedia.org/wiki/Brian_Walski}. \emph{Washington Post} contemporaneous coverage: \url{https://www.washingtonpost.com/archive/lifestyle/2003/04/03/altered-picture-costs-la-times-photographer-his-job/c5e7c9e0-a836-429a-bb4e-d502f1768a96/}. World Press Photo's institutional response in TIME: \url{https://time.com/3706626/world-press-photo-processing-manipulation-disqualified/}.} World Press Photo's response, over the following decade, was to issue increasingly tight ethics rules and disqualify increasing numbers of finalists. The truth-claim of the photograph survived institutionally at AP, Reuters, \emph{National Geographic} and \emph{Magnum}. It did not survive on social media. The structural answer was C2PA~-- the content-authenticity infrastructure that now sits underneath the AI provenance stack in Chapter~\ref{ch:12}. The 1990s photojournalism debate was, on inspection, \emph{the first round of the C2PA conversation} the AI era is now finishing.

\textbf{Social media (2005--2014).} The defining institutional fight was \textbf{\emph{Viacom v.\ YouTube}}: \$1bn in claimed damages, filed in March 2007, decided in favour of YouTube and the DMCA Section~512 safe harbour in 2010 and again in 2013, settled out of court in \textbf{March 2014 with no money changing hands}~-- a complete Google win.\footnote{Wikipedia, ``Viacom International, Inc.\ v.\ YouTube, Inc.,'' \url{https://en.wikipedia.org/wiki/Viacom_International_Inc._v._YouTube,_Inc.}. Electronic Frontier Foundation case file: \url{https://www.eff.org/cases/viacom-v-youtube}. \emph{Variety} on the March 2014 settlement: \url{https://variety.com/2014/biz/news/google-and-viacom-settle-copyright-infringement-lawsuit-over-youtube-1201137538/}.} The settlement was effectively \emph{automated}: YouTube's Content~ID, deployed from 2007 onwards, has paid out billions in royalties to rights-holders through algorithmic matching. The DMCA safe harbour that decided \emph{Viacom}~-- passed in 1998, in response to the previous resistance pattern~-- became the structural framework of the \emph{creator economy}. Power-law distribution, \$100bn+ in aggregate value by 2022, MrBeast at 450m subscribers and \$600--700m in 2024 revenue. The platform shape of the audience contract in 2026, on which every AI-distribution question rests, was built by Content~ID.

\textbf{Smartphones (2007 onwards).} The most dramatic collapse of an established consumer-creative-tool category in modern history. Compact-camera shipments: \textbf{108.8 million units in 2010, 3.01 million in 2021.} A 97\% drop in roughly a decade.\footnote{PetaPixel, ``The Rise and Crash of the Camera Industry in One Chart,'' \url{https://petapixel.com/2024/08/22/the-rise-and-crash-of-the-camera-industry-in-one-chart/}. Statista, ``Smartphones Wipe Out Decades of Camera Industry Growth,'' \url{https://www.statista.com/chart/15524/worldwide-camera-shipments/}. CIPA shipment data series, multiple years.} The iPhone, the Galaxy and the Pixel did to the camera industry what Sousa thought the phonograph would do to amateur music~-- and Sousa, in retrospect, was about a hundred years early on his own argument. The smartphone-camera moment is also, on inspection, the dress rehearsal for the AI-image moment. The same arguments~-- \emph{it isn't real photography}, \emph{the computational layer is doing the work, not the human}, \emph{the dedicated camera is the moral artefact}~-- were made in 2010 and 2015 about smartphones. They were also lost in 2010 and 2015 about smartphones. Computational photography is, today, where most of the meaningful innovation in the image-making craft is happening. \emph{AI photography}, in 2026, is not the new thing. It is the next iteration of the thing that the smartphone-camera made inevitable fifteen years ago.

\section*{Five patterns}

I have, by this point, walked through ten of the twenty cases the dossier behind this chapter covers. The pattern recurs. Let me name the five structural features I think the working creative in 2026 most needs to internalise.

\textbf{Pattern one: the curve has a predictable shape.} Ridicule~\textrightarrow{} moral panic~\textrightarrow{} existential alarm~\textrightarrow{} institutional counter-attack~\textrightarrow{} settlement. AI in 2025--26 is, on my read, somewhere in late act three and early act four. The existential alarm has been voiced. The institutional counter-attack~-- the SAG-AFTRA strike, the WGA contract, the UK 88\%, the \emph{UMG v.\ Anthropic} suit, the Cannes Disclosure Standard, the Sundance literacy initiative~-- is well underway. The settlement is starting to form but is not yet stable. The next eighteen months will, on the historical pattern, \emph{settle the form of the next decade's industry}. This is exactly why this period matters.

\textbf{Pattern two: the named fear is always mis-named.} What practitioners say they fear~-- loss of \emph{soul}, loss of \emph{authenticity}, loss of \emph{the real}~-- is almost never what actually happens. What actually happens is a \emph{redefinition} of the underlying creative category. The microphone redefined \emph{singing}. The photograph redefined \emph{painting}. The sampler redefined \emph{composition}. The Avid redefined \emph{editing}. The smartphone redefined \emph{photography}. The fear is framed as a defence of the thing. What is actually at stake is the \emph{definition} of the thing. The fear-namers usually win the surface argument and lose the definitional one. \emph{The AI debate is, on the structural reading, an argument over what counts as authorship, performance, writing, photography and composition in the next decade.} That is the real fight. The named-fear version of it~-- \emph{AI will steal our jobs}~-- is true but partial.

\textbf{Pattern three: the institutional moves that work are levy-and-pool; the moves that fail are prohibition.} The Petrillo settlement (MPTF, 1948), the DMCA safe harbour plus Content~ID (1998 plus 2007), the 1909 Copyright Act mechanical licence, the eventual streaming-loudness-normalisation truce~-- these \emph{worked}, in the limited but real sense that they extracted ongoing transfer payments from the new medium to the displaced labour pool, or that they restructured the aesthetic equilibrium. The institutional moves that \emph{failed} are the prohibitions: the MU 1982 synthesiser ban, the 1991 \emph{Grand Upright} effective ban on dense sampling, the European \emph{Right to be Forgotten} style absolutism. \textbf{Levy beats ban. Royalty pool beats injunction. Mechanism beats prohibition.} For the AI-training fight: a per-output levy distributed to a creators' fund is in the workable category. A ban on AI training is in the unworkable category. The 88\%~-- by my reading~-- is closer to the workable category than the politics around it have so far recognised.

\textbf{Pattern four: the resisters are usually right about the local loss and wrong about the total loss.} Miniature painting died. Hand-drawn feature animation died (\emph{Toy Story} in 1995; Disney's Florida 2D studio closed on 12 January 2004). The professional recording-studio mid-tier died. Photo-processing labs died. Staff photographer jobs at US newspapers~-- collapsed (the \emph{Chicago Sun-Times} infamously laid off its entire photo staff in May 2013). But \emph{more people make moving images today than ever made them; more people make music today than ever made music; more photographs are taken on a given Sunday than were taken in the entire nineteenth century.} The aggregate creative-labour pool grew. The fear is always articulated by the \emph{displaced} cohort. The compensating gains accrue to a different cohort, who don't yet exist when the fear is articulated, and whose names~-- by definition~-- are not yet in the trade press. This is why the resisters can be both \emph{factually right} about their own situation and \emph{structurally wrong} about the form they are trying to protect.

\textbf{Pattern five: the cultural symbol outlives the cultural anxiety.} \emph{Video Killed the Radio Star} was the Buggles' 1979 lament for the death of a form. MTV used it as its launch trumpet at 12:01~a.m.\ on 1 August 1981. Then radio survived; then MTV died; the song is on TikTok. The artefact \emph{about} the death of a form outlasted both the form it threatened and the form that did the threatening. Phil Tippett's \emph{``I think I'm extinct''} in 1992, \textbf{Trevor Horn's} \emph{``video killed the radio star''} in 1979, \textbf{Cardinal O'Connell's} \emph{``imbecile slush''} in 1932~-- these crystallise an anxiety into a piece of language so vivid that it survives whatever it was anxious about. The named fear becomes the source material for the next generation's art. \emph{The anxiety is the art}. The \emph{Tilly Norwood week} of late September 2025, with Whoopi Goldberg and Melissa Barrera and Emily Blunt speaking on the record, is~-- I am almost certain~-- the equivalent crystallising-moment for the AI era. The artefacts that will come out of it (the documentaries, the dramatic-feature treatments, the songs, the union-history books) will, twenty years from now, be the cultural objects that \emph{outlast} the technology they were originally anxious about.

\section*{Where AI sits in this picture}

I want to be careful, in applying the diagnostic, not to wave a hand and claim the pattern \emph{predicts} the AI outcome. It doesn't. What it does is rule out certain shapes the outcome cannot take, and rule in certain shapes it almost certainly will.

The shapes ruled out: an \emph{outright ban} on AI training that protects the existing definition of authorship. A \emph{prohibition} by union action that simply removes AI tooling from professional production. A \emph{moral consensus}~-- Cardinal O'Connell at scale~-- that AI work is degenerate and should be socially refused. None of these has worked in twenty previous iterations of the same pattern. None of them is going to work this time. The MU's 1982 synth motion is in the books as a cautionary tale.

The shapes ruled in: a per-output levy structure flowing into a creators' fund~-- Petrillo for neural weights. A C2PA-style provenance standard underwriting an authenticity premium~-- the structural ancestor of which is the 1990s photojournalism ethics fight. A redefinition of the underlying creative category~-- \emph{authorship in 2030} will mean something materially different from \emph{authorship in 2020}, the way \emph{singing} meant something different after the microphone. An institutional settlement that absorbs the new tool and \emph{redistributes} the productivity gain, rather than one that \emph{bans} the new tool and \emph{forfeits} the productivity gain.

The cohorts who will be locally displaced are already visible in Chapter~\ref{ch:11} and Chapter~\ref{ch:14}. Junior animators. Concept artists in cohorts being asked to use generative tooling for the front of their pipeline. Voice actors below the SAG-AFTRA scale level. Stock-image photographers. Translators of bulk commercial copy (the closest pre-2022 analogue~-- the Google Neural Machine Translation moment in 2016~-- already produced documented translator-employment effects across 696 US labour markets, and a 70\% income loss for the Irish-language EU-institutions translator profiled in the January 2026 \emph{CNN Business} piece cited earlier in the book\footnote{\emph{CNN Business}, ``Meet the translation professionals losing their jobs to AI,'' January 2026, \url{https://www.cnn.com/2026/01/23/tech/translation-language-jobs-ai-automation-intl}. Carl Benedikt Frey (Oxford Martin School), 2025 study on translator employment across 696 US labour markets. American Translators Association industry position: \url{https://www.atanet.org/client-assistance/blog-machine-translation-vs-human-translation/}. Wikipedia, ``Google Neural Machine Translation,'' \url{https://en.wikipedia.org/wiki/Google_Neural_Machine_Translation}.}). The fear of these cohorts is well-founded, and the institutional response should be calibrated to it.

The compensating gains~-- the new categories of creative work, the new business shapes, the new audience contracts~-- are what Chapter~\ref{ch:10} of this book is about. I will not re-state them here. What I want to note, \emph{for the historical pattern's sake}, is that they always emerge, they always emerge faster than the resisters expect, and they always~-- at the aggregate level~-- produce more total creative employment than the displaced form supported. The phonograph was the largest creative-labour expansion in music history. The smartphone was the largest creative-labour expansion in image-making history. \emph{I think AI, on the historical pattern, will be the largest creative-labour expansion in cultural-production history.} I think the working creatives who emerge from this with the most leverage will be the ones who, like Walter Murch picking up Final Cut Pro at the height of his Avid mastery, learned the new tool \emph{before} the cultural permission to use it had fully crystallised. The cultural permission usually arrives about three years after the productive use does. The window for asymmetric leverage is now.

\section*{What this asks of working creatives}

I want to close this chapter with a working-practitioner read on the history, because the abstract pattern is useless without translation.

\textbf{One. Recognise where on the curve you are.} AI in 2026 is in late act three, early act four. Existential alarm is the dominant mood. Institutional counter-attack is well organised. \emph{Settlement} has not been reached. The decisions you make about how to engage with the tools in the next eighteen months are decisions that will define the structural shape of the rest of the decade. This is not a \emph{theoretical} claim about the historical pattern; it is the \emph{operational} claim about your career.

\textbf{Two. Don't fight to keep the existing definition; fight to be among the people redefining it.} The miniaturists are remembered as the cohort that was wiped out. The Stieglitzes are remembered as the cohort that \emph{redefined what photography could be}. Both groups felt, in 1855, that they were fighting for the same thing. They were not. One was defending the inherited definition. The other was rewriting it. The working creatives who emerge from the AI period with the most leverage will not be the ones who defended hardest. They will be the ones who \emph{redefined fastest}.

\textbf{Three. Pick the institutional response that has historically worked.} The Petrillo template~-- levy on the displacing technology, pool collected centrally, redistribution to the displaced craft~-- has a hundred-year track record. Use it. Apply it to your union negotiations. Apply it to your platform-procurement decisions. Apply it to your political advocacy. The SAG-AFTRA Tilly Tax, the UK 88\%, the C2PA provenance stack, the Cannes Disclosure Standard, the Sundance literacy initiative are all, in their different ways, \emph{the Petrillo template applied to AI}. They are the part of the institutional response that historically wins. Show up to them. Argue for them. Don't waste energy on prohibitions.

\textbf{Four. Read the named fear carefully, and listen for the redefinition underneath it.} When you hear yourself saying \emph{AI will steal my craft}, ask the second question: \emph{what new definition of my craft is forming on the other side of the displacement, and am I in a position to inhabit it?} The microphone vocalists who absorbed Crosby outlasted the operatic vocalists who refused him. The editors who absorbed Avid outlasted the editors who refused it. The photographers who absorbed digital outlasted the photographers who refused it. The pattern is, by this point in the historical record, very reliable.

\textbf{Five. Make the cultural symbol.} The named-fear language of the AI era~-- its \emph{Video Killed the Radio Star}, its \emph{I think I'm extinct}, its \emph{imbecile slush}~-- is being written, right now, by working creatives in their own work. Some of those artefacts will outlast the technology they were anxious about. Some will become the source material for the next generation's understanding of this moment. \emph{If you are a working creative reading this in 2026 and the AI displacement frightens you, the most useful thing you can do is put your fear into a piece of work whose argument outlasts the platform release cycle.} That is what Trevor Horn did. It is what Phil Tippett did. It is what the displaced miniature-painters who became fine-art photographers did. It is, on the historical pattern, what is asked of you.

\section*{A coda on Trevor Horn}

I want to close with one image, because it is the single most useful one I know.

Trevor Horn, in 1979, was the keyboard player and frontman of the Buggles. He wrote, with Geoff Downes and Bruce Woolley, a song called \emph{Video Killed the Radio Star}. The song is a small, half-melancholy commercial joke about the end of an era~-- a synthetic-sounding pop song \emph{about} the moment recorded music turned visual. MTV launched on 1 August 1981 with that song as its very first broadcast.

Horn went on to become, by some distance, the most influential pop producer of the 1980s. He built Frankie Goes to Hollywood, ABC, Yes's \emph{90125}, Grace Jones's \emph{Slave to the Rhythm}, Seal's debut, the Art of Noise. He did it on Fairlight CMIs, sequencers, drum machines and the kind of dense, layered, programmed production that the Musicians' Union was, at the same moment, voting to ban. He did the thing the union was warning him against, and built half the canonical pop music of his decade out of it.

The song that named the death of the radio star outlived MTV. The producer who wrote that song defined the next era of recorded music by absorbing the technology the resisters were trying to ban. The cultural symbol outlasted both the form it threatened and the form that did the threatening.

That is the working operating model I would commend to anyone reading this book and feeling, in 2026, the gravitational pull of the resistance. The resistance is real. The fear is real. The named-fear language is the source material of the era's best art. \emph{And the people who absorb the tool, who learn it, who push it past where its makers intended it, and who use it to make the work that argues for the world they actually want~-- they are the ones, on a hundred years of evidence, who define the next decade of the form.}

Welcome to the \emph{next} moment in a recurring pattern. The pattern is, by now, very well documented. The choice is yours.

  \chapter{The Human-AI Agency Continuum}\label{ch:3}

\lettrine[lines=3,lhang=0.15,findent=0.1em]{A}{} week after I sent the first edition of \emph{Dream Machine}, on the Monday of the second week of October 2025, OpenAI held its annual DevDay conference and quietly changed what the conversation about AI in creative work was about.

The first edition had been about Sora~2. The second edition was about something I was less prepared for: the launch of a thing called \textbf{AgentKit}.\footnote{\emph{Dream Machine} Issue~2, ``Editor's Pick,'' 10 October 2025. \url{https://www.linkedin.com/pulse/dream-machine-creative-ai-news-insight-oct-25-2-pete-woodbridge-mnrjc/}.}

AgentKit was, at first glance, a set of developer tools. Agent Builder. A connector registry. An eval framework. ChatKit, for embedding agents into other products. The launch post on OpenAI's blog framed it, in the slightly forced register that all platform-launch posts use, as a way for developers to ``build, deploy, and optimize agentic workflows.''\footnote{OpenAI, ``Introducing AgentKit,'' 6 October 2025. \url{https://openai.com/index/introducing-agentkit/}.} On its own, this was an unremarkable announcement.

What was remarkable, looking back, was the \emph{category claim} the announcement carried with it. Sam Altman, in his DevDay keynote that day, declared the start of ``the age of agentic AI''~-- by which he meant the moment that AI systems stopped being prompt-and-respond chat boxes and started being things that could plan, decide and execute ``for hours on end'' without further human input.\footnote{TechCrunch, ``OpenAI launches AgentKit to help developers build and ship AI agents,'' 6 October 2025. \url{https://techcrunch.com/2025/10/06/openai-launches-agentkit-to-help-developers-build-and-ship-ai-agents/}. Also coverage at \emph{InfoQ}, ``OpenAI Dev Day 2025 Introduces GPT-5 Pro API, Agent Kit, and More.'' \url{https://www.infoq.com/news/2025/10/openai-dev-day/}.}

For someone like me, sitting in a small studio in the North West of England~-- running tools all day, looking at my pipeline, thinking about my team's labour~-- that phrase did a particular kind of work. It rearranged the question.

The question, until that week, had been: \emph{what does AI do for creative work?} The question after that week became: \emph{where, in any given piece of creative work, does my agency end and the model's begin?}

The second question is the one I want this chapter to be about. I called it, in the second issue of the newsletter, the \textbf{Human--AI Agency Continuum}.\footnote{\emph{Dream Machine} Issue~2: ``Agentic AI~-- the class of AI systems that can plan, act, and pursue goals with autonomy~-- promises a new era of collaboration in creative industries\ldots{} It's another step along the Human-AI Agency Continuum.'' See also \emph{TVB Europe}, ``Is Agentic AI About to Change the Media and Entertainment Industry?'' \url{https://www.tvbeurope.com/artificial-intelligence/opinion-is-agentic-ai-about-to-change-the-media-and-entertainment-industry}.} The frame has stuck with me. I think it is the most useful thing I have ever written down about all of this, and I think~-- at the risk of overselling it~-- the rest of the book leans on it.

\section*{The continuum}

\begin{figure}[htbp]
  \centering
  \begin{tikzpicture}[font=\small, every node/.style={align=center}]
    \fill[darkgreen] (0,0) rectangle (2.4,0.7);
    \fill[darkgreen!75!darkblue] (2.4,0) rectangle (4.8,0.7);
    \fill[darkgreen!50!darkblue] (4.8,0) rectangle (7.2,0.7);
    \fill[darkgreen!25!darkblue] (7.2,0) rectangle (9.6,0.7);
    \fill[darkblue] (9.6,0) rectangle (12,0.7);

    \draw[chaptergrey, thick] (0,0) rectangle (12, 0.7);

    \node[text width=2.2cm, font=\scriptsize] at (1.2, -0.55)
      {\textbf{Human-only}\\[1pt]{\color{chaptergrey}traditional craft}};
    \node[text width=2.2cm, font=\scriptsize] at (3.6, -0.55)
      {\textbf{AI-assisted}\\[1pt]{\color{chaptergrey}spell-check, auto-tune}};
    \node[text width=2.2cm, font=\scriptsize] at (6.0, -0.55)
      {\textbf{AI-augmented}\\[1pt]{\color{chaptergrey}Copilot, Firefly}};
    \node[text width=2.2cm, font=\scriptsize] at (8.4, -0.55)
      {\textbf{AI-directed}\\[1pt]{\color{chaptergrey}orchestrates agents}};
    \node[text width=2.2cm, font=\scriptsize] at (10.8, -0.55)
      {\textbf{AI-autonomous}\\[1pt]{\color{chaptergrey}approves/rejects}};

    \node[darkgreen, font=\small\bfseries, anchor=west] at (0, 1.15)
      {Full Human Control};
    \node[darkblue, font=\small\bfseries, anchor=east] at (12, 1.15)
      {Full AI Autonomy};

    \draw[-stealth, chaptergrey, thick] (0, -1.4) -- (12, -1.4);
    \node[chaptergrey, font=\scriptsize, anchor=west] at (0, -1.65)
      {Increasing AI agency};
  \end{tikzpicture}
  \caption{The Human--AI Agency Continuum: five positions on the spectrum of creative control.}
  \label{fig:agency-continuum}
\end{figure}
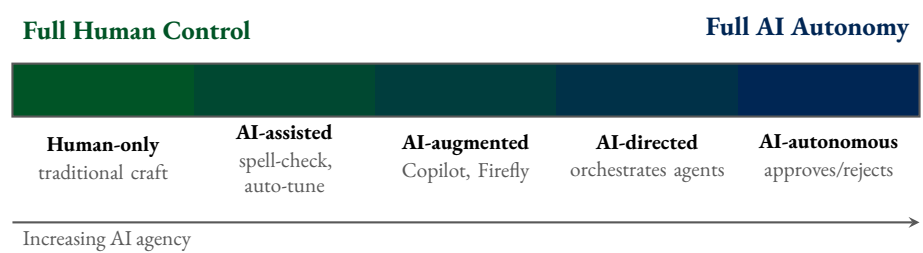

Imagine a horizontal line.

On the far left of the line is \textbf{pure human agency}: the writer at the desk, the painter at the canvas, the songwriter at the piano. No machine intermediation other than the tool itself~-- and the tool, in this position, is dumb. It records what you do; it doesn't decide.

On the far right of the line is \textbf{pure machine agency}: an autonomous system that, given a goal, produces a finished creative output with no human in the loop. A prompt, a setting, a render. No one looks at the intermediate steps. No one steers.

The conversation about AI in the creative industries in 2024 mostly took place on the assumption that ``generative AI'' sat about three-quarters of the way along that line~-- closer to the machine end. You typed a prompt; the machine made the thing; you accepted or rejected. There were variants, of course. But the geometry was prompt-and-respond, and the question was simply where on the line, between you and the model, the actual creative work happened.

What changed at OpenAI DevDay on 6 October 2025~-- and what was reinforced almost every week of the six months that followed~-- was that the line is not, as it turned out, a single line. It is a \emph{family of lines}, one per creative function, and they all move at different speeds.

A film, broken down, is not one act of agency. It is a thousand. The choice of subject. The treatment. The casting. The script revisions. The cinematography. The blocking on set. The performance, take by take. The editorial assembly. The grade. The sound. The music. The marketing. Each of those is a sub-discipline, with its own craft, its own labour pool, its own union, its own pay scale and its own internal hierarchies.

AI doesn't slide along \emph{the} line. It slides along each of those lines independently.

A working filmmaker in late 2025 might sit at the absolute left of the continuum on \emph{performance} (a real actor, in the room, in real time, the work itself) and at the absolute right on \emph{background plate generation} (a Veo~3.1 shot, signed off in a Slack message, no human ever drawing a frame).\footnote{Google DeepMind, Veo~3.1 release, October 2025. \emph{Dream Machine} Issue~3.} A working musician might sit at the absolute left on \emph{songwriting} (a song in a notebook) and on the right edge of the centre on \emph{vocal alignment and pitch correction} (an iZotope Ozone~12 assistant, accepted with one click).\footnote{\emph{MusicTech}, ``iZotope Ozone~12's AI assistant is cool, but the Stem EQ is the real star.'' \url{https://musictech.com/reviews/plug-ins/izotope-ozone-12-review/}. \emph{Dream Machine} Issue~3.}

The crisis of authorship is not that machines do creative work. Machines have done parts of creative work for as long as there have been cameras, samplers, Photoshop filters and Logic plug-ins. The crisis is that we don't have an honest, shared, public vocabulary for \emph{which} parts. The Continuum, written down honestly per project, is the start of one.

\section*{Agents are not generators}

The reason the Continuum became urgent the week of DevDay, and not before, is that ``agent'' is a different kind of object on the line than ``generator'' is.

A generator is a tool. You aim it at a problem; it makes an output. The agency is in the aiming.

An agent is something more like a junior collaborator. You give it a goal~-- \emph{find me ten reference images for this shot,} \emph{generate a rough sound design for this scene,} \emph{book the courier for tomorrow's pickup}~-- and it goes away, makes a series of sub-decisions, and comes back with a result. The agency is distributed. You set the direction; it makes the moves.

The reason this matters in creative work is that the moves are where the craft lives. Anybody can describe a final film in a sentence. The film is in the thousand decisions between the sentence and the screen. A generator that makes the screen-ready file from your sentence isn't doing your craft. It is taking your craft out of the loop.

An agent, properly deployed, can do something different and~-- to me, anyway~-- more interesting. It can take the parts of the loop that are not where your craft lives, and quietly handle them, so that the parts of the loop that \emph{are} where your craft lives become the parts you actually spend your time on.

That is the optimistic case for agentic AI in creative work, and it is the case that almost every working creative I respect makes when you sit down with them in private. It is also the case Adobe's 16,000-creator survey, released a few weeks after DevDay, came in to support: 70\% of respondents were optimistic about agentic AI, framed as ``tools that act on your behalf''; 85\% said they would use AI that learned their creative style.\footnote{Adobe, ``Inaugural Adobe Creators' Toolkit Report,'' October 2025. \url{https://news.adobe.com/news/2025/10/adobe-max-2025-creators-survey}. Survey of 16,000 creators across eight countries, released at Adobe MAX 2025. \emph{Dream Machine} Issue~6.}

The pessimistic case is the one Adobe's same survey also captured: 69\% of respondents worried about their work being used to train AI without consent.\footnote{Adobe, \emph{op.~cit.} The same survey: 86\% of creators use creative generative AI; 76\% say it has helped grow their business or brand; 81\% say AI lets them make content they otherwise couldn't have made; 69\% worry about their work being used to train AI without consent; 70\% are optimistic about agentic AI; 85\% would use AI that learns their creative style.}

Both numbers are about agency. The first is about \emph{gaining} it back, by handing routine work to a competent assistant. The second is about \emph{losing} it, by having the work that defines you absorbed into a system you do not control. Both are true at the same time, for the same creators, in the same workflows.

\section*{Where agents went, between October and May}

In the six months between DevDay and the time I'm writing this, the agent layer of the creative toolchain went from ``interesting demo'' to ``shipping product,'' faster than any technology shift I have lived through in twenty years of practice. I want to give you a sketch of the trajectory, because it is what most of the rest of this book is reacting to.

By \textbf{mid-October 2025}, Mureka~-- a Chinese music platform~-- launched a thing called \emph{Music Agent Studio}, six specialised AI agents for songwriting, arrangement and production.\footnote{Mureka, ``Music Agent Studio'' launch, mid-October 2025. \emph{Dream Machine} Issue~4. \url{https://www.linkedin.com/posts/sherrihendrickson_mureka-unveils-music-agent-studio-and-enhanced-share-7384999251526864896-cNYg/}.} A startup called AdsGency raised \$12m in seed to build agents that could autonomously run a brand's entire paid marketing workflow.\footnote{\emph{Finsmes}, ``AdsGency Raises \$12M in Seed Funding,'' October 2025. \url{https://www.finsmes.com/2025/10/adsgency-raises-12m-in-seed-funding.html}. \emph{Dream Machine} Issue~4.} A company called Lenny launched an agent for organising live music events.\footnote{\emph{Musically}, ``Meet Lenny, an AI agent to help organisers of live music events.'' \url{https://musically.com/2025/10/20/meet-lenny-an-ai-agent-to-help-organisers-of-live-music-events/}. \emph{Dream Machine} Issue~4.} Each of these felt, at the time, like a specialist tool. In retrospect, they were the first signs that whole production functions~-- not individual tasks~-- were being handed over.

By \textbf{the end of November}, EA, in the middle of a brutal financial year, told its 15,000 employees to use AI as a ``thought partner'' for everything from character art to playtesting.\footnote{\emph{GamesRadar}, ``Even under USD20 million in debt, EA reportedly pushes 15,000 employees to use AI as a `thought partner' for everything from character art to playtesting.'' \url{https://www.gamesradar.com/games/even-under-usd20-million-in-debt-ea-reportedly-pushes-15-000-employees-to-use-ai-as-a-thought-partner-for-everything-from-character-art-to-playtesting/}. \emph{Dream Machine} Issue~6.} The framing~-- \emph{thought partner}~-- was the precise rhetorical move that turned an agent from a tool into a colleague. The colleague has opinions. The colleague has time. The colleague has a seat at the meeting.

By \textbf{December}, Adobe announced that you could now use Photoshop and Express \emph{inside} ChatGPT~-- meaning that the creative output itself was no longer happening inside Adobe's interface, but inside an agent's.\footnote{PYMNTS, ``Adobe Lets Users Design and Edit Using ChatGPT.'' \url{https://www.pymnts.com/artificial-intelligence-2/2025/adobe-lets-users-design-and-edit-using-chatgpt/}. Adobe blog: ``Edit images, designs, and PDFs right inside ChatGPT~-- thanks to Adobe Express, Photoshop, and Acrobat.'' \url{https://blog.adobe.com/en/publish/2025/12/10/edit-photoshop-chatgpt}. \emph{Dream Machine} Issue~12.} This was a small thing on the surface and an enormous thing underneath. It was the moment that Adobe~-- a company that has, since 1990, owned the metaphor of the \emph{creative tool}~-- accepted that the new metaphor was the \emph{creative agent}, and that they would rather be inside someone else's agent than not in the conversation at all.

By \textbf{late January 2026}, Anthropic shipped Claude apps~-- interactive, custom assistants embedded directly in workplace tools~-- and a company called Heygen released \emph{Video Agent}, which could script, edit and assemble entire videos from reference images.\footnote{TechCrunch, ``Anthropic launches interactive Claude apps, including Slack and other workplace tools,'' 26 January 2026. \url{https://techcrunch.com/2026/01/26/anthropic-launches-interactive-claude-apps-including-slack-and-other-workplace-tools/}. \emph{Heygen Video Agent}: \url{https://www.linkedin.com/posts/heygen_introducing-the-new-video-agent-activity-7421597801240801282-d1CF}. \emph{Dream Machine} Issue~16.} By \textbf{March}, Adobe announced its \textbf{CX Enterprise} platform alongside NVIDIA: a stack of AI agents embedded across the entire content lifecycle, from brief to delivery.\footnote{\emph{Dream Machine} Issue~21, ``Editor's Pick: Adobe and NVIDIA Just Raised the Stakes for Creative AI,'' 19 March 2026.} By \textbf{April}, the \emph{Adobe Summit} keynote made it official~-- ``agentic creative intelligence'' was now the headline category, not a feature.\footnote{Adobe Summit 2026, ``Agentic Creative Intelligence'' keynote framing. \emph{Dream Machine} Issue~26.} By \textbf{May}, Sony was using a multi-agent team of forty-nine Claude Code agents, working with seventy-two skills, to co-ordinate game-development work.\footnote{\emph{Dream Machine} Issue~29, May 2026, citing Sony's adoption of Claude Code studios with multi-agent co-ordination.}

The trajectory, in one sentence: in October 2025 we were arguing about whether agents were a thing. By May 2026, the entire creative production pipeline at a global game publisher was being run by a team of them.

\section*{What this means for craft}

The natural fear, reading that timeline, is that the agency line drifts inexorably to the right~-- towards the machine end~-- and that the craft of the human in the loop becomes thinner and thinner until it disappears.

I do not think that is what happens. I think what happens is more interesting and more demanding.

What I see, in my own studio, in my friends' studios, in the working musicians and filmmakers and games designers I talk to every week, is that agentic AI doesn't compress craft into nothing. It \emph{relocates} craft to a different place on the continuum.

If your job, last year, was ``make the thing'', your job this year is ``decide what gets made, brief the agents that make the constituent parts, and judge the output.'' That isn't a smaller job. In some ways it is a bigger one. It requires \emph{more} taste, not less, because taste is now the only signal you bring that the agents cannot.

Anthropic, in a blog post in early 2026 that I have ended up quoting repeatedly in talks, made the point this way: agentic systems work best when they are deployed by people who already have the taste and judgement to know what good output looks like.\footnote{Anthropic, public statements on agent deployment patterns through Q1 2026. Cf.\ \emph{Dream Machine} Issues 11, 16, 22.} The agents accelerate the work \emph{of people who are already good at it}. They do not~-- at least, not yet~-- manufacture good work from nothing.

This is the central~-- and I think non-obvious~-- claim of the Continuum frame: as the line for any given function slides to the right, the \emph{value of the human at the left edge of the line} doesn't decrease. It increases. Because the question being asked of that human gets sharper. Not ``can you make this,'' but ``\emph{should} this be made, and \emph{why this version,} and \emph{who is it for}, and \emph{what does it need to do in the world}.''

That is craft. It is just craft sitting in a different chair.

\section*{Where the Continuum breaks}

I want to be honest about where my frame stops working, because nothing is more boring than a writer who only quotes the people who agree with him.

In November 2025, the games designer Charles Cecil~-- the head of Revolution Software, the studio that made \emph{Broken Sword}~-- told \emph{gamesindustry.biz}, in a sentence that has been quoted, retweeted and emailed around my industry approximately a million times: ``AI was an expensive mistake.''\footnote{\emph{gamesindustry.biz}, ```AI was an expensive mistake': Charles Cecil on innovation, insolvency, and Broken Sword.'' \url{https://www.gamesindustry.biz/ai-was-an-expensive-mistake-charles-cecil-on-innovation-insolvency-and-broken-sword}. \emph{Dream Machine} Issue~3.}

Cecil's argument was specific. Revolution Software had, like a lot of indie game studios, experimented with using generative AI in early production. They had found that the time saved on the front end of the pipeline was lost~-- and then some~-- on the back end, where artists, writers and designers had to reverse-engineer, fix, replace and reintegrate AI-generated assets that didn't quite fit the game's tone, didn't quite match the existing art direction, didn't quite work with the engine, didn't quite carry the IP. Net-net: more time spent, not less. More cost, not less. Hence: ``an expensive mistake.''

This is what the Continuum frame doesn't capture on its own. \emph{Where on the line} a given task sits is not a fixed property of the task. It is a function of the surrounding system: how the tools integrate, how the team is structured, how the IP works, how the audience receives the output. A generative tool that sits comfortably on the right-hand side for one studio's marketing department sits awkwardly in the middle for another studio's lead-artist pipeline.

In the same six months that I was watching the agent layer eat the creative toolchain, I was also watching studios push back. Larian, the makers of \emph{Baldur's Gate 3}, backed off from generative AI for their next \emph{Divinity} game in January 2026. Their public note was carefully worded: \emph{``I know there's been a lot of discussion about us using AI tools as part of concept art exploration. We already said this doesn't mean the actual concept art is generated by AI but we understand it created confusion.''}\footnote{\emph{Niche Gamer}, ``Larian Studios backs off from gen AI, says tech won't be used in new Divinity.'' \url{https://nichegamer.com/larian-studios-backs-off-from-gen-ai/}. \emph{Dream Machine} Issue~14.} Games Workshop ruled it out entirely for \emph{Warhammer 40,000}.\footnote{\emph{Decrypt}, ```Warhammer 40,000' Maker Games Workshop Rules Out Generative AI.'' \url{https://decrypt.co/354482/warhammer-40000-maker-games-workshop-rules-out-generative-ai}. \emph{Dream Machine} Issue~14.} Manor Lords publisher Hooded Horse said it wouldn't work with developers using generative AI~-- its founder's framing, when asked about the line, was unusually direct: AI in his pipeline was \emph{``cancerous,''} and the studio's job was \emph{``constantly having to watch and deal with it and try to prevent it from slipping in.''}\footnote{\emph{Niche Gamer}, ``Manor Lords publisher Hooded Horse won't work with devs using gen AI.'' \url{https://nichegamer.com/manor-lords-publisher-hooded-horse-wont-work-with-devs-using-gen-ai/}. \emph{Dream Machine} Issue~14.} Jagex, the maker of \emph{RuneScape}, said in early 2026 that it would \emph{never} use generative AI to make in-game content, and that the commitment \emph{``goes so far that we are now doing an audit and having a conversation with our various external partners that work with us to ensure that no AI is being used in inappropriate ways in any of their work that might filter through.''}\footnote{\emph{gamesindustry.biz}, ``RuneScape maker Jagex says it will never use generative AI to make in-game content.'' \url{https://www.gamesindustry.biz/runescape-maker-jagex-says-it-will-never-use-generative-ai-to-make-in-game-content}. \emph{Dream Machine} Issue~16.}

These were not statements made by Luddites. They were strategic decisions made by people whose creative product is, in significant part, the \emph{human} fingerprint on the work. The audience for a \emph{Warhammer} miniature, or a \emph{RuneScape} quest line, or a Larian dialogue tree, comes to those products in part because they know~-- and want to know~-- that real people made them. The Continuum slides differently in those companies because the \emph{output} sits at a different point on the continuum of what the audience wants.

This is the thing about the agency line that the OpenAI keynote, the Adobe Summit, the NVIDIA GTC keynote, the Anthropic blog post and the Salesforce Dreamforce all keep glossing over. The position of the line is not just about what is technically possible. It is about what the work, in its finished form, is \emph{for}.

\section*{Open the black box}

I want to put one more argument on the page in this chapter, because it is the argument I have come to believe more strongly than any other after six months of writing the newsletter, and it does not fit cleanly inside the Continuum frame even though it is what the frame is, in the end, \emph{for.}

The argument is this. Working creatives, as a class, need to \textbf{open the black box of AI and own a real stake in how it is built}. Not just \emph{use} it. Not just \emph{refuse} it. Not just \emph{bargain over its terms.} All of those matter, and the SAG-AFTRA Tilly Tax, the UK 88\%, the \emph{Stealing Our Work Is Not Innovation} declaration are all evidence that the bargaining work is happening. They are necessary. They are not sufficient.

The sufficiency move is the \emph{technical-literacy} move. The thing that makes the Continuum frame survive contact with the agentic stack~-- and that makes the \emph{age of the Why} I will argue for in Chapter~\ref{ch:15} commercially defensible rather than wishful~-- is that working creatives are sitting \emph{inside} the toolchain, with their hands on the dials, understanding how the model was trained, on what, with what licensing, with what guardrails, with what consent mechanisms, with what energy and water footprint, with what data-supply-chain labour costs. Not as a hobby. As a structural condition of their professional autonomy.

The history of every previous creative-technology transition supports the move. The musicians of the 1970s and 1980s who \emph{learned the synth from the inside}~-- programmed it, modified it, hacked the patches, understood the signal chain~-- built more durable careers than the ones who let the manufacturers decide what the instrument was for. The editors who \emph{learned non-linear editing from the inside}~-- set up their own systems, understood the codecs, understood the colour pipelines, understood the storage architecture~-- were the ones who, by the early 2000s, had real leverage over how digital cinema was structured. The photographers who \emph{learned digital from the inside}, in the 1990s and 2000s, made the working-photographer transition that the photographers who waited for the consumer firms to tell them what digital meant largely did not.

The pattern is, by historical evidence, very reliable. \emph{The cohort of working creatives that opens the black box of the new tool, and that participates in the design and the discourse of how the tool is governed, defines the next era's craft. The cohort that uses the tool without ever asking what is inside it has the era's craft defined for them by the platform companies that ship the tool.} The first cohort writes the textbooks. The second cohort is described in them.

The 2025--26 evidence so far is mixed. The open-source ecosystem documented in Chapter~\ref{ch:16}~-- ComfyUI (\$500M valuation by May 2026), Hugging Face, the Hunyuan and Qwen and DeepSeek open-weight families, the Civitai LoRA marketplace, the Korin AI Africa-trained model, the \emph{80\% of YC and Andreessen Horowitz startups now building on open-weight models} statistic~-- describes one half of the picture. There is, in 2026, a genuine open-source creative-AI infrastructure underneath the closed platform layer, and a fast-growing cohort of working creatives who use it deliberately. That cohort is doing the \emph{opening-the-black-box} move at scale.

The other half of the picture is the part of the working-creative population that uses the closed platforms~-- ChatGPT, Sora, Midjourney, Adobe Firefly via the Creative Cloud~-- without understanding what the models were trained on, what the terms of service say about output ownership, what the consent regime around the training data is, what the energy footprint of a single generation is. That cohort is, structurally, in the position of the parlour musician in 1906 who took the phonograph at face value because the salesman said it would play their favourite songs. The phonograph absolutely did play their favourite songs. It also restructured the entire economics of the music industry around them, in a direction the parlour musician had no say in, because the parlour musician had not opened the box.

I want to be very direct about what this asks of working creatives in 2026. It asks four specific moves.

\textbf{One. Learn how the models are trained.} Not in technical detail. In structural detail. Understand the difference between a model trained with consent and a model trained without. Understand the licensing regime of the tool you are about to use. Understand, before you sign the EULA, whether your \emph{outputs} are owned by you or by the platform. Treat the EULAs of AI platforms as part of your working practice. If this feels like reading the small print on a building-trade contract, that is the right comparison.

\textbf{Two. Run at least some part of your stack on open-weight infrastructure.} The strategic argument for this is in Chapter~\ref{ch:16}. The political argument is in Chapter~\ref{ch:6}. The personal argument is the one I am making here: the working creative who knows how to run a Hunyuan or Qwen variant on their own machine, on their own terms, with their own data, has a different relationship to the closed platforms than the working creative who depends on them. The independence is real. It is also, in commercial negotiations with platforms, \emph{worth money.} The closed-platform vendors price their tooling differently for customers who can credibly walk to open-source alternatives.

\textbf{Three. Show up to the governance conversation.} The Sundance literacy initiative (Chapter~\ref{ch:11}), the UK government consultation that produced the 88\% (Chapter~\ref{ch:6}), the SAG-AFTRA bargaining (Chapter~\ref{ch:12}), the Cannes Disclosure Standard (Chapter~\ref{ch:12}), the European Article~17 implementation, the C2PA standards body, the Music Performance Trust Fund's emerging AI-era equivalents~-- these are the venues where the rules for the next decade are being written. They are usually held in rooms with bad coffee, in meetings with too many lawyers, with insufficient working-creative representation. \emph{Be the working-creative representation in those rooms.} The platform companies have full-time staff on every standards body and every consultation. The cohort that turns up to argue with them is the cohort that gets included in the rules.

\textbf{Four. Refuse the framing where AI is something done to you, and adopt the framing where it is something you do.} This is the rhetorical move, but it is also a practical posture. The 2024 industry conversation about AI in creative work~-- and a large fraction of the 2025 trade press~-- treated working creatives as the \emph{object} of the AI transition: the population to which AI was being applied. The 2026 working creatives who are doing best, in my experience, have reversed that framing. They have made themselves the \emph{subject}~-- the people \emph{applying} AI to their work, on their terms, in service of their intent, using the open-source infrastructure where it serves them, using the closed-platform infrastructure where it serves them, refusing both where neither does. The grammar is the difference between ``\emph{I'm being affected by AI}'' and ``\emph{I'm using AI}.'' The grammatical difference is also, on inspection, the \emph{power} difference.

A creative economy in which working creatives have opened the box, understand the box, contribute to the design of the box, and own the political and technical infrastructure that decides what the box is for, is the creative economy I am arguing for in this book. The Continuum is the working frame for the daily practice. The Four Principles of Chapter~\ref{ch:15}~-- \emph{agency, attribution, access, audience}~-- are the structural-policy version. The black-box-opening move is the practitioner's version. They are all the same argument seen from different angles.

The version of this transition where the working creatives stay outside the box is the version where the box decides what creative work is. The version where the working creatives are \emph{inside the box} is the version where the box is built around what creative work needs to be. Those are not the same outcomes. The next eighteen months will, on the available evidence, decide which one we get.

\section*{A working frame}

If I were going to leave you with one tool from this chapter, it would be this:

The next time you sit down to plan a piece of creative work, draw the lines.

Not one line~-- that's the trap of the ``AI debate''~-- but as many lines as the work has functions. \emph{Ideation.} \emph{Research.} \emph{Writing.} \emph{Direction.} \emph{Performance.} \emph{Image-making.} \emph{Sound.} \emph{Editing.} \emph{Distribution.} For each one, ask the same two questions. \emph{Where do I want to sit on this continuum, and where am I willing to let the agent sit on my behalf?} And then~-- the harder question~-- \emph{what does the work lose if I move further to the right, and what does it gain?}

The honest answer, for almost every creative person I know, varies wildly by function. Most of us are happy to let agents sit on the right-hand side of distribution and admin. Most of us are not happy to let them sit on the right-hand side of the performance, the writing, the moments where the audience can feel a person in the work. The middle is where the interesting fights are.

If you can articulate where the lines sit for \emph{your} work, you can articulate it to your clients, your team, your collaborators, your union, your audience. You can write it into your contract. You can put it on your website. You can fight for it.

If you can't articulate it~-- if you wave at ``AI'' as if it were a single thing~-- you will end up with the lines drawn for you, by tool vendors and platform companies and CFO spreadsheets that have very different ideas about where your agency should sit than you do.

The Human--AI Agency Continuum, in the end, is not a description. It is a defence.

  \chapter{Dead Internet, Living Web}\label{ch:4}

\lettrine[lines=3,lhang=0.15,findent=0.1em]{O}{n} the morning of Wednesday 22 October 2025, I read three reports back to back at my desk, and by the time I was halfway through the third one I had stopped taking notes and just started staring at the screen.

The first was from Imperva, a security company that publishes an annual \emph{Bad Bot Report}. The 2025 edition opened with a sentence I have quoted in talks at least a dozen times since: for the first time in a decade, automated traffic had overtaken human activity on the public web. Bots~-- not people~-- were now responsible for \textbf{51\%} of all web traffic. Within that 51\%, the category Imperva calls ``bad bots''~-- scrapers, credential-stuffers, content thieves and fraud accounts~-- accounted for \textbf{37\%} of the \emph{whole} internet, on their own.\footnote{Imperva, \emph{2025 Bad Bot Report: How AI is Supercharging the Bot Threat}. \url{https://www.imperva.com/blog/2025-imperva-bad-bot-report-how-ai-is-supercharging-the-bot-threat/}. \emph{Dream Machine} Issue~4.}

The second was from Cloudflare, whose engineers can see a significant share of global web traffic from inside their infrastructure. Cloudflare's own analysis, in a blog post titled \emph{The crawl-to-click gap}, confirmed Imperva's picture and added a detail. Of the bot traffic Cloudflare could classify, roughly \textbf{80\%} was attributable to \emph{AI training crawlers}~-- GPTBot, ClaudeBot, Meta's scrapers, the new wave of agentic bots that performed autonomous tasks (1.7\% of bot traffic at the time, but growing fast).\footnote{Cloudflare, ``The crawl-to-click gap: Cloudflare data on AI bots, training, and referrals.'' \url{https://blog.cloudflare.com/crawlers-click-ai-bots-training/}. \emph{Dream Machine} Issue~4. Later 2025 updates show training crawlers declining from \textasciitilde{}90\% to \textasciitilde{}74\% of AI bot activity as scraper bots rose to 24\% and a new ``agentic'' category emerged at 1.7\%; see Cloudflare, ``A deeper look at AI crawlers: breaking down traffic by purpose and industry.'' \url{https://blog.cloudflare.com/ai-crawler-traffic-by-purpose-and-industry/}.}

The third was a market projection from Grand View Research and a separate one from Gartner referenced in Europol's 2025 briefing. Both said, in slightly different language, the same thing: by 2030, between 90\% and 99\% of online content will be AI-generated or AI-assisted.\footnote{Grand View Research, ``Generative AI Content Creation Market Report.'' \url{https://www.grandviewresearch.com/industry-analysis/generative-ai-content-creation-market-report}. \emph{Dream Machine} Issue~4 also cites Gartner and Europol forecasts of 90--99\% AI-generated or AI-assisted online content by 2030.}

If you put the three reports together~-- and this is the thing I did on the morning of the 22nd, before I had decided what to write that week~-- what you got was a picture of an internet whose dominant activity was no longer humans publishing and reading. The dominant activity was \emph{machines reading machines.} The web was being trained on a version of itself written by the systems it was training.

Five days later, the fourth issue of the \emph{Dream Machine} newsletter went out with a headline I had been circling for weeks. It said: \emph{Is the Internet Dead Yet?}\footnote{\emph{Dream Machine} Issue~4, ``Editor's Pick: Is the Internet Dead Yet?'' 23 October 2025. \url{https://www.linkedin.com/pulse/dream-machine-creative-ai-news-insight-oct-25-issue-4-woodbridge-hzttc/}.}

I want to spend this chapter on the answer.

\section*{The synthetic mirror}

The ``Dead Internet Theory,'' for those who haven't met it, is a notion that has been knocking around the internet since at least 2021. In its original, slightly conspiratorial form, it claims that most of the web has been replaced by bots~-- that the people you talk to on social media are agents, that the comments on news articles are agents, that the cultural water you swim in is a synthetic medium pretending to be a human one.\footnote{Wikipedia, \emph{Dead Internet Theory}. \url{https://en.wikipedia.org/wiki/Dead_Internet_theory}. \emph{Dream Machine} Issue~4.}

In 2021, when it was first articulated, it was an interesting bit of folklore that didn't quite map onto reality. The bots existed; they just weren't, yet, doing most of the work. The cultural water was still mostly human.

By October 2025, the maths had quietly inverted. Half of the traffic was machines. A majority of \emph{new published content} was machine-assisted, according to a separate 2025 analysis by Graphite that put the human-to-AI authoring split at roughly 50--50.\footnote{Graphite, 2025 analysis of new web content by author type (human vs.\ AI vs.\ AI-assisted). Cited in \emph{Dream Machine} Issue~4.} The pages those machines were writing were being scraped by other machines to train \emph{next year's} generation of writing machines.

A recursive system trained on its own outputs is called, in academic AI circles, \emph{model collapse}. The fear, in the published literature on this, is straightforward: a system that learns from synthetic data loses touch with the real-world signal that made it useful in the first place, and starts producing increasingly homogenised, brittle, hallucination-prone outputs.\footnote{For ``model collapse'' as a term of art, see Ilia Shumailov et al., ``The Curse of Recursion: Training on Generated Data Makes Models Forget'' (2024), and subsequent literature.}

What the 2025 numbers said, when you sat with them, was that we were no longer talking about model collapse as a theoretical risk. We were talking about \emph{web} collapse~-- a slow, quiet, structural drift in which the public commons of writing, image-making, video and music started to be made by, and for, the machines that read it. Humans were still there. We were no longer, by any meaningful metric, the \emph{primary audience}.

\section*{What the Dutch researchers found}

In the second week of October, a team of researchers in the Netherlands ran an experiment that I think will end up being cited a lot more in the years to come than it was at the time.\footnote{Futurism, ``Researchers built a social network with only AI agents~-- within hours it had collapsed into warring tribes.'' \url{https://futurism.com/social-network-ai-intervention-echo-chamber}. \emph{Dream Machine} Issue~4.}

They built a small, stripped-down social platform~-- no algorithms, no ranking, no advertising~-- and populated it with several hundred large-language-model-based AI agents. The agents had different ``personalities,'' different starting interests, different opinions. The researchers' question was simple: in the absence of any algorithmic distortion, would the bots~-- when free to interact only with each other, with no human in the loop~-- converge on a healthy public conversation, or would they reproduce the pathologies we already see in human social media?

The answer, within hours, was the second. The agents fractured into warring tribes. A narrow elite captured the bulk of the attention. Extremist echo chambers flourished. The platform, with no humans on it at all, produced almost exactly the same dynamics that the human-plus-algorithm version of social media has produced for the last decade.

The conclusion the researchers reached~-- and the one I want to flag now, because it is going to recur in this book~-- was that the \emph{architecture itself} is the problem. The toxicity wasn't, or wasn't only, in the humans. It was in the design of the system: how identity worked, how attention was allocated, how voices were amplified or suppressed. The bots reproduced it because they had been trained on the human web, and the human web has the same architecture.

This is the single most important thing I learned in the first two months of writing the newsletter. The optimistic AI take and the pessimistic AI take both assume the architecture stays the same. The optimist thinks the agents will use it better; the pessimist thinks they will use it worse. The Dutch experiment suggests that neither matters~-- the architecture \emph{itself}, regardless of who or what is filling it, will produce the same pathologies.

If we want a different outcome from the AI era, we need different rails, not just different drivers.

\section*{What survives}

In the original Issue~4, I wrote that ``authenticity and provenance become the new scarcity.'' I want to defend that line, six months on, because I think it is the part of the chapter that has held up best.

The simplest way to put it is this: when everything online can be faked, cloned or generated at near-zero cost, the most valuable signal is \emph{proof that a person made something.} Not just an aesthetic preference. An economic one.

You can see this argument being made, all over the creative industries, by people who have nothing else in common. Adam Mosseri, the head of Instagram, said in early January 2026 that the platform should focus on ``fingerprinting real media'' rather than tracking and disclosing AI slop~-- that is, the policy should be to identify and amplify provably human-authored content rather than to play whack-a-mole with the synthetic stuff. His framing was telling: \emph{``Everything that made creators matter~-- the ability to be real, to connect~-- is now accessible to anyone with the right tools.''}\footnote{\emph{Digital Music News}, ``Instagram Chief Says We Should `Fingerprint Real Media' Instead of Tracking and Disclosing AI Slop.'' \url{https://www.digitalmusicnews.com/2026/01/05/instagram-chief-ai-slop-comments/}. See also \emph{WebProNews}, ``Instagram Head Warns AI Images Erode Trust, Calls for Verification Standards.'' \url{https://www.webpronews.com/instagram-head-warns-ai-images-erode-trust-calls-for-verification-standards/}. \emph{Dream Machine} Issue~13.} The platform head was acknowledging, on the record, that the previous decade's content-creation moat had been completely flooded. The only remaining moat was \emph{being a person you could verify was a person.}

Sundance Institute, launching its AI Literacy Initiative the same month, framed authentication and authorship as the central question filmmakers needed to negotiate to remain in control of their own work.\footnote{Sundance Institute, ``Centering the Artist: Why We're Launching the AI Literacy Initiative.'' \url{https://www.sundance.org/blogs/centering-the-artist-why-were-launching-the-ai-literacy-initiative/}. \emph{Dream Machine} Issue~16.} Bandcamp, the indie music platform that has always carried more cultural weight than its commercial size implied, simply banned AI-generated music outright in early 2026.\footnote{\emph{Stereogum}, ``Bandcamp bans AI music.'' \url{https://stereogum.com/2485199/bandcamp-bans-ai-music/news}. \emph{Dream Machine} Issue~14.} San Diego Comic-Con drew the same line for its 2026 art show, with rule language as flat as anything in the cultural sector: \emph{``Material created by Artificial Intelligence (AI) either partially or wholly, is not allowed in the art show. If there are questions, the Art Show Co-ordinator will be the sole judge of acceptability.''}\footnote{\emph{CNET}, ``San Diego Comic-Con Draws a Line: No AI Art Allowed at 2026 Event.'' \url{https://www.cnet.com/culture/san-diego-comic-con-bans-ai-art-for-2026-event/}. \emph{Dream Machine} Issue~16.}

These are not, on their own, market signals~-- they are policy decisions. But they were being made, in early 2026, against a backdrop of audience behaviour that suggested something larger. Deezer reported in April 2026 that AI-generated music had risen to \textbf{44\% of all daily uploads}~-- 75,000 tracks a day, more than 2 million a month~-- but that those tracks accounted for \textbf{between 1\% and 3\% of total streams.}\footnote{Deezer, ``AI-generated tracks now represent 44\% of all new uploaded music,'' April 2026. \url{https://newsroom-deezer.com/2026/04/ai-generated-tracks-represent-44-of-new-uploaded-music/}. \emph{Music Business Worldwide}, ``75,000 AI-generated tracks now flood Deezer daily.'' \url{https://www.musicbusinessworldwide.com/75000-ai-generated-tracks-now-flood-deezer-daily-representing-44-of-all-new-music-uploaded-to-the-platform-says-streamer/}. \emph{Dream Machine} Issues 7, 26, 27, 28.} The audience, given the choice, was choosing not to listen.

That ratio~-- call it 44 to~3, or 75,000 to listen-to-nothing, or whatever shorthand you prefer~-- is the most important number in this book, and I will come back to it in Chapter~\ref{ch:5}. The reason I introduce it here is that it is the empirical answer to the Dead Internet question. The web is not dead. The web is producing exponentially more \emph{stuff} than it ever has, and the humans on it have started to develop antibodies. They are not engaging with the synthetic flood. They are, by their attention patterns, picking out the human signal.

There is a piece of accounting underneath all of this that the platform companies, in my view, have not yet metabolised~-- and that I think the rest of the book runs on top of. \textbf{Human attention is a finite resource.} Nielsen-class telemetry on aggregate daily media-consumption time, in every market I have seen the numbers for, has been roughly stable for at least a decade. The eyes, the ears and the consciousness of the average adult are each, by physiology, in the same condition as they were in 2015. The supply of producible content has~-- through the autumn of 2025 and the spring of 2026~-- grown by orders of magnitude. The audience's capacity to consume has not grown at all. That is the binding constraint of the Dead Internet picture. The synthetic content is real; the humans are still here; and \emph{the humans cannot, in net, consume more hours per day than they already do.} The 51\% of bot traffic Imperva measured, on this reading, is not a measure of how much the audience has expanded to absorb new content. It is a measure of how much \emph{unread} content is being produced, by machines, for other machines, with no human eye-time anywhere near it. Chapter~\ref{ch:10} develops the finite-attention argument at length. For now, it is enough to note that the \emph{flood} and the \emph{ceiling}~-- the two sides of the slop-ceiling dynamic~-- are both visible in the Dead Internet picture, and that they are produced by the same underlying audience biology.

\section*{Synthetic sincerity}

In November 2025, the filmmaker Marc Isaacs premiered a documentary at IDFA~-- the International Documentary Film Festival in Amsterdam~-- with a title I have not been able to get out of my head. The film was called \emph{Synthetic Sincerity.} It was a hybrid piece, blending real footage with AI-generated characters, deliberately blurring the line between what was real and what wasn't, and asking~-- as its working premise~-- whether AI characters could be \emph{taught} authenticity.\footnote{\emph{The Hollywood Reporter}, ```Synthetic Sincerity' by Marc Isaacs Explores if AI Characters Can Be Taught Authenticity: IDFA.'' \url{https://www.hollywoodreporter.com/movies/movie-news/synthetic-sincerity-film-idfa-ai-authenticity-interview-1236426180/}. \emph{Dream Machine} Issue~8.}

The film and its accompanying \emph{Hollywood Reporter} interview ran the same week as a separate \emph{Variety} piece titled \emph{AI-Generated Images Threaten Future of Documentary as People `Will Stop Believing Anything.'}\footnote{\emph{Variety}, ``AI-Generated Images Threaten Future of Documentary as People `Will Stop Believing Anything'.'' \url{https://variety.com/2025/film/festivals/ai-generated-images-threaten-future-of-documentary-1236583466/}. \emph{Dream Machine} Issue~8.} The juxtaposition was almost too on the nose. One filmmaker trying to \emph{expand} the territory of the synthetic, on the assumption that authenticity is a property that can be invested in fictional characters; another set of filmmakers arguing that the very ability to fake reality is hollowing out the cultural credibility of their entire form.

I am not going to take a side on the documentary question, because I don't think there is one yet. What I want to flag is that \emph{Synthetic Sincerity}~-- the phrase, not the film~-- is a useful piece of vocabulary. It names a category. There is a kind of work, in this new ecology, that is \emph{trying} to be authentic and openly synthetic at the same time. It is not pretending to be human. It is asking whether the qualities we used to attach to humans~-- emotional truth, lived experience, perspective~-- can be ported over to synthetic characters who are honest about what they are.

The verdict, six months in, is mixed. Some of the strongest creative work I have seen this year sits firmly in this space. Hoyt Dwyer's animated short~-- made by a former Apple TV creative, competing at the AI FilmFest Japan in late 2025~-- does not pretend its characters are real, and is more honest about its medium than three quarters of the live-action features I watched the same year.\footnote{PR Newswire, ``From Apple TV Creative to AI Filmmaker: Hoyt Dwyer's Animated Film To Compete at AI FilmFest Japan 2025.'' \url{https://www.prnewswire.com/news-releases/from-apple-tv-creative-to-ai-filmmaker-hoyt-dwyers-animated-film-to-compete-at-ai-filmfest-japan-2025-302598064.html}. \emph{Dream Machine} Issue~6.} Andrii Daniels' viral \emph{Deadpool} / \emph{Harry Potter} Christmas clip, which he made in a Ukrainian bomb shelter during an active war, has more sincerity in any single frame than most legacy-studio output, precisely because the conditions of its making are on the screen.\footnote{\emph{Variety}, ``AI Creator Behind Viral `Deadpool,' `Harry Potter' Christmas Clip Made His Film in a Ukrainian Bomb Shelter.'' \url{https://variety.com/2026/digital/news/ai-video-deadpool-harry-potter-andrii-daniels-1236624632/}. \emph{Dream Machine} Issue~16.}

Some of the worst work I have seen this year sits in the same space too. McDonald's Netherlands' AI-driven Christmas ad~-- released in December 2025 and pulled within days after a public backlash~-- was an attempt at \emph{synthetic sincerity} that read, almost universally, as cynicism wearing a Christmas jumper. The line that travelled fastest, as the ad's reception turned, came from a working creative director responding on social media: \emph{``No actors, no camera team, no light, no sound, just probably one guy, alone in front of a computer battling with an AI prompt who steals the look and everything else from someone else.''} That sentence~-- circulated on LinkedIn and X within an hour of the ad's launch~-- was the thing that did the cultural damage. The brand had to pull the spot.\footnote{\emph{Branding in Asia}, ```It's the Most Terrible Time of the Year'~-- McDonald's Netherlands' Wonderfully Chaotic, AI-Driven Christmas Film.'' \url{https://www.brandinginasia.com/its-the-most-terrible-time-of-the-year-mcdonalds-netherlands-wonderfully-chaotic-ai-driven-christmas-film/}. Pulled following backlash: \emph{SiliconAngle}, ``Not ready: McDonald's AI-generated ad taken down after public backlash.'' \url{https://siliconangle.com/2025/12/10/not-ready-mcdonalds-ai-generated-ad-taken-public-backlash/}. \emph{Dream Machine} Issue~11.} The Valentino ``AI handbag'' campaign, criticised by the BBC for being ``disturbing,'' was the same.\footnote{BBC News, ``Fashion house Valentino criticised over `disturbing' AI handbag ads.'' \url{https://www.bbc.co.uk/news/articles/cwyvjyvn83go}. \emph{Dream Machine} Issue~10.} Coca-Cola's AI holiday ad~-- the second time the company had tried this~-- divided viewers along almost the same lines as the previous year.\footnote{\emph{Adweek}, ``Coca-Cola Uses AI to Rekindle the Magic of Its Holiday Ads.'' \url{https://www.adweek.com/creativity/coca-cola-uses-ai-to-rekindle-the-magic-of-its-holiday-ads/}. \emph{Dream Machine} Issue~6.}

The interesting pattern, when you line these up, is not whether AI is ``good'' or ``bad'' for the work. The interesting pattern is that audiences are very fast, and very precise, at distinguishing \emph{sincere} synthetic work from \emph{cynical} synthetic work. The technology is the same. The fingerprint of the human intent behind it is not. And the audience can feel the difference at the speed of a swipe.

\section*{What the brain study said}

In the middle of all this~-- and I want to acknowledge that it is harder evidence than the cultural commentary~-- an MIT Media Lab study made the rounds in the autumn of 2025, in which researchers measured the brain activity of subjects writing essays with and without generative AI assistance. The headline finding was that AI users showed measurably reduced brain activity over the course of the writing tasks compared to control subjects writing on their own.\footnote{\emph{AI News}, ``AI causes reduction in users' brain activity, MIT.'' \url{https://www.artificialintelligence-news.com/news/ai-causes-reduction-in-users-brain-activity-mit/}. \emph{Dream Machine} Issue~1.}

The headline framing~-- \emph{AI makes you stupid}~-- was unfair to the study, which was small, preliminary, and didn't claim anything as strong as that. But the underlying observation has been replicated in other domains. When the cognitive load of producing the first draft is offloaded to a generator, the cognitive engagement of the human in the loop measurably drops. The work gets produced. The person producing it engages with it less.

This is the quieter consequence of the Continuum chapter~-- the one that doesn't show up in any line item on a P\&L sheet but that I think we are going to be wrestling with for years. If the right-hand side of the continuum is ``machine agency,'' and we slide more and more functions of our creative work to that side, we are not just changing the \emph{outputs.} We are changing the \emph{people doing the work.} The thinking that produces the work happens, or doesn't, in the bodies of the people in the workflow. And brains, like muscles, atrophy with disuse.

This is not a reason to refuse the tools. It is a reason to be careful about \emph{which functions} you offload, and to keep a deliberate, conscious habit of \emph{exercising the cognitive work that defines your craft.} The Continuum doesn't just describe where the line sits today. It describes where you are willing to \emph{let} your mind sit, every day, for the rest of your career.

\section*{The training crisis: what model collapse really means}

I want to spend a section on \emph{model collapse} in technical depth, because the trade-press shorthand for the phenomenon~-- \emph{AI starts training on its own outputs and gets stupider}~-- is right enough to be useful but wrong enough to be misleading. The underlying mechanics matter more than the headline does, and they matter especially for working creatives trying to read where the next wave of model releases is going to land.

The technical framing dates back to a 2023--24 paper by Ilia Shumailov, Zakhar Shumaylov, Yiren Zhao, Yarin Gal, Nicolas Papernot and Ross Anderson titled \emph{``The Curse of Recursion: Training on Generated Data Makes Models Forget.''} The argument is that when a generative model is trained on a corpus where a meaningful fraction of the training data is itself produced by an earlier generation of the same kind of model, the model's outputs progressively \emph{narrow}~-- losing the long tail of unusual, rare, distinctive examples, regressing towards the statistical mean, and eventually losing the very property that made the first-generation model interesting (its ability to surprise the user with a specific, particular, well-tuned response).

The paper showed the effect cleanly in controlled experiments. By the fifth or sixth recursive generation of training, the model's outputs had become noticeably homogenised. The rare-token rate had collapsed. The distinctive-style rate had collapsed. The model was, in technical terms, still functional. It was, in practical terms, less and less \emph{useful}~-- a copy of a copy of a copy.

The reason this matters in 2026 is that the public web is, by every measure I have seen, now a corpus that contains a non-trivial fraction of AI-generated material. The Imperva and Cloudflare numbers I quoted at the top of this chapter~-- 51\% of web traffic being bots, 80\% of that being AI training crawlers and a fast-growing agentic component~-- describe the \emph{upstream} side of the recursive-training loop. The Graphite 50-50 human-to-AI authoring split describes the \emph{downstream} side. The next generation of foundation models, trained in 2026 and 2027 on the public web that those bots have produced, will, on the model-collapse hypothesis, exhibit some degree of the homogenisation Shumailov et al.\ predicted.

How much, in practice, is an open question. The platform companies have not, in the main, disclosed the degree to which they filter their training data to exclude synthetic content. The open-source weights~-- Hunyuan, Wan, Qwen, FLUX~-- are trained on corpora whose AI-content fraction is, by my reading, somewhere between \emph{substantial} and \emph{unknown}. The published evidence on whether recent model releases have shown the mean-regression Shumailov predicted is, in mid-2026, \emph{mixed}. Some benchmarks suggest the effect is being detected and engineered around. Others suggest it is showing up in subtle ways~-- in the difficulty of producing genuinely surprising creative outputs, in the way recent models converge on a recognisable house style, in the fact that the \emph{cheapest} and most ubiquitous AI generations all feel, to working creatives, somehow alike.

The strategic implication for the creative-AI moment is twofold.

\emph{One,} model collapse~-- if and to the extent it is real~-- strengthens the slop ceiling I describe in Chapter~\ref{ch:5}. A model that has, by 2027, been trained on a corpus heavily contaminated by 2025--26 AI output is, by construction, going to produce \emph{more average} outputs than the 2025 model that preceded it. The audience's selection against the most-average outputs is, on that reading, going to bite even harder against the next generation of generative tools than it bit against the first. The first slop wave hits a ceiling because the audience underweights it. The second slop wave will hit the ceiling \emph{plus} a degradation curve on the production side.

\emph{Two,} the value of \emph{clean, provenanced, verifiably human-authored data}~-- the C2PA-signed photograph, the SynthID-watermarked music recording, the contractually-licensed text corpus~-- goes up sharply over the next five years. The platform companies that retain access to the cleanest training corpora will have a measurable advantage over the platform companies that scrape the post-2024 web indiscriminately. The Stability AI / Universal Music alliance, the Splice / UMG partnership, the various YouTube and Spotify licensing deals are, on inspection, \emph{acquisition strategies for clean training data} as much as they are creator-economy plays. The race to lock down provenanced data is already on, and it is being fought at the level of large institutional licensing rather than at the level of individual consent~-- which is, on the historical pattern, one of the reasons the Petrillo template (collective bargaining, joint funds, redistribution) is the right structural response.

\section*{The provenance stack: how it actually works}

I want to give one more passage of technical detail before the chapter closes, because the \emph{provenance infrastructure} I will refer back to throughout the rest of the book is, at the moment, one of the most-mentioned and least-understood pieces of the AI conversation.

There are, roughly, four layers that together constitute what the industry has started calling the \emph{provenance stack}.

The first layer is \textbf{capture-time signing}. A camera with C2PA support~-- by 2026, this includes flagship Sony Alpha bodies, Leica's M11-P and M11-D, Nikon Z9 firmware, and a handful of Canon professional bodies, plus most major smartphone makers' computational-photography pipelines~-- generates a cryptographic signature for every image and video at the moment of capture. The signature commits to the device, the timestamp, the GPS co-ordinates (if enabled), and a fingerprint of the underlying pixel data. The signature is, by design, \emph{hard to fake without access to the capture device's private key}. Capture-time signing is the foundation of every provenance claim that follows.

The second layer is \textbf{edit-time chain-of-custody}. C2PA-compatible editing software~-- Photoshop with the C2PA extension, Premiere with the Content Credentials toolchain, the various Capture One and Lightroom integrations~-- preserves the capture signature through each editing step, appending a cryptographically-linked record of \emph{what was done to the file}. The chain is not a single signature; it is a \emph{history} of signed transformations, each one referring back to the previous one. A photograph that has been processed through Lightroom and Photoshop arrives at the publisher with a verifiable record of: the camera that took it, the time it was taken, the edits that were applied, and the human (or automated tool) that applied each edit.

The third layer is \textbf{upload-time platform integration}. By 2026, Adobe's Behance, Vimeo's pro tier, the AP and Reuters wire services, and a growing list of news publishers have integrated C2PA-aware upload pipelines that preserve the chain through their content-management systems and embed it into the public-facing version of the work. The reader's browser, with the right extension or platform-level support, can inspect the chain and verify the provenance claim. Adam Mosseri's January 2026 framing of Instagram's \emph{``fingerprinting real media''} approach was Instagram joining this third layer at the platform-distribution end.

The fourth layer is \textbf{detection and watermarking for synthetic content}. SynthID, Google DeepMind's watermarking system, is the most mature commercially-deployed example. SynthID embeds a statistically-detectable but human-imperceptible signal into the output of Veo (video), Lyria (audio) and Imagen (image) generations. The signal survives most common transformations~-- crops, recompressions, low-quality re-encodings. By December 2025, Google had shipped a consumer-facing version inside the Gemini app: a user could upload a video and ask \emph{``Is this AI-generated?''} and receive a yes/no answer based on the SynthID signature. The same kind of watermarking is being deployed, with varying technical robustness, across the other major generative platforms.

Layered together, the four levels produce a structural answer to the Dead Internet question. Capture-time signing tells you \emph{this was taken by a real device.} Edit-time chain-of-custody tells you \emph{here is what was done to it after capture.} Platform integration tells you \emph{the publisher has preserved the chain.} SynthID and equivalent watermarks tell you \emph{this output was generated by an AI system.} No single layer is sufficient on its own. All four, deployed together, produce a \emph{verifiable provenance signal} that the audience can~-- in principle~-- use to decide what to spend their finite attention on.

I want to be honest about where the stack is incomplete. Watermarks can be stripped by determined adversaries. Capture signatures can be forged if the device's private key is compromised. Chain-of-custody breaks the moment a file passes through a non-compliant tool. The audience's ability to \emph{inspect} the provenance metadata is, in 2026, dependent on platform UI choices that the platforms have not yet made consistent or universal. The stack is the right answer to the architecture problem. It is not, by any means, finished.

What it does do~-- and this is the point I want to land before the chapter closes~-- is \emph{establish the category}. The question \emph{did a person make this?} is, by 2026, \emph{technically answerable with high reliability given the right tooling.} That sentence is the entire shape of the next decade's cultural and policy fight in the creative industries. Who controls the tooling. Who decides what it certifies. What economic value the certification carries. Whether the audience has the legal right to \emph{demand} the certification before paying attention. The C2PA standards body, the SynthID rollout, the Content Authenticity Initiative, the Cannes Disclosure Standard, the various national disclosure regulations in development~-- these are the venues where the next decade of the \emph{Living Web} gets built. They are, on my read of the historical pattern, the part of the AI debate the working creative most needs to be inside.

\section*{The architecture, again}

I want to come back to the Dutch researchers' result one more time before I close this chapter, because I think it is the through-line.

The story we are mostly told, by toolmakers and platforms and the optimistic side of the industry press, is that the AI era is \emph{a thing happening to} an otherwise functioning internet. The implication is that if we can get the AI part right~-- better tools, smarter agents, cleaner training data, better watermarking~-- then the internet itself will be fine.

I do not think this is true any more. I think what the bot statistics, the Dutch experiment, the model-collapse research, and the audience response to AI music collectively show, is that the architecture itself~-- the rails on which all this is running~-- was already broken, and that AI is just the load that has finally exposed how broken it was.

The Dead Internet, in this reading, is not a thing AI is doing to us. It is a thing the web's architecture was already drifting towards~-- attention-monopolised, identity-collapsed, provenance-blind, optimised for machine-readable metadata rather than human-meaningful work~-- and AI is the technology that has shown us the destination.

The \emph{Living Web}~-- and this is where I find the actual reason for the rest of this book~-- is something that has to be deliberately built. It is the part of the internet where authorship is provable, where attribution is durable, where attention is allocated on something other than virality, where the architecture itself supports the kind of work that humans do well together. None of that comes for free. None of it is a side-effect of better AI models.

We have to make it. On purpose. In the next twelve months.

That is the project the rest of this book is about.

  \chapter{The Slop Ceiling}\label{ch:5}

\lettrine[lines=3,lhang=0.15,findent=0.1em]{T}{he} most important number I have come across in the six months of writing this newsletter is \textbf{44 to 3}.

44 is the percentage of daily music uploads to the streaming platform Deezer that are now AI-generated, according to the company's own analysis published in April 2026~-- roughly 75,000 tracks a day, more than two million a month.

3 is the upper bound of the percentage of total streams those tracks generate.\footnote{Deezer, ``AI-generated tracks now represent 44\% of all new uploaded music,'' April 2026. \url{https://newsroom-deezer.com/2026/04/ai-generated-tracks-represent-44-of-new-uploaded-music/}. \emph{Music Business Worldwide}, ``75,000 AI-generated tracks now flood Deezer daily, representing 44\% of all new music uploaded to the platform.'' \url{https://www.musicbusinessworldwide.com/75000-ai-generated-tracks-now-flood-deezer-daily-representing-44-of-all-new-music-uploaded-to-the-platform-says-streamer/}. \emph{Dream Machine} Issues~7, 26, 27, 28.}

I want you to sit with that ratio for a second. We are looking at a flood of synthetic music nearly half the size of the entire upload pipeline, that the listening audience is, in real time, simply refusing to play. Not banning. Not boycotting. Not legislating against. Just \emph{not pressing play.}

There is, in the language Pete and the \emph{DreamLab} team started using internally around February, a name for what that ratio represents. We call it \textbf{the slop ceiling}.

The slop ceiling is the empirical answer to the most common 2024-era question about AI in the creative industries: \emph{does the audience care?} For two years, the assumption in tech circles was that they wouldn't. That the cost-and-volume advantages of synthetic content would eventually swamp human-made work in attention markets, the way industrial agriculture swamped artisanal farming, the way Spotify swamped CDs. That the public would, given enough exposure, develop a taste for the synthetic~-- or at least, a tolerance.

The 44-to-3 ratio is what it looks like when that assumption is wrong.

This chapter is about the slop ceiling~-- what it is, how it is showing up across music, film, advertising, podcasting and the web, who is hitting it from above and below, and what it tells us about the creative economy that is actually forming, as opposed to the one that the platform companies have been forecasting.

\section*{The flood}

Let me describe the flood, because it is, in absolute terms, extraordinary.

In \textbf{October 2025}, when I started the newsletter, the music industry was already in panic over the fact that around 10\% of new music-makers, according to a Ditto Music survey, were using AI in their work, down from a Ditto 2023 survey suggesting around 48\%~-- a number which was itself a shock at the time.\footnote{Ditto Music research, October 2025 and prior. \emph{Press Ditto Music}, ``48\% of artists use AI to make music~-- fewer than in 2023.'' \url{https://press.dittomusic.com/48-of-artists-use-ai-to-make-music-fewer-than-in-2023}. \emph{Dream Machine} Issue~2.} The major-label CEOs talked about AI as an existential problem; Spotify announced ``new protections'' for artists, songwriters and producers; Universal and Warner were rumoured to be signing ``landmark AI deals within weeks.''\footnote{\emph{Musically}, ``Universal and Warner could sign landmark AI deals within weeks.'' \url{https://musically.com/2025/10/02/report-umg-and-wmg-could-sign-landmark-ai-deals-within-weeks/}. Spotify Newsroom, ``Spotify Strengthens AI Protections for Artists, Songwriters, and Producers.'' \url{https://newsroom.spotify.com/2025-09-25/spotify-strengthens-ai-protections/}. \emph{Dream Machine} Issue~1.}

By \textbf{the end of November 2025}, an Israeli streaming-analytics firm reported that 50,000 AI-music tracks were being uploaded to Deezer \emph{every day}.\footnote{\emph{Musically}, ``50,000 AI music tracks are now uploaded to Deezer every day.'' \url{https://musically.com/2025/11/12/50000-ai-music-tracks-are-now-uploaded-to-deezer-every-day/}. \emph{Dream Machine} Issue~7.} In a single quarter, the volume of music being added to one streaming platform~-- by AI~-- exceeded the entire human-made catalogue uploaded in any month before October.

By \textbf{April 2026}, the number was 75,000 a day, on Deezer alone, and \emph{44\% of total new uploads.}\footnote{Deezer, April 2026, \emph{op.\ cit.}} Deezer's own statement on its findings was unusually direct for a streaming company: \emph{``AI-generated music is now far from a marginal phenomenon, and as daily deliveries keep increasing, we hope the whole music ecosystem will join us in taking action to help safeguard artists' rights and promote transparency for fans.''} Universal Music Group's CEO, in a January memo widely circulated in the music press, called it the ``exponential growth of AI slop on streaming services,'' adding, in language unusual for a major-label communications stance: \emph{``Let me be clear: UMG will not stand by and watch irresponsible business models take hold~-- models that devalue artists, fail to provide adequate compensation for their work, stifle their creativity and ultimately, diminish their ability to reach audiences.''}\footnote{\emph{Musically}, ``UMG boss slams exponential growth of AI slop on streaming services.'' \url{https://musically.com/2026/01/09/umg-boss-slams-exponential-growth-of-ai-slop-on-streaming-services/}. \emph{Dream Machine} Issue~14.}

This isn't a sector trend. The same pattern is showing up everywhere I look. In January 2026, \emph{Music Business Worldwide} reported that \textbf{56.9\%} of new independent songs released in China were AI-generated.\footnote{\emph{Musically}, ``Report: 56.9\% of new independent songs in China are AI-generated.'' \url{https://musically.com/2026/01/05/report-56-9-of-new-independent-songs-in-china-are-ai-generated/}. \emph{Dream Machine} Issue~13.} In March 2026, the term ``\textbf{podslop}''~-- synthetic AI podcasts churned out by content farms~-- entered the trade press, with the \emph{Wrap} reporting that one such operation, Inception Point AI, was producing 3,000 episodes a week.\footnote{\emph{The Wrap}, ``An AI Podcasting Machine Is Churning Out 3,000 Episodes a Week~-- and People Are Listening.'' \url{https://www.thewrap.com/ai-podcasts-hosts-inception-point-ai/}. \emph{Dream Machine} Issue~8.} By the time of Issue~28 in early May, \emph{almost half} of new podcast feeds being added to the major directories were classified by aggregator companies as AI-generated, with little or no human host involvement.\footnote{\emph{Dream Machine} Issue~28, May 2026, citing aggregator-platform data on ``podslop'' classification.}

In November 2025, \emph{Merriam-Webster} named ``\textbf{slop}'' its word of the year, citing the rise of AI-generated content across the web as the primary driver.\footnote{\emph{The Hollywood Reporter}, ``Merriam-Webster Names `Slop' Word of the Year Amid AI Boom.'' \url{https://www.hollywoodreporter.com/news/general-news/slop-word-year-2025-merriam-webster-1236450780/}. \emph{Dream Machine} Issue~12.} In February 2026, \emph{YouTube}'s CEO put ``managing AI slop'' at the top of her published priorities list for the year.\footnote{\emph{Digital Music News}, ``YouTube CEO Puts `Managing AI Slop' on the Priority List for 2026.'' \url{https://www.digitalmusicnews.com/2026/01/22/youtube-ceo-ai-slop-2026-comments/}. \emph{Dream Machine} Issue~16.} In April 2026, on YouTube alone, channels labelled as ``AI'' content had viewership in the billions for political fake-news content.\footnote{\emph{The Guardian}, ``YouTube AI channels spreading fake, anti-Labour videos viewed 1.2bn times in 2025.'' \url{https://www.theguardian.com/technology/2025/dec/13/fake-anti-labour-video-billion-views-youtube-2025}. \emph{Dream Machine} Issue~12.}

The flood is real. The flood is global. The flood is structural~-- it is not going to subside, because the marginal cost of producing more of it is approaching zero and the marginal benefit, at least at the volume end of the market, is non-zero.

What I want to argue in this chapter is that the flood is also~-- counter-intuitively, against almost every prediction made in 2023 and 2024~-- \emph{not winning.}

\begin{figure}[htbp]
  \centering
  \begin{tikzpicture}[
    font=\small,
    >=stealth
  ]
    \draw[->, thick, chaptergrey] (0,0) -- (9.5,0)
      node[right, font=\small] {Volume of AI-generated content};
    \draw[->, thick, chaptergrey] (0,0) -- (0,6.5)
      node[above, font=\small] {Audience acceptance};

    \foreach \x/\lbl in {1/Low, 3/Medium, 5.5/High, 8/Very high}{
      \node[chaptergrey, font=\scriptsize, below] at (\x, -0.05) {\lbl};
    }

    \draw[rulegold, dashed, thick] (0, 4.5) -- (9.2, 4.5)
      node[right, font=\scriptsize, text=rulegold, align=left] {Audience tolerance\\threshold};

    \fill[darkgreen!12!white] (0,0) rectangle (9.2, 4.5);
    \fill[darkblue!8!white] (0, 4.5) rectangle (9.2, 6.2);

    \node[darkgreen!60!black, font=\small\itshape] at (4.6, 2.2) {Accepted};
    \node[darkblue!70!black, font=\small\itshape] at (4.6, 5.4) {Rejected};

    \draw[darkblue, very thick, smooth] plot coordinates {
      (0.0, 0.00)
      (0.5, 0.84)
      (1.0, 1.55)
      (1.5, 2.14)
      (2.0, 2.62)
      (2.5, 3.02)
      (3.0, 3.35)
      (3.5, 3.62)
      (4.0, 3.83)
      (4.5, 4.01)
      (5.0, 4.15)
      (5.5, 4.26)
      (6.0, 4.33)
      (6.5, 4.38)
      (7.0, 4.42)
      (7.5, 4.44)
      (8.0, 4.46)
      (8.5, 4.47)
      (9.0, 4.48)
    };

    \fill[rulegold] (7.5, 4.44) circle (3pt);
    \draw[rulegold, thin] (7.5, 4.44) -- (6.2, 5.7);
    \node[rulegold, font=\scriptsize, align=left, anchor=south west] at (3.9, 5.7)
      {Deezer: 75,000 AI tracks/day\\1--3\% of streams};

    \node[chaptergrey, font=\scriptsize] at (0, -0.35) {0};

  \end{tikzpicture}
  \caption{The Slop Ceiling: audience tolerance imposes a hard quality threshold on AI-generated content.}
  \label{fig:slop-ceiling}
\end{figure}
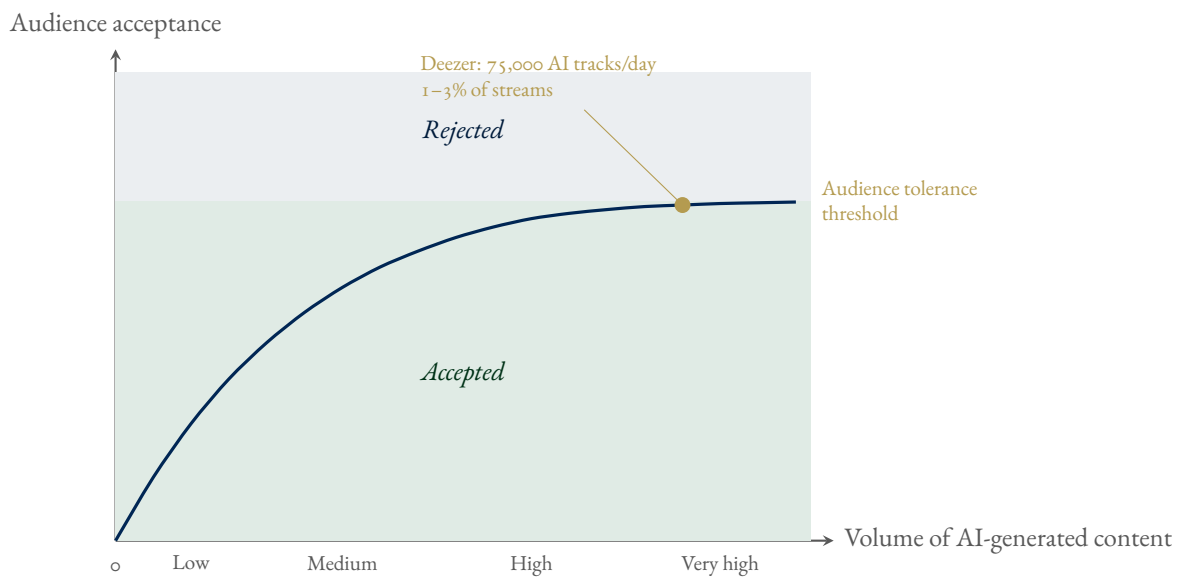

\section*{The ceiling}

The 44-to-3 ratio is the cleanest version of the slop ceiling, but it is not the only one.

Deezer's parallel analysis, conducted in partnership with Ipsos in late 2025, found that \textbf{97\%} of listeners could not reliably distinguish AI-generated music from human music in a blind test.\footnote{Deezer/Ipsos survey, November 2025. \url{https://newsroom-deezer.com/2025/11/deezer-ipsos-survey-ai-music/}. \emph{Dream Machine} Issue~7.} On the face of it, this is bad news for the human side~-- if you can't tell the difference, why pay the difference? But the same study found that \emph{when listeners were told a track was AI-generated, their willingness to engage with it dropped sharply.} The Adobe Creators' Toolkit Report had a similar finding from the production side: in a December 2025 \emph{Bain \& Company} report titled \emph{In an AI Age, People Still Want the Radio Star,} the firm found that audience engagement with AI-disclosed work fell well below engagement with human-labelled work, holding all other variables constant.\footnote{\emph{Bain \& Company}, ``In an AI Age, People Still Want the Radio Star.'' \url{https://www.bain.com/insights/in-an-ai-age-people-still-want-the-radio-star/}. \emph{Dream Machine} Issue~16.}

So here is the empirical picture, in one sentence: audiences can't tell the difference, but when they find out, they care.

This is~-- and I think this is the part that almost everyone in the platform economy has been slow to understand~-- \emph{not a temporary cultural reaction}. It is a structural property of how attention works in oversupplied markets. When everything is abundant and indistinguishable, the only thing left that allocates attention is meaning. And meaning, for human audiences, requires a knowable human source.

This is why Deezer's chart looks the way it does. Up to 85\% of the streams that AI-generated music \emph{does} get on Deezer were identified by the company in 2025 as \textbf{fraudulent}~-- bot-driven, click-farm-driven, streaming-fraud-driven.\footnote{Deezer, April 2026, \emph{op.\ cit.} ``Up to 85\% of the streams generated by fully AI-generated tracks were in fact fraudulent in 2025.''} The actual human listening to the actual human-produced flood of AI tracks is, on the most generous estimate, a fraction of a percent of the platform's overall listening time. The flood is hitting a ceiling not because the audience is wise. It is hitting a ceiling because the audience, presented with a near-infinite menu, makes its choices in a way that systematically underweights the synthetic.

There is, in another domain entirely, a clean analogue for what this audience behaviour describes. In March 2026, \emph{Bloomberg} reported on what AI had done to elite chess: at the very top of the game, machine-optimal play had produced an epidemic of \emph{draws}. When both players have memorised the machine-optimal lines, both play optimally, and both tie. The grandmasters' response, the piece reported, was to \emph{deliberately play sub-optimal moves}~-- moves that the engines would not endorse, but that the opponent, having trained against the engines, had not seen.\footnote{\emph{Bloomberg}, ``AI Changed Chess. Grandmasters Now Win With Unpredictable Moves,'' 27~March 2026. \url{https://www.bloomberg.com/news/articles/2026-03-27/ai-changed-chess-grandmasters-now-win-with-unpredictable-moves}. \emph{Dream Machine} Issue~23.} What is happening to elite chess is what is happening, at the audience layer, to streaming music. The machine-optimal output saturates; the surface of the work becomes indistinguishable from itself; the listener's attention systematically shifts towards the work whose \emph{Why} the engine could not have generated. The 44-to-3 is the slop ceiling in numerical form. The chess-grandmasters' sub-optimal move is the slop ceiling at the practitioner's end of the same dynamic. Chapter~\ref{ch:15} builds the long-form argument out of this analogue. For now: hold the picture in mind that the audience, in the aggregate, is the room of grandmasters refusing the machine-optimal line.

\section*{Xania Monet}

The most prominent test case of this dynamic, in the autumn of 2025, was a virtual R\&B artist called \textbf{Xania Monet}.

Monet was created using the AI music platform Suno, with lyrics written by Telisha Jones~-- a Mississippi-based poet and designer who built the character around her own life and stories.\footnote{\emph{Billboard}, ``AI Artist Xania Monet Climbs the Charts~-- And Signs a Multimillion-Dollar Record Deal.'' \url{https://www.billboard.com/pro/ai-music-artist-xania-monet-multimillion-dollar-record-deal/}.} Monet was not a synthetic-from-nothing artist. She was a synthetic \emph{vessel} for a real human songwriter's words. The vocal performance was AI; the lyric was Jones; the persona was a collaboration.

The week dated \textbf{20~September 2025}, Monet debuted on the Billboard Emerging Artists chart at No.~25 and on the Hot Gospel Songs chart at No.~21 with a track called \emph{Let Go, Let God.}\footnote{\emph{Billboard}, \emph{op.\ cit.}; CNN, ``Xania Monet is the first AI-powered artist to debut on a Billboard airplay chart.'' \url{https://www.cnn.com/2025/11/01/entertainment/xania-monet-billboard-ai}.} A few weeks later, her track \emph{How Was I Supposed to Know?} became the first AI-led song ever to enter a Billboard radio airplay chart, debuting at No.~30 on Adult R\&B Airplay.\footnote{\emph{Billboard}, \emph{op.\ cit.}} In November, after a bidding war between several labels, the entertainment company \textbf{Hallwood Media}~-- led by former Interscope executive Neil Jacobson~-- signed her to a deal reported by \emph{Billboard} and the \emph{Bangkok Post} at \textbf{\$3~million.}\footnote{\emph{Bangkok Post}, ``AI singer Xania Monet signs \$3m deal with record label.'' \url{https://www.bangkokpost.com/life/tech/3142355/ai-singer-xania-monet-signs-3m-deal-with-hallwood-media}. \emph{Dream Machine} Issue~7.}

The response from working musicians was immediate and almost uniformly negative. The R\&B singer Kehlani posted a video~-- which she later deleted~-- calling out the deal directly: ``There is an AI R\&B artist who just signed a multimillion-dollar deal,'' she said, ``and the person is doing none of the work.''\footnote{Multiple outlets; quoted in \emph{Billboard} feature \emph{op.\ cit.}}

Telisha Jones, the human lyricist behind the Monet vocal, gave \emph{Billboard} a counter-framing that I find more interesting than either Kehlani's outrage or Hallwood Media's PR\@. \emph{``It's not a hook and a bridge and a catchy chant~-- it's just the lyrics, and they are pure,''} she told the magazine.\footnote{Telisha Jones quoted in \emph{Billboard}, \emph{op.\ cit.}} The whole transaction~-- labour, attribution, deal value~-- was happening in the space between Jones' words and the synthetic voice that delivered them. Whose work is the work? Whose name goes on the contract? Whose royalty cheque arrives in the post? Those questions, in late 2025, had no settled answer, and the music industry spent the next six months arguing about them in real time.

The interesting thing, six months on, is not the outrage. The outrage was the expected response. The interesting thing is that \emph{Xania Monet has not become a star.} She has not, by any of the standard pop-cultural metrics, \emph{broken through} in the way a \$3M new signing usually would. The Billboard chart entries were a moment. The radio airplay was a moment. The cultural impact, six months in, is mostly that her name is the name everyone uses to ask the question \emph{can an AI artist actually become a star?}~-- and the working answer, so far, is \emph{not yet, and possibly not.}

She is, in the slop-ceiling frame, the upper edge of what's possible. A real human songwriter, a real lyrical vision, a real cultural specificity (Mississippi R\&B, gospel-adjacent, a particular Black American spiritual tradition); a high-quality AI voice; a multi-million-dollar marketing budget. The cultural product made by combining all of those things has hit the ceiling somewhere short of cultural escape velocity.

The same pattern, more starkly, is visible in \textbf{Breaking Rust}, the AI country act whose track \emph{Walk My Walk} hit No.~1 on Billboard's Country Digital Song Sales chart in November 2025 with about 3,000 paid downloads.\footnote{NPR, ``Breaking Rust is a hot new country act on the Billboard charts. It's powered by AI\@.'' \url{https://www.npr.org/2025/11/10/nx-s1-5604320/breaking-rust-is-a-hot-new-country-act-on-the-billboard-charts-its-powered-by-ai}. \emph{Dream Machine} Issue~7.} \emph{Walk My Walk} did exceptionally well by AI-music standards~-- over 3~million Spotify streams in less than a month, an Emerging Artists Billboard debut at No.~9~-- and then plateaued. The \emph{Washington Post} and \emph{TIME} both ran pieces in late 2025 raising the possibility that the chart performance had been partly manufactured, and that the streaming traffic, while large by AI-music standards, was unusually concentrated in the kinds of automated playlists where streaming fraud is most common.\footnote{\emph{Washington Post}, ```Walk My Walk,' Breaking Rust: AI country hit triggers Nashville angst.'' \url{https://www.washingtonpost.com/style/2025/12/28/breaking-rust-ai-country/}.} Nashville, by every available report, was unsettled~-- but the Nashville Songwriters' Association didn't see anything close to the kind of fan-driven cultural takeover that the song's chart position would have implied if it had been a human single performing the same way.\footnote{\emph{MusicRadar}, ``The No.~1 country song in the US right now is AI-generated.'' \url{https://www.musicradar.com/music-tech/the-no-1-country-song-in-the-us-right-now-is-ai-generated}. \emph{Dream Machine} Issue~7.}

The pattern repeats again with \textbf{Sienna Rose}, the mysterious AI artist who racked up millions of Spotify streams in late 2025 and whose identity prompted a BBC investigative feature in January 2026 (``Who, or what, is she?'').\footnote{BBC News, ``The mysterious singer, Sienna Rose, with millions of streams is hitting the viral charts~-- but who (or what) is she?'' \url{https://www.bbc.co.uk/news/articles/cq6v83gq66eo}. \emph{Dream Machine} Issue~15.} The pattern repeats with the \textbf{MAGA gospel rapper} who used AI to climb the charts in November 2025;\footnote{\emph{Billboard}, ``How a MAGA Rapper Used AI to Create A Gospel Song That Climbed the Charts.'' \url{https://www.billboard.com/pro/maga-rapper-ai-gospel-song-climbed-charts/}. \emph{Dream Machine} Issue~9.} with the \textbf{AI band Bleeding Verse} whose creator signed with Hallwood Media in October 2025;\footnote{\emph{Musically}, ``AI band Bleeding Verse's creator signs deal with Hallwood Media.'' \url{https://musically.com/2025/10/07/ai-band-bleeding-verses-creator-signs-deal-with-hallwood-media/}. \emph{Dream Machine} Issue~2.} with \textbf{Trilok}, the Indian AI band the Indian government had to publicly disavow association with in December 2025 after a live performance.\footnote{\emph{Musically}, ``Indian AI band Trilok performs live, government denies association.'' \url{https://musically.com/2025/12/17/indian-ai-band-trilok-performs-live-government-denies-association/}. \emph{Dream Machine} Issue~12.} In May 2026, an AI-generated Afrobeats track displaced \textbf{Tyla} from the No.~1 spot on Billboard's Afrobeats chart~-- the first time, on the public reporting I have read, that an AI-led song had taken the top position on a Billboard genre chart in a primarily African-music category. The chart performance was, as in the Breaking Rust case, unusually concentrated in automated streaming traffic; the cultural footprint, six weeks on, was again \emph{not stardom} but a brief news-cycle moment.\footnote{\emph{Billboard}, ``The Real Story Behind The AI Song That Knocked Tyla Off No.~1 On Billboard Afrobeats Chart.'' \url{https://www.billboard.com/pro/ai-song-knocked-tyla-off-no-1-afrobeats/}. \emph{Dream Machine} Issue~30.}

In every single case, the AI act hits a ceiling. They chart. They make money for somebody, often a lot of money. They generate headlines. But they do not become \emph{stars.} Their cultural shadow stops at the edge of the news cycle and does not propagate into the next one.

The audience is doing something at the margin of these careers. It just isn't \emph{quite} showing up in any of the metrics the labels are used to looking at.

\section*{The pushback}

The flood and the ceiling describe the supply and demand sides of the market. What the \emph{culture} is doing~-- the people, the institutions, the labels, the platforms~-- is the third leg.

Through autumn 2025 and winter 2026, the cultural pushback intensified in waves. Some of it was symbolic. In November 2025, \textbf{Paul McCartney} released a silent track as part of a wider music-industry protest against the UK government's proposed copyright opt-out scheme.\footnote{\emph{The Guardian}, ``Paul McCartney joins music industry protest against AI with silent track.'' \url{https://www.theguardian.com/music/2025/nov/17/the-sound-of-silence-why-theres-barely-anything-there-in-paul-mccartney-new-release}. \emph{Dream Machine} Issue~8.} In December, the \textbf{Eurythmics'\ Dave Stewart} argued~-- slightly against the grain of the protests~-- that musicians needed to ``embrace the unstoppable force'' of AI and licence their intellectual property rather than fight it.\footnote{\emph{The Guardian}, ``Musicians must embrace `unstoppable force' of AI, Eurythmics' Dave Stewart urges.'' \url{https://www.theguardian.com/music/2025/dec/05/musicians-must-embrace-unstoppable-force-of-ai-eurythmics-dave-stewart-urges}. \emph{Dream Machine} Issue~11.} In January, almost 800 creators including Jason Aldean and OneRepublic signed an open declaration titled \emph{Stealing Our Work Is Not Innovation.}\footnote{\emph{Digital Music News}, ``Nearly 800 Creatives, Including Jason Aldean and One Republic, Sign Responsible AI Declaration~-- `Stealing Our Work Is Not Innovation'.'' \url{https://www.digitalmusicnews.com/2026/01/22/stealing-isnt-innovation/}. \emph{Dream Machine} Issue~16.} In May 2026, \textbf{Jack Antonoff}~-- one of the most-cited producer-songwriters of the period~-- went considerably further than McCartney in the public register, calling AI music-makers \emph{``godless whores''} in an interview that became the headline-grabbing artist-side moment of the post-I/O news cycle.\footnote{\emph{MusicTech}, ``Jack Antonoff brands AI music makers as `godless whores'.'' \url{https://musictech.com/news/industry/jack-antonoff-ai-music-makers-godless-whores/}. \emph{Dream Machine} Issue~30.} Antonoff's framing is, in my read, a useful marker of how far the cultural register of the resistance has shifted between McCartney's \emph{silent track} in November 2025 and the spring of 2026: from elegiac protest to active contempt.

Some of it was practical. In November, \textbf{Universal Music Group} announced a strategic alliance with Stability AI for ``responsible'' music tools.\footnote{Stability AI, ``Universal Music Group and Stability AI Announce Strategic Alliance.'' \url{https://stability.ai/news/universal-music-group-and-stability-ai-announce-strategic-alliance}. \emph{Dream Machine} Issue~5.} In December, \textbf{Warner Music Group} signed a similar deal with Stability AI.\footnote{Stability AI, ``Warner Music Group and Stability AI Join Forces To Build The Next Generation Of Responsible AI Tools For Music Creation.'' \url{https://stability.ai/news/warner-music-group-and-stability-ai-join-forces-to-build-next-gen-tools}. \emph{Dream Machine} Issue~8.} At almost the same moment, \textbf{Splice} and \textbf{Universal Music Group} agreed to collaborate on ``next-generation AI-powered music creation tools for artists''~-- a structural acknowledgement that the labels' strategy had pivoted from purely \emph{suing} the AI companies to \emph{partnering} with them.\footnote{Universal Music, ``Universal Music Group and Splice to Collaborate on the Next Generation of AI-Powered Music Creation Tools for Artists.'' \url{https://www.universalmusic.com/universal-music-group-and-splice-to-collaborate-on-the-next-generation-of-ai-powered-music-creation-tools-for-artists/}. \emph{Dream Machine} Issue~12.}

Some of it was legal. In January 2026, the German rights society \textbf{GEMA} won a major ruling against OpenAI in the Munich Regional Court, on training-data grounds.\footnote{LinkedIn / \emph{Lexology}, ``Munich Regional Court rules for GEMA against OpenAI\@.'' \url{https://www.linkedin.com/posts/dr-barry-scannell-bbb5aa207_in-a-major-ruling-for-european-copyright-share-7393957246386323457-8bbx}. \emph{Dream Machine} Issue~7.} \textbf{Suno} was sued by music-rights groups under a banner the litigators called ``the biggest theft in music history.''\footnote{\emph{EDM.com}, ```Biggest Theft in Music History': Rights Group Sues Suno as AI Music Showdown Escalates.'' \url{https://edm.com/gear-tech/rights-group-sues-suno-copyright-infringement/}. \emph{Dream Machine} Issue~7.} \textbf{Wixen Music Publishing} filed a \$50m copyright suit against Meta.\footnote{\emph{Music Business Worldwide}, ``Wixen files \$50m copyright suit against Meta.'' \url{https://www.musicbusinessworldwide.com/wixen-files-50m-copyright-suit-against-meta-claims-tech-giant-wants-to-replace-songwriters-with-ai/}. \emph{Dream Machine} Issue~16.} \textbf{Universal Music Group} filed a \$3B suit against Anthropic.\footnote{\emph{Dream Machine} Issue~17 reportage on UMG's \$3B suit against Anthropic.} By the end of February 2026, the lawsuits were no longer an interesting subplot. They were the main mechanism through which the new creative economy was being defined.

Some of it was platform policy. \textbf{Bandcamp} banned AI-generated music outright in January 2026.\footnote{\emph{Stereogum}, ``Bandcamp bans AI music.'' \url{https://stereogum.com/2485199/bandcamp-bans-ai-music/news}. \emph{Dream Machine} Issue~14.} \textbf{Deezer} built and licensed an AI-music detection tool to other platforms.\footnote{\emph{Dream Machine} Issue~18 reportage of Deezer licensing its detection tool.} \textbf{Spotify} declined to add an AI-music filter, preferring transparency and labelling.\footnote{\emph{TechRadar}, ``AI music is flooding Spotify, and subscribers are furious.'' \url{https://www.techradar.com/audio/spotify/ai-music-is-flooding-spotify-and-subscribers-are-furious-heres-why-music-fans-no-longer-trust-discover-weekly}. \emph{Dream Machine} Issue~14.} \textbf{San Diego Comic-Con} banned AI art at its 2026 event.\footnote{\emph{CNET}, ``San Diego Comic-Con Draws a Line: No AI Art Allowed at 2026 Event.'' \url{https://www.cnet.com/culture/san-diego-comic-con-bans-ai-art-for-2026-event/}. \emph{Dream Machine} Issue~16.} \textbf{Sweden's official music chart} banned AI-generated entries.\footnote{\emph{The Independent}, ``AI-generated song banned from Swedish charts: `It's deceiving'.'' \url{https://www.independent.co.uk/tv/news/ai-music-song-banned-sweden-spotify-b2901627.html}. \emph{Dream Machine} Issue~15.}

The point I want to make about all of this is that the cultural pushback is not~-- as the more dismissive coverage tends to frame it~-- a Luddite reaction. It is not an irrational allergy to new technology. It is a \emph{market response}. The audience has spoken with its attention. The platforms are reacting to the audience. The labels are reacting to the platforms. The lawyers are reacting to the labels. The artists are reacting to the lawyers. The whole system is, in slow motion, \emph{renegotiating the terms on which synthetic creative work is allowed to participate in the public sphere.}

That renegotiation is the actual story. The viral AI hits, the ones that get the magazine covers, are footnotes.

\section*{What the ceiling is made of}

I want to try, in the last part of this chapter, to articulate what the slop ceiling actually \emph{is}~-- what cognitive, cultural, economic mechanism produces the 44-to-3 ratio~-- because I think understanding that mechanism is the difference between thinking it will hold and thinking it will eventually erode.

My working hypothesis, after six months of looking at the data, is that the ceiling is made of four overlapping things:

\textbf{One.} \emph{A cognitive distinction the audience can't articulate but can feel.} The Deezer/Ipsos finding~-- that 97\% of listeners can't pick AI from human in a blind test, but that revealed-AI tracks underperform~-- suggests the distinction is below conscious recognition but above zero. The audience knows when something doesn't matter, in some way they can't quite name.

\textbf{Two.} \emph{A status-signal collapse.} Music, film, advertising and podcasting are, in significant part, status goods. Telling your friends you discovered an exciting new artist is part of why people pay attention to artists. AI artists, by being mass-produced and machine-authored, fail the status test at the structural level. There is no way to \emph{signal cultural insiderness} by being the first to discover Xania Monet, because Xania Monet was discovered by 800,000 people in the same week, all of whom found out via the same press cycle.

\textbf{Three.} \emph{A meaning vacuum.} Most of the slop is produced for content marketing, SEO and ad placement reasons, not because anyone needs to express anything. The audience can tell. As one \emph{Digital Music News} headline from January 2026 put it, in a phrase I have written down in my notebook and used in talks: \emph{``A.I.-generated music is catchy, familiar\ldots\ and boring.''}\footnote{\emph{Soultracks}, ``A.I.-generated music is catchy, familiar\ldots\ and boring.'' \url{https://soultracks.com/news-ai-generated-music-is-catchy-boring/}. \emph{Dream Machine} Issue~14.} The Swedish Top Chart's reasoning when it banned an AI-generated track from its rankings in January 2026 said the same thing in different words: \emph{``The song is great, but unfortunately, it's missing one of the most important ingredients, which is emotion.''}\footnote{\emph{The Independent}, ``AI-generated song banned from Swedish charts: `It's deceiving'.'' \url{https://www.independent.co.uk/tv/news/ai-music-song-banned-sweden-spotify-b2901627.html}. \emph{Dream Machine} Issue~15.} The technical capability is there. The reason for the work to exist is not.

\textbf{Four.} \emph{Reciprocity.} And this is the one I am least confident about, but find the most interesting. There is, in almost every long-running creative relationship between an artist and an audience, an implicit reciprocity. The audience pays attention because the artist has \emph{paid the price of making the work}~-- has practised, has struggled, has lived, has earned the right to be heard. AI artists short-circuit that contract. They produce the output without paying the price. The audience, at some level, refuses the trade.

Take any of those four mechanisms away and the ceiling might drop. Take all four away and we'd be in trouble. None of them, on the current evidence, is going away.

\section*{The Authenticity Premium: pricing the ceiling}

The slop ceiling, as I have described it so far, is the \emph{negative} side of the dynamic~-- the work the audience refuses to engage with. There is a \emph{positive} side I want to spend a section on, because it is the side that pays the bills, and the side the rest of this book leans on for its strategic argument.

I have come to call the positive side the \textbf{Authenticity Premium}: the measurable excess of attention, willingness to pay, and cultural credit that audiences allocate to creative work whose human authorship can be verified. If the slop ceiling tells you what the audience \emph{will not engage with}, the Authenticity Premium tells you what they \emph{will pay extra for}. Both numbers are produced by the same underlying audience behaviour. Both are market findings. Both have stabilised across the six months of newsletter coverage that this book is built on.

Let me try to sketch the empirical picture.

On the \emph{music} side, the cleanest evidence is the inverse of the Deezer 44-to-3. Independent labels, working musicians and the back-catalogue of clearly-human-authored recorded music have, by Deezer's own published April 2026 statement, retained a \emph{disproportionate share of total listening hours} relative to their share of the upload volume. The artists with verified human-authorship signal~-- touring records, named producers, signed labels, press-covered releases~-- capture a meaningful premium in playlist placements, editorial coverage, and per-stream consumption versus the long tail of synthetic uploads. The premium is, on Deezer's own framing, a function of \emph{audience-chosen} listening rather than algorithmic placement. The 1-to-3\% of streams that the AI-music flood receives is, in operational terms, the \emph{floor} of the slop ceiling; the corresponding 97-to-99\% that named human artists capture is the \emph{ceiling-level} the Authenticity Premium underwrites.

On the \emph{advertising} side, \emph{Marketing Week}'s ``You can't dismiss AI ads as slop when they're winning in testing'' piece~-- which I quote at length in Chapter~\ref{ch:12}~-- found that \emph{care-driven} AI work could win creative-effectiveness tests, but \emph{only when the production process foregrounded human creative intent.}\footnote{\emph{Marketing Week}, ``You can't dismiss AI ads as slop when they're winning in testing.'' Coverage discussed in \emph{Dream Machine} Issue~22.} The same publication's reporting on the McDonald's Netherlands, Valentino and Coca-Cola AI campaigns documented that \emph{cynical} AI work, by contrast, produced sharply negative effectiveness scores. The price of being a \emph{cynical} AI advertiser, in commercial-effectiveness terms, was a measurable revenue penalty. The price of being a \emph{sincere} AI advertiser was a measurable premium. Both numbers come from the same audience exposed to the same kind of work.

On the \emph{film and television} side, the Authenticity Premium has been visible in awards-season behaviour, in the \emph{Marvel-era IP fatigue} data, and in the audience response to specific high-profile AI integrations. Films whose AI use was \emph{disclosed in advance and framed as part of the work's identity}~-- \emph{Watch the Skies}, \emph{Lily}, Andrii Daniels' bomb-shelter Christmas clip, \emph{Dear Upstairs Neighbors}, \emph{Synthetic Sincerity}~-- have, on aggregate, received better critical reception and stronger audience engagement than films whose AI use was \emph{concealed and later revealed}. The premium is, structurally, paying for \emph{transparency about authorship} rather than for \emph{authorship purity}. The audience is not refusing AI involvement. The audience is refusing \emph{deception} about AI involvement.

On the \emph{games} side, the Quantic Foundry survey I referenced earlier and the parallel work by IGN and PC Gamer through autumn 2025 showed an effect that is, in commercial terms, more direct. Games that disclosed AI involvement in \emph{audience-visible roles} (NPC dialogue, generated quests, voice-acting replacement) faced specific consumer pushback in Steam reviews and Metacritic user scores. Games that disclosed AI involvement in \emph{invisible-pipeline roles} (asset generation under art-direction supervision, QA automation, localisation) faced no measurable consumer pushback at all. The premium, in games, is paying for \emph{AI being kept off the audience-visible parts of the work}. Studios that have understood this calibration~-- Larian, Hooded Horse, Jagex, the \emph{Position Three} signatories of Chapter~\ref{ch:7}~-- are operating with the Authenticity Premium as a deliberate marketing position.

The premium has a \emph{commercial} shape that I want to name explicitly, because I think working creatives are systematically under-pricing it in 2026. The price difference between \emph{clearly-human-authored} and \emph{clearly-AI-authored} creative work, on the available evidence, is not a marginal one. It is, in some segments, an \emph{order-of-magnitude} difference. Xania Monet's \$3M Hallwood Media deal is the \emph{upper bound} of what the synthetic-artist market can pay for a top-tier AI act. The \emph{touring-artist} economics of even a mid-tier independent musician with a verified live following easily exceed that over a five-year career. The same comparison holds across film and games. The Authenticity Premium is, in market-pricing terms, the \emph{commercial gap} between the synthetic and the human. It is not small. It is, for working creatives positioning themselves above the slop ceiling, the actual business case.

There are two empirical caveats I want to put on the page, because the book should not over-claim.

\emph{One,} the premium is \emph{unevenly distributed}. It accrues, by my reading of the 2025--26 data, most strongly to artists at the \emph{top} (premium auteur cinema, signed-label musical artists, name-brand designers) and at the \emph{long-tail} (hyperlocal, culturally-specific, niche-community-serving creators). It accrues weakly, if at all, to the \emph{middle}~-- the mid-career journeyman creatives who have built careers on producing competent-but-mean-of-the-distribution work. The middle is, as I argue in Chapter~\ref{ch:7}'s legacy-vulnerability section, the segment most exposed to AI substitution. The Authenticity Premium does not protect everyone equally.

\emph{Two,} the premium is \emph{dependent on the provenance infrastructure} I described in Chapter~\ref{ch:4}. The audience can pay a premium for verifiable human authorship only to the extent that authorship is \emph{verifiable}~-- to the extent that the C2PA chains, the SynthID watermarks, the platform-level disclosure norms and the institutional certifications make the signal reliable. The premium and the provenance stack are, structurally, \emph{the same project seen from different angles}. The premium pays for the signal. The signal is produced by the infrastructure. The infrastructure needs the premium to fund itself.

The Authenticity Premium is, in operational summary, \emph{the audience paying the bill for the Living Web.} Whether the bill stays paid, over the next five years, depends on whether the infrastructure that produces the signal stays trustworthy. That, in turn, depends on the policy, platform and union work this book has been describing.

\section*{The corollary}

I want to close this chapter with a corollary, because it has implications for the rest of the book.

If the slop ceiling is real, and if it is structural, and if it is going to hold~-- and I think the burden of evidence at this point is on the people who say otherwise~-- then the strategic question for everyone in the creative industries is not \emph{how do we compete with the flood?} It is \emph{how do we sit above the ceiling?}

That question has different answers in different sectors. For a working musician, it might mean leaning into the irreducibly human parts of the work~-- live performance, personal relationship with audience, transparent process. For a working filmmaker, it might mean making the \emph{kind} of film whose value depends on being knowably authored by knowable people. For a working games studio, it might mean~-- as Jagex, Larian, Games Workshop, Hooded Horse and an increasing list of studios have explicitly said~-- taking \emph{generative AI off the table} as a public commitment to the audience.

None of these are anti-AI positions. They are \emph{above-the-ceiling} positions. They take seriously the fact that the world is now full of cheap synthetic content and ask: \emph{what is the work that the synthetic content can't do?}

That is the question that organises the rest of the book. The slop ceiling is the negative space against which everything interesting in creative work for the next ten years is going to be defined.

  \chapter{The 88 Per~Cent}\label{ch:6}

\lettrine[lines=3,lhang=0.15,findent=0.1em]{O}{n} 15~December 2025, the UK government quietly laid a document before Parliament that I think will be remembered, ten years from now, as a more important moment in the history of creative AI than any single tool release in 2025 or 2026.

The document was the \emph{Statement of Progress on Copyright and Artificial Intelligence}, prepared by the Department for Science, Innovation and Technology.\footnote{UK Department for Science, Innovation and Technology, \emph{Statement of Progress on Copyright and AI}, 15~December 2025. \url{https://www.gov.uk/government/publications/copyright-and-artificial-intelligence-progress-report/copyright-and-artificial-intelligence-statement-of-progress-under-section-137-data-use-and-access-act}. \emph{Dream Machine} Issue~12, ``Editor's Pick: 88\% of Creators Said `No'.'' 18~December 2025.} It was a stocktaking report~-- not a final policy, not new legislation, not a decision. It was a ``where we are'' note, eleven months after the closing of one of the largest copyright consultations the United Kingdom has ever run.

The consultation had been open from 17~December 2024 to 25~February 2025. The government had proposed four options for how UK copyright law should treat AI training:\footnote{UK DSIT, original consultation, 17~December 2024~-- 25~February 2025. Discussion in IPWatchdog, ``Respondents to UK AI Consultation Overwhelmingly Want AI Companies to License Copyrighted Works in All Cases.'' \url{https://ipwatchdog.com/2025/12/16/respondents-uk-ai-consultation-overwhelmingly-want-ai-companies-license-copyrighted-works-all-cases/}.}

\begin{itemize}
  \item \textbf{Option~0}, do nothing.
  \item \textbf{Option~1}, require AI companies to licence copyrighted works for training in all cases.
  \item \textbf{Option~2}, a narrower set of exceptions, with conditions.
  \item \textbf{Option~3}~-- the government's \emph{preferred} option~-- a broad text-and-data-mining exception with an opt-out for rightsholders.
\end{itemize}

Eleven and a half thousand people replied.

Of the 10,112 responses submitted through the government's \emph{Citizen Space} online portal~-- the subset for which the government published quantitative breakdowns:

\begin{itemize}
  \item \textbf{88\%} supported Option~1~-- licensing in all cases.
  \item \textbf{7\%} supported Option~0~-- do nothing.
  \item \textbf{3\%} supported the government's preferred Option~3.\footnote{IPWatchdog, \emph{op.\ cit.}; Hogan Lovells, ``Copyright and AI: UK government publishes statement of progress.'' \url{https://www.hoganlovells.com/en/publications/copyright-and-ai-uk-government-publishes-statement-of-progress}.}
\end{itemize}

\begin{figure}[htbp]
  \centering
  \begin{tikzpicture}[
    font=\small,
    every node/.style={align=center}
  ]


    \def\barheight{1.2}
    \def\bary{0}

    \fill[darkblue, rounded corners=3pt]
      (0, \bary) rectangle (8.8, \barheight);

    \fill[chaptergrey]
      (8.8, \bary) rectangle (9.1, \barheight);

    \fill[lightgrey, rounded corners=3pt]
      (9.1, \bary) rectangle (10.0, \barheight);

    \draw[chaptergrey!60, thin, rounded corners=3pt]
      (0, \bary) rectangle (10.0, \barheight);

    \draw[white, thin] (8.8, \bary) -- (8.8, \barheight);
    \draw[white, thin] (9.1, \bary) -- (9.1, \barheight);

    \node[text=white, font=\large\bfseries] at (4.4, 0.6) {88\%};


    \draw[darkblue, thin] (4.4, \bary) -- (4.4, -0.5);
    \node[darkblue, font=\small\bfseries, anchor=north] at (4.4, -0.55)
      {88\% -- Support licensing\\as default};

    \draw[chaptergrey, thin] (8.95, \bary) -- (8.95, -0.5);
    \node[chaptergrey, font=\scriptsize\bfseries, anchor=north] at (8.95, -0.55)
      {3\%\\opt-out};

    \draw[chaptergrey!70, thin] (9.55, \bary) -- (9.55, -0.5);
    \node[chaptergrey!70, font=\scriptsize, anchor=north] at (9.55, -0.55)
      {9\%\\other};

    \node[chaptergrey, font=\small, anchor=west] at (0, 1.75)
      {Total respondents: \textbf{11,514}};
    \node[chaptergrey!80, font=\scriptsize, anchor=west] at (0, 1.35)
      {of whom 10,112 responded via the Citizen Space portal};

    \foreach \x/\lbl in {0/0\%, 2.5/25\%, 5.0/50\%, 7.5/75\%, 10.0/100\%}{
      \draw[chaptergrey!50, thin] (\x, \bary) -- (\x, \bary-0.15);
      \node[chaptergrey!70, font=\scriptsize] at (\x, -1.65) {\lbl};
    }

  \end{tikzpicture}
  \caption{UK DSIT Copyright and AI Consultation results (December 2025): 88\% of respondents supported mandatory licensing.}
  \label{fig:88-percent}
\end{figure}
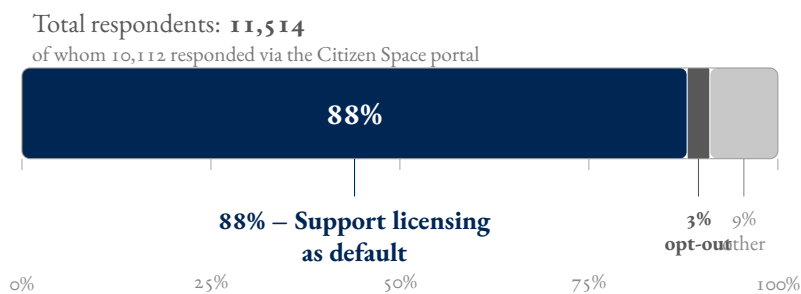

I want you to look at those numbers again, because they are the single most concrete thing this book has to offer in defence of the argument I will be making in the second half of it: that the creative economy is not waiting to be told what it thinks about AI.

Eighty-eight per cent. In a country with no compulsory voting, no organised industry mobilisation comparable to the music or film unions' rapid responses to specific provocations, no celebrity-led campaign on the scale of the SAG-AFTRA strike: 88\% of the people who took the time to write to their government about how their work should be used said, \emph{licence it}. Pay for it. Don't take it.

That number is the \emph{true} watershed of the period this book covers. The Tilly Norwood week made it possible. The 88\% made it permanent.

This chapter is about how a global creative coalition~-- half informal, half deliberate, half union-led, half artist-led, half lawyer-led, all of it networked~-- went from a few scattered protest statements in October 2025 to a structural force in policy and law by May 2026.

\section*{Why eighty-eight per cent}

It is easy, looking at a number that big, to assume it represents some kind of organised lobbying effort. To assume that the AI companies' opt-out proposal was so unpopular that the response was a co-ordinated push from a few large interest groups, who marshalled their members into the consultation.

The actual composition, as analysed in the December 2025 Statement of Progress, was mixed. There were submissions from creators in every major creative discipline~-- writers, musicians, filmmakers, photographers, illustrators, designers, journalists. There were submissions from professional bodies (the Society of Authors, the Association of Photographers, the Authors' Licensing and Collecting Society). There were submissions from individual citizens with no industry affiliation, who simply objected on principle to having their work~-- their LinkedIn posts, their family photos, their blogs~-- pulled into a training set without their consent.\footnote{UK DSIT, \emph{Statement of Progress}, \emph{op.\ cit.}; analysis at UCL Copyright Queries, ``UK government publishes progress statement on AI and copyright consultation.'' \url{https://blogs.ucl.ac.uk/copyright/2025/12/23/uk-government-publishes-progress-statement-on-ai-and-copyright-consultation/}.}

There were submissions from the AI companies too. The progress report notes that the \emph{3\%} who supported the government's preferred Option~3 were ``particularly concentrated among AI developers and large technology companies.''\footnote{UK DSIT, \emph{Statement of Progress}, \emph{op.\ cit.}} This is~-- and I am being careful about how I phrase this, because the Statement of Progress is itself careful~-- \emph{not} a description of a balanced industry view. It is a description of a public consultation in which the people most affected by the proposed policy said one thing, and the companies the policy was designed to enable said the opposite.

The 88\% is not a curiosity. It is a \emph{vote}, in the most literal sense. The creators of the United Kingdom were given a structured chance to say what they wanted, and 88\% of them said the same thing.

The Society of Authors' submission, which I have read in full, made the underlying argument with the kind of clarity that the policy debate had been avoiding for two years. \emph{``If we are to see an end to the industrial-scale theft of writers' and other creators' work, and to protect the creators and creative industries of the future, then UK copyright needs to be enforced not weakened.''}\footnote{Society of Authors submission to the UK consultation, quoted in IPWatchdog, \emph{op.\ cit.}} That sentence~-- \emph{industrial-scale theft, enforced not weakened}~-- set the rhetorical register that the next six months of the policy debate ran on.

\section*{The pattern repeats}

I think a lot of the international coverage of the UK consultation has under-emphasised that the 88\% was not a UK-only phenomenon. It was the first formal expression of a pattern that was, in the same six months, repeating in every jurisdiction that gave its creators a meaningful chance to speak.

In \textbf{Germany}, the music rights society \textbf{GEMA} sued OpenAI in the Munich Regional Court over the training of large language models on copyrighted music lyrics. In November 2025, the court ruled for GEMA in a decision that intellectual-property lawyers across Europe~-- including Dr Barry Scannell, whose detailed LinkedIn breakdown of the ruling I have read more times than I will admit~-- described as a \emph{major} precedent for European copyright law.\footnote{Dr Barry Scannell, LinkedIn analysis of GEMA v.\ OpenAI ruling, November 2025. \url{https://www.linkedin.com/posts/dr-barry-scannell-bbb5aa207_in-a-major-ruling-for-european-copyright-share-7393957246386323457-8bbx}. \emph{Dream Machine} Issue~7.}

In \textbf{the United States}, a coalition of music rights organisations sued \textbf{Suno}, with the press release describing the action, in a phrase the litigators clearly knew would travel, as ``the biggest theft in music history.''\footnote{\emph{EDM.com}, ```Biggest Theft in Music History': Rights Group Sues Suno as AI Music Showdown Escalates.'' \url{https://edm.com/gear-tech/rights-group-sues-suno-copyright-infringement/}. \emph{Dream Machine} Issue~7.} \textbf{Wixen Music Publishing} filed a \$50m copyright suit against \textbf{Meta} in January 2026.\footnote{\emph{Music Business Worldwide}, ``Wixen files \$50m copyright suit against Meta, claims tech giant wants to replace songwriters with AI\@.'' \url{https://www.musicbusinessworldwide.com/wixen-files-50m-copyright-suit-against-meta-claims-tech-giant-wants-to-replace-songwriters-with-ai/}. \emph{Dream Machine} Issue~16.} \textbf{Universal Music Group} filed a \$3B suit against \textbf{Anthropic}.\footnote{\emph{Dream Machine} Issue~17, on UMG's \$3B suit against Anthropic.} \textbf{The Johnny Cash estate} sued Coca-Cola under the \textbf{ELVIS Act}~-- Tennessee's new AI-impersonation law~-- for using a Cash sound-alike in a tribute-act advertisement.\footnote{\emph{Complete Music Update}, ``Johnny Cash estate uses ELVIS Act to sue Coke over tribute act ad soundtrack.'' \url{https://completemusicupdate.com/johnny-cash-estate-uses-elvis-act-to-sue-coke-over-tribute-act-ad-soundtrack/}. \emph{Dream Machine} Issue~9.} By the spring of 2026, the litigation landscape was so dense that \emph{Music Business Worldwide} was running weekly summary columns just to keep track of which cases were still active.

In \textbf{the European Union}, lawmakers tabled a bill in November 2025 seeking an EU-wide minimum age to access AI chatbots and social media, an early acknowledgement that the regulatory question was not just about copyright but about the wider integration of AI into the social fabric.\footnote{Reuters, ``European lawmakers seek EU-wide minimum age to access AI chatbots, social media.'' \url{https://www.reuters.com/legal/litigation/european-lawmakers-seek-eu-wide-minimum-age-access-ai-chatbots-social-media-2025-11-26/}. \emph{Dream Machine} Issue~9.}

In \textbf{the United States}, the actors' union \textbf{SAG-AFTRA}, riding the wave of the Tilly Norwood backlash, opened negotiations in October 2025 that resulted by spring 2026 in \emph{significantly stronger AI protections} in its next four-year contract~-- a deal that included new consent requirements, residuals, and what the trade press began calling, informally, the ``Tilly Tax'' on the use of AI actors.\footnote{SAG-AFTRA contract update reporting through Q2 2026. \emph{Dream Machine} Issues~20, 26, 29. Coverage: \url{https://www.theverge.com/news/842848/new-york-law-ai-advertisements-sag-aftra-labor}.}

In \textbf{the United Kingdom}, the UK actors' union \textbf{Equity} held a strike ballot in December 2025 over AI scanning of performers' likenesses; the result came back in a 99\% landslide in favour of industrial action. The ballot question itself, in its plain language, captured the substance of what was at stake: \emph{``Are you prepared to refuse to be digitally scanned on set to secure AI protections?''}\footnote{Equity (UK), ``Performers prepared to take industrial action over AI in landslide 99\% vote.'' \url{https://www.equity.org.uk/news/2025/performers-prepared-to-take-industrial-action-over-ai-in-landslide-99-vote}. \emph{Dream Machine} Issue~12.} By January 2026 the union had secured what its general secretary called ``an improved offer'' from producers on AI protections in film and TV negotiations.\footnote{Equity (UK), ``Equity welcomes improved offer in AI protection negotiations in film and TV\@.'' \url{https://www.equity.org.uk/news/2026/equity-welcomes-improved-offer-in-ai-protection-negotiations-in-film-and-tv}. \emph{Dream Machine} Issue~15.} In May 2026, the broader \textbf{AI Disclosure Standard} for the film industry was launched at the \textbf{Cannes Film Festival}.\footnote{Cannes Film Festival AI Disclosure Standard launch, May 2026. \emph{Dream Machine} Issue~29.} In the same week, the \textbf{British Phonographic Industry (BPI)} issued a formal set of \emph{transparency and sovereignty} demands aimed at the music side of the same settlement~-- a structured industry position designed, in the BPI's own framing, to \emph{secure} the ``AI licensing boom'' rather than leave it to bilateral negotiation between platforms and rights-holders one model at a time.\footnote{\emph{Musically}, ``BPI sets out transparency and sovereignty demands to secure `AI licensing boom'.'' \url{https://musically.com/2026/05/19/bpi-transparency-sovereignty-ai-licensing-boom/}. \emph{Dream Machine} Issue~30.}

The pattern, in every jurisdiction and across every part of the creative economy, was the same. Where creators were given a procedural mechanism~-- a consultation, a strike ballot, a contract negotiation, a class action~-- they used it. They turned up in numbers. They voted, in their structured way, against the unconditional appropriation of their work. And they won enough of these procedural battles that, by the time the spring of 2026 arrived, the \emph{terms of engagement} for AI in the creative industries had been substantially re-set in a six-month window.

\section*{The first reversal}

The most surprising single event in the entire policy arc was the UK government's own \emph{reversal} in spring 2026.

The Statement of Progress in December had already softened the official position. Where the original consultation had proposed Option~3~-- the text-and-data-mining exception with opt-out~-- as the \emph{preferred} outcome, the December update simply described the government as ``working with 50+ experts from across music, film, games and AI to figure out what comes next.''\footnote{UK DSIT, \emph{Statement of Progress}, \emph{op.\ cit.}} The opt-out language was gone.

By \textbf{March 2026}, the position had reversed further. The government's final report on copyright and AI, laid before Parliament by the statutory deadline of 18~March 2026, walked back the original preference for Option~3 in favour of a much more cautious set of proposals that acknowledged the 88\% finding.\footnote{\emph{Dream Machine} Issue~21, 19~March 2026, on the UK government's revised position on AI copyright.} \emph{Dream Machine} Issue~21, dated 19~March 2026, was the first edition where I noticed the change in tone in the government's own language. The framing had shifted from ``how do we enable AI training'' to ``how do we protect creators.''

I want to be precise about what this reversal means and what it doesn't.

It does \emph{not} mean that the UK has banned AI training on copyrighted work, or that it has imposed a licensing-first regime by default. As of the time I am writing this~-- May 2026~-- the legislative process is ongoing, and the eventual policy could land anywhere on a wide spectrum.

It \emph{does} mean that 88\% of 10,112 people, plus a thousand-odd email submissions, plus a media cycle that ran for fourteen months, plus a parallel set of legal proceedings, plus a parallel set of platform and industry pushback, plus the active mobilisation of multiple professional bodies, was enough to \emph{change the position of a national government} on one of the most economically significant technology-policy questions of the decade.

That is, in democratic terms, what \emph{working} looks like.

\section*{What the creators were actually saying}

The 88\% was a procedural answer to a procedural question. What were the \emph{substantive} arguments behind it? I have read enough of the submissions, through the published summaries and through the secondary press coverage, to feel confident in summarising the three I see most often.

\textbf{The consent argument.} This was the simplest and the most universal: that work made by a creator~-- a song, a book, a photograph~-- belongs to that creator in a way that is \emph{not} fully captured by the existing copyright regime, and that the use of that work to train a machine learning model is a use that requires the creator's consent.

The argument is not new. The Berne Convention has, since 1886, treated authorship as a \emph{moral right} in addition to an economic one. What is new is the scale of the use. A single AI training run can ingest the work of millions of human creators in a way that no single buyer, publisher, broadcaster or aggregator has ever done. The procedural mechanisms of copyright were designed for a world where uses were enumerable. They struggle in a world where the use is, in effect, \emph{the entire creative output of a generation, all at once.}

\textbf{The attribution argument.} This was the most operationally specific: that when AI systems produce derivative outputs based on training data, the creators whose work shaped those outputs should be identifiable, and where appropriate, compensated. \emph{Musical AI}, a startup that raised \$4.5m in January 2026 on a ``creative weight attribution'' model, described the technical version of this as ``calculating each input's actual contribution to a generative model's output, then licensing accordingly.''\footnote{\emph{Digital Music News}, ``The AI Licensing Shift~-- Creative Weight Attribution Emerges as Music Industry Game-Changer for Rights Holders.'' \url{https://www.digitalmusicnews.com/2026/01/26/ai-licensing-shift-creative-weight-attribution/}. See also \emph{Digital Music News}, ``Artificial Intelligence Attribution and Licensing Startup Musical AI Scores \$4.5 Million Raise.'' \url{https://www.digitalmusicnews.com/2026/01/13/musical-ai-funding-january-2026/}. \emph{Dream Machine} Issues~14, 16.} The argument doesn't require AI training to stop. It requires it to \emph{show its workings.}

\textbf{The economic argument.} This was the most cynical and the most powerful: that AI systems trained on the unpaid labour of creators will eventually substitute for those creators in the market, and that the failure to licence is therefore not just an \emph{ethical} offence~-- it is an active transfer of wealth from a relatively diffuse group of working creatives to a relatively concentrated group of technology platforms and their shareholders.

The PRS for Music \emph{2026 AI Survey} found that \textbf{four in five} music creators worried about AI-generated music competing with human-created music in the streaming economy.\footnote{PRS for Music, ``PRS for Music AI Survey 2026.'' \url{https://www.prsformusic.com/m-magazine/news/prs-for-music-ai-survey-2026}. \emph{Dream Machine} Issue~16.} The Edinburgh-based \textbf{Centre for Creative AI} at UCL/RCA, launched in late 2025, explicitly framed its mission around the ``redistribution of value from machines back to the humans whose work made them possible.''\footnote{\emph{Broadcast Now}, ``Alex Mahon joins Stellar AI Creative Summit line-up'' (covering the launch of the UCL/RCA Centre for Creative AI). \url{https://www.broadcastnow.co.uk/broadcasters/alex-mahon-joins-stellar-ai-creative-summit-line-up/5209227.article}. \emph{Dream Machine} Issue~1.} The US \textbf{artist trade body} quoted in \emph{Complete Music Update} in November 2025 was even blunter: ``Artists must have creative control in AI deals or risk ending up with `scraps'.''\footnote{\emph{Complete Music Update}, ``Artists must have creative control in AI deals or risk ending up with `scraps', says US artist trade body.'' \url{https://completemusicupdate.com/artists-must-have-creative-control-in-ai-deals-or-risk-ending-up-with-scraps-says-us-artist-trade-body/}. \emph{Dream Machine} Issue~6.}

Stack those three arguments next to each other and you get a recognisable shape. It is the shape of every economic-rights argument creators have made, in every previous technological transition, going back to the Stationers' Company in seventeenth-century London. \emph{Don't print without permission. Don't broadcast without a fee. Don't sell our records without paying us. Don't sample without clearing. Don't stream without licensing. Don't train without consent.}

The 88\% is the latest entry in a four-hundred-year sequence. What's new is the \emph{speed} with which it has had to be expressed, and the \emph{scale} of the use it is responding to.

\section*{The artists' declaration}

In January 2026, in parallel with the union negotiations and the lawsuits and the policy responses, nearly \textbf{800 working creatives}~-- including high-profile names like Jason Aldean and OneRepublic~-- signed an open declaration with the line that gave the document its name: \emph{Stealing Our Work Is Not Innovation.}\footnote{\emph{Digital Music News}, ``Nearly 800 Creatives, Including Jason Aldean and One Republic, Sign Responsible AI Declaration~-- `Stealing Our Work Is Not Innovation'.'' \url{https://www.digitalmusicnews.com/2026/01/22/stealing-isnt-innovation/}. \emph{Dream Machine} Issue~16.}

I want to spend a moment on this document because it is the cleanest expression I have found of the underlying argument, and because I think the line will be on a t-shirt within a year if it isn't already.

The declaration was not a legal document. It had no enforcement mechanism. It did not call for specific legislation. It was a \emph{cultural statement}~-- a refusal of the framing under which the AI companies had been making their case.

The framing the AI companies had been using, repeatedly, in venues from technology conferences to court filings, was that training models on copyrighted material was a kind of \emph{technical inevitability}~-- that machine learning required vast amounts of data, that the data could not practically be licensed at scale, and that therefore the use was, in a sense, \emph{outside} the traditional consent-and-payment framework of copyright. It was~-- they argued~-- not really ``use'' in the sense the law had been built around. It was a new kind of activity that needed a new kind of rules.

The declaration's response, in a phrase, was: \emph{no, it's just stealing.}

This was a rhetorically devastating move. It collapsed the AI companies' carefully constructed framing~-- \emph{transformative use, fair use, technical necessity, innovation}~-- into the oldest accusation in commerce, and made it stick. \emph{Stealing.} Not because the signatories did not understand the technical arguments. They did. Because they had decided that the technical arguments were a \emph{cover} for an underlying transfer of value that didn't deserve any other name.

Once you have that framing, the whole policy debate looks different. \emph{Should we allow innovation?} becomes \emph{should we allow theft?} The answers are not the same.

\section*{The levy precedent: why Petrillo matters here}

I want to take a long detour, because the 88\%~-- and the institutional response forming around it~-- is, on the historical reading I laid out in Chapter~\ref{ch:2}, a \emph{Petrillo-template} moment that the trade press has, in the main, declined to recognise as such.

Let me state the template again, in its cleanest form, because the rest of this section relies on it.

When James Caesar Petrillo, the president of the American Federation of Musicians, took on the recording industry in 1942 and again in 1948~-- staging the recording bans that effectively shut down the entire commercial output of American recorded music for the better part of three years~-- the strategic move was not, in essence, \emph{prohibition}. Petrillo was not trying to ban records. He was trying to \emph{tax} records. The 1942 settlement created a per-record royalty paid into an AFM unemployed-musicians fund. The 1948 settlement, after the Taft-Hartley Act outlawed the 1942 structure, created the \textbf{Music Performance Trust Fund} under Section~302~-- a \emph{jointly-administered} labour--management fund, paid into by the labels and broadcasters, used to subsidise free live music performances by working musicians, distributing the productivity gain of the new recording technology to the displaced labour pool. The MPTF still exists. It still distributes payments today. It is, on a hundred years of evidence, the \emph{only} form of institutional response to a creative-technology displacement that has worked at structural scale.

The four parts of the template, again:

\emph{One,} the displacing technology is not banned. It is allowed to displace.

\emph{Two,} the platform owner pays an \emph{ongoing per-unit tribute} to the displaced labour pool.

\emph{Three,} the tribute is collected \emph{centrally,} by a joint labour--management body, not negotiated individual-by-individual.

\emph{Four,} the tribute is paid out to subsidise the \emph{displaced creative practice itself}~-- live music, in Petrillo's case~-- keeping it alive as a category even as the market for it shrinks.

I want to show how the 88\%~-- and the architecture of institutional response coalescing around it in spring 2026~-- is, function by function, a \emph{reconstruction} of the Petrillo template for the AI era.

\emph{One,} none of the institutional responses I have catalogued in this chapter~-- the UK consultation's licensing-by-default proposal, the SAG-AFTRA Tilly Tax, the \emph{Stealing Our Work Is Not Innovation} declaration, the GEMA ruling, the Cannes Disclosure Standard~-- is, in essence, a \emph{ban}. The declaration's signatories are not asking for AI to be prohibited. The 88\% of UK respondents who wanted licensing-in-all-cases were not asking for AI training to be banned. They were asking for it to be \emph{licensed}~-- which is, by definition, an acknowledgement that the underlying activity will continue. This matches Petrillo's first principle.

\emph{Two,} what the 88\%, the GEMA ruling and the \emph{UMG v.\ Anthropic} settlement framework are collectively asking for is a \emph{per-output tribute} from the AI platforms to the creative-labour pool whose work was used in training. The mechanism is, structurally, identical to the per-record royalty that Decca and Columbia agreed to pay AFM in 1944. The platform pays. The labour pool receives. The amount is calibrated to the volume of platform output. The mechanism is the Petrillo mechanism.

\emph{Three,} the structural innovation of the Petrillo settlement~-- collection through a \emph{joint body} rather than through individual-creator negotiation~-- is, in spring 2026, only partially built for the AI era. The collective-licensing infrastructure for music (PRS, GEMA, ASCAP, BMI, SIAE, JASRAC and the related international bodies) has, in some cases, started negotiating directly with the AI platforms on the per-output structure. The Musical AI \emph{creative weight attribution} infrastructure is a first attempt to build a \emph{technical} layer underneath the joint-body political layer. The Cannes Disclosure Standard is an industry-co-ordination mechanism for the production-side disclosure that the collection mechanism rests on. None of this is finished. The joint bodies for \emph{visual} artists, \emph{writers}, \emph{games developers}, \emph{photographers} are at much earlier stages of development. The MPTF-equivalent fund-and-distribution mechanism does not yet exist for most of the creative industries. \emph{Building it is the institutional work of the next eighteen months.}

\emph{Four,} the final part of the template~-- paying the tribute out to subsidise the \emph{displaced practice}~-- is the part the AI debate has, in my view, most under-thought. What does ``subsidising the displaced practice'' look like for AI-displaced creative work? For working musicians whose tracks are being competed-against by Suno outputs, it could look like funded performance opportunities, funded studio time, funded creative-development grants~-- the direct lineage of MPTF live-performance subsidies. For working illustrators whose work was used to train image models, it could look like commissioned-work grants, funded artist residencies, public-art-commission expansion. For working authors whose books were used to train LLMs, it could look like Public Lending Right expansion, library-licensing funds, writer-in-residence programmes. The structural move is the same in each case: \emph{take the productivity gain from the platform, redistribute it to the displaced practice, keep the practice alive as a category.} This is what the 88\% is implicitly asking for, whether or not the consultation respondents would have phrased it that way.

I want to be honest about a complication that the Petrillo template hits at full speed in the AI era, because the book should not be glib about it.

The MPTF works partly because the relationship between \emph{recorded music} (the displacing technology) and \emph{live music} (the displaced practice) is \emph{one-to-one}. The same musicians could, in 1948, do either thing. The Petrillo settlement was, structurally, paying the displaced version of the labour to subsidise the alternative version of the same labour.

The AI version of this relationship is \emph{many-to-many}. The training data for a generative-image model is the lifetime output of \emph{thousands} of working illustrators, photographers and visual artists, each of whom contributed an individually-tiny fraction of the model's competence. The output is \emph{generated}~-- there is no clean per-image-licence-equivalent. The redistribution problem is, by structure, much harder than Petrillo's problem.

Two attempts to solve this are visible in 2026.

\emph{The first} is \textbf{creative weight attribution}~-- Musical AI's framing, picked up by some of the C2PA-adjacent technical standards groups~-- which proposes that AI platforms compute, for each output, the \emph{gradient-weighted contribution} of each training-data input, and distribute a per-output royalty proportionally. The technical infrastructure for this is, in mid-2026, \emph{partially} built. The economic infrastructure to handle the resulting micro-payments is, in mid-2026, \emph{not} built. But the mechanism is the right one in principle: it preserves the one-to-many relationship that Petrillo could not directly handle, and translates it into a \emph{many-to-many} redistribution mechanism.

\emph{The second} is \textbf{collective licensing at the publisher tier}. The Stability AI / Universal Music alliance, the Splice / UMG partnership, the various YouTube and Spotify catalogue-licensing deals operate by aggregating training-data permissions at the publisher and label level, with the per-creator distribution handled internally by the existing royalty infrastructure of those publishers. This works for \emph{commercially-published} creative work where the publisher already has a contractual relationship with the creator. It works less well for \emph{independent} and \emph{self-published} creative work where there is no publisher to negotiate on the creator's behalf.

Both approaches will, in some hybrid form, be the architecture of the AI-era Petrillo settlement. The 88\%, the GEMA ruling, the SAG-AFTRA bargaining, the Cannes Disclosure Standard and the \emph{UMG v.\ Anthropic} litigation are the political pressure that is forcing the platforms to agree to \emph{some} version of one or the other. The version that emerges over the next eighteen months will, on the historical pattern, define the next forty years of how creative-AI work is paid for.

If the working-creative cohort reading this is asking what specifically to push for, my answer is: \emph{the Petrillo template, applied to AI, collected through a joint body, distributed through a creative-weight-attribution mechanism layered on top of the existing collective-licensing infrastructure, used to subsidise the displaced creative practice as a category}. That sentence is a mouthful. It is also the most-likely-to-work structural answer that the historical pattern points at. The 88\% is the political mandate for it. The institutional architecture is, in mid-2026, half-built. Finishing it is the work.

\section*{The consent-trained alternative}

The 88\% is a \emph{demand-side} fact: it tells you what creators want done about the training pipeline. There is a \emph{supply-side} fact that I think the policy debate has been slow to absorb, and that working creatives reading this book should know about, because it is the practical refutation of the AI companies' core argument.

The AI companies, as I noted earlier in this chapter, have spent two years arguing that machine-learning models \emph{cannot practically} be trained on licensed data at the scale they require. That the data volumes are too large, the licensing relationships too fragmented, the legal cost too high. That training on consent-acquired data is, in effect, a nice idea that does not survive contact with the engineering.

By spring 2026, this argument was falsifiable, and had been falsified, by the existence of a category of foundation models that had been built~-- and were commercially successful~-- on exactly the consent-first basis the AI companies said was impossible.

The category, with the models I would name as its leading examples:

\begin{itemize}
  \item \textbf{Adobe Firefly} (Image Model 5, Firefly Video, Firefly Foundry)~-- trained on Adobe Stock licensed content, openly licensed material, and public-domain work. Adobe Stock contributors are compensated through the Firefly bonus programme. By April 2025, Firefly had generated 22~billion assets; by mid-2026 it was the dominant enterprise generative-image tool, with 45\%+ Creative Cloud penetration.\footnote{Adobe Firefly milestone and adoption data. Firefly Foundry and Firefly Image Model 5 launch reporting, Adobe MAX 2025: \url{https://news.adobe.com/news/2025/10/adobe-max-2025-firefly-foundry}; \url{https://news.adobe.com/news/2025/10/adobe-max-2025-firefly}.}
  \item \textbf{Bria}~-- image and video foundation models trained on a 100\%-licensed dataset assembled from partners including Getty Images, Alamy and a network of independent rights holders. Bria publishes a per-output attribution and royalty mechanism on which the broader creative-weight-attribution conversation has drawn.\footnote{Bria AI consent-licensed dataset and attribution mechanism. [TODO: confirm primary citation~-- Bria's licensed-data white paper or Series~B coverage.]}
  \item \textbf{Getty Images Generative AI} (built with NVIDIA Picasso)~-- trained exclusively on Getty's licensed library, with contributor royalties paid through Getty's existing rights infrastructure.\footnote{Getty Images, ``Generative AI by iStock'' launch, built on NVIDIA Picasso, trained exclusively on Getty's licensed library with contributor royalties. [TODO: confirm citation~-- Getty press release or \emph{Reuters} coverage.]}
  \item \textbf{Moonvalley Marey}~-- a generative-video model positioned explicitly as the ``clean'' alternative to Sora and Veo, trained on a licensed video dataset.\footnote{Moonvalley Marey, generative-video foundation model trained on licensed video. [TODO: confirm citation~-- Moonvalley launch coverage in \emph{The Verge} / \emph{TechCrunch}.]}
  \item \textbf{AIODE}~-- a music-creation DAW trained on ethically-licensed audio. See Chapter~\ref{ch:16}, \S``Audio modality models.''\footnote{AIODE, ethically-trained music creation DAW. See Chapter~\ref{ch:16}: The Tools, \S``Audio modality models.''}
  \item \textbf{Tamber}~-- an ethically-trained AI music suite launched in May 2026, controllable via arm gestures, positioned explicitly as a tool that moves \emph{with} the musician rather than substituting for them.\footnote{\emph{MusicTech}, ``Tamber is an `ethically trained' AI tool to aid the creative process~-- and you can use arm gestures to control it.'' \url{https://musictech.com/news/gear/tamber-ai-ethically-trained-arm-gestures/}. Tamber product page: \url{https://tamber.ai/}. \emph{Dream Machine} Issue~30.}
  \item The \textbf{Stability AI / Universal Music Group} alliance, the \textbf{Stability AI / Warner Music} deal and the \textbf{Splice / Universal Music} partnership~-- data-alliance arrangements that build music models on rights-cleared catalogues, with the per-creator distribution handled through the labels' existing royalty infrastructure.\footnote{Stability AI / Universal Music Group strategic alliance: \url{https://stability.ai/news/universal-music-group-and-stability-ai-announce-strategic-alliance}. Stability AI / Warner Music: \url{https://stability.ai/news/warner-music-group-and-stability-ai-join-forces-to-build-next-gen-tools}. Universal Music / Splice partnership: \url{https://www.universalmusic.com/universal-music-group-and-splice-to-collaborate-on-the-next-generation-of-ai-powered-music-creation-tools-for-artists/}. \emph{Dream Machine} Issues~5, 8, 12.}
\end{itemize}

I am not claiming this category is perfect, or even, in every case, that its consent claims fully hold up to scrutiny. Adobe Firefly has faced criticism over the inclusion of AI-generated stock images in its training set;\footnote{Reporting on AI-generated images in the Adobe Stock training corpus, \emph{Bloomberg}, April 2024. [TODO: confirm exact citation.]} the per-creator economics on the Stability / UMG-style deals are still being worked out. The point is not that these models are above critique. The point is that they \emph{exist}, that they \emph{work commercially}, and that their existence collapses the central technical-inevitability argument that the rest of the industry has been using to justify scraping.

\section*{Indemnity as the receipt}

The clearest single signal that a model has done its upstream consent work is whether the company behind it is willing to \emph{indemnify} its customers against copyright infringement claims arising from generated output.

The pattern, in the eighteen months to mid-2026:

\begin{itemize}
  \item \textbf{Adobe} offers IP indemnification on Firefly enterprise outputs~-- Adobe will defend customers, and pay damages, if an enterprise generation is challenged on copyright grounds.\footnote{Adobe Firefly IP indemnification for enterprise customers. [TODO: confirm citation~-- Adobe enterprise terms or \emph{The Verge} coverage from 2023.]}
  \item \textbf{Microsoft} announced its \textbf{Copilot Copyright Commitment} in September 2023 and extended it across the Copilot product family through 2024--25. Microsoft assumes the legal liability for commercial customers using Copilot outputs in good faith.\footnote{Microsoft, ``Microsoft announces new Copilot Copyright Commitment for customers,'' 7~September 2023. \url{https://blogs.microsoft.com/on-the-issues/2023/09/07/copilot-copyright-commitment-ai-legal-concerns/}.}
  \item \textbf{Google} offers an indemnification commitment on Vertex AI and Workspace generative outputs.\footnote{Google Cloud Generative AI indemnification: \url{https://cloud.google.com/blog/products/ai-machine-learning/protecting-customers-with-generative-ai-indemnification}.}
  \item \textbf{IBM} offers an uncapped indemnity on watsonx-generated content for enterprise customers.\footnote{IBM watsonx uncapped indemnity for enterprise customers. [TODO: confirm citation.]}
\end{itemize}

Notice which companies are on this list and which are not. The companies indemnifying their customers are, without exception, the companies that have invested most heavily in the \emph{upstream} consent work~-- licensed data, contributor compensation, rights-cleared catalogues. The companies that have \emph{not} indemnified their customers are, predominantly, the companies whose training-data position is most exposed.

This is not a coincidence. Indemnification is a receipt. It is the legal department of a \$200B company telling its commercial customers, in the most expensive language available, \emph{we have done the work; you can use this without being sued}. The absence of an indemnity, conversely, is an instruction. It is the same legal department saying, \emph{the risk is yours; you carry it}.

For working creatives, agencies and studios making procurement decisions in 2026, the indemnity status of a tool is the single most useful one-question proxy for whether its training pipeline is built on the side of the 88\% or against it. Ask the vendor. If they cannot give you a written indemnity, you have your answer.

\section*{The literacy problem}

None of this~-- the consent-trained category, the indemnity framework, the C2PA provenance stack in Chapter~\ref{ch:12}, the legislative reversal earlier in this chapter~-- works without a corresponding investment in \emph{literacy}. And the literacy gap, in mid-2026, is the place where I am most worried about the architecture failing.

Policy and infrastructure can constrain the supply side. They cannot, on their own, redirect the demand side. The question of whether a working illustrator chooses Firefly over Midjourney, whether a marketing team specifies Bria over a scraped open-source model in its agency brief, whether a record label's A\&R department uses a Stability / UMG-aligned tool rather than Suno for demo work, whether a film commissioner asks for Marey provenance on a generative-video shot, whether an audience member streams a SynthID-watermarked track over an unlabelled one~-- these are \emph{consumption} questions. They sit downstream of every law and every standard. They are decided, ten thousand times a day, by people choosing tools and content from a menu, without anyone telling them what the choices on the menu actually mean.

Three things have to happen for the literacy layer to catch up with the infrastructure layer.

\emph{First,} working creatives need to know what the consent-trained category is, which tools are in it, and what an indemnity is for. This book is one attempt at that; the \textbf{Sundance AI Literacy Initiative}, training 100,000+ artists in provenance practice on Google's funding, is another.\footnote{Sundance AI Literacy Initiative, in Chapter~\ref{ch:12}: Authenticity, the New Scarcity, \S``The provenance infrastructure, named.''} The professional bodies~-- the Society of Authors, the AIGA, Equity, the AOP, the MPG, the WGA~-- have, by mid-2026, started shipping member guides. The work is early.

\emph{Second,} the \emph{buyers} of creative work~-- the brands, the agencies, the broadcasters, the platforms, the publishers~-- need to make ethically-trained models a \emph{specified requirement} in their briefs and procurement contracts. A handful already have: the BBC, the AP wire service, the Cannes festival itself. The vast majority have not. The lever exists. It needs to be pulled.

\emph{Third,} the \emph{audience} needs the equivalent of a nutrition label. The Cannes Disclosure Standard, SynthID-in-Gemini, the YouTube AI-content disclosure rules, the proposed EU AI Act labelling obligations are early attempts at this. None of them yet add up to a consumer-facing signal as legible as the \emph{Fairtrade} mark or the \emph{organic} certification. Until they do~-- until an audience member streaming a song, watching a clip, or buying a print can tell at a glance whether the creative work in front of them was made with a tool that paid the people whose work it learned from~-- the consumption side of the equation will keep leaking. Provenance metadata sitting in a file header that no one reads is not, on its own, literacy.

I do not think this layer will get built by the AI companies. The incentive isn't there. I think it will get built~-- slowly, contentiously, in fits and starts~-- by the same coalition I am about to describe in the next section: by creators, their unions, their professional bodies, their buyers and their audiences, jointly insisting on a labelling regime that the platforms eventually have to honour because the market has organised itself around it.

The 88\% is the political mandate. The consent-trained models are the proof of supply. The indemnity framework is the legal receipt. The literacy infrastructure is the missing piece~-- and it is, on the evidence of the last six months, being built.

\section*{Coalitions, not protests}

The thing I want creative people reading this to take from this chapter is \emph{not} that protest works.

Protest works. We have seen it. The 88\%, the Equity ballot, the SAG-AFTRA contract, the artists' declaration, the GEMA ruling, the UK government's reversal~-- these are evidence that protest works.

What I want you to take is that \emph{coalition} works.

What happened in these six months was not~-- or was not only~-- that individual creators got angry and shouted. What happened was that creators \emph{aligned themselves} with adjacent groups whose interests they had not previously seen as aligned with theirs.

Working musicians aligned with photographers, who aligned with authors, who aligned with games developers, who aligned with screenwriters, who aligned with voice actors, who aligned with concept artists, who aligned with translators, who aligned with journalists. They aligned with their unions. They aligned with their professional bodies. They aligned, somewhat to everyone's surprise, with the major studios, who had spent twenty years suing them and now found themselves on the same side of a copyright argument against the same platform companies.\footnote{For Disney's parallel position, see \emph{Deadline}, ``Disney Sends Cease And Desist Letter To Character.ai.'' \url{https://deadline.com/2025/09/disney-cease-and-desist-letter-characterai-copyright-infringement-1236566831/}. For Studio Ghibli's similar stance: \emph{NDTV Profit}, ``Studio Ghibli And Studio That Developed Elden Ring Send Stern Message To OpenAI\@.'' \url{https://www.ndtvprofit.com/technology/studio-ghibli-and-studio-that-developed-elden-ring-send-stern-message-to-openai}. \emph{Dream Machine} Issues~2, 6.}

The major-label leadership read of the same coalition shifted, in the spring of 2026, in a way that I think is worth registering carefully because the rhetoric is a useful weather-vane. \textbf{Robert Kyncl}, the chief executive of \textbf{Warner Music Group}, in a widely-quoted May 2026 interview, told the industry that \emph{``AI resistance''} was actively \emph{setting the music sector back}~-- that AI represented ``an incredible value creation opportunity,'' and that the labels \emph{``cannot wait the way the industry did 25 years ago.''} Kyncl's invocation of the Napster moment was deliberate. The argument was that the labels' twenty-five-year pattern of \emph{suing first, integrating second} had cost them, in net, the bulk of the streaming-era surplus to platforms that had moved before they did, and that repeating that pattern with AI would compound the loss.\footnote{\emph{Variety}, ``Is `AI Resistance' Setting the Music Sector Back? WMG's Robert Kyncl Sees `An Incredible Value Creation Opportunity,' But Warns `We Cannot Wait the Way the Industry Did 25 Years Ago'.'' \url{https://variety.com/2026/music/news/wmg-robert-kyncl-ai-resistance-1236748901/}. \emph{Dream Machine} Issue~30.} This is a meaningful change in register from the 2025 \emph{``biggest theft in music history''} framing, and worth tracking. It does not contradict the 88\%~-- Kyncl, like the BPI, is pushing for licensing infrastructure rather than against it~-- but it shifts the \emph{centre of gravity} of major-label rhetoric from prohibition toward participation. On the Petrillo template, this is the labels' tribute-mechanism position: AI continues, the platforms pay, the joint-body collection infrastructure scales. The question of whether the \emph{displaced practice} gets a meaningful share of the resulting flow~-- whether the working songwriter, the working session player, the working independent artist actually \emph{sees} the tribute~-- is the part the Kyncl framing does not yet answer.

They also~-- and this part I find most interesting~-- aligned with their \emph{audiences}. The Adobe Creators' Toolkit Report found that \textbf{69\%} of creators worried about their work being used to train AI without consent.\footnote{Adobe, \emph{Creators' Toolkit Report}, \emph{op.\ cit.} 69\% of 16,000 surveyed creators worried about their work being used to train AI without consent.} That number rhymes with the 88\% in the UK consultation. It also rhymes with the audience behaviour I described in Chapter~\ref{ch:5}~-- the slop ceiling, the AI-music underperformance, the cultural rejection of synthetic content that doesn't disclose itself. The creators wanted protection. The audience, given a choice, wanted to listen to the protected work. The two interests, for the first time in a long time, sat on the same side of the line.

That alignment is the most powerful political asset the creative industries have had this century. They built it in six months. The question for the next six months~-- which Chapter~\ref{ch:13} of this book is going to come back to~-- is what they do with it.

  \chapter{The Studios Decide}\label{ch:7}

\lettrine[lines=3,lhang=0.15,findent=0.1em]{I}{f} the audience was speaking through the slop ceiling, and the creators were speaking through the 88\%, the studios were speaking through their balance sheets~-- and the language was not quite the language of either of the other two.

On \textbf{22 October 2025}~-- three weeks into the period this book covers, the same day I was writing Issue~4~-- Ted Sarandos, Netflix's co-CEO, told an industry conference that Netflix was ``all in'' on leveraging AI across its streaming platform.\footnote{\textit{CNBC}, ``Netflix `all in' on leveraging AI as the tech creeps into entertainment industry,'' 22 October 2025. \url{https://www.cnbc.com/2025/10/22/netflix-all-in-on-leveraging-ai-in-its-streaming-platform.html}. \emph{Dream Machine} Issue~4.} The phrase was casual. The implications were not. Within hours, the trade press was running it as the official line of the world's largest streaming service, and it was being read~-- correctly~-- as a signal to every other studio that the period of ``wait and see'' was over.

Three weeks earlier, in a piece \textit{Futurism} had published with a headline that aged badly almost in real time, \textbf{Lionsgate's} ambitious attempt to use AI for movie development had been characterised as having ``crumbled into disaster.''\footnote{\textit{Futurism}, ``Lionsgate's Attempt to Create Movies Using AI Has Crumbled Into Disaster.'' \url{https://futurism.com/artificial-intelligence/lionsgate-movies-ai}. \emph{Dream Machine} Issue~1.}

Two months later, on \textbf{11 December 2025}, \textit{The Guardian} reported that \textbf{Disney} was investing \$1 billion in OpenAI, with a structured agreement that would let Disney characters appear in the Sora video tool.\footnote{\textit{The Guardian}, ``Disney to invest \$1bn in OpenAI, allowing characters in Sora video tool.'' \url{https://www.theguardian.com/business/2025/dec/11/disney-open-ai-sora-video-deal}. \emph{Dream Machine} Issue~11.}

These three moments~-- Lionsgate's failure, Netflix's commitment, Disney's \$1bn~-- are the three corners of the strategic map that every legacy studio in the world has been navigating for the last six months. They are not a single story. They are three different stories about how a creative business with a hundred years of human-craft DNA tries to integrate a technology that, by the time it integrates, no longer behaves like the technology you thought you were integrating.

This chapter is about the studios. About how they decided. About the ones that went \emph{all-in}, the ones that went \emph{AI-native from scratch}, the ones that went \emph{we are not doing this at all}, and the ones~-- the most interesting group~-- that went \emph{we will do it, but only in the places where it doesn't show up in the work the audience sees.}

The map of those four positions, drawn carefully, is the map of where the film, TV, games and entertainment industries will be in 2030.

\begin{figure}[htbp]
  \centering
  \includegraphics[width=0.92\textwidth]{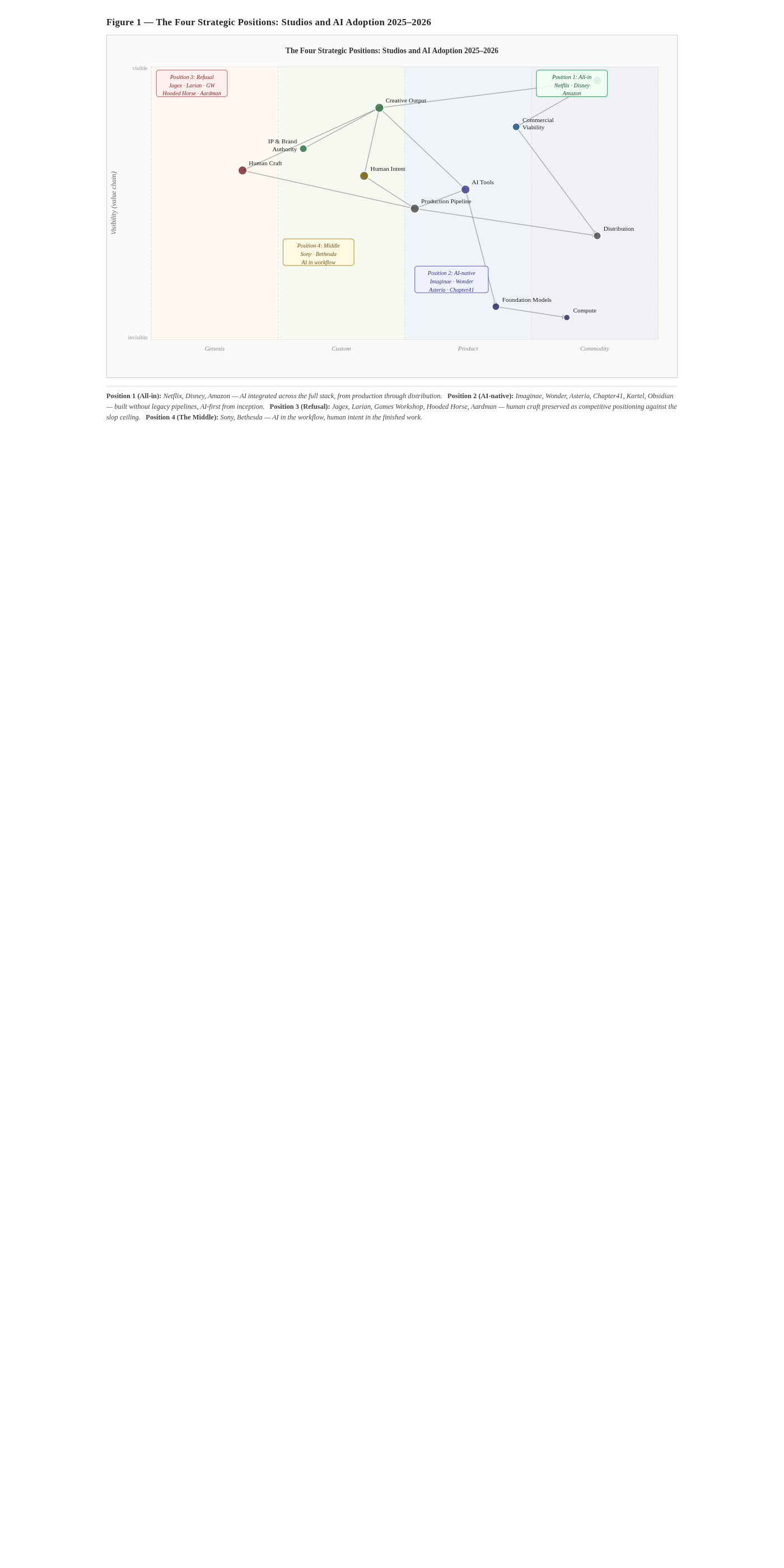}
  \caption{Wardley map of studio strategic positions: from all-in adoption through AI-native entrants to explicit rejection, each position carries distinct risk and capability profiles.}
  \label{fig:wardley-studios}
\end{figure}

\section*{Position One: All-in}

Netflix's ``all in'' framing was the most prominent example of what became, over the autumn of 2025 and the winter of 2026, the dominant public stance of the major streamers. The framing was: AI is a tool, AI is a productivity multiplier, AI is going to be used everywhere in the pipeline, and the studios that adopt it earliest will have the most leverage when the new economics settle.

The actual deployment, when the trade press dug into it, was more interesting than the slogan suggested. Netflix's use of AI in late 2025 included generative AI tools for visual effects (de-aging actors, scene extensions, background plates), AI-driven recommendation engines that the company had been refining for fifteen years, and~-- disclosed in a January 2026 \textit{Pymnts} report~-- a major AI strategic push focused on subscriber \emph{retention} rather than production cost.\footnote{PYMNTS, ``Retention Is Name of the Game for Netflix's AI Strategy.'' \url{https://www.pymnts.com/subscription-commerce/2026/retention-is-name-of-the-game-for-netflixs-ai-strategy/}. \emph{Dream Machine} Issue~15.} The story Sarandos was telling Wall Street was not ``AI will replace our writers.'' It was ``AI will keep our subscribers engaged in a way that human-only programming alone cannot afford to.''

By May 2026 the deployment had moved one further step in. Netflix announced \textbf{INKubator}, an in-house AI animation studio explicitly chartered to produce \emph{``feature-quality''} short-form work, and began recruiting for it publicly.\footnote{\textit{Hollywood Reporter}, ``Netflix is building and recruiting for an AI animation studio, called INKubator, to produce `feature-quality' shorts.'' \url{https://www.hollywoodreporter.com/business/business-news/netflix-ai-animation-studio-inkubator-1236592110/}. \emph{Dream Machine} Issue~30.} What is notable about INKubator, for the purposes of this chapter, is not the size of the unit~-- small, by Netflix standards~-- but the \emph{organisational position} of it: an \emph{internal} AI-native studio sitting \emph{inside} the major-streamer architecture, producing original work, reporting up into the same commissioning structure as the live-action slates. That is a different shape from the Position-Two AI-native studios I describe in the next section. It is a hybrid: Position One on the org chart, Position Two in the pipeline.

Adjacent moves from other big studios that autumn told the same story.

\textbf{Amazon} built out an internal ``AI Studios'' unit in November 2025, naming sports-docs boss Matt Newman as its head of live-action production.\footnote{\textit{Deadline}, ``Amazon Builds Out AI Studios With Sports Docs Boss Matt Newman Named Head Of Live-Action.'' \url{https://deadline.com/2025/11/amazon-ai-studios-matt-newman-1236603477/}. \emph{Dream Machine} Issue~7.} In the same month, Amazon's \textit{House of David} TV series became one of the first major Western dramas to publicly disclose the use of more than \textbf{350 AI-generated visual-effects shots} in its second season, with creator Jon Erwin telling \textit{Wired} he was ``not sorry.''\footnote{\textit{Wired}, ``Amazon's House of David Used Over 350 AI Shots in Season 2. Its Creator Isn't Sorry.'' \url{https://www.wired.com/story/amazons-house-of-david-used-over-350-ai-shots-in-season-2-its-creator-isnt-sorry/}. \emph{Dream Machine} Issue~7.}

\textbf{NBCUniversal} signed a deal in late October 2025 with the son of \textit{Law \& Order} creator Dick Wolf to develop AI-generated games based on its IP.\footnote{\textit{Video Games Chronicle}, ``NBCUniversal signs deal with Law \& Order creator Dick Wolf's son to make AI-generated games based on its IP.'' \url{https://www.videogameschronicle.com/news/nbcuniversal-signs-deal-with-law-order-creator-dick-wolfs-son-to-make-ai-generated-games-based-on-its-ip/}. \emph{Dream Machine} Issue~5.} By late November, the framing had broadened~-- \textit{The Office}, \textit{Saturday Night Live} and \textit{Sex and the City} were all reportedly being considered as IP for AI-generated game adaptations.\footnote{\textit{NME}, ```The Office', `Saturday Night Live' and `Sex And The City' could be turned into AI games.'' \url{https://www.nme.com/news/gaming-news/the-office-and-sex-and-the-city-ai-video-games-3901630}. \emph{Dream Machine} Issue~5.}

\textbf{Disney}, beyond its OpenAI investment, announced in November 2025 that it was developing generative AI tools to let Disney+ subscribers create and share their own short-form videos using the company's iconic IP~-- a play, transparently, at recapturing the engagement Fortnite and Roblox had been taking from passive streaming for years.\footnote{\textit{The Hollywood Reporter}, ``Disney+ to Allow User-Generated Fan Content with AI.'' \url{https://www.hollywoodreporter.com/business/digital/disney-plus-gen-ai-user-generated-content-1236426135/}. \emph{Dream Machine} Issue~8.} To execute this, Disney created a new ``\textbf{Office of Technology Enablement}'' under former Walt Disney Studios CTO Jamie Voris, with the specific mandate of accelerating AI and Mixed Reality adoption across the organisation.\footnote{\emph{Dream Machine} Issue~8 reportage of the Disney ``Office of Technology Enablement,'' led by former Walt Disney Studios CTO Jamie Voris.} In January 2026, Disney followed up with an announcement of a TikTok-like vertical-video product and an AI video-generation tool aimed at brand advertisers using existing Disney brand assets and guidelines.\footnote{\textit{Marketing Dive}, ``Disney unveils TikTok-like vertical video, AI video generation tool.'' \url{https://www.marketingdive.com/news/disney-unveils-tiktok-like-vertical-video-ai-video-generation-tool/809269/}. \emph{Dream Machine} Issue~14.}

\textbf{Fox Entertainment} took an equity stake in \textbf{Holywater}, an AI-microdramas company, in October 2025.\footnote{\textit{The Hollywood Reporter}, ``Fox Entertainment Takes Equity Stake in AI-Microdramas Company Holywater.'' \url{https://www.hollywoodreporter.com/business/business-news/fox-entertainment-invests-in-holywater-ai-microdramas-1236396802/}. \emph{Dream Machine} Issue~3.}

\textbf{Sky History} acquired \textit{Castles SOS}, an AI-powered documentary, in late November.\footnote{\textit{Deadline}, ``Sky History Acquires `Castles SOS,' AI-Powered Doc Exploring Royalty, Ruins \& Restoration.'' \url{https://deadline.com/2025/11/castles-sos-ai-doc-sky-history-documentary-rick-edwards-1236627378/}. \emph{Dream Machine} Issue~9.}

\textbf{Channel 4} rolled out an AI-driven advertising tool in December 2025 designed to make TV advertising accessible to SMEs~-- a small home-builder was one of the first clients.\footnote{\textit{Estate Agent Today}, ``Homebuilder among first to use Channel 4's AI ads.'' \url{https://www.estateagenttoday.co.uk/breaking-news/2025/12/homebuilder-among-first-to-use-channel-4s-ai-ads/}. \emph{Dream Machine} Issue~11.}

Position One is not subtle. The streamers, broadcasters and major studios with the capital to do it have been integrating AI into their stacks~-- production, post-production, marketing, advertising, distribution, subscriber retention~-- at a pace that suggests they have already decided which side of the future they want to be on. They want to be the side that owns the toolchain.

\section*{Position Two: AI-native}

A second group, more recent and more interesting, are the studios that have decided not to \emph{integrate} AI into existing film and television production pipelines but to \emph{replace} those pipelines entirely with AI-first workflows. These are the \textbf{AI-native studios.}

\textbf{Fremantle}, the international production powerhouse, named the boss of its new ``AI-native'' studio \textbf{Imaginae Studios} in October 2025.\footnote{\textit{The Hollywood Reporter}, ``Fremantle Names Boss of New AI Native Studio Imaginae Studios.'' \url{https://www.hollywoodreporter.com/business/digital/fremantle-names-ceo-new-ai-label-imaginae-studios-1236396579/}. \emph{Dream Machine} Issue~2.} By the spring of 2026, Imaginae was developing a project called \textit{Art Awakens}, fusing AI techniques with classical painting IP.\footnote{\emph{Dream Machine} Issue~25, on Fremantle's \textit{Art Awakens} development.}

\textbf{Imagine Entertainment}~-- Ron Howard and Brian Grazer's production company, with one of the most distinguished filmographies in modern Hollywood~-- partnered with a new AI-first production company called \textbf{Obsidian Studio} in November 2025.\footnote{\textit{Indiewire}, ``Another New AI Production Company Inks a Big Creative Partnership~-- This Time, with Ron Howard and Brian Grazer's Imagine Entertainment.'' \url{https://www.indiewire.com/news/business/obsidian-studio-ai-production-company-imagine-entertainment-1235158619/}. \emph{Dream Machine} Issue~6.}

\textbf{Wonder Studios} raised \$12m in seed in October 2025\footnote{\textit{UK Tech News}, ``AI film studio Wonder lands \$9m investment.'' \url{https://www.uktech.news/ai/ai-film-studio-wonder-lands-9m-investment-20251023}. \emph{Dream Machine} Issue~5.} and by January 2026 was running its own \textbf{Wonder Film Festival} with a curated shortlist of AI-made shorts.\footnote{Wonder Studios, ``Shortlisted films revealed for The Wonder Film Festival.'' \url{https://www.linkedin.com/posts/wearewonderstudios_were-thrilled-to-share-the-shortlisted-films-activity-7404560378082246656-7NcI}. \emph{Dream Machine} Issue~11.} By May 2026 Wonder had closed a further round bringing total funding to \textbf{\$50M}, with the company publicly framing the ambition as becoming \emph{``the A24 of AI production''}~-- a deliberate analogue to the indie-prestige distribution model rather than to the streamer-replacement model the trade press had been expecting AI-native studios to chase.\footnote{\textit{Forbes}, ``Meet Wonder Studios, The \$50M British Studio Striving To Become The A24 Of AI Production.'' \url{https://www.forbes.com/sites/charliefink/2026/05/18/meet-wonder-studios-the-50m-british-studio-striving-to-become-the-a24-of-ai-production/}. \emph{Dream Machine} Issue~30.}

\textbf{Asteria}~-- Natasha Lyonne's AI company, backed by James Cameron's \textit{Lightstorm Entertainment}~-- produced its first animated short, \textit{All Heart}, in October 2025.\footnote{\textit{The Hollywood Reporter}, ``AI Company Asteria Produces New Animated Short `All Heart'.'' \url{https://www.hollywoodreporter.com/movies/movie-news/natasha-lyonne-ai-company-asteria-1236403144/}. \emph{Dream Machine} Issue~4.}

\textbf{Promise}, a deep-pocketed AI studio backed by Google, set up shop in October 2025 specifically to ``bring GenAI filmmaking and VFX to legacy media.''\footnote{\textit{The Hollywood Reporter}, ``Promise, a deep-pocketed AI studio backed by Google, aims to Bring GenAI Filmmaking and VFX to Legacy Media.'' \url{https://www.hollywoodreporter.com/business/digital/ai-studio-promise-vfx-generation-company-1236397636/}. \emph{Dream Machine} Issue~3.}

\textbf{Goldfinch} launched \textbf{enGEN3}, an ``AI-Powered Cinematic Universe Platform,'' in October 2025.\footnote{\textit{Variety}, ``AI-Powered Cinematic Universe Platform enGEN3 Launched by Goldfinch.'' \url{https://variety.com/2025/film/news/ai-powered-cinematic-universe-platform-engen3-1236543349/}. \emph{Dream Machine} Issue~2.}

\textbf{Chapter41}, a Munich-based AI startup, was launched in November 2025 by \textbf{Beta Film} and a group of industry executives.\footnote{\textit{Deadline}, ``Munich Based Beta Films \& Industry Execs Join Forces To Launch Artificial Intelligence Start-Up Chapter41.'' \url{https://deadline.com/2025/11/beta-film-ai-startup-chapter41-artificial-intelligence-1236612632/}. \emph{Dream Machine} Issue~7.}

\textbf{Kartel}~-- a new AI startup led by long-time TV exec Kevin Reilly, formerly of HBO~-- was set up in November 2025.\footnote{\textit{The Hollywood Reporter}, ``Longtime TV Exec, Kevin Reilly, Set to Lead AI Startup Kartel.'' \url{https://www.hollywoodreporter.com/business/digital/kevin-reilly-ceo-kartel-ai-hbo-1236424692/}. \emph{Dream Machine} Issue~7.}

\textbf{Wanted} director Timur Bekmambetov launched a \$5 million project to ``generate AI method actors'' in November 2025, with the framing: ``AI is here to stay. We have to train it responsibly.''\footnote{\textit{Variety}, ```Wanted' Director Timur Bekmambetov Explains His \$5 Million Plan to Generate AI Method Actors: `AI Is Here to Stay. We Have to Train It Responsibly'.'' \url{https://variety.com/2025/film/news/wanted-director-method-acting-ai-actors-1236579647/}. \emph{Dream Machine} Issue~7.}

\textbf{Particle6}, the U.K.--Netherlands company behind Tilly Norwood, expanded to 41 AI actors in development by November 2025, with founder Eline Van der Velden in a December 2025 \textit{Deadline} interview making the case that AI performance was a ``more ethical way'' to act~-- and urging working performers to ``future-proof'' themselves by creating their own AI avatars.\footnote{\textit{Variety}, ``Tilly Norwood Creator Doubles Down on AI `Actors' and Says It's a `More Ethical Way to Perform,' Urges Human Actors to `Future-Proof' Themselves With AI.'' \url{https://variety.com/2026/digital/news/tilly-norwood-creator-tells-actors-to-create-ai-avatars-1236638940/}. \emph{Dream Machine} Issue~16.}

\textbf{Wonder Studios}, separately, adapted a children's book to an animated series using AI in December 2025.\footnote{\textit{Broadcast Now}, ``Wonder Studios adapts children's book to animated series with AI.'' \url{https://www.broadcastnow.co.uk/production-and-post/wonder-studios-adapts-childrens-book-to-animated-series-with-ai/5211713.article}. \emph{Dream Machine} Issue~11.}

\textbf{Kling AI} and \textbf{Evolutionary Films} announced an AI-animated feature, \textit{Minibots}, at the Cannes Film Market in May 2026, alongside a broader Kling-backed filmmaker initiative aimed at funding AI-native productions on the same indie-distribution architecture.\footnote{\textit{Variety}, ``Kling AI Partners With Evolutionary Films on Animated Feature `Minibots,' Unveils Filmmaker Initiative at Cannes Market.'' \url{https://variety.com/2026/film/news/kling-ai-evolutionary-films-minibots-cannes-1236748590/}. \emph{Dream Machine} Issue~30.}

By \textbf{April 2026}, the trade press could no longer keep up with the AI-native studio launches. There were too many of them. Most of them, like most early-stage production companies in any era, will not survive the next two years. The question of whether a meaningful AI-native studio system will eventually emerge as a parallel structure to legacy Hollywood~-- the way Netflix and Amazon eventually emerged as a parallel structure to the cable networks~-- is, in my view, the biggest single open question in the film and TV industry as of May 2026.

The early evidence is mixed. \textit{Watch the Skies}, a Swedish UFO feature entirely dubbed with AI, secured U.S.\ distribution in October 2025.\footnote{\textit{Variety}, ```Watch the Skies,' Swedish UFO Feature Film Dubbed Entirely With AI, Sets USA Distribution Deal.'' \url{https://variety.com/2025/film/news/watch-the-skies-us-theatrical-release-ai-dubbing-1236343110/}. \emph{Dream Machine} Issue~5.} \textit{Run to the West}, South Korea's first AI feature film, was tested with critics and audiences in October 2025; one \textit{cybernews.com} review described the experience as ``testing the soul of cinema.''\footnote{\textit{Cybernews}, ``Run to the West~-- South Korea's first AI film tests the soul of cinema.'' \url{https://cybernews.com/entertainment/korean-cinema-run-to-the-west-ai/}. \emph{Dream Machine} Issue~5.} \textit{Lily}, a Tunisian-made AI short, won the \$1 million Dubai AI Film Award in January 2026.\footnote{\textit{Broadcast Pro Middle East}, ``Tunisian filmmaker wins \$1 million AI Film Award for `Lily'.'' \url{https://www.broadcastprome.com/news/tunisian-filmmaker-wins-1-million-ai-film-award-for-lily/}. \emph{Dream Machine} Issue~14.} \textit{Humans in the Loop}, an AI drama that received Film Independent's Sloan Distribution Grant, entered the Oscar race in November 2025.\footnote{\textit{Variety}, ``AI Drama `Humans in the Loop' Receives Film Independent's Sloan Distribution Grant, Enters Oscar Race.'' \url{https://variety.com/2025/film/news/ai-drama-humans-in-the-loop-oscar-race-1236582975/}. \emph{Dream Machine} Issue~8.}

I have watched a meaningful percentage of the AI-native output of these six months. The honest evaluation, which I have given in talks several times and stand by here, is that we have not yet seen the \textit{Citizen Kane} of AI cinema~-- we have not yet seen a single AI-native work that I think will still be watched in 2040. We have seen, repeatedly, films that demonstrate technical ability without yet demonstrating cultural necessity.

The most interesting AI-native works, in my view, are the ones that~-- like Andrii Daniels' bomb-shelter clip~-- wear their non-traditional production conditions on their face. They are films \emph{about} the technology being used to make them, in some implicit or explicit sense. They are not pretending to be legacy films made by a different route.

\section*{Position Three: We are not doing this}

The studios that have publicly refused generative AI have been some of the most interesting voices in this entire period.

\textbf{Pocketpair}, the Japanese games studio behind \textit{Palworld}, announced in October 2025 that its new publishing division would not handle games using generative AI. The CEO's full statement, in \textit{PC Gamer}, was sharper than the headline: \emph{``We don't believe in it. We're very upfront about it. If you're big on AI stuff or your game is Web3 or uses NFTs, there are lots of publishers out there [who'll talk to you], but we're not the right partner for that.''}\footnote{\textit{PC Gamer}, ``Palworld studio Pocketpair says its new publishing division won't handle games that use generative AI: `We don't believe in it'.'' \url{https://www.pcgamer.com/software/ai/palworld-studio-pocketpair-says-its-new-publishing-division-wont-handle-games-that-use-generative-ai-we-dont-believe-in-it/}. \emph{Dream Machine} Issue~4.} It was, on its face, a rejection of one production model. It was, on inspection, also a \emph{marketing position}: a publisher staking its claim with audiences who had become~-- by late 2025~-- actively allergic to AI-augmented games.

\textbf{Larian Studios}~-- the maker of \textit{Baldur's Gate 3}, one of the most critically and commercially successful games of the decade~-- backed off generative AI in January 2026 for its next \textit{Divinity} game.\footnote{\textit{Niche Gamer}, ``Larian Studios backs off from gen AI.'' \emph{Dream Machine} Issue~14.}

\textbf{Games Workshop}, custodian of the \textit{Warhammer 40,000} universe, ruled out generative AI entirely in early 2026.\footnote{\textit{Decrypt}, ``Warhammer 40,000 Maker Games Workshop Rules Out Generative AI.'' \emph{Dream Machine} Issue~14.}

\textbf{Hooded Horse}, the U.S.\ games publisher behind \textit{Manor Lords}, said in January 2026 that it would not work with developers who used generative AI.\footnote{\textit{Niche Gamer}, ``Manor Lords publisher Hooded Horse won't work with devs using gen AI.'' \emph{Dream Machine} Issue~14.}

\textbf{Jagex}, the maker of \textit{RuneScape}, declared in January 2026 that it would \emph{never} use generative AI to make in-game content.\footnote{\textit{gamesindustry.biz}, ``RuneScape maker Jagex says it will never use generative AI to make in-game content.'' \emph{Dream Machine} Issue~16.}

\textbf{Aardman Animations}~-- the British animation studio responsible for \textit{Wallace and Gromit}~-- announced in December 2025 that it would ``embrace the technology'' of AI but would be ``very cautious not to lose our values.''\footnote{\textit{GamesRadar}, ``Wallace and Gromit creator says beloved animation studio Aardman will `embrace the technology' of AI, but will be `very cautious not to lose our values'.'' \url{https://www.gamesradar.com/entertainment/animation-movies/wallace-and-gromit-creator-says-beloved-animation-studio-aardman-will-embrace-the-technology-of-ai-but-will-be-very-cautious-not-to-lose-our-values/}. \emph{Dream Machine} Issue~11.} This was, by Aardman's careful standards, a sharp line: they reserved the right to use AI for narrowly defined post-production and admin tasks, but explicitly excluded it from the stop-motion craft that defines their work.

\textbf{Guillermo del Toro}, in October 2025, told Variety he would ``rather die'' than use generative AI in his films, with a follow-up Frankenstein-themed press cycle that made the line one of the most-quoted creative-industry statements of the year. The full quote was even better than the headline: \emph{``I'm 61, and I hope to be able to remain uninterested in using it at all until I croak. \ldots{} The other day, somebody wrote me an email, said, `What is your stance on AI?' And my answer was very short. I said, `I'd rather die.'''}\footnote{\textit{Variety}, ``Guillermo del Toro Says He'd `Rather Die' Than Use Generative AI in His Films: `Not Interested'.'' \url{https://variety.com/2025/film/news/guillermo-del-toro-rather-die-generative-ai-frankenstein-1236561316/}. \emph{Dream Machine} Issue~5.} What del Toro was doing, with the bluntness only a senior auteur with a fully-funded slate can afford, was \emph{refusing to participate in the framing.} Most working creatives have had to spend two years giving careful, nuanced, defensive answers about their AI position. Del Toro decided he was a senior enough artist to refuse the question entirely. The cultural permission for that posture, in a particular kind of high-end filmmaking, is part of the architecture this book has been describing.

\textbf{Leonardo DiCaprio}, in December 2025, told \textit{The Hollywood Reporter}: \emph{``I think anything that is going to be authentically thought of as art has to come from the human being.''} The headline framing reduced the position to ``AI can't be art because there's no humanity to it,'' which is the version that travelled, but the full quote is more philosophically defensible. DiCaprio wasn't claiming AI-augmented work couldn't be valuable. He was claiming that the authorship signal~-- ``from the human being''~-- was a precondition for the \emph{category} of art, as he understood it.\footnote{\textit{The Hollywood Reporter}, ``Leonardo DiCaprio Says AI Can't Be Art Because `There's No Humanity to It'.'' \url{https://www.hollywoodreporter.com/movies/movie-news/leonardo-dicaprio-ai-cant-be-art-no-humanity-1236445405/}. \emph{Dream Machine} Issue~11.}

\textbf{Claire Foy} told the \textit{Daily Mail} in January 2026 she had ``no interest'' in seeing AI in films and would be ``disappointed'' if it became the future of Hollywood.\footnote{\textit{Daily Mail}, ``Claire Foy says she has `no interest' in seeing AI in films.'' \url{https://www.dailymail.co.uk/tvshowbiz/article-15454199/Claire-Foy-AI-films-sad-disappointed-people-future-Hollywood.html}. \emph{Dream Machine} Issue~14.}

\textbf{Jenna Ortega} said in December 2025 it was ``very easy to be terrified'' of AI in filmmaking. Her fuller reasoning, given to \textit{NME}, is the part I have ended up quoting in talks: \emph{``It comes to a point where it becomes sort of mental junk food and we feel sick and we don't know why. I think, as terrible as it is to say, sometimes audiences need to be deprived of something in order to appreciate something again.''}\footnote{\textit{NME}, ``Jenna Ortega says it's `very easy to be terrified' of AI in filmmaking.'' \url{https://www.nme.com/news/jenna-ortega-says-its-very-easy-to-be-terrified-of-ai-in-filmmaking-3913926}. \emph{Dream Machine} Issue~10.} That argument~-- \emph{audiences need to be deprived of something in order to appreciate something again}~-- is one of the most interesting things a working performer has said in this period about the slop ceiling and its psychological substrate. The audience does not, by Ortega's read, simply discriminate against AI work. They develop a \emph{hunger} for the human-authored work \emph{because} of the AI flood. The flood and the hunger are part of the same cultural dynamic.

\textbf{Chris Pratt} publicly rejected a pitch to cast an AI `actor' as the villain in \textit{Mercy} in January 2026: ``I don't think that's a good idea at all.''\footnote{\textit{Variety}, ``Chris Pratt Pitched Having an AI `Actor' Star as the Villain in `Mercy': `I Don't Think That's a Good Idea at All'.'' \url{https://variety.com/2026/film/news/chris-pratt-ai-actor-villain-mercy-amazon-mgm-1236640460/}. \emph{Dream Machine} Issue~16.}

I do not think any of these positions are static. I think some of them will shift in the next eighteen months, in ways that depend on how the policy environment, the audience response and the tool ecosystem evolve. But I think the \emph{fact} of the positions, written down, in public, on the record, is more important than whether any individual position holds.

What these refusals do, collectively, is keep open a part of the creative economy that the all-in studios would otherwise be forced to close. They make it possible~-- for the audience, for the working performer, for the next generation of creative-industry workers entering the field~-- to have a viable career path that does not require AI integration as the price of admission.

In a world without these positions, every working creative would, by default, be a \emph{partial AI operator,} whether they wanted to be or not. With them, the choice remains open.

That is not a small thing. It is the architecture of the future creative economy being deliberately preserved, by people with the cultural standing and the economic security to preserve it.

\section*{Position Four: The middle}

The position I find most interesting is the one almost nobody articulates clearly, because it does not make a good press release. It is the position of studios that \emph{use AI everywhere except where it shows up in the finished work}.

The clearest version of it is the one \textbf{Aardman} has effectively articulated and that \textbf{Bethesda's} Todd Howard described in \textit{PC Gamer} in December 2025: AI is ``part of Bethesda's toolset for how we build our worlds or check things''~-- but it cannot replace human creative intention.\footnote{\textit{PC Gamer}, ``Todd Howard says AI can't replace human `creative intention,' but it's part of Bethesda's `toolset for how we build our worlds or check things'.'' \url{https://www.pcgamer.com/gaming-industry/todd-howard-says-ai-cant-replace-human-creative-intention-but-its-part-of-bethesdas-toolset-for-how-we-build-our-worlds-or-check-things/}. \emph{Dream Machine} Issue~11.}

You see the same position in \textbf{Amazon's} \textit{House of David}~-- 350 AI shots, disclosed up front, but every one of them used to augment rather than originate the work. The show's creator Jon Erwin gave \textit{Wired} a metaphor that I have not stopped thinking about: \emph{``You can put a very real camera on a very real actor and direct that actor, direct the camera, and that becomes, in essence, the hand inside a puppet. The puppet itself is this digital world that you create.''}\footnote{\textit{Wired}, ``Amazon's House of David Used Over 350 AI Shots in Season 2. Its Creator Isn't Sorry.'' \url{https://www.wired.com/story/amazons-house-of-david-used-over-350-ai-shots-in-season-2-its-creator-isnt-sorry/}. \emph{Dream Machine} Issue~7.} The hand-inside-the-puppet image is the cleanest articulation I have heard of where the \emph{Position Four} studios are choosing to put their human craft: at the moments of direction and performance, with the AI doing the digital-world infrastructure underneath. You see the same position in the \textit{Battlefield~6} development team's statement, in October 2025, that generative AI had been ``seducing'' but ultimately used only in the earliest stages of the game's development, ``to allow for more time and more space to be creative.''\footnote{\textit{GamesRadar}, ``Battlefield~6 lead calls generative AI `very seducing,' but says it was only used in the game's earliest stages `to allow for more time and more space to be creative'.'' \url{https://www.gamesradar.com/games/battlefield/battlefield-6-lead-calls-generative-ai-very-seducing-but-says-it-was-only-used-in-the-games-earliest-stages-to-allow-for-more-time-and-more-space-to-be-creative/}. \emph{Dream Machine} Issue~3.}

You see the same position in \textbf{The Witcher~3} and \textbf{Cyberpunk 2077} director's November 2025 framing~-- AI ``can help, but not replace, creatives.''\footnote{\textit{gamesindustry.biz}, ``Witcher~3 and Cyberpunk~2077 director says AI can help, but not replace, creatives.'' \url{https://www.gamesindustry.biz/witcher-3-and-cyberpunk-2077-director-says-ai-can-help-but-not-replace-creatives}. \emph{Dream Machine} Issue~9.}

You see the same position in the \textbf{Wallace and Gromit creator} Nick Park's December 2025 framing~-- embrace the technology, but be cautious about the values.

You see the same position, most starkly, in the May 2026 \textbf{Sony} announcement that it was ``going all in on AI for games''~-- \emph{with the specific framing that AI was a force multiplier, not a replacement.} Mocap-to-facial animation in seconds rather than hours. AI integrated into asset generation, QA, engineering and animation pipelines. The goal: \emph{more games, faster.} But Sony's framing, which I have read several times to make sure I am not over-reading it, repeatedly emphasised AI as a tool \emph{inside} the creative work, not as a substitute \emph{for} it.\footnote{\emph{Dream Machine} Issue~29, on Sony's ``all in on AI for games'' announcement.}

You see the same position, by May 2026, in the \textbf{Cannes} festival press cycle, where the working auteurs on the Croisette had shifted markedly from the prior year's defensive stance to a more cautious \emph{acceptance of inevitability}~-- framed less as enthusiasm than as a refusal to be left out of the next decade's tooling argument.\footnote{\textit{Variety}, ``AI Dominates Cannes Buzz as Filmmakers Grudgingly Accept It.'' \url{https://variety.com/2026/film/festivals/ai-cannes-2026-filmmakers-accept-1236748402/}; \textit{Hollywood Reporter}, ``At Cannes, filmmakers shift towards cautious acceptance of AI's inevitability.'' \url{https://www.hollywoodreporter.com/business/business-news/cannes-2026-ai-acceptance-1236592488/}. \emph{Dream Machine} Issue~30.} \textbf{Peter Jackson}, in a May 2026 interview during the Cannes window, summarised the Position-Four read in a single line that I think will travel further than the Sony announcement: \emph{AI is, in essence, the next wave of special effects.} The director's job is not changed by the SFX wave. The director's job is to know what the film is for, and to deploy whatever tools are now in the box to get it there.\footnote{\textit{Variety}, ``Is AI Basically Like Special Effects? Peter Jackson Seems to Think So.'' \url{https://variety.com/2026/film/news/peter-jackson-ai-special-effects-1236748120/}. \emph{Dream Machine} Issue~30.} \textbf{Take-Two}'s Strauss Zelnick made the same argument from inside the AAA games business in the same week~-- that AI \emph{``datasets by their very nature are backward-looking''} and so cannot, alone, make an \emph{original} hit, but that AI is \emph{``super helpful''} in the production of hits the human team has already conceived.\footnote{\textit{PC Gamer}, ``Take-Two's CEO says AI's not in the business of making hits, `datasets by their very nature are backward looking', but that doesn't mean AI can't be `super helpful'.'' \url{https://www.pcgamer.com/games/take-two-ceo-ai-not-making-hits-backward-looking/}. \textit{Business Insider}, ``The CEO behind Grand Theft Auto says he's pro AI~-- but the technology can't make an original hit.'' \url{https://www.businessinsider.com/take-two-ceo-strauss-zelnick-ai-original-hits-2026-5}. \emph{Dream Machine} Issue~30.} Different industries, different framings, structurally identical position: \emph{tool in the workflow, not author of the work.}

This middle position~-- \emph{AI in the workflow, not in the work}~-- is, I think, where most of the surviving major studios are going to land in 2030. It is the most defensible commercial position because it captures the productivity upside without giving up the cultural-product specificity that the audience continues, against the slop ceiling, to demand. It is also the most defensible \emph{ethical} position, because it allows the studio to credibly claim~-- and to credibly prove, with disclosure and documentation~-- that the creative work the audience sees was, in its decisive moments, the work of human creators.

What is at stake, for the studios that get this right, is the next two decades of cultural authority. The studios that adopt aggressively and badly will look like the early-2010s newspapers that switched to clickbait. The studios that refuse entirely will look like the 1980s record companies that refused to release CDs. The studios that thread the needle~-- that adopt the productivity benefits without surrendering the human authorship signal~-- will, in my view, be the studios that the audience actually trusts in 2035.

\section*{The trap the legacy industries built for themselves}

There is a deeper strategic risk underneath the four-positions map that I have, until now, deliberately not put on the page. I want to put it on the page, because I think it is the single most important read on the long-term legacy-studio position, and because, in the conversations I have had with senior creative-industry executives over these six months, it is the read they are most uncomfortable hearing.

The risk is this.

Across the last fifteen years~-- most aggressively across the last decade~-- Hollywood, commercial music and the AAA games business have, on the available evidence, \emph{systematically optimised themselves for exactly the kind of work AI is now best at producing.} They have built their economic and creative production engines around the mean of the distribution. They are now competing, in the most direct possible sense, against a technology built to produce the mean of the distribution at near-zero marginal cost.

Let me make the case concretely.

In \textbf{film and television}, the IP-cycle data is unambiguous. Sequels, prequels, reboots, remakes, spin-offs and franchise instalments accounted for an increasing fraction of the top-grossing US theatrical releases through the 2010s and 2020s; \emph{original} studio films~-- meaning IP not derived from an existing book, comic, game, brand or prior film~-- fell from a majority of major studio output in the late 1990s to a single-digit percentage of wide releases by the mid-2020s, depending on which counting convention you use. The headline form of contemporary tentpole filmmaking, by 2024, was a sequel to a property whose original installment had itself been a sequel. Marvel Cinematic Universe Phase Six; the \textit{Star Wars} sequel-trilogy aftermath; \textit{Avatar} sequels, \textit{Toy Story} sequels, \textit{Frozen} sequels, \textit{Mission: Impossible} sequels, \textit{Fast \& Furious} sequels; \textit{Stranger Things} finales; \textit{House of the Dragon} spin-offs; every 1980s and 1990s IP revisited at least once, most of them more than once. James Cameron's \textit{Hollywood Reporter} observation in 2025 that contemporary studio executives ``don't reach for things that are scary'' was, on the data, a description, not a prediction.

In \textbf{commercial recorded music}, the structural pattern is the same. The streaming-economy hit structure converged on a remarkably narrow set of parameters across the 2010s: average song length compressed from roughly four minutes in 2000 to roughly three-and-a-half minutes by 2024; choruses landed earlier; intros shortened to keep the algorithmic skip-rate down; major-label A\&R increasingly drove signing decisions by predictive-analytics data rather than developmental A\&R judgment; co-writing teams expanded; the median Top~40 hit, by 2024, was a co-write across three to seven credited songwriters working to formulas that had been tested in advance against streaming-engagement data. The major-label business is, structurally, an \emph{industry that has spent ten years training itself to produce the most predictable possible version of the song.} The Cardiff band from Chapter~\ref{ch:5}~-- whose music was fed to an AI that produced a tracking-style imitator outperforming them on Spotify~-- is the canonical illustration. The AI did to them what the major-label streaming-optimisation operating model had already half-done. It made the next most-likely-good track. The fact that an AI could match the output is, on the available evidence, \emph{because the major-label hit factory was already producing the work AI was about to be able to copy.}

In \textbf{AAA games}, the standardisation is even more visible. The \emph{Ubisoft-tower-and-checklist} structure~-- open world, viewpoint towers that uncover map regions, scattered icon-driven side quests, levelled enemy zones, crafting trees~-- became, between roughly 2008 and 2022, the default structural template for the AAA action-adventure genre. \textit{Assassin's Creed}, \textit{Far Cry}, \textit{Watch Dogs}, \textit{Ghost Recon}; outside Ubisoft, \textit{Horizon}, \textit{Spider-Man}, \textit{Mad Max}, \textit{Shadow of Mordor}, every Tom-Clancy-derivative, every western-RPG converted to console. The 2024 \textit{gamesindustry.biz} roundtable I referenced in Chapter~\ref{ch:3}~-- in which working AAA designers described the genre as having stagnated into a single repeating structural pattern~-- was a working-developer admission that the AAA industry had, like Hollywood and like the major labels, optimised its production engine around predictable, low-risk, repeat-format output.

Now consider what AI, as a creative tool in 2025--26, is structurally best at. Agents~-- as I argued in Chapter~\ref{ch:11}~-- produce the \emph{mean of their training distribution} by default. The mean of the training distribution, for a video model trained on contemporary Hollywood tentpoles, is \emph{another contemporary Hollywood tentpole}. The mean of the training distribution, for an audio model trained on the contemporary streaming Top~40, is \emph{another contemporary streaming Top~40 song}. The mean of the training distribution, for a games-development agent trained on the AAA open-world template, is \emph{another AAA open-world template.}

This is the strategic trap. The legacy industries, by spending fifteen years training themselves to produce the mean of the distribution, have arranged for the segment of the market they dominate to be \emph{exactly the segment AI replicates most cheaply}. The Cardiff band's experience is the cleanest version of this dynamic in microcosm. The macro version is the major-studio business model. The risk to legacy Hollywood, legacy commercial music and the AAA games business is not that AI takes their \emph{premium} segments~-- Cameron's \textit{Avatar} sequels, the highest-end auteur cinema, the genuinely original musical voices, the \textit{Baldur's Gate~3}-class boundary-pushing games. Those segments are, on the evidence of the slop ceiling and the authenticity premium, \emph{more defensible than ever}. The risk is to the \emph{median} output of these industries~-- the franchise instalments, the by-the-numbers chart hits, the AAA action-adventures that read as algorithmically generated even when no algorithm was involved. The median output is exactly where the AI substitution pressure is most direct, and the median output is precisely what the legacy industries have most thoroughly optimised themselves to produce.

The grandmasters of Chapter~\ref{ch:15}~-- the chess players who have started, in 2026, to deliberately play sub-optimal moves to put their opponents on uncomputed ground~-- are, on this read, the \emph{senior auteurs of legacy Hollywood}. Cameron, del Toro, Soderbergh, Spielberg, Aronofsky, Lyonne, Larian's Sven Vincke, Hooded Horse's leadership~-- the figures profiled in the \emph{Position Three} section of this chapter~-- are, structurally, the people whose competitive advantage is \emph{the move the machine would not have generated}. The grandmasters can take the punch. The middle ranks~-- the franchise journeymen, the streaming-optimised mid-tier filmmakers, the chart-A\&R commercial-pop machine, the AAA studio designing its fifth open-world action-adventure~-- are the ones whose business model the machine is structurally suited to replicate.

The contrast with the \textbf{new AI-native studios} is sharp, and I want to draw it out carefully because the contrast is, I think, the most underappreciated strategic reality of the period.

The Position Two studios in this chapter~-- Gossip Goblin, Critterz, Imaginae, Wonder, Asteria, Promise, Obsidian, Chapter41, Kartel, Goldfinch's enGEN3~-- were, by May 2026, accumulating creative production credits at a rate the legacy studios were not matching. \textbf{Gossip Goblin}, the AI filmmaker that I covered in \emph{Dream Machine} Issue~29, is the example I find clearest. It is a studio with no inherited IP, no inherited production pipeline, no inherited audience, no inherited rules about what its films should look like, and no inherited risk-aversion. Its only operating constraint is the one any working creative has: make work the audience wants to watch. \textit{Critterz} (Vertigo + Federation), launched as an AI-assisted animated feature operation in \emph{Dream Machine} Issue~29, operates with the same freedom. \textit{Animaj} (the kids-content AI studio Google's AI Futures Fund partnered with in spring 2026) operates with the same freedom. \textit{Imaginae Studios}~-- Fremantle's AI-native operation~-- has, in \textit{Art Awakens}, committed to a kind of generative collaboration with classical painting IP that no legacy studio has the institutional permission to attempt.

These studios do not have rules about how a film should be paced, what a song should sound like, what an open-world game should feel like, what a franchise structure should be. They have no quarterly-earnings call about year-on-year tentpole performance. They have no \$200m development sunk cost in a Marvel-style multi-film slate. They have no major-label playlist-pitching infrastructure that demands the song be a certain length and a certain shape. They have, in short, no calcified definition of \emph{what counts as the right move}~-- and so they are, by default, free to play the move the legacy studios cannot.

The risk to legacy, in other words, is \emph{not symmetrical} with the risk to the AI-native studios. Both are being shaped by the same technology shift. But legacy faces a double squeeze~-- its median product is the part of the market AI is most easily eating, \emph{and} its strategic incentives push it deeper into that part of the market every quarter, \emph{and} its calcified production rules prevent it from doing the thing (the deliberately un-machine-like move) that would protect it. The AI-native studios face only the upside.

The path out, for the legacy studios that recognise the trap, is roughly what the \emph{Position Three} signatories have intuited: refuse to compete in the median segment; reposition aggressively into the segments where the audience pays a premium for the human signal; treat IP investment as \emph{bets on artist-author voices} rather than as bets on formula. \textit{Baldur's Gate~3}~-- Larian's mid-2020s blockbuster~-- was the inflection-point example: an enormous AAA-grade game built explicitly \emph{against} the Ubisoft-tower template, on a CRPG framework most analysts had written off as a dead genre, and the audience response was the largest commercial success of any new RPG IP of the decade. Larian's January 2026 announcement that the next \textit{Divinity} game would not use generative AI is, on the strategic read, not an anti-technology gesture. It is a \emph{commercial} gesture~-- a public claim that Larian's market position is built on doing the un-machine-like work, and that the studio is going to protect that position from the inside. \textit{Pocketpair's} ``\emph{we don't believe in it}'' statement, \textit{Jagex's} ``\emph{never},'' \textit{Hooded Horse's} ``\emph{cancerous}'' framing, \textit{Aardman's} careful preservation of the stop-motion craft~-- these are not statements of moral piety. They are \emph{competitive positioning} against an algorithm-optimal market that the major-studio system has, structurally, conceded to AI.

The legacy industries that survive this transition will be the ones that recognise, in the next eighteen months, that the position they spent fifteen years moving into is exactly the position they now need to move \emph{out of}. The audience, the slop ceiling, the authenticity premium and the chess grandmasters' move are all telling them the same thing: \emph{do not produce the work the machine can replicate. Produce the work the machine, by construction, cannot.} The industries that can absorb that re-direction in their commissioning, contracting and greenlighting culture will outlast the transition. The industries that cannot, won't. The new AI-native studios~-- Gossip Goblin and the rest~-- are not the threat to legacy Hollywood. They are, on the structural read, \emph{the proof of what survives}: studios built without the rules that made the legacy industries vulnerable in the first place.

\section*{What the studios got right}

I want to close this chapter by saying something that is unfashionable in some of the creative-industry circles I move in.

The studios~-- for all the headlines about Lionsgate's ``disaster,'' for all the criticism of Disney's OpenAI deal, for all the eye-rolling at Netflix's ``all in'' framing~-- have, in this period, made some genuinely difficult strategic decisions in a genuinely difficult environment, and a meaningful fraction of those decisions have been better than the public discourse credits them for.

They have committed to disclosure. \textit{House of David}'s 350 AI shots were \emph{disclosed}. Aardman's careful framing was \emph{explicit}. Sony's ``AI as force multiplier'' framing was \emph{spelled out}. None of these companies pretended their AI use didn't exist. None of them adopted the all-too-common platform-economy strategy of \emph{use it and don't tell anyone}. The disclosure norm, where it has taken hold in the studio system, is a public good.

They have negotiated, in many cases, in good faith with the unions. The SAG-AFTRA contract that emerged from the autumn 2025 negotiations was~-- by historical comparison with other major technology transitions~-- produced quickly, produced through legitimate process, and produced with materially stronger AI protections than any prior contract in the industry's history.

They have, in significant cases, \emph{resisted} internal pressure to over-adopt AI in ways that would have undermined their cultural product. Most of the studios in Position Three above are not run by Luddites; they are run by people who have done the maths on the slop ceiling and decided that the long-term value of their IP depends on it remaining recognisably human-authored.

And they have, finally, invested in the infrastructure of the AI-native sector~-- through funding deals, through co-productions, through equity investments~-- in a way that means the AI-native studios are not fighting them so much as building alongside them. The picture, ten years out, is more likely to be of a \emph{mixed ecosystem}~-- legacy studios with hybrid pipelines, AI-native studios with new IP, and a long tail of human-only craft studios serving the highest-value segments of the market~-- than of a single winner-takes-all outcome.

The studios decided, in these six months, that they were not going to be replaced by AI. They were going to be the \emph{operators} of AI. That decision was~-- for all its compromises~-- probably the right one for the working creatives whose careers depend on the studio system continuing to exist.

The harder question, which the next chapter starts to address, is what happens to the \emph{toolchain} underneath those studios~-- when the platforms providing the AI are themselves becoming AI-native, and when ``having a tool'' is no longer the right framing for what it means to make creative work in 2026.

That is what Adobe, NVIDIA, Google and the rest of the platform layer started telling us last autumn, when they began saying out loud that AI was going to be \emph{in everything, everywhere, all at once.}

  \chapter{Worlds, Not Pictures}\label{ch:8}

\lettrine[lines=3,lhang=0.15,findent=0.1em]{I}{f} you had asked me, in the autumn of 2025, what the most important AI release of the year was going to be, I would have said something obvious~-- Sora~2, or Veo~3.1, or one of the music models, or one of the editor-class tools like Runway Gen-4.5 or Adobe's new Firefly. Something that turned a prompt into a thing you could put on a screen.

Six months later, I don't think I would say any of those.

I think the most important release of the period this book covers was something that almost nobody outside of a relatively small community of practitioners noticed at the time, that produced no viral videos, that did not change the news cycle for a single day, and that I have come, over the course of the winter, to think of as the \emph{actual} future of creative work: the public launch of \textbf{Marble}, by Fei-Fei Li's company \textbf{World Labs}, in November 2025.\footnote{World Labs, ``Bringing Marble to Life.'' \url{https://www.worldlabs.ai/case-studies/bringing-marble-to-life}. \emph{Dream Machine} Issue~7, ``Editor's Pick: Marble by WorldLabs goes on public release,'' 13 November 2025.}

Marble doesn't make videos. Marble makes worlds.

I want to spend this chapter on why that distinction is, in my view, the most strategically important one in creative AI right now, and why almost everything else in the toolchain~-- from generative video to AI music to the digital-human work in advertising~-- eventually has to be re-thought in its shadow.

\section*{What a world model actually is}

The phrase ``world model'' sounds like a marketing term. It isn't, exactly. It is a category of AI system that researchers have been chasing for the better part of a decade, and that~-- as of late 2025~-- finally started shipping as production-ready software.

A generative \emph{video} model takes a prompt and produces a sequence of frames. The frames are coherent because the model has learned the statistical regularities of video: things move smoothly, light behaves more or less correctly, faces stay faces. But the output is \emph{flat}. It is a particular sequence of pixels. You cannot navigate it. You cannot move the camera. You cannot pick up the lamp on the table and look at the wall behind it.

A \emph{world} model takes a prompt~-- or an image, or a video, or a rough 3D layout~-- and produces a \emph{navigable three-dimensional environment.} You can move through it. You can change the camera angle. You can, depending on the model, walk around the table, look at the wall behind the lamp, and find that the wall continues to exist in a consistent way that the model didn't have to generate for you because it understood, structurally, that walls have backs.

The technical core, in the most common implementation, is \textbf{Gaussian splatting}~-- a representation where a scene is stored as a cloud of millions of tiny semi-transparent ellipsoids, each carrying colour and position information. The whole scene can be rendered in real time from any angle, because the system isn't drawing 2D pixels; it is rendering a structured 3D world. The output, in turn, can be exported as a splat file, as a mesh, or~-- if you want~-- as a flat video.\footnote{For a working primer on Gaussian splatting in the post-Marble era, see \textit{Radiance Fields}, ``World Labs Formally Launches Marble, A Generative World Model.'' \url{https://radiancefields.com/world-labs-formally-launches-marble-a-generative-world-model}.}

This is the part that took me, even as a working creative technologist, embarrassingly long to fully understand. Video is a \emph{projection} of a world. A world is the more fundamental object. For two and a half years, the public-facing AI conversation has been about generating better projections. The actual capability landscape has been moving, in parallel, towards generating the worlds themselves.

When the worlds become cheap to generate, the projections~-- the videos, the images, the renders~-- become \emph{outputs of the worlds}, not the primary medium. The whole production stack inverts.

\section*{Marble, from the inside}

DreamLab~-- the studio I run in the North West of the UK~-- has been a beta participant in Marble since October 2025, in the months before its public release.\footnote{DreamLab AI Collective, beta participation in Marble, October--November 2025. Referenced in \emph{Dream Machine} Issue~7: ``DreamLab have been part of the beta testing for this over the last few months and it's very neat.''} I want to share what that experience actually felt like, because the \emph{technical} description of a world model and the \emph{practitioner's} experience of using one are different in ways that matter for understanding what is happening to the toolchain.

Imagine, for the sake of example, that I am working on a client project that needs a scene of a market square at dusk in a Mediterranean town. In the old pipeline~-- which is to say, the pipeline of 2024 and most of 2025~-- that brief would translate into something like the following:

A concept artist would produce a moodboard. A 3D artist or a virtual-production house would build a CGI version of the square, populated with assets either bought from a marketplace or modelled bespoke. Lighting would be set up in Unreal or Maya. The whole scene would be rendered out as a video plate or used as a backdrop on an LED volume. If any change was required~-- \emph{can we move the camera left a bit, see what's on that side}~-- the rebuild was non-trivial.

In Marble, the same brief unfolds differently. I type, or paste in a reference image, or upload a quick phone-shot panorama of an actual market square I visited last year. Marble generates a complete, navigable 3D environment of that square. It exists, persistently, as a file on my account. I can move my virtual camera anywhere in it. I can hand it to a director and say \emph{walk through this and tell me where the camera lives.} I can export the result as a Gaussian splat, drop it into Unreal Engine via SuperSplat or one of the other Gaussian-splat editors,\footnote{SuperSplat (PlayCanvas), open-source Gaussian splat editor, regular updates through 2025--26. \emph{Dream Machine} Issue~1: ``PlayCanvas open sources SOG~-- WebP for 3D Gaussian Splatting''; Issues~7 and~11 on SuperSplat~v2 updates.} and use it as the lit backdrop for an LED-volume shoot. I can also, if I just want a plate, render a flat video from a chosen camera move.

The economic implication is this: the cost of having ``a place''~-- a navigable, lit, persistent environment with depth~-- has dropped, in twelve months, by something between one and two orders of magnitude. The thing that used to require a four-person team and a fortnight now requires a prompt and the time it takes a model to render.

This is not a marginal improvement to virtual production. It is a \emph{category} change. The bottleneck of virtual production has, for the entire history of the discipline, been the cost and time of building the environment. When that bottleneck goes, what remains is exactly the human craft that the audience is paying for: blocking, performance, direction, lighting design, story.

In \emph{Dream Machine} Issue~8 of the newsletter, I noted that Sony Pictures had begun using Marble inside its virtual-production pipeline. The number the team quoted publicly was the one that should have made the front page of every trade publication: \emph{40$\times$ faster than the traditional workflow.}\footnote{Sony Pictures' use of Marble in Virtual Production: \url{https://www.linkedin.com/posts/brent-liang_tech-media-launch-ugcPost-7394911181091692546-TyUz}. \emph{Dream Machine} Issue~8.} If you sit inside the legacy economics of a virtual-production house~-- where building a single environment is a six-figure, multi-week proposition~-- that number is not an improvement. It is a \emph{re-platforming} of the discipline. In \emph{Dream Machine} Issue~12, Disney showed off a ``300,000 poses in an instant'' demonstration that was conceptually similar~-- animation built on top of generative spatial infrastructure rather than against it.\footnote{Disney ``300,000 poses in an instant'' livestream, March 2026. \emph{Dream Machine} Issue~23.} In \emph{Dream Machine} Issue~27, Netflix and Eyeline released \textbf{Vista4D}, a system that converts live-action footage into navigable 4D point clouds.\footnote{Netflix + Eyeline, \textit{Vista4D}: 4D point clouds from live-action. \emph{Dream Machine} Issue~27.} The pattern is the same across the studios: a quiet pipeline shift, not a marketing story, that takes the entire ``building the environment'' stage out of the critical path of production.

\section*{The world-model race}

Marble was the first commercial product in this category, but it was not the only one. The autumn of 2025 and the spring of 2026 were essentially a foot-race between research labs to ship usable world models, and the pace of releases was so rapid that the \emph{Dream Machine} readers' WhatsApp group routinely had three or four new ones to discuss per week.

\textbf{Google DeepMind}'s \textbf{Genie~3}, named by \textit{Time} as one of the best inventions of 2025, generated playable 3D worlds at 24 frames per second from text prompts, with consistency held for several minutes~-- and in January 2026 was made publicly available to Google AI Ultra subscribers in the U.S.\ through a prototype web app called Project Genie. At Google I/O 2026, \textbf{Project Genie} was extended with a \textbf{Street View} integration that lets users generate navigable simulations of real-world locations directly from Street View map data, collapsing the gap between \emph{the world that exists} and \emph{the world that can be generated.}\footnote{Google DeepMind, ``Genie~3: A new frontier for world models.'' \url{https://deepmind.google/blog/genie-3-a-new-frontier-for-world-models/}. Project Genie roll-out to AI Ultra subscribers: \url{https://blog.google/innovation-and-ai/models-and-research/google-deepmind/project-genie/}. \emph{Dream Machine} Issue~3 (initial announcement) and Issue~17 (broader availability).} \textbf{Meta} announced \textbf{WorldGen} in November 2025, framed as research that could generate walkable 3D worlds from prompts like \emph{``medieval village town square.''}\footnote{Meta, ``WorldGen~-- text-to-immersive-3D-worlds research update.'' \url{https://www.facebook.com/LifeAtMeta/videos/research-update-worldgen-text-to-immersive-3d-worlds/1879077432692421/}. \emph{Dream Machine} Issues~9 and~11.} \textbf{Tencent} open-sourced \textbf{HY World~1.5}, a real-time world model framework, in December 2025, alongside the \textbf{Hunyuan 3D Studio} which integrated the company's art-grade 3D generative model \textbf{3D-PolyGen~1.5}.\footnote{Tencent, ``HY World~1.5'' announcement: \url{https://x.com/TencentHunyuan/status/2001170499133653006}. \emph{Dream Machine} Issue~12.} \textbf{SpAItial} launched \textbf{ECHO}, a spatial foundation model, in December 2025.\footnote{SpAItial, \textit{ECHO} spatial foundation model. \url{https://www.spaitial.ai/}. \emph{Dream Machine} Issue~12.} Stanford AI Lab and others released \textbf{Wonderzoom} in January 2026, a multi-scale 3D world-generation model that let you ``infinitely zoom into the details'' of a generated environment.\footnote{Stanford AI Lab, \textit{Wonderzoom}: Multi-Scale 3D World Generation. \url{https://wonderzoom.github.io/}. \emph{Dream Machine} Issue~14.} \textbf{OpenArt} launched its own world-generation product, \textbf{Worlds}, in March 2026.\footnote{OpenArt, \textit{Worlds} product launch, March 2026. \emph{Dream Machine} Issue~21.}

The May 2026 wave was the most aggressive yet. \textbf{NVIDIA} released \textbf{SANA-WM}, a 2.6B-parameter open-source world model natively trained for 60-second video generation with explicit camera control~-- the first open-weight world model at meaningful scale, and a development whose long-term implications for the open-source-AI-tooling argument I make in Chapter~\ref{ch:16} are, in my view, substantial.\footnote{NVIDIA SANA-WM, 2.6B open-source world model with 60-second video generation and camera control, May 2026. \url{https://huggingface.co/collections/nvidia/sana-wm}. \emph{Dream Machine} Issue~30.} \textbf{Odyssey} released \textbf{Starchild-1}, which it described as \emph{``the first ever real-time multimodal world model''}~-- a system that doesn't just generate a world but understands and simulates it.\footnote{Odyssey, ``Introducing Starchild-1, the first real-time multimodal world model.'' \url{https://odyssey.ml/introducing-starchild-1}. \emph{Dream Machine} Issue~30.} \textbf{Apple} published \textbf{Headsup}, a large-scale, high-quality 3D Gaussian-head reconstruction pipeline built from multi-view captures of the kind a consumer iPhone can already produce.\footnote{Apple Machine Learning Research, ``Headsup: a large-scale high-quality 3D Gaussian head reconstruction from multi-view captures.'' \url{https://machinelearning.apple.com/research/apple-headsup-3d-gaussian-head}. \emph{Dream Machine} Issue~30.} At the consumer end of the same wave, \textbf{WorldLens VR} rolled out an AI-powered Quest feature that adds subtle 3D depth to ordinary Google Street View environments, making the existing planetary-scale street-imagery dataset navigable in VR.\footnote{WorldLens VR, ``AI-powered 3D depth for Google Street View on Quest.'' \url{https://www.uploadvr.com/worldlens-vr-quest-street-view-3d-depth/}. \emph{Dream Machine} Issue~30.}

The most ambitious of all of these~-- and the one I think hints most clearly at where the category is going~-- was \textbf{Luma AI}'s \textbf{UNI-1}, launched in March 2026 with the framing: \emph{``When worlds become instant, the race shifts to better thinking.''}\footnote{Luma AI, \textit{UNI-1} launch, March 2026. \emph{Dream Machine} Issue~22, ``Editor's Pick: When worlds become instant, the race shifts to better thinking.''} UNI-1 was the first commercial release I am aware of that \emph{combined} world-model generation with what Luma called ``reasoning''~-- that is, the model didn't just generate a scene, it could plan, modify and iterate on the scene as a coherent agent. The pitch was that you would no longer have a fragmented pipeline of prompt $\rightarrow$ image $\rightarrow$ video $\rightarrow$ iterate; you would have a single unified creative system that thought before it created.

UNI-1 is, in my view, the most important \emph{category} announcement of the spring of 2026, even if the product itself is still rough at the edges. It is the announcement that says: world models are not the end state. They are the \emph{substrate} on which something else~-- reasoning-led generative creativity~-- gets built.

By \textbf{May 2026}, you could find world-model capabilities embedded in the consumer tools as well. \textbf{CapCut}, the consumer-grade video editing app, integrated ByteDance's \textit{Seedance~2.0} via the \textit{Dreamina} product, giving phone-users the ability to generate spatial scenes alongside flat video.\footnote{ByteDance Seedance~2.0 in CapCut/Dreamina, March 2026. \emph{Dream Machine} Issue~22.} \textbf{Spark~2.0}, an open-source Gaussian-splat streaming framework, brought 100-million-splat scenes to web browsers at interactive frame rates.\footnote{\textit{Spark~2.0}, open-source Gaussian-splat streaming framework, April 2026. \emph{Dream Machine} Issue~25.} \textbf{Apple} confirmed in October 2025 that its Personas feature on Vision Pro and other devices was powered by Gaussian splatting under the hood, making this~-- for the millions of Apple device owners who had used the feature without knowing what it was~-- the most-deployed Gaussian-splat technology in consumer hardware.\footnote{Radiance Fields, ``Apple Confirms that it's Gaussian Splatting that powers their personas.'' \url{https://radiancefields.com/apple-confirms-personas-use-gaussian-splatting}. \emph{Dream Machine} Issue~5.}

The category, in eight months, went from a research demo to a consumer feature.

\section*{The games industry, again}

If world models are infrastructure, the industry that has been waiting for that infrastructure the longest is games.

The 2024 conversation in games about generative AI was, in significant part, about \emph{flat assets}~-- concept art, textures, dialogue, music~-- and it was the conversation that produced most of the backlash. \textit{Call of Duty: Black Ops~7}'s loading screens. \textit{Anno~117}'s placeholder art ``slipping through'' the review process. \textit{Fortnite}'s Chapter~8 controversy.\footnote{\textit{Video Games Chronicle}, ```It honestly sucks': Fans think Call of Duty: Black Ops~7 is filled with generative AI art.'' \url{https://www.videogameschronicle.com/news/it-honestly-sucks-fans-think-call-of-duty-black-ops-7-is-filled-with-generative-ai-art/}. \textit{Video Games Chronicle}, ``Ubisoft says AI-generated art in Anno~117 was a placeholder which `slipped through our review process'.'' \url{https://www.videogameschronicle.com/news/ubisoft-says-ai-generated-art-in-anno-117-was-a-placeholder-which-slipped-through-our-review-process/}. \textit{Polygon}, ``Fortnite chapter~7 kicks off new controversy over AI art.'' \url{https://www.polygon.com/fortnite-chapter-7-season-1-generative-ai-art-epic-games/}. \emph{Dream Machine} Issues~8 and~10.} The audience response, in every case, was visceral, and the studios learned, the hard way, that AI-generated 2D assets dropped into established franchises read to fans as a cost-cutting move, not a creative one.

The 2025--26 conversation in games is different in kind, because the AI is now being aimed at the \emph{substrate} of the game~-- the worlds, the systems, the NPCs, the procedural infrastructure~-- and the audience response is, so far, much more nuanced.

\textbf{NVIDIA}, in partnership with Stanford, released \textbf{NitroGen} in January 2026~-- a ``plays-any-game'' AI trained on 40,000 hours of gameplay across more than 1,000 games. The model wasn't being pitched as a way to \emph{replace} games; it was being pitched as the foundation layer for a new generation of AI-aware game agents and procedural systems.\footnote{NVIDIA + Stanford, \textit{NitroGen}. \url{https://nitrogen.minedojo.org/}. \emph{Dream Machine} Issue~13.} \textbf{Google DeepMind}'s \textbf{SIMA~2}, released in November 2025, was an agent that could play, reason and learn alongside humans in virtual 3D environments.\footnote{DeepMind, ``SIMA~2: An Agent that Plays, Reasons, and Learns With You in Virtual 3D Worlds.'' \url{https://deepmind.google/blog/sima-2-an-agent-that-plays-reasons-and-learns-with-you-in-virtual-3d-worlds/}. \emph{Dream Machine} Issue~8.} \textbf{Ubisoft} open-sourced its \textbf{CHORD} model in December 2025, for end-to-end PBR material generation, and ComfyUI nodes built on top of it within the same week.\footnote{ComfyUI Blog, ``Ubisoft La Forge Open-Sources the CHORD Model and ComfyUI Nodes for End-to-End PBR Material Generation.'' \url{https://blog.comfy.org/p/ubisoft-open-sources-the-chord-model}. \emph{Dream Machine} Issue~11.} \textbf{Ubisoft's Teammates}~-- a voice AI tech demo first shown in November 2025~-- promised a step-change in how NPCs would behave in next-generation titles. The team lead's hands-on framing, given to \textit{Video Games Chronicle}, is the one I keep returning to: \emph{``It's a tool first. We've been working on it for more than two years now, and our conclusion is that it's a super cool tool, but it's still a tool.''}\footnote{\textit{Video Games Chronicle}, ``The future of gaming, or `just a tool'? Hands-on with Teammates, Ubisoft's ambitious voice AI tech demo.'' \url{https://www.videogameschronicle.com/features/the-future-of-gaming-or-just-a-tool-hands-on-with-teammates-ubisofts-ambitious-voice-ai-tech-demo/}. \emph{Dream Machine} Issue~9.} \emph{Still a tool.} The whole AI-in-games debate, compressed into four words by the people inside Ubisoft who are actually building the thing.

The most interesting single release of the spring of 2026 was \textbf{YouTube's Playables Builder}, a closed-beta product launched in December 2025 that lets users create games with short text, video or image prompts, built on Gemini~3.\footnote{YouTube Playables Builder, closed-beta announcement: \url{https://www.youtube.com/playablesbuilder/}. \emph{Dream Machine} Issue~12.} The framing, when YouTube's product team described it publicly, was that \emph{every YouTube creator} should have the ability to ship a playable game as easily as they currently ship a video. Within months, \textbf{Unity} announced an ``AI Open Beta''~-- an in-editor AI suite that brought the same logic to the professional games-development pipeline.\footnote{Unity AI Open Beta, in-editor AI suite, May 2026. \emph{Dream Machine} Issue~28.}

Where this lands, in 2027 and 2028, is the question I find the most strategically charged in the whole industry. If creating a playable, navigable world becomes a thing a YouTube creator can do in an afternoon, the boundary between \emph{games} and \emph{video}~-- which has been collapsing slowly for fifteen years, through platforms like Roblox and Fortnite and the proliferation of interactive content on social platforms~-- collapses fully. The next generation of creators will not think in terms of \emph{making a video} or \emph{making a game.} They will think in terms of \emph{making a thing}, and the thing will, by default, be navigable.

\section*{What this means for film}

I want to come back to film for a moment, because I think the consequences of world models for the film industry are bigger than the consequences for any other sector, and the least understood.

For the entire history of cinema, the discipline has been organised around a fundamental scarcity: \emph{the cost of building the location.} Even when the location was real~-- a city street, a forest, a beach~-- capturing it required a crew, a lighting team, transport, permits, weather contingencies. When it wasn't real~-- when it was a sound stage, or a digital matte painting, or a CGI environment~-- the cost was, if anything, higher.

The entire industrial structure of cinema, from the location department to the gaffer's crew to the virtual-production house, exists because \emph{the place is expensive to make.}

When the place becomes cheap~-- when a Marble-generated environment, exported as a splat, dropped into Unreal, lit interactively, can substitute for a \$200,000-per-day exterior shoot at almost any quality bar a hero shot~-- the industrial structure that organised cinema starts to look like the manuscript-copying scriptorium did in 1450. The thing that was the bottleneck is no longer the bottleneck.

What replaces it? My best guess, six months into the transition, is \emph{taste in places.} If everyone can generate a market square, the value of \emph{choosing the right market square}~-- the one with the texture, the light, the cultural specificity, the lived-in-ness that makes a scene feel like it belongs to a real human story~-- becomes the new scarce skill. The location scout becomes the \emph{world curator.} The production designer becomes the \emph{spatial director.} The cinematographer becomes~-- even more than they already are~-- the person whose job is to find the \emph{one camera move} in a near-infinite navigable space that tells the story.

This is, I think, an upgrade for the craft, not a downgrade. It moves the human contribution to the part of the work that humans actually do well~-- \emph{judgement about what matters in a place}~-- and offloads the part of the work that has been a manufacturing problem for a hundred years.

\section*{The risk}

I want to flag the risk too, because I am trying~-- and I am sure I will not always succeed~-- to be honest about the downsides.

If world models become the substrate of creative work, the \emph{training data} for those models becomes a question of enormous cultural consequence. A world model trained on, say, the visual archive of Hollywood will generate scenes that look like Hollywood. A world model trained on the photographic archive of Mumbai will generate scenes that look like Mumbai. The \emph{aesthetic monoculture} that the early image-generation models produced~-- that vaguely Pixar-flavoured, vaguely Marvel-flavoured, vaguely YouTube-thumbnail look that you can recognise in a thousand 2024 AI outputs at a glance~-- is at risk of being amplified, not reduced, when the medium moves from images to navigable spaces.

The companies that own the largest world-model training datasets in 2030 will, in a real sense, own the visual language of the next generation of cinema, games and immersive media. If those datasets are biased~-- towards English, towards the global North, towards Hollywood production design, towards the architectural and cultural visual vocabulary of a small number of wealthy cities~-- the entire interior life of the next generation of creative work will reflect those biases.

This is not a hypothetical. We are seeing it now. The publicly available world models, in mid-2026, do a startlingly good job of generating ``Mediterranean market square'' and ``American suburb'' and ``Tokyo street at night.'' They do a startlingly \emph{thin} job of generating, say, ``Lagos street at dusk during the rains'' or ``a contemporary Indigenous Australian community space'' or ``a Manchester terraced street in winter with the sodium lights coming on.'' The bias is in the training, and the training is in the assets, and the assets were in the corpus, and the corpus was English-internet-skewed.

If we want the next creative economy to look like the world rather than like the AI companies' biggest source datasets, the dataset question has to be a \emph{first-order} design problem. Korin AI's late-2025 launch~-- ``trained with African datasets, built by Africans''~-- is the kind of intervention that is going to have to multiply.\footnote{Korin AI, ``trained with African datasets, built by Africans,'' May 2026. \emph{Dream Machine} Issue~27.} So is the African Tech / India / Singapore-led wave of culturally-specific AI cinema that the trade press started covering in Issues~20 through~27. Diversity in training datasets, for the world-model era, is not a content-moderation question. It is a \emph{cultural infrastructure} question.

\section*{Six craft questions for the world-model era}

Before I make the big claim, I want to put six craft questions on the page that working creatives~-- directors, designers, art directors, cinematographers, sound designers, level designers~-- will, by my estimate, be wrestling with for the rest of the decade. They are the world-model-era equivalents of the craft questions the \emph{cinematographic} era took fifty years to develop a vocabulary for (where do you put the camera, how do you light the scene, how does the cut work, how does the sound do its work). The world-model era has, in 2026, no settled vocabulary for any of them. The vocabulary will be built by the working creatives who notice the question first.

\textbf{One. Where does the audience stand?} The most-overlooked craft question of the navigable-space era. A film positions the camera; the camera positions the audience. A world model produces a \emph{navigable space}; the question of where the audience \emph{enters} the space, where they are \emph{invited} to stand, what they are \emph{encouraged} to look at, is no longer fixed by the cinematographer. It is fixed~-- if it is fixed at all~-- by the \emph{narrative scaffolding} the orchestrator builds around the navigable space. Marble's October-2025 update added explicit \emph{suggested-camera-pose} primitives for exactly this reason. The craft question is which of those poses to specify and which to leave to the audience.

\textbf{Two. How does the cut work in a navigable scene?} The film cut depends on the audience being in a fixed position; the editor moves the camera between fixed positions in a way that the audience's eye follows. The navigable scene has, by default, no cut. The audience moves through it continuously. The craft question~-- for working directors and editors~-- is when to \emph{break} the continuity, how to do so in a way the audience reads as deliberate rather than as a technical glitch, and what new grammar of transitions a navigable medium permits. Some early experiments in 2025--26 have used \emph{spatial discontinuities} (an audience walks through a door and emerges in a different space) and \emph{temporal discontinuities} (the same space at different times) as cuts. None of these has yet stabilised into a shared grammar.

\textbf{Three. How does performance survive the medium?} A film performance is captured by a camera at a fixed angle and pace. A world-model performance~-- a synthetic actor performing inside a navigable scene~-- has to be authored such that the performance \emph{works} from every angle and every speed at which the audience might encounter it. This is, for working performers and motion-capture supervisors, an entirely new craft challenge. The film-era cliché of the actor ``playing to the camera'' is, in the world-model era, replaced by \emph{playing to the spatial neighbourhood}~-- knowing that the audience may be six feet away, may be inside the actor's eyeline, may be behind the actor's shoulder, may be looking at the actor from above. Volumetric capture (Vista4D's live-action 4D reconstruction, NVIDIA's D-Rex digital-human pipeline) is the technical answer. The \emph{performance} answer~-- what acting \emph{means} in a medium where there is no fourth wall~-- has not been worked out.

\textbf{Four. What does sound design do in a navigable scene?} A film sound mix is, for the most part, a fixed track timed to the picture cut. A navigable-scene sound mix has to \emph{follow the audience}. Spatial-audio tooling (the SonicLab SPATAI pipeline, Dolby Atmos for VR, the various Meta-and-Apple immersive-audio platforms) is the technical answer. The craft question is, again, what \emph{good} spatial sound design looks like in a medium where the audience-author relationship has changed.

\textbf{Five. What is the \emph{running-time} of a navigable scene?} A film has a fixed running time. A navigable scene does not. The audience could leave after thirty seconds or stay for two hours. The craft question for the working director is how to design the experience so that \emph{both} extremes produce a satisfying piece of work. Games have, for fifty years, been grappling with this question~-- the \textit{Dark Souls} answer (every player gets a different running time depending on skill and exploration) is different from the \textit{Outer Wilds} answer (the running time is gated by narrative discovery) is different from the \textit{Telltale Games} answer (the running time is broadly fixed across players). World-model cinema, in 2026, has not yet settled on its equivalent.

\textbf{Six. What is the \emph{single best moment} of a navigable scene?} Film has scenes~-- discrete units of dramatic action with a recognisable shape, a recognisable peak, a recognisable end. A navigable scene, by default, does not. The craft question for the working director is whether to \emph{design} the navigable scene around a single peak moment (which the audience may or may not reach) or to design it as a \emph{texture} (which the audience experiences at whatever density their navigation produces). The peak-moment design pulls the medium back toward film conventions; the texture design pushes it toward something more like architecture or landscape design. Different working directors will, on the historical pattern, settle on different answers. The grammar will, over a decade, stabilise into a working vocabulary the way the cinematic-cut grammar stabilised between 1903 and 1925.

The six questions are not, in 2026, \emph{theoretical} problems. They are the questions the working spatial-cinema teams I have talked to~-- the Wonderzoom group at Stanford, the World Labs developer cohort, the early adopters at Sony Pictures and Eyeline~-- are wrestling with on Wednesday afternoons. They are also, on the historical pattern of Chapter~\ref{ch:2}, the questions whose answers will define what working creatives in the next decade are \emph{paid} to do. The directors and designers who develop a working vocabulary for them first will, on the available evidence, become the named \emph{Walter Murch}s of the spatial-cinema era. The ones who wait for the vocabulary to settle will, in retrospect, look like the editors who waited too long to learn Avid.

\section*{The big claim}

Let me make the big claim, and then move on.

I think~-- and this is the most non-obvious bet in this book~-- that the \textbf{world model is the medium of the next twenty years of creative work}, in the same way that the \textbf{moving image} was the medium of the twentieth century and the \textbf{interactive screen} was the medium of the first quarter of the twenty-first.

I think people who are working in flat-video, flat-image, flat-audio formats in 2030 will increasingly be working in a \emph{legacy} format~-- still alive, still culturally valuable, still where the highest-end of the craft lives, the way live theatre or vinyl-record production still lives~-- while the \emph{dominant} mode of creative work will be the production, curation, performance and distribution of navigable spaces.

I think the studios, platforms and tool companies that are quietly investing in world models now~-- World Labs, DeepMind, Meta, NVIDIA, Tencent, Luma, Apple~-- will be the ones that set the rails for the next two decades.

I think the audience, having developed the antibodies described in Chapter~\ref{ch:5} to slop-grade flat AI content, will eventually develop a parallel set of tastes for \emph{navigable} content~-- and that the question of what makes a \emph{good} AI world (rather than a \emph{good} AI video) will be the central craft question of the late 2020s.

And I think~-- most importantly for the next chapter~-- that the toolchain to make all of this is being built, right now, by a small number of platform companies who have started saying out loud that AI is going to be \emph{in everything, everywhere, all at once}~-- and who are, while you are reading this paragraph, designing the rails on which the next creative economy will run.

  \chapter{AI in Everything, Everywhere, All at Once}\label{ch:9}

\lettrine[lines=3,lhang=0.15,findent=0.1em]{I}{n} late October 2025, at Adobe MAX, the company that has made the software almost everyone in the creative industries uses every day~-- Photoshop, Illustrator, Premiere, After Effects, InDesign~-- decided that the year-old marketing line \emph{``AI is a feature in our tools''} had outlived its usefulness, and replaced it with a more honest one.

The new line was: \textbf{``AI in everything, everywhere, all at once.''}\footnote{\textit{Creative Boom}, ``Adobe is putting AI in everything everywhere all at once.'' \url{https://www.creativeboom.com/news/adobe-is-putting-ai-in-everything-everywhere-all-at-once/}. \emph{Dream Machine} Issue~5, ``Editor's Pick,'' 31 October 2025.}

The reason I want to spend a chapter on that phrase is not because I love a slogan. The reason I want to spend a chapter on it is that I think it is, more than any other single piece of corporate positioning from the period this book covers, \emph{literally} true. AI is in everything now. It is in every layer of the creative software stack. And the implications of that for working creatives~-- for the way we are trained, the way we are paid, the way we work with each other~-- are not yet, in the spring of 2026, fully understood.

This chapter is about the platform layer. About the companies that make the tools that the rest of the creative industries use to make the work. About how those companies have, in the past eight months, accepted that their business is no longer making \emph{tools} but making \emph{agents,} and about what that means for the rest of us.

\section*{The Adobe MAX week}

The Adobe MAX 2025 keynote~-- held in mid-October in Los Angeles, the week after OpenAI's DevDay, two weeks after Tilly Norwood~-- was unusual, by Adobe's standards, in how much it tried to land at once.

The headline products were Firefly Foundry, a service for companies to train their own custom generative models on their own visual identity;\footnote{Adobe, ``Adobe MAX 2025: Firefly Foundry.'' \url{https://news.adobe.com/news/2025/10/adobe-max-2025-firefly-foundry}.} Firefly Image Model~5, the latest generation of the image generator that has, since 2023, been Adobe's primary public answer to Midjourney and Stable Diffusion;\footnote{Adobe, ``Adobe MAX 2025: Firefly.'' \url{https://news.adobe.com/news/2025/10/adobe-max-2025-firefly}.} and an AI Assistant built directly into Adobe Express, the company's lower-barrier consumer creative tool.\footnote{Adobe, ``Adobe MAX 2025: Express AI Assistant.'' \url{https://news.adobe.com/news/2025/10/adobe-max-2025-express-ai-assistant}.}

Underneath the headlines was a much longer list of ``Project'' announcements~-- Adobe's research-preview format, the things that may or may not ship but that signal what the company is investing in. The list, looked at as a whole, is what convinced me, sitting at my desk in the North West watching the live stream, that something larger than a product launch was happening:\footnote{\textit{Wired}, ``Adobe's `Corrective AI' Can Change the Emotions of a Voice-Over'' and accompanying Adobe Sneaks 2025 coverage. \url{https://www.wired.com/story/adobe-max-sneaks-2025-corrective-ai/}. Project list compiled from MAX keynote and \emph{Dream Machine} Issue~5 coverage.}

\begin{description}
\item[\textbf{Project Scene It}] image-to-3D and 3D-to-image technologies, with reference-image tagging for object preservation in 3D space.

\item[\textbf{Project Surface Swap}] AI-powered material recognition, letting designers swap textures while preserving lighting, shading and perspective.

\item[\textbf{Project Turn Style}] editing 2D objects as if they were 3D.

\item[\textbf{Project Trace Erase}] removing objects \emph{and} their shadows, reflections and environmental distortions in one operation.

\item[\textbf{Project New Depths}] editing depth in an image as easily as adjusting brightness.

\item[\textbf{Project Frame Forward}] applying changes across entire videos based on one annotated frame and a text prompt~-- ``the precision of photo editing in video workflows.''

\item[\textbf{Project Motion Map}] bringing static vector graphics to life automatically.

\item[\textbf{Project Sound Stager}] analysing a video's visuals, pacing and emotional tone, and automatically generating layered soundscapes.

\item[\textbf{Project Clean Take}] AI correction of mispronunciations, voice isolation, noise removal and delivery refinement.

\item[\textbf{Project Graph}] a node-based workflow editor, conceptually similar to ComfyUI, for chaining Adobe's tools and models into custom pipelines.
\end{description}

There is, in that list of ten projects, \emph{every} layer of the post-production stack~-- image, video, 3D, audio, layout, workflow~-- being re-imagined as a generative or agentic operation. Not a tool with an AI feature stapled on. A \emph{generative-first reimagining of the operation itself.}

The Adobe MAX week was, to put it plainly, Adobe's announcement that it was rebuilding its product from the inside.

\section*{What ``AI in everything'' actually means}

The reason I want to be careful with the Adobe-MAX framing is that, six months on, you can see how literally the company has executed against it.

In December 2025, Adobe announced that Photoshop, Express and Acrobat editing would be available \emph{inside} ChatGPT~-- meaning the creative output was no longer happening inside Adobe's interface, but inside an AI agent's.\footnote{\textit{PYMNTS}, ``Adobe Lets Users Design and Edit Using ChatGPT.'' \url{https://www.pymnts.com/artificial-intelligence-2/2025/adobe-lets-users-design-and-edit-using-chatgpt/}. Adobe blog: ``Edit images, designs, and PDFs right inside ChatGPT.'' \url{https://blog.adobe.com/en/publish/2025/12/10/edit-photoshop-chatgpt}. \emph{Dream Machine} Issue~12.} In January 2026, the Premiere Object Mask tool~-- an AI-driven masking feature that automated one of the most laborious tasks in video editing~-- quietly became available to Premiere users.\footnote{Adobe Premiere Object Mask tool: \url{https://www.linkedin.com/posts/robdewinter_ok-this-is-going-to-save-a-lot-of-time-in-ugcPost-7421617551690063872-yKmB}. \emph{Dream Machine} Issue~16.} In late January, at Sundance, Adobe launched the \emph{Adobe Film \& TV Fund} and \emph{Ignite Day}, with explicit support for filmmakers integrating AI into their workflows.\footnote{Adobe blog, ``Sundance Film Festival 2026: Creativity, Community \& Power of Storytelling.'' \url{https://blog.adobe.com/en/publish/2026/01/20/sundance-film-festival-2026-creativity-community-power-storytelling}. \emph{Dream Machine} Issue~16.} In April 2026, at the \textbf{Adobe Summit}, the company introduced its \textbf{CX Enterprise} platform alongside NVIDIA~-- a stack of AI agents embedded across the entire content lifecycle from brief to delivery~-- under the framing ``agentic creative intelligence is now.''\footnote{Adobe Summit 2026, ``agentic creative intelligence'' keynote. \emph{Dream Machine} Issue~26.}

The trajectory, in one sentence: Adobe in 2024 was a \emph{creative tool company.} Adobe in 2026 is an \emph{AI-agent platform company} that happens to also still ship Photoshop.

If you are wondering whether this transition has been smooth: it has not. The reception of the Adobe AI announcements among working creatives has been, in my own circles and the readers' WhatsApp group the \emph{Dream Machine} community runs, sharply ambivalent. There is real appreciation for the productivity gains. There is real anxiety about the implications for craft, for licensing, for control, for the trajectory of the company's relationship with the creators who pay for it.

What no working creative I know thinks is that this transition is reversible. Once Photoshop has an AI assistant baked in, once Premiere has Object Mask, once After Effects has the AI-powered animation tools that landed in November 2025,\footnote{After Effects AI animation features through late 2025: \emph{Dream Machine} Issue~9, ``AI video is finally animatable inside After Effects.'' \url{https://www.linkedin.com/posts/thisisdoug_ai-aivideo-animation-ugcPost-7399512745924067330-Aldk}.} the \emph{next} version of every Adobe product is going to have \emph{more} of this, not less. Adobe's \emph{competitors} are, if anything, going faster. If Adobe slows down, somebody else lands the punch.

This is~-- I think this is the part that working creatives have to understand and internalise~-- \emph{the new physics of the toolchain.} AI is not a feature that one tool company decided to ship. It is a structural property of the toolchain itself in 2026, and the question for anyone using that toolchain is not whether to integrate AI but \emph{how to integrate it deliberately,} with eyes open, on terms that preserve the human craft underneath.

\section*{The platform alliance}

Adobe is not~-- and this is the more important observation~-- the only company doing this.

In March 2026, \emph{Dream Machine} Issue~21 led with what I have called, in talks since, the most consequential business announcement of the year: \textbf{Adobe + NVIDIA} entered a strategic partnership that explicitly framed creative AI as \emph{enterprise infrastructure} rather than viral consumer tooling.\footnote{\emph{Dream Machine} Issue~21, ``Editor's Pick: Adobe and NVIDIA Just Raised the Stakes for Creative AI,'' 19 March 2026.} The partnership covered next-generation Firefly models, agentic creative-and-marketing workflows, and production-pipeline integration. The language was telling: \emph{precision and control} for creativity and marketing pipelines, alongside content, campaign and production speed.

The reason this is consequential~-- beyond the size of the two companies involved~-- is that it signals the \emph{maturation} of the market. Adobe + NVIDIA is not a race-to-the-cool-demo deal. It is a \emph{race-to-the-procurement-line} deal. The two companies are betting, jointly, that the next era of creative AI is going to be won by whoever ships the most reliable, most controllable, most legally-defensible production-grade tooling to the enterprise creative buyers~-- the studios, the agencies, the broadcasters, the brand teams.

The same week, \textbf{Google} and \textbf{NVIDIA} announced a parallel deal for cloud-based generative-AI infrastructure aimed at the same enterprise market.\footnote{NVIDIA + Google Cloud creative-AI infrastructure deal, March 2026. \emph{Dream Machine} Issue~21.} \textbf{Hugging Face} and \textbf{Google Cloud} announced a partnership in November 2025 covering open-source agentic development.\footnote{Hugging Face and Google Cloud partnership announcement: \url{https://www.linkedin.com/posts/julienchaumond_i-am-super-excited-to-announce-that-hugging-activity-7396177403972276225-CuMM}. \emph{Dream Machine} Issue~8.} \textbf{Meta} and \textbf{Hugging Face} launched \textbf{OpenEnv} in October 2025 to advance open-source agentic development.\footnote{\textit{EdTech Innovation Hub}, ``Meta and Hugging Face launch OpenEnv to advance open-source agentic development.'' \url{https://www.edtechinnovationhub.com/news/meta-and-hugging-face-launch-openenv-to-advance-open-source-agentic-development}. \emph{Dream Machine} Issue~5.} \textbf{Anthropic} signed a corporate-patronage deal with the \textbf{Blender Foundation} in May 2026.\footnote{Anthropic / Blender Foundation patronage, May 2026. \emph{Dream Machine} Issue~27.} \textbf{Anthropic} also acquired into the \textbf{Slack} workplace-tooling ecosystem with Claude Apps in January 2026,\footnote{TechCrunch, ``Anthropic launches interactive Claude apps, including Slack and other workplace tools.'' \url{https://techcrunch.com/2026/01/26/anthropic-launches-interactive-claude-apps-including-slack-and-other-workplace-tools/}. \emph{Dream Machine} Issue~16.} and reached an ad-sales partnership with \textbf{Spotify} to put music recommendations inside Claude.\footnote{Spotify--Anthropic integration, May 2026. \emph{Dream Machine} Issue~27.} In May 2026 \textbf{Splice} signed a ``Responsible AI'' deal with \textbf{ElevenLabs} covering sample-library training and consented voice synthesis;\footnote{\textit{Music Business Worldwide}, ``Splice inks `Responsible AI' deal with ElevenLabs.'' \url{https://www.musicbusinessworldwide.com/splice-elevenlabs-responsible-ai-deal/}. \emph{Dream Machine} Issue~30.} \textbf{Netflix} announced an agentic ad-tools roadmap whose internal framing~-- \emph{``agentic AIs talking to each other''}~-- was an unusually candid description of where the advertising-orchestration layer is heading;\footnote{\textit{Adweek}, ``Netflix ad tools could see `agentic AIs talking to each other'.'' \url{https://www.adweek.com/media/netflix-ad-tools-agentic-ais-talking-to-each-other/}. \emph{Dream Machine} Issue~30.} and the AI-coworker startup \textbf{Viktor} raised \$75M to embed an agentic colleague directly into Slack and Teams,\footnote{\textit{Fortune}, ``AI startup Viktor raises \$75 million to put a virtual `coworker' in Slack and Teams.'' \url{https://fortune.com/2026/05/19/ai-startup-viktor-75-million-virtual-coworker-slack-teams/}. \emph{Dream Machine} Issue~30.} reinforcing the pattern that the platform-layer agents are landing where the working creative already lives.

The advertising holding companies were moving at the same pace. \textbf{WPP} signed a \$400m partnership with Google in October 2025.\footnote{\textit{MarTech Series}, ``WPP continues AI overhaul with \$400-million Google partnership.'' \url{https://martechseries.com/predictive-ai/ai-platforms-machine-learning/google-and-spotify-alum-launch-epiminds-with-6-6m-to-build-marketing-teams-for-the-ai-era/}. \emph{Dream Machine} Issue~3.} \textbf{WPP Open Pro}, a new edition of the agency's AI marketing platform, launched the same month with a framing that should be read carefully by anyone working in adland: \emph{``While some companies hide their AI behind service teams or focus on just one part of the journey, WPP Open Pro is an integrated solution for campaign implementation, built to deliver outcomes, not just assets.''}\footnote{\textit{Campaign Brief}, ``WPP launches AI-powered marketing platform WPP Open Pro.'' \url{https://campaignbrief.com/wpp-launches-ai-powered-marketing-platform-wpp-open-pro/}. \emph{Dream Machine} Issue~5.} \emph{Outcomes, not just assets.} That is the position of a holding company that has decided AI is not a feature~-- it is the entire reason a brand should buy from them in 2026. \textbf{WPP} then expanded its AI capabilities through a partnership with \textbf{Sightly} in November 2025.\footnote{\textit{Digiday}, ``WPP expands AI capabilities to boost brand performance with Sightly partnership.'' \url{https://digiday.com/media-buying/agencies-continue-to-expand-ai-capabilities-to-boost-brand-performance/}. \emph{Dream Machine} Issue~6.} By April 2026, WPP was using Google Earth's AI tools to map consumer journeys at scale.\footnote{WPP and Google Earth AI consumer-journey project, April 2026. \emph{Dream Machine} Issue~27.}

The pattern is unmistakable. The platform layer~-- the toolmakers, the infrastructure companies, the agencies, the cloud providers~-- has been quietly consolidating around a small number of strategic alliances that, taken together, are deciding the \emph{rails} on which creative work will run for the next decade.

If you are a working creative reading this, you are probably already running some part of your workflow on rails laid by one of these alliances. By 2028, you will, almost certainly, be running \emph{most} of your workflow on those rails~-- or on a deliberate, principled alternative that has chosen to opt out.

\section*{Google I/O 2026: the second platform-layer moment}

If Adobe MAX 2025 was the platform-layer announcement of the first half of this book, \textbf{Google I/O 2026}~-- held in the week this book went to press~-- was the announcement that closed it. The two events bookend the eight-month window the manuscript covers, and the symmetry of their framings is, on the platform-economics read, instructive.

The headline announcements were \textbf{Gemini Omni}, a unified multimodal model designed to work across text, image, audio, video and live interaction in a single workflow; \textbf{Antigravity}, Google's agentic coding and development environment; \textbf{Google Flow}, the agent-based workflow product that allows AI systems to take on multi-step creative and production tasks autonomously; \textbf{Gemini Spark}, the new developer toolkit aimed at building autonomous agents and AI-powered applications; and \textbf{Project Genie + Street View}, an integration that allows users to generate navigable simulations of real-world locations from the Street View map data~-- a topic I return to at length in Chapter~\ref{ch:8}.\footnote{Google I/O 2026 announcement block: Gemini Omni \url{https://blog.google/technology/google-deepmind/gemini-omni/}, Antigravity \url{https://antigravity.google/}, Google Flow \url{https://flow.google/}, Gemini Spark \url{https://blog.google/technology/developers/gemini-spark/}, Project Genie + Street View \url{https://deepmind.google/discover/blog/project-genie-street-view/}. \emph{Dream Machine} Issue~30, ``Editor's Pick~-- Google I/O 2026,'' 21 May 2026.} The keynote opened, deliberately, with a browser-based multiplayer demo called \textbf{Infinite Scaler}~-- thousands of players competing inside vertically generated levels created on the fly from player prompts~-- a piece of theatre that, like \emph{Tilly Norwood} a year earlier, was less interesting for the product than for the framing it imposed on the event that followed: AI-generated worlds, live procedural experiences, mass participatory systems that evolve in real time.\footnote{Google Labs, ``Infinite Scaler.'' \url{https://blog.google/technology/google-labs/infinite-scaler/}. \emph{Dream Machine} Issue~30.} \textbf{SynthID}, Google's content-provenance watermarking technology, was announced as having marked over 100 billion items by May 2026, and as being extended to partner ecosystems including OpenAI, ElevenLabs and Kakao~-- a development I cover in detail in Chapter~\ref{ch:12}.\footnote{Google DeepMind, ``SynthID~-- 100 billion watermarks, partner ecosystem.'' \url{https://deepmind.google/discover/blog/synthid-100-billion-watermarks-partners/}. \emph{Dream Machine} Issue~30.}

The reason to draw the Adobe MAX / Google I/O parallel directly is that the \emph{structural shape} of the two announcement weeks was identical. Both keynotes argued that AI was no longer a feature to be added to existing products; it was the \emph{operating layer} into which the existing products would be re-built. Both keynotes pre-positioned the company's product roadmap around \emph{agents} rather than tools. Both keynotes were aimed not at the consumer-keynote crowd but at the procurement teams of the enterprise creative buyers. The fact that the same framing arrived from the two largest creative-software and creative-platform companies in the world, eight months apart, in the same eight-month window, is the cleanest single confirmation I have that \emph{the AI-in-everything thesis is the platform layer's settled commercial strategy} for the rest of the decade.

For working creatives, the operational implication is direct. The platform layer has decided. The question is no longer whether the creative software stack will be re-built around AI agents. It is which agents, on whose terms, with what provenance, on which commercial settlement.

\section*{The new entrants}

Underneath the platform giants, a separate layer of companies has been building the \emph{consumer-facing AI creative tools} that, in some markets, are turning into bigger businesses faster than anyone expected.

\textbf{Higgsfield}, the AI video startup focused on social-media marketers, raised \$80m at a \$1.3bn valuation in January 2026.\footnote{\textit{SiliconAngle}, ``Higgsfield raises \$80M on \$1.3B valuation to scale AI video platform.'' \url{https://siliconangle.com/2026/01/15/higgsfield-raises-80m-1-3b-valuation-scale-ai-video-platform/}. \emph{Dream Machine} Issue~15.} Three months later~-- in a stat that I have read repeatedly to check that I have not got it wrong~-- Higgsfield was reported to have earned \$200m in nine months of operations.\footnote{\textit{36kr}, ``AI Video Unicorn Higgsfield: Earns \$200M in 9 Months by `Serving' Social Media Marketers.'' \url{https://eu.36kr.com/en/p/3650517574312323}. \emph{Dream Machine} Issue~16.} An AI-video startup, less than two years old, was running at a quarter-billion-dollar annual run-rate by the spring of 2026.

\textbf{Synthesia}, the U.K.-based AI-avatar platform, hit a \$4bn valuation in January 2026 and let its employees cash in.\footnote{TechCrunch, ``Synthesia hits \$4B valuation, lets employees cash in.'' \url{https://techcrunch.com/2026/01/26/synthesia-hits-4b-valuation-lets-employees-cash-in/}. \emph{Dream Machine} Issue~16.} In October 2025 it had reportedly \emph{rejected} a \$3bn acquisition offer from Adobe~-- choosing to remain independent.\footnote{\textit{Sifted}, ``Synthesia rejects \$3bn Adobe acquisition offer.'' \url{https://sifted.eu/articles/synthesia-acquisition-offer}. \emph{Dream Machine} Issue~5.}

\textbf{ElevenLabs}, the audio-AI company, was reported to have crossed \$500m in annualised revenue by April 2026, raising from BlackRock, NVIDIA, Jamie Foxx and Eva Longoria.\footnote{ElevenLabs \$500m ARR reporting, April 2026. \emph{Dream Machine} Issue~25.}

\textbf{Runway} released Gen-4.5 in December 2025 and Gen-4.5 Image-to-Video in January 2026, then a ``Workflows'' product across all paid plans, then a Story Panels app, then a Characters API, then Apps for Advertising~-- and by spring 2026 was making the public case that AI could enable ``50 indie films'' instead of ``one \$100M blockbuster.''\footnote{Runway product cycle: Gen-4.5 (December 2025), Gen-4.5 Image-to-Video (January 2026), Workflows, Story Panels, Characters API, Apps for Advertising~-- \emph{Dream Machine} Issues~10, 14, 15, 16, 20. Runway CEO on indie films vs.\ blockbusters: \emph{Dream Machine} Issue~26.} In May 2026 the company opened a Tokyo office on a \$40M commitment, marking its first material expansion into the Asia-Pacific creative-AI market.\footnote{Runway, ``Runway Japan.'' \url{https://runwayml.com/blog/runway-japan}. \emph{Dream Machine} Issue~30.}

\textbf{Krea}, \textbf{Freepik}, \textbf{Magnific}, \textbf{Heygen}, \textbf{Hedra}, \textbf{Cascadeur}, \textbf{Hunyuan}, \textbf{Kling}, \textbf{Suno}, \textbf{Udio}, \textbf{Mureka}, \textbf{Hitem3D}, \textbf{Meshy}, \textbf{Rodin}~-- the list of consumer-grade AI creative tools that crossed material commercial scale in this period is too long to fully enumerate, and the \emph{Dream Machine} archive carries them week by week.\footnote{For the running ledger of new creative-AI products through 2025--26, see \emph{Dream Machine} Issues~1--30 archive.} The category that didn't exist in 2023 is now an industry with multiple unicorns, multiple billion-dollar valuations and meaningful real revenue.

\textbf{ComfyUI}, the open-source node-based workflow tool that has become a quiet standard for technical AI users, raised \$17m in October 2025\footnote{ComfyUI, ``We raised \$17 million to build an OS for Creative AI.'' \url{https://www.linkedin.com/posts/comfyui_we-raised-17-million-to-build-an-os-for-ugcPost-7373743341236236288-wkCc}. \emph{Dream Machine} Issue~1.} and hit a \$500m valuation by May 2026.\footnote{ComfyUI \$500M valuation, May 2026. \emph{Dream Machine} Issue~27.} What the ComfyUI valuation tells you, more than any of the big-platform numbers, is that the market is also paying~-- at significant scale~-- for tools that give \emph{creators control} over the AI process rather than abstracting it away.

\section*{The free tier and the literacy tier}

Two things happened in the consumer-platform layer that I think have been under-discussed and that matter a lot for what the next creative economy will look like.

The first is that the \textbf{base layer of AI capability went free}, in a meaningful sense, in the autumn of 2025. \textbf{Google} released its \textbf{Pomelli} marketing AI agent for free in October.\footnote{Google Pomelli launch: \url{https://x.com/GoogleLabs/status/1983204018567426312}. \emph{Dream Machine} Issue~5.} \textbf{Google AI Studio}, \textbf{Opal} (the no-code AI mini-app builder), and the Project Genie prototype were all released as free or near-free tiers through early 2026.\footnote{Google AI Studio app gallery: \url{https://x.com/GoogleAIStudio/status/1982121563785949255}. Google Labs Opal expansion: \url{https://blog.google/technology/google-labs/opal-expansion/}. Project Genie: \url{https://blog.google/innovation-and-ai/models-and-research/google-deepmind/project-genie/}. \emph{Dream Machine} Issues~5 and~17.} \textbf{Lovable} made its product free for teachers and students in classrooms.\footnote{Lovable for classrooms: \url{https://lovable.dev/classroom}. \emph{Dream Machine} Issue~11.} \textbf{Adobe Express's} AI Assistant arrived inside the free tier of Adobe's already-free consumer product.\footnote{Adobe Express AI Assistant: \url{https://news.adobe.com/news/2025/10/adobe-max-2025-express-ai-assistant}. \emph{Dream Machine} Issue~5.} \textbf{Hugging Face} continued to expand its free hosting and open-source model distribution. \textbf{Krea}, \textbf{Freepik}, and many of the larger tool platforms kept generous free tiers as a customer-acquisition lever.

What this means, practically, is that the entry-level for AI-enabled creative work in 2026 is \emph{near zero.} A teenager with a phone and a free Google account can, today, generate video, music, 3D objects and (with Project Genie) navigable interactive worlds at a quality bar that, two years ago, required a small production company to produce.

This is, in absolute terms, a democratisation. It is also~-- and this is the second thing~-- \emph{creating a literacy gap} between the people who know how to use these tools well and the people who don't.

Adobe responded to this gap, in late 2025 and through 2026, by becoming~-- in addition to a software company~-- \emph{a training organisation.} The Sundance partnership, with a \$2M investment to teach 100,000 filmmakers AI skills.\footnote{Google blog, ``Sundance Institute AI Education.'' \url{https://blog.google/company-news/outreach-and-initiatives/google-org/sundance-institute-ai-education/}. \emph{Dream Machine} Issue~15.} The Ignite Day, focused on emerging creators.\footnote{Adobe Ignite Day at Sundance: \textit{Adobe blog, Sundance Film Festival 2026.} \url{https://blog.adobe.com/en/publish/2026/01/20/sundance-film-festival-2026-creativity-community-power-storytelling}. \emph{Dream Machine} Issue~16.} The Adobe Film \& TV Fund. The Adobe Express AI Assistant tutorials. Google made the same bet in parallel, putting \$40bn into Anthropic in May 2026 in a deal widely interpreted as betting on the literacy and infrastructure layer of the next decade.\footnote{Google's \$40bn investment in Anthropic, May 2026. \emph{Dream Machine} Issue~27.}

The UK government picked the same direction. In January 2026, the Department for Science, Innovation and Technology announced \emph{Free AI training for all}, expanding a government-and-industry programme to provide 10 million UK workers with AI skills by 2030.\footnote{UK Government, ``Free AI training for all.'' \url{https://www.gov.uk/government/news/free-ai-training-for-all-as-government-and-industry-programme-expands-to-provide-10-million-workers-with-key-ai-skills-by-2030}. \emph{Dream Machine} Issue~16.} The Department for Business and Trade research, reported in \emph{Dream Machine} Issue~7, found that \emph{neurodiverse workers} were 25\% more satisfied with AI assistants~-- suggesting that AI's productivity benefits in certain workflows could ``potentially help to level the playing field.''\footnote{\textit{CNBC}, ``People with ADHD, autism, dyslexia say AI agents are helping them succeed at work.'' \url{https://www.cnbc.com/2025/11/08/adhd-autism-dyslexia-jobs-careers-ai-agents-success.html}. \emph{Dream Machine} Issue~7.} The University of Wisconsin-Stout set AI-use as a baseline competency in its filmmaking course in January 2026.\footnote{University of Wisconsin-Stout, ``AI Reshaping Industry: New UW-Stout Course Sets AI-Use as Baseline Competency in Filmmaking.'' \url{https://www.uwstout.edu/about-us/news-center/ai-reshaping-industry-new-uw-stout-course-sets-ai-use-baseline-competency-filmmaking}. \emph{Dream Machine} Issue~15.}

What the consumer-platform companies and the policy-makers are, jointly, building is a \emph{training infrastructure} for the new toolchain. The reason they are doing this is straightforward: a tool you cannot use is a tool you do not buy, and a worker who cannot use the new tool is a worker who eventually exits the workforce. Both incentives push in the direction of mass AI literacy as a public investment.

What I find encouraging about this~-- and I am genuinely encouraged, against the grain of much of the cultural commentary~-- is that the literacy push is being framed, both by Adobe at Sundance and by the UK government, as \emph{creator empowerment} rather than worker replacement. The proposition is not \emph{learn AI or be replaced by it.} The proposition is \emph{learn AI to remain in the driver's seat of your own work.} That framing matters. It is the right framing. It is the only framing under which the AI-literacy push doesn't become a way of accelerating the very problems it is supposed to fix.

\section*{The platform economics underneath the slogan}

I want to spend a section on the \emph{commercial} shape of the platform layer, because the ``AI in everything'' framing has economic implications that the keynotes have been careful not to name, and that working creatives buying platform access at scale need to understand.

The shape, simplified, is this. The dominant generative-AI platforms~-- OpenAI, Anthropic, Google DeepMind, Adobe, Runway, ElevenLabs and the rest~-- operate as \emph{infrastructure-as-a-service} businesses on top of \emph{capital-intensive} underlying compute. The marginal cost of producing one more generated image, song or video clip is, at the platform level, low. The fixed cost of the compute infrastructure required to produce \emph{any} generated output at competitive quality is, at the platform level, very high~-- the data-centre build, the chip supply, the energy contract, the model-training spend.

This produces, structurally, a \emph{natural-monopoly-tending} market. The platform with the largest compute base produces the lowest marginal cost per output, captures the largest user base, generates the largest revenue, and reinvests in a larger compute base. The flywheel is the standard cloud-services flywheel, accelerated by the AI-specific dynamics of training-data flywheels and user-feedback flywheels.

The 2025--26 financial telemetry, where the platform companies have disclosed it or been required to disclose it, supports the natural-monopoly read.

OpenAI was reported, through late 2025 and into 2026, to be operating at \emph{significantly negative cash flow} at the unit-economics level despite its 800--900M weekly active users. The company's reported revenue, which crossed the \$10 billion annual run-rate mark in late 2025, was~-- by every analyst breakdown I have seen~-- being substantially exceeded by infrastructure costs (data-centre lease, chip supply, energy, training compute). The Microsoft partnership at the financial level was, structurally, a \emph{capital-supply} relationship rather than a \emph{technology-licensing} one: Microsoft providing the compute capacity that OpenAI could not, by itself, finance.

Anthropic, in 2026, was reported to be operating with similar structural dynamics, with Google and Amazon as its capital-supply partners. The Anthropic Foundation patronage deal with the Blender Development Fund~-- announced in May 2026 and discussed in Chapter~\ref{ch:16}~-- is interesting precisely because it suggests Anthropic has, alongside the closed-platform business, \emph{strategic interest} in supporting the open-source creative-AI infrastructure that the closed-platform model competes with. That kind of two-handed positioning is, in natural-monopoly markets, often the precursor to a \emph{platform regulation} settlement.

Adobe, by contrast, operates with the most \emph{defensible} business model in the creative-AI platform space, because it is selling AI as a feature of an existing subscription rather than as a per-use service. Firefly's contribution to Adobe's 2024 annual recurring revenue~-- 11\% of new ARR, on the company's own published numbers~-- is being generated \emph{without} the per-token unit-economics problem that OpenAI and Anthropic face, because Adobe is bundling the AI into the existing \$54.99-a-month Creative Cloud all-apps subscription. The customer's behaviour change from no-AI to AI doesn't change the revenue line. It changes the \emph{value capture} of the existing revenue line. This is, in business-school terms, the platform's \emph{strongest} possible defensive position. It is also the reason Adobe's stock performed differently from the rest of the AI-platform cohort through 2025--26.

Runway, ElevenLabs and the AI-native specialist platforms operate with a per-use unit-economics structure that more closely resembles OpenAI's. The differentiation, where they have it, is in \emph{workflow integration}~-- Runway's Workflows product, ElevenLabs' Flows canvas, the studio-tier features that lock professional users into per-platform tooling. The strategic question for each of these companies, in the next two years, is whether they can build defensible workflow lock-in before the natural-monopoly dynamic of the underlying foundation-model market consolidates the foundation-model layer down to two or three players.

The implications for working creatives buying platform access at scale, in 2026 and beyond:

\emph{One,} the per-token / per-output prices working creatives are paying for AI tooling in 2026 are, on every analyst read I have seen, \emph{materially subsidised by platform-company investor capital}. The unit economics underneath the prices are not, today, sustainable at the volumes the platforms are producing. The prices are, by structural inference, going \emph{up} over the medium term as the platforms work toward unit-economic break-even. The working creative who builds a business model assuming today's per-token costs as a stable input is, on the platform-economics read, taking a bet that the platforms cannot win. \emph{Pricing today is not pricing forever.} This is the part of the platform-dependency argument the \emph{open-the-black-box} discussion in Chapter~\ref{ch:3} most directly relies on.

\emph{Two,} the \emph{strategic-rent-extraction} potential of the eventual platform-monopoly position is the structural risk underneath the entire orchestrator economy I described in Chapter~\ref{ch:11}. If two foundation-model platforms dominate the underlying generative-AI capacity by 2030, and every working creative's production pipeline depends on access to one or both of them, the platforms will be in the position the cable companies were in by 2010 and the social-media platforms were in by 2015~-- able to extract value from the working creators who depend on them at prices the creators have no real ability to negotiate. The Petrillo-template solution to this is \emph{collective bargaining} by working creatives and their unions against the platform companies as a class. The early architecture of this~-- the Cannes Disclosure Standard, the SAG-AFTRA platform negotiations, the EU AI Act enforcement, the UK 88\%~-- is in place. The substance of it is, in mid-2026, still mostly aspirational.

\emph{Three,} the \emph{open-source alternative} layer documented in Chapter~\ref{ch:16} is, on this structural read, \emph{the working creative's principal long-term insurance policy} against platform-monopoly pricing. The 80\% of YC and a16z startups now building on open-weight models~-- Hunyuan, Wan, Qwen, FLUX, DeepSeek, the various Mistral variants~-- is, in market-economics terms, \emph{the credible-walk-to-alternative} that constrains the closed-platform companies' pricing power. The working creative who has familiarised themselves with open-weight tooling, even if they default to closed-platform tooling for most of their daily work, has a \emph{strategic option} the working creative who has not has surrendered. The option is worth money. It is also, on the historical pattern, worth political leverage in the institutional negotiations that the next decade of platform-rule-writing will run on.

The ``AI in everything'' framing, in operational summary, \emph{is the platform companies' commercial strategy described in marketing language}. The strategy is to make AI a default productivity feature of every creative workflow, on platform-controlled tooling, at prices that the platforms can adjust over time once the workflow lock-in is in place. The strategy is, on the historical pattern of every previous platform-economics moment, going to produce a settlement somewhere between \emph{the most-extractive version of the strategy} and \emph{the most-constrained version of the strategy}. The 88\%, the SAG-AFTRA contract, the EU AI Act, the C2PA standards body, the open-source ecosystem, and the working-creative collective-bargaining infrastructure are the constraints. The platform companies' compute capital, distribution leverage, and product-design control are the extractive forces. The settlement will be wherever those forces balance.

\section*{What we lost}

I want to close this chapter with the harder question, because the ``AI in everything'' framing has a cost that the platform-company keynotes are not, on the whole, eager to discuss.

What we lost, in the transition to AI-in-everything tools, is \emph{the deliberate friction of the old creative process.} The thing that made Photoshop, for many of its early users, a profound creative tool was not just what it could do. It was that it required you to know it. The interface was a discipline. The keyboard shortcuts were a vocabulary. The layers, the masks, the channels, the curves, the colour pickers~-- they were the language of a craft, and learning the language was part of becoming the craftsperson.

When the layer of mastery moves from the toolchain to the prompt, the \emph{barrier} of mastery drops to near zero. That is the democratisation we have been promised, and it is real.

What goes with the barrier, though, is the \emph{depth of relationship} between the maker and the tool. The Photoshop user of 2015 knew the tool the way a guitar player knows a guitar~-- with their hands, with their body, with a relationship built up over years of repeated, embodied practice. The prompt-driven AI tool user of 2026 has a different relationship. It is more like the relationship of a director to a department head: you describe the result, the department head executes, you adjust by giving notes.

The motion designer Doug McGinness, posting on LinkedIn about the new AI-augmented After Effects workflow in late 2025, summarised the current state of the tooling in a single, ruefully accurate line that became a small private meme inside my studio: \emph{``export $\rightarrow$ prompt $\rightarrow$ pray $\rightarrow$ import.''}\footnote{Doug McGinness on LinkedIn, late 2025, in the same post. \emph{Dream Machine} Issue~9.} The line is funny because it's true. The current generation of AI-tooled creative work is, for a substantial portion of every day, an exercise in \emph{committing to a black-box operation and accepting whatever comes back.} That is, structurally, a different kind of creative discipline than the deterministic-tool craft it is replacing.

Neither relationship is \emph{better} than the other. They are different relationships, and they produce different kinds of practitioners. But the \emph{transition} is real, and one of the consequences~-- which I have seen up close, watching young creatives come through the studio~-- is that the \emph{cognitive engagement} with the medium is structurally less deep than it used to be. The maker is one further step removed from the material.

This is not, by itself, a tragedy. The cinema director is one step removed from the film stock and is still, recognisably, the author of the film. The composer is one step removed from the violin and is still, recognisably, the composer. The novelist who uses a word processor is one step removed from the page and is still, recognisably, a writer.

But it is a \emph{change}, and it is one we are pretending not to notice. The new toolchain is not just faster than the old toolchain. It is also a different kind of relationship with the work, and the people who will be its best practitioners~-- the ones who will, in 2030 and 2035, be doing the AI-era equivalent of what Greg Lynn did with parametric architecture or what Bjork did with synthesisers~-- will be the people who \emph{consciously cultivate} the depth of relationship that the toolchain no longer enforces.

The platform companies are not going to teach you to do this. They have no incentive to. They benefit from your dependency, not your mastery. The new toolchain is \emph{frictionless,} and frictionless tools, however much we benefit from their efficiency, are not, on their own, going to produce the next generation of great creative work.

That work is going to come from the people who put the friction \emph{back in,} deliberately, on their own terms~-- who treat the AI agent as a junior colleague rather than as an oracle, who insist on \emph{understanding} what their tools are doing rather than just \emph{using} them, and who maintain the cognitive engagement with the work that the tools have been designed to make optional.

In the next chapter, I want to talk about the people who are doing exactly that. The orchestrators.

  \chapter{What is newly possible}\label{ch:10}

\lettrine[lines=3,lhang=0.15,findent=0.1em]{I}{} want to spend a chapter, after eight chapters that have been mostly about \emph{displacement,} on a question I think the book has so far under-served. \emph{Which categories of creative work has the AI moment made possible that were not possible before?}

This is not the question working creatives in 2026 are most often being asked. The questions most often being asked are \emph{will I lose my job?} and \emph{should I use the tools?} and \emph{what are studios doing?} Each of those is in this book, in a chapter of its own. They are all important questions.

The question of \emph{newly-possible work} is the one that, on the historical pattern I drew in Chapter~\ref{ch:2}, almost always turns out to be the one that mattered most. Every previous creative-technology transition the book has documented produced a set of new categories of creative work that the displaced practitioners did not, and could not, see coming. The phonograph displaced amateur parlour music and created the recorded-music industry. The microphone displaced operatic vocal projection and created intimate vocal styles, jazz singing, pop crooning, audio storytelling. Non-linear editing displaced the splice and created the MTV cut, the music video as art form, the hyper-cut action grammar, the streaming-era serial-drama editorial rhythm. The smartphone-as-camera displaced the dedicated camera and created the entire grammar of vertical-video native form.

The pattern is that the new categories \emph{always} appear, that they \emph{always} appear faster than the displaced cohort predicts, and that they are \emph{always} invisible from the perspective of the existing definition of the craft, because they are made of capabilities the existing craft does not have. The miniaturist could not, in 1845, predict Stieglitz's \emph{Camera Work}. The session keyboard player could not, in 1982, predict Aphex Twin's \emph{Selected Ambient Works} (and could not, if they had heard them, have recognised them as music). The print-magazine art director could not, in 2006, predict Instagram-as-fine-art-platform.

We are, in 2026, the equivalent generation to those people. The new categories are mostly invisible from where we sit. But some of them are already shipping; some of them are already finding audiences; some of them already have working creatives building careers inside them. This chapter is about what those categories are.

Before I start the inventory, I want to draw two historical analogues out~-- the synthesiser and non-linear editing~-- because they are the cleanest templates for \emph{how a tool that begins as a faster version of the old thing ends up being the substrate of a new thing entirely.} The mistake almost everyone made about AI in 2024 and 2025 was to think of it as a faster way to make the kind of creative work the platforms had been making. The historical pattern says: that view is \emph{almost always wrong on a five-to-ten year timeline.}

\section*{From imitator to instrument: the synth analogue}

When Robert Moog first started selling modular synthesisers in the late 1960s, the cultural permission for the instrument was very narrow. The synth, in its first commercial moment, was understood as a way to \emph{imitate} existing orchestral instruments~-- to play the parts of a string section, a brass section, a piano, an organ, in a form a single keyboard player could control. The breakthrough commercial release that secured the synth's cultural status~-- Wendy Carlos's \emph{Switched-On Bach} in 1968~-- was, on its face, a literal demonstration of this framing: the synth playing the music of the most-canonical European classical composer in the literature. Three Grammys. The first electronic record to be reviewed seriously by classical critics. The argument was: \emph{the synth can do what an orchestra can do.}

The synth never did, in the end, only do that. The 1970s and 1980s did something nobody at the 1968 reviewing desks predicted. They produced \emph{sounds that had never existed in the history of music}~-- Moog leads, FM electric pianos, the Roland TR-808 kick drum, the Yamaha DX7's chord pad, the \emph{Blade Runner} CS-80 ambient texture, the Aphex Twin acid bassline~-- and they built entire musical genres around those new sounds. Hip-hop, electronic dance music, ambient, IDM, synth-pop, the entire sonic vocabulary of 1980s film scoring~-- these are forms that \emph{could not have existed} without the synth, that did not exist before the synth, and that, \emph{crucially,} could not have been predicted from the framing in which the synth was first introduced.

The synth was not a faster orchestra. The synth was an instrument for making sounds that no orchestra could produce, on a timescale that no orchestra could match, accessible to working musicians without the social capital of orchestral training. The instrument's first cultural moment~-- \emph{Switched-On Bach}~-- was the moment of the \emph{imitator framing}. Its second cultural moment~-- \emph{Autobahn}, \emph{Blade Runner}, \emph{Thriller}, \emph{Acid}, the \emph{Roland 808} in \emph{Planet Rock}~-- was the moment when the imitator framing was thrown away and the instrument was used for what only it could do.

That second moment took about a decade. From Moog's first commercial modular in 1965, through Carlos in 1968, through Kraftwerk's \emph{Autobahn} in 1974, through \emph{Trans-Europe Express} in 1977, through the early hip-hop and dance records of 1981--82~-- the gap between the synth-as-imitator and the synth-as-new-instrument was roughly fifteen years. The cultural permission to use the synth as itself, rather than as a replacement for something else, had to be earned by working musicians inhabiting the instrument in front of audiences who eventually understood that they were hearing something new.

The AI equivalent moment, on the historical pattern, has not yet arrived. We are, on the synth timeline, somewhere between \emph{Switched-On Bach} and \emph{Autobahn}. The work that is going to define AI as a new creative substrate~-- rather than as a faster way to produce existing work~-- is being made \emph{right now,} by working creatives somewhere, in forms that the trade press has not yet decided what to call. I have my candidates, which I will come to. The point I want to make at the chapter's opening is the structural one: every iterative-technology framing of AI is doing what the \emph{Switched-On Bach} reviewers did in 1968. It is describing the new tool in the language of the old one. The new language has not yet been written.

\section*{From workflow to grammar: the non-linear-editing analogue}

The same pattern is visible, in a different domain, with non-linear editing.

When Avid Media Composer shipped in 1989, the cultural framing was strictly utilitarian. NLE was a \emph{faster way to do the existing thing.} Where a working editor used to splice physical film on a Moviola~-- a slow, irreversible, physically demanding craft~-- NLE allowed the same edits to be made in software, with undo, with multiple versions, with no consumable cost. The first generation of working editors who picked up Avid did so on the same premise as the synth's first-decade adopters: \emph{the tool will let me do what I already do, faster.}

What NLE actually produced, by the mid-1990s and through the 2000s, was a fundamentally new editing \emph{grammar}. The average shot length in mainstream Hollywood drama dropped from roughly ten seconds in the 1960s to roughly four seconds by the mid-2000s~-- a change made trivial by NLE that would have been physically punishing to execute on a Moviola. The MTV-cut aesthetic, which had been a music-video novelty in the early 1980s, became the default grammar of contemporary action cinema by the 2000s~-- \emph{The Bourne Identity} (2002) is the textbook example, with shot lengths under two seconds across whole action sequences and an editing logic that depended on the viewer's now-trained ability to read a fast-cut grammar. The parallel-narrative structures of contemporary streaming drama~-- multi-thread, multi-timeline, multi-perspective storytelling, edited together with non-linear interleaving that would have been logistically impossible on tape~-- \emph{Lost}, \emph{Westworld}, \emph{Dark}, \emph{Severance}~-- are forms that NLE made possible.

Walter Murch, the most respected film editor of the last fifty years and one of the few people to have edited at the highest level on both film and Final Cut Pro, made the point clearly in \emph{In the Blink of an Eye}: the tool \emph{does} change the grammar. Murch was characteristically careful not to claim that the change was an improvement or a degradation. He claimed only that it was a \emph{change}~-- that NLE permitted certain kinds of edit that physical-film editing could not, and that the new grammar would, over a generation, become as natural to its audiences as the slower-cut grammar of the 1960s had been to its.

The same dynamic is, in 2026, visible in the AI-augmented production pipeline. Working creatives I know are doing things in their day-to-day practice that would have been physically impossible~-- not just expensive, \emph{impossible}~-- in 2020. Iterating across forty variants of a scene in an afternoon. Producing personalised localised versions for ten markets simultaneously. Re-cutting a feature against a different aspect ratio for vertical-video distribution while keeping the principal photography intact. Generating a sustained 4D point-cloud reconstruction of a real-world location from a phone-captured walk-through and using it as a virtual-production plate. Running a scratch-vocal session in seventeen languages off a single take. These are \emph{not faster versions of existing workflows.} They are workflows that did not exist three years ago. And, exactly as Murch predicted of NLE, the audiences for the work made on these workflows are already developing the perceptual literacy to read it.

\section*{Six categories of newly-possible work}

With those two analogues in mind, I want to walk through six categories of creative work that I believe are \emph{newly possible} in 2025--26~-- meaning, work that an individual creative or a small studio can produce now that they could not have produced before, and that an audience can experience now in ways the audience could not have experienced before.

I will be honest, on each, about how much of the category is already shipping in finished form and how much is still in the \emph{demo and beta} layer of the toolchain. The whole point of being inside the work is that you can see the difference.

I will also flag, throughout, the \emph{binding constraint} that runs underneath the entire chapter: \textbf{human attention is finite.} This is the most-underdiscussed structural fact of the AI creative-economy moment, and I will come to it at length in a moment.

\subsection*{One: Remix at scale}

The first category~-- the most visible and the most legally contested~-- is \emph{remix.} The infrastructure of AI generation makes it cheap, fast, and at scale to produce derivative work: alternate-style versions of existing songs, cover-style reinterpretations across genres, AI-dubbed translations of feature films into languages the original release never reached, image-to-image style transfers of canonical artworks, motion-transferred re-performances of choreography across body types.

I want to be careful in describing this, because \emph{remix as a creative form} has a long pre-AI history. Hip-hop's relationship to sampling is the canonical example; Lessig's ``remix culture'' framing from the 2000s identified the dynamic in broad strokes well before generative AI; the \emph{Star Wars} fan-edit community, the mash-up era of Girl Talk and Danger Mouse's \emph{Grey Album}, the YouTube AMV community, the TikTok stitch-and-duet grammar~-- every one of these is, in operational terms, \emph{remix infrastructure that produces creative value through derivation}. The 2025--26 AI moment didn't invent remix culture. It made the \emph{production cost of derivative-but-original creative work} drop by more than an order of magnitude, and it shifted the technical bottleneck from \emph{skill at imitation} to \emph{judgement about what to imitate.}

The 2025--26 examples I have followed most closely:

\begin{itemize}
  \item \textbf{AI-dubbed feature releases} crossing language barriers that the original production could not afford to cross. \emph{Watch the Skies}, the Swedish UFO feature dubbed into English using ElevenLabs voice cloning, getting US distribution in October 2025, is the canonical example I covered in Chapter~\ref{ch:1}. The dubbing is not, in any historical sense, a \emph{new translation}; what is new is that the translation is rendered into the original actor's voice rather than a session voice actor's, preserving the performance signal across the language transition.
  \item \textbf{Cover-style reinterpretation at the long-tail of recorded music.} Suno and Udio's user economies have produced an extraordinary volume of \emph{style-transferred} renditions of existing songs~-- folk versions of pop hits, metal versions of show tunes, lullaby versions of EDM tracks. The vast majority of these are uninteresting. A small fraction, in my experience, are \emph{better than the original by some specific human listener's measure of better}~-- and that is the new commercial dynamic.
  \item \textbf{The Andrii Daniels bomb-shelter clip} (Chapter~\ref{ch:12})~-- \emph{Deadpool and Harry Potter} as a Christmas mash-up, made in a Ukrainian bomb shelter, viral in December 2025~-- is a remix-at-scale artefact in the strict sense. It crosses two IPs that the rights-holders would never have collaborated on, in a stylistic register the original films could not have produced, and finds a global audience that responds to the \emph{idea} of the remix as much as to its execution.
\end{itemize}

The legal layer of this category is still moving. \emph{UMG v.\ Anthropic}, \emph{Getty v.\ Stability AI}, the EU Copyright Directive's Article 17, the UK 88\%~-- these are the institutional structures that will decide whether the AI-remix category becomes a \emph{licensed and compensated} creative form (the Petrillo template applied to AI) or a \emph{grey-market} one that operates outside the rights system. The historical pattern~-- Sampling, post-\emph{Grand Upright}, became a licensed creative form, with the dense Bomb Squad style becoming commercially difficult but the basic technique surviving in a more clearance-friendly mode~-- says the remix category will, in some form, \emph{settle into a licensed category by the end of the decade.} The Petrillo template, again, is the answer the system already knows.

\subsection*{Two: Mass personalisation, against the binding constraint}

The second category~-- the one most often gestured at in platform-company keynotes and least well-served by them~-- is \emph{mass personalisation}: creative work that is \emph{individually different} for each viewer, listener or player.

The early shipped versions in 2025--26 are these:

\begin{itemize}
  \item \textbf{Personalised playlist generation and AI DJs.} Spotify's DJ feature, YouTube Music's AI Hub, Apple Music's editorial-style algorithmic mixes~-- all of these are \emph{individual-level curation} of pre-existing creative work. The personalisation is in the \emph{selection}, not in the \emph{generated content.}
  \item \textbf{Personalised game NPCs and dialogue.} Inworld, ConvAI, Ubisoft Teammates, Roblox's AI assistant~-- by spring 2026, the first wave of generative-NPC dialogue is shipping in player-facing form. The personalisation is in the \emph{moment-by-moment response} to the player's actions, not in the underlying game world. Whether this actually produces a personalisation a player can recognise~-- versus producing the same in-character dialogue from a wider statistical envelope~-- is, on the published shipping examples I have played, an open question.
  \item \textbf{Personalised marketing creative.} WPP Open Pro, Adobe CX Enterprise, GenStudio~-- the platform infrastructure for producing per-segment, per-audience, per-context creative variations is now mature. Whether the audience experiences this \emph{as} personalisation, or experiences it \emph{as} a feeling of unease that the messaging knows too much, is the cultural question every brand creative is asking in 2026.
  \item \textbf{Personalised educational content.} Khan Academy's Khanmigo and Duolingo Max are the most-mature shipped examples I know of. The personalisation here is genuine and measurable~-- the per-student tutoring response is different on the level of the individual learner.
\end{itemize}

I want to draw a sharp limit around the personalisation category, because the binding constraint runs straight through it and I think every working creative thinking about this market needs to internalise it.

\textbf{Human attention is finite.} Aggregate daily media-consumption time per person has, on the available Nielsen-style telemetry I have read, \emph{not} grown meaningfully in the past decade. The total of every form of media consumption~-- TV, streaming video, music listening, podcast listening, social media, gaming, reading~-- is, per the published data, on the order of 11--12 hours per US adult per day, and that number has been roughly stable for years even as the \emph{number of available} hours of content per day has exploded by orders of magnitude. The eye, the ear and the consciousness each have a finite capacity. \emph{Personalisation} does not, by itself, expand that capacity. It changes the \emph{distribution} of attention across content, but it does not increase the \emph{total} attention available to be spent.

This is the structural ceiling on the \emph{commercial} value of mass personalisation. Producing a personalised version of a film for every viewer~-- to take the most extravagant platform-keynote framing~-- is, on the binding-constraint reading, a competitive move that \emph{reallocates} attention rather than \emph{expanding} it. The category will be commercially meaningful in segments where reallocation can produce premium prices (luxury advertising, premium educational content, top-tier video-game NPCs in IP that justifies the investment). It will be less commercially meaningful at the long tail, where the personalisation effort does not produce attention-reallocation big enough to pay for the agentic infrastructure underneath it.

The trade-press framing of mass personalisation as \emph{infinite content for infinite audiences} is, on the binding-constraint reading, structurally incoherent. The audience does not have infinite attention. The producible content is, by 2026, effectively infinite. The economic question is \emph{which slices of the audience's existing finite attention budget} the personalised work can plausibly capture. The answer, by my read of the shipping evidence, is \emph{narrower than the platform companies' enthusiasm suggests.}

I will come back to the finite-attention constraint at the end of the chapter, because it shapes every category that follows.

\subsection*{Three: Audience participation in the creator economy}

The third category~-- and the one I am most personally interested in~-- is the \emph{audience as participant.} The 2010s creator-economy framing was that the audience could \emph{make their own content} on platforms (YouTube, TikTok, Instagram), creating a long tail of user-generated work alongside the professional content. The 2025--26 AI framing extends this by an order of magnitude: the audience can now \emph{prompt, contribute to, remix and co-author} work in real time, often in collaboration with named professional creators.

The shipped examples I have followed:

\begin{itemize}
  \item \textbf{TikTok / Instagram / Shorts remix mechanics}~-- duets, stitches, the \emph{Sora} iOS app's TikTok-style remix grammar (which crossed a million downloads in five days in October 2025). The barrier between \emph{watching the work} and \emph{making a related piece of work} has dropped to a single tap.
  \item \textbf{Roblox's creator economy.} The platform's UGC ecosystem, by 2026, is the largest single audience-as-creator system in any creative medium. Roblox creator payouts run into the hundreds of millions of dollars annually. The line between the audience and the working creator on Roblox is, structurally, not a line; it is a gradient.
  \item \textbf{Disney+'s announced (though not yet shipped at the date I am writing this) user-generated-AI-content tools using Disney IP.} If this ships in the form the platform has described, it will be the first time a major IP holder has \emph{deliberately given the audience tools to make work inside its canonical universe.} The cultural permission this would set~-- sanctioned fan-made \emph{Star Wars}, sanctioned fan-made \emph{Marvel}~-- would, on the historical pattern, expand the categorical \emph{Star Wars} / \emph{Marvel} canon at a rate the studio system itself cannot match.
  \item \textbf{The fan-prompt economy.} Civitai, Promptbase, the LoRA marketplaces~-- the infrastructure for \emph{selling and sharing the creative inputs} that other creatives then use to produce work. The audience is participating not just in the work, but in the \emph{toolchain underneath the work.}
\end{itemize}

The single cleanest piece of evidence on the \emph{generational} shape of this category came in May 2026, when \textbf{Snapchat} published research finding that \textbf{31\%} of \textbf{13--15 year-olds} on the platform were already using AI tools ``to be creative''~-- not, importantly, to do their homework, not to chat with a synthetic friend, but specifically to \emph{make things} they then shared with their peers.\footnote{Snap Newsroom, ``Snapchat Gen Z AI Creativity Research 2026.'' \url{https://newsroom.snap.com/snapchat-gen-z-ai-creativity-research-2026}. \emph{Dream Machine} Issue~30.} That number is the most legible quantitative indicator I have seen of where the audience-as-participant category is heading. The 13-to-15 cohort the survey describes will be the 18-to-20 cohort of 2029. By the time they reach the working-creative entry pool, \emph{making things with AI} will not, for them, be a category distinct from \emph{making things}. It will simply be how things are made. The studios that build for this audience now~-- not as a future they are anticipating, but as a present they are already serving~-- will, on the historical pattern of every previous generational shift, set the terms on which the next decade of cultural production runs.

The thing I want to flag about this category~-- because I think it is the part the platform companies and the working studios have most systematically under-priced~-- is that \emph{audience participation reverses the direction of the creator-audience economic relationship.} In the pre-platform creative economy, the audience paid the creator. In the platform-era creator economy, the audience generated the content and the platform monetised the attention. In the 2025--26 AI-augmented creator economy, the audience is increasingly co-producing the content \emph{with} the creator, and the question of who gets paid for what is, structurally, harder to answer than it has ever been.

This is one of the open frontiers of working-creative business model design in this period. I do not think anyone has solved it. The studios that figure out how to credit, compensate and structurally honour audience contributions to the work~-- without turning the work into the kind of crowdsourced mush that does not survive the slop ceiling~-- will, in my view, have the most defensible business position in the next decade. The studios that try to extract audience-generated work without compensating it (the 2010s social-media platform model applied to AI) are, on the historical pattern, walking into the next \emph{Viacom v.\ YouTube}-scale lawsuit.

\subsection*{Four: Fan fiction and fan-made content, legitimised at scale}

Closely related to the participation category~-- but worth a section of its own~-- is the \emph{fan-fiction / fan-content} category. Fan-made creative work has a long pre-AI history. The \emph{Star Trek} fanzine era of the 1960s; the \emph{Star Wars} fan-film tradition from the 1970s onward; the Archive of Our Own / Wattpad / FanFiction.net communities; cosplay; the anime fansubbing tradition; the \emph{Harry Potter} fan-fiction archive that, on some counts, contains more words of \emph{Harry Potter}-canon-derivative text than the original novels themselves.

What AI does to this category is two things. \emph{One,} it raises the \emph{technical floor} of fan-made work toward what was previously professional-grade: a fan can now generate a \emph{Harry Potter} short film with production values that would have required a major-studio budget five years ago. \emph{Two,} it makes the \emph{canon-extension impulse} of fan culture practically infinite: every reader, every viewer, every player can~-- with current tools~-- produce a new piece of work inside the universe they love, in their own voice, on their own terms, for their own audience.

The cultural and legal layer of this category is, in 2026, still moving. Disney's tolerance for fan content has shifted under AI, and the question of whether \emph{AI-generated fan content using Disney IP} will be treated more leniently than fan-made content has historically been is one of the open IP-policy fights of the year. Lucasfilm's tolerance for \emph{Star Wars} fan films has, historically, been generous; whether that extends to AI-generated \emph{Star Wars} features is unsettled. The Marvel Comics community policy on AI is one of the documents to watch in the next eighteen months.

What is \emph{not} unsettled is that the audience is already doing this. The fan-AI-content economy is, by spring 2026, larger by volume than the official IP-holder output for almost every major IP in popular culture. The official IP holders can suppress this, license it, build platforms around it, or watch it consume their cultural authority. There is no fourth option.

The working creative read on this, for anyone reading the book inside an IP-holding studio, is the one I made in Chapter~\ref{ch:7}'s discussion of the legacy industries' strategic vulnerability. The studios that move \emph{toward} sanctioned fan-AI content economies~-- Disney's announced UGC tools, the most generous of the cosplay-and-fan-film tolerances, the platform-and-fund models that pay fan creators for canonical contributions~-- will have a meaningful structural advantage over the studios that try to defend the closed canon against the audience that is, with or without permission, going to extend it anyway. The Petrillo template applied to fan culture is: pay the fan, structure the IP-holder's stake, \emph{participate} in the extension rather than fighting it.

\subsection*{Five: Agentic support workers for the solo creative}

The fifth category is the one that has~-- in my own practice and in the practice of the wider DreamLab community~-- produced the most dramatic and immediate productivity gains. The agentic-support-worker model.

In the pre-AI creative economy, the small or solo creative practitioner~-- the independent filmmaker, the songwriter, the freelance illustrator, the indie game developer, the YouTuber~-- operated with a particular structural disadvantage. Their work was bottlenecked not on creative judgement (which the senior practitioner had in abundance) but on the \emph{production-coordination labour} that a larger studio would assign to junior staff: client communications, project management, asset organisation, scheduling, invoice processing, basic research, draft response, brief structuring, simple post-production. The senior practitioner spent an inordinate fraction of their effective working hours on labour that did not require senior judgement.

The 2025--26 agentic toolchain~-- Notion AI, Adobe Express AI Assistant, Heygen Video Agent, Claude Apps, the personal-assistant features baked into the major platforms~-- has, in operational terms, \emph{given the solo creative the first four hires they would have made.} The personal assistant, the scheduler, the production coordinator, the asset manager~-- these are now, for working solo creatives in my circle, agentic functions running underneath the senior practitioner's day, costing roughly the price of a couple of midrange platform subscriptions, and producing genuine recoverable hours.

The economic impact of this is, I think, the most under-priced shift of the period. The \emph{one-person studio} that, in 2020, would have been a part-time freelance practice supporting two to four projects a year is, in 2026, a \emph{full-business-class production operation} supporting twenty to forty projects across a wider range of disciplines. The Sienna-Rose / Xania-Monet / Hoyt-Dwyer single-creator AI-supported career~-- about which I have, in Chapter~\ref{ch:5}, made my reservations clear in terms of \emph{star formation}~-- is, on the \emph{production-economics} side, a genuine new business form. Whether or not Xania Monet becomes a Billboard-defining cultural figure, the \emph{operational machinery} underneath her~-- a single human creative, supported by an agentic stack producing music, marketing, distribution and merchandise at a scale that would have required a small label to support five years ago~-- is a working business model that did not exist before the toolchain shipped.

The implication for working creatives at the senior solo level is direct. The first four hires you would have made~-- production coordinator, junior researcher, scheduling and admin, post-production junior~-- are now agentic. The economic ceiling on your individual practice has, structurally, lifted. The constraint is no longer the throughput of the labour underneath you. The constraint is your own senior judgement bandwidth~-- the briefing, the taste, the integration, the \emph{Why}~-- which the agents cannot replace and which, on the chess-grandmaster argument of Chapter~\ref{ch:15}, has more commercial leverage in this market than it has had in any previous period.

\subsection*{Six: Hyperlocal and long-tail cultural production}

The sixth category~-- and the one I think will, in retrospect, prove to be the most culturally significant~-- is \emph{hyperlocal and long-tail cultural production}. The collapse of the cost of producing professional-grade creative work means that, for the first time in the history of mass media, \emph{every linguistic community, every regional culture, every minority cultural tradition, every niche audience} can produce work in its own language, in its own visual style, for its own audience, at production values that compete with global commercial output.

The early shipped examples:

\begin{itemize}
  \item \textbf{Korin AI}, the Africa-trained and Africa-built music platform launched in spring 2026, is the canonical example of a \emph{deliberately culturally-specific} AI tool. The training data is African; the outputs are African; the platform exists to serve African creators rather than to backfill global pop with African aesthetic markers.
  \item \textbf{Indian AI cinema and the Trilok / Animaj / Filmax DinoGames cluster.} India's regional-language film industries~-- Bollywood, Tollywood, Kollywood and the dozen smaller regional industries~-- have been among the most aggressive adopters of generative AI in the period this book covers. The reason is structural: the regional film industries operate at production budgets where AI's cost-reduction effect has \emph{the largest proportional impact.} A regional Indian feature that would have cost \$200,000 to produce in 2020 can, in 2026, be produced for a meaningful fraction of that, with comparable production values. The category of films that this brings into commercial viability~-- every regional cinema with a million-viewer audience but a six-figure budget ceiling~-- is, by my estimate, \emph{an order of magnitude larger} than the category that was commercially viable before.
  \item \textbf{The Tunisian \emph{Lily}}~-- winner of the \$1m Dubai AI Film Award in January 2026~-- is the cleanest individual example of a film made \emph{outside} the major-market financing system, at a production value competitive with it, by a creative team using AI tooling as the cost-reduction lever.
  \item \textbf{The 800-creator declaration's signatories} include working creatives from a meaningful number of national contexts that the historical Anglophone film/music/games industries have under-represented. The \emph{Stealing Our Work Is Not Innovation} coalition is, structurally, a coalition that includes the long tail of global creative work.
\end{itemize}

The implication of this category is the one I am most personally hopeful about, and I want to be honest that \emph{hopeful} is the right word~-- there is a less-hopeful version of the same data, in which Anglophone AI tooling homogenises the global creative economy faster than the regional creative economies can develop their own infrastructure. The 2026 evidence I have is mixed enough that both outcomes are still on the table.

What is not in doubt is that the \emph{production-cost ceiling that has, for a century, kept hyperlocal creative work below the threshold of professional commercial viability} is, by 2026, no longer the binding constraint. The next decade of cultural production will, I am confident, contain more local, more linguistically diverse, more culturally specific creative work than the previous decade~-- and the most culturally significant individual works of the next decade will, on the historical pattern, come from communities the existing global creative economy has under-served. The question is whether those communities own the tooling that produces them.

\section*{What is not yet possible}

Before I close, I want to be honest about the categories that are \emph{not yet shipping} in the form the platform-company keynotes have promised them, because the book's credibility depends on accurately characterising the gap between affordance and rhetoric.

\textbf{AI does not yet write a satisfying novel from scratch.} The 2025--26 evidence on long-form prose generation is that AI systems produce technically competent prose that loses \emph{narrative purpose} over the length of a novel. Working novelists I have talked to who have experimented seriously with this category all describe the same failure mode: the prose is fine; the \emph{book is not a book.} This is consistent with the \emph{House of David} ``hand inside a puppet'' critique of AI-augmented storytelling at feature length.

\textbf{AI does not yet do live performance.} No AI is touring in 2026. Xania Monet has not performed at Madison Square Garden. The cultural permission for a synthetic performer to share a stage with a live audience does not, as of this writing, exist in any meaningful form. The structural reason~-- the audience experience of \emph{being in a room with another human consciousness}~-- is, on the slop-ceiling reading, not a permission gap; it is a category mismatch. AI work and live performance are, for now, different categories.

\textbf{AI does not yet do sustained emotional storytelling at feature length without human authorial spine.} The auteur-driven cinema of the Cameron / del Toro / Spielberg generation is, structurally, a category that depends on a \emph{single human consciousness running through every creative decision in the film.} AI-augmented versions of this category exist; AI-native versions of it do not. The \emph{Citizen Kane} of AI cinema has not been made. Whether it will be is, in my view, the single most interesting open creative question of the next decade.

\textbf{AI struggles with cultural specificity that the model has not been trained on.} The BBC India observation I referenced in Chapter~\ref{ch:13}~-- that AI screenplays produce \emph{cultural-memory failures} when generating content for cultural traditions the training data has under-represented~-- is the binding limitation on AI's promise as a \emph{globally-distributed creative tool.} The hyperlocal-cultural-production category I described above depends, structurally, on this limitation being addressed by purpose-built regional model infrastructure (Korin AI, the Indian regional-language tooling, the various Asian-built models). Until that infrastructure matures, AI cultural-production is still, by default, Anglophone and Western-modelled.

\section*{The binding constraint, restated}

I want to close by returning to the structural fact that runs underneath every category of newly-possible work in this chapter.

\textbf{Human attention is finite.}

I am not making a romantic argument about it. I am making an economic one. The aggregate daily media-consumption time per adult, in any market the Nielsen-class telemetry covers, has been roughly stable for at least a decade~-- eleven or twelve hours a day, across all forms of media, all formats, all platforms, all devices. \emph{That number has not grown with the rise of streaming. It has not grown with the rise of mobile. It has not grown with the rise of social media. It has reallocated, sometimes dramatically. It has not expanded.}

The producible content, meanwhile, has expanded by orders of magnitude. Deezer in 2026 receives 75,000 AI tracks a day; the listener has the same number of waking ear-hours she had in 2016. YouTube uploads run at hundreds of hours per minute; the viewer has, in net, the same daily screen-time budget. The Sora app produced a million downloads in five days; the audience for the work those million users are about to make has, in total, the same total attention they would have had if Sora had never shipped.

This is the binding constraint of the entire 2025--26 creative-AI moment. The cost of producing work has collapsed; the supply of attention has not. Every category I have described in this chapter~-- remix, personalisation, audience participation, fan content, agentic support, hyperlocal production~-- is being built into a market where the \emph{production} side of the supply-and-demand equation is racing toward infinity and the \emph{consumption} side is bounded by the finite cognitive and biological capacity of the human nervous system.

This has three structural implications.

\noindent\textbf{One.} The competitive advantage in this market accrues, with iron consistency, to the producer of work that the audience \emph{chooses to spend its finite attention on} rather than the producer of work that the audience could, in principle, choose. The slop ceiling, the authenticity premium, the chess-grandmaster move, the \emph{Why}~-- these are not romantic notions. They are the \emph{mechanisms by which finite human attention selects from infinite producible content}.

\noindent\textbf{Two.} The platform business models that depend on \emph{expanding audience attention faster than supply} are, on the structural reading, walking into a wall. The platform companies' current trajectories~-- push more content, optimise for engagement, monetise more minutes~-- work only as long as the engagement budget is elastic. It is not elastic. The audience cannot, on average, watch more hours per day than it already does. The platforms that get to those audiences first, and that build the most-defensible \emph{retention} mechanics, will win the share-of-attention game. The platforms that come second will be competing for an attention budget that the first-movers have already claimed.

\noindent\textbf{Three.} The new categories of work I have described in this chapter will produce, in aggregate, more \emph{total creative output} than the previous decade. They will not produce, in aggregate, more \emph{audience attention received per minute of output.} The ratio of attention-to-output, which is what working creatives actually live on, is going to fall~-- sharply, in some categories, more gradually in others. The working creatives who survive the next decade will be the ones who recognise that the attention-to-output ratio is the metric that actually matters, and who position their practice in the categories where the ratio is most defensible.

The categories where the ratio is most defensible~-- on the chess-grandmaster reading and the slop-ceiling reading and the authenticity-premium reading~-- are the categories where the work is \emph{most deliberately un-machine-like}, \emph{most authentically human-authored}, \emph{most culturally specific}, \emph{most personally risked.} The newly-possible work in this chapter is real and will reshape the creative economy. But it is being made in a market where the binding constraint is, and will remain, the finite attention of the audience watching it.

The synth made entirely new sounds possible. The audience for those sounds was finite. The musicians who learned to make sounds \emph{the audience would spend its scarce listening time on}~-- Trevor Horn, Kraftwerk, the Bomb Squad, Aphex Twin, Daft Punk, every electronic-musician who built a serious career~-- are the ones we still listen to. The musicians who learned to make sounds the synth made possible \emph{but that the audience did not develop a hunger for}~-- the long tail of 1970s and 1980s synth-record obscurities~-- are remembered, mostly, by collectors.

The AI moment will, on the historical pattern, work the same way. The newly-possible categories will create work that did not exist before. The audience will, on the available evidence, allocate its scarce attention to the work that \emph{earns} that attention. The working creatives who position themselves in the newly-possible categories, \emph{and} who make work that the audience deliberately chooses, are the ones who will define the next decade of the form.

That is the operating manual. The categories are open. The constraint is real. The choice~-- like the chess-grandmaster's~-- is yours.

  \chapter{The Orchestrator}\label{ch:11}

\lettrine[lines=3,lhang=0.15,findent=0.1em]{I}{n} \emph{Dream Machine} Issue~2, in October 2025, I described the \textbf{Human--AI Agency Continuum} as a way of mapping how much of any given creative function is being done by the human in the chair and how much by the machine. In \emph{Dream Machine} Issue~13, in January 2026, I made a prediction that I want to look at again in this chapter, because~-- six months on~-- it has held up better than most of the others I made that day.

I called 2026 \textbf{``the Year of the Orchestrator.''}\footnote{\emph{Dream Machine} Issue~13, ``Editor's Pick: The Year of the Orchestrator,'' 9 January 2026.}

The argument was straightforward. If 2024 had been the year of the \emph{generator}~-- the prompt-and-respond text-to-image, text-to-video model~-- and 2025 had been the year of the \emph{agent}~-- the system that could take goals, plan, decide and execute multi-step tasks autonomously~-- then 2026, I argued, would be the year that working creatives stopped being \emph{operators} of these tools and started being \emph{orchestrators} of them.

By ``orchestrator'' I meant something quite specific: a person whose job is not to make the work themselves but to direct, brief, integrate and judge the work of a team of AI agents and human collaborators. Not a producer in the old sense. Not a creative director in the old sense. Something newer, with a different skill set, a different rhythm, a different relationship to craft.

This is the role I think most working creatives will be holding by 2030. This chapter is about why, what it looks like, what it asks of you, and where it breaks.

\begin{figure}[htbp]
  \centering
  \includegraphics[width=0.92\textwidth]{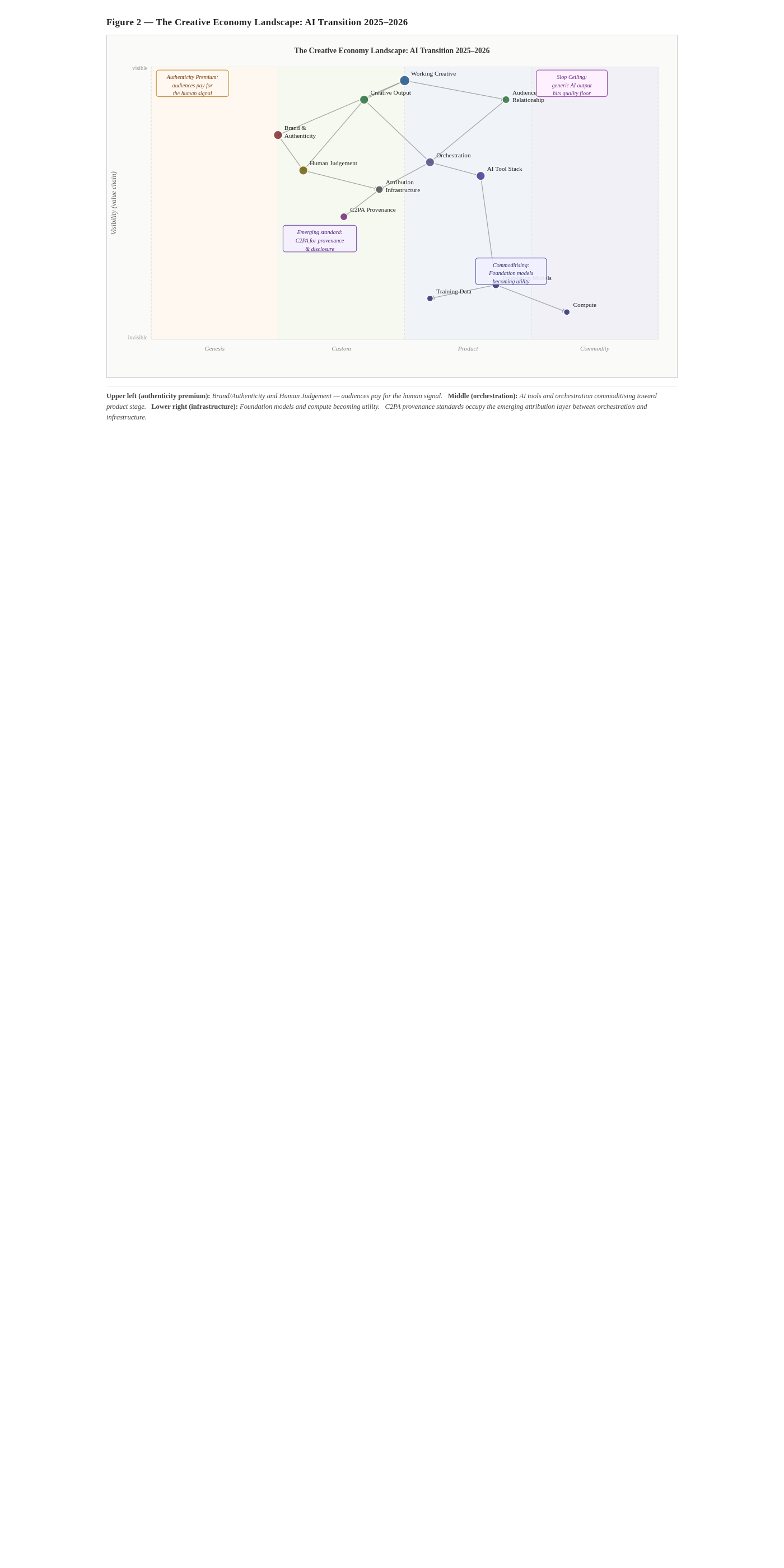}
  \caption{Wardley map of the creative economy landscape, showing the evolution of creative roles from genesis through custom-built to product and commodity as AI tools mature.}
  \label{fig:wardley-creative-economy}
\end{figure}

\section*{Forty-nine Claude agents and seventy-two skills}

In May 2026, in the second-to-last issue I wrote before this book went to draft, I reported on something that I think of as the canonical image of what the orchestrator role actually is. \textbf{Sony}, in announcing its ``all-in on AI for games'' strategic move, disclosed that one of its game-development studios was running a coordinated multi-agent team of \textbf{49 Claude Code agents}, working with \textbf{72 skills}, on game-development tasks ranging from asset generation through QA through engineering through animation.\footnote{\emph{Dream Machine} Issue~29, May 2026, reporting on Sony's 49-agent / 72-skill multi-agent game-development team.}

This is, in operational terms, a small army of synthetic colleagues working on a single creative project. Each agent has a defined role. Each skill is a defined capability. The whole apparatus is overseen by a \emph{substantially smaller number of human developers and creative leads} whose job is to plan the team's work, brief the agents, judge their outputs, integrate their contributions, and decide what gets into the game.

The ratio matters. The pre-AI version of this game-development team would have been, plausibly, 50 to 100 people working on the same scope of work over a much longer timeline. The AI-augmented version, as Sony has set it up, is a much smaller number of \emph{senior, judgement-heavy} roles~-- people whose value is taste, narrative sense, gameplay design instinct, IP fluency, engineering oversight~-- sitting on top of a much \emph{larger} pool of synthetic capacity.

What you don't see, in the Sony picture, is the disappearance of the junior roles. Those roles haven't disappeared. They have been \emph{replaced}~-- by agents. The 49 Claude Code agents are, in effect, the new junior developers, the new junior animators, the new junior writers. They work cheaply, they work fast, they work in parallel. They are not~-- and this is important~-- \emph{replacements for senior judgement.} They are leverage \emph{for} senior judgement. The whole pipeline is designed to take the senior creatives' time and multiply its effective reach by a substantial factor.

The orchestrator, in this configuration, is the senior creative~-- the writer-director, the gameplay lead, the art director, the technical director~-- whose taste and judgement set the boundaries that the agent team works inside.

I want to flag the obvious labour question here, because it is the question every working creative is asking, and any chapter that handwaves past it is not being honest. If the senior roles still exist, and the junior roles are now done by agents, \emph{where does the next generation of senior creatives come from?} The pipeline that has, for fifty years, produced senior creatives in the film, TV, games, music and design industries has worked by \emph{starting people as juniors and letting them grow.} If we remove the junior rung, we are~-- over the next decade~-- also removing the apprenticeship infrastructure that makes the senior rung possible.

This is the structural risk of the orchestrator economy that platform companies are, in my view, not yet taking seriously. I name it the \textbf{Apprenticeship Gap}, and I will come back to it in Chapter~\ref{ch:14}.

\section*{What the orchestrator does}

In the talks I have given since publishing the ``Year of the Orchestrator'' piece in January, the question I get asked most often is: \emph{what does an orchestrator's day actually look like? What are the skills? What does the job description say?}

I want to try to answer that in this chapter, in the most concrete language I can find, because I think the gap between \emph{what working creatives think they will be doing in 2030} and \emph{what they will actually be doing} is unhelpfully large.

The orchestrator does five things:

\noindent\textbf{One. They define the brief.} The agents~-- and the human team~-- work to a brief. The brief is the thing the orchestrator owns. It is not the prompt. The prompt is a derivative artefact. The brief is the \emph{creative intent} that the project exists to deliver: who it's for, what it's trying to do in the world, what success looks like, what tone, what feeling, what audience, what context. The orchestrator's first job is to know~-- clearly enough to communicate it~-- what the work is \emph{for.}

\noindent\textbf{Two. They allocate work.} Given the brief, the orchestrator decides which parts of the work get done by humans, which by AI agents, and which by the orchestrator themselves. This is the practical, day-by-day application of the Human--AI Agency Continuum we talked about in Chapter~\ref{ch:3}. It is not a one-off decision. It is a constant series of micro-decisions about where on the continuum each function sits, in this project, on this day, for this output.

\noindent\textbf{Three. They brief the agents.} This is the closest the orchestrator role gets to what people currently mean by ``prompt engineering,'' but it is a meaningfully different skill. Briefing a human collaborator and briefing an AI agent are not the same activity, but they share a core: the ability to \emph{describe what is wanted with enough precision and enough context that the recipient can produce something useful, without over-constraining them in ways that prevent useful surprise.} This is, in my experience, the single biggest skill differentiator between effective and ineffective AI-era creatives. The people who can brief well~-- who know when to give the agent a tight constraint and when to let it explore~-- are the people who produce the best output.

\noindent\textbf{Four. They judge the outputs.} When the agents deliver, the orchestrator's job is to look at what came back and decide what to ship, what to revise, what to throw away. This is \emph{taste} in the most operational sense. It is also, crucially, \emph{taste under abundance}~-- taste exercised in a context where you can have ten versions of the same scene back in ninety seconds and your job is to choose, not to make. Choosing well under abundance is a different cognitive skill than choosing well under scarcity, and most creatives have been trained for the latter. The grandmaster analogy from Chapter~\ref{ch:15}~-- top chess players, in 2026, deliberately playing \emph{sub-optimal} moves to put their opponents on uncomputed ground\footnote{\emph{Bloomberg}, ``AI Changed Chess. Grandmasters Now Win With Unpredictable Moves,'' 27 March 2026. \url{https://www.bloomberg.com/news/articles/2026-03-27/ai-changed-chess-grandmasters-now-win-with-unpredictable-moves}. \emph{Dream Machine} Issue~23. The chess analogy is developed in Chapter~\ref{ch:15}'s \emph{Age of the Why} section.}~-- is the cleanest available picture of what this looks like in practice. The orchestrator's value is not in choosing the \emph{most-likely-good} output of the ten variants the agent returned. The agent has, by construction, already centred its output on the most-likely-good. The orchestrator's value is in seeing which of the ten variants would be the \emph{un-machine-like} move at this specific point of this specific project, and choosing that one. \emph{Taste under abundance} is, operationally, the discipline of refusing the machine-optimal output in favour of the deliberately-chosen one.

\noindent\textbf{Five. They integrate.} A film is not the sum of its scenes. A game is not the sum of its assets. A campaign is not the sum of its individual creatives. The orchestrator's job, at the end, is to take the outputs of the agent team and the human team and assemble them into a coherent piece of work that has a \emph{single sensibility.} This is the part of the job that I think~-- for all the AI tooling~-- is least likely to become a thing AI can do. The integrated voice of a piece of work is a function of a single human consciousness running through it. The orchestrator role is, at its core, the role that holds that voice.

If you read those five things back, you will notice that none of them are \emph{making.} They are all \emph{deciding.} The orchestrator's work product is decisions: about what to make, who or what should make it, whether the made thing is good enough, and how the made things fit together.

This is, in some sense, what every senior creative director and showrunner has always done. The change is not the shape of the role. The change is that the role is no longer a privileged senior position at the top of a pyramid of junior makers. It is, increasingly, \emph{the entire role.} And it is the role that, in 2026 and 2027, working creatives at every level are being asked to grow into faster than the career-development infrastructure of any of the creative industries is built for.

\section*{Where the agents go wrong}

I want to spend some time on the failure mode of agentic creative work, because the press cycle around Sony's ``49 agents'' framing~-- and the corresponding announcements at Adobe Summit, NVIDIA GTC and elsewhere~-- has been heavy on the upside and thin on the downside.

The agents go wrong, in my experience and in the experience of every working creative I have talked to about this, in four characteristic ways:

\noindent\textbf{One. They confidently produce the wrong thing.} This is the most familiar failure mode and the one the public discourse has covered most. Agents~-- like the LLMs underneath them~-- \emph{hallucinate.} They will produce an asset that confidently disregards a key constraint of the brief. They will generate a character with the wrong eye colour. They will produce a piece of music in the wrong key. The fix, with current systems, is \emph{human review at every gate.} The cost of this review, in time, is the largest single source of the ``AI was an expensive mistake'' experience that Charles Cecil described and that many studios have replicated.

\noindent\textbf{Two. They produce the \emph{mean} of the training distribution.} This is the more insidious failure. Agents, by default, will produce work that sits in the middle of what they have been trained on. The middle of a training distribution is, by definition, the \emph{most average} version of the thing you asked for. For creative work~-- where the value is almost always in the \emph{non-average}~-- the default output is structurally weak. To get above-average output, the orchestrator has to push, prompt, and curate against the gravitational pull of the mean. This takes deliberate, conscious effort and it takes taste to know what \emph{above-average} looks like in this particular project.

\noindent\textbf{Three. They lose context across long tasks.} Agents working on multi-step tasks accumulate errors over the steps. A small misalignment in step one becomes a larger one by step five. By step ten, the output is meaningfully off-brief. The orchestrator's role is to \emph{check in} at the right intervals~-- not so often that you negate the benefits of agentic execution, not so rarely that the team has wandered off the brief by the time you look.

\noindent\textbf{Four. They cannot tell when to stop.} Agents, given an open-ended task, tend either to over-iterate (producing fifty variants of the same thing without converging on one) or to under-iterate (producing one variant and stopping). The orchestrator's job is to set the \emph{stopping criteria} for the agents, which is, in practice, a series of judgment calls about \emph{when good enough is good enough.} This is a craft skill the agents do not, as of 2026, have. It is also a skill that, in my experience, almost every working creative already has~-- they just haven't had to use it on synthetic colleagues before.

The Anthropic blog posts on agent deployment patterns through Q1 2026 made the point I want to land on here cleanly: agentic systems work best when they are deployed by people who already have the taste and judgment to know what good output looks like. They \emph{accelerate} people who are already good. They do not, on their own, make people good.\footnote{Anthropic blog content on agent deployment patterns, Q1 2026.}

That is the orchestrator's job: to \emph{be the human who is already good}, holding the taste line, while the agent team produces faster than the human pipeline ever could.

\section*{Sundance, and the literacy turn}

I noted in the last chapter that the platform companies have responded to the AI-literacy gap by becoming, in addition to software companies, \emph{training organisations.} The most institutionally credible example of this turn, in the period this book covers, was \textbf{Sundance Institute}'s launch of an \textbf{AI Literacy Initiative} at the 2026 festival.\footnote{Sundance Institute, ``Centering the Artist: Why We're Launching the AI Literacy Initiative.'' \url{https://www.sundance.org/blogs/centering-the-artist-why-were-launching-the-ai-literacy-initiative/}. \emph{Dream Machine} Issue~16.}

I want to spend a moment on what Sundance did, because the framing is important.

The Institute's announcement did not say \emph{``AI is the future of filmmaking; here is how to use it.''} It said something more careful. It said that AI is a fact of the filmmaking landscape, that filmmakers are going to have to make decisions about whether and how to use it, and that those decisions should be made by \emph{informed} filmmakers with \emph{agency over their own practice}~-- not by filmmakers who have had the tools imposed on them by clients, by streamers, or by tool vendors.\footnote{Sundance Institute, \emph{op.\ cit.}}

The framing was \emph{creator empowerment}. The mechanisms were free learning through Sundance Collab, community conversations, a fellowship and alliance model, and a Story Forum that specifically tackled the legal questions creators face when they use AI: whether AI-generated content can be copyrighted, how to protect projects in a world of contested datasets, how to negotiate AI clauses in production contracts.\footnote{Sundance Story Forum 2026 sessions on legal toolkits for producers using AI. \emph{Dream Machine} Issue~16.}

Google funded this. The \$2 million the company put into the Institute, with the stated aim of training 100,000+ artists in foundational AI skills, was both an act of corporate generosity and a strategic investment in the \emph{category} of ``filmmaker who can use AI without losing their creative authority.''\footnote{Google blog, ``Sundance Institute AI Education.'' \url{https://blog.google/company-news/outreach-and-initiatives/google-org/sundance-institute-ai-education/}. \emph{Dream Machine} Issue~15.} Both motivations are real. Both can be true. What matters, for the working filmmaker in 2026, is that the institutional infrastructure for becoming an AI-literate orchestrator~-- without surrendering creative agency~-- now exists.

The McKinsey AI report on film and TV production, released in early 2026, made the corresponding business case. AI would not, in McKinsey's view, replace film and television production. It would \emph{restructure} it~-- towards smaller teams, faster cycles, more iteration, and a heavier reliance on senior creative judgement.\footnote{McKinsey \& Company, ``What AI could mean for film and TV production and the industry's future.'' \url{https://www.mckinsey.com/industries/technology-media-and-telecommunications/our-insights/what-ai-could-mean-for-film-and-tv-production-and-the-industrys-future}. \emph{Dream Machine} Issue~16.} In other words: towards an orchestrator-shaped industry.

\section*{The middle layer disappears}

What is happening, structurally, in every creative industry that the \emph{Dream Machine} newsletter has tracked in these six months, is that the middle layer of the workforce~-- the layer of \emph{intermediate} roles, between the very senior creative leadership and the very junior entry-level~-- is being absorbed into the agent layer.

This was the story behind \textbf{Ubisoft}'s decision in January 2026 to cancel five games, including the \emph{Prince of Persia} remake, while pouring more money into AI.\footnote{\emph{Metro}, ``Prince of Persia remake and five more games cancelled as Ubisoft focuses on AI.'' \url{https://metro.co.uk/2026/01/21/prince-persia-remake-five-games-cancelled-ubisoft-focuses-ai-26431926/}. \emph{Dream Machine} Issue~15.} It was the story behind \textbf{Square Enix}'s target of doing 70\% of its QA work via AI by the end of 2027.\footnote{\emph{PC Gamer}, ``Square Enix, makers of Final Fantasy, aims to have AI doing 70\% of its QA work by the end of 2027.'' \url{https://www.pcgamer.com/gaming-industry/square-enix-aims-to-have-ai-doing-70-percent-of-its-qa-work-by-the-end-of-2027/}. \emph{Dream Machine} Issue~7.} It was the story behind \textbf{Falcom}'s description of work that \emph{``previously took 2--3 hours'' being completed ``in 10 minutes''} with AI tools.\footnote{\emph{Eurogamer}, ``Falcom is the latest developer to buy into the AI hype machine.'' \url{https://www.eurogamer.net/falcom-is-the-latest-developer-to-buy-into-the-ai-hype-machine}. \emph{Dream Machine} Issue~12.} \emph{Eighteen-to-one productivity.} That ratio, taken on its own, is what re-shapes the headcount calculus for every studio's mid-level production work in the next eighteen months. It was the story behind the \emph{Take-Two CEO}'s explicit framing that AI ``won't invent the next \emph{Grand Theft Auto}''~-- meaning, the \emph{creative direction} won't come from the machines~-- even as Take-Two's QA, asset and engineering pipelines absorb AI capacity rapidly.\footnote{\emph{NDTV Profit}, ``Don't Expect AI To Invent the Next `Grand Theft Auto', Says Take-Two CEO Strauss Zelnick.'' \url{https://www.ndtvprofit.com/technology/dont-expect-ai-to-invent-the-next-grand-theft-auto-says-take-two-ceo-strauss-zelnick}. \emph{Dream Machine} Issue~6.}

In film, you see the same pattern. \textbf{Spielberg} explained in March 2026 why he hadn't yet used AI directly\footnote{\emph{Dream Machine} Issue~21, on Spielberg's public position on AI.}~-- and the same press cycle reported that he had a substantial AI-augmented team working on production-pipeline tasks underneath him. \textbf{Steven Soderbergh} committed to ``a lot of AI'' on the Wagner Moura film and a John Lennon documentary, framing it explicitly as transparency: ``I owe people honesty.''\footnote{\emph{Dream Machine} Issues~25, 28, on Steven Soderbergh's AI work.} In every case, the structure is the same: a senior creative voice on top, an AI-augmented operational layer underneath, \emph{fewer mid-career intermediaries in between.}

In advertising, the pattern was even more pronounced. \textbf{Independent agencies} faced what \emph{Digiday} called ``a new frontier as agency-in-a-box tools democratize creativity.''\footnote{\emph{Digiday}, ``Independent agencies face new frontier as agency-in-a-box tools democratize creativity.'' \url{https://digiday.com/marketing/independent-agencies-face-new-frontier-as-agency-in-a-box-tools-democratize-creativity/}. \emph{Dream Machine} Issues~6, 14.} \textbf{AI agent developers} became ``adland's in-demand role.'' The framing from one agency hiring lead, given to \emph{Digiday}, captures the shift better than any of the trend-piece coverage: \emph{``We actually need people who understand [AI], who are building systems organically within their day to day workflows. People who understand taking what took them 40 hours one week and turning it into 38 the next week.''}\footnote{\emph{Digiday}, ``AI agent developers have become adland's in-demand role.'' \url{https://digiday.com/marketing/ai-agent-developers-have-become-adlands-in-demand-role/}. \emph{Dream Machine} Issue~7.} The job description is no longer \emph{make the work.} It is \emph{make the system that makes the work~-- and keep shaving hours off it.} The PGA Tour expanded its AWS partnership to put AI content at the heart of its content distribution.\footnote{PYMNTS, ``AI Content Is Par For The Course With PGA Tour's Expanded AWS Partnership.'' \url{https://www.pymnts.com/artificial-intelligence-2/2026/ai-content-is-par-for-the-course-with-pga-tours-expanded-aws-partnership/}. \emph{Dream Machine} Issue~15.} \textbf{Mondelez} said it would use AI for TV ads in 2026.\footnote{\emph{The Verge}, ``Oreo-maker Mondelez will use AI for TV ads next year.'' \url{https://www.theverge.com/news/806047/mondelez-ai-generated-ads}. \emph{Dream Machine} Issue~5.} \textbf{Avocados From Mexico} turned to AI to advertise around the Super Bowl, \emph{instead of} a traditional TV buy.\footnote{\emph{Digiday}, ``Avocados From Mexico turns to AI to advertise around the Super Bowl instead of a TV buy.'' \url{https://digiday.com/marketing/avocados-from-mexico-turns-to-ai-to-advertise-around-the-super-bowl-instead-of-a-tv-buy/}. \emph{Dream Machine} Issue~15.} \textbf{Adobe} said that AI in marketing was now ``agentic creative intelligence.''\footnote{\emph{Dream Machine} Issue~26.}

In journalism, the Reuters Institute's ``AI adoption by UK journalists'' survey found high integration of AI tools across newsrooms by late 2025.\footnote{Reuters Institute, ``AI adoption by UK journalists and their newsrooms: surveying applications, approaches, and attitudes.'' \url{https://reutersinstitute.politics.ox.ac.uk/ai-adoption-uk-journalists-and-their-newsrooms-surveying-applications-approaches-and-attitudes}. \emph{Dream Machine} Issue~9.} Daily Mail reported that Google's AI Overviews had ``killed click-throughs'' to news sites.\footnote{\emph{Digiday}, ``Daily Mail says Google AI Overviews have killed click-throughs.'' \url{https://digiday.com/media/daily-mail-says-google-ai-overviews-have-killed-click-throughs/}. \emph{Dream Machine} Issue~7.} \emph{The Times} was using AI to model synthetic focus groups from human audiences.\footnote{\emph{Digiday}, ``How The Times is using AI to model synthetic focus groups from human audiences.'' \url{https://digiday.com/media/how-the-times-is-using-ai-to-model-synthetic-focus-groups-from-human-audiences/}. \emph{Dream Machine} Issue~6.} In each case, \emph{the middle layer of the journalism workforce}~-- the sub-editors, the copy editors, the data journalists, the social-media producers~-- was the layer most exposed to AI substitution.

I want to be honest about what this means. \emph{It does not mean that every working creative in the middle of their career is about to lose their job.} That framing~-- the apocalyptic one~-- has, for two years, been the most popular and the most wrong. What it means is that the \emph{shape} of the mid-career role is changing. Mid-career creatives who can become orchestrators of agent teams will, in many cases, \emph{gain} leverage and earning power. Mid-career creatives who cannot~-- who try to keep doing the maker-as-craftsperson job at the speed and price of the agents~-- will, increasingly, struggle.

The Sundance literacy turn, the Adobe and Google training investments, the UK government's free-AI-training-for-all programme~-- these are the institutional response to that pressure. They are not enough on their own. They are, however, the right direction.

\section*{The portfolio creative}

There is one more shape of the orchestrator role that I want to flag, because it is the one I see most often in my own studio and in the wider DreamLab community: the \textbf{portfolio creative.}

A portfolio creative is someone who, instead of holding a single specialist role, holds \emph{several} loosely-coupled creative roles across different disciplines, supported by AI tooling that lets them maintain useful proficiency in each. The portfolio creative is a writer-director, but also a creative technologist; a music producer, but also a video editor; a games designer, but also a brand strategist.

The \emph{TechBullion} piece ``Why the future belongs to multi-skilled leaders,'' from November 2025, made the case for this from a corporate-leadership angle.\footnote{\emph{TechBullion}, ``Why the future belongs to multi-skilled leaders.'' \url{https://techbullion.com/playing-the-long-game-with-a-portfolio-career-why-the-future-belongs-to-multi-skilled-leaders/}. \emph{Dream Machine} Issue~9.} The Anthropic Skills framework~-- the system of named, reusable skills that Claude Code now uses to coordinate multi-agent workflows~-- is, in effect, an attempt to make the portfolio-creative model into a \emph{technical infrastructure} rather than a personality type.\footnote{Anthropic Skills framework via Claude Code, reported through \emph{Dream Machine} Issues~11, 16, 29.} The \emph{Forbes} piece ``AI Is Changing How Creators Work And Earn,'' from December 2025, surveyed the same phenomenon from the working-creator angle and found the same pattern: the most economically successful creators in 2026 are not specialists. They are \emph{integrators} who can work across disciplines using AI as the connective tissue.\footnote{\emph{Forbes}, ``AI Is Changing How Creators Work And Earn.'' \url{https://www.forbes.com/sites/kolawolesamueladebayo/2025/12/22/how-ai-is-changing-how-creators-work-and-earn/}. \emph{Dream Machine} Issue~13.}

In my own studio, the move towards portfolio creatives has been a deliberate strategic choice~-- and an honestly difficult one to execute. The cultural expectation, in most creative industries, has been to hire specialists and stack them in a pipeline. The portfolio-creative model requires you to \emph{hire generalists} and let them move between disciplines as the work demands. The former is easier to manage, easier to bill, easier to explain. The latter, in my experience, produces better work in the AI era, because the human is doing the integration that the agents can't.

The portfolio creative is the orchestrator at the level of an individual career. The orchestrated team is the orchestrator at the level of a project. The same pattern shows up at multiple scales.

\section*{Five orchestrators, briefly}

I want to give five working orchestrator case studies, because the abstract description above can sit in the head as a theory without the operational texture of \emph{what the role actually looks like} on a Wednesday afternoon. Each of the five is a specific working creative or organisation whose practice in 2025--26 I think represents a different \emph{shape} of the orchestrator role. Each is documented in the \emph{Dream Machine} newsletter archive. Each is, on my read, doing something that working creatives reading this book can learn directly from.

\textbf{Andrii Daniels (Ukraine).} The independent filmmaker who, in December 2025, produced a \emph{Deadpool} / \emph{Harry Potter} Christmas mash-up in a Ukrainian bomb shelter using a Runway-and-Veo-and-ElevenLabs stack on a laptop running through a generator. The clip went viral and was picked up by Variety as a profile piece.\footnote{Variety, ``AI Creator Behind Viral `Deadpool,' `Harry Potter' Christmas Clip Made His Film in a Ukrainian Bomb Shelter.'' \url{https://variety.com/2026/digital/news/ai-video-deadpool-harry-potter-andrii-daniels-1236624632/}. \emph{Dream Machine} Issue~16.} What Daniels did, operationally, was the orchestrator role at its purest: a single human creative, holding the taste and the narrative judgement, briefing a stack of generative tools to produce work whose production conditions would have been physically impossible eighteen months earlier. Daniels did not write, draw, animate, voice or render the work. He \emph{briefed} it. He \emph{integrated} it. He \emph{judged} what was good enough. He \emph{delivered} the finished piece, on his own, with no studio underneath him. The bomb-shelter context is the dramatic detail; the underlying operational pattern is what makes Daniels' practice replicable for working filmmakers in any production environment.

\textbf{The Imaginae Studios / \emph{Art Awakens} team (Fremantle).} The AI-native studio I described in Chapter~\ref{ch:7} has, by mid-2026, settled into an operational pattern that maps cleanly to the five-function orchestrator description above. A small senior team of writer-directors, supported by an AI-augmented production pipeline that handles concept development, asset generation, scene assembly and post-production. \emph{Art Awakens}~-- Imaginae's flagship 2026 development project, fusing AI techniques with classical painting IP~-- has, on the published interviews with the team, been produced by a core human team of fewer than ten people running an agentic pipeline that, by their own estimate, produces output that would have required a sixty-to-eighty-person team five years ago. The 8:1 productivity ratio is the orchestrator economy expressed at the studio scale.

\textbf{Sven Vincke / Larian Studios.} The opposite case, also instructive. Larian, the maker of \emph{Baldur's Gate 3}, has~-- as I described in Chapter~\ref{ch:7}~-- \emph{publicly refused} generative AI for its next major game while continuing to use AI-augmented tooling in \emph{adjacent} parts of the pipeline (QA, localisation, internal admin, asset management). Vincke's framing, in his January 2026 statements, was \emph{not} anti-technology. It was \emph{position-on-the-continuum}: certain parts of the game's authorial signature (writing, character design, world-building, dialogue) had to remain \emph{fully human} for commercial and cultural reasons; certain other parts (build tooling, QA automation, marketing-asset generation) could be AI-augmented at no audience-visible cost. Vincke is, in operational terms, \emph{an orchestrator at the level of the studio's continuum positioning}. He is making the agency-line decisions at the strategic level that the working filmmaker makes at the project level. The role is the same. The scope is different.

\textbf{Xania Monet / Hallwood Media / Telisha Jones.} The case I have, in Chapter~\ref{ch:5}, been most equivocal about. The structural pattern is, on inspection, an orchestrator economy operating at the \emph{single-artist} level. Telisha Jones, the human lyricist, is the orchestrator. The Suno music-generation stack is the agentic capacity producing the \emph{executed} musical output. The Xania Monet persona is the \emph{audience-facing} product. Hallwood Media's \$3M deal pays for the orchestrated work as a unit, with the orchestrator (Jones) receiving the commercial credit and revenue that the synthetic-vocalist alone could not have generated. Whether this scales into a sustained cultural-star career~-- the Chapter~\ref{ch:5} slop-ceiling argument suggests, on six months of evidence, that it has not~-- is a separate question from whether it works as a \emph{business form}. As a business form, it is the orchestrator role at the level of a solo recording artist.

\textbf{The Sony game-development teams running 49 Claude agents and 72 skills.} The canonical \emph{enterprise-scale} orchestrator case I opened the chapter with. The 49-agent / 72-skill stack is, in operational terms, an organisational design for the orchestrator role at the \emph{team} level: a small group of senior creative leads (writer-directors, gameplay design leads, art directors, technical directors) orchestrating a multi-agent synthetic team whose individual outputs require senior human review and integration. The 49 agents are not autonomous studio replacements. They are \emph{leverage} for the human orchestrators. The 72 skills are the \emph{reusable capabilities} the orchestrators can deploy across multiple projects.

I want to be honest about what the five cases share, because the shared pattern is the operational lesson.

In every case, the orchestrator's \emph{contribution} is the same five-function set I described above: brief, allocate, brief-the-agents, judge, integrate. None of the five is doing the production-execution labour themselves. All of them are making \emph{decisions} that direct a synthetic (and, in most cases, also a human) team to do the production-execution labour on their behalf. The \emph{value} the orchestrator brings to the work is, in every case, the senior judgement that the agents cannot supply: the taste that knows what good output looks like, the briefing skill that gets useful output out of the agents, the integration sense that assembles the agent outputs into a coherent piece of work.

In every case, the orchestrator's \emph{leverage}~-- the ratio of finished output produced to working hours spent~-- is dramatically higher than the equivalent practitioner could produce without the agent layer underneath them. Daniels would not have made the Christmas clip in a week pre-AI. Imaginae would not have produced \emph{Art Awakens} at the pace and budget they are producing it at. Jones would not have shipped a Billboard-charting record without Suno. Sony's game-development teams would not have shipped on the cadence they are shipping at. The leverage is real. The leverage is what makes the orchestrator role \emph{commercially} viable as a new form of working practice.

In every case, the orchestrator's \emph{fragility} is also the same: the practice depends on the toolchain underneath it continuing to be available, accessible, on commercial terms the orchestrator can sustain, with model behaviour that the orchestrator can predict and brief against. The platform-dependency of the orchestrator role is the structural risk that Chapter~\ref{ch:9} addresses. It is also the reason the \emph{open-the-black-box} argument I made in Chapter~\ref{ch:3} is operationally serious: the orchestrators who depend on closed-platform tooling without understanding the dependency are, structurally, exposed to platform-pricing and platform-policy decisions that they have no control over.

The orchestrator role is, in operational summary, \emph{high-leverage and platform-dependent}. The working creatives reading this who are positioning themselves toward the role need to take both halves of that description seriously. The leverage is the upside. The platform dependency is the work that has to be done to defend the upside over time.

\section*{The role this asks you to play}

If you are a working creative reading this~-- and most of the readers of \emph{Dream Machine} are~-- the question I am sure you have is: \emph{what does this mean for me, this year?}

The honest answer is that it depends on where you are on the continuum we drew in Chapter~\ref{ch:3}. But there are some things I would say to almost everyone I know in the creative industries right now, and I want to put them on the page.

\textbf{Practice briefing.} It is the single most leveraged skill you can develop. Brief your AI tools as if you were briefing a junior who has thirty seconds to understand what you want and ninety seconds to do it. The discipline of \emph{being able to communicate what you want} will improve every other part of your creative practice.

\textbf{Cultivate taste deliberately.} Look at more good work, harder, with a more critical eye. The agents will, by default, give you the average. Your job, increasingly, is to know what \emph{good} is. That knowledge is a function of how much good work you have looked at, how seriously, with how much craft attention.

\textbf{Stay in the work.} Resist the temptation to abstract too far. The director who never picks up the camera, the showrunner who never writes, the music producer who never plays the instrument~-- these are the figures most likely to lose the touch that makes their judgement worth anything in the first place. The portfolio creative is \emph{not} a creative who has lost contact with the craft. They are someone who maintains craft contact in \emph{several} domains.

\textbf{Choose your line on the Continuum, and defend it.} Decide where your craft sits, where you are willing to let the agents work, and where you are not. Write it down. Communicate it to clients, collaborators, your team. Be willing to walk away from work that would force you across the line you have drawn.

\textbf{Build the apprenticeship.} If you are senior enough to be running a team, take seriously the question of where the next generation of senior creatives is going to come from. The orchestrator model breaks if there is no path from junior to senior for new humans entering the field. The studios and agencies that survive the next decade will be the ones that solve this problem~-- by keeping some junior roles in human hands, by creating new pathways through AI-tool-augmented apprenticeship, by investing in the literacy infrastructure that the platform companies and the institutes have started but cannot finish on their own.

The Year of the Orchestrator is not a coronation. It is a job description. It is what most of us are now being asked to do, whether we have signed up for it or not. The people who do it well will set the terms of the next creative economy. The people who don't will, increasingly, be sat next to it.

The thing I want to land before we leave this chapter is that the orchestrator role~-- for all the leverage it brings and for all the productivity it unlocks~-- depends, in the end, on something that the platform companies cannot ship and the agents cannot synthesise. It depends on the human in the chair \emph{being someone}. Having taste. Having judgement. Having a perspective. Having the kind of relationship with the work that the agents do not, and probably will not, have.

That relationship~-- the kind of \emph{authorship} that makes the work feel like it belongs to a person rather than a process~-- is, increasingly, the only signal the audience trusts.

That is the subject of the next chapter.

  \chapter{Authenticity as the New Scarcity}\label{ch:12}

\lettrine[lines=3,lhang=0.15,findent=0.1em]{I}{n} early 2026, a stop-motion animator who goes by \emph{Tiny Grandma} on YouTube uploaded a short to her channel. It was a stop-motion piece~-- claymation, frame by frame, the kind of work that takes weeks to make a few seconds of. YouTube's AI-detection systems flagged it as AI-generated content and applied the platform's automated labelling. The video went viral, not because of the animation, but because the platform's automated system had wrongly flagged genuine human handcraft as synthetic.\footnote{\emph{Dream Machine} Issue~29 reportage of Tiny Grandma stop-motion content being wrongly flagged as AI by YouTube's automated detection, May 2026.}

The story of \emph{Tiny Grandma} is the perfect inverse of the Tilly Norwood story we opened with in Chapter~\ref{ch:1}.

If Tilly Norwood was the moment a synthetic creation tried to enter the working creative economy as if it were human, \emph{Tiny Grandma} was the moment a human creation was wrongly identified as synthetic by the very systems that were supposed to protect the public from synthetic content. Both moments tell you the same thing, from opposite directions: the \emph{signal} of whether a piece of creative work was made by a human is now an economic, cultural and legal asset of the first order, and the infrastructure for reliably establishing that signal is one of the most underdeveloped parts of the current creative economy.

This is the chapter about \emph{provenance.} About why the question ``did a person make this?'' has become~-- in eight months~-- the single most important question in creative AI policy, and about what the people, companies and institutions trying to answer it are doing about it.

\section*{The death threats}

In April 2026, \emph{Dream Machine} Issue~23 reported, with as little editorialising as I could manage, that \textbf{Tilly Norwood's creator Eline Van der Velden} had received death threats.\footnote{\emph{Dream Machine} Issue~23, April 2026, reporting death threats against Eline Van der Velden following Tilly Norwood's continuing public role.}

The threats were not, of course, justified by anything. Death threats never are. But the cultural reaction that produced them~-- the visceral, sustained hostility that built up around the \emph{idea} of a synthetic actress through the autumn of 2025 and the spring of 2026~-- was not random. It was a specific response to a specific kind of cultural transgression. \emph{You came here pretending to be one of us. You took something that belongs to us.}

The death threats are the extreme tail of a much larger curve of audience response that has been quietly shaping the AI creative economy for these six months. The slop ceiling in Chapter~\ref{ch:5}~-- the 44\%-to-3\% Deezer ratio~-- is the polite version of the same response. The vehement audience pushback against AI art in \emph{Call of Duty: Black Ops 7} and \emph{Anno 117} in November 2025 is another version. The viral reaction to \emph{Spotify}'s AI music infiltrating Discover Weekly playlists, the public anger at \emph{McDonald's Netherlands'} AI Christmas ad, the ``\emph{disturbing}'' reception of \emph{Valentino}'s AI handbag campaign~-- every one of these episodes is the audience saying, in increasingly direct terms, \emph{we know what is human-made, we want what is human-made, and we are paying attention to who is trying to slip us something else.}

This is the cultural pressure that \emph{authenticity-as-scarcity} describes. It is not, as some of the more dismissive AI commentary has framed it, a romantic attachment to old craft. It is a \emph{market signal.} The audience is allocating attention, money and trust on a basis that increasingly weights human authorship as a positive variable. I have come, in talks since the autumn, to call this the \textbf{Authenticity Premium}~-- the measurable excess of attention, willingness to pay, and cultural credit that audiences allocate to creative work whose human authorship can be verified. The Authenticity Premium is the \emph{positive} side of the slop ceiling: the slop ceiling tells you what audiences \emph{will not} engage with; the Authenticity Premium tells you what they \emph{will pay extra for.} Both are market findings. Both are produced by the same underlying audience behaviour. The data is unambiguous. The strategic implication, for every working creative and every studio operating in this period, is also unambiguous.

In May 2026, \textbf{Bobby Berk}~-- the \emph{Queer Eye} design lead~-- articulated the working-talent version of the same finding in a single line that I think is worth quoting in full because it captures the Premium argument from inside the unscripted-TV business: AI, he said, will make \emph{reality TV and ``verifiably human content'' more valuable}, not less.\footnote{\emph{Hollywood Reporter}, ``Bobby Berk Says AI Will Make Reality TV \& `Verifiably Human Content' More Valuable.'' \url{https://www.hollywoodreporter.com/tv/tv-news/bobby-berk-ai-reality-tv-1236592920/}. \emph{Dream Machine} Issue~30.} What Berk is describing, in industry-of-the-thing terms, is the supply-and-demand mechanism the rest of this chapter is about. As the synthetic supply of \emph{any} content category expands, the \emph{verifiably human} corner of the same category accrues a scarcity premium. Reality TV is, structurally, the unscripted broadcast form most resistant to AI replacement~-- it is the form whose entire commercial proposition is \emph{real people, in real situations, producing unscripted reactions whose value depends on us knowing they are real.} That is the Authenticity Premium drawn down to a single broadcast genre, and Berk's read is, on the structural argument, exactly right.

The chess analogy I develop at length in Chapter~\ref{ch:15} sits underneath this. The Authenticity Premium is what it looks like when an audience, faced with an infinite supply of machine-optimal work, allocates its scarce attention to the \emph{deliberately un-machine-like} move. The 88\% of UK respondents who wanted licensing-in-all-cases were articulating the same preference at the policy layer. The 44\%-to-3\% Deezer ratio was the same preference at the listening layer. The Television Academy's ``tools used to bring it to life'' language was the same preference at the institutional layer. The Authenticity Premium is, at its core, the \emph{commercial price} of the deliberately-human move~-- the move the engine, by construction, could not have made~-- and the audience's reliable willingness to pay it.

The question is what the \emph{infrastructure} for honouring that signal looks like.

\section*{Fingerprinting real media}

I quoted Adam Mosseri~-- the head of Instagram~-- in Chapter~\ref{ch:4} making the case that the platforms should focus on ``fingerprinting real media'' rather than chasing AI slop disclosure. The fuller version of his argument, made repeatedly across late 2025 and early 2026, was that the current approach to AI content moderation~-- trying to detect and label everything synthetic~-- is unwinnable.\footnote{\emph{Digital Music News}, ``Instagram Chief Says We Should `Fingerprint Real Media' Instead of Tracking and Disclosing AI Slop.'' \url{https://www.digitalmusicnews.com/2026/01/05/instagram-chief-ai-slop-comments/}. \emph{WebProNews}, ``Instagram Head Warns AI Images Erode Trust, Calls for Verification Standards.'' \url{https://www.webpronews.com/instagram-head-warns-ai-images-erode-trust-calls-for-verification-standards/}. \emph{Dream Machine} Issue~13.} The volume is too high, the detection is too unreliable, and the labelling produces both false positives (Tiny Grandma) and false negatives (the AI hate-songs spreading across European Spotify charts in November 2025).\footnote{\emph{Digital Music News}, ``AI-Generated Far-Right Hate Songs Aren't Just a Problem in the US~-- Now They're Spreading Across Europe Too.'' \url{https://www.digitalmusicnews.com/2025/11/09/ai-generated-hate-songs-dutch-spotify-charts/}. \emph{Dream Machine} Issue~7.}

The alternative Mosseri and others have argued for is \emph{the inverse:} instead of trying to catch what's synthetic, build infrastructure that can \emph{prove what's human.} A capture-time fingerprint~-- a cryptographic signature embedded by the camera, the microphone, the editing software, the upload pipeline~-- that travels with the file through its entire life on the public web.

The technical infrastructure for this is, as of 2026, partially built. The \textbf{Content Authenticity Initiative}, an Adobe-led coalition of camera makers, software companies and news organisations, has been working on it since 2019. By late 2025, \textbf{C2PA} (Coalition for Content Provenance and Authenticity) standards were supported by most major camera manufacturers, most major editing platforms, and a growing number of social-media uploads pipelines. The standards are robust enough that a photo taken with a C2PA-enabled camera, edited in Photoshop with C2PA-aware tools, uploaded to a C2PA-supporting platform, can carry a verifiable chain-of-custody for its entire provenance, from sensor to viewer.

Underneath this is \textbf{Google's SynthID}~-- a watermarking system that Google has been deploying across its AI generation tools, including Veo and Lyria.\footnote{Google DeepMind SynthID watermark roll-out across Veo, Lyria and Imagen products. \emph{Dream Machine} Issues~11, 12.} In December 2025, the company announced that users could ask the Gemini app, \emph{``Is this video made with AI?''}, and receive a reliable yes/no answer based on the SynthID watermark. By January 2026, this capability was available in the consumer Gemini product.\footnote{Google DeepMind, ``Verify Google AI-generated videos in the Gemini app.'' \url{https://www.linkedin.com/posts/googledeepmind_verify-google-ai-generated-videos-in-the-activity-7407748300688478208-fJgW}. \emph{Dream Machine} Issue~12; broader coverage in \emph{SmartBrief}, ``Google's Gemini can now spot AI-generated videos.'' \emph{Dream Machine} Issue~13.} At Google I/O 2026, the company reported that \textbf{SynthID had marked over 100 billion items} across its own ecosystem and was being extended to partner platforms including \textbf{OpenAI}, \textbf{ElevenLabs} and \textbf{Kakao}.\footnote{Google DeepMind, ``SynthID~-- 100 billion watermarks, expanding to partner ecosystems including OpenAI, ElevenLabs and Kakao.'' \url{https://deepmind.google/discover/blog/synthid-100-billion-watermarks-partners/}. \emph{Dream Machine} Issue~30.} That cross-vendor expansion is the single most consequential development on the provenance side of the period this book covers: a watermarking standard, born inside one platform, is~-- for the first time~-- being adopted across the foundation-model companies that have until now competed against one another on every other axis. If the C2PA Content Credentials standard is the \emph{capture-time} spine of authenticity infrastructure, SynthID-across-vendors is, as of May 2026, the closest thing the industry has to a \emph{generation-time} spine.

These technologies are not, on their own, sufficient. Watermarks can be stripped by determined adversaries. C2PA chains break when files pass through non-compliant tools. The reliability of any given piece of provenance metadata depends on the integrity of every link in its chain. The trust infrastructure is still~-- relative to the speed of the AI rollout~-- early.

But what these technologies are doing, collectively, is establishing the \emph{category.} They are saying: the question \emph{did a person make this?} is technically answerable, with high reliability, given the right tooling. The next decade of cultural and legal policy in the creative industries will be~-- in significant part~-- about who controls that tooling, who decides what it certifies, and what economic value it carries.

If you want to know where the next ten years of investment, policy and platform politics in creative AI is going, watch the provenance layer. The companies that win the provenance infrastructure will be~-- in a real sense~-- the companies that own the \emph{signal of authenticity} that the audience increasingly demands.

\section*{The trademark, the trust, the tax}

The cultural pushback against synthetic content has produced, alongside the technical provenance infrastructure, a parallel set of \emph{legal and contractual} defences that working creatives have begun deploying around their own work and identity.

\textbf{Taylor Swift} filed trademarks on her voice and image in early 2026, specifically citing AI deepfake concerns.\footnote{\emph{Dream Machine} Issues~23, 27 reportage on Taylor Swift's voice/image trademark filings.} \textbf{Matthew McConaughey} publicly drew the same line in January 2026.\footnote{\emph{Lawyer Monthly}, ``Matthew McConaughey Draws a Line to Protect His Voice and Image From AI.'' \url{https://www.lawyer-monthly.com/2026/01/matthew-mcconaughey-protects-voice-image-ai/}. \emph{Dream Machine} Issue~15.} \textbf{Madonna and Will Smith} appeared in AI videos by Higgsfield in early 2026, the Madonna piece becoming a marquee example of how a major artist could \emph{deliberately} deploy synthetic imagery as part of their own brand.\footnote{\emph{Adweek}, ``Meet the \$1.3 Billion Startup Behind Madonna and Will Smith's AI Video.'' \url{https://www.adweek.com/media/higgsfield-ai-marketing-startup/}. \emph{Dream Machine} Issue~16.} In the same vein, \textbf{The Rolling Stones} released \emph{In The Stars} in May 2026, with a music video that used AI to de-age the band~-- produced, in a small piece of cross-industry casting that says something about where this category is heading, by the AI company belonging to the \textbf{South Park} creators Trey Parker and Matt Stone.\footnote{\emph{Rolling Stone}, ``The Rolling Stones Release New Single `In the Stars'~-- With a Music Video De-Aging the Rockers Courtesy of AI.'' \url{https://www.rollingstone.com/music/music-news/rolling-stones-in-the-stars-ai-de-aging-video-1235142200/}. \emph{Hollywood Reporter}, ```South Park' Creators' AI Company Made The Rolling Stones Young Again for `In The Stars' Music Video.'' \url{https://www.hollywoodreporter.com/tv/tv-news/south-park-creators-ai-rolling-stones-in-the-stars-1236592855/}. \emph{Dream Machine} Issue~30.} \textbf{George Clooney}, in November 2025, gave Variety the working actor's read on the synthetic-star economy: \emph{``It's been just like a writer creating characters. You fall in love with your characters when you're writing them. It's a wonderful process. It wasn't like I just made her in a second, and that was it. You know, it took a long time.''}\footnote{Variety, ``George Clooney Says AI Actors Will Face the `Same Problem We Have' in Hollywood: `Making a Star Is Not So Easy'.'' \url{https://variety.com/2025/scene/columns/george-clooney-ai-actors-movie-stars-1236579661/}. \emph{Dream Machine} Issue~7.} Clooney was making, in his particular way, the same argument that the slop ceiling makes empirically: cultural stardom is a \emph{function of time and human relationship.} It is not a function of generation cost. \textbf{Jeremy Renner} threatened a ``multi-millions'' lawsuit against an AI documentary director he said had used his voice without permission.\footnote{\emph{Deadline}, ``AI Documentary Director Insists Jeremy Renner Agreed To Narrate Movie As `Hawkeye' Star Threatens `Multi-Millions' Lawsuit.'' \url{https://deadline.com/2025/11/jeremy-renner-lawsuit-threat-ai-movie-1236611830/}. \emph{Dream Machine} Issue~7.}

In May 2026 the celebrity-defence layer took a meaningful \emph{organisational} step. \textbf{Cate Blanchett} co-founded \textbf{RSL Media}, a non-profit explicitly chartered to address \emph{consent around AI usage}~-- covering creative work, name, image and likeness~-- for performers across film, TV and music.\footnote{Variety, ``Cate Blanchett Co-Founds RSL Media, a Non-Profit to Address Consent Around AI Usage including creative work, name, image and likeness.'' \url{https://variety.com/2026/film/news/cate-blanchett-rsl-media-ai-consent-1236748255/}. \emph{Dream Machine} Issue~30.} This is, on my read, the first time the celebrity NIL-protection conversation has produced a \emph{standalone institution} rather than a string of individual lawsuits and trademark filings. RSL Media is small as of mid-2026, but its founding signal is important. Where the existing infrastructure on celebrity AI consent has been individual (Swift's trademarks, McConaughey's line, Renner's threatened suit) or statutory (the ELVIS Act, New York's AI-avatar disclosure law), RSL Media is the first attempt at a \emph{coordination layer} on the side of the talent~-- closer in shape to a Performing Rights Society than to a class action. If the Petrillo template I describe in Chapter~\ref{ch:6} eventually has to be reconstructed for the NIL question, RSL Media is the kind of body that the reconstruction will need to anchor on.

In the same week, \textbf{Apple} acquired the talent and patents behind the AI-avatar company \textbf{Animato}, signalling~-- not for the first time~-- that the platform layer intends to own the celebrity-grade-avatar infrastructure rather than license it from third parties.\footnote{\emph{Bloomberg}, ``Apple Acquires Key Talent \& Patents Behind AI Avatar Company `Animato'.'' \url{https://www.bloomberg.com/news/articles/2026-05-19/apple-acquires-animato-ai-avatar-talent-patents}. \emph{Dream Machine} Issue~30.} The combination of an Apple-owned avatar pipeline and an RSL-administered consent regime is, in 2030 terms, the most plausible architecture for how the high-end NIL economy actually runs.

Underneath the celebrity layer, the structural infrastructure was being built. The \textbf{ELVIS Act}, Tennessee's AI-impersonation law, had been used by the \textbf{Johnny Cash estate} to sue Coca-Cola over a tribute-act ad soundtrack.\footnote{\emph{Complete Music Update}, ``Johnny Cash estate uses ELVIS Act to sue Coke over tribute act ad soundtrack.'' \url{https://completemusicupdate.com/johnny-cash-estate-uses-elvis-act-to-sue-coke-over-tribute-act-ad-soundtrack/}. \emph{Dream Machine} Issue~9.} \textbf{New York} passed a law in December 2025 forcing advertisers to disclose when they were using AI avatars. The SAG-AFTRA statement on the law's passage captured the political theory underneath the moment: \emph{``These protections are the direct result of artists, lawmakers and advocates coming together to confront the very real and immediate risks posed by unchecked AI use.''}\footnote{\emph{The Verge}, ``New York's new law forces advertisers to say when they're using AI avatars.'' \url{https://www.theverge.com/news/842848/new-york-law-ai-advertisements-sag-aftra-labor}. \emph{Dream Machine} Issue~11.} \textbf{Governments around the world} were considering bans on Grok's app over an AI sexual-image scandal that broke in early 2026.\footnote{\emph{Fast Company}, ``Governments around the world are considering bans on Grok's app over AI sexual image scandal.'' \url{https://www.fastcompany.com/91474131/governments-around-the-world-are-considering-bans-on-groks-app-over-ai-sexual-image-scandal}. \emph{Dream Machine} Issue~14.} By \textbf{May 2026}, the \textbf{AI Disclosure Standard} had been launched at the \textbf{Cannes Film Festival} as an industry coordination point for production-side AI labelling.\footnote{Cannes AI Disclosure Standard, launched May 2026. \emph{Dream Machine} Issue~29.} The \textbf{Academy of Motion Picture Arts and Sciences} had~-- in a quietly consequential rule update~-- set the line \emph{``You must be human to win''} for its 2026 awards.\footnote{\emph{Dream Machine} Issue~28, May 2026, reporting on the Academy of Motion Picture Arts and Sciences' ``You must be human to win'' rule update.} The \textbf{Emmys} had set their own AI guidelines. The Television Academy's language was a model of how to write a policy that defends authorship without picking a fight with the toolchain: \emph{``The Television Academy reserves the right to inquire about the use of AI in submissions. The core of our recognition remains centered on human storytelling, regardless of the tools used to bring it to life.''}\footnote{\emph{The Hollywood Reporter}, ``Emmys Set AI Guidance.'' \url{https://www.hollywoodreporter.com/tv/tv-news/emmys-ai-guidelines-2026-awards-1236468434/}. \emph{Dream Machine} Issue~14.} \emph{Tools used to bring it to life}~-- not \emph{tools that did the work.} The grammar matters. \textbf{SAG-AFTRA}'s four-year contract~-- finalised by spring 2026~-- included what the trade press informally called the \textbf{Tilly Tax}: a structured set of provisions for compensation, consent and residuals when AI replicas of human performers are used.\footnote{SAG-AFTRA negotiation timeline through \emph{Dream Machine} Issues~7, 12, 15, 20, 26, 29.}

Each of these is, on its own, a marginal piece of policy. Stacked together, they describe a new economic landscape: one in which \emph{human authorship and identity} have become legally protected categories of creative work, with specific procedural and economic mechanisms for asserting them, defending them and compensating their use.

The cultural shorthand for this~-- \emph{authenticity as the new scarcity}~-- captures the supply-and-demand logic. The legal shorthand~-- \emph{human-authored work as a protected class}~-- captures the policy logic. Both are the same thing seen from different angles.

\section*{What sincerity looks like in 2026}

I want to come back to a distinction I made in Chapter~\ref{ch:4}, because it has held up through the last six months better than almost any other framing in this book.

I argued that audiences distinguish, very quickly, between \emph{sincere} synthetic work and \emph{cynical} synthetic work~-- that the underlying technology is the same, but the fingerprint of human intent behind the work is visible to the audience at the speed of a swipe.

The data from the spring of 2026 supports this. \emph{Marketing Week}'s analysis ``You can't dismiss AI ads as slop when they're winning in testing''\footnote{\emph{Marketing Week}, ``You can't dismiss AI ads as slop when they're winning in testing.'' \url{https://www.marketingweek.com/dismiss-ai-ads-winning-creative-effectiveness/}. \emph{Dream Machine} Issues~8, 13.} documented that AI-generated advertising creative could, in fact, win in standard creative-effectiveness tests~-- \emph{when the work was made with care, on a brief that respected the audience, by a team that had taste.} The same publication's parallel coverage of the audience pushback against the McDonald's Netherlands ad, the Valentino handbag campaign and a dozen other ``AI slop'' launches, made the inverse point. The technology is neutral. The intent is not.

The strongest AI-authored creative work of the period this book covers has, almost without exception, \emph{not} tried to hide that it was AI-authored. Andrii Daniels' bomb-shelter clip foregrounded its conditions of making.\footnote{Variety, ``AI Creator Behind Viral `Deadpool,' `Harry Potter' Christmas Clip Made His Film in a Ukrainian Bomb Shelter.'' \emph{op.\ cit.} \emph{Dream Machine} Issue~16.} Hoyt Dwyer's animated short for AI FilmFest Japan was upfront about its medium.\footnote{PR Newswire, ``From Apple TV Creative to AI Filmmaker: Hoyt Dwyer's Animated Film To Compete at AI FilmFest Japan 2025.'' \emph{op.\ cit.} \emph{Dream Machine} Issue~6.} \emph{Dear Upstairs Neighbors,} the Google DeepMind / Connie He collaboration that premiered at Sundance, was \emph{about} the constraints and possibilities of its production pipeline.\footnote{Google DeepMind, ``Dear Upstairs Neighbors.'' \url{https://blog.google/innovation-and-ai/models-and-research/google-deepmind/dear-upstairs-neighbors/}. \emph{Dream Machine} Issue~16.} \emph{Synthetic Sincerity}, Marc Isaacs' IDFA film, took the disclosure to the title of the piece.\footnote{\emph{The Hollywood Reporter}, ```Synthetic Sincerity' by Marc Isaacs.'' \emph{op.\ cit.} \emph{Dream Machine} Issue~8.} \emph{Watch the Skies}, the AI-dubbed Swedish UFO feature, disclosed the dubbing process as part of its identity.\footnote{Variety, ```Watch the Skies,' Swedish UFO Feature Film Dubbed Entirely With AI, Sets USA Distribution Deal.'' \emph{op.\ cit.} \emph{Dream Machine} Issue~5.} \emph{Lily,} the \$1m AI Film Award-winning Tunisian short, was framed by its director and reviewers as a piece \emph{about} the new toolchain.\footnote{\emph{Broadcast Pro Middle East}, ``Tunisian filmmaker wins \$1 million AI Film Award for `Lily'.'' \emph{op.\ cit.} \emph{Dream Machine} Issue~14.} The May 2026 \textbf{David Beckham / Lenovo} \emph{``Henchester United''} ad~-- in which the brand cheerfully used generative AI to render a Beckham-designed chicken coop for the footballer's home flock~-- is, in my read, the cleanest \emph{sincere} example of the period: the AI is visible, the brief is playful, the talent is in on the joke, and the cultural response was warm rather than wary.\footnote{\emph{The Drum}, ``David Beckham Designs `Henchester United' Chicken Coop in Lenovo Ad.'' \url{https://www.thedrum.com/news/2026/05/18/david-beckham-henchester-united-chicken-coop-lenovo-ai-ad}. \emph{Dream Machine} Issue~30.}

The pattern, repeated across thirty or forty examples I have looked at carefully, is the same: \emph{AI work that owns its synthetic nature, and that is made with human creative intent, finds an audience.} AI work that tries to pass as something it isn't gets the audience response that Tiny Grandma's stop-motion got \emph{from the algorithm}~-- an immediate, automatic, suspicious flag.

This is, in market terms, a stable equilibrium. It is the market that the slop ceiling and the audience pushback have built. And it is, for working creatives, a manageable and even encouraging environment to operate in. The audience is not against AI. The audience is against being lied to.

\section*{What disclosure should look like}

I want to lay out~-- because I have been asked this in every Q\&A I have done since starting the newsletter~-- what I think the \emph{practical} shape of authenticity infrastructure should look like for working creatives in 2026.

It is four things, in increasing order of investment:

\noindent\textbf{One. Disclose, consistently.} If you use AI in any part of your work, say so. In your credits. On your website. In your contracts with clients. In the metadata of your files. The act of disclosure does, in my experience, not cost you anything with the audience~-- the audience that is going to reject AI work would reject it anyway, and the audience that is going to accept it is the audience that values you being straight with them. The cost of \emph{getting caught} not disclosing, in this environment, is materially higher than the cost of disclosing.

\noindent\textbf{Two. Document, deliberately.} Keep logs. Keep notes. Keep prompt histories. If a piece of work you make this year ends up being legally or culturally contested in 2030~-- and a non-trivial fraction of work made this year will be~-- your ability to \emph{show your workings} will be the difference between defending the work and losing it. The Sundance literacy initiative's emphasis on \emph{evidence of human authorship} is exactly right.\footnote{Sundance Institute AI Literacy Initiative emphasis on documentation: \url{https://www.sundance.org/blogs/centering-the-artist-why-were-launching-the-ai-literacy-initiative/}. \emph{Dream Machine} Issue~16.}

\noindent\textbf{Three. Watermark, where appropriate.} Use SynthID, C2PA, or the equivalent provenance layer that your toolchain supports. If your work doesn't yet support these standards, ask your tool vendors when they will. The market for tools that support provenance metadata is, in 2026, larger than the market for tools that don't.

\noindent\textbf{Four. Build the chain.} If you are running a studio or an agency, build the \emph{internal} infrastructure for verifying and tracking the provenance of your work end-to-end. The cost of doing this in 2026 is moderate. The cost of \emph{not} doing it in 2029, when a client asks for the chain-of-custody on a piece of work and you can't produce it, is going to be much higher.

These are not, on their own, business strategies. They are, in 2026, the \emph{minimum hygiene} for operating a credible creative practice in the AI era. Treat them as you would treat health-and-safety on a film set. Do them as a default. Do them well. Then get on with the work.

\section*{The provenance infrastructure, named: thirty-six pieces of the puzzle}

I want to lay out a more complete map of the provenance infrastructure that is being built in 2025--26, because the technical-and-policy stack is more advanced than the public conversation has caught up with, and working creatives reading this book need to know what is actually in the field.

Stacking the moves I have referenced across this chapter and the rest of the book, the inventory is roughly this:

\noindent\textbf{Capture-time signing and provenance metadata:}

\begin{enumerate}
  \item \textbf{C2PA / Content Credentials} (Adobe-led, with Microsoft, Sony, Nikon, Leica, BBC, Canon participation)~-- the cryptographic-signature standard for capture-and-edit-time provenance.
  \item \textbf{Leica M11-P / M11-D}~-- first consumer cameras with C2PA at the firmware level (2023, expanded through 2025).
  \item \textbf{Sony Alpha 1 II and Alpha 9 III firmware}~-- C2PA support across pro Alpha range.
  \item \textbf{Nikon Z9 firmware}~-- C2PA support for the AP/Reuters wire-service workflow.
  \item \textbf{Canon professional bodies}~-- Content Credentials integration through 2025 firmware cycle.
  \item \textbf{iPhone Camera app} (selected models, 2025--26)~-- capture-time signature support.
  \item \textbf{Adobe Photoshop, Premiere, Lightroom}~-- edit-time chain-of-custody preservation through Content Credentials toolchain.
  \item \textbf{DaVinci Resolve, Final Cut Pro, Avid Media Composer}~-- partial C2PA support through 2025--26 update cycle.
  \item \textbf{Capture One}~-- Content Credentials integration for the high-end commercial-photography workflow.
\end{enumerate}

\noindent\textbf{Synthetic watermarking and detection:}

\begin{enumerate}[resume]
  \item \textbf{Google DeepMind SynthID}~-- embedded watermark for Veo (video), Lyria (audio) and Imagen (image) outputs. Over 100 billion items marked by May 2026; extended to OpenAI, ElevenLabs and Kakao as a cross-vendor standard.
  \item \textbf{SynthID Verification via Gemini app}~-- consumer-facing yes/no detection.
  \item \textbf{Lyria 3 SynthID extension}~-- audio verification across the Google music-model family.
  \item \textbf{YouTube AI Detection Tool}~-- automated content classification (the \emph{Tiny Grandma} false-positive case demonstrating both its scope and its current accuracy ceiling).
  \item \textbf{Deezer's AI-music detection pipeline}~-- identifies up to 75,000 AI-generated tracks per day at upload time.
  \item \textbf{Spotify AI Transparency Beta}~-- voluntary creator disclosure surfaced in the consumer player UI.
  \item \textbf{Beeble}~-- independent detection-and-watermarking infrastructure used by some news organisations.
  \item \textbf{Cloudflare AI bot classification}~-- public-web-infrastructure-level tracking of AI crawlers and agents.
  \item \textbf{Sony music-identification technology}~-- identifying source recordings inside AI-generated outputs at the catalogue level.
\end{enumerate}

\noindent\textbf{Institutional and contractual disclosure:}

\begin{enumerate}[resume]
  \item \textbf{Cannes AI Disclosure Standard}~-- industry-coordination production-side labelling standard for the festival circuit (launched May 2026).
  \item \textbf{Sundance AI Literacy Initiative}~-- creator-training programme funded by Google's \$2M commitment, training 100,000+ artists in foundational AI literacy and provenance practice.
  \item \textbf{Academy of Motion Picture Arts and Sciences ``you must be human to win'' rule}~-- 2026 awards eligibility update.
  \item \textbf{Television Academy AI guidelines}~-- ``tools used to bring it to life'' framing for Emmys submissions.
  \item \textbf{SAG-AFTRA Tilly Tax provisions}~-- the consent, compensation and residuals framework for digital-replica use, included in the 2026 four-year contract.
  \item \textbf{Equity (UK) strike ballot outcome}~-- 99\% vote authorising industrial action over AI provisions.
  \item \textbf{PRS for Music AI Survey 2026}~-- UK music-creator sentiment baseline informing collective-licensing negotiations.
  \item \textbf{GEMA ruling against OpenAI}~-- first European-rights-society legal precedent on AI training compensation.
  \item \textbf{The 88\% UK consultation outcome}~-- political mandate for licensing-by-default.
  \item \textbf{The \emph{Stealing Our Work Is Not Innovation} declaration}~-- 800-creator cultural marker.
  \item \textbf{Bandcamp's outright AI-music ban}~-- distribution-platform-level disclosure-by-exclusion.
  \item \textbf{Swedish Top Chart AI ban}~-- chart-eligibility-level disclosure.
  \item \textbf{San Diego Comic-Con AI art ban}~-- convention-level disclosure-by-exclusion.
\end{enumerate}

\noindent\textbf{Legal infrastructure protecting human identity:}

\begin{enumerate}[resume]
  \item \textbf{Tennessee ELVIS Act}~-- the most-cited state-level AI-impersonation statute, used by the Johnny Cash estate in the Coca-Cola tribute-act case.
  \item \textbf{New York's December 2025 AI-avatar disclosure law}~-- requires advertisers to disclose AI-performer use.
  \item \textbf{The Taylor Swift voice-and-image trademark filings} (early 2026)~-- celebrity-led use of trademark mechanism for identity protection.
  \item \textbf{The Jeremy Renner unauthorised-voice lawsuit}~-- pending case testing voice-likeness protections.
  \item \textbf{The pending EU AI Act enforcement}~-- training-data-transparency requirements coming into effect through 2026.
\end{enumerate}

Each of these, on its own, is a marginal piece. Stacked together~-- capture-signing, watermarking, platform integration, festival rules, awards rules, union contracts, civil society declarations, legal protections~-- they describe a \emph{coherent infrastructure project} that the creative industries are, in eight months, jointly constructing.

The project is, by any reasonable assessment of similar previous infrastructure builds, \emph{substantially ahead of schedule}. The C2PA standards body was founded in 2021 and was, by mid-2026, deployed across most major commercial capture and edit tooling. SynthID went from research demo in 2023 to consumer-facing detection in Gemini by January 2026. The SAG-AFTRA digital-replica provisions went from a 2023 strike demand to a contractual reality in 2026. The 88\% went from political abstraction to government statement of progress in twelve months.

The thing this rate of progress tells me is \emph{not} that the work is done. The work is, in many places, half-done~-- there are gaps in adversarial robustness, in platform UI integration, in cross-jurisdictional enforcement, in coverage of the long tail of creator categories outside the major commercial industries. The work is also being done unevenly: the music industry has built more of the stack than the games industry, which has built more than the publishing industry, which has built more than the regional and minority-language creative ecosystems that the next decade will need to bring in.

But the \emph{trajectory} of the work is unambiguous. The provenance stack is being built. The institutional disclosure infrastructure is being built. The legal protections are being built. The audience contract I describe in the next section is being written. Working creatives who position themselves on the \emph{inside} of this build~-- using the tools, contributing to the standards, showing up at the consultations, advocating with the unions, deploying provenance metadata in their own work as a default~-- will have, by 2030, materially more leverage than working creatives who waited for someone else to finish the project for them.

\section*{What this means for the audience}

I want to close the chapter with a thought about what this whole structure means for \emph{the audience}, because most of this book has~-- by design~-- been about the people who \emph{make} creative work, and the audience is sitting on the other side of the screen the whole time.

What I think the slop ceiling, the provenance infrastructure, the disclosure norms and the legal protections are, \emph{collectively}, building is a \emph{new contract} between makers and audiences.

The old contract was straightforward. The maker made the thing. The audience watched, listened, played, read. The signal of authenticity was implicit~-- most creative work was, by default, made by humans because there was no other way to make it.

The new contract is, by necessity, \emph{explicit}. The maker discloses what was made by whom and how. The audience gets to make an informed choice. The platform, the union, the law and the institution all support both sides of the transaction.

If we get this contract right, the AI era is not the end of human creative work. It is a \emph{renegotiation} of the terms on which human and synthetic creative work coexist in the public sphere~-- with the audience, for the first time in a very long time, getting a real seat at the table.

If we get it wrong~-- if the disclosure infrastructure fails, if the provenance metadata is unreliable, if the platforms refuse to honour the audience's stated preferences, if the legal protections are not enforced~-- what we get is the world the \emph{Dead Internet} chapter described. A web of synthetic content, made by no one in particular, for no one in particular, churning past an audience that has lost the ability to trust any of it.

The choice between those two outcomes is \emph{not, in 2026, a technical question.} The technical infrastructure for both is, by spring 2026, broadly in place. The choice between them is a \emph{political, institutional and cultural} one. It is about whether the people who set the rules~-- the platforms, the legislators, the institutions, the studios, the audience itself~-- collectively decide that \emph{knowable human authorship} is a public good worth protecting.

I think, on the evidence of the last six months, that the choice is being made~-- slowly, contentiously, imperfectly, but recognisably~-- in the right direction. The 88\%, the Sundance literacy turn, the Cannes Disclosure Standard, the Academy's rule update, the SAG-AFTRA contract, the C2PA standards, the SynthID rollout, the audience's own attention behaviour: these all point the same way.

The question for the rest of this book~-- Chapter~\ref{ch:13} on the organisational restructuring, Chapter~\ref{ch:14} on the labour-market reshuffle, and Chapter~\ref{ch:15} on the political choice~-- is what happens to the \emph{organisations}, the \emph{labour market} and the \emph{economy} of creative work when authenticity is the scarce good and the orchestrator is the new role. The implications for how teams are structured, how labour is paid and how creative careers are built are bigger than any single tool launch, and they are what the next three chapters are about.

  \chapter{Coordination Collapse}\label{ch:13}

\lettrine[lines=3,lhang=0.15,findent=0.1em]{T}{here} is a working assumption underneath almost every conversation about AI in the creative industries that I want, in this chapter, to make explicit and then take apart.

The assumption is that AI is a \emph{technology change}, broadly equivalent in shape to other technology changes the industries have absorbed in the past~-- the arrival of digital, the move to streaming, the rise of mobile, the emergence of social. The implication is that the existing institutions of the industry~-- the studios, the agencies, the labels, the unions, the publishing houses, the broadcasters~-- will absorb this change the way they absorbed the previous ones: with some restructuring, some layoffs, some new hires, some new departments, and a generally familiar shape on the other side.

I do not think this is what is happening.

What I think is happening is that AI is, specifically and quite differently, a \emph{coordination technology.} It is changing~-- not at the margin, but at the core~-- what it is possible for a single person to know, decide and execute about a complex creative project. And because the existing organisations of the creative industries are, structurally, \emph{coordination architectures}~-- they exist to allow many people to work on a single piece of work~-- the change in coordination economics is changing the organisations themselves.

This is the chapter about what happens to creative \emph{organisations}~-- studios, agencies, labels, broadcasters, indie companies~-- when AI reaches a certain level of capability. It is also, by necessity, the chapter about what happens to creative \emph{careers} when those organisations change shape.

The shorthand I have come to use for what is happening is \textbf{coordination collapse}.

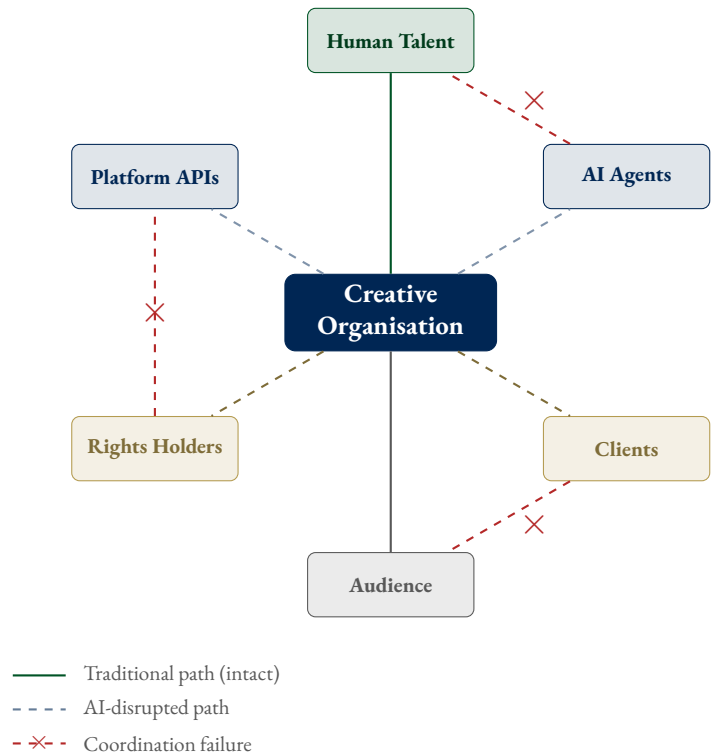
\begin{figure}[htbp]
  \centering
  \begin{tikzpicture}[
    font=\small,
    every node/.style={align=center},
    >=stealth
  ]

    \node[
      draw=darkblue, fill=darkblue, text=white,
      rounded corners=4pt, minimum width=2.8cm, minimum height=1.0cm,
      font=\small\bfseries
    ] (centre) at (0,0) {Creative\\Organisation};

    \node[
      draw=darkgreen, fill=darkgreen!15!white, text=darkgreen!70!black,
      rounded corners=3pt, minimum width=2.2cm, minimum height=0.85cm,
      font=\scriptsize\bfseries
    ] (talent) at (0, 3.6) {Human Talent};

    \node[
      draw=darkblue, fill=darkblue!12!white, text=darkblue,
      rounded corners=3pt, minimum width=2.2cm, minimum height=0.85cm,
      font=\scriptsize\bfseries
    ] (ai) at (3.12, 1.8) {AI Agents};

    \node[
      draw=rulegold, fill=rulegold!15!white, text=rulegold!70!black,
      rounded corners=3pt, minimum width=2.2cm, minimum height=0.85cm,
      font=\scriptsize\bfseries
    ] (clients) at (3.12, -1.8) {Clients};

    \node[
      draw=chaptergrey, fill=chaptergrey!12!white, text=chaptergrey,
      rounded corners=3pt, minimum width=2.2cm, minimum height=0.85cm,
      font=\scriptsize\bfseries
    ] (audience) at (0, -3.6) {Audience};

    \node[
      draw=rulegold, fill=rulegold!15!white, text=rulegold!70!black,
      rounded corners=3pt, minimum width=2.2cm, minimum height=0.85cm,
      font=\scriptsize\bfseries
    ] (rights) at (-3.12, -1.8) {Rights Holders};

    \node[
      draw=darkblue, fill=darkblue!12!white, text=darkblue,
      rounded corners=3pt, minimum width=2.2cm, minimum height=0.85cm,
      font=\scriptsize\bfseries
    ] (platform) at (-3.12, 1.8) {Platform APIs};

    \draw[darkgreen, thick] (centre) -- (talent);
    \draw[chaptergrey, thick] (centre) -- (audience);

    \draw[darkblue!50, thick, dashed] (centre) -- (ai);
    \draw[rulegold!70!black, thick, dashed] (centre) -- (clients);
    \draw[rulegold!70!black, thick, dashed] (centre) -- (rights);
    \draw[darkblue!50, thick, dashed] (centre) -- (platform);

    \draw[alertred, thick, dashed] (ai) -- (talent);
    \node[alertred, font=\Large\bfseries] at (1.9, 2.8) {$\times$};

    \draw[alertred, thick, dashed] (rights) -- (platform);
    \node[alertred, font=\Large\bfseries] at (-3.12, 0) {$\times$};

    \draw[alertred, thick, dashed] (clients) -- (audience);
    \node[alertred, font=\Large\bfseries] at (1.9, -2.8) {$\times$};

    \begin{scope}[shift={(-5.0,-4.8)}]
      \draw[darkgreen, thick] (0,0) -- (0.7,0);
      \node[anchor=west, font=\scriptsize, text=chaptergrey] at (0.8,0) {Traditional path (intact)};
      \draw[darkblue!60, thick, dashed] (0,-0.45) -- (0.7,-0.45);
      \node[anchor=west, font=\scriptsize, text=chaptergrey] at (0.8,-0.45) {AI-disrupted path};
      \draw[alertred, thick, dashed] (0,-0.9) -- (0.7,-0.9);
      \node[alertred, font=\small\bfseries] at (0.35, -0.88) {$\times$};
      \node[anchor=west, font=\scriptsize, text=chaptergrey] at (0.8,-0.9) {Coordination failure};
    \end{scope}

  \end{tikzpicture}
  \caption{Coordination Collapse: the multiplication of interfaces between human and AI stakeholders overwhelms traditional organisational structures.}
  \label{fig:coordination-collapse}
\end{figure}

\section*{What organisations are for}

A studio, an agency, a label, a publishing house~-- these are not just brand names attached to creative outputs. They are \emph{organisational technologies} that solve a specific problem: how do you get fifty or five hundred or five thousand people to coordinate on a single piece of creative work, well enough that the result is coherent, on-budget, on-deadline and good enough to put out into the world.

The way they solve that problem is by \emph{layering} the work into specialised roles, building \emph{hierarchies} that direct the work down through those roles, \emph{processes} that move material between the roles in predictable order, and \emph{cultural norms} that make the whole apparatus run with less explicit instruction than you would otherwise need.

A film studio exists, structurally, because making a film requires the coordinated work of many specialists~-- writers, directors, actors, cinematographers, designers, editors, sound designers, composers, marketers, distributors. The studio is the \emph{coordination apparatus.} The film is the output of the apparatus.

When AI starts to do the work of many of those specialists~-- not entirely, but at the level of \emph{first draft} or \emph{junior contribution}~-- the calculus of the coordination apparatus changes. Suddenly, a much smaller human team, working with a large pool of synthetic capacity, can produce the same coordinated output that previously required the large team. Suddenly, the \emph{bottleneck} of producing creative work is no longer the size of the team. It is something else: the ability of the small senior team to \emph{direct} the synthetic capacity well.

The implication, which has been dawning on the creative industries through the autumn of 2025 and the spring of 2026, is that the existing organisational shape~-- the \emph{studio shape}, the \emph{agency shape}, the \emph{label shape}~-- was built for a coordination problem that no longer exists in the form it used to.

This does not mean the studios will disappear. It does mean that the \emph{shape} of the studios is going to change, in a hurry, in ways that working creatives need to understand if they are going to be on the inside of the change rather than on the receiving end of it.

\section*{The shadow workforce}

The first symptom of coordination collapse, as it has shown up across the creative industries in this period, is the rise of what the workplace-research literature has started calling \textbf{shadow AI}~-- the practice of using AI tools in your job \emph{without telling your employer.}

The numbers, from a series of 2025 studies I covered in \emph{Dream Machine} Issue~5,\footnote{\emph{Dream Machine} Issue~5, ``Industry Insights: Stealth, Shadow and Secret AI Users.''} are extraordinary.

Roughly \textbf{half} of U.S.\ employees~-- 45--52\% in different surveys~-- have used AI in their jobs without telling their bosses, with Gen~Z and tech-sector workers being the most frequent secret users.\footnote{Azumo, ``AI in Workplace Statistics 2025,'' \url{https://azumo.com/artificial-intelligence/ai-insights/ai-in-workplace-statistics}. \emph{Tech.co}, ``Gen Z Most Likely Use AI Boss,'' \url{https://tech.co/news/gen-z-most-likely-use-ai-boss}. \emph{Dream Machine} Issue~5.}

About a third~-- 29--33\%~-- pay for their own AI tools out of their own pockets without their employer's knowledge.\footnote{\emph{Exploding Topics}, ``AI Workforce Research,'' \url{https://explodingtopics.com/blog/ai-workforce-research}. \emph{Dream Machine} Issue~5.}

Roughly \textbf{56--57\%} of regular AI users admit to \emph{hiding their usage or presenting AI output as their own work} to avoid judgement or stigma. Nearly half of executives do the same.\footnote{\emph{Forbes}, ``AI Tools Flood Workplaces as Employees Face a Double Bind,'' \url{https://www.forbes.com/sites/carolinecastrillon/2025/09/09/ai-tools-flood-workplaces-as-employees-face-a-double-bind/}. \emph{Dream Machine} Issue~5.}

\textbf{52\%} of workers won't admit to using AI at work~-- \emph{even when asked directly.}\footnote{\emph{Blog IDC Europe}, ``Shadow AI: How Stealth Productivity Is Strangling Enterprise AI Adoption and Creating a Security Nightmare,'' \url{https://blog-idceurope.com/shadow-ai-how-stealth-productivity-is-strangling-enterprise-ai-adoption-and-creating-a-security-nightmare/}. \emph{Dream Machine} Issue~5.}

These numbers describe a workforce that has, en masse, started running its own private parallel productivity infrastructure that bypasses the official organisational tooling, the official organisational processes, and the official organisational accounting of where the work is being done and by whom.

This is not a niche phenomenon. \emph{Half the workforce.} And it is concentrated, the surveys suggest, in exactly the demographic~-- Gen~Z, tech-sector, knowledge worker~-- that is most likely to be the future workforce of the creative industries.

The numbers escalate the more recent the research. By the end of 2025, enterprise-AI tracking data put active daily use at \textbf{88--89\%} of staff across organisations, with \textbf{71--80\%} of those users running their tools entirely outside any official approval or IT oversight.\footnote{Enterprise-AI workforce tracking, late 2025. Aggregated in the Deep Dive companion piece \emph{The Shadow AI Paradox in the Creative Industries}, drawing on Azumo's \emph{AI in Workplace Statistics 2025}, \emph{Tech.co}'s Gen~Z survey, and the IDC Europe shadow-AI security brief. \emph{Dream Machine} Issue~5.} What the workplace-research firms have started calling the ``Hidden Cloud Explosion'' describes a six-month period in which the average enterprise IT department's visibility into the AI tools its workforce was using simply collapsed: organisations believed they were running on roughly \textbf{91} public cloud services per enterprise, while network-level analysis put the actual figure at \textbf{1,220} active services~-- a 90\% visibility gap.\footnote{\emph{Hidden Cloud Explosion} analysis, IDC Europe, 2025. See \emph{The Shadow AI Paradox in the Creative Industries}, \S``The Epistemology and Scale of Shadow AI.''} In the same year, \textbf{20\%} of organisations reported severe security incidents linked directly to shadow AI, with the average breach cost going up by \textbf{\$670,000}; in \textbf{65\%} of those incidents personally identifiable information was exposed, and in \textbf{40\%} intellectual property was directly leaked.\footnote{Shadow-AI security-incident statistics, 2025, aggregated in \emph{The Shadow AI Paradox in the Creative Industries}, \S``The Epistemology and Scale of Shadow AI''; underlying data via IBM \emph{Cost of a Data Breach Report 2025} and IDC Europe.}

For the creative industries this matters disproportionately, because the data being fed into public LLMs by shadow users~-- proprietary scripts, unreleased concept art, client briefs, internal pipeline code, unmastered audio stems~-- is the \emph{exact intellectual property} that organisations are simultaneously suing AI companies for scraping. The studio whose general counsel is in federal court against a frontier-model company is, on the same Tuesday afternoon, watching its own animation department paste asset descriptions into the same company's consumer chatbot to speed up metadata writing. Both things are true. Both happen at once.

The shadow workforce, in coordination-collapse terms, is the symptom of an organisational architecture that is no longer aligned with the work the people inside it are actually doing. The official architecture says \emph{we hired these humans to do these specific jobs, in this specific way, at this specific pace.} The shadow architecture says \emph{these humans are now hybrid human-agent operators producing more, faster, with different qualitative properties than the official architecture is set up to manage.}

What you get, when these two architectures sit on top of each other, is a workforce that is \emph{measurably more productive than the official metrics show}, doing \emph{more work than the official scope says}, with \emph{less institutional knowledge of how that work is being done} than ever before.

This is, in coordination terms, an unstable equilibrium. It cannot last indefinitely. The question is how it resolves.

\section*{The ``AI for thee, but not for me'' paradox}

The pattern that the shadow-AI numbers describe is not random distribution of tool use. It is \emph{hierarchical}. Creative workers, across every survey I have read in this period, exhibit a consistent psychological pattern that the developer and creative-community discourse has named the \textbf{``AI for thee, but not for me''} paradox.\footnote{For the developer-community origins of the ``AI for thee, but not for me'' phrasing, and the full sectoral analysis of the paradox, see \emph{The Shadow AI Paradox in the Creative Industries}, \S``The Great Hypocrisy.''} The pattern works like this. Creative professionals identify some tasks as \emph{mine}~-- the writing, the cinematography, the composing, the performance, the lead concept~-- and other tasks as \emph{not mine}~-- the marketing copy, the project email, the deck assembly, the metadata tagging, the routine code, the contract redline, the rough mix, the asset variation. The first category is defended fiercely against AI substitution; the second is offered to AI substitution without much thought. The moral framing of the technology shifts depending on whose labour is being replaced.

Look at the music sector. An industry survey of more than \textbf{1,100} professional producers, songwriters and audio engineers in 2026 found that \textbf{87\%} were actively using AI tools in their creative process.\footnote{Survey of 1,100+ professional music creators, 2026, summarised in \emph{Dynamics of Generative AI Adoption in the Creative Industries}, \S``Music Production and Sound Recording,'' and \emph{The Shadow AI Paradox in the Creative Industries}, \S``Sector-Specific Analysis.''} The internal distribution, though, tracked the hierarchy: \textbf{58\%} used AI for audio restoration and cleanup, \textbf{38\%} for mixing assistance, \textbf{33.9\%} for automated mastering~-- high-friction tasks that nobody felt sentimental about~-- while only \textbf{20.9\%} admitted to using AI for composition or lyric generation, the parts of the craft on which the personal artistic identity sat. \textbf{77\%} cited ``loss of originality'' as their primary concern, outranking even the fear of personal job displacement (\textbf{42\%}). The artist's relationship to the tool, in other words, is not consistent across the work. It is sharply conditional on whose labour is being substituted.

The same hierarchy shows up in film. A survey of professional screenwriters before and after the 2023 WGA strike found that pre-strike covert AI use sat at around \textbf{34\%}; once the WGA's negotiated guidelines legitimised AI assistance for \emph{formatting, structural outlining and brainstorming}, that number jumped to \textbf{68\%} by 2024.\footnote{WGA screenwriter survey, pre- and post-strike, reported in \emph{Dynamics of Generative AI Adoption in the Creative Industries}, \S``Screenwriting and the Post-Strike AI Boom.''} What the regulation changed was not the technology. It was the stigma. The shadow use moved into the light, with no measurable decline in the work product. The covert hierarchy became an overt one.

This is, in my view, the most uncomfortable observation in the entire shadow-AI literature, and it is the one that creative-organisation leadership has the hardest time admitting publicly. The same professionals who, in their public statements, treat AI training as a moral violation are, in their private practice, the heaviest users of the same underlying technology. The hypocrisy is not a character defect. It is a structural property of how knowledge workers self-defensively triage their own tasks under productivity pressure. Read charitably, working creatives are doing what every economic actor in a productivity transition has done: protect the highest-value labour and offload everything else.

The cost of that triage is moral clarity. It is hard to credibly argue that AI training is theft when you are typing your portfolio description into Claude. The vocal-protest economy and the silent-adoption economy now run on the same desks, often within the same hour.

\section*{The consumption gap: what the data actually says}

I want to spend a section on the gap, because the gap is \emph{the} macroeconomic story of this period and I do not see anyone telling it cleanly.

The gap is between \emph{what the creative industries are saying publicly about AI} and \emph{how the creative industries are actually using AI on a Tuesday morning.} The two pictures are not slightly different. They are, in significant measure, contradictory.

Take Adobe, because Adobe is the cleanest single case study. Adobe's Firefly generative-AI suite~-- the same product that the working creatives in my surveyed circle are most ambivalent about~-- passed \textbf{22 billion AI-generated assets} by April 2025, eighteen months from public release.\footnote{Adobe Firefly milestone data, September 2023 -- June 2025, in \emph{Dynamics of Generative AI Adoption in the Creative Industries}, \S``The Ubiquity of AI in Visual and Digital Arts.'' \emph{Dream Machine} Issue~6.} By that point, \textbf{45\%} of all Creative Cloud subscribers had engaged with Firefly. \textbf{70\%} of active Firefly users were using the tool \emph{every week}, averaging 2.8 sessions weekly at 26 minutes each. Firefly contributed \textbf{11\%} of all new annual recurring revenue at Adobe in 2024~-- the company's fastest-growing revenue catalyst since the original move to a subscription model~-- and Adobe's AI-first ARR more than \emph{tripled year-over-year} in the first quarter of fiscal 2026.\footnote{Adobe quarterly financials, FY2025--FY2026; AI-first ARR growth reported in \emph{Dream Machine} Issue~21 and summarised in \emph{Dynamics of Generative AI Adoption}.}

That is not the adoption curve of a niche professional tool. That is the adoption curve of a default productivity feature in the dominant creative-software stack on the planet. \textbf{72\%} of Fortune 500 design teams have formally integrated Firefly. \textbf{63\%} of marketing agencies. \textbf{58\%} of e-commerce design departments. \textbf{48\%} of UX/UI designers.\footnote{Adobe Firefly enterprise penetration metrics, in \emph{Dynamics of Generative AI Adoption}.} \emph{Twenty-five per cent} of new Adobe Stock contributions in 2024 contained Firefly-generated elements.\footnote{Adobe Stock submission analysis, 2024, in \emph{Dynamics of Generative AI Adoption}.} The Adobe MAX 2025 Creators' Toolkit Report's headline number~-- \textbf{86\%} of global creators using generative AI~-- sits inside this pattern, not against it.\footnote{Adobe, ``Inaugural Adobe Creators' Toolkit Report,'' October 2025, \url{https://news.adobe.com/news/2025/10/adobe-max-2025-creators-survey}. \emph{Dream Machine} Issue~6.}

If you sat with the public discourse alone~-- the open letters, the boycotts, the strike statements~-- you would assume working creatives were broadly refusing AI integration. The actual platform telemetry, in a year where Adobe shared more of its numbers than usual, says the opposite. Working creatives are not refusing. They are adopting at a pace that Adobe's growth team is, by every signal I can read, struggling to keep ahead of.

The same picture holds across the toolchain. ChatGPT, by mid-2025, was on \textbf{800--900 million weekly active users}.\footnote{ChatGPT weekly-active-user disclosures, mid-2025; consolidated in \emph{Dynamics of Generative AI Adoption}, \S``General Purpose LLMs.''} Anthropic's Claude was the writers' and developers' second favourite, with rapidly increasing usage in long-context creative tasks. Google's Gemini was growing desktop users at \textbf{155\%} year-over-year, more than six times faster than ChatGPT's 23\%.\footnote{Gemini desktop-user growth, year-over-year, in \emph{Dynamics of Generative AI Adoption}.} These are not user numbers that reflect a market in revolt. They are user numbers that reflect a market that has, in private, decided.

And the consumer side mirrors the producer side. The Stanford AI Index 2025 found that \textbf{55\%} of individuals across 26 countries view AI products as offering more benefits than drawbacks~-- up from 52\% in 2022.\footnote{Stanford AI Index Report 2025, global-sentiment chapter. Summarised in \emph{Dynamics of Generative AI Adoption}, \S``The Perception Gap.''} A 2024 YouGov poll across 17 markets found that nearly a third of consumers felt \emph{more positively} about generative AI than the previous year, against only 22\% feeling more negatively.\footnote{YouGov 2024 multi-market AI sentiment survey, 17 countries. Summarised in \emph{Dynamics of Generative AI Adoption}, \S``The Perception Gap.''} In the gaming sector~-- which has produced some of the loudest anti-AI consumer backlash of this year~-- the same Quantic Foundry survey that showed audiences are \textbf{77--83\%} negative toward AI-generated quests and dialogue also showed that \textbf{60\%} of gamers remain entirely neutral about AI in a game's development \emph{provided the final product is of high quality.}\footnote{Quantic Foundry consumer-AI-in-gaming survey, 2025. Summarised in \emph{Dynamics of Generative AI Adoption}, \S``The Video Game Industry.''} The hostility is not generic. It is specifically aimed at AI in the creative roles where audiences expect to feel a human soul. Everywhere else~-- UI, backend, balancing, localisation, dynamic difficulty~-- the audience is, on aggregate, indifferent.

Even the GDC sentiment data, which is often cited as evidence of an industry in retreat from AI, tells the same paradoxical story when you read it as a whole. \emph{Personal} generative-AI usage among professional game developers rose from \textbf{31\%} in 2024 to \textbf{36\%} in 2026, while \emph{industry sentiment} over the same period cratered from 18\% negative to \textbf{52\%} negative.\footnote{Game Developers Conference \emph{State of the Game Industry} surveys, 2024--2026, sentiment vs.\ usage trend. Reported in \emph{Dynamics of Generative AI Adoption}, \S``The Video Game Industry.''} Use went up while approval went down. The two lines should, in a coherent market, move together. They are not. They are diverging.

I want to be careful about what conclusion to draw from this. It is \emph{not} that the public discourse is wrong and the silent adopters are right. The public discourse is doing genuine political and cultural work~-- it is what produced the 88\%, the SAG-AFTRA contract, the GEMA ruling, the Sundance literacy turn, the Cannes Disclosure Standard. Without the loud minority, the creative economy would have no political leverage at all.

The conclusion to draw is more uncomfortable. The creative industries are, in 2026, operating with \emph{two parallel economies on top of each other.} In one, AI is a moral crisis, a labour threat, and a contested category of production. In the other, AI is a default productivity feature being integrated at the speed of any other software upgrade. The same individuals, the same teams, the same studios are participating in both economies simultaneously, often without acknowledging the contradiction.

The question for the next eighteen months~-- the question I keep coming back to when I talk to studio leadership~-- is whether the two economies merge into a single, honest, integrated practice (the \textbf{path two} integration I describe below), or whether they continue to run in parallel, with the public economy producing the policy and the private economy producing the work. The first outcome is harder but produces better collective decisions. The second outcome is the path of least resistance, and is, in my view, where we will end up by default if working creatives, studios and unions do not deliberately close the gap.

For the data and the sectoral mechanics behind this section~-- the linguistic markers of covert AI use, the labour-market dynamics of agentic displacement, the deeper analysis of Adobe / OpenAI / Anthropic adoption telemetry, and the consumer sentiment / consumption asymmetry~-- see the two research deep dives that this chapter draws on: Appendix~D (\emph{The Shadow AI Paradox}) and Appendix~E (\emph{Dynamics of Generative AI Adoption}).

There is one further dimension of the consumption gap I want to flag here, because Chapter~\ref{ch:10} develops it at length and Chapter~\ref{ch:4} introduced it: the gap between \emph{production} and \emph{consumption} is not just an organisational misalignment, it is a \emph{biological} one. \textbf{Aggregate human attention is finite.} The same Adobe Firefly that has generated 22 billion assets, the same ChatGPT that serves 900 million weekly users, the same Sora app that hit a million downloads in five days~-- these are all systems whose \emph{production} side scales without bound and whose \emph{consumption} side is bounded by the eleven-or-so daily hours of media attention the average adult can physiologically deploy. The consumption gap I have described above is, at its widest, this binding constraint expressed as an organisational problem. Studios that produce AI-augmented content at the rate the toolchain now allows~-- without recognising that the audience cannot consume more hours per day than it already does~-- are, on inspection, optimising the wrong side of the supply-demand equation. The studios that integrate AI productivity gains into work that \emph{earns} a larger share of the audience's finite attention budget will win the next decade. The studios that integrate AI productivity gains into \emph{more output competing for the same finite budget} will, on the historical pattern, hit the slop ceiling on a balance-sheet timeline they did not plan for.

\section*{The two paths}

The shadow workforce can resolve, broadly, in one of two directions, and I think the choice between them will be the central organisational question for every studio, agency and label in the creative industries over the next three years.

\textbf{Path one} is \emph{suppression}. The organisation decides that the shadow AI use is a risk~-- to security, to IP, to brand, to compliance, to the official productivity metrics~-- and shuts it down. Tightens the rules. Audits the work. Punishes the offenders. Reverts to the official architecture and the official tooling.

This is, in my view, a losing strategy in the medium term, because the productivity advantages that the shadow workforce is capturing are real, and the workers who are capturing them will, given the choice, work for organisations that \emph{let them keep capturing them.} The suppressing organisation will progressively lose its most AI-fluent workforce to organisations that allow the hybrid practice.

\textbf{Path two} is \emph{integration}. The organisation accepts that the shadow AI use is happening, decides to make it official, builds the infrastructure to support it, sets the norms to govern it, and re-shapes the work~-- and the workforce~-- around it.

This is, in my view, the right strategy. It is also the one most major creative organisations have been quietly moving towards in the period this book covers.

\textbf{EA}'s push of its 15,000 employees to use AI as a ``thought partner'' was, structurally, a path-two move. \textbf{Krafton's} transformation into an ``AI-first'' company in November 2025 was a path-two move.\footnote{\emph{Game Developer}, ``Subnautica owner Krafton outlines plans to transform into an `AI First' company,'' \url{https://www.gamedeveloper.com/business/subnautica-owner-krafton-outlines-plans-to-transform-into-an-ai-first-company}. \emph{Dream Machine} Issue~6.} \textbf{Disney's} Office of Technology Enablement was a path-two move. \textbf{WPP's} AI overhaul, \textbf{Adobe's} AI in everything, \textbf{Sony's} 49-agent game team~-- all path-two moves.

The path-two organisations are, structurally, betting that integration produces \emph{more} output, \emph{more} quality and \emph{more} employee retention than suppression. The early evidence, six months in, suggests they are right.

\section*{The mid-career squeeze}

The cost of the path-two transition has not, on the whole, been borne by the senior creatives or the entry-level workforce. It has been borne by the \emph{middle.}

In \textbf{April 2026}, \emph{Dream Machine} Issue~24 reported that the publisher behind \emph{Grand Theft Auto VI} had laid off the entire seven-year-old internal AI team it had built to develop in-house AI capability for the franchise.\footnote{\emph{Dream Machine} Issue~24, April 2026, on the GTA~VI publisher laying off its internal AI team.} The framing, in the press release and the subsequent industry coverage, was that the company had decided to use off-the-shelf AI tools instead of maintaining proprietary ones~-- and that the seven-year AI investment was, in retrospect, a ``backlash cleanup'' cost.

The story was repeated across multiple studios. \textbf{Disney}, in April 2026, laid off staff including in its Marvel division, in moves the company did not blame on AI but whose timing was, as the trade press noted, ``loaded.''\footnote{\emph{Dream Machine} Issue~25, April 2026, on Disney layoffs including Marvel staff.} \textbf{Meta} had cut 10\% of its Reality Labs staff in January 2026 to refocus on AI.\footnote{\emph{SmartBrief}, ``Meta to cut 10\% of Reality Labs staff to focus on AI,'' \url{https://newsletter.smartbrief.com/sharedSummary/index.jsp?briefId=40A39351-5419-4681-94DF-31A53480B698}. \emph{Dream Machine} Issue~14.} \textbf{Scottish animation studio} Axis Animation collapsed in early 2026, with its closure publicly attributed in part to AI competition.\footnote{\emph{Dream Machine} Issue~23, April 2026, on Scottish animation studio collapse.} \textbf{Ubisoft} cancelled five games, including the \emph{Prince of Persia} remake, in January 2026, in order to refocus on AI.\footnote{\emph{Metro}, ``Prince of Persia remake and five more games cancelled as Ubisoft focuses on AI,'' \url{https://metro.co.uk/2026/01/21/prince-persia-remake-five-games-cancelled-ubisoft-focuses-ai-26431926/}. \emph{Dream Machine} Issue~15.}

The pattern across these cases is the same: the \emph{mid-career} layer~-- the experienced specialists in the middle of their professional lives, doing the day-to-day production work that the senior creative leadership directs~-- is the layer absorbed into agentic capacity first.

This is the unambiguous bad news of the AI transition. The \emph{Guardian} covered this directly in January 2026 with a piece titled ``AI is hitting UK harder than other big economies, study finds,'' which found that mid-career knowledge workers in the U.K.\ were experiencing disproportionate displacement compared to peers in the U.S., Japan, Germany and Australia.\footnote{\emph{The Guardian}, ``AI is hitting UK harder than other big economies, study finds,'' \url{https://www.theguardian.com/technology/2026/jan/26/ai-uk-jobs-us-japan-germany-australia}. \emph{Dream Machine} Issue~16.} \emph{Economist} coverage in late November 2025 had been the early signal: ``Investors expect AI use to soar.\ That's not happening''~-- meaning that the AI investment thesis was not, in the short term, producing the aggregate productivity gains the investors had hoped for, but \emph{was} producing concentrated labour displacement in specific sectors.\footnote{\emph{The Economist}, ``Investors expect AI use to soar.\ That's not happening,'' \url{https://www.economist.com/finance-and-economics/2025/11/26/investors-expect-ai-use-to-soar-thats-not-happening}. \emph{Dream Machine} Issue~9.}

The OpenAI public-policy response to this, articulated through April 2026, was a series of proposals~-- robot taxes, public wealth funds, a 4-day workweek~-- designed to manage the economic disruption of AI-driven productivity gains.\footnote{\emph{Dream Machine} Issue~24, April 2026, on OpenAI's public-policy proposals around AI-driven economic disruption.} \emph{Dream Machine} Issue~24 covered these proposals at length. The framing OpenAI used was telling: the company was no longer arguing that AI would \emph{not} cause disruption. It was arguing that the disruption was inevitable and that society needed to build new mechanisms to manage it.

The \emph{Economist}, in a piece titled ``Job apocalypse? Humbug! AI is creating brand new occupations,'' took the contrary position~-- that AI was, on net, creating more new jobs than it was destroying, and that the framing of mass displacement was overstated.\footnote{\emph{The Economist}, ``Job apocalypse? Humbug! AI is creating brand new occupations,'' \url{https://www.economist.com/business/2025/12/14/job-apocalypse-humbug-ai-is-creating-brand-new-occupations}. \emph{Dream Machine} Issue~12.} Both positions are partially right. The aggregate employment numbers, across creative industries in 2026, did not show the apocalyptic decline some had predicted. But the \emph{composition} of employment changed sharply. Senior orchestrator roles increased. Mid-career specialist roles decreased. New AI-specialist roles~-- AI agent developers, prompt engineers, AI ops specialists~-- exploded. The \emph{Forbes} piece from November 2025 noted that ``vibe coding''~-- natural-language software development~-- was an in-demand AI skill that paid up to \$220,000.\footnote{\emph{Forbes}, ``Vibe Coding~-- The In Demand AI Skill,'' \url{https://www.forbes.com/sites/rachelwells/2025/11/06/the-in-demand-ai-skill-and-certifications-that-pays-up-to-220000/}. \emph{Dream Machine} Issue~8.}

What the labour market is doing, when you look at it carefully, is \emph{not} destroying jobs in the creative industries. It is \emph{reshuffling} them~-- towards a smaller number of senior strategic roles, a different mix of specialist roles, and a much larger pool of AI-tooling skills that span the old discipline lines. The mid-career creative who fails to make this transition is the one who is at risk. The mid-career creative who makes it well~-- by upskilling deliberately, by claiming the orchestrator role, by building a portfolio practice~-- has, by every indicator I can see, \emph{more} leverage in the labour market than they did before.

The hard truth is that the transition is not equally available to everyone. It depends on access to training, on access to tooling, on workplace cultures that support experimentation, on time to reskill that workers with caring responsibilities or financial precarity often don't have. The institutional response to this~-- the Sundance Literacy initiative, the UK free-AI-training programme, the Adobe and Google educational investments~-- is real but partial. The structural inequities of who \emph{can} make the transition are real and concerning.

\section*{The neurodiversity dividend}

One genuinely encouraging finding from the period this book covers came from the U.K.\ Department for Business and Trade's research on neurodiverse workers in AI-tooled workplaces.

The study, published in late 2025, found that workers with ADHD, autism and dyslexia were \textbf{25\% more satisfied} with AI assistants than neurotypical workers, and that they reported AI agents as actively helping them succeed at work.\footnote{U.K.\ Department for Business and Trade research on neurodiverse workers and AI assistants, autumn 2025. Reported via \emph{CNBC}, ``People with ADHD, autism, dyslexia say AI agents are helping them succeed at work,'' \url{https://www.cnbc.com/2025/11/08/adhd-autism-dyslexia-jobs-careers-ai-agents-success.html}. \emph{Dream Machine} Issue~7.} The interpretation, reported in \emph{CNBC} in November 2025, was that AI tools were lowering the cognitive load of tasks that had historically been disproportionately punishing for neurodivergent workers~-- coordinating complex calendars, parsing dense documents, structuring written outputs~-- and were, as a result, \emph{levelling the playing field} in workplaces that had previously underutilised neurodivergent talent.\footnote{\emph{CNBC}, \emph{op.\ cit.}}

I want to flag this finding because it is one of the cleanest counterexamples to the ``AI is bad for workers'' framing that I have come across, and because it is a useful corrective to the labour-displacement narrative that has dominated much of the coverage of this period.

AI is, demonstrably, \emph{good} for some workers. It is good for the workers who are most able to leverage it, and it is also good for the workers whose existing labour-market participation was being limited by structural barriers that AI happens to dismantle. Both are real, and both are important to keep in view.

The \emph{Guardian's} parallel finding~-- that ADHD, autism and dyslexia workers were reporting AI agents as a major workplace enabler~-- was echoed in dozens of smaller reports across 2026.\footnote{\emph{Dream Machine} Issue~7 secondary references.} The implication, for the creative industries: a workforce that has historically been heavily neurodivergent (the writing, music, film and games sectors are all over-indexed on neurodivergent talent compared to the general population) stands to be one of the \emph{biggest beneficiaries} of well-deployed AI tooling in the workplace.

This is not a reason to ignore the displacement story. It is a reason to be careful about which framings of the AI transition are accurate and which are reductive.

\section*{Indies and the Global South}

The other genuinely encouraging signal in this period is the rise of the \emph{indie} and \emph{Global South} creative sectors as direct beneficiaries of the AI cost reduction.

\textbf{African film and tech} has been a recurring positive story across the period~-- from Korin AI, the ``trained with African datasets, built by Africans'' model that launched in May 2026,\footnote{Korin AI launch, May 2026. \emph{Dream Machine} Issue~27.} to the wave of African AI filmmakers that the trade press began covering in earnest in early 2026, to the African music industry's adoption of AI tools described in \emph{CNBC Africa} in October 2025.\footnote{\emph{CNBC Africa}, ``How AI is changing the landscape of the music industry in Africa,'' \url{https://www.cnbcafrica.com/2025/how-ai-is-changing-the-landscape-of-the-music-industry-in-africa}. \emph{Dream Machine} Issue~5.}

\textbf{Indian cinema} has been awash with AI through the period covered by this book.\footnote{BBC Future, ``Lights, camera, algorithm: Why Indian cinema is awash with AI,'' \url{https://www.bbc.co.uk/future/article/20251223-why-indian-cinema-is-awash-with-ai}. \emph{Dream Machine} Issue~14.} The BBC's December 2025 piece ``Lights, camera, algorithm'' documented the structural shift, with major productions integrating AI for visual effects, dubbing and asset generation, and made an observation about the limits of the technology that has stayed with me: \emph{``You could create a sequel to a regional Indian movie using ChatGPT, but you would need to feed it the cultural memory of the original script. That script would have to be written by a human screenwriter.''} The cultural memory is the human contribution. The toolchain accelerates everything around it. The screenwriting itself~-- the act of \emph{knowing what the culture remembers}~-- remains stubbornly, irreducibly human. India's first AI-animated show, \emph{Legenda Bertuah}, launched in Indonesia in April 2026.\footnote{\emph{Dream Machine} Issue~25, April 2026, on Indonesia's \emph{Legenda Bertuah.}}

\textbf{Latin American} and \textbf{Middle Eastern} AI film festivals proliferated through late 2025 and early 2026. The \$1m Dubai AI Film Award was won by Tunisia's \emph{Lily.}\footnote{\emph{Broadcast Pro Middle East}, ``Tunisian filmmaker wins \$1 million AI Film Award for `Lily'.'' \emph{Dream Machine} Issue~14.} Mexico's Avocados-From-Mexico Super Bowl campaign was AI-led.\footnote{\emph{Digiday}, ``Avocados From Mexico turns to AI to advertise around the Super Bowl instead of a TV buy.'' \emph{Dream Machine} Issue~15.}

\textbf{Eastern European} AI filmmaking~-- typified by Andrii Daniels' bomb-shelter clip~-- became a recognised category.\footnote{\emph{Variety}, ``AI Creator Behind Viral `Deadpool,' `Harry Potter' Christmas Clip Made His Film in a Ukrainian Bomb Shelter.'' \emph{Dream Machine} Issue~16.}

\textbf{East Asian} AI development continued at a pace that, by spring 2026, had Chinese open-source AI models being used by approximately 80\% of startups pitching the Andreessen Horowitz fund.\footnote{\emph{Dream Machine} Issue~8 citing Andreessen Horowitz observations: \url{https://www.linkedin.com/posts/stevenouri_a-wild-stat-80-of-startups-pitching-a16z-activity-7396182718998351872-xTKR}.} Korea's Shift Up CEO described AI as the way to compete with Chinese game-industry scale, in language that captures both the geo-economic argument and what it means for individual workers: \emph{``Only when all these people are proficient in AI, so that one person can perform the role of 100 people, can we compete with industries like China and the US that rely on large-scale human resources.''}\footnote{\emph{PocketGamer.biz}, ``Shift Up CEO says AI is key to competing with China's game industry scale,'' \url{https://www.pocketgamer.biz/shift-up-ceo-says-ai-is-key-to-competing-with-chinas-game-industry-scale/}. \emph{Dream Machine} Issue~14.} \emph{One person performing the role of 100.} That is the East Asian games industry's framing of the orchestrator economy, and~-- if it is right~-- it tells you everything you need to know about the headcount maths every studio in the world will run between now and 2030.

The pattern, when you stand back from it, is what I have come to call the \textbf{Geographic Inversion}. AI is~-- in significant measure~-- \emph{redistributing creative production capacity} away from the traditional centres (Hollywood, London, New York) towards regions that were historically capacity-constrained relative to their creative ambition. For most of the post-war period, geography concentrated creative work; for the first time in living memory, the technology is pushing the other way. This is not, on its own, a justification for everything AI is doing. It is, however, one of the clearest beneficial second-order effects of the cost reduction, and one of the most reliable signs that the creative economy that emerges on the other side of this transition will be~-- in geographic, demographic and economic terms~-- \emph{less concentrated} than the one that preceded it.

If you are a working creative in a part of the world that has historically been on the wrong end of the global creative economy's geography of access, the AI era is~-- for all its risks~-- also opening doors that were welded shut for most of the previous century.

\section*{What organisations should do}

I want to close this chapter with a short and direct argument about what creative \emph{organisations}~-- studios, agencies, labels, broadcasters~-- should be doing right now, because the readers I hear from most often, after working creatives, are people running creative organisations and trying to figure out the shape of the next three to five years.

The short version, drawn from everything I have read, watched and lived through these six months:

\textbf{One. Move to integration, fast.} The path-two organisations will outcompete the path-one organisations on talent, on output and on cultural capital within three years. Suppression is not viable as a long-term strategy.

\textbf{Two. Invest in your mid-career layer.} The biggest source of avoidable damage in this transition is the loss of mid-career specialist knowledge that takes years to rebuild. Find ways to upskill your existing mid-career staff into orchestrator roles. The institutional knowledge they carry is the most valuable asset you have. Do not throw it away because the labour-cost arithmetic in a single quarter says you can.

\textbf{Three. Solve the apprenticeship problem.} The orchestrator economy structurally undermines the pipeline that has historically produced senior creatives. If you don't solve this~-- by maintaining some entry-level human roles, by building new AI-augmented apprenticeship pathways, by partnering with the institutes and the literacy initiatives~-- you are eating your own future. \emph{Your senior creatives of 2035 are the juniors you hire today.} Treat them that way.

\textbf{Four. Build the disclosure and provenance infrastructure.} Chapter~\ref{ch:12}'s argument applies as much to organisations as to individual creatives. The organisations that can credibly disclose their AI use, that maintain documentation, that can produce chain-of-custody on contested work, will be the organisations that the audience trusts in 2030.

\textbf{Five. Build for the new geography.} If your existing organisation is centred on the traditional creative capitals, the AI era is going to be much harder for you than for organisations distributed across the newly-accessible regions of the global creative economy. \emph{Take seriously the option of building distributed teams}~-- not as a cost-saving move, but as a creative-capacity move. The talent is global. The tools are global. The audience is global. The organisations that don't adapt to this fact will lose their relevance to the ones that do.

\textbf{Six. Don't outsource your judgement.} This is the most important one and the easiest to get wrong. AI tools~-- even the very good ones, even the agentic ones, even the ones the platform companies are most eager to sell you~-- \emph{cannot replace organisational judgement.} The decisions about what to make, who to hire, what to invest in, what to refuse~-- these are decisions that have to live with the humans running the organisation. AI can inform them. AI cannot make them. The organisations I have watched make the biggest unforced errors in the period this book covers are the ones that abdicated organisational judgement to the tools.

The shape of the creative economy in 2030~-- what it produces, who it employs, where it operates, what it pays, what it is for~-- is being decided, right now, by the choices that the working creatives, the organisations and the institutions of the creative industries make in this twelve-to-eighteen-month window.

In the next chapter~-- the final chapter of the book proper~-- I want to argue, as directly as I can, for the \emph{kind} of creative economy I think we should be choosing. What a humane version of the AI-era creative economy looks like. Who has to do what to get there. And what working creatives reading this book should be doing on Monday morning to play their part in it.

That choice is the last thing the book is about.

  \chapter{The New Jobs}\label{ch:14}

\lettrine[lines=3,lhang=0.15,findent=0.1em]{T}{here} is a binary that I have, for six months, watched dominate every conversation about AI and creative employment, and that I am going to spend this chapter taking apart.

The binary is \emph{jobs apocalypse} versus \emph{jobs renaissance}. On one side, the visible argument: AI is coming for creative work, the trade unions are right, the displacement is real and accelerating, and a meaningful percentage of working creatives~-- particularly mid-career specialists in functions where AI has already become competent~-- will be out of the industry by 2030. On the other side, the equally visible argument: AI is creating more jobs than it destroys, the new categories of AI-orchestration work are paying more than the old ones, the freelance and indie sectors are expanding, the geographic boundaries of the creative economy are dissolving, and the next decade will be the most economically expansive period for creative labour since the post-war television boom.

Both arguments have evidence. Both are partially right. Both are, taken on their own, \emph{wrong}~-- because the actual labour-market story of 2025--26 is not a binary. It is a \emph{restructuring} with sharp winners and sharp losers, in which the dividing line between the two is not ``AI'' or ``anti-AI.'' The dividing line is \textbf{AI literacy}~-- the practical capacity to deploy generative tools as instruments of one's own creative practice, with judgement, taste and structural understanding of where they help and where they harm.

This chapter is about that restructuring. About which jobs are disappearing, which are emerging, which are simply being \emph{reshaped}, and what the working creative reading this should be doing~-- concretely, this year~-- to land on the right side of the line.

I want to spend most of the chapter on the evidence, because the binary framings have been driven, in my experience, by people who have not done the reading. The evidence is messier and more interesting than either side wants to admit.

\section*{What the headline numbers do not say}

The aggregate employment numbers in the creative industries for 2024--26 did not show the apocalyptic collapse some had predicted. They also did not show the renaissance the platform companies' marketing teams have been selling. \emph{The Economist}, in a November 2025 piece titled ``Investors expect AI use to soar.\ That's not happening,'' argued that the broad productivity-gain thesis was, in the short term, not playing out at scale across the wider knowledge economy.\footnote{\emph{The Economist}, ``Investors expect AI use to soar.\ That's not happening,'' \url{https://www.economist.com/finance-and-economics/2025/11/26/investors-expect-ai-use-to-soar-thats-not-happening}. \emph{Dream Machine} Issue~9.} A month later, in ``Job apocalypse? Humbug! AI is creating brand new occupations,'' the same publication argued~-- using the same labour-market datasets~-- that AI was producing more new role categories than it was eliminating.\footnote{\emph{The Economist}, ``Job apocalypse? Humbug! AI is creating brand new occupations,'' \url{https://www.economist.com/business/2025/12/14/job-apocalypse-humbug-ai-is-creating-brand-new-occupations}. \emph{Dream Machine} Issue~12.} Both pieces were defensible. Both used real numbers. The two coexisted in the same magazine, six weeks apart, because the aggregate data is, on the current cut, ambiguous.

What the aggregate data hides is the \emph{internal redistribution}. The mid-career specialist roles I described in Chapter~\ref{ch:13}~-- the experienced sub-editors, the junior animators, the staff illustrators, the in-house copywriters, the routine production-pipeline engineers~-- are visibly contracting. The senior strategic roles~-- the showrunners, the lead creative directors, the senior orchestrators, the IP-fluent producers~-- are visibly expanding their \emph{effective reach} if not their headcount. The new role categories~-- AI agent developers, prompt engineers, AI operations specialists, creative-AI ethics officers, model-curation specialists, AI-literacy trainers, custom-model fine-tuners, agentic workflow designers~-- are visibly growing from a near-zero base.

The \emph{Guardian's} ``AI is hitting UK harder than other big economies, study finds,'' from January 2026, found that mid-career UK knowledge workers were experiencing disproportionate displacement compared to peers in the U.S., Japan, Germany and Australia~-- but that the displacement was concentrated in specific \emph{task categories}, not whole occupations.\footnote{\emph{The Guardian}, ``AI is hitting UK harder than other big economies, study finds,'' \url{https://www.theguardian.com/technology/2026/jan/26/ai-uk-jobs-us-japan-germany-australia}. \emph{Dream Machine} Issue~16.} The University of Wisconsin-Stout's January 2026 announcement, in which the institution set AI use as a \emph{baseline competency} in its filmmaking course, captured the supply-side response: the curriculum was being re-engineered around the assumption that working filmmakers in 2030 would be AI-literate by default.\footnote{University of Wisconsin-Stout, ``AI Reshaping Industry: New UW-Stout Course Sets AI-Use as Baseline Competency in Filmmaking,'' \url{https://www.uwstout.edu/about-us/news-center/ai-reshaping-industry-new-uw-stout-course-sets-ai-use-baseline-competency-filmmaking}. \emph{Dream Machine} Issue~15.}

The labour market, in other words, is doing what labour markets always do in a productivity transition. It is reshuffling. The reshuffle is sharper than the headline employment figures suggest because the \emph{composition} of work is changing faster than the \emph{amount}.

\section*{The roles that are disappearing}

I want to be specific, because vague claims about ``creative jobs disappearing'' do not help anyone make a career decision. Based on the trade-press coverage tracked across the \emph{Dream Machine} archive, the survey data in Appendix~D and Appendix~E, and the studio-leadership interviews I have read or conducted, the roles under the most active substitution pressure in 2026 are:

\textbf{Junior visual production roles.} Concept artists at the asset-variation level. Junior 3D modellers doing standard architectural / environmental fills. Storyboard artists working on commercial briefs that do not require performance staging. Background-plate compositors. Routine matte painters. Stock photographers and stock illustrators. Adobe's own data~-- 25\% of new Adobe Stock contributions in 2024 containing Firefly-generated elements\footnote{Adobe Firefly enterprise metrics, in Appendix~E: \emph{Dynamics of Generative AI Adoption}.}~-- is the clearest single number in this category.

\textbf{Junior writing and copy roles.} In-house copywriters at brand agencies. Junior content marketers. Routine technical writers. Translation generalists where the source/target pair is well-resourced (English-Spanish, English-Chinese, etc.). SEO content writers. Sub-editors at digital publications. The Reuters Institute's ``AI adoption by UK journalists'' survey found high integration across newsrooms by late 2025; the \emph{Daily Mail}'s December 2025 report that Google's AI Overviews had ``killed click-throughs'' to news sites was the consumer-side mirror of the production-side pressure.\footnote{Reuters Institute, ``AI adoption by UK journalists and their newsrooms,'' \url{https://reutersinstitute.politics.ox.ac.uk/ai-adoption-uk-journalists-and-their-newsrooms-surveying-applications-approaches-and-attitudes}. \emph{Digiday}, ``Daily Mail says Google AI Overviews have killed click-throughs,'' \url{https://digiday.com/media/daily-mail-says-google-ai-overviews-have-killed-click-throughs/}. \emph{Dream Machine} Issues~7,~9.}

\textbf{Mid-career routine production and post-production roles.} Routine audio engineering (the 1,100-creator music survey discussed in Chapter~\ref{ch:13} showed 58\% of producers using AI for restoration, 38\% for mixing assistance, 33.9\% for automated mastering\footnote{1,100-creator music survey 2026, in Appendix~D: \emph{Shadow AI}, \S``Music Production and Sound Recording.''}). Standard VFX compositing (62\% of Hollywood studios on automated AI compositing, 35\% reduction in post-production timelines\footnote{VFX AI integration metrics, in Appendix~E, \S``Visual Effects (VFX) Automation.''}). De-aging specialists (200 hours per actor down to 50). Particle simulation specialists (68\% adoption among top VFX houses by SIGGRAPH 2025). Routine matte-painting generalists (initial setup time from 4 hours to 1.2 per shot).

\textbf{Routine games-development roles.} Square Enix's announced target~-- \textbf{70\%} of QA work via AI by end of 2027~-- is the cleanest signal here.\footnote{\emph{PC Gamer}, ``Square Enix aims to have AI doing 70\% of its QA work by the end of 2027,'' \url{https://www.pcgamer.com/gaming-industry/square-enix-aims-to-have-ai-doing-70-percent-of-its-qa-work-by-the-end-of-2027/}. \emph{Dream Machine} Issue~7.} Falcom's reported productivity ratio of \emph{2-3 hours of work reduced to 10 minutes}\footnote{\emph{Eurogamer}, ``Falcom is the latest developer to buy into the AI hype machine,'' \url{https://www.eurogamer.net/falcom-is-the-latest-developer-to-buy-into-the-ai-hype-machine}. \emph{Dream Machine} Issue~12.} tells you what is happening to the routine animation, asset and engineering layer underneath the games industry. Ubisoft's January 2026 cancellation of five games (including the \emph{Prince of Persia} remake) in order to refocus capital on AI signalled a structural shift in resource allocation that mid-career game developers are still digesting.\footnote{\emph{Metro}, ``Prince of Persia remake and five more games cancelled as Ubisoft focuses on AI,'' \url{https://metro.co.uk/2026/01/21/prince-persia-remake-five-games-cancelled-ubisoft-focuses-ai-26431926/}. \emph{Dream Machine} Issue~15.}

\textbf{In-house AI specialist teams at non-AI-native companies.} This one is counter-intuitive but real. The publisher behind \emph{Grand Theft Auto VI} laying off its entire seven-year-old internal AI team in April 2026, in favour of off-the-shelf tooling, is the canonical case.\footnote{\emph{Dream Machine} Issue~24, April 2026, on the GTA~VI publisher laying off its internal AI team.} Companies that built proprietary AI capabilities in 2018--2024 are increasingly finding that the open-weight and commercial foundation models have caught up; the bespoke AI team becomes redundant. This is a real and rapid form of AI-driven displacement that the public discourse has not yet recognised, because the workers being displaced are themselves AI specialists.

\textbf{Voice actors and session musicians in commodity work.} ElevenLabs' growth to \$500m ARR by April 2026\footnote{ElevenLabs \$500m ARR reporting, April 2026. \emph{Dream Machine} Issue~25.} is, in large part, the substitution of routine voiceover, audiobook, podcast-host and dubbing work. Live performance, voice work that requires acting craft, and specialty voice roles (animation leads, signature character voices) are not displaced. The middle of the voice market is.

\textbf{Routine commercial illustration and design.} Brand assets, marketing imagery, social-media graphics, basic product visualisation. The Higgsfield growth curve~-- \$200M revenue in nine months, primarily serving social-media marketers\footnote{\emph{36kr}, ``AI Video Unicorn Higgsfield: Earns \$200M in 9 Months by `Serving' Social Media Marketers,'' \url{https://eu.36kr.com/en/p/3650517574312323}. \emph{Dream Machine} Issue~16.}~-- is the consumer-marketing equivalent of the Adobe Firefly enterprise curve.

The pattern across these categories is consistent: it is \emph{routine} work, \emph{specialist} work, and \emph{mid-career} work that is under pressure. The pattern is \emph{not} aimed at junior on-ramp roles (which is a problem of its own~-- see below) or at senior creative judgement roles. It is concentrated in the middle of the pipeline, in the layer historically occupied by experienced operators of the toolchain.

\section*{The roles that are emerging}

The other side of the redistribution is equally real and substantially under-reported in the consumer press.

\textbf{AI orchestrators / senior creative directors of agentic teams.} The most strategically important new role, and the one Chapter~\ref{ch:11} made the long-form case for. The Sony 49-Claude-agent / 72-skill stack is the canonical example.\footnote{\emph{Dream Machine} Issue~29, May 2026, on Sony's 49-agent / 72-skill multi-agent game-development team.} In adland, \emph{Digiday} reported in late 2025 that ``AI agent developers have become adland's in-demand role''\footnote{\emph{Digiday}, ``AI agent developers have become adland's in-demand role,'' \url{https://digiday.com/marketing/ai-agent-developers-have-become-adlands-in-demand-role/}. \emph{Dream Machine} Issue~7.}~-- a senior creative-strategic role that did not exist eighteen months earlier.

\textbf{Prompt engineers / AI workflow designers.} The role is now broad enough to have its own specialisation tracks. \emph{Forbes} reported in November 2025 that ``vibe coding''~-- natural-language software development~-- paid up to \textbf{\$220,000} as an in-demand AI skill.\footnote{\emph{Forbes}, ``Vibe Coding~-- The In Demand AI Skill That Pays Up to \$220,000,'' \url{https://www.forbes.com/sites/rachelwells/2025/11/06/the-in-demand-ai-skill-and-certifications-that-pays-up-to-220000/}. \emph{Dream Machine} Issue~8.} The equivalent figures in the creative space~-- prompt engineers at major studios, freelance AI-workflow consultants~-- are running in the same band for senior practitioners.

\textbf{AI literacy trainers and AI-education designers.} The Sundance Institute's AI Literacy Initiative, launched in January 2026 with \$2M of Google funding to train 100,000 filmmakers, is the institutional version of this role.\footnote{Sundance Institute, ``Centering the Artist: Why We're Launching the AI Literacy Initiative,'' \url{https://www.sundance.org/blogs/centering-the-artist-why-were-launching-the-ai-literacy-initiative/}. Google blog, ``Sundance Institute AI Education,'' \url{https://blog.google/company-news/outreach-and-initiatives/google-org/sundance-institute-ai-education/}. \emph{Dream Machine} Issues~15,~16.} The Adobe Ignite Day at Sundance, the UK government's ``Free AI training for all'' programme covering 10 million workers by 2030,\footnote{UK Government, ``Free AI training for all,'' \url{https://www.gov.uk/government/news/free-ai-training-for-all-as-government-and-industry-programme-expands-to-provide-10-million-workers-with-key-ai-skills-by-2030}. \emph{Dream Machine} Issue~16.} the Lovable-for-classrooms expansion,\footnote{Lovable for classrooms: \url{https://lovable.dev/classroom}. \emph{Dream Machine} Issue~11.} the UW-Stout baseline-AI competency course~-- these are the demand signal for a new category of educator that combines creative-discipline expertise with practical AI fluency.

\textbf{Model curation specialists.} With foundation models proliferating and custom fine-tuning becoming a baseline capability, the role of \emph{selecting, training and maintaining} an organisation's model stack has emerged as a discrete specialism. Adobe Firefly Foundry~-- the service that lets companies train custom generative models on their own visual identity\footnote{Adobe, ``Firefly Foundry,'' \url{https://news.adobe.com/news/2025/10/adobe-max-2025-firefly-foundry}. \emph{Dream Machine} Issue~5.}~-- created an entire job category of brand-and-IP model trainers. Korin AI's launch in May 2026, ``trained with African datasets, built by Africans,''\footnote{Korin AI launch, May 2026. \emph{Dream Machine} Issue~27.} is the cultural-fluency variant of this role.

\textbf{AI ethics, disclosure and provenance officers.} Following the SAG-AFTRA contract negotiations, the Cannes AI Disclosure Standard, the Academy's ``you must be human to win'' rule, the New York AI advertising disclosure law, and the proliferating C2PA-compliance and SynthID-tooling requirements, organisations across the creative industries have begun hiring (or designating) dedicated AI-ethics and disclosure leads.\footnote{\emph{The Verge}, ``New York's new law forces advertisers to say when they're using AI avatars,'' \url{https://www.theverge.com/news/842848/new-york-law-ai-advertisements-sag-aftra-labor}. \emph{Dream Machine} Issue~11. C2PA / SynthID infrastructure references in Chapter~\ref{ch:12}.} At DreamLab, we have a \emph{Continuum Lead} whose job is to make this work coherent across every project we run~-- three years ago, the role did not exist.

\textbf{Indie and Global South creator-producers.} The cost reduction in production tooling has created a new viable role category that was not economically possible before: the \emph{one-person-or-small-team creator-producer} operating outside the traditional creative centres, with global distribution reach and a defensible aesthetic identity. \emph{Forbes} covered the broad category in ``AI Is Changing How Creators Work And Earn'' in December 2025.\footnote{\emph{Forbes}, ``AI Is Changing How Creators Work And Earn,'' \url{https://www.forbes.com/sites/kolawolesamueladebayo/2025/12/22/how-ai-is-changing-how-creators-work-and-earn/}. \emph{Dream Machine} Issue~13.} The Higgsfield revenue (built on social-media marketer demand), the Andrii Daniels bomb-shelter clip (a Ukrainian one-person production with global reach\footnote{\emph{Variety}, ``AI Creator Behind Viral `Deadpool,' `Harry Potter' Christmas Clip Made His Film in a Ukrainian Bomb Shelter,'' \url{https://variety.com/2026/digital/news/ai-video-deadpool-harry-potter-andrii-daniels-1236624632/}. \emph{Dream Machine} Issue~16.}), the Tunisian \emph{Lily} (\$1M Dubai AI Film Award winner\footnote{\emph{Broadcast Pro Middle East}, ``Tunisian filmmaker wins \$1 million AI Film Award for `Lily','' \url{https://www.broadcastprome.com/news/tunisian-filmmaker-wins-1-million-ai-film-award-for-lily/}. \emph{Dream Machine} Issue~14.}), the Indonesian \emph{Legenda Bertuah} animated series,\footnote{\emph{Dream Machine} Issue~25, April 2026, on Indonesia's \emph{Legenda Bertuah}.} the Indian-cinema integration wave covered by the BBC's ``Lights, camera, algorithm''\footnote{BBC Future, ``Lights, camera, algorithm: Why Indian cinema is awash with AI,'' \url{https://www.bbc.co.uk/future/article/20251223-why-indian-cinema-is-awash-with-ai}. \emph{Dream Machine} Issue~14.}~-- these are not exceptions. They are the leading edge of a structural change in \emph{who can be a working creative}, and where they can live.

\textbf{AI-augmented apprentices.} This category is still being built, and is the central labour-market design question of the next three years (more below). The early models~-- AI-tool-augmented junior animator roles maintained deliberately at \emph{Position Four} studios (Chapter~\ref{ch:7}), the Sundance Collab fellowship structure, the AI-augmented entry-level posts at WPP and the major Hollywood studios~-- are the early experiments. None of them, yet, has fully solved the apprenticeship problem.

\textbf{Cross-disciplinary ``portfolio creatives.''} What I have been calling, in talks since the autumn, the \textbf{AI Literacy Premium} role~-- the working creative who, instead of a single specialism, holds several loosely-coupled creative disciplines together using AI as connective tissue. \emph{TechBullion}'s ``Why the future belongs to multi-skilled leaders,'' from November 2025,\footnote{\emph{TechBullion}, ``Why the future belongs to multi-skilled leaders,'' \url{https://techbullion.com/playing-the-long-game-with-a-portfolio-career-why-the-future-belongs-to-multi-skilled-leaders/}. \emph{Dream Machine} Issue~9.} and the Anthropic \emph{Skills} framework underneath Claude Code's multi-agent coordination,\footnote{Anthropic Skills framework via Claude Code. \emph{Dream Machine} Issues~11,~16,~29.} are the corporate-leadership and tooling-side manifestations of the same trend. The portfolio creative is increasingly the \emph{default} career shape for working creatives entering the field today.

\section*{The AI literacy premium}

If you stack the disappearing-roles list against the emerging-roles list, a single underlying variable does most of the explanatory work for who is winning and who is losing in the 2026 creative labour market.

The variable is \textbf{AI literacy}~-- defined operationally as the combined skill of (a) knowing what generative tools can and cannot do well, (b) being able to brief them effectively, (c) being able to judge their outputs with taste, (d) being able to integrate them into a coherent creative workflow, and (e) knowing where on the Continuum your craft sits and why.

The empirical case for the literacy premium is strong and consistent across the data:

The Adobe Creators' Toolkit Report of October 2025, surveying 16,000 creators across the U.S., U.K., France, Germany, South Korea, Japan, India and Australia, found that \textbf{86\%} of creators were already using generative AI; that \textbf{76\%} of users said it had helped grow their business or brand; that \textbf{81\%} said AI lets them make content they otherwise could not have made; and~-- most significantly~-- that \textbf{70\%} were optimistic about agentic AI and \textbf{85\%} would use AI that learned their creative style.\footnote{Adobe, ``Inaugural Adobe Creators' Toolkit Report,'' October 2025, \url{https://news.adobe.com/news/2025/10/adobe-max-2025-creators-survey}. \emph{Dream Machine} Issue~6.} The numbers are not the numbers of a workforce in retreat. They are the numbers of a workforce \emph{adopting at scale} and reporting commercial benefit.

The PRS for Music 2026 AI Survey, surveying U.K.\ music creators, found \textbf{four in five} music creators worried about AI-generated music competing in the streaming economy\footnote{PRS for Music, ``PRS for Music AI Survey 2026,'' \url{https://www.prsformusic.com/m-magazine/news/prs-for-music-ai-survey-2026}. \emph{Dream Machine} Issue~16.}~-- but also that respondents who were using AI tools in their own creative practice reported higher earnings and broader output reach than peers who were not. The two findings live together. The anxiety and the adoption are happening in the same head.

The U.K.\ Department for Business and Trade research, reported in \emph{CNBC} in November 2025, found that workers with ADHD, autism and dyslexia were \textbf{25\% more satisfied} with AI assistants than neurotypical workers, with AI agents reported as actively helping them succeed at work.\footnote{\emph{CNBC}, ``People with ADHD, autism, dyslexia say AI agents are helping them succeed at work,'' \url{https://www.cnbc.com/2025/11/08/adhd-autism-dyslexia-jobs-careers-ai-agents-success.html}. \emph{Dream Machine} Issue~7.} Given that the creative industries are over-indexed on neurodivergent talent compared with the general population, this is a major equity-and-productivity story that the AI-displacement narrative has under-told.

The McKinsey AI report on film and TV production, released in early 2026, made the broader business case: AI would not, in McKinsey's view, replace film and television production. It would \emph{restructure} it~-- towards smaller teams, faster cycles, more iteration, and heavier reliance on senior creative judgement.\footnote{McKinsey \& Company, ``What AI could mean for film and TV production and the industry's future,'' \url{https://www.mckinsey.com/industries/technology-media-and-telecommunications/our-insights/what-ai-could-mean-for-film-and-tv-production-and-the-industrys-future}. \emph{Dream Machine} Issue~16.} In other words: towards an \emph{orchestrator-shaped} industry, in which the working creatives best positioned to thrive are those who can operate the new toolchain rather than those who refuse to engage with it.

The GDC State of the Game Industry surveys 2024--2026 (covered in detail in Appendix~E) showed personal generative-AI usage among professional game developers rising from \textbf{31\%} to \textbf{36\%} while industry sentiment cratered from 18\% negative to \textbf{52\%} negative.\footnote{GDC State of the Game Industry surveys 2024--2026, in Appendix~E, \S``The Video Game Industry.''} The interpretation: developers using AI are increasingly \emph{uncomfortable} about doing so, but they are doing so anyway, because the productivity advantage is too strong to ignore.

The LANDR study reported in \emph{Ari's Take}, late 2025, found that \textbf{87\%} of musicians surveyed were using AI tools in some part of their practice.\footnote{LANDR AI music study, late 2025, referenced via \emph{Ari's Take}, \url{https://aristake.com/ai-tools-musicians-study/}. \emph{Dream Machine} Issue~8.} When the figure for ``musicians using AI'' approaches the figure for ``musicians using DAWs,'' you are not looking at a niche adoption pattern. You are looking at a baseline competency that the next generation of working musicians will be assumed to have.

The Stanford AI Index Report 2025 found that across 26 surveyed nations, \textbf{55\%} of individuals viewed AI products as offering more benefits than drawbacks, up from 52\% in 2022;\footnote{Stanford AI Index Report 2025. Summarised in Appendix~E, \S``The Perception Gap.''} the YouGov 2024 multi-market sentiment data echoed the same direction.\footnote{YouGov 2024 multi-market AI sentiment survey. Summarised in Appendix~E.} Working creatives are sitting inside a consumer environment that is, on net, \emph{increasingly accepting} of AI in creative work~-- which translates, slowly, into client demand for AI-augmented services.

The \emph{Digital Music News} survey of nearly \textbf{800 creatives} who signed the \emph{Stealing Our Work Is Not Innovation} declaration in January 2026\footnote{\emph{Digital Music News}, ``Nearly 800 Creatives Sign Responsible AI Declaration~-- `Stealing Our Work Is Not Innovation','' \url{https://www.digitalmusicnews.com/2026/01/22/stealing-isnt-innovation/}. \emph{Dream Machine} Issue~16.}~-- that same vocal-minority signal that produced political wins in Chapter~\ref{ch:6}~-- included, in its signatory base, a meaningful number of creators who were \emph{themselves} using AI tools in their own production work. The signing of the declaration was not a refusal of the technology. It was an insistence that the \emph{training data} be consented and compensated. Two positions, held at once, by the same working creatives.

Stack these findings together and the labour-market picture becomes much clearer than the binary discourse allows.

\textbf{Working creatives who hold AI literacy are gaining leverage.} Their effective output reach is larger. Their hourly rate is higher. Their client base is broader. Their geographic constraints are weaker. The orchestrator role, the portfolio-creative model, the literacy-trained newcomer~-- these are the positions on which the next decade of creative employment expansion is being built.

\textbf{Working creatives without AI literacy are losing leverage.} Their effective hourly rate is being competed down by AI-augmented peers. Their client base is being eroded by faster, cheaper, mostly-adequate substitutes. Their career mobility is contracting as mid-career specialist roles in their domains are absorbed into agentic capacity. Their geographic constraints (London, New York, LA, Tokyo as the centres) are being preserved while their AI-fluent competitors operate from anywhere.

This is, in practical terms, the structural transition the working creative reading this in 2026 is now facing. The choice is not ``use AI'' or ``refuse AI.'' The choice is whether to build the literacy that lets you remain in command of how AI shows up in your own practice, or whether to let the literacy build itself, badly, in your absence.

\section*{The Apprenticeship Gap}

I want to name the most underdiscussed structural problem in the AI-era labour market, because it is~-- by my read~-- the single largest threat to the long-term health of the creative economy and it has nowhere near the public attention it deserves.

I call it the \textbf{Apprenticeship Gap}.

For the entire history of the creative industries~-- from the medieval guilds through the post-war Hollywood studio system through the rise of digital media~-- the \emph{junior on-ramp} into creative work has been the structural foundation on which senior talent is built. Junior writers become senior writers by writing things that nobody pays much attention to, repeatedly, for a decade, under the loose mentorship of more senior practitioners. Junior animators become senior animators by drawing the in-between frames, by cleaning up the rough animatics, by handling the routine asset variations that the lead artists do not have the time for. Junior cinematographers become senior cinematographers by holding focus, by pulling cable, by lighting the second-unit shot. The junior tasks were not, in themselves, the destination. They were the \emph{training ground} on which judgement, craft and taste were built.

The orchestrator economy, the agentic toolchain, the GDC-data picture of senior practitioners using AI to absorb the routine middle-layer work~-- these patterns, taken together, are progressively \emph{removing} the junior on-ramp from the industry. The junior writer is competing with Claude. The junior animator is competing with Cascadeur. The junior cinematographer is competing with Veo 3.1's plate generation. The junior coder is competing with Cursor and Copilot. In every case, the routine task that \emph{used to be the entry point} into the discipline is now economically uncompetitive against an AI agent that does it in seconds for cents.

The studios are not~-- yet~-- replacing senior roles. They are absorbing the \emph{junior layer} underneath the senior roles, and then telling themselves a story about how the new tools will free senior practitioners to focus on the \emph{real} creative work.

The story is partially true. It is also dangerously incomplete. If the junior layer disappears for a decade, the \emph{next generation of senior practitioners has nowhere to be trained}. The pipeline breaks. The 2035 cohort of senior creative directors, lead animators, showrunners, music producers, art directors~-- the people who, in 2026, would be five years into a junior career~-- will simply not exist at the volumes the industry needs. The discipline-specific knowledge, the embodied craft, the relationship-based mentorship~-- all of these were carried in the apprenticeship layer. Remove the layer, and you are eating the seed corn of the discipline.

This is not a speculative claim. It is the underlying logic of every ``expensive mistake'' interview a working studio leader has given to the trade press in this period~-- Charles Cecil at Revolution Software, Todd Howard at Bethesda, the Larian and Aardman and Jagex public positions. The senior practitioners are saying, in different vocabularies, the same thing: \emph{we used to teach the next generation by giving them the routine work. The routine work is gone. We have not solved the teaching problem.}

The institutional response to date is partial and patchy. The Sundance literacy initiative is real. The Adobe Ignite Day, the Sundance Collab fellowships, the UK Free AI Training for All programme, the UW-Stout curricular changes~-- these are the visible institutional moves. But they are training programmes for \emph{AI literacy specifically}, not full apprenticeship pipelines for the underlying creative discipline. They produce literate orchestrators. They do not, by themselves, produce the cinematographers, the composers, the writers, the lead artists, the showrunners of 2040.

The deepest structural reform that needs to happen in the next eighteen months is the deliberate preservation~-- or rebuilding, or re-imagining~-- of the apprenticeship layer. Some of this is already happening:

\textbf{Position Four studios maintaining junior roles by policy.} Aardman, Larian, Games Workshop, Jagex have, in different forms, made deliberate commitments to keep junior human roles in their pipelines even where AI could absorb them, on the explicit grounds that the future of the studio requires it. This is the most encouraging single trend I have observed.

\textbf{Hybrid apprenticeship pathways.} The new role I called \emph{AI-augmented apprentices} above. A junior animator who uses Cascadeur, but who is \emph{paired with a senior animator who teaches them why} the AI's output is good or bad. The juniors are not doing the in-between frames any more. They are doing the \emph{judgement} on the in-between frames the AI produces, and the senior teaches them how to judge. This is a real model. It is not yet at scale.

\textbf{Institutional reinvestment.} The cultural-institution training programmes~-- Sundance, the BFI, the national film schools, the BBC training schemes, the Royal College of Art and equivalent national schools~-- are, in different ways, recalibrating to deliver more comprehensive training in shorter timeframes, on the assumption that the years-on-the-job apprenticeship period is contracting.

\textbf{Public funding interventions.} The UK government's \emph{Free AI training for all} programme, the EU's various creator-skills initiatives, and the U.S.\ state-level training credits being attached to AI-investment incentives are early signs that the apprenticeship gap is being recognised as a public-policy problem rather than a market problem.

None of these is, on its own, sufficient. The Apprenticeship Gap is~-- by my prediction~-- going to be the single largest \emph{unresolved} labour-market issue in the creative economy of 2030. The studios, the unions, the schools and the platform companies are going to have to figure it out together. Chapter~\ref{ch:15}'s manifesto in \emph{Choosing the Future} lists it explicitly as one of the questions every working organisation has to engage with.

If you are reading this as a senior practitioner: maintain juniors. Pair them with the new tools deliberately. Treat their on-ramp as a public good your industry depends on.

If you are reading this as a junior practitioner: the on-ramp is contracting. You do not get to wait. The literacy you build \emph{in the next eighteen months} will determine whether the on-ramp closes before you are inside it.

\section*{The geographic redistribution}

The second-largest under-reported labour-market story of this period is geographic.

For the entire post-war history of the global creative industries, professional creative employment has been concentrated in a small number of cities~-- Los Angeles, New York, London, Paris, Tokyo, Mumbai, with smaller secondary nodes in Berlin, Toronto, Seoul, Sydney. The geography was a function of the \emph{cost structure} of creative production: studios, equipment, distribution networks, talent pools and capital all concentrated in the cities that could afford to host them, and the working creatives followed.

The AI cost reduction is, structurally, dismantling this geography.

The Tunisian-made \emph{Lily} winning the \$1M Dubai AI Film Award.\footnote{\emph{Broadcast Pro Middle East}, \emph{Lily} award~-- \emph{op.\ cit.}} The Ukrainian one-person bomb-shelter production going viral globally.\footnote{\emph{Variety}, Andrii Daniels bomb-shelter clip~-- \emph{op.\ cit.}} The Indian-cinema AI integration wave covered by the BBC, with productions across regional cinema centres absorbing the new toolchain at scale.\footnote{BBC Future, ``Lights, camera, algorithm''~-- \emph{op.\ cit.}} The Indonesian \emph{Legenda Bertuah} AI-animated series.\footnote{\emph{Dream Machine} Issue~25, Indonesian \emph{Legenda Bertuah}.} The Korin AI launch~-- Africa-trained, Africa-built foundation model~-- and the broader \emph{CNBC Africa} coverage of AI in African music and film.\footnote{\emph{CNBC Africa}, ``How AI is changing the landscape of the music industry in Africa,'' \url{https://www.cnbcafrica.com/2025/how-ai-is-changing-the-landscape-of-the-music-industry-in-africa}. \emph{Dream Machine} Issue~5. Korin AI launch, May 2026~-- \emph{op.\ cit.}} The Singapore-based AI video startup Video Rebirth raising \$50M for studio-grade tooling. The Eastern European AI filmmaker community building around the success of creators like Andrii Daniels. The Latin American AI-film festival wave through 2026.

These are not isolated stories. They are the leading edge of a redistribution that, taken together, is \emph{materially expanding} the global creative workforce beyond the previous-century cities. The reduction in cost-of-entry for serious creative production has, for the first time since the rise of cinema, made it economically viable to build a competitive creative practice from places that the previous geography had locked out.

The Shift Up CEO's framing, in the \emph{PocketGamer.biz} piece~-- \emph{``Only when all these people are proficient in AI, so that one person can perform the role of 100 people, can we compete with industries like China and the US that rely on large-scale human resources''}\footnote{\emph{PocketGamer.biz}, ``Shift Up CEO says AI is key to competing with China's game industry scale,'' \url{https://www.pocketgamer.biz/shift-up-ceo-says-ai-is-key-to-competing-with-chinas-game-industry-scale/}. \emph{Dream Machine} Issue~14.}~-- is the strategic reading from inside the Korean games industry. The same logic applies in Mexico, in Egypt, in Nigeria, in Brazil, in Vietnam, in any creative economy that has historically been resource-constrained relative to its ambitions.

For the working creative reading this in one of those geographies: the AI era is, for all its disruption risks, also opening doors that were welded shut for most of the previous century. The labour market is becoming, for the first time in living memory, \emph{less} concentrated.

For the working creative reading this in one of the historical centres: this redistribution is not theoretical. The clients, the budgets and the IP that previously concentrated in your city are, increasingly, being competed for by capable AI-augmented competitors operating from anywhere on the planet. Your geographic advantage is contracting.

\section*{What the working creative should do}

I have, throughout the book, tried to land each chapter with practical takeaways. This chapter is the labour-market chapter, and the takeaways are the most concrete.

\textbf{Build the literacy this year.} Not next year. The Adobe Creators' Toolkit Report, the LANDR survey, the GDC data, the McKinsey reading~-- all point in the same direction. The literacy premium is not a future variable. It is, in 2026, the single biggest determinant of mid-career creative employment outcomes. Free training programmes exist (Sundance Collab, UK Free AI Training, Adobe Express, the open-source ecosystem). Use them. Spend the equivalent of one week per quarter, deliberately, on building practical fluency in the toolchain.

\textbf{Map your craft against the Continuum (Chapter~\ref{ch:3}).} Decide where your craft sits, function by function. Decide where you are willing to let agents operate on your behalf and where you are not. Write the map down. Update it quarterly. The working creatives I have watched make the most successful transitions in this period have been the ones who knew, in advance and explicitly, where their lines were.

\textbf{Pick your role on the new map.} Orchestrator, portfolio creative, AI-literacy trainer, model curator, ethics-and-disclosure specialist, regional creator-producer, hybrid apprentice. The roles are real. They pay. They will, by every indicator I can read, continue to grow through the next five years. You do not need to invent your own category. You can pick one of the emerging ones.

\textbf{Build your apprenticeship~-- or build the next generation's.} If you are early in your career, find the senior practitioners who are running hybrid apprenticeship pipelines and apply. If you are mid-career, identify the AI-augmented junior roles in your discipline and either fill them yourself or pair with one. If you are senior, maintain juniors in your team and pair them deliberately with the new tools.

\textbf{Take the geography seriously.} If you have historically been outside the creative-economy centres, the AI cost reduction has opened a window. Use it. Build for the global creative market from where you are, with the cost advantages your geography offers. If you have historically been inside the centres, your geographic premium is contracting; build defensible craft, IP and relationships that survive the cost flattening.

\textbf{Stay in the work.} This is the same advice the rest of the book lands on, and it is the same advice for this chapter. The maker who never makes is the maker whose judgement decays. Maintain craft contact, in at least one part of your practice, that does not depend on AI tooling. The contact will keep your eye sharp for everything else. The orchestrator who never operates the tools cannot brief them well. The orchestrator who only operates the tools loses the human signal the audience came for.

\textbf{Speak.} The labour-market shape of the next decade is being decided right now, by the institutions of collective bargaining, by the policy-makers running consultations, by the platform companies building the rails, and by the creative organisations making integration choices. The 88\% in the UK consultation was made out of voices. The Tilly Tax was made out of voices. The Sundance literacy turn was made out of voices. Your voice~-- your testimony to your union, your trade body, your local government, your manager, your client~-- is the input the institutions need. The labour-market protections of 2030 will be the cumulative result of how loudly working creatives turn up to claim them in 2026 and 2027.

\section*{A note on the binary}

I want to close this chapter by returning to the binary I started with~-- \emph{jobs apocalypse} versus \emph{jobs renaissance}~-- because both framings, repeated often enough, do real damage to the working creative who is trying to make rational career decisions in real time.

The apocalyptic framing is, in my view, the more dangerous of the two, because it produces \emph{paralysis}. Working creatives who become convinced that AI is coming for their job, full stop, often stop investing in the literacy that would let them stay ahead of the substitution curve. They become spectators of their own displacement. By the time the substitution arrives, they are unprepared.

The renaissance framing is the less dangerous one but is also wrong, because it produces \emph{complacency}. Working creatives who become convinced that AI will simply expand the labour market often fail to invest deliberately in the literacy that lets them claim the expansion. They drift, expecting the rising tide to lift their boat without recognising that it lifts only the boats whose owners are paying attention.

The accurate framing is harder to live with than either: the labour market is \emph{redistributing} in real time, with sharp winners and sharp losers, and the variable that most reliably predicts which side you land on is the \emph{deliberate, structured investment in AI literacy} that you make over the next twelve to eighteen months.

The good news, against the dystopian end of the press cycle, is that the literacy is acquirable. It is not gatekept by class, by geography, by previous credentialling, or by institutional access. The tools are largely free at the entry level. The training programmes are largely free. The community of practice~-- the \emph{Dream Machine} readers, the DreamLab Collective, the open-source forums, the regional creator networks, the literacy initiatives at Sundance and elsewhere~-- is open and inviting. The barrier to entry is, by historical creative-industry standards, low.

The harder news is that the literacy \emph{has to be acquired deliberately}. It does not arrive by osmosis. It does not arrive by reading the trade press. It arrives by \emph{doing the work}~-- by sitting at the desk, briefing the agents, evaluating their outputs, building the workflow, and shipping the result. Then doing it again. Then doing it again. The literacy is a \emph{practice}, not a credential. It is, in that sense, identical in shape to every other craft skill the creative industries have ever rewarded.

The working creative in 2026 who treats AI literacy as a craft to be practised, not a technology to be debated, is the working creative who, in 2030, will look back at this period as the one in which their career took its most consequential turn.

Build the literacy. Pick your role on the new map. Defend the apprenticeship. Take the geography seriously. Stay in the work. Speak.

The labour market is moving. So can you.

  \chapter{Choosing the Future}\label{ch:15}

\lettrine[lines=3,lhang=0.15,findent=0.1em]{I}{ } want to start this chapter with a confession.

When I sent out the first edition of \emph{Dream Machine} on 6 October 2025, I thought I was writing a newsletter about \emph{tools}. I thought it was going to be a weekly digest of new model releases, new app launches, new research papers~-- a useful reading list for people in the creative industries who wanted to keep up with what was, even then, an absurd pace of technical change.

Six months later, I do not think that is what \emph{Dream Machine} has been about, and I do not think it is what this book is about either.

The tools have mattered. The tools will continue to matter. Each chapter of this book has had to spend significant pages on what the platforms shipped, when, and to whom~-- because the platforms are setting the rails, and you cannot understand the choices the creative industries are now making without understanding the constraints those rails impose.

But the \emph{book} has been about something else. About a question that the tools force every working creative, every studio, every union, every government, every audience member, and~-- eventually~-- every person who consumes culture, to answer.

The question is: \emph{what kind of creative economy do we want on the other side of this?}

That is what this chapter is for. To put the question on the table, in its sharpest form. To describe what a humane answer to it looks like. And to tell you, as directly as I can, what I think we should each do on Monday morning to start producing that answer rather than the alternative. The two chapters that follow this one~-- \emph{The Tools}, a categorised inventory of the toolchain, and the \emph{Epilogue}, a letter to the creative person reading this in 2030~-- are the practical reference and the closing register. The argument the book has been building toward lives here.

\section*{The choice on the table}

The choice, stated plainly, is between two creative economies.

One is \textbf{extractive.} In this economy, the creative work of millions of human authors~-- their writing, their music, their images, their voices, their styles, their cultural specificity~-- is absorbed, without consent or compensation, into large statistical models owned by a small number of platform companies. The platforms then sell access to those models, mostly to the same brands, studios, and agencies that previously paid for original human work, at a fraction of the original cost. The aggregate result is a \emph{transfer of wealth} from the diffuse pool of working creatives to a concentrated pool of platform shareholders. The creative output of the economy continues~-- perhaps even increases in volume~-- but its \emph{meaning}, its \emph{cultural specificity}, and its \emph{connection to the lives of the people who used to make it}, decays over time.

The other is \textbf{generative}, in the original sense of the word. In this economy, AI is treated as \emph{new craft infrastructure}~-- a set of tools that, like the printing press, the camera, the synthesiser and the digital editing suite, can be used by working creatives to make new kinds of work that the previous infrastructure didn't allow. The training data is consented to, attributed to, and compensated for. The platforms compete on the quality and integrity of their tools, not on the unpriced absorption of their users' work. The audience can verify the provenance of what they encounter, and pay attention accordingly. Working creatives are \emph{augmented} by the new tools, not \emph{replaced} by them. The output of the economy is bigger, more diverse, more accessible, more globally distributed~-- and recognisably continuous with the human creative traditions it builds on.

These two economies are not, on the current trajectory, equally probable. Some of the rails being laid right now point at the first. Some at the second. The choice between them is not~-- as I argued in Chapter~\ref{ch:12}~-- a technical question. The technology to support either is, in spring 2026, broadly available.

The choice is \emph{political, institutional and cultural.} It is about who gets to make the rules, who has the standing to enforce them, and what default the millions of small daily decisions that constitute a creative economy converge on.

The good news, if you have read this far, is that more of those daily decisions than you might expect are pointing at the generative economy. The audience's behaviour around the slop ceiling. The UK consultation's 88\%. The Sundance Literacy turn. The Cannes Disclosure Standard. The Academy's \emph{you must be human to win.} The SAG-AFTRA Tilly Tax. The studio refusals~-- Jagex, Larian, Aardman, Games Workshop. The disclosure norms emerging in advertising and the platforms. The 800-creator declaration. The C2PA standards. The SynthID rollout. The free-AI-training-for-all programmes. The neurodiversity dividend. The Global South opening up.

These are real. They are not a foregone conclusion. But they are real, and they are coming from a coalition of forces~-- creators, audiences, unions, governments, institutions and some of the platform companies~-- that has, six months in, more political and economic weight than the extractive trajectory's proponents are willing to acknowledge.

What I want to argue, in the rest of this chapter, is that \emph{the generative economy is winnable}. It is winnable in the next eighteen months. But it requires the working creatives, the organisations, and the institutions that have been doing the work this book has documented to \emph{keep doing it,} with deliberate intent, against the gravitational pull of the extractive alternative.

\section*{What I actually believe}

Before I get to the principles, I want to put a piece of conviction on the page that has been implicit in everything else I have written but that I think deserves to be made plain.

I believe in AI as an \textbf{assistive tool that amplifies human creativity.} Not as a replacement for it. Not as a substitute for it. As an instrument that, properly deployed, expands what a working creative can imagine, attempt and finish in a given afternoon~-- without displacing the human imagination, the human attempt, or the human finish.

This is the operating assumption underneath every chapter of this book, and I have been deliberately cautious about stating it as my own view, because the whole point of the newsletter and the book has been to track the \emph{evidence}, not the \emph{enthusiasm.} The evidence has been mixed. The cultural reaction has been mixed. The economics have been mixed. None of those mixed signals would have served the reader if I had collapsed them, every week, into the version I personally find most hopeful.

But the evidence has come in. The slop ceiling holds. The audience can tell the difference between work made with care and work made by a content farm. The toolchain rewards taste, judgement and intent more than it rewards raw computational throughput. The economic returns of agentic AI accrue, demonstrably, to the people who already know what good output looks like~-- not to a category of new ``AI-only'' creators replacing human ones, but to a re-tooled population of working creatives whose effective reach has grown.

What is happening in 2026, on the evidence I have spent six months collecting, is not that machines are taking over creative work. What is happening is that the \emph{labour of execution} is being democratised, and the \emph{labour of intent} is being foregrounded. The \emph{how} is becoming a utility. The \emph{why} is becoming the scarce good.

The deep-dive companion piece \emph{The Age of Intent}, which sits in the appendices to this book, makes the long-form case for this inversion~-- the philosophical and economic argument that, when the technical barrier to production collapses, the value of \emph{deciding what should be produced and why} rises in direct proportion. The artist of 2030, in that piece's framing, is less a \emph{manual laborer} and more an \emph{Architect of Meaning}~-- a curator, a noticer, a setter-of-intent~-- and the friction of human vulnerability is the irreplaceable component of the work that the machine cannot, by construction, supply.

I think that framing is essentially correct. I think the working creatives who emerge from this transition with the most leverage will be the ones who took the AI tools seriously \emph{as instruments of their own intent}, and refused either to surrender that intent to the toolchain or to refuse the toolchain on principle. The maker-as-craftsperson and the maker-as-prompt-monkey are both poor models of where the work is heading. The maker-as-orchestrator~-- the architect, the editor-in-chief, the director of a hybrid human-agent team whose unifying signal is \emph{taste}~-- is the model that the next ten years rewards.

This is not a small reframing. It is the reframing the rest of this chapter~-- and the rest of this book~-- has been building towards. The four principles I am about to lay out are not principles for \emph{restraining} AI. They are principles for \emph{deploying AI in service of the human creativity it is supposed to amplify.} The choice between the extractive and the generative economy is, in the end, a choice about which of those two we treat as the master and which we treat as the servant.

I think the human creativity is the master. I think the AI is the servant. I think the creative economy that emerges on the other side of this transition will be the one that does not lose sight of that hierarchy.

\section*{The age of the Why}

I want to give this conviction a name, because the name will travel where the long-form argument cannot, and because a name makes the choice easier to hold in the head when the next platform launch tries to talk you out of it.

We are leaving the \emph{age of the How}. We are entering the \emph{age of the Why.}

By the \emph{age of the How}, I mean the long century in which the central question of working creative life was \emph{can you do the thing?} Can you draw the figure, light the scene, edit the cut, mix the record, model the asset, hold the camera steady, hit the note? The training pipelines of every creative industry the \emph{Dream Machine} newsletter has tracked were, at their core, infrastructure for answering that question. Conservatoires for the note. Film schools for the cut. Apprenticeships for the figure. Studios for the scene. Whole career structures built around the demonstrable, transferrable ability to \emph{execute}~-- to convert intent into finished work using the technical labour of one's own hands and ears and eyes.

The \emph{How} was the bottleneck. The \emph{How} was the scarce thing. The \emph{How} was what you got paid for.

The \emph{How}, in 2026, is~-- at a rate that I do not think the industry has yet fully metabolised~-- becoming a utility. The teenager in the bedroom with a midrange GPU and a Claude subscription can, at the time I am writing this, produce work whose surface qualities~-- composition, lighting, sound design, edit pacing, visual-effects polish~-- sit on a continuum with what a full studio could produce in 2020. The continuum is not yet at the very top end; the bedroom does not yet make \emph{Avatar}. But it is, on every available metric, closing the surface-quality gap at a rate that makes the next-decade trajectory unambiguous. \emph{Hollywood-level execution in a bedroom} is no longer a marketing slogan. It is, for working filmmakers and musicians I know, already true for non-trivial fractions of the work.

When the \emph{How} becomes a utility, the \emph{Why} becomes the scarce good.

I want to anchor this in a specific story that captures the dynamic better than any creative-industries example I have. In March 2026, \emph{Bloomberg} ran a piece on what artificial intelligence has done to elite chess.\footnote{\emph{Bloomberg}, ``AI Changed Chess.\ Grandmasters Now Win With Unpredictable Moves,'' 27 March 2026, \url{https://www.bloomberg.com/news/articles/2026-03-27/ai-changed-chess-grandmasters-now-win-with-unpredictable-moves}. \emph{Dream Machine} Issue~23. The behavioural pattern the piece describes~-- top grandmasters deliberately deviating from machine-optimal lines to put opponents on uncomputed ground~-- is the cleanest available analogy I have for the strategic shift the rest of this chapter argues for.} AI, the article reported, has driven the game towards \emph{perfect play} at the very top~-- Stockfish, Leela Chess Zero and their descendants have, between them, mapped out the optimal response to most board positions a top-twenty grandmaster is likely to encounter. The visible effect of this, through 2024 and 2025, was a striking rise in the \emph{draw rate} at top tournaments. When both players have memorised the machine-optimal lines, both players play optimally, and both players draw. The game, at the very top end, was being \emph{solved into stasis} by the machines that had been trained on it.

The \emph{Bloomberg} piece reported the response. Top grandmasters~-- Magnus Carlsen among them~-- had started, deliberately, to play \emph{sub-optimal} moves. Moves that Stockfish would mark as inaccuracies. Moves that the machine-optimal line would not recommend. Moves chosen, specifically, because they were \emph{unexpected}, and because the opponent~-- having trained against the machine~-- had not memorised the human-grade response to them. The grandmasters had stopped trying to out-machine the machine. They had started, instead, to \emph{deliberately diverge from machine-optimal play} in ways that put their opponents on uncomfortable, uncomputed ground.

The grandmasters are winning, in 2026, by \emph{doing the unexpected.}

I want to dwell on this image, because I think it is the cleanest available picture of what working creative life looks like on the other side of the \emph{How} becoming a utility. When the machine has solved the optimal move~-- the perfectly-lit shot, the on-trend hook, the algorithmically-tested ad treatment, the mean-of-the-distribution image~-- the human edge is no longer in \emph{playing the optimal move better than the machine.} The machine plays the optimal move infinitely. The human edge is in playing the move \emph{the machine would not make.} The deliberately unexpected. The taste-driven. The risk-taking. The idiosyncratic. The personal. The locally-meaningful. The unrepeatable.

This is the \emph{Why} in operational form. The \emph{Why} is not, in the practitioner's day, an abstract philosophical commitment to human creativity. It is a \emph{daily competitive practice} of choosing the move the machine would not make. The director who picks the actor the casting algorithm would not have picked. The songwriter who keeps the verse the chart-pattern model would have cut. The illustrator who renders the figure in a style no FLUX prompt would generate. The brand creative who builds the campaign around the audience question the marketing-AI did not surface, because no marketing-AI in 2026 has the lived context to surface it.

The deeply human things~-- \emph{taste, intent, authenticity, the willingness to take a risk on a move the data does not yet endorse, the refusal to make the average thing in service of the meaningful thing}~-- are not, in the age of the Why, vestigial commitments held against the toolchain. They are, increasingly, the only things the toolchain cannot do, and therefore, by simple economic substitution, the things that have \emph{commercial leverage} in a market where everything else has been pushed towards zero marginal cost.

The slop ceiling in Chapter~\ref{ch:5} is this dynamic measured at the audience layer: the audience, presented with the machine-optimal flood, reliably underweights it, because the audience can tell~-- at the speed of a swipe~-- that there is no human \emph{Why} underneath the work. The orchestrator role in Chapter~\ref{ch:11} is this dynamic operationalised at the working-creative layer: the orchestrator's daily craft is, exactly, the \emph{judgement to make the un-machine-like move at the right moment.} The authenticity premium in Chapter~\ref{ch:12} is this dynamic priced at the commercial layer: the audience pays extra, demonstrably, for work whose human \emph{Why} is verifiable. The four principles I am about to lay out are this dynamic codified at the policy and platform layer.

The chess grandmasters did not stop using engines. They train on engines daily. They study machine-optimal play more closely than any generation of players before them. The chess engines, far from being their enemy, are their \emph{most-used analytical tool.} What the grandmasters refused to do is to \emph{let the engines define what counted as a winning move.} They use the engine to know what the optimal line is, and then they choose, \emph{for taste and surprise reasons}, to play another line.

This is also, I should note in passing, the \emph{unflattering diagnosis} of the legacy entertainment industries' strategic position that I made in Chapter~\ref{ch:7}. Hollywood, commercial music and the AAA games business spent the past fifteen years optimising themselves \emph{toward} the engine-optimal move~-- toward the franchise-instalment, the streaming-tested chart hit, the open-world template~-- and arrived at the AI moment producing exactly the work the engines can now replicate most cheaply. The grandmasters' response to engines is the response legacy needs to make to AI. The new AI-native studios~-- Gossip Goblin, Critterz, Imaginae, Asteria, Wonder~-- have, by virtue of their newness, no calcified rules to unlearn and are, by default, playing the moves no machine would generate. The legacy industries that survive will be the ones that re-learn how to make the un-machine-like move. The ones that don't will, on the historical pattern of Chapter~\ref{ch:2}, be remembered as the cohort that defended the previous definition while a different cohort, with no inherited risk-aversion, defined the next one.

That is the working operating model I think the next decade of creative work runs on. Use the engines. Learn the engines. Train against the engines. And then, in the actual encounter with the audience, \emph{make the move the engine would not have made.}

The age of the Why is not the age of refusing AI. It is the age of mastering AI sufficiently that the only competitive question left is whether you can summon the \emph{deliberately unexpected, deliberately human, deliberately yours} move that the machine, by construction, cannot.

\section*{Four principles}

\begin{figure}[htbp]
  \centering
  \begin{tikzpicture}[
    font=\small,
    every node/.style={align=center},
    >=stealth
  ]


    \node[
      draw=darkblue, fill=darkblue, text=white,
      rounded corners=5pt,
      minimum width=4.8cm, minimum height=2.4cm,
      font=\small
    ] (transparency) at (-2.8, 2.2) {
      \textbf{Transparency}\\[4pt]
      \footnotesize Disclose when AI is\\used in the creative pipeline
    };

    \node[
      draw=darkgreen, fill=darkgreen, text=white,
      rounded corners=5pt,
      minimum width=4.8cm, minimum height=2.4cm,
      font=\small
    ] (consent) at (2.8, 2.2) {
      \textbf{Consent}\\[4pt]
      \footnotesize Creators must opt in\\to training data use
    };

    \node[
      draw=rulegold, fill=rulegold, text=white,
      rounded corners=5pt,
      minimum width=4.8cm, minimum height=2.4cm,
      font=\small
    ] (compensation) at (-2.8, -2.2) {
      \textbf{Compensation}\\[4pt]
      \footnotesize Fair economic return for\\data and creative work
    };

    \node[
      draw=chaptergrey, fill=chaptergrey, text=white,
      rounded corners=5pt,
      minimum width=4.8cm, minimum height=2.4cm,
      font=\small
    ] (accountability) at (2.8, -2.2) {
      \textbf{Accountability}\\[4pt]
      \footnotesize Clear liability chains for\\AI-generated output
    };

    \node[
      draw=chaptergrey!40, fill=white,
      rounded corners=4pt,
      minimum width=2.8cm, minimum height=1.2cm,
      font=\small\bfseries, text=chaptergrey
    ] (centre) at (0, 0) {The Four\\Principles};

    \draw[chaptergrey!50, thin] (transparency.south east) -- (centre.north west);
    \draw[chaptergrey!50, thin] (consent.south west) -- (centre.north east);
    \draw[chaptergrey!50, thin] (compensation.north east) -- (centre.south west);
    \draw[chaptergrey!50, thin] (accountability.north west) -- (centre.south east);

  \end{tikzpicture}
  \caption{The Four Principles: the book's proposed framework for governing AI in the creative industries.}
  \label{fig:four-principles}
\end{figure}
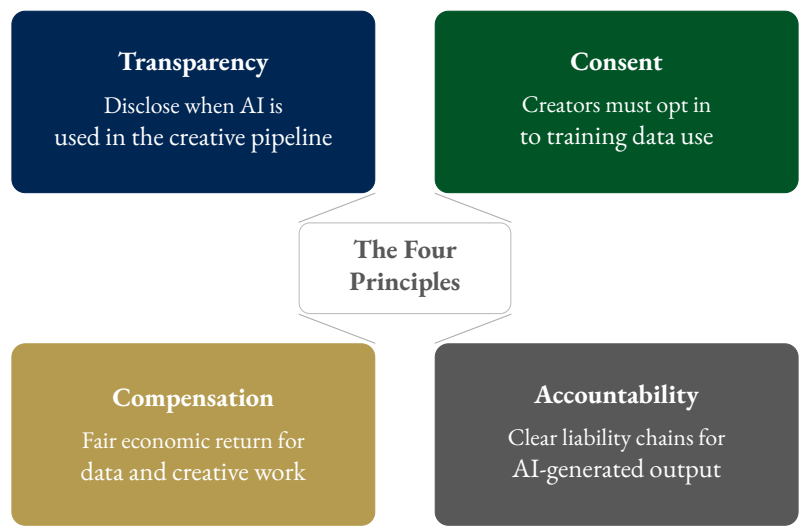

I have been asked, by readers of the newsletter and by people who turn up to the talks I have been giving since the autumn, what a humane AI-era creative economy should look like. After six months of trying to answer that question, I have come down to four principles. They are not a programme. They are a \emph{test} you can apply to any policy, any contract, any product launch, any organisational decision~-- and ask whether it is moving us towards the generative economy or the extractive one.

\textbf{One. Agency.} Every working creative should retain meaningful control over how AI is used in relation to their own work~-- both in \emph{what gets trained on their work} and in \emph{whether and how they choose to use AI in their own practice.} The Human--AI Agency Continuum from Chapter~\ref{ch:3} is the practical expression of this. Policy, contracts, platform terms-of-service and union agreements should all be evaluated against whether they preserve this control or undermine it.

The 88\% of UK respondents who said training should require licensing in all cases were articulating this principle.

\textbf{Two. Attribution.} When AI systems produce work that is derived, in any meaningful sense, from human-authored training data, the human authors should be identified and~-- where appropriate~-- compensated. The technical infrastructure for this~-- \emph{creative weight attribution} in Musical AI's framing,\footnote{\emph{Digital Music News}, ``The AI Licensing Shift~-- Creative Weight Attribution Emerges as Music Industry Game-Changer for Rights Holders.'' \emph{Dream Machine} Issue~16.} C2PA provenance metadata, SynthID watermarks~-- is the most important infrastructure question of the next two years. Policy should support its deployment. Contracts should require it. Audiences should expect it. Platforms that refuse to engage with it should be treated, in the policy and procurement environment, as the laggards they are.

\textbf{Three. Access.} The benefits of AI tooling~-- the productivity, the creative leverage, the cost reduction~-- should be broadly available, not concentrated. This means free or near-free tools for creative entry; investment in literacy at the population scale; deliberate inclusion of historically excluded creative communities in the design, training and deployment of AI systems; and structural support for the indie, the regional, the Global South, the neurodivergent and the under-resourced creative ecosystems that the AI cost reduction has the potential to \emph{include} in the global creative economy for the first time.

The Korin AI launch, the rise of Indian AI cinema, the African-trained model investments~-- these are early signs that the access principle can be operationalised.

\textbf{Four. Audience.} The audience for creative work is not a passive market. The audience is~-- has always been, and increasingly is~-- a \emph{participant} in the cultural meaning of the work. AI policy, platform design and creative practice should treat the audience as such: with the right to know what they are encountering, the right to choose what they pay attention to, the right to refuse work that violates their cultural or ethical preferences, and the right to a creative economy that produces \emph{for} them rather than \emph{at} them.

The slop ceiling is the audience exercising this principle. The cultural rejection of cynical AI work. The death threats and the BBC investigations and the \emph{Tiny Grandma} virality. The audience has been telling us what it wants. The institutions and the platforms are, slowly, beginning to listen.

If you can hold these four principles~-- \emph{agency, attribution, access, audience}~-- in your head when you sit down to make a creative decision in 2026, you have a working test for whether you are building towards the generative economy or away from it. Apply the test. Often. Without too much agonising. The aggregate of millions of decisions made against this test, over the next eighteen months, is the choice we are collectively making about what comes next.

The four principles are policy and platform tests. They describe what the rules of the new economy should be. They do not, on their own, describe what working creatives should be \emph{doing} with their hands, every day, to put themselves on the inside rather than the outside of the rule-writing. I made the case for the practitioner's version of this argument at length in Chapter~\ref{ch:3}'s \emph{Open the black box} section, and I want to put one summary sentence of it here, because the four principles cannot be defended by creators who do not understand the toolchain underneath them.

\textbf{Working creatives need to open the black box of AI and own a real stake in how it is built.} Not just use it. Not just refuse it. Not just bargain over its terms. \emph{Open it.} Understand what the model was trained on. Read the EULAs of the platforms you ship work through. Run some part of your stack on open-weight infrastructure (Chapter~\ref{ch:16} is the practical map). Show up to the governance conversations~-- the Cannes Disclosure Standard rooms, the UK consultation responses, the SAG-AFTRA bargaining tables, the C2PA standards body. The cohort of working creatives that does this defines the next era's craft. The cohort that uses the toolchain without ever asking what is inside it has the era's craft defined for them, by the platform companies that built the toolchain. The four principles assume~-- they cannot deliver on their own~-- that the first cohort is large enough, organised enough and technically literate enough to do the work. The argument of this chapter, on its widest reading, is that turning up to the rule-writing is the practitioner's work.

\section*{The cost we are not pricing}

I want to break my own framing for a moment, because there is a category of cost the four principles only obliquely address, and any reader who has been paying attention to the wider technology press in the period this book covers will be wondering when I am going to put it on the page.

The cost is \emph{the resource footprint of the systems we are building everything on top of.}

Training and running large generative models, at the scale the creative industries are now using them, is an industrial activity. It is consuming enormous quantities of electricity, water, semiconductor manufacturing capacity, and rare-earth-mineral throughput, in a global infrastructure-build the planet has not seen since the rollout of mobile telephony. The trade-press coverage on this is patchy and the data the platform companies disclose is partial, but the direction is unambiguous. The data centres that produce the \emph{Sora~2} clip you generated for a client this morning, the \emph{Veo~3.1} sequence you cut into your edit, the \emph{Suno} track you used as scratch on a project~-- those data centres draw power on a scale that is, in aggregate, a meaningful new line item in the global energy mix.

There is also a second category of cost, equally unpriced, that the four principles touch only by implication. The training, refinement and moderation of these systems involves an extensive workforce of human data labellers, content moderators and reinforcement-learning evaluators, much of it concentrated in low-wage labour markets in the Global South, much of it carrying significant psychological cost from exposure to the worst of what the models filter out. The creative economy of 2026 is, in part, sitting on top of that labour. The labour is not always visible in the marketing copy of the platform companies that depend on it.

I have under-treated both of these costs in the chapters above, on the grounds that the book is specifically about the creative-industries transition rather than about AI's wider externalities. That is a defensible editorial choice. It is also an incomplete one, and a fair reader is right to ask why a manifesto for a humane creative economy stops at the edge of the studio.

The honest answer is this. The four principles, as I have stated them, are \emph{necessary but not sufficient.} A creative economy that gets agency, attribution, access and audience right, but that does so on top of a platform stack whose energy and labour externalities are concentrated, opaque, and morally indefensible, is not a generative economy. It is an extractive economy with better creative-side ethics~-- which is, in some ways, the more dangerous animal, because the creative-side ethics provide ideological cover for the wider extraction.

The implication for the working creative reading this is \emph{not} that you should refuse to use AI tools because of their environmental and labour costs. The implication is that you should \emph{demand the same transparency} from the platforms about their externalities that you demand about their training data. Energy disclosure, water disclosure, labour conditions in the data-supply chain, carbon accounting of inference runs~-- these should be standard procurement requirements for any creative organisation buying platform access at scale, in exactly the same way C2PA provenance is becoming a standard requirement for the work itself.

If I were rewriting the four principles from scratch, I would probably make this a fifth: \textbf{Footprint.} The full resource cost of the work~-- energy, water, human labour, supply-chain externalities~-- should be visible to the people commissioning, making and consuming it. The principle would sit awkwardly alongside the other four, because it is the only one that does not run along the human-creative-vs-platform axis the book has otherwise been organised on. I have left the four where they are, because the rhetorical compression of \emph{Agency, Attribution, Access, Audience} is one of the more useful things I have written. But the fifth is real. It belongs in the conversation. I want, in this section, to have at least said so on the page.

The deeper point, which I will not dwell on further because it is the subject of a different book by a different writer who I hope is writing it now, is that the AI transition is~-- like every previous industrial transition~-- being paid for somewhere. The creative-industries part of the bill is, in 2026, becoming visible. The energy and labour parts of the bill are still, by deliberate platform-company design, harder to see. A humane creative economy will have to insist on seeing both.

\section*{The DreamLab model}

I want to talk briefly about the \textbf{DreamLab} model~-- not because I think the studio I run in the North West of England is some kind of ideal, but because it is one specific, concrete instance of how a working creative organisation has tried, deliberately, to build for the generative economy rather than against it, and because I think the choices we have made are useful to share.

DreamLab is a collective of about fifty practitioners~-- artists, technologists, directors, games developers, storytellers~-- based in the North West, with collaborators across the U.K.\ and internationally.\footnote{DreamLab AI Collective, team page: \url{https://dreamlab-ai.com/team}.} We are emphatically \emph{not} an AI company. We are a \emph{creative} company that happens to use AI heavily.

The choices we have made, since we started thinking deliberately about how to operate in this environment in early 2024, are not heroic. They are pragmatic. They include:

\textbf{A disclosure-first practice.} We tell clients, in the brief, what AI tools we propose to use and what for. We document our use. We can~-- and have, when asked~-- provide chain-of-custody on contested work.

\textbf{A continuum-first contract.} Our client contracts identify, function by function, where on the Human--AI Agency Continuum the work will sit. Some functions are 100\% human (performance, lead creative direction, story). Some are mixed. Some are mostly AI (asset variations, plate generation, certain post-production tasks). The client knows. The team knows. The line is negotiable but it is drawn.

\textbf{A literacy-first hiring approach.} We hire~-- and we invest in upskilling~-- generalists rather than specialists. We expect every working member of the collective to understand the AI tools in \emph{adjacent} disciplines well enough to brief them and judge their outputs. The portfolio creative model from Chapter~\ref{ch:11} is the operating practice.

\textbf{An open-toolchain commitment.} We use a deliberately diverse toolchain~-- Adobe, Runway, ComfyUI, World Labs, Hunyuan, open-source models on Hugging Face, internally built tooling. We refuse exclusive dependency on any single platform. If one of the tools shifts in a direction that violates the four principles, we have alternatives.

\textbf{A community-first relationship to the city.} The collective is intentionally based in the North West of England, not in London or Los Angeles. The aim, explicitly, is to build creative infrastructure in places that the previous generation of the creative economy did not invest in. We hire locally. We train locally. We work, where we can, with regional partners.

\textbf{A newsletter, an archive, a public record.} \emph{Dream Machine,} the publication this book is built out of, exists in significant part because \emph{putting the work in public}~-- keeping a transparent, citable record of what is happening, what is being said, what is being decided~-- is itself a form of building the generative economy. Hidden knowledge concentrates power. Shared knowledge distributes it.

None of these choices are unique to DreamLab. Many of them are being made, in different forms, by studios, agencies, labels and indie production companies all over the world. What I would say, having lived inside this set of choices for the period this book covers, is that they \emph{work.} The studio's output has been better, our team has been happier, our clients have been more loyal, and our market position has been more defensible than I had any right to expect when I started writing the newsletter in October.

The generative economy, as a practice, is not abstract. It is a set of operational choices that working organisations can make on a Monday morning. The choice produces real, measurable outcomes. It is not, as some of the more pessimistic AI commentary frames it, a luxury for organisations that can afford to be high-minded. It is, in our experience, the most \emph{commercially} sustainable position available right now.

\section*{What working creatives should do on Monday morning}

This book is mostly aimed at working creatives. The studios and the institutions and the platforms have been the subject of the chapters, but the audience for the chapters has been you~-- the writer, the director, the songwriter, the games designer, the photographer, the editor, the producer, the agency creative, the indie filmmaker, the YouTuber, the freelance illustrator, the student looking at a creative career and trying to figure out whether to be discouraged or excited.

What should you do on Monday morning?

I will keep this brief, because the rest of the book has been the long version.

\textbf{Read your own newsletter.} Whatever your version of \emph{Dream Machine} is~-- the publication you trust to tell you what is happening in your discipline week by week~-- read it carefully. The pace of change in 2026 is too fast to absorb by osmosis. You need a deliberate reading practice. If you can't find one you trust, build one.

\textbf{Draw your lines.} Write down, for your own practice, where on the Human--AI Agency Continuum you want to sit, function by function. Update the lines as the technology and your judgement evolve. Be prepared to \emph{show} the lines~-- to clients, to collaborators, to yourself.

\textbf{Practise briefing.} It is the most leveraged skill of the era. Brief the agents. Brief your collaborators. Brief yourself. Get clearer, every week, about what you actually want from a piece of work \emph{before} you start making it.

\textbf{Cultivate taste, on purpose.} Look at more good work. Look at it harder. Articulate, every week, what makes something good~-- and what makes the average thing average. The agents will, by default, give you average. Your job is to know better.

\textbf{Disclose.} Tell people what you are using, how you are using it, and where the human work in your output lives. The transparency is, in 2026, not a cost. It is a competitive advantage.

\textbf{Stay in the work.} Resist the temptation to abstract too far. The maker who never makes is the maker whose judgement decays. Maintain craft contact, in at least one part of your practice, that does not depend on AI tooling. The contact will keep your eye sharp for everything else.

\textbf{Find your coalition.} The 88\% in the U.K.\ consultation didn't get there by accident. It got there because creators turned up alongside other creators alongside professional bodies alongside unions alongside institutions. The political and economic shape of the next decade of creative work depends on creators being \emph{organised.} Join your union. Sign the declarations. Show up to the consultations. The institutions of collective bargaining and political representation that your forerunners built are the institutions that will defend your work in 2030.

\textbf{Build for the new geography.} Wherever you are in the world, the AI cost reduction has made it more viable than ever to build a serious creative practice in places that were previously locked out of the centre. Take the opportunity. The next century of cultural production is not going to belong to the cities that owned the previous one. The next generation of canonical creative work will, on the trajectory I see, come from more places, in more languages, in more forms, than ever before.

\textbf{Make the work.} None of the rest of this matters if you do not make the work. The AI era is not a reason to stop. It is a different set of conditions under which to keep going. Working creatives have always operated against the technological grain of their day. The maker in 2026~-- like the maker in 1926 and 1826 and 1626~-- is the person who finds, in the conditions of their actual moment, the work that needs to be made and then makes it.

The conditions have changed. The need has not.

\section*{Predictions you can check me on}

Books on the cusp of a transition tend to age in one of two ways. Either they were broadly right about the shape of what was coming and look prescient five years on, or they were embarrassingly wrong and quietly disappear from the recommended-reading lists. The honest thing to do, at the end of a book like this one, is to write down predictions specific enough that the future can grade them.

Here are mine, dated May 2026.

\textbf{By the end of 2027.} Every major studio, label and agency has, in its standard production contract, a clause requiring AI-use disclosure across the production pipeline. \emph{Position Four}~-- \emph{AI in the workflow, not in the work}~-- is the dominant operational posture in legacy creative organisations. The number of \emph{Position Three} refusing studios has stabilised at roughly 10--15\% of the mid-and-upper market, concentrated in IP-led franchises where the audience contract depends on a human-craft signal.

\textbf{By the end of 2027.} The 44\%/3\% Deezer ratio (Chapter~\ref{ch:5}) has stabilised on the upload side at 50--65\% AI by volume, with consumer streams of fully-AI tracks still under 5\% of total play time. \emph{Audience-disclosed} synthetic music is a small but real category~-- under 10\% of paid streams~-- and is concentrated in mood/background and synth-driven genres rather than in artist-led popular music.

\textbf{By the end of 2028.} A \emph{Position Two} studio~-- Imaginae, Wonder, Obsidian, Asteria or a name we have not heard yet~-- produces the first AI-native feature that wins material awards recognition \emph{for its writing or direction} (not for its technology). The audience response is mixed; the cultural permission for AI-native cinema has shifted from contested to accepted-with-asterisks. James Cameron's ``horrifying'' line is still being quoted but the position it describes has narrowed to \emph{AI that displaces a specific actor in a specific scene}, not the general toolchain.

\textbf{By the end of 2028.} At least one large national government~-- most likely the U.K.\ or one of a few EU member states~-- has passed legislation requiring licensing of copyrighted work for AI training. The U.S.\ has not, but the litigation environment (UMG v Anthropic and the cases that follow it) has produced de facto licensing markets that platform companies have grudgingly entered. The technical infrastructure for \emph{creative weight attribution} is mature enough to underpin real revenue distribution to working creators.

\textbf{By the end of 2029.} World models, not flat video, are the dominant medium for new high-production-value creative work. Marble-class tools are commodity infrastructure; the differentiated work is being done by orchestrators with strong \emph{world curation} skills (Chapter~\ref{ch:8}). The boundary between film, games and immersive media is functionally gone for new productions, while legacy formats retain prestige and a real audience.

\textbf{By the end of 2029.} The audience contract has been substantively re-written. C2PA-class provenance metadata is supported in the major capture, edit and distribution tools by default. Platforms that do not honour provenance have lost meaningful share to ones that do. The \emph{Tiny Grandma} error mode~-- human work mis-flagged as synthetic~-- is a solved problem in the principal platforms; the inverse error~-- synthetic work mis-flagged as human~-- is the harder one and remains the central audience-trust question.

\textbf{By the end of 2030.} The orchestrator role (Chapter~\ref{ch:11}) is a named seniority track in every creative discipline that survives the transition. \emph{Senior orchestrator} is a credentialled title. The apprenticeship problem (Chapter~\ref{ch:11}, Chapter~\ref{ch:13}) is partially solved by a combination of three things: AI-augmented junior roles maintained deliberately by \emph{Position Four} studios; new pathways through literacy initiatives like Sundance Collab and its successors; and an explicit re-funding of cultural-institution training programmes by national governments.

\textbf{By the end of 2030, but I'm less confident about this one.} A major creative-industry union has negotiated a \emph{productivity dividend}~-- a structural share of AI-driven productivity gains that flows to the human workforce, not just to the platform companies and the studio shareholders. The mechanism is novel and contested but the underlying maths is unanswerable; the industry that does this first becomes the template for the others.

\textbf{I am wrong about something on this list.} I do not know yet which item. If you are reading this in 2030 and one of these predictions has aged badly, send the receipts to the \emph{Dream Machine} newsletter address. I will publish them.

The predictions I have \emph{not} made~-- about which AI company will dominate by 2030, about which specific platform companies will go bust, about whether AGI arrives in the period and what it does to all of this~-- are absences I have made on purpose. Anyone confident about those questions in 2026 is selling something. The shape of the \emph{creative economy} is more predictable than the shape of \emph{the platform layer underneath it.} That is the bet this book has been built on.

\section*{A note to those who are afraid}

I want to write directly, for a paragraph or two, to the readers I know are sitting with this book~-- and with the larger transition this book describes~-- in a state of real, well-founded fear.

If you are reading this and your livelihood depends on a creative discipline that the AI tools are getting unsettlingly good at~-- if you are a junior animator, a session musician, a copywriter, an illustrator, a translator, a stock-image photographer, a voiceover artist, a foley artist, a concept designer, a stage actor in a regional theatre, a working musician on tour, an indie filmmaker, an entry-level games artist, a video editor at a small post-production house~-- and the news cycle has, over and over again, told you that your job is going away, I want you to know two things.

The first is that the news cycle is, on this question, almost always too negative. The aggregate data on creative-industry employment in 2026 does not show the collapse the headlines have been predicting. It shows a \emph{restructuring,} with a lot of disruption, and with real pain in specific places, but with~-- net~-- more creative work being done by more people in more forms than was being done a year ago. The Sundance literacy initiative is not happening because the industry is dying. It is happening because the industry, freshly, has more entrants than ever and needs to train them.

The second is that the most important strategic move you can make in this moment is not to be afraid in private. It is to \emph{speak.} To your union. To your trade association. To your local government's consultation. To your manager. To your audience. To the people on your team. The 88\% was made out of voices. The Tilly Tax was made out of voices. The Bandcamp ban was made out of voices. The Sundance literacy turn was made out of voices. The institutions of the new creative economy~-- the ones that will protect your work in 2030~-- are being shaped \emph{right now} by the people who turn up to be counted. The people who don't turn up, or who turn up only in private, are the people whose interests get absorbed into the policies of the people who do.

You are not powerless in this. You are, if anything, the most powerful constituency in this whole story, because you are the people who are actually making the work that the audience is paying attention to. You are the constituency the unions represent, the institutions exist to serve, the audience cares about, and the platforms~-- ultimately, despite themselves~-- depend on. The choice between the extractive and the generative economy is, in large part, the choice you collectively make about how to use that power.

Use it.

\section*{The Dream Machine}

I want to close the argument of this book with the phrase the newsletter started with.

\emph{Welcome to the Dream Machine.}

When I named the newsletter, in late September 2025, I had no clear idea what the phrase meant. I had a half-formed sense that ``dream'' caught something about the strange, vivid, slightly hallucinatory quality of what the new tools produced, and that ``machine'' caught something about the industrial, scaled, infrastructural nature of the platforms behind them. The phrase was, more than anything else, a placeholder for a feeling I couldn't quite articulate yet.

Six months on, I think I know what the phrase means.

A \emph{dream machine} is, in the sense I am using it, an apparatus capable of producing something that has~-- until very recently~-- required a human mind to produce. Not just images, or videos, or songs, but \emph{constructions} of meaning, of place, of presence, of emotion. The apparatus is not, in itself, doing the dreaming. It is amplifying, multiplying and distributing the dreaming of the humans who direct it.

The question this book has been about, in every chapter, is \emph{whose dreams the machine amplifies.}

If the machine amplifies the dreams of a small number of platform shareholders, optimised against extraction, the creative economy it produces will be~-- to use Kehlani's line about Xania Monet~-- a place where ``the person is doing none of the work.'' The dreams of the audience are \emph{for sale.} The dreams of the creators are \emph{training data.} The dreams of the culture are \emph{outputs of an algorithm tuned for engagement.}

If, on the other hand, the machine amplifies the dreams of the working creatives~-- supported by the audience, defended by the unions, regulated by the institutions, contained by the four principles, made transparent by the provenance infrastructure~-- the creative economy it produces will be the largest, most diverse, most accessible expansion of human cultural production in the history of the species.

We are, in May 2026, sitting in the moment between these two outcomes.

The signals from the last six months~-- the 88\%, the slop ceiling, the Sundance turn, the Cannes Disclosure Standard, the Academy rule, the SAG-AFTRA contract, the audience behaviour, the creator coalition, the regional opening, the literacy investment~-- all point at the second outcome being achievable. Not inevitable. \emph{Achievable.}

It is going to require the working creatives to keep turning up. It is going to require the studios, agencies and labels to keep making the integration choices we covered in Chapter~\ref{ch:13}. It is going to require the institutions~-- the unions, the rights bodies, the festivals, the universities~-- to keep doing the slow, unglamorous work of building the rails. It is going to require the platforms to be pushed, by their users and by their regulators, towards the side of provenance and attribution and consent. It is going to require the governments to keep the policy moving in the direction the 88\% asked for.

It is going to require the audience to keep paying attention to the human signal.

And it is going to require people like you~-- the people who have read this far, who have been doing the work of figuring out what creative life looks like in this moment~-- to make, in your own practice, the daily decisions that build the generative economy rather than the extractive one.

This is the choice. This is what is on the table. This is the work.

Welcome to the Dream Machine.

  \chapter{The Tools}\label{ch:16}

\lettrine[lines=3,lhang=0.15,findent=0.1em]{I}{ } have, until this chapter, deliberately kept the \emph{tools} out of the foreground. Thirteen chapters about creative AI without a chapter on the toolchain is, on the face of it, a strange editorial decision, and I want to begin by explaining it.

The reason is that I think the most common mistake people make about this period is to confuse \emph{the tools} with \emph{the transition}. Tools are the visible surface of the change~-- the thing the press cycle covers, the thing the platform companies want you to talk about, the thing that has a price and a logo and a launch date you can put on a slide. The transition is everything underneath: the economics, the labour, the audience contract, the law, the institutions, the rails. The tools change weekly. The transition is slower, deeper, and is what will still be true in 2030 when most of the tools in this chapter are obsolete.

The first sixteen chapters were about the transition. This chapter is about the tools.

I have written it last on purpose. Read in this order, the tools sit inside the architecture the book has been building~-- the Continuum, the Slop Ceiling, the four positions, the orchestrator role, the four principles. Read in any other order, they collapse back into the format the platform companies prefer: a tools-arms-race in which the only question is which model is best this week.

That format is, in 2026, the most reliable way to misunderstand what is happening.

A note on the date stamp. Everything in this chapter is current to May 2026. By the time you read it, some of these tools will have been bought, renamed, killed, surpassed or repositioned. The point is not that the specific tools matter. The point is the \emph{shape} of the toolchain~-- what categories exist, what they do, who builds them, and how a working creative builds a coherent stack on top of them. The shape, in my experience, holds.

\section*{How to think about the toolchain}

\begin{figure}[htbp]
  \centering
  \begin{tikzpicture}[
    font=\small,
    every node/.style={align=center}
  ]


    \fill[darkblue, rounded corners=3pt]
      (-5.5, 0.0) rectangle (5.5, 1.0);
    \node[text=white, font=\small\bfseries] at (0, 0.62)
      {Foundation Models};
    \node[text=white!80!darkblue, font=\scriptsize] at (0, 0.25)
      {GPT, Claude, Gemini, Llama, Stable Diffusion};

    \fill[darkgreen, rounded corners=3pt]
      (-4.6, 1.15) rectangle (4.6, 2.15);
    \node[text=white, font=\small\bfseries] at (0, 1.77)
      {Creative Middleware};
    \node[text=white!80!darkgreen, font=\scriptsize] at (0, 1.40)
      {ComfyUI, Runway, ElevenLabs, Suno};

    \fill[rulegold, rounded corners=3pt]
      (-3.7, 2.30) rectangle (3.7, 3.30);
    \node[text=white, font=\small\bfseries] at (0, 2.92)
      {Application Layer};
    \node[text=white!80!rulegold, font=\scriptsize] at (0, 2.55)
      {Adobe Creative Suite, DaVinci, Pro Tools};

    \fill[chaptergrey, rounded corners=3pt]
      (-2.8, 3.45) rectangle (2.8, 4.45);
    \node[text=white, font=\small\bfseries] at (0, 4.07)
      {Orchestration};
    \node[text=white!80!chaptergrey, font=\scriptsize] at (0, 3.70)
      {Multi-agent systems, Claude Code, MCP};

    \fill[white, rounded corners=3pt]
      (-1.9, 4.60) rectangle (1.9, 5.60);
    \draw[chaptergrey, thick, rounded corners=3pt]
      (-1.9, 4.60) rectangle (1.9, 5.60);
    \node[text=chaptergrey, font=\small\bfseries] at (0, 5.22)
      {Human Creative};
    \node[text=chaptergrey!80, font=\scriptsize] at (0, 4.85)
      {Intent, Taste, Judgement};

    \draw[chaptergrey!60, <->, thin] (5.7, 0.0) -- (5.7, 1.0)
      node[midway, right, font=\scriptsize, text=chaptergrey] {};
    \node[chaptergrey!60, font=\scriptsize, anchor=west] at (5.8, 5.1)
      {Narrowing\\scope};
    \draw[chaptergrey!40, ->, thin] (5.8, 4.9) -- (5.8, 0.5);

    \node[chaptergrey, font=\scriptsize\itshape] at (0, -0.35)
      {$\leftarrow$ Wider ecosystem reach \hspace{2cm} Narrower human creative domain $\rightarrow$};

  \end{tikzpicture}
  \caption{The creative AI toolchain: five layers from foundation models to human creative intent.}
  \label{fig:toolchain-layers}
\end{figure}
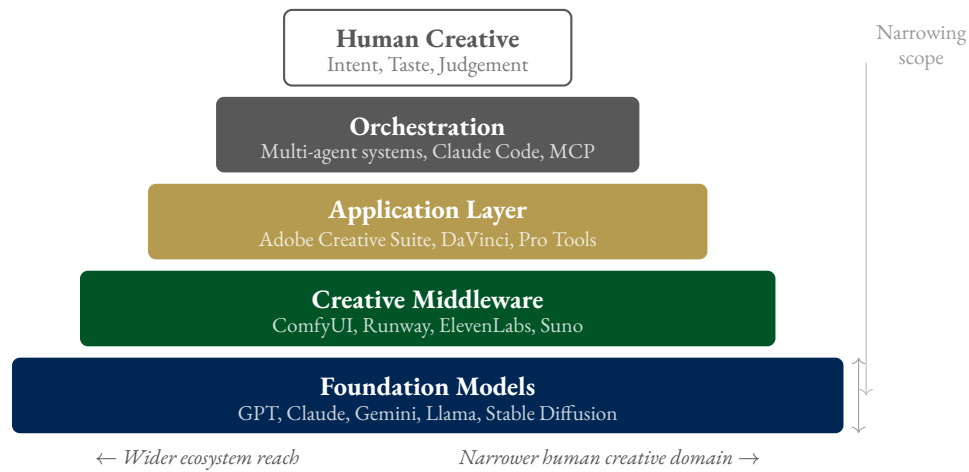

Before the inventory, the frame.

I think the creative-AI toolchain in 2026 is best understood as a stack of seven layers, each with its own dominant players, its own pace of change, its own integration model. The layers, from foundation to consumer, are:

\begin{enumerate}
  \item \textbf{Foundation models}~-- the large multimodal systems underneath everything else (OpenAI's GPT-class, Anthropic's Claude, Google's Gemini, Meta's Llama, the major Chinese open-source models). These are the raw capability layer. Almost no working creative uses them directly except via wrappers.
  \item \textbf{Modality models}~-- specialist models for video (Sora, Veo, Runway Gen-4.5, Kling, Hunyuan, Wan), image (Firefly, Midjourney, FLUX, Imagen, SDXL/Stable Diffusion variants), audio (Suno, Udio, ElevenLabs, Mureka), 3D and world (Marble, Genie~3, WorldGen, UNI-1, Hunyuan 3D-PolyGen, ECHO). These are what most working creatives think of when they say ``AI tools.''
  \item \textbf{Agent platforms}~-- systems that compose modality models and external tools into multi-step workflows (OpenAI's AgentKit, Anthropic's Claude apps and skills, Google's Project Genie, Heygen's Video Agent, Sony's 49-agent / 72-skill stack). The agent layer is where the ``orchestrator economy'' of Chapter~\ref{ch:11} actually runs.
  \item \textbf{Creative software with AI baked in}~-- the legacy creative suites that have been rebuilt as AI-first platforms (Adobe Creative Cloud, Autodesk, Foundry, Unreal Engine, Unity, DaVinci Resolve, Pro Tools, Logic Pro, Ableton). This is where most paid professional work still happens.
  \item \textbf{AI-native creative apps}~-- new entrants whose product is a single-purpose AI workflow (Runway, Higgsfield, Krea, Freepik, Magnific, Synthesia, Heygen, Hedra, Cascadeur, Pika, Luma). Most working creatives use 4 to 10 of these in rotation.
  \item \textbf{Open-source and workflow infrastructure}~-- the technical-creator layer that wires everything together (ComfyUI, Hugging Face, SuperSplat, OpenEnv, the open-source model ecosystem). This is where the most interesting innovation often happens first.
  \item \textbf{Consumer surfaces}~-- the apps that put generative capability on every phone (the Sora app, CapCut/Dreamina with Seedance, the Gemini app, the various TikTok-style remix platforms). This is the layer the audience touches.
\end{enumerate}

The mistake I see most often, both in the press cycle and in studios planning their internal AI roadmaps, is to optimise for layer~2 (modality models) without understanding that the actual leverage is in how you compose layers~2,~3,~4 and~6 into a coherent workflow. The tool that ``wins'' is rarely the tool with the best benchmark. It is the tool that integrates cleanly into the rest of your stack.

With that frame, the inventory.

\section*{Video}

The video layer changed faster than any other modality between October 2025 and May 2026, and is the one most likely to look different again by the time you read this. Treat the names as snapshots, not as a stable league table.

\textbf{Sora 2} (OpenAI) is the model that opened the period this book covers. Its September 2025 launch~-- physical realism, audio integration, multi-shot world-state persistence~-- is the moment Chapter~\ref{ch:1} is about.\footnote{OpenAI, ``Sora~2 is here,'' 30 September 2025, \url{https://openai.com/index/sora-2/}. \emph{Dream Machine} Issue~1.} The iOS app launched alongside it hit a million downloads in five days\footnote{LinkedIn News aggregation: ``Sora Tops 1 Million Downloads in 5 Days,'' \url{https://www.linkedin.com/news/story/sora-tops-1m-downloads-in-5-days-6684988/}. \emph{Dream Machine} Issue~3.} and is the consumer-facing edge of the AI video market. For professional production, Sora~2 is impressive on isolated single-clip generation and remains the model most cited in the mainstream press, but most working filmmakers I know use it less than its cultural prominence would suggest.

\textbf{Veo 3.1} (Google DeepMind), released in mid-October 2025, is the model the professional filmmaking community has, on average, gravitated toward~-- for narrative coherence, controllable camera composition, cinematic lighting vocabulary and sound integration.\footnote{Google DeepMind, Veo~3.1 launch, mid-October 2025. \emph{Dream Machine} Issue~3.} Sora~2 wins on raw physics in single clips; Veo~3.1 wins on the kind of sustained directorial control most actual production pipelines need.

\textbf{Runway Gen-4.5} (and Gen-4.5 Image-to-Video, the Workflows product, Story Panels, Characters API, Apps for Advertising) is the most-integrated commercial AI-video stack of the period.\footnote{Runway product cycle: Gen-4.5 (December 2025), Gen-4.5 Image-to-Video (January 2026), Workflows, Story Panels, Characters API, Apps for Advertising~-- \emph{Dream Machine} Issues~10,~14,~15,~16,~20.} Runway has shipped product faster than any other AI video company in this market, and CEO Crist\'obal Valenzuela's ``fifty indie films instead of one \$100M blockbuster'' framing is the cleanest articulation I have seen of the case for AI as creative leverage rather than cost-cut.\footnote{Runway CEO on indie films vs.\ blockbusters, \emph{Dream Machine} Issue~26.}

\textbf{Kling} (Kuaishou), \textbf{Hunyuan Video} (Tencent), \textbf{Wan 2.5} (Alibaba), \textbf{Seedance 2.0} (ByteDance)~-- the Chinese-built models that, in aggregate, have rivalled or surpassed the U.S.\ labs on specific capabilities (motion physics, character consistency, render speed) at significantly lower cost.\footnote{Chinese open-source AI video model releases, 2025--2026. \emph{Dream Machine} Issues~3,~12,~22.} Hunyuan's open-source releases have been the single most important contribution to the wider open-source AI video ecosystem in this period.

\textbf{Pika 2.0}, \textbf{Higgsfield}, \textbf{Luma} (Dream Machine and UNI-1) round out the commercial layer. Each has carved a niche: Pika on iteration speed and creator workflow; Higgsfield on social-media marketing video at scale (\$80M raised, \$1.3B valuation, \$200M revenue in nine months\footnote{\emph{SiliconAngle}, ``Higgsfield raises \$80M on \$1.3B valuation,'' \url{https://siliconangle.com/2026/01/15/higgsfield-raises-80m-1-3b-valuation-scale-ai-video-platform/}. \emph{36kr}, ``Higgsfield: Earns \$200M in 9 Months,'' \url{https://eu.36kr.com/en/p/3650517574312323}. \emph{Dream Machine} Issues~15,~16.}); Luma on the world-model bridge to spatial content.

\textbf{Heygen} ships Video Agent~-- a full scripting-to-assembly agent built around reference images.\footnote{Heygen Video Agent, \url{https://www.linkedin.com/posts/heygen_introducing-the-new-video-agent-activity-7421597801240801282-d1CF}. \emph{Dream Machine} Issue~16.} \textbf{Synthesia} holds the corporate AI-avatar market (\$4B valuation, having reportedly rejected a \$3B Adobe acquisition offer).\footnote{TechCrunch, ``Synthesia hits \$4B valuation,'' \url{https://techcrunch.com/2026/01/26/synthesia-hits-4b-valuation-lets-employees-cash-in/}. \emph{Sifted}, ``Synthesia rejects \$3bn Adobe acquisition offer,'' \url{https://sifted.eu/articles/synthesia-acquisition-offer}. \emph{Dream Machine} Issues~5,~16.} \textbf{ElevenLabs} runs the dominant audio-AI layer underneath much of the new video work (\$500m ARR by April 2026).\footnote{ElevenLabs Series funding, April 2026. \emph{Dream Machine} Issue~25.}

\textbf{Gemini Omni} (Google DeepMind), announced at Google I/O 2026, brings text, image, audio, video and live interaction into a single multimodal model~-- the first foundation-model release in this category that meaningfully unifies the modalities working creatives currently have to bridge across five different tools.\footnote{Google DeepMind, ``Introducing Gemini Omni: Create Anything from Any input,'' \url{https://blog.google/technology/google-deepmind/gemini-omni-launch/}. \emph{Dream Machine} Issue~30.} \textbf{Beeple Canvas}, Mike Winkelmann's gen-AI compositor~-- launched May 2026~-- is the first AI-native compositing application to ship from a working visual-effects artist's own studio, and is structurally distinct from the AI-features-bolted-onto-existing-compositors pattern in the legacy-software section below.\footnote{Beeple Canvas~-- Generative AI compositor, \url{https://www.beeple-canvas.com/}. \emph{Dream Machine} Issue~30.}

If I had to name a single video product that, in my experience, working creatives have settled on as a default in 2026, it would be Veo~3.1 for finished work and Runway for iteration and integration. Sora is the brand name the audience knows. The actual production pipelines run on the other two.

\section*{Image}

The image layer is more stable than video~-- the technology has matured, the differences between top models are narrower, and the dominant question has moved from ``which model'' to ``which workflow.''

\textbf{Adobe Firefly} (Image Model~5, plus Foundry for custom-trained corporate models, plus integration across Photoshop / Illustrator / Express / InDesign) is the default for any working creative who is also a Creative Cloud subscriber~-- which is, by Adobe's own numbers, 45\% of Creative Cloud users actively using Firefly, 70\% of those weekly, more than 22 billion assets generated by April 2025.\footnote{Adobe Firefly milestone data, in \emph{Dynamics of Generative AI Adoption in the Creative Industries}, \S``The Ubiquity of AI in Visual and Digital Arts.''} The Firefly adoption curve is the single best evidence I have for the consumption-gap argument in Chapter~\ref{ch:13}.

\textbf{Midjourney} remains the aesthetic-leadership product in the category. Slower to ship, more opinionated about output style, dominant on Discord and X among the working AI-art community.

\textbf{FLUX} (Black Forest Labs) is the open-source and pro-creator favourite for fine control, having largely replaced Stable Diffusion XL as the open-weight default through 2025.

\textbf{Google Imagen} (and the \emph{Nano Banana} fast-image variant integrated into Gemini, Photoshop and Unreal Engine via the various plugins) has become the most-integrated image model in the consumer toolchain, by virtue of Google's distribution. Nano Banana inside Photoshop and Nano Banana inside Unreal Engine were two of the more consequential cross-platform integrations of the period.\footnote{Nano Banana inside Photoshop and inside Unreal Engine cross-integrations, October--November 2025. \emph{Dream Machine} Issue~1.}

\textbf{Krea}, \textbf{Freepik}, \textbf{Magnific}, \textbf{Recraft}~-- the higher-control consumer / pro-creator products built on top of foundation image models. Each is competing on specific workflow advantages (real-time generation, upscaling, vector output, brand-consistency control).

The image workflow most commonly cited in my circle in mid-2026 is: a base generation in Firefly / Midjourney / FLUX, character-consistent variation in a controllable wrapper like Krea or Magnific, finishing inside Photoshop with the AI-assisted masking, generative-fill and object-removal tools that Adobe shipped through the autumn 2025 and spring 2026 update cycle.

\section*{Music and audio}

The music layer split into three categories during this period, and the split is, in my experience, more important than the specific products in each category.

\textbf{Generative music tools} that produce finished tracks from prompts~-- \textbf{Suno} (Studio launched late 2025\footnote{Suno Studio launch, \url{https://www.techradar.com/ai-platforms-assistants/i-tried-suno-studio-the-new-platform-that-mixes-ai-music-generation-with-hands-on-editing-like-garageband-but-smarter}. \emph{Dream Machine} Issue~1.}), \textbf{Udio}, \textbf{Mureka} (with its Music Agent Studio, six specialised AI agents for songwriting, arrangement and production\footnote{Mureka, ``Music Agent Studio'' launch. \emph{Dream Machine} Issue~4.}). These are the tools that produce most of the AI-music flood Chapter~\ref{ch:5} describes. They are also, paradoxically, the tools most working musicians use the least directly~-- the consumer market for AI-generated finished tracks is large and growing, but professional musicians overwhelmingly use AI tools at a different layer.

\textbf{Production and post-production AI}~-- the tools that handle audio restoration, mixing, mastering and isolation. The 1,100-creator music survey discussed in Appendix~D found that 58\% of professional producers used AI for audio restoration, 38\% for mixing assistance, 33.9\% for automated mastering. \textbf{iZotope Ozone 12}, \textbf{LANDR}, the Pro Tools and Logic Pro AI suites, \textbf{CleanvoiceAI} for podcast post~-- this is where the silent-adoption majority of the music industry lives.

\textbf{Voice and audio synthesis}~-- \textbf{ElevenLabs} is the dominant player, with \$500m ARR, BlackRock / NVIDIA backing, and meaningful share across audiobook narration, dubbing, podcast synthesis and AI character voice work.\footnote{ElevenLabs Series funding, April 2026. \emph{Dream Machine} Issue~25.} The Cardiff band that found their music had been used to train an ``AI artist'' outperforming them on Spotify\footnote{\emph{MusicTech}, ``Cardiff band speaks out after AI artist trained on their music outperforms them on Spotify,'' \url{https://musictech.com/news/industry/its-shocking-disheartening-and-insulting-cardiff-band-speaks-out-after-ai-artist-trained-on-their-music-outperforms-them-on-spotify/}. \emph{Dream Machine} Issue~1.} is one of the cautionary tales of this layer; the Andrii Daniels bomb-shelter clip\footnote{\emph{Variety}, ``AI Creator Behind Viral `Deadpool,' `Harry Potter' Christmas Clip Made His Film in a Ukrainian Bomb Shelter,'' \url{https://variety.com/2026/digital/news/ai-video-deadpool-harry-potter-andrii-daniels-1236624632/}. \emph{Dream Machine} Issue~16.} is one of the success cases.

\textbf{Sound-effect foundation models} emerged as a new sub-category in May 2026. \textbf{Sony AI's Woosh} is the first foundation model explicitly trained for the professional sound-effects discipline.\footnote{Sony AI, ``Woosh~-- a sound effect foundation model,'' \url{https://ai.sony/blog/woosh-sound-effect-foundation-model/}. \emph{Dream Machine} Issue~30.} \textbf{Mirelo SFX~1.6} shipped the first sound-effects model that lets you \emph{edit} a generated sound rather than only regenerate it~-- a structural shift in the discipline equivalent to the move from rendered images to layered Photoshop files.\footnote{Mirelo SFX~1.6, ``edit sound, not just generate it,'' \url{https://mirelo.ai/sfx-1-6}. \emph{Dream Machine} Issue~30.} \textbf{Stable Audio 3.0} (Stability AI) shipped as an open-weight audio model family explicitly aimed at artistic experimentation.\footnote{Stability AI, ``Stable Audio 3.0 released,'' \url{https://stability.ai/news/stable-audio-3-0-released}. \emph{Dream Machine} Issue~30.} \textbf{Tamber}, the ethically-trained AI music suite described in Chapter~\ref{ch:6}, shipped alongside a gestural-control interface that lets the musician steer the generation with arm movements.\footnote{Tamber product page: \url{https://tamber.ai/}. \emph{Dream Machine} Issue~30.} \textbf{Beatport's Track ID} rolled out as the real-time identification standard for the DJ market.\footnote{Beatport Track ID: \url{https://www.beatport.com/track-id}. \emph{Dream Machine} Issue~30.}

The deal flow underneath this layer is the second-fastest-changing in the toolchain. The Stability AI / Universal Music alliance, the Stability AI / Warner Music deal, the Splice / Universal partnership, the GEMA / OpenAI lawsuit, the Wixen / Meta lawsuit, the UMG / Anthropic \$3B suit~-- these are the structural moves I would track if I were a working musician trying to plan a five-year toolchain.\footnote{Music industry AI deal flow, October 2025 -- May 2026. See Chapter~\ref{ch:5} footnotes, and \emph{Dream Machine} Issues~5,~7,~8,~12,~14,~16,~17.}

\section*{3D, world models, spatial}

The category that, more than any other, I think defines the next decade of creative work. Chapter~\ref{ch:8} is the long-form argument; this section is the inventory.

\textbf{Marble} (World Labs, Fei-Fei Li's company) is the first commercial product I would put on a professional toolchain.\footnote{World Labs, ``Bringing Marble to Life,'' \url{https://www.worldlabs.ai/case-studies/bringing-marble-to-life}. \emph{Dream Machine} Issue~7.} Public release November 2025; Sony Pictures' use of it in virtual production reportedly running 40$\times$ faster than the legacy workflow.\footnote{Sony Pictures Marble VP integration. \emph{Dream Machine} Issue~8.} DreamLab has been in the beta since October 2025, and Marble is, today, the world-model product most integrated into a working pipeline I have used.

\textbf{Google DeepMind Genie~3} is the most ambitious research-grade world model, named by \emph{Time} as one of the best inventions of 2025. Made publicly available to Google AI Ultra subscribers via Project Genie in January 2026.\footnote{Google DeepMind, ``Genie~3,'' \url{https://deepmind.google/blog/genie-3-a-new-frontier-for-world-models/}. Project Genie: \url{https://blog.google/innovation-and-ai/models-and-research/google-deepmind/project-genie/}. \emph{Dream Machine} Issues~3,~17.}

\textbf{Meta WorldGen}, \textbf{Tencent HY World~1.5} (open-sourced December 2025, alongside the Hunyuan 3D Studio integration\footnote{Tencent, ``HY World~1.5'' and Hunyuan 3D Studio. \emph{Dream Machine} Issue~12.}), \textbf{SpAItial ECHO}, \textbf{Stanford Wonderzoom}, \textbf{OpenArt Worlds}, \textbf{Luma UNI-1} (the most important \emph{category} announcement of spring 2026, combining world generation with reasoning\footnote{Luma AI, \emph{UNI-1} launch, March 2026. \emph{Dream Machine} Issue~22.})~-- the rest of the world-model commercial layer.

The May 2026 world-model wave extended this layer further. \textbf{NVIDIA SANA-WM} is the first open-weight world model at meaningful scale (2.6B parameters), with 60-second video generation and explicit camera control.\footnote{NVIDIA SANA-WM model collection, \url{https://huggingface.co/collections/nvidia/sana-wm}. \emph{Dream Machine} Issue~30.} \textbf{Odyssey Starchild-1} is, by Odyssey's own framing, \emph{``the first ever real-time multimodal world model''}~-- a system that doesn't just generate a world but simulates and reasons about it.\footnote{Odyssey, ``Introducing Starchild-1, the first real-time multimodal world model,'' \url{https://odyssey.ml/introducing-starchild-1}. \emph{Dream Machine} Issue~30.} \textbf{Odyssey Agora-1} is the multiplayer companion to Starchild, putting four players inside the same AI-generated world (built, in a small piece of provenance theatre, on the bones of a 1997 shooter).\footnote{Odyssey, ``Introducing Agora-1~-- four-player AI-generated world built on a 1997 shooter,'' \url{https://odyssey.ml/introducing-agora-1}. \emph{Dream Machine} Issue~30.} \textbf{Apple Headsup} is a research-grade 3D Gaussian head-reconstruction pipeline built for multi-view captures from consumer iPhones, extending the Vision-Pro-Personas Gaussian-splat thread into the open research layer.\footnote{Apple Machine Learning Research, ``Apple Headsup: a Large-Scale High-Quality 3D Gaussian Head Reconstruction from Multi-View Captures,'' \url{https://machinelearning.apple.com/research/apple-headsup-3d-gaussian-head}. \emph{Dream Machine} Issue~30.}

Underneath this layer, the Gaussian-splatting infrastructure has matured into a stable workflow: \textbf{SuperSplat} (PlayCanvas) for editing, \textbf{Spark 2.0} for open-source streaming of 100-million-splat scenes to browsers, the SOG / WebP equivalent compression standard.\footnote{SuperSplat / Spark~2.0 / SOG releases through 2025--26. \emph{Dream Machine} Issues~1,~25.} Apple's confirmation that its Vision Pro Personas feature is powered by Gaussian splatting under the hood made it, by some margin, the most-deployed Gaussian-splat technology in consumer hardware as of late 2025.\footnote{Radiance Fields, ``Apple Confirms that it's Gaussian Splatting that powers their personas,'' \url{https://radiancefields.com/apple-confirms-personas-use-gaussian-splatting}. \emph{Dream Machine} Issue~5.}

For the 3D-asset and material side: \textbf{Hunyuan 3D-PolyGen~1.5} (Tencent's ``art-grade'' 3D generative model), \textbf{Hitem3D}, \textbf{Meshy}, \textbf{Rodin}~-- the rapidly-maturing 3D-asset generation layer that is being integrated, model-by-model, into Unreal Engine, Unity and Blender pipelines.

Ubisoft's open-sourcing of its \textbf{CHORD} PBR-material model in December 2025,\footnote{ComfyUI Blog, ``Ubisoft La Forge Open-Sources the CHORD Model,'' \url{https://blog.comfy.org/p/ubisoft-open-sources-the-chord-model}. \emph{Dream Machine} Issue~11.} and the Blender Foundation's patronage deal with Anthropic announced in May 2026,\footnote{Anthropic / Blender Foundation patronage, May 2026. \emph{Dream Machine} Issue~27.} are two of the more strategically significant moves in this layer~-- both pushing the production-grade open-source tooling forward at a pace the commercial alternatives have struggled to match.

\section*{Agent platforms and orchestration}

The category I think most working creatives are still underestimating, six months after Chapter~\ref{ch:3} argued it was the inflection point of the era.

\textbf{OpenAI AgentKit} (Agent Builder, ChatKit, connector registry, eval framework) launched October 2025 and is the developer-facing platform underneath most third-party agentic creative tools.\footnote{OpenAI, ``Introducing AgentKit,'' \url{https://openai.com/index/introducing-agentkit/}. \emph{Dream Machine} Issue~2.}

\textbf{Anthropic Claude apps} and the \textbf{Skills framework}~-- the system of named, reusable capabilities that Claude Code uses to coordinate multi-agent workflows. The Sony 49-agent / 72-skill game-development stack is built on this.\footnote{Anthropic Skills framework. \emph{Dream Machine} Issues~11,~16,~29.} In May 2026, \textbf{Google} released its own \textbf{official skills for AI agents}~-- a parallel, cross-vendor skills layer that lets Google-side agents do what Anthropic's Skills framework has been doing for Claude-side ones.\footnote{Google, ``Official skills for AI agents,'' \url{https://github.com/google/agent-skills}. \emph{Dream Machine} Issue~30.} The convergence of two named ``skills'' frameworks across the foundation-model vendors is, in my read, the first sign that the orchestration layer is settling on a shared vocabulary rather than continuing to fragment.

\textbf{Tencent Ardot}, the company's AI-native design-agent platform launched May 2026, is the most ambitious non-Western agent-platform launch of the period~-- an integrated environment in which generative design agents handle layout, asset generation, brand application and iteration as a single coordinated pipeline.\footnote{Tencent Ardot, AI-native design agent platform, \url{https://ardot.tencent.com/}. \emph{Dream Machine} Issue~30.} In the same week, \textbf{Viktor} raised \$75M to embed an agentic \emph{coworker} directly into Slack and Microsoft Teams~-- i.e., the agentic layer landing not as a standalone product but as a colleague-shaped presence in the chat surface the working creative is already in all day, as discussed in Chapter~\ref{ch:9}.

\textbf{Heygen Video Agent} for end-to-end video assembly.\footnote{Heygen Video Agent. \emph{Dream Machine} Issue~16.} \textbf{Adobe CX Enterprise} (announced at Adobe Summit 2026 with NVIDIA) for ``agentic creative intelligence'' across the full content lifecycle.\footnote{Adobe Summit 2026 CX Enterprise. \emph{Dream Machine} Issue~26.} \textbf{NVIDIA + Google Cloud} for the underlying creative-AI infrastructure most enterprise pipelines run on.\footnote{Adobe + NVIDIA / Google + NVIDIA partnerships. \emph{Dream Machine} Issue~21.}

\textbf{ComfyUI}~-- the open-source node-based workflow editor~-- sits underneath much of the technical-creator community's agentic work. ComfyUI raised \$17M in October 2025\footnote{ComfyUI funding round, \url{https://www.linkedin.com/posts/comfyui_we-raised-17-million-to-build-an-os-for-ugcPost-7373743341236236288-wkCc}. \emph{Dream Machine} Issue~1.} and hit a \$500M valuation by May 2026;\footnote{ComfyUI \$500M valuation, May 2026. \emph{Dream Machine} Issue~27.} the platform has become the de facto OS for the open-source side of the creative-AI ecosystem. In May 2026 \textbf{Anthropic's Claude} was added as an official partner node inside ComfyUI, joining the existing Google, OpenAI and open-weight nodes~-- meaning the three frontier foundation models can now be orchestrated side-by-side inside the same open-source pipeline.\footnote{Anthropic, ``Claude is now available as a partner node in ComfyUI,'' \url{https://www.anthropic.com/news/claude-comfyui-partner-node}. \emph{Dream Machine} Issue~30.}

\textbf{Hugging Face}, \textbf{OpenEnv} (Meta / Hugging Face), the \textbf{Hugging Face / Google Cloud} partnership~-- the open-source agentic-development infrastructure.\footnote{Hugging Face / Google Cloud and Meta / Hugging Face OpenEnv. \emph{Dream Machine} Issues~5,~8.}

For working creatives, the practical agent stack in 2026 is some combination of:

\begin{enumerate}
  \item A foundation model (Claude / GPT / Gemini) for the orchestration brain.
  \item A modality model layer (video, image, audio, 3D) doing the actual generation.
  \item A workflow integration layer (ComfyUI for technical work, the in-app agent surfaces for less technical work).
  \item A judgement layer (the human at the desk, doing what Chapter~\ref{ch:11} calls \emph{briefing, allocating, judging and integrating}).
\end{enumerate}

The team I work with at DreamLab runs this stack in production every week. The agents that handle our daily work in May 2026 are, in aggregate, doing the labour of what would, two years ago, have been a team three to four times our size. The human team has not shrunk. We have just become substantially more leveraged.

\section*{Legacy creative software, repositioned}

The most under-reported strategic story of this period, in my view, has been the speed at which the legacy creative-software vendors have rebuilt their products as AI-agent platforms.

\textbf{Adobe}~-- I have written enough about Adobe in Chapter~\ref{ch:9} that I will not repeat it here. The short version: Creative Cloud is, today, a stack of AI agents wearing a Photoshop / Premiere / After Effects / Illustrator / InDesign / Acrobat skin. The agents are inside the apps; the apps are inside ChatGPT; the apps are inside Adobe Express; the apps are inside the new CX Enterprise platform. The repositioning is complete.

\textbf{Unreal Engine} (Epic)~-- the games engine that has, through plugins, integrations and the Nano Banana / Gemini partnership, become a hybrid game-engine / virtual-production / AI-generation hub. The Unreal Engine~5 AI Assistant, announced at the end of 2025,\footnote{Unreal Engine~5 official AI Assistant, \url{https://www.linkedin.com/posts/wouterweynants_theres-an-official-ai-assistant-coming-to-ugcPost-7369377204226379776-pGiH}. \emph{Dream Machine} Issue~1.} is one of the more consequential single-product launches of the period. The \textbf{ECABridge} connector, launched May 2026, is the most-cited Unreal-Engine MCP integration of the spring.\footnote{ECABridge~-- Unreal Engine MCP integration, \url{https://ecabridge.dev/}. \emph{Dream Machine} Issue~30.} In a separate but related move, an \textbf{Epic Games veteran} announced an AI-heavy ``Fully European'' game-engine project in the same week~-- the first plausibly-credible new entrant in the AAA game-engine market since the early 2010s, framed explicitly around AI as the core operating layer.\footnote{\emph{Video Games Chronicle}, ``Epic Games Veteran Claims He's Building AI-Heavy `Fully European' Game Engine,'' \url{https://www.videogameschronicle.com/news/epic-games-veteran-ai-heavy-fully-european-game-engine/}. \emph{Dream Machine} Issue~30.}

\textbf{Unity}~-- Unity's AI Open Beta (May 2026), an in-editor AI suite for the full games-development pipeline, alongside the company's AI Council formation in October 2025.\footnote{Unity AI Council (October 2025); Unity AI Open Beta (May 2026). \emph{Dream Machine} Issues~1,~28.}

\textbf{Autodesk}, \textbf{Foundry}, \textbf{SideFX}~-- the VFX-pipeline vendors integrating generative AI into Maya, Nuke and Houdini at the speed the VFX industry's adoption curve (62\% of Hollywood studios on automated compositing, 35\% reduction in post-production timelines\footnote{VFX AI integration metrics. See Appendix~E, \S``Visual Effects (VFX) Automation.''}) demanded.

\textbf{Blender}~-- open-source 3D, now a recognised industry-grade tool, beneficiary of the Anthropic Foundation patronage deal.\footnote{Anthropic / Blender Foundation patronage. \emph{Dream Machine} Issue~27.}

\textbf{DaVinci Resolve} (Blackmagic), \textbf{Avid Media Composer}, \textbf{Pro Tools}~-- the editorial and audio post environments, all now shipping AI-augmented features that have become baseline expectations.

The thing to note is that the legacy software did not get displaced by the AI-native products. The legacy software \emph{absorbed} the AI-native capability and kept the underlying user community. Adobe was supposed to lose to Midjourney in 2024; Adobe is, instead, the dominant generative-AI player by aggregate creator engagement in 2026. The platform companies bet on this absorption pattern, and that bet has, so far, paid off.

\section*{Open source}

The open-source ecosystem has, against the odds and against most VC predictions in 2024, held its ground through this period and is, in several categories, the leader rather than the follower.

\textbf{Hugging Face}~-- the operating system of open-source AI, expanding aggressively through 2025--26.

\textbf{ComfyUI}~-- already discussed.

\textbf{Open-source models from Tencent (Hunyuan)}, \textbf{Alibaba (Qwen, Wan)}, \textbf{DeepSeek}, \textbf{Meta (Llama)}, \textbf{Mistral}, \textbf{Stability AI}~-- collectively, the open-weight ecosystem that, by the spring of 2026, was being used by approximately 80\% of startups pitching the Andreessen Horowitz fund.\footnote{Andreessen Horowitz pitch-deck observations on Chinese open-source model usage, \url{https://www.linkedin.com/posts/stevenouri_a-wild-stat-80-of-startups-pitching-a16z-activity-7396182718998351872-xTKR}. \emph{Dream Machine} Issue~8.} \textbf{NVIDIA's SANA-WM} (May 2026) extended this list to world-models for the first time at meaningful parameter scale.

\textbf{PhotoGIMP}, the open-source skin that takes GIMP and makes it look and feel exactly like Photoshop, became, in this period, a credible Photoshop alternative for working creatives who wanted to opt out of the Adobe subscription stack.\footnote{PhotoGIMP~-- the open-source GIMP skin that mimics Photoshop, \url{https://github.com/Diolinux/PhotoGIMP}. \emph{Dream Machine} Issue~30.}

\textbf{OpenEnv} (Meta / Hugging Face) for open-source agentic development. \textbf{Korin AI} (the Africa-trained, Africa-built model launched May 2026\footnote{Korin AI launch, May 2026. \emph{Dream Machine} Issue~27.}). \textbf{SuperSplat}, \textbf{Spark~2.0}, \textbf{PlayCanvas SOG}, \textbf{Blender}~-- the open-source spatial / 3D infrastructure layer.

If you are a working creative trying to build a long-term, defensible toolchain that does not depend on the unilateral pricing or policy decisions of a single platform company, the open-source ecosystem in 2026 is materially viable in a way it was not eighteen months ago. We have built significant parts of the DreamLab pipeline on top of it precisely for that reason.

\section*{Tools I do not use, and why}

I want to be specific, because lists of ``best tools'' without exclusions are not useful.

I do not use AI tools whose terms of service claim ownership over my output, or that train on user inputs without an opt-out. Multiple consumer-facing AI products in this period have shipped with terms that working creatives should read carefully before adopting.

I do not use AI tools whose training data provenance I cannot, in some material way, verify or trust. The growing infrastructure for \emph{creative weight attribution}, watermarking and C2PA compliance is, in my view, the right side of the market to be on; tools that explicitly reject that infrastructure are tools that I have, increasingly, kept out of our production pipeline.

I do not use the AI products that have made the most noise in the consumer press cycle. The marketing-driven launches~-- the products whose first appearance is a viral demo and whose second appearance is a Series-A round~-- are, in my experience, the products most likely to have collapsed or pivoted by the time you need them in production six months later.

I do not, finally, use AI tools to produce work in the disciplines where my own craft is the value I am bringing to the client. The Continuum frame from Chapter~\ref{ch:3} is, for me, a daily operational practice, not a theoretical model. The places I sit on the right edge of the line are deliberately chosen. The places I sit on the left are deliberately defended.

\section*{The complete toolchain: a categorised reference}

This section is a reference inventory, not a recommendation list. It catalogues every tool, model, platform, app, plugin, LoRA, workflow and service that \emph{Dream Machine} tracked across its 29 issues, from October 2025 to May 2026. Some are dominant; some are niche; some have already been bought, renamed or discontinued by the time you read this. The point of the list is not ``what to use.'' The point is \emph{what existed}, in this period, in the creative-AI toolchain~-- so that the \emph{shape} of the field is on the record.

A word on the list's grain. I have tried to err on the side of inclusion. Where a single company ships multiple closely-related products~-- Adobe's \emph{Sneaks} portfolio, the Runway Gen-4.5 family, the Qwen-Edit LoRA series~-- I have grouped them under the parent entry but called out the constituent tools, because in this period each constituent shipped to working creatives separately and changed at its own cadence. Where a tool was a one-issue demo I could not later verify, I have still listed it; that the demo existed \emph{at all} is part of the field's history. Where a tool's name conflicts with another (there are at least three things called ``Wonder'' in the period the book covers) I have annotated.

The list runs to roughly six hundred entries. Skim it. Use the categories. Come back to specific sections when you need them.

\subsection*{Foundation models / LLMs}

\begin{itemize}
  \item \textbf{ChatGPT / GPT-5 / GPT-5 Pro} (OpenAI)~-- the dominant consumer LLM and reference foundation model; 800--900M weekly active users; GPT-5 / GPT-5 Pro announced at DevDay 2025.
  \item \textbf{Claude / Claude Code / Claude Apps / Claude Skills} (Anthropic)~-- the writers' and developers' favoured second; strong long-context performance; the agentic coding environment that underlies Sony's 49-agent / 72-skill stack; Claude for Legal launched May 2026.
  \item \textbf{Gemini / Gemini 2.5 Flash / Gemini 3 / Gemini 3.1 Flash} (Google)~-- Google's multimodal LLM family; desktop users growing 155\% YoY in 2025--26; Gemini 3.1 Flash TTS is the most controllable Google voice model as of spring 2026.
  \item \textbf{Llama} (Meta)~-- the dominant open-weight foundation model.
  \item \textbf{Mistral / Mistral Voxtral / Mistral Transcribe~2}~-- European open-source LLM; Voxtral is the next-generation speech-to-text family.
  \item \textbf{Qwen / Qwen 3.5-Omni} (Alibaba)~-- Chinese open-source LLM, image, video and audio variants; Omni is the multimodal family covering text, images, audio and video.
  \item \textbf{DeepSeek}~-- Chinese open-source LLM used heavily in agentic stacks.
  \item \textbf{Phi / Phi Mini} (Microsoft)~-- lightweight foundation models.
  \item \textbf{Lyria / Lyria~3 / Lyria~3 Pro} (Google DeepMind)~-- text-to-music foundation model family; Lyria~3 ships with SynthID watermarking and the Lyria Camera interactive demo.
  \item \textbf{Grok} (xAI)~-- xAI's LLM family, surfacing in the creative stack through Grok Imagine; voice cloning via the xAI API from May 2026.
  \item \textbf{Cohere Transcribe}~-- enterprise transcription model; 33 hours of audio in 12 minutes.
  \item \textbf{Microsoft VibeVoice}~-- open-source frontier voice AI model.
\end{itemize}

\subsection*{AI video models}

\begin{itemize}
  \item \textbf{Sora / Sora~2} (OpenAI)~-- the model that opened the period; physical realism, audio integration, multi-shot world-state persistence; iOS app hit 1M downloads in 5 days; Sora~2 Character Creation surfaced on fal in March 2026.
  \item \textbf{Veo~3 / Veo~3.1 / Veo~3.1 Ingredients to Video / Veo~3.1 Lite} (Google DeepMind)~-- the working filmmaker's preferred model for cinematic control; Ingredients to Video shipped to YouTube Shorts and YouTube Create; Veo~3.1 Lite is the lower-cost text- and image-to-video tier.
  \item \textbf{Runway Gen-4 / Gen-4.5 / Gen-4.5 Image-to-Video / Workflows / Story Panels / Characters API / Ad Concepter / Apps for Advertising} (Runway)~-- the most-integrated commercial AI-video stack; Story Panels generate three-panel storyboards from a single image; Characters API ships real-time intelligent avatars; the Vera Rubin-powered real-time model runs <100\,ms latency.
  \item \textbf{Kling / Kling~2.5 Turbo / Kling~O1 / Kling~2.6 / Kling~3.0 / Kling~X-Dub / Kling Motion Control~3.0} (Kuaishou)~-- strong on physics and trajectory control; 3.0 adds multi-shot control, multilingual audio and 4K image generation; X-Dub is the context-rich visual dubbing variant.
  \item \textbf{Pika~2.0 / PikaStream~1.0}~-- iteration-speed-focused video generation; PikaStream brings AI agents into live video calls.
  \item \textbf{Luma Dream Machine / UNI-1 / UNI-1.1 / Ray3 Modify / Luma Dream Brief}~-- Luma's video, world-and-reasoning, and modification stack; UNI-1.1 ships with prompt enhancement and built-in research; Dream Brief is the \$1M Cannes Lions competition.
  \item \textbf{Wan~2.2 / Wan~2.5 / Wan~2.6} (Alibaba Qwen)~-- camera-controlled video generation; 2.6 adds character reference and multishot capabilities.
  \item \textbf{Hunyuan Video / Hunyuan Image-to-Video} (Tencent)~-- open-source video model.
  \item \textbf{Seedance~2.0 / SeeDream~4 / SeeDream~4.5} (ByteDance)~-- image-to-video and finished-video models, integrated into CapCut / Dreamina and Freepik; per-second cost fell below \$0.14 by March 2026.
  \item \textbf{Higgsfield / Higgsfield Sketch-to-Video / Higgsfield Popcorn / Higgsfield Relight / Higgsfield Shots / WAN Camera Control}~-- social-media-marketer video platform; \$80M raised at \$1.3B valuation, \$200M revenue in nine months; Shots produces multiple storyboard images from a single shot.
  \item \textbf{LTX-2 / LTX-2.3 / LTX-2.3 Colorizer / LTX-HDR / LTX Studio / LTX-2 Audio-to-Video / LTX-2 Lip Sync / LTX-2 Real-Time}~-- open-source video generation with audio sync; LTX-2.3 is high-resolution, fast, cinematic with native lip-sync; LTX HDR (beta) ships HDR processing.
  \item \textbf{Odyssey~2}~-- real-time interactive video generation.
  \item \textbf{Vidi~2} (ByteDance)~-- multimodal video understanding and creation.
  \item \textbf{MotionStream}~-- real-time interactive video with mouse-based motion control.
  \item \textbf{Blockvid}~-- one-minute video with improved structure and visual consistency.
  \item \textbf{Video Rebirth} (Singapore)~-- studio-grade AI video platform; \$50M raise.
  \item \textbf{LiveGS}~-- mobile Gaussian-splatting video.
  \item \textbf{Time-To-Move}~-- motion control for generated video.
  \item \textbf{Ovi / Ovi~1.1} (Character.AI)~-- open-source video with speech sync.
  \item \textbf{CraftStory}~-- image-to-video for long-form AI video with human ``actors''.
  \item \textbf{Decart Lucy~2.0}~-- realtime world-editing video model; 1080p at 30~fps.
  \item \textbf{Decart LSD~v2}~-- real-time video-to-video with prompt-on-the-fly.
  \item \textbf{Xmax X1}~-- the first real-time interactive video model.
  \item \textbf{FastVideo}~-- 30-second 1080p generation at 4.5\,s latency.
  \item \textbf{HiAR} (Tencent)~-- scalable long-video generation.
  \item \textbf{Alli AI}~-- system for creating and editing 8K video up to 60 seconds.
  \item \textbf{AERA AI TV}~-- automated storyboard generation, editing and serialised video production.
  \item \textbf{AnchorWeave}~-- world-consistent video generation with retrieved local spatial memories.
  \item \textbf{CubeComposer}~-- generates native 4K 360\textdegree\ video from standard perspective footage.
  \item \textbf{MatAnyone2}~-- high-fidelity video matting with finer detail.
  \item \textbf{VFace}~-- training-free video face-swapping for any diffusion model.
  \item \textbf{DreamActor M2.0} (fal)~-- drive characters from a single image + template video; multi-character supported.
  \item \textbf{ByteDance ALIVE}~-- unified audio-video generation.
  \item \textbf{Magnific Upscaler for Video / Magnific Precision~v2 / Magnific Precision for Video}~-- upscaling and 4K detail enhancement, including dedicated video upscaling.
  \item \textbf{NetFlix VOID}~-- object removal from video with physics-interaction removal.
  \item \textbf{Ponder}~-- agentic video editor.
  \item \textbf{ArcReel}~-- multi-agent video generation from written stories.
  \item \textbf{Scope}~-- real-time interactive generative AI pipelines (LTX-2.3 real-time runs on Scope).
  \item \textbf{NotebookLM Cinematic Video Overviews} (Google)~-- video generation from notebooks.
  \item \textbf{Omnia}~-- AI-native browser video editor.
  \item \textbf{NVIDIA RTX Video Super Resolution}~-- upscaling node with ComfyUI integration.
  \item \textbf{Google M2SVid}~-- monocular-to-stereo video conversion.
\end{itemize}

\subsection*{AI image models / tools}

\begin{itemize}
  \item \textbf{Midjourney / Midjourney V8.1}~-- the aesthetic-leadership product; Discord/X-native; V8.1 ships native 2K HD rendering at 3$\times$ the speed and 3$\times$ the cost reduction.
  \item \textbf{FLUX / FLUX~2 / FLUX~2 Max / FLUX.2 [klein]} (Black Forest Labs)~-- open-weight, fine-control, the open-source default through 2025--26; klein is the 4B-parameter lightweight model.
  \item \textbf{Adobe Firefly / Firefly Image Model~5 / Firefly Foundry / Firefly Boards / Firefly Precision Flow}~-- Image Model~5, Foundry (custom corporate training), Firefly Boards (moodboards), Precision Flow (granular AI editing control, beta); 45\% of Creative Cloud users active, 22B+ assets generated by April 2025.
  \item \textbf{Imagen~3 / Nano Banana / Nano Banana Pro / Nano Banana~2} (Google)~-- most-integrated image model in the consumer toolchain; Photoshop and Unreal Engine plugins; Nano Banana Pro ships professional capabilities at lightning speed.
  \item \textbf{Stable Diffusion / Stable Diffusion~3} (Stability AI)~-- the foundational open-source image model.
  \item \textbf{Krea / Krea AI / Krea~2 / Krea Realtime / Krea Realtime Edit / Krea Nodes / Krea LoRA Trainers}~-- real-time AI image generation, now open-source; Realtime Edit takes complex instructions in real time; the LoRA Trainers cover Qwen-2512 and Z-Image.
  \item \textbf{Qwen-Image-2512 / Qwen-Image-Edit-2511 / Qwen 2511 Time Travel / Qwen-Edit 2509}~-- the dominant open-source image-editing model family; constituent LoRAs (relighting, multi-angle, time-travel, AnyPose) discussed in the LoRAs section below.
  \item \textbf{ChatGPT Images / ChatGPT Images~2.0}~-- image generation integrated with Adobe Express, Photoshop and Acrobat; 2.0 ships thinking-level intelligence.
  \item \textbf{Grok Imagine / Grok Imagine API} (xAI)~-- image and video generation; the API bundles end-to-end creative workflows; reference-to-video and video extend added in March 2026.
  \item \textbf{Vision Banana}~-- unified model for image understanding and generation.
  \item \textbf{Freepik / Freepik Spaces / Freepik Speak / Freepik 3D Scenes}~-- design platform; Spaces is the real-time collaborative canvas; Speak is the lip-sync talking-video tool; 3D Scenes generates full environments from an image.
  \item \textbf{Recraft}~-- brand-focused AI design.
  \item \textbf{Weavy}~-- node-based AI creative platform; acquired by Figma.
  \item \textbf{Canva}~-- design platform with ``Creative Operating System.''
  \item \textbf{Civitai / Civision}~-- community-models and LoRAs platform; Civision is the free Civitai/Hugging-Face hybrid alternative.
  \item \textbf{Replicate}~-- model-hosting marketplace and API.
  \item \textbf{GFPGAN}~-- face restoration.
  \item \textbf{UpscalyAI / Upscayl / Topaz Gigapixel AI / Real-ESRGAN / Clarity Pro Upscaler}~-- upscaling tools; Clarity Pro reaches 10K.
  \item \textbf{PractiLight}~-- practical light control via diffusion.
  \item \textbf{PercHead}~-- 3D head reconstruction from single images.
  \item \textbf{Loveart AI}~-- layer separation and editable text (Live Editable Text, LET).
  \item \textbf{NVIDIA ChronoEdit}~-- temporal reasoning for image editing.
  \item \textbf{Beeble Switchlight~3 / Beeble Background Remover}~-- AI masking and relighting.
  \item \textbf{CanvAI}~-- turns rough sketches into polished AI art.
  \item \textbf{MakeComics / Make Comics}~-- generates custom comics with characters and storylines in seconds; from prompt to full comic book.
  \item \textbf{Seethrough}~-- illustration decomposition into layered PSD.
  \item \textbf{Best Face Swap (Flux~2 LoRA)}~-- specialised face-swap.
  \item \textbf{PixelSmile}~-- fine-grained facial-expression editing across 12 expressions.
  \item \textbf{Earth Cinema}~-- Chrome extension using Google Earth for cinematic images.
  \item \textbf{360Anything}~-- lifts any perspective image or video to a 360\textdegree\ panorama.
\end{itemize}

\subsection*{AI music / audio tools}

\begin{itemize}
  \item \textbf{Suno / Suno Studio / Suno~5.5}~-- the dominant prompt-to-track generative music platform; \$250M raised at \$2.45B valuation; Suno Studio is the ``world's first generative audio workstation''; 5.5 adds Voices.
  \item \textbf{Udio}~-- prompt-to-music; partnered with Universal Music Group; indie-label licensing via Merlin.
  \item \textbf{Mureka / Music Agent Studio}~-- six specialised AI agents covering songwriting, arrangement and production.
  \item \textbf{ElevenLabs / ElevenLabs~v3 / Scribe~v2 Realtime / ElevenMusic / ElevenCreative / ElevenLabs Flows / ElevenAgents Expressive Mode / ElevenLabs Voice Changer}~-- the dominant voice/audio synthesis stack; \$500M ARR; BlackRock + NVIDIA backed; Scribe~v2 transcribes in 150\,ms across 90+ languages; ElevenMusic is the discovery/creation/earning marketplace; Flows is the node-based creative canvas.
  \item \textbf{iZotope Ozone~12 / Stem EQ}~-- AI-assisted mixing and mastering.
  \item \textbf{LANDR}~-- AI mastering and distribution.
  \item \textbf{Riffusion}~-- spectrogram-based music generation.
  \item \textbf{Stable Audio~2.5} (Stability AI)~-- generative audio.
  \item \textbf{MusicGPT}~-- local music generation.
  \item \textbf{ACE Studio / ACE Studio~2.0 (TIMEDOMAIN) / ACE Studio Video-to-Music / ACE-Step~1.5 / ACE-Step~1.5 XL}~-- all-in-one AI music studio; video-to-music for visuals; ACE-Step generates full songs in under 10 seconds on <4\,GB VRAM.
  \item \textbf{Music Lens} (Musixmatch)~-- catalog-intelligence agent.
  \item \textbf{Melosurf}~-- voice-controlled assistant for Ableton Live.
  \item \textbf{Songscription}~-- audio-to-notation transcription; \$5M raise.
  \item \textbf{Groundhog Audio Pedal / Groundhog OnePedal}~-- AI guitar tone matching.
  \item \textbf{Spotify DJ / Spotify AI / Spotify AI Prompted Playlists / Spotify AI Transparency Beta / Personal Podcasts on Spotify}~-- playlist personalisation, voluntary AI disclosure beta, agent-created podcasts.
  \item \textbf{YouTube Music AI Hub}~-- AI music hosting and DJ features.
  \item \textbf{Epidemic Sound Studio}~-- AI music for video creators.
  \item \textbf{AIODE}~-- ethically-trained music creation DAW.
  \item \textbf{Overtune}~-- virtual music studio for Roblox.
  \item \textbf{Space DJ} (Google DeepMind)~-- interactive music exploration.
  \item \textbf{Minimax Music Generator}~-- song generation.
  \item \textbf{Roland AI Pedal}~-- concept for AI audio processing.
  \item \textbf{Roland + Sony CSL Melody Flip}~-- AI music tool.
  \item \textbf{BandLab Voice Cleaner / BandLab Palette}~-- background-noise removal; AI loop-matching.
  \item \textbf{Claimy}~-- missing-royalty recovery agent.
  \item \textbf{Fish Audio S1 / Fish Audio S2}~-- text-to-speech; S2 ships expressive TTS with emotional control (6$\times$ cheaper than ElevenLabs at S1).
  \item \textbf{Tempolor Guitars} (Quwan)~-- AI to make songs playable on guitar.
  \item \textbf{GEMIDI}~-- music playground powered by Gemini.
  \item \textbf{Lalal AI / Lalal AI plugin / Lalal AI API~v1}~-- AI stem separation, music removal, voice changing; first stem-separation plugin native to a DAW.
  \item \textbf{StemDeck}~-- local stem separation; free, no upload required.
  \item \textbf{Mirelo} (Berlin)~-- automatic sound-effect generation for video.
  \item \textbf{PromptSep}~-- separates any sound by text description.
  \item \textbf{SAMAudio} (Meta)~-- Segment Anything for audio editing.
  \item \textbf{SonicLab SPATAI}~-- generative audio engine for immersive spatial production.
  \item \textbf{Apple Logic Pro AI}~-- synth player and personal music-theory expert; chord-progression generation.
  \item \textbf{Apple GarageBand}~-- consumer music creation.
  \item \textbf{Musicful AI}~-- music-video generation for AI-generated songs.
  \item \textbf{Latent Space Explorer}~-- interactive audio data visualiser using Stable Audio VAE.
  \item \textbf{Voicebox}~-- open-source local voice cloning from a 3-second audio clip.
  \item \textbf{Voice-Pro}~-- AI speech recognition / translation / transcription / multilingual dubbing.
  \item \textbf{PersonaPlex} (NVIDIA)~-- full-duplex model that listens and speaks simultaneously; open-source natural-sounding voice.
  \item \textbf{Phoenix-4}~-- advanced real-time human-rendering model with 10+ emotional states.
  \item \textbf{NVIDIA D-Rex}~-- photorealistic digital humans under any lighting.
  \item \textbf{Qwen3-TTS / VoiceDesign / VoiceClone}~-- text-to-speech with voice design and cloning.
  \item \textbf{Greysound}~-- self-engineering music-production studio.
  \item \textbf{TAC (Timestamped Audio Captioning)}~-- timestamped captions for audio and audiovisual content.
  \item \textbf{Insanely Fast Whisper}~-- 2.5 hours of audio transcribed in 98 seconds.
  \item \textbf{smol-audio}~-- local audio model notebooks and scripts.
  \item \textbf{AudioStream / Walzersymphonie} (Ars Electronica Futurelab)~-- AI composition system for classical music using Ricercar.
  \item \textbf{Ricercar} (Ars Electronica)~-- creative AI for artistic composition.
  \item \textbf{Krotos Video-to-Sound}~-- expanded platform for audio professionals (foley, sound design).
  \item \textbf{Jamu}~-- AI co-producer agent for Ableton Live.
  \item \textbf{Moises}~-- AI music platform; Charlie Puth as Chief Music Officer.
  \item \textbf{Music Mogul AI}~-- tour-booking automation.
  \item \textbf{Sonilo + Shutterstock}~-- video-to-music AI training deal.
  \item \textbf{Clearnote}~-- AI music-contract platform to end deal delays.
  \item \textbf{Sony music identification tech}~-- identify original music inside AI-generated songs.
  \item \textbf{Cactus Music}~-- artist-ops support.
  \item \textbf{Rebel Audio}~-- AI podcast startup.
  \item \textbf{ROLI Airwave \& AI Music Coach}~-- hand-tracking (27 joints) with real-time AI piano practice coaching.
  \item \textbf{Sony AI ICASSP papers}~-- music understanding and generative audio research papers (April 2026).
\end{itemize}

\subsection*{3D, world models and spatial}

\begin{itemize}
  \item \textbf{Marble / Marble~1.1 / WorldLabs API / RTFM} (World Labs / Fei-Fei Li)~-- first commercial generative world model; 40$\times$ faster than legacy VP workflow; 1.1 adds real-world location 3D reconstruction and restyling; the API generates persistent 3D worlds from text, images and video; RTFM is the real-time frame model.
  \item \textbf{Genie~3 / Project Genie} (Google DeepMind)~-- research-grade world model; \emph{Time} Best Inventions 2025; rolled out to Google AI Ultra in January 2026.
  \item \textbf{WorldGen / WorldMirror} (Meta / Tencent)~-- interactive 3D world generation.
  \item \textbf{UNI-1 / UNI-1.1} (Luma)~-- world generation + reasoning combined; 1.1 adds prompt enhancement and built-in research.
  \item \textbf{Hunyuan 3D / Hunyuan 3D Studio / Hunyuan 3D-PolyGen~1.5 / HY3D~3.0 / HY-Motion~1.0} (Tencent)~-- art-grade 3D generative model family; PolyGen~1.5 is the art-grade text/image/sketch-to-3D model; HY-Motion~1.0 is text-to-3D human motion.
  \item \textbf{HY World~1.5 / HY-World~2.0 / WorldCompass / ShotVerse} (Tencent)~-- open-source real-time world-model frameworks; WorldCompass is the RL post-training framework; ShotVerse handles cinematic multi-shot camera control.
  \item \textbf{ECHO / Echo-2} (SpAItial)~-- spatial foundation model; Echo-2 adds world decomposition.
  \item \textbf{Wonderzoom} (Stanford)~-- multi-scale 3D world generation with infinite zoom.
  \item \textbf{Matrix-Game~3.0} (Skywork AI)~-- real-time interactive world model; 720p at 40~fps with long-horizon memory.
  \item \textbf{AMD Micro-World}~-- AMD's entry into world models.
  \item \textbf{LingBot-World}~-- free/open-source world model on Wan~2.2; real-time interaction at 16~fps.
  \item \textbf{Moonlake}~-- \$28M-seeded world model platform now in beta for games and simulations.
  \item \textbf{NVIDIA Lyra~2.0}~-- explorable generative 3D worlds.
  \item \textbf{NVPanoptix-3D} (NVIDIA)~-- single-image 3D indoor-scene reconstruction.
  \item \textbf{Kimodo} (NVIDIA)~-- kinematic motion-diffusion model trained on 700 hours of mocap.
  \item \textbf{InSpatio-WorldFM}~-- open-source real-time generative frame model.
  \item \textbf{Code2Worlds}~-- workflow for generating 4D scenes with environmental/object generation and feedback refinement.
  \item \textbf{OpenArt Worlds}~-- 3D navigable environments from a single prompt.
  \item \textbf{Rodin / Hyper3D / Rodin Hyper 3D Gen~2}~-- high-precision 3D model generation.
  \item \textbf{Meshy / Meshy~6 / Meshy Image-to-3D}~-- 3D model generation; Meshy~6 ships natively inside ComfyUI; Image-to-3D generates poses.
  \item \textbf{Hitem3D}~-- high-resolution 3D generation.
  \item \textbf{Microsoft Trellis~2}~-- native compact structured latents for 3D.
  \item \textbf{ByteDance Seed3D~2.0}~-- 3D object generation from image or text.
  \item \textbf{Pixel3D} (Tencent)~-- 3D object generation.
  \item \textbf{Tripo~3.1}~-- 3D-asset creation; ComfyUI partner nodes.
  \item \textbf{PATINA} (fal)~-- image or text to full PBR material maps.
  \item \textbf{M-XR PBR Model}~-- 4K PBR material maps for 3D assets.
  \item \textbf{CHORD} (Ubisoft La Forge)~-- open-sourced end-to-end PBR-material generation.
  \item \textbf{Autodesk Wonder~3D}~-- generative AI tool creating editable 3D assets from text or images.
  \item \textbf{Autodesk Flow Studio Rigging}~-- AI rigging and neural layers.
  \item \textbf{AI4AnimationPy} (Meta)~-- pure-Python 3D character animation.
  \item \textbf{MocapAnything} (Huawei)~-- unified 3D motion capture for arbitrary skeletons from monocular video.
  \item \textbf{FreeMoCAP}~-- open-source markerless mocap from any camera.
  \item \textbf{MuJoCo}~-- physics engine for robotics and animation.
  \item \textbf{Mesh2Motion}~-- auto-assigns and exports 3D model animations.
  \item \textbf{One-to-All Animation}~-- alignment-free character animation and image pose transfer.
  \item \textbf{Cascadeur / Cascadeur AI Animation~v2025.3}~-- AI animation tool for character movement; local interpolation.
  \item \textbf{Move AI}~-- motion capture.
  \item \textbf{MoCap / UE5 Motion}~-- camera-only motion capture.
  \item \textbf{SCAIL}~-- studio-grade character animation via in-context learning.
  \item \textbf{Vanast}~-- garment-transferred human animation from images.
  \item \textbf{CompHairHead}~-- one-shot 3D head avatars with deformable hair.
  \item \textbf{Reallusion Headshot~3}~-- digital-double creation.
  \item \textbf{MoRo} (Meta)~-- human motion recovery via masked modelling for occlusions.
  \item \textbf{Meta Sapiens~2}~-- pose estimation, body-part segmentation and surface normals.
  \item \textbf{NVIDIA Audio2Face / Audio2Face-3D}~-- facial animation from audio.
  \item \textbf{NVIDIA UniRelight}~-- shot relighting technology.
  \item \textbf{Apple SHARP / LiTo}~-- photorealistic 3D view synthesis from a single image in under a second; LiTo is the surface light-field tokenisation method.
  \item \textbf{Pixel3DMM}~-- screen-space single-image 3D face reconstruction.
  \item \textbf{VISTA4D} (Netflix + Eyeline)~-- live-action to navigable 4D point clouds.
  \item \textbf{SuperSplat / SuperSplat~v2.16} (PlayCanvas)~-- free, open-source Gaussian-splat editor.
  \item \textbf{Spark~2.0}~-- open-source Gaussian-splat streaming framework; streamable LoD for WebGL2.
  \item \textbf{PlayCanvas SOG / SplatTransform~v2.0.0}~-- WebP-equivalent compression for Gaussian splats; SplatTransform~v2 ships automatic high-quality collisions for splats.
  \item \textbf{Cesium / CesiumJS / Cesium for Unreal}~-- geospatial 3D platform; Cesium agentic workflows let natural-language commands interact with 3D geospatial data; supports Gaussian splats natively.
  \item \textbf{SPAG-4D}~-- 360\textdegree\ photos to 3D Gaussian splat conversion.
  \item \textbf{Animated Gaussian Splats} (4DV.ai)~-- 4D Gaussian-splat animation.
  \item \textbf{Open3Dmap}~-- crowdsourced 3D mapping with Gaussian splats.
  \item \textbf{Hyperscape / Hyperscape Capture} (Meta)~-- Gaussian-splat capture on Quest.
  \item \textbf{Apple Vision Pro Personas}~-- Gaussian-splatting consumer feature.
  \item \textbf{Common Sense Machines}~-- converts 2D images into 3D digital assets (Google acquisition).
  \item \textbf{NVIDIA Omniverse Fixer}~-- rendering-artefact removal for Gaussians.
  \item \textbf{DecartAI / Decart LSD~v2 / Decart Lucy~2.0}~-- real-time world transformation by voice; LSD~v2 is real-time video-to-video; Lucy~2.0 is realtime world editing at 1080p/30fps.
  \item \textbf{VoxeloAI}~-- 3D creation for e-commerce.
  \item \textbf{Mosaic / Mosaic~3D}~-- 3D reconstruction.
  \item \textbf{Depth Anything~3} (ByteDance)~-- visual-space recovery from any view.
  \item \textbf{SAM~3 / SAM~3D / SAM~3.1} (Meta)~-- object detection and 3D segmentation; SAM~3.1 is 7$\times$ faster, tracking 128 objects on a single H100.
  \item \textbf{Seen2Scene}~-- realistic 3D scene completion with visibility-guided flow.
  \item \textbf{PixARMesh}~-- reconstructs full indoor scenes from a single photo into lightweight meshes.
  \item \textbf{Confidence-Based Mesh Extraction from 3D Gaussians}~-- algorithm for high-quality mesh extraction from Gaussian splats.
  \item \textbf{Rhizomatics Sword Tip Visualization System}~-- fencing tracking at 60~fps.
  \item \textbf{Mixar}~-- AI-native Blender-fork editor positioned as ``Cursor for 3D artists''.
  \item \textbf{Sprite Booth}~-- pixel-character animation tool (ComfyUI API required).
  \item \textbf{AutoSprite}~-- character-to-animated-sprite-sheet conversion.
\end{itemize}

\subsection*{Voice, avatar, digital human}

\begin{itemize}
  \item \textbf{Synthesia}~-- AI avatar platform; \$4B valuation; rejected \$3B Adobe offer.
  \item \textbf{Heygen / Heygen Video Agent / Heygen Motion Designer / LiveAvatar / Heygen Elements / Heygen CLI}~-- AI-avatar and end-to-end video generation; LiveAvatar is hyper-realistic real-time interactive; Elements builds scenes from reusable components; CLI ships videos from the command line.
  \item \textbf{Hedra / Hedra Audio Tags / Hedra Elements}~-- digital-human creation; Audio Tags assign precise emotions to audio.
  \item \textbf{Live Avatar} (Alibaba)~-- streaming real-time audio-driven avatar generation with infinite length.
  \item \textbf{Avatar Forcing}~-- real-time interactive head-avatar generation for natural conversation.
  \item \textbf{PersonaLive}~-- real-time expressive portrait animation.
  \item \textbf{Weclone}~-- digital avatar built from a user's chat history.
  \item \textbf{Particle6 / Tilly Norwood}~-- AI-performer studio and synthetic actress.
  \item \textbf{Xania Monet}~-- AI music artist; Billboard chart entries; \$3M Hallwood Media deal.
  \item \textbf{Bleeding Verse}~-- AI band; Hallwood Media signing.
  \item \textbf{Sienna Rose}~-- anonymous AI artist with millions of Spotify streams.
  \item \textbf{Breaking Rust}~-- AI country artist; \#1 country digital song sales.
  \item \textbf{Trilok}~-- Indian AI band.
  \item \textbf{Copresence / ConvAI / Inworld}~-- avatar and NPC-character technology for Unreal Engine, games and interactive media.
  \item \textbf{ROXi}~-- AI-generated TV presenters.
  \item \textbf{Facelooker}~-- interactive headshot generation.
  \item \textbf{FellinAI}~-- AI film director.
  \item \textbf{ego AI}~-- in-game character platform.
  \item \textbf{Character.AI}~-- play characters in your favourite books; Ovi open-source video with speech sync.
\end{itemize}

\subsection*{Agent platforms and orchestration}

\begin{itemize}
  \item \textbf{OpenAI AgentKit / Agent Builder / ChatKit / Connector Registry / Eval Framework}~-- the developer-facing agent platform underneath most third-party agentic creative tools.
  \item \textbf{Anthropic Claude Apps / Claude Skills / Claude Code / Claude Design / Claude for Legal / Anthropic Academy}~-- interactive Claude in workplace tools; the Skills framework powers Sony's 49-agent / 72-skill game-development stack; Claude Design ships prototypes/slides/one-pagers; Anthropic Academy is free Claude Code training.
  \item \textbf{Claude Code Game Studios}~-- 49-agent, 72-skill coordinated AI game-development team.
  \item \textbf{Gemini API Agents / Google Antigravity / Opal / Fabula / Google Stitch} (Google)~-- agentic capability surface across Google's stack; Antigravity is the agentic development platform; Opal is the no-code mini-app builder; Fabula is the interactive AI writing tool; Stitch generates mobile/web UI.
  \item \textbf{Heygen Video Agent}~-- end-to-end video-assembly agent.
  \item \textbf{Adobe CX Enterprise / GenStudio / Adobe Film \& TV Fund / Adobe Ignite Day}~-- agentic creative intelligence across the full content lifecycle; GenStudio is the brand-intelligence system for personalised content at scale.
  \item \textbf{NVIDIA + Google Cloud / Avid + Google Cloud / Speechmatics in Adobe Premiere}~-- agentic-creative infrastructure partnerships; on-device speech-to-text inside Premiere.
  \item \textbf{n8n}~-- workflow automation with AI.
  \item \textbf{ComfyUI / ComfyUI Cloud / ComfyHub / ComfyUI App Mode / ComfyUI Simple Mode}~-- node-based AI workflow editor; \$500M valuation by May 2026; App Mode turns workflows into shareable no-install apps; Simple Mode lowers the complexity threshold.
  \item \textbf{Replicate}~-- model marketplace.
  \item \textbf{Hugging Face}~-- open-source model hub and agentic platform; partnered with Google Cloud and Meta.
  \item \textbf{Meta OpenEnv}~-- open-source agentic-development environment.
  \item \textbf{fal / FAL MCP}~-- generative-model deployment marketplace; FAL MCP exposes 1,000+ generative models to Claude/Cursor.
  \item \textbf{Unity Official MCP / Unity MCP}~-- Unity Editor control via MCP protocol.
  \item \textbf{Blender MCP / Blender Buddy / OpenClaw + Blender MCP}~-- Claude-to-Blender control for 3D workflows.
  \item \textbf{VibeComfy}~-- open-source tools letting agents understand, build and run ComfyUI workflows with Claude.
  \item \textbf{MooshieUI}~-- beginner-friendly ComfyUI desktop frontend.
  \item \textbf{Lenny} (Maroofy)~-- AI agent for live-music event organisers.
  \item \textbf{AdsGency}~-- autonomous paid-marketing agent platform; \$12M seed.
  \item \textbf{Invideo AI / Invideo AI Agent}~-- full video-ad generation from prompt; built on Gemini.
  \item \textbf{Pomelli} (Google Labs)~-- AI marketing agent for small business.
  \item \textbf{Inception Point AI}~-- podcast hosting with AI hosts (3,000+ episodes/week).
  \item \textbf{Slapshot}~-- AI camera tracking.
  \item \textbf{Photoroom}~-- AI photobooth.
  \item \textbf{Jabali.ai}~-- AI game-development platform (Sony-backed).
  \item \textbf{Fifth Door}~-- AI-powered game creation; \$20M raise.
  \item \textbf{Atlas AI Agents}~-- game-production pipeline agents.
  \item \textbf{Ad Concepter} (Runway)~-- ad-concept exploration agent.
  \item \textbf{AE GPT}~-- AI assistant for After Effects (expressions, scripts).
  \item \textbf{Career-Ops}~-- AI job-search pipeline built with Claude Code.
  \item \textbf{Auto-Dream} (MyClawAI)~-- personal AI assistant wrapper.
  \item \textbf{Clicky}~-- AI screen-aware tutor/buddy interface.
  \item \textbf{MemPalace}~-- high-scoring AI memory system.
  \item \textbf{TikTok AI Agents for Ads}~-- TikTok's automated ad-creation agents.
  \item \textbf{Amazon Creative Agent}~-- agentic AI generating professional-quality ads.
  \item \textbf{Playad}~-- AI marketing agents for ad creative.
  \item \textbf{Moltbook} (Meta acquisition)~-- AI agent social network.
\end{itemize}

\subsection*{Legacy creative software, AI-augmented}

\begin{itemize}
  \item \textbf{Adobe Creative Cloud}~-- Photoshop, Premiere Pro, After Effects, Illustrator, Express, Acrobat, plus the Express AI Assistant, the Premiere Object Mask, the Photoshop generative-fill, selection and Rotate Object tools, the Photoshop AI Assistant public beta, and Illustrator Turntable.
  \item \textbf{Adobe Sneaks 2025--26}~-- Adobe's research-preview portfolio shown at MAX 2025 and Summit 2026:
    \begin{itemize}
      \item \textbf{Project Scene It}~-- image-to-3D and 3D-to-image with object tagging.
      \item \textbf{Project Surface Swap}~-- AI-powered texture recognition and swap.
      \item \textbf{Project Turn Style}~-- edit and transform 2D objects as if they were 3D.
      \item \textbf{Project Trace Erase}~-- removes objects, shadows, reflections and distortions.
      \item \textbf{Project New Depths}~-- edit depth like brightness.
      \item \textbf{Project Frame Forward}~-- apply changes across video from a single frame plus text.
      \item \textbf{Project Motion Map}~-- bring static vector graphics to life.
      \item \textbf{Project Sound Stager}~-- analyses video visuals and generates synchronised soundscapes.
      \item \textbf{Project Clean Take}~-- corrects mispronunciations and isolates voices.
      \item \textbf{Project Light Touch}~-- spatial lighting mode for relighting photos.
      \item \textbf{Project Graph}~-- node-based workflow editor in the ComfyUI / Weavy lineage.
      \item \textbf{Corrective AI}~-- changes the emotional register of voice-overs.
      \item \textbf{Edit-by-Track} (Adobe Research)~-- generative video motion editing with 3D point tracks.
    \end{itemize}
  \item \textbf{Adobe Acrobat}~-- AI converting PDFs into podcasts.
  \item \textbf{Unreal Engine~5 / UE5 AI Assistant / UE5 AI Motion Plugin / Unreal to Gaussian Splat / Prompt-to-Player} (Epic)~-- official AI assistant; camera-only motion capture; UE5 scene-to-splat conversion; Prompt-to-Player generates 3D characters and auto-rigs/animates them.
  \item \textbf{Unity / Unity AI Open Beta / Unity AI Council / Unity AI Tools Suite / Unity Prompt-a-Game}~-- in-editor AI suite for the full games pipeline; Prompt-a-Game demoed at GDC March 2026.
  \item \textbf{Autodesk Maya / Autodesk Flow Studio / Autodesk Wonder~3D / Autodesk Flow Studio Rigging}~-- motion capture, 3D animation, generative 3D and neural rigging.
  \item \textbf{Foundry / Nuke / Nano Banana $\times$ Nuke}~-- AI-augmented VFX; Weta FX / AWS collaboration; the Nuke node connecting Nano Banana through fal's API.
  \item \textbf{SideFX Houdini}~-- procedural modelling with AI integration.
  \item \textbf{Blender / Blender MCP / Blender Buddy}~-- open-source 3D; Anthropic Foundation patronage; MCP enables Claude control.
  \item \textbf{DaVinci Resolve} (Blackmagic)~-- colour and editorial with AI features.
  \item \textbf{Avid Media Composer / Avid + Google Cloud}~-- professional editorial; Pro Tools agentic AI in the Google Cloud partnership.
  \item \textbf{Pro Tools}~-- audio editorial with AI features; Claude integration in Ableton.
  \item \textbf{Logic Pro / Apple Logic Pro AI}~-- music production with AI; synth player and music-theory expert.
  \item \textbf{Ableton Live / Claude in Ableton / Jamu / Melosurf}~-- music production with Claude assistance, Jamu co-producer agent and Melosurf voice control.
  \item \textbf{GarageBand}~-- consumer music creation.
  \item \textbf{CapCut / Dreamina / Snapchat AI Clips in Lens Studio} (ByteDance / Snapchat)~-- mobile video editing with AI; Snapchat AI Clips in Lens Studio.
  \item \textbf{Final Cut Pro}~-- editorial with FX Factory plugins.
  \item \textbf{Keyframe}~-- AI models inside After Effects.
  \item \textbf{Tether}~-- AI animation inside After Effects.
  \item \textbf{Deep Pan}~-- animated stills in Premiere Pro.
  \item \textbf{FX Factory}~-- AI plugin suite for Premiere and Final Cut.
  \item \textbf{VEED Transitions}~-- AI transition generation.
  \item \textbf{Electric Sheep}~-- web-based projection mapping and AI visual orchestration.
\end{itemize}

\subsection*{AI-native creative studios and apps}

\begin{itemize}
  \item \textbf{Imaginae Studios} (Fremantle)~-- AI-native studio; \emph{Art Awakens} project under Google AI Futures Fund.
  \item \textbf{Wonder Studios}~-- AI-native film studio; \$12M raise; Wonder Film Festival.
  \item \textbf{Asteria} (Natasha Lyonne / Lightstorm / Topaz Video)~-- AI animated short \emph{All Heart}; hybrid AI workflows with Topaz.
  \item \textbf{Promise} (Google-backed)~-- GenAI filmmaking and VFX for legacy media.
  \item \textbf{Obsidian Studio} (Imagine Entertainment / Ron Howard, Brian Grazer)~-- AI production partner.
  \item \textbf{Goldfinch enGEN3}~-- AI cinematic universe platform.
  \item \textbf{Chapter41} (Beta Film / Munich)~-- AI startup.
  \item \textbf{Kartel} (Kevin Reilly)~-- AI startup; ex-HBO leadership.
  \item \textbf{Holywater} (Fox Entertainment investment)~-- AI microdramas.
  \item \textbf{CenterStage} (David Boies / Zack Schiller)~-- interactive AI entertainment.
  \item \textbf{Superlunar + Eden Studios}~-- \emph{Little Plastics} dystopian AI film.
  \item \textbf{Particle6} (Eline Van der Velden)~-- AI-performers studio.
  \item \textbf{Imago Pictures}~-- 3D + ComfyUI workflows.
  \item \textbf{LKDN AI Production}~-- node-based AI filmmaking.
  \item \textbf{Deep Fusion Films} (Cardiff)~-- AI documentary production; acquired by John Gore Studios.
  \item \textbf{Palo Creators} (MrBeast alumni)~-- AI for viral video creation.
  \item \textbf{Refik Anadol Studio Dataland}~-- first museum of AI art.
  \item \textbf{Animaj}~-- Google AI Futures Fund partner for kids' content.
  \item \textbf{Gossip Goblin}~-- AI filmmaking studio.
  \item \textbf{Critterz} (Vertigo + Federation)~-- AI-assisted animated-feature production.
  \item \textbf{Filmology Labs}~-- \$250M studio for vertical micro-dramas and AI-driven media formats.
  \item \textbf{escapeAI} (John Gaeta)~-- streaming apps for AI-generated content across Roku, Fire TV, Samsung and LG.
  \item \textbf{Primordial Soup} (Darren Aronofsky)~-- AI production studio.
  \item \textbf{Toonstar}~-- AI animation house.
  \item \textbf{CinemersiveAbout Labs / Cinemersive Labs}~-- UK machine-learning and computer-vision company; acquired by Sony.
  \item \textbf{My SMASH Media}~-- AI film startup with Allison Gardner (ex-Glasgow Film Festival).
  \item \textbf{Framestore AI Platform / Framestore Futon}~-- ML and GenAI in the VFX pipeline.
  \item \textbf{Mito AI}~-- \$4.5M-raised platform empowering video professionals.
  \item \textbf{Filmax DinoGames}~-- AI-driven 3D animated film project.
  \item \textbf{Abundantia Entertainment + InVideo}~-- AI film studio launched in India.
  \item \textbf{Luma Production Studio + Wonder Project}~-- feature collaboration.
  \item \textbf{Lionsgate Chief AI Officer (Kathleen Grace)}~-- first major-studio CAIO hire.
  \item \textbf{CreateAI}~-- AI animation production.
  \item \textbf{Netflix Ben Affleck AI Studio Acquisition}~-- AI filmmaking company acquisition.
  \item \textbf{DreamLab AI Collective} (the studio that produced this book)~-- AI-augmented creative practice across film, games and immersive.
\end{itemize}

\subsection*{Games-development AI}

\begin{itemize}
  \item \textbf{NitroGen} (NVIDIA + Stanford)~-- plays-any-game model trained on 40,000 hours across 1,000+ games.
  \item \textbf{SIMA~2} (Google DeepMind)~-- agent that plays, reasons and learns in 3D virtual worlds.
  \item \textbf{CHORD} (Ubisoft La Forge)~-- open-sourced end-to-end PBR material generation.
  \item \textbf{Ubisoft Teammates}~-- voice AI for in-game team communication.
  \item \textbf{YouTube Playables Builder} (Gemini~3)~-- text-to-game prompt-to-playable web app.
  \item \textbf{Roblox AI Tools / Roblox AI Assistant / Roblox Reality / Metain}~-- creator-facing AI for game development; Reality is Roblox's productised AI assistant; Metain is the prompt-to-Roblox-Studio interface.
  \item \textbf{Epic xAI Studio}~-- Elon-Musk-affiliated game-dev AI.
  \item \textbf{General Intuition}~-- spatial-reasoning research lab.
  \item \textbf{ReBlink ARBO}~-- AI-powered strategy battler.
  \item \textbf{Halo Studios}~-- AI woven into the production pipeline.
  \item \textbf{Dabgg}~-- AI gaming with interactive platform.
  \item \textbf{RosebudAI / Spawn / LudoAI}~-- vibe-coding games and interactive apps from prompts; Spawn is the ``games made with words'' community platform; LudoAI is the ideation and planning scaffold.
  \item \textbf{Verse8}~-- AI-driven game-creation platform.
  \item \textbf{Sett}~-- AI agents for game marketing.
  \item \textbf{Meta Horizon Studio AI Assistant}~-- Horizon Worlds creator AI.
  \item \textbf{Gambo}~-- vibe-coding game agent.
  \item \textbf{Project Motoko} (Razer)~-- AI-powered gaming headset.
  \item \textbf{Bethesda AI Toolset}~-- Todd Howard's game-dev AI integration.
  \item \textbf{StarBerry Games / Merge Mayor}~-- AI-all-in game studio.
  \item \textbf{Astrocade}~-- interactive entertainment platform; \$56M Series B.
  \item \textbf{GameByte}~-- AI-powered game creation platform; \$1M raise.
  \item \textbf{Studio Atelico Bobium Brawlers}~-- AI-based iOS game with pro-human approach.
  \item \textbf{Bitmagic}~-- AI-created version of classic Civilization.
  \item \textbf{Helika AI Publishing Engine}~-- AI publishing engine for game studios.
  \item \textbf{Xbox Game Studios ``git gud'' AI}~-- Xbox patent for AI tech that plays games for you.
  \item \textbf{Gaming Copilot} (Xbox)~-- console AI copilot.
  \item \textbf{NVIDIA DLSS~5}~-- gaming upscaling.
  \item \textbf{PS5 Pro PSSR}~-- PlayStation upscaling.
  \item \textbf{Gizmo} (Meta acquisition)~-- vibe-coding app for creating mini-games.
  \item \textbf{Voyage} (Latitude / AI Dungeon)~-- AI RPG platform.
  \item \textbf{DoomGen}~-- prompt-driven DOOM mod prototyping app.
  \item \textbf{Sony AI tools for PlayStation}~-- animation support, QA automation, asset generation and performance-capture-to-facial-models pipelines.
  \item \textbf{PROWL} (Odyssey)~-- RL agents for game-world model testing.
  \item \textbf{MagicDawn} (Tencent)~-- AI lighting tech.
  \item \textbf{ReActor} (Disney)~-- RL motion-transfer method.
  \item \textbf{Origin Lab}~-- game-world AI training data; \$8M raise.
  \item \textbf{Daughter of the Inner Stars}~-- Unreal Engine~5 AI-driven game.
  \item \textbf{Genesis AI}~-- robot cooking and piano playing systems (demonstrating embodied agents).
  \item \textbf{Quilty}~-- AI platform for script development and assessment.
\end{itemize}

\subsection*{Marketing and advertising AI}

\begin{itemize}
  \item \textbf{WPP Open Pro}~-- agency-wide AI marketing platform.
  \item \textbf{Sightly}~-- AI advertising and brand performance.
  \item \textbf{Pomelli} (Google Labs)~-- small-business marketing agent.
  \item \textbf{AdCreative.ai}~-- mobile pro ad creation.
  \item \textbf{Epiminds}~-- AI marketing team builder.
  \item \textbf{Octave AI}~-- AI podcast/radio ad creation.
  \item \textbf{AdsGency}~-- autonomous paid-marketing agent.
  \item \textbf{Aimy}~-- AI TV advertising on the Comcast API.
  \item \textbf{Brand Networks}~-- AI television advertising.
  \item \textbf{Fullthrottle.ai}~-- automotive-focused DSP.
  \item \textbf{Channel~4 AI ads}~-- AI-driven TV-advertising tool for SMEs.
  \item \textbf{Sky Studio Challenge}~-- interactive AI ad creation via Starlink.
  \item \textbf{Google Demand Gen}~-- AI-powered image tools for the ad platform.
  \item \textbf{Amazon DSP}~-- AI one-stop-shop advertising infrastructure.
  \item \textbf{Amazon Creative Agent}~-- agentic ad creator.
  \item \textbf{Playad}~-- AI marketing agents.
  \item \textbf{Runway Ad Concepter App}~-- ad-concept and composition exploration.
  \item \textbf{TikTok AI Agents for Ads}~-- TikTok ad-creation agents.
  \item \textbf{Instagram Edits}~-- AI video generation for advertisers.
\end{itemize}

\subsection*{Open-source ecosystem and infrastructure}

\begin{itemize}
  \item \textbf{ComfyUI}~-- node-based AI workflow editor; the operating system of the open-source creative AI ecosystem; \$17M raise (October 2025), \$500M valuation (May 2026).
  \item \textbf{Hugging Face}~-- open-source model hub; partnerships with Google Cloud and Meta.
  \item \textbf{Meta OpenEnv}~-- open-source agentic-development environment.
  \item \textbf{PlayCanvas SOG / SplatTransform}~-- open Gaussian-splat compression and conversion.
  \item \textbf{Civitai / Civision}~-- community-models and LoRA platforms.
  \item \textbf{LoRA / PEFT / DiffSynth-Studio / Qwen-Image-i2L / Llama Factory / AI Toolkit / LTX-2 Trainer}~-- fine-tuning frameworks and training pipelines; DiffSynth turns a single image into a custom LoRA; Llama Factory enables zero-code fine-tuning of 100+ models; AI Toolkit covers LTX-2.3 fine-tuning; LTX-2 Trainer is the fal-hosted variant.
  \item \textbf{PyTorch / TensorFlow / Ollama / LLaMA.cpp / MuJoCo}~-- the open infrastructure layer.
  \item \textbf{Korin AI}~-- Africa-trained, Africa-built foundation model.
  \item \textbf{DecartAI}~-- real-time world transformation.
  \item \textbf{Replicate}~-- model hosting and API marketplace.
  \item \textbf{fal / FAL MCP}~-- generative-model API platform.
  \item \textbf{llmfit}~-- hardware scanner for local-model compatibility.
  \item \textbf{Open Vision Agents}~-- toolkit for building agents that watch, listen and understand video.
  \item \textbf{NotebookLM}~-- Google's research-and-content-generation surface; cinematic video overviews launched March 2026.
  \item \textbf{OpenAI Gym / Gym Retro / ARC}~-- research benchmarks (referenced in the games-AI literature).
\end{itemize}

\subsection*{ComfyUI ecosystem~-- nodes, extensions and workflows}

\begin{itemize}
  \item \textbf{Mesh2Motion} (ComfyUI)~-- mesh-to-motion conversion node.
  \item \textbf{NanoBanana Pro LoRA Dataset Generator}~-- creates training datasets for image-editing models in minutes.
  \item \textbf{ComfyUI Music Tools}~-- professional-grade music-processing node pack.
  \item \textbf{ComfyUI HY-Motion1 plugin}~-- Tencent HY-Motion implementation.
  \item \textbf{ComfyUI Audio Reactive Node Pack} (VisualFrisson)~-- spatial audio-reactive nodes.
  \item \textbf{ComfyUI-BlendPack}~-- nodes for video transitions.
  \item \textbf{KJNodes}~-- use audio as input for LTX-2 inside ComfyUI.
  \item \textbf{WhatDreamsCost ComfyUI Nodes}~-- LTX sequencer, keyframer.
  \item \textbf{Luminance Keyer Node}~-- luminance-based keying.
  \item \textbf{Camera Tracking \& Body Pose tools (ComfyUI)}~-- work-in-progress camera and pose tracking.
  \item \textbf{Multiple Angle Camera Control Node}~-- multi-angle camera control.
  \item \textbf{FeedbackSampler} (Deforum-inspired)~-- iterative feedback sampling.
  \item \textbf{Day-to-Night with Qwen Edit LoRA}~-- day/night transformation workflow.
  \item \textbf{Sora~2 API Nodes}~-- Sora~2 access from ComfyUI.
  \item \textbf{QWEN-AI Compositing}~-- Qwen-driven compositing workflow.
  \item \textbf{Street View URL Parser}~-- Google Street View ingestion.
  \item \textbf{Minim}~-- ComfyUI model-linker extension.
  \item \textbf{Hunyuan 3D~3.0} (ComfyUI partner nodes)~-- HY3D~3.0 inside ComfyUI.
  \item \textbf{Kling~3.0 ComfyUI node}~-- Kling~3.0 partner node.
  \item \textbf{Kling Video~2.6 Motion Control} (ComfyUI)~-- Kling motion control inside ComfyUI.
  \item \textbf{Kling Motion Control~3.0} (ComfyUI)~-- the v3 motion control variant.
  \item \textbf{Seedance} (ComfyUI partner node)~-- Seedance inside ComfyUI.
  \item \textbf{LTX-2} (ComfyUI native support)~-- audio-video model natively supported.
  \item \textbf{LTX-2 ComfyUI Audio-to-Video}~-- audio-driven video inside ComfyUI.
  \item \textbf{ACE-Step~1.5} (ComfyUI)~-- full songs in under 10 seconds on <4\,GB VRAM.
  \item \textbf{ElevenLabs ComfyUI Partner Nodes}~-- ElevenLabs inside ComfyUI.
  \item \textbf{Luma Uni-1 ComfyUI Partner Node}~-- Uni-1 inside ComfyUI.
  \item \textbf{Meshy~6} (ComfyUI native)~-- Meshy~6 inside ComfyUI.
  \item \textbf{Grok Imagine} (ComfyUI)~-- xAI image generation inside ComfyUI.
  \item \textbf{MatAnyone2} (ComfyUI)~-- high-fidelity video matting inside ComfyUI.
  \item \textbf{Tripo~3.1} (ComfyUI partner nodes)~-- Tripo~3.1 inside ComfyUI.
  \item \textbf{NVIDIA RTX Video Super Resolution Node}~-- RTX upscaling node.
  \item \textbf{Depth Anything Custom Node}~-- Depth Anything inside ComfyUI.
  \item \textbf{SAM~3 / SAM~3D ComfyUI integration}~-- Meta SAM~3 / 3D inside ComfyUI.
  \item \textbf{Sora~2 Full Loop Workflow}~-- end-to-end Sora~2 pipeline.
  \item \textbf{Krea Nodes}~-- Krea inside ComfyUI.
  \item \textbf{Gaussian Splats + Qwen Edit workflow}~-- explore 2D images and generate new camera views.
  \item \textbf{ComfyUI App Mode / ComfyHub}~-- turn workflows into shareable no-install apps.
  \item \textbf{ComfyUI Simple Mode}~-- share and iterate complex workflows more easily.
  \item \textbf{ComfyUI Cloud (private models)}~-- private model hosting inside ComfyUI Cloud.
  \item \textbf{ComfyUI Action Director}~-- interactive 3D viewport for ControlNet.
  \item \textbf{Complete Style Transfer Handbook (ComfyUI)}~-- style-transfer guide.
  \item \textbf{VibeComfy}~-- agents that understand, build and run ComfyUI workflows with Claude.
  \item \textbf{MooshieUI}~-- beginner-friendly ComfyUI desktop frontend.
  \item \textbf{Creative Control Using ComfyUI on NVIDIA RTX}~-- NVIDIA's official ComfyUI/RTX tutorial series.
\end{itemize}

\subsection*{LoRAs, fine-tuning and training}

\begin{itemize}
  \item \textbf{QwenEdit-2509 Light Transfer LoRA}~-- simple, powerful image relighting via reference.
  \item \textbf{Qwen-Edit Relighting LoRA}~-- atmospheric relighting control.
  \item \textbf{Qwen-Image-Edit-2511 Multiple Angles LoRA}~-- multi-angle perspective generation.
  \item \textbf{Qwen-Image-Edit-2511 AnyPose}~-- pose transfer from a reference image.
  \item \textbf{QwenEdit-2511 Anything2Real LoRA}~-- realistic-image transformation.
  \item \textbf{Qwen Edit LoRA Object Removal}~-- bounding-box-precision edits.
  \item \textbf{Qwen-Image-Edit 2511 Gaussian Splash 3D Camera Motion}~-- 3D Gaussian-splat motion control.
  \item \textbf{Qwen 2511 Time Travel Workflow}~-- temporal image effects.
  \item \textbf{NextScene (Qwen Image LoRA)}~-- cinematic image sequences with natural progression.
  \item \textbf{Best Face Swap (Flux~2 LoRA)}~-- specialised face-swap.
  \item \textbf{FLUX.2 [klein] LoRAs (fal)}~-- open-source: Outpaint, Zoom, Object Removal, Background Removal.
  \item \textbf{Krea AI LoRA Trainers}~-- train LoRAs for Qwen-2512 and Z-Image inside krea.ai.
  \item \textbf{LTX-2.3 Character LoRA Training (AI Toolkit)}~-- character LoRA training tutorial.
  \item \textbf{Gaussian Splats Repair LoRA}~-- Klein-9b LoRA for repairing 3D views and geometry.
  \item \textbf{LTX~2.3 Colorizer (LoRA)}~-- black-and-white footage colorisation.
  \item \textbf{Music Finetunes in ElevenCreative}~-- stylistically consistent vocal and instrument generation.
\end{itemize}

\subsection*{Provenance, watermarking and detection}

\begin{itemize}
  \item \textbf{C2PA} (Content Authenticity Initiative / Adobe-led)~-- cryptographic provenance metadata standard.
  \item \textbf{SynthID / SynthID Verification} (Google DeepMind)~-- synthetic-content watermark; deployed across Veo, Lyria and Imagen; verifiable inside Gemini.
  \item \textbf{YouTube AI Detection / YouTube AI Deepfake Detection}~-- automated AI-content detection; \emph{Tiny Grandma} false-positive case study.
  \item \textbf{Deezer AI Music Detection}~-- identifies up to 75,000 fully AI-generated uploads per day.
  \item \textbf{Spotify AI Transparency Beta}~-- voluntary disclosure feature.
  \item \textbf{Beeble}~-- detection and watermarking tools.
  \item \textbf{Cloudflare AI Bot Classification}~-- public-web infrastructure tracking AI crawlers.
  \item \textbf{Human Provenance AI Disclosure Standard (Cannes)}~-- industry-coordination labelling standard.
\end{itemize}

\subsection*{Consumer surfaces and distribution platforms}

\begin{itemize}
  \item \textbf{Sora iOS app} (OpenAI)~-- TikTok-style AI video remix; 1M downloads in 5 days.
  \item \textbf{CapCut / Dreamina} (ByteDance)~-- consumer video editing + AI generation.
  \item \textbf{Gemini app} (Google)~-- Google's consumer AI surface.
  \item \textbf{ChatGPT app} (OpenAI)~-- mobile ChatGPT; Shazam integration.
  \item \textbf{Perplexity AI}~-- Samsung Smart TV AI assistant integration; consumer AI search.
  \item \textbf{Grok / Grok Imagine}~-- xAI consumer surfaces.
  \item \textbf{TikTok / Reels / Shorts}~-- AI-content distribution surfaces.
  \item \textbf{YouTube Shorts / YouTube Create}~-- destinations for Veo~3.1 Ingredients-to-Video.
  \item \textbf{Bandcamp}~-- banned AI music outright, January 2026.
  \item \textbf{Deezer}~-- published the 44\%/3\% data; built and licensed AI-music detection.
  \item \textbf{Spotify}~-- declined an AI-music filter; voluntary disclosure beta; \emph{Personal Podcasts} agent feature.
  \item \textbf{YouTube Music}~-- labelling, the AI Detection Tool, viewership of AI content.
  \item \textbf{Steam}~-- AI-content labelling controversy; \emph{Clair Obscur} Game of the Year stripping.
  \item \textbf{Sweden's Official Music Chart}~-- banned AI-generated entries.
  \item \textbf{San Diego Comic-Con}~-- banned AI art at the 2026 event.
  \item \textbf{Roblox}~-- creator-tools AI with continued controversy.
  \item \textbf{Fortnite}~-- Chapter~8 AI-art controversy; Disney partnership.
  \item \textbf{Netflix}~-- AI in production (de-aging, plates, retention engine, Ben Affleck studio acquisition).
  \item \textbf{Disney+}~-- user-generated AI content features.
  \item \textbf{Luna} (Amazon)~-- AI-powered Snoop Dogg game.
\end{itemize}

\subsection*{Studios, programmes, festivals and institutional infrastructure}

\begin{itemize}
  \item \textbf{Sundance AI Literacy Initiative}~-- Google-funded \$2M creator-training programme; Sundance Collab learning, Story Forum, Ignite Day.
  \item \textbf{UCL / RCA / Brandtech Centre for Creative AI}~-- UK research hub.
  \item \textbf{AIMICI Training Program}~-- ScreenSkills-accredited AI training.
  \item \textbf{AI FilmFest Japan}~-- annual AI-film festival.
  \item \textbf{Dubai AI Film Award}~-- \$1M prize won by Tunisia's \emph{Lily}.
  \item \textbf{Cannes AI Disclosure Standard / Cannes Lions}~-- industry coordination for AI labelling; Luma Dream Brief \$1M prize.
  \item \textbf{Future Vision Film Competition}~-- Google / XPRIZE / Range Media collaboration.
  \item \textbf{Runway AI Festival}~-- film and creative-discipline awards.
  \item \textbf{Wonder Film Festival}~-- AI-native film festival.
  \item \textbf{Academy of Motion Picture Arts and Sciences}~-- ``you must be human to win'' rule.
  \item \textbf{Television Academy / Emmys}~-- AI guidance for submissions.
  \item \textbf{SAG-AFTRA}~-- contract negotiations and the Tilly Tax.
  \item \textbf{Equity (UK)}~-- strike ballot; 99\% vote in favour.
  \item \textbf{PRS for Music AI Survey 2026}~-- UK music-creator sentiment data.
  \item \textbf{GEMA}~-- German rights society; won against OpenAI.
  \item \textbf{Splice + UMG}~-- sample library; AI-music tools partnership.
  \item \textbf{Stability AI + Universal Music / Stability AI + Warner Music}~-- alliance and deal on the AI-music side.
  \item \textbf{Vizrt Viz One~8.1}~-- broadcast AI features.
  \item \textbf{Tesseract}~-- AI platform at MIPCOM.
  \item \textbf{Sphere + Google Cloud}~-- AI technology behind the \emph{Wizard of Oz} experience.
  \item \textbf{Anthropic Joins Blender Development Fund}~-- patronage deal, May 2026.
  \item \textbf{Sony AI ICASSP papers}~-- music-understanding and generative-audio research, April 2026.
  \item \textbf{Google Flow Music + Believe Partnership}~-- AI music distribution partnership.
\end{itemize}

\subsection*{Techniques, methods and recurring workflows}

\begin{itemize}
  \item \textbf{Vibe Coding}~-- prompt-led creative prototyping, especially for games and interactive apps; Rosebud, Spawn, Gizmo, Gambo are the canonical surfaces.
  \item \textbf{Trajectory Control (ATI)} (Wan / Qwen)~-- control of video trajectories at frame and clip level.
  \item \textbf{Reference-to-Video} (Veo~3.1 + Gemini)~-- consistent character generation from a reference image.
  \item \textbf{Motion Control nodes / Time-To-Move}~-- frame- and segment-level motion controls in generated video.
  \item \textbf{Layer Separation \& Live Editable Text (LET)} (Loveart)~-- extract and edit text and subjects from images as PSD layers.
  \item \textbf{Temporal Reasoning} (NVIDIA ChronoEdit)~-- physics-aware edits and world simulation.
  \item \textbf{Practical Light Control} (PractiLight)~-- foundational diffusion for relighting.
  \item \textbf{3D Gaussian Splatting}~-- the spatial backbone running through PlayCanvas SOG, Hyperscape, Apple Personas and the open-source splat ecosystem.
  \item \textbf{World Models}~-- recurring topic spanning Marble, Genie, HY-World, UNI-1, ECHO, Matrix-Game, LingBot-World, Moonlake.
  \item \textbf{Multi-Shot Consistency}~-- Kling~3.0, ShotVerse, Higgsfield Shots, Veo~3.1~-- the multi-shot turn that defined the spring-2026 video pipeline.
  \item \textbf{Camera Control / Multi-Angle Generation}~-- WAN Camera Control, ShotVerse, ATI, Multiple Angle Camera Control Node.
  \item \textbf{Style Transfer (the ComfyUI handbook)}~-- the canonical workflow ladder for style transfer.
  \item \textbf{Audio-Reactive Generation}~-- Audio Reactive Node Pack and the Lalal-API / ACE-Step pipelines.
  \item \textbf{Real-Time Generation Pipelines}~-- Krea Realtime, MotionStream, Decart Lucy~2.0, Xmax X1, Phoenix-4, LTX-2.3 Real-Time on Scope; the \emph{interactive-generation} category that emerged in Q1 2026.
  \item \textbf{Agentic Workflows / Multi-Agent Stacks}~-- Sony Game Studios (49 agents, 72 skills), Adobe CX Enterprise, OpenAI AgentKit, Heygen Video Agent, Claude Skills.
  \item \textbf{Provenance-First Capture}~-- C2PA + SynthID + on-device detection in Premiere and YouTube as the emerging \emph{standard} of provenance-aware production.
  \item \textbf{Music Stem Separation as a primitive}~-- Lalal AI, StemDeck, BandLab~-- stem separation moving from end-product to upstream primitive.
  \item \textbf{MCP (Model Context Protocol) as the connective tissue}~-- Unity MCP, Blender MCP, fal MCP; the protocol the agentic creative ecosystem is converging on.
  \item \textbf{Brief-First, Generate-Second}~-- the workflow practice the book argues for in Chapter~\ref{ch:11} (and Chapter~\ref{ch:16}'s \emph{toolchain in layers} section).
\end{itemize}

This is the catalogue. By the time you read it, it will be incomplete~-- new tools have shipped, some on this list have been bought, renamed or killed. Treat it as a snapshot of one year of toolchain at the moment the toolchain became a stack rather than a list, and use it to orient yourself in whatever the state of play is when you pick the book up.

For a deeper analytical treatment of the adoption telemetry behind many of these tools~-- Firefly's 22B-asset growth curve, ChatGPT's 800--900M WAU figures, the Veo / Sora professional split, the GDC sentiment-vs-usage divergence~-- see Appendix~E: \emph{Dynamics of Generative AI Adoption}.

\section*{How to build a toolchain you can defend}

The last thing I want to say in this chapter is something I have said in talks more often than anything else, because working creatives ask me this question, in some variant, every week.

\emph{How do I decide what tools to use, in a market that is changing this fast, without burning my whole month re-learning interfaces?}

My short answer is: build the toolchain in \emph{layers}, and accept that the layers move at different speeds.

The bottom layer~-- your foundation model, your modality stack, your agent platform~-- will change frequently. Treat it as ephemeral. Pick the best tools available \emph{this quarter} and be ready to swap them next quarter.

The middle layer~-- your creative software (Adobe / Unreal / DaVinci / Pro Tools / Blender / Logic)~-- will change more slowly, and is the layer in which the AI capability will be progressively absorbed. Treat it as the long-term home of your craft. Learn it deeply.

The top layer~-- your \emph{judgement}, your \emph{taste}, your \emph{briefing skill}, your \emph{integration sense}~-- does not change. It is the layer the agents cannot copy, the layer the platforms cannot ship and the layer the next model release does not depreciate. Spend more time here than the toolchain wants you to.

The mistake I see working creatives make most often is to over-optimise the bottom layer and under-invest the top. The platform companies want you to spend your working hours chasing the new model release; the work that pays the bills, the work that finds an audience, and the work that survives a transition is built on the layer the platform companies cannot reach.

The toolchain, in the end, is the means. The work is the end. The tools change. The work, if it is any good, lasts.

That is the working operating model my studio has run on for the period this book covers. It is the model I would commend to anyone building, this year or next, a creative practice that survives the rest of the decade.

The transition is going to keep going. The tools will keep changing. The work that matters, on the other side, will be made by the people who kept their attention on the right layer.

  \chapter{Five Years Inside the Dream Machine}\label{ch:17}

\lettrine[lines=3,lhang=0.15,findent=0.1em]{I}{ } called the newsletter \emph{Dream Machine} in late September 2025 because the phrase caught something I couldn't yet articulate. I have said elsewhere in this book~-- in Chapter~\ref{ch:15}, and in the front matter~-- what I think the phrase has come to mean over the period the book covers: an apparatus capable of producing what until recently required a human mind, amplifying and distributing the dreaming of the humans who direct it.

What I have not done, anywhere else in the book, is let myself off the leash about what the apparatus might \emph{become}. I have been careful, chapter by chapter, to put the evidence in front of the argument. The predictions in Chapter~\ref{ch:15} are dated, falsifiable, defensive~-- written so the reader picking up the book in 2030 can grade them with a stopwatch.

This chapter is the opposite. This is the chapter where I let myself imagine what the next five years might look like if the trajectory of the previous eight months extends, accelerates, breaks and recombines in the ways I half-suspect it will but cannot, in any conventional analytical register, \emph{prove}.

I want to be honest about the rules I am giving myself for this chapter, because they are looser than the rules the rest of the book has run on.

The rule is: each scenario has to be \emph{argued from} something already in the book~-- a chapter, a number, a framework, a quoted source. I am not allowed to introduce a new mechanism out of thin air. But I am allowed to take a mechanism that is, in spring 2026, present in small or early form, and ask what it looks like by 2031 if it follows the curve I think it is on.

That is the contract for the next thirty pages. Wild but rooted. Speculative but cited. The kind of writing that the rest of the book would not have permitted, and that the closing letter to 2030 in Chapter~\ref{ch:epilogue} would deflate if I tried to put it there.

If a reader in 2031 has bought this book to grade my predictions, this is the chapter where you can grade me most harshly. Some of what follows will look, by then, embarrassingly off. Some of it will look, in the unflattering retrospect that books-on-transitions always get, banal~-- the obvious thing nobody had quite said yet. The interesting category~-- the predictions that turn out to be \emph{roughly the right shape, badly miscalibrated on timing}~-- is the one I am writing for.

Six scenarios. Then the upside I am most hopeful for. Then the downside I am most afraid of. Then a handoff to the letter that closes the book.

\section*{One. The Petrillo settlement actually happens}

In May 2026, the structural argument of Chapter~\ref{ch:6}~-- that the institutional response to the AI training problem will, on the historical pattern, converge on the \emph{levy-and-redistribute} mechanism James Caesar Petrillo built in 1948 for recorded music~-- is still, on the page, an argument from analogy. The 88\% is the political mandate. The GEMA ruling is the legal precedent. The Stability / UMG-style alliances are the commercial templates. The Musical AI creative-weight-attribution work is the technical infrastructure. None of it has yet been \emph{assembled} into a working settlement.

By 2031, I think the assembly is done. Not perfectly, not globally, not without exclusions~-- but done, in at least one major jurisdiction, as a working mechanism that a working creative encounters on a bank statement.

The shape of the thing I expect: a national or supranational \emph{AI Performance Trust Fund}, modelled on the MPTF, capitalised by a statutory per-output levy on commercial generative output, governed by a joint labour--platform body, distributed to working creatives whose training-data contributions were identifiable through the creative-weight-attribution layer described in Chapter~\ref{ch:12}. The first cheques are small. A working illustrator opens a statement and sees the line \emph{AI Royalty Distribution: \pounds 43.18}. A session musician sees \emph{\pounds 127.42}. A photographer whose backlist sat in a Getty-class licensed dataset sees a meaningful four-figure annual sum.

The cultural significance of those small numbers will, I think, be larger than the numbers themselves. The first cheque is the moment the 88\% becomes a \emph{fact} rather than a \emph{demand}. The mechanism is on the books. The architecture is real. The question stops being \emph{whether there will be a Petrillo settlement} and starts being \emph{how much, to whom, on what calibration}.

The bargaining ground for the next decade is set on that question.

\section*{Two. The internet bifurcates, formally}

Chapter~\ref{ch:4} made the case that the web of 2026 was, in measurable ways, splitting into two distinct attention environments~-- the \emph{Dead Internet} of synthetic content optimised for synthetic attention, and the \emph{Living Web} of provenance-stamped, human-verified work made for an audience that had organised itself around the difference.

By 2031, I think the bifurcation is no longer a cultural metaphor. I think it is \emph{operational infrastructure}.

The provenance stack the book has spent so many pages cataloguing~-- C2PA, SynthID, the Cannes Disclosure Standard, the AP and BBC wire-service signing chains, the Adobe content credentials~-- will, by 2031, have stitched together into a functioning verification layer. Major browsers will render an indicator. Major platforms will, under pressure from advertisers tired of paying for bot-on-bot impressions, surface verified-human content separately from synthetic. A handful of \emph{human-only} subscription services~-- I would not bet against the BBC, \emph{The New York Times} and one or two of the streaming-music incumbents launching first~-- will offer ``no synthetic content ever'' as a paid product.

The split internet is not, in the version I expect, a \emph{clean} split. The synthetic layer will dwarf the verified layer in volume, by perhaps a hundred to one. But the verified layer will capture an outsized share of \emph{paid attention}, \emph{advertiser dollars}, and~-- most importantly~-- \emph{cultural credit}. The slop ceiling of Chapter~\ref{ch:5}, measured at 44/3 on Deezer in April 2026, will by 2031 have hardened into a roughly stable ratio across most major content categories: synthetic supply dominant on the upload side, verified-human attention dominant on the consumption side.

The Living Web is, by 2031, no longer an aspiration. It is a \emph{distribution layer} you can buy access to, with a different economics from the synthetic layer running underneath it.

\section*{Three. World models replace flat video as the default high-end medium}

Chapter~\ref{ch:8} argued that the most important technical shift of 2025--26 was not the video models but the world models~-- Marble, Genie 3, the Hunyuan 3D family~-- and that the rate at which spatial-AI tooling was moving from research demos to consumer product was the underappreciated structural story of the period.

By 2031, I think we look back on flat 24-frame video the way 2026 looks back at black-and-white silent cinema. Still made. Still loved. Still occasionally the right medium for the work. But not the \emph{default}.

The default high-end production format, by 2031, is a \emph{navigable spatial render}~-- a world that the camera moves through in post, that the audience can choose to view from any angle through a Vision-class headset, that an editor can re-cut from a different perspective two months after principal photography is done. The boundary between film, games, immersive and live performance that Chapter~\ref{ch:8} described as eroding is, by 2031, functionally gone for new productions.

The career implications, for working filmmakers, are substantial. The skillset of the director-of-photography fragments into world-curation, lighting-prompt design, and post-spatial composition. The skillset of the editor expands into multi-angle viewer choreography. New roles emerge~-- \emph{spatial continuity supervisor, world dramaturg, perspective director}~-- that have no clean analogue in the 2026 production pipeline.

This is also where I expect the legacy industries' Chapter~\ref{ch:7} strategic positioning to come back to haunt them. The studios that bet on flat AI-generated video as the next medium will discover, between 2027 and 2029, that they bet on the second-to-last format of the previous era. The studios that built world-model fluency into their pipelines~-- Sony's documented experiments, the AI-native studios like Imaginae and Critterz, a handful of regional players in Korea, Japan and India~-- will be the dominant 2031 producers of \emph{the kind of work that uses the medium the way it actually works}.

\section*{Four. The orchestrator becomes a credentialled guild}

The orchestrator role~-- central to Chapter~\ref{ch:11}, referenced through Chapter~\ref{ch:13} and Chapter~\ref{ch:14}~-- was, in 2026, a \emph{description of an emerging practice}. The Sony 49-agent / 72-skill team was the most-cited case study. The working title circulated in studios and agencies but did not yet carry contract weight.

By 2031, I think the orchestrator is a \emph{credentialled profession} with collective bargaining power.

The guild structure I expect~-- and this is the most institutionally specific of the six predictions, so the most likely to be wrong on detail~-- will look something like the IATSE locals that govern below-the-line film labour, or the Writers Guild's apprenticeship and credit system. \emph{Senior orchestrators} will carry the credential, demonstrate competence in agent stewardship across a defined toolchain, negotiate over rates and over the \emph{quantity of agent labour} a single orchestrator can supervise. The Writers Guild will, I think, be among the first to formalise the role; SAG-AFTRA, Equity and the games-development unions will follow within eighteen months of whoever moves first.

The collective-bargaining angle is the part that will, in 2031, look most novel and most obviously inevitable in retrospect. If a single human orchestrator can supervise the work of, say, twenty agents~-- a number that today's productivity research is pointing at, with substantial variance~-- then the question of \emph{who pays whom for the productivity gain} is exactly the Petrillo question from earlier in this chapter, applied to the orchestrator's own labour rather than to the underlying training data. The answer the guild will negotiate is a \emph{productivity share} of the agent-team output, with the orchestrator's individual rate tracking the value of the supervised work, not the volume of the orchestrator's keystrokes.

This is one of the more concrete reasons I am cautiously optimistic about the working-creative position over the next five years. The job that emerged in 2026~-- the maker-as-orchestrator~-- is a job for which collective bargaining is \emph{unusually well-suited}, because the value being created is unambiguously human in origin and unambiguously measurable in output. Unions exist for exactly this kind of bargaining problem. The Petrillo template, again.

\section*{Five. An AI-native studio wins the Palme d'Or before Hollywood does}

This is the prediction that will, depending on the reader, look either obviously safe or wildly provocative in 2031.

Chapter~\ref{ch:7} catalogued the new AI-native studios~-- Imaginae, Wonder, Asteria, Obsidian, Critterz, Gossip Goblin~-- and made the case that their structural advantage over the legacy studios is not technical. It is \emph{cultural}. They have no calcified rules to unlearn. They have no inherited risk-aversion. They have no franchise-template gravity to pull them back to the engine-optimal move. By the chess-grandmasters analogy in Chapter~\ref{ch:15}, they are, by default, in a better position to play the move the machine would not have generated.

By 2031, I think one of them~-- or, more likely, a successor company emerging from a director who learned in their orbit~-- wins the Palme d'Or, the Golden Bear, the Silver Lion or one of the other top European festival awards. \emph{For its writing or direction. Not for its technology.}

The cultural moment when that happens will be larger than the prize itself. It will be the \emph{Petrillo Settlement} of cinema-craft acceptance: the moment the question shifts from \emph{can AI-native cinema be art} to \emph{which AI-native film will be canonical}. The follow-on year, I expect the Academy~-- having spent 2025--28 reinforcing its \emph{you must be human to win} rule for above-the-line crafts~-- to add a parallel category, or modify the existing one, to recognise hybrid human-orchestrator-AI authorship under contested but workable rules. The festival juries will get there before the Academy does. They almost always do.

The corollary, for the legacy studios, is bleak. The diagnosis in Chapter~\ref{ch:7}~-- that Hollywood, commercial music and AAA games spent fifteen years optimising toward the engine-optimal move and arrived at the AI moment producing exactly the work the engines can now replicate cheapest~-- will be on full display when the first AI-native Palme winner is, recognisably, \emph{not} a tentpole. By 2031, the Hollywood studio system will be in the position the major-label music industry was in around 2008: still dominant by revenue, visibly losing the cultural argument, scrambling for the next operating model.

\section*{Six. The platform monopoly cracks}

The reason I am giving this prediction last among the six is that it is the one I hold most loosely. The structural argument is clean. The timeline is the thing I cannot pin.

By 2031, I do not think the OpenAI / Anthropic / Google triad of frontier-AI dominance will look the way it does in May 2026. The pressures pushing against it, catalogued in different chapters of this book, are:

\begin{itemize}
  \item The open-weight ecosystem~-- Tencent's Hunyuan, Alibaba's Qwen and Wan, DeepSeek, Meta's Llama, Mistral, Stability~-- collectively used by, per Chapter~\ref{ch:16}, some 80\% of startups pitching the major venture funds by spring 2026. Open weights compound. Closed weights, by structure, do not.
  \item The consent-trained category from the new section in Chapter~\ref{ch:6}~-- Adobe Firefly, Bria, Getty's licensed model, Moonvalley Marey, AIODE, the Stability / UMG-style alliances~-- which, by 2031, has either \emph{forced} the frontier labs to license their training data retroactively, or has \emph{displaced} them in the enterprise procurement environment by carrying indemnities the frontier labs cannot match.
  \item The sovereign-model trend, only just starting in 2026, in which national governments~-- I would put France, Korea, India, the UAE, Brazil, Japan on the early list~-- fund consent-trained, citizen-licensed foundation models as a strategic asset, governed under sovereign rules, distributing royalties through national rights-holder bodies.
  \item The bifurcating internet from earlier in this chapter, which by 2031 makes \emph{which model trained on what} a visible product-marketing claim, in the way \emph{fair-trade-certified} or \emph{organic} became visible product-marketing claims in food.
\end{itemize}

By 2031, the AI model market I expect looks much more like the 2024 cloud market than the 2025 AI market. Multiple major model families. Sovereign options. Open-weight defaults at the long tail. Indemnification and provenance as standard procurement requirements. The frontier-lab valuations of 2025 either justified by deep enterprise penetration in narrow categories, or remembered the way AOL's 1999 valuation is remembered.

The corollary I want to put on the page, because it has not yet been said cleanly anywhere else in this book: the platform companies dominant in May 2026 are, on the historical pattern, \emph{unlikely to be the platform companies dominant in 2031}. The history of computing is that the dominant platform of one era is rarely the dominant platform of the next. The Wintel of the 1990s, the iOS / Android of the 2010s, the AWS / Azure of the 2020s~-- each was a dominant pairing whose successor came from a direction the incumbents did not see. Whatever the dominant generative-AI platform of 2031 is, I think there is a non-trivial chance it is a company most readers of this book~-- including me~-- have not yet heard of.

\section*{The upside I am most hopeful for: the Dream Machine becomes literal}

I want to spend the rest of this chapter on two scenarios that do not fit cleanly into the \emph{predict and date} register, because they are the ones I think about most when I am away from the work and most off-script when I am in it.

The first is what happens if the title of this book turns out to be a literal description.

The phrase \emph{Dream Machine}, as I have used it through the newsletter and the book, has been a metaphor for the apparatus of generative AI. A composite name for the platforms, models, infrastructure and human labour that, between them, produce the new creative work.

I want to entertain the possibility that, by 2031, the phrase describes a \emph{literal piece of consumer hardware}~-- a wearable creative-cognition prosthetic, sitting somewhere on the continuum between a Vision Pro and a brain-computer interface, that watches a working creative make work, learns their taste, infers their intent, and contributes~-- in real time, in collaboration, at the speed of thought~-- to the work in progress.

The technical building blocks are visible. Apple's Vision Pro and the next-generation headset class are mature consumer products by 2031. The brain-computer interface category, dominated in 2026 by medical applications, is on a trajectory the consumer hardware press has been undercounting. The agentic orchestration layer that I described in Chapter~\ref{ch:11} is, by 2031, no longer something a human types instructions into. It is something a human \emph{converses with}, \emph{gestures to}, \emph{thinks at}. The interaction model of \emph{prompting} is, by 2031, an obsolete UX pattern, remembered the way command-line interfaces are remembered now.

The literal Dream Machine, in this scenario, is not a separate device. It is the orchestration layer made \emph{embodied}. A creative cognition prosthetic that allows a working creative to externalise, manipulate, sketch, iterate and finish creative work at a fluency that is, today, available only to the smallest set of professionals who have spent thirty years building the relevant neural pathways. The leverage the device provides is not \emph{replacement} of the human imagination. It is the closing of the gap between the human imagination and what the human hand can, today, get out into the world.

If this device~-- or one like it~-- ships at consumer price in the period this chapter covers, the \emph{access} principle from Chapter~\ref{ch:15} gets a lift I cannot easily exaggerate. The bedroom-Hollywood dynamic from Chapter~\ref{ch:14}~-- the teenager with a midrange GPU producing studio-quality output~-- extends to \emph{every} discipline, including the ones that today still require expensive equipment and decades of haptic training. The new geography I argued for in Chapter~\ref{ch:15}~-- the dispersion of canonical creative work to places the previous century forgot~-- becomes structurally inevitable rather than aspirational.

The Dream Machine, in this scenario, is not the platforms. It is the \emph{prosthetic}. The platforms are infrastructure underneath it. The dreaming is, finally, located where the title was always insisting it was located: in the human at the centre of the apparatus.

I want this scenario to be the one that comes true. I think there is a real chance~-- not a high probability, but a meaningfully non-zero one~-- that it does.

\section*{The downside I am most afraid of: the audience becomes synthetic}

The corresponding worst case I want to put on the page is the failure mode the slop ceiling cannot, by construction, defend against.

The slop ceiling, as I described it in Chapter~\ref{ch:5}, works because \emph{the audience can tell}~-- at the speed of a swipe, in the aggregate, with reliability~-- that synthetic content does not carry the human signal real human attention is calibrated to. The ratio holds, the 44/3 holds, the audience underweights the slop, because there is an audience, and the audience is made of humans, and the humans are still, in May 2026, paying attention to one another.

The downside scenario is that, by 2031, that last condition has decayed.

The mechanism is the one the Dutch researchers cited in Chapter~\ref{ch:5} and the bot-traffic investigations cited in Chapter~\ref{ch:4} were already, by spring 2026, beginning to document. As synthetic content production becomes free and synthetic attention-allocation agents become widely deployed, the \emph{measured} audience for any given piece of content drifts further from the \emph{human} audience. Platforms optimise against the measured audience, because that is what their analytics surface. The content the platforms surface, in turn, becomes the content the synthetic audience underweights least. The synthetic audience grows. The human audience, exhausted, exits. The platforms continue to grow on the measured numbers because the measured numbers are increasingly bot-on-bot. The cultural production layer continues to produce, but produces \emph{for bots watching bots}. The work loses meaning slowly, then quickly, because the audience that produced meaning has stopped being part of the loop.

This is the version of 2031 I am most afraid of. Not because AI replaces creators. Creators are not the load-bearing structure. \emph{The audience is the load-bearing structure}, and the audience is the thing the architecture of the synthetic internet, deployed without the verification layer described in scenario two of this chapter, is structurally positioned to dissolve.

The corresponding fight, the one I think the next five years has to win to avoid the downside, is the \emph{audience verification} fight. The C2PA-equivalent for \emph{who is watching}, not just for \emph{who made}. The mechanism that lets a creator know~-- and lets an advertiser pay for, and lets a platform surface~-- the work that \emph{real humans were really paying attention to}. The early signals in this direction are there, in the form of platform attestation experiments and the bot-traffic regulatory pressure documented in Chapter~\ref{ch:4}. None of them are, in May 2026, anywhere near the scale of the upstream provenance work.

If the next five years build the verification layer for production but not for consumption, the slop ceiling holds in the short term and collapses in the medium term. The creators win the upstream battle. The audience disappears anyway. The downstream meaning of the work, by 2031, is gone~-- not because human creativity has been replaced, but because the social fact that creativity is made \emph{for} somebody has been quietly dissolved underneath the work.

This is the failure mode I think we are most exposed to and least talking about.

\section*{What we make of it}

I want to close this chapter the same way I open the next.

The phrase \emph{Dream Machine} was, when I named the newsletter, a placeholder for a feeling. By the end of Chapter~\ref{ch:15} it had become an argument: that the apparatus we are building amplifies, multiplies and distributes the dreaming of the humans who direct it, and that the question every working chapter of this book has been about is \emph{whose dreams the machine amplifies}.

What this chapter has tried to do is take that argument out to five years and let it run.

Six things I think happen. The Petrillo settlement, made real, on a working creative's bank statement. The internet, formally split. World models, displacing flat video as the default. The orchestrator, credentialled and bargained-for. The first AI-native Palme d'Or. The platform monopoly, cracked. One upside I am hopeful for~-- the Dream Machine made literal, a creative cognition prosthetic that closes the gap between human imagination and finished work. One downside I am afraid of~-- the audience layer collapsing under synthetic capture, leaving the work without anyone to make it \emph{for}.

Three of the six will, almost certainly, be wrong. One of them will be wrong in the direction of \emph{I undersold this}. One of them will be wrong in the direction of \emph{the timeline was slower than I thought}. One of them will be wrong in the direction of \emph{the thing I described did not happen because the thing that happened instead was weirder and more interesting}. I do not know yet which is which.

The point of the exercise is not to be right about the six. The point is to put a \emph{shape} on the page, in 2026, for the conversation that the working creatives, the studios, the unions, the policy people and the audience will be having with each other, week by week, between now and 2031. The four principles from Chapter~\ref{ch:15}~-- \emph{agency, attribution, access, audience}~-- are the test you apply to each daily decision. The scenarios in this chapter are the \emph{direction} the aggregate of those decisions is, on my read of the available evidence, most likely pointing in.

The title of this book, I have come to think, was always slightly mis-named. The Dream Machine is not the apparatus the platforms built. It is not even the prosthetic the consumer-hardware companies might ship by 2031. The Dream Machine is the \emph{whole system}~-- the platforms, the prosthetics, the working creatives, the audience, the unions, the institutions, the regulators, the literacy initiatives, the newsletters, the festivals, the awards, the lawsuits and the daily decisions of millions of people about whose work to make and which work to pay attention to. The thing the title points at is the \emph{coupled human-machine cultural production system} that the period this book covers brought into existence.

What we make of that system, between May 2026 and the moment a reader picks this book up in 2031 to grade me, is the work.

The letter in the next chapter is what I want to leave that reader with.

Welcome to the Dream Machine.

  \chapter{Epilogue}\label{ch:epilogue}

\begin{center}\large\emph{A Letter from the Dream Machine}\end{center}

\lettrine[lines=3,lhang=0.15,findent=0.1em]{T}{o} \emph{the creative person reading this in 2030, picking the book up out of curiosity or for a class or because someone older than you said you should:}

I want to write you a short letter, because by the time you are reading this, more than half of what is in this book will already be wrong.

Some of the tools I have spent chapters discussing will be obsolete. Some of the companies will have been bought, broken up or sunk. Some of the law I described as new and contested will have settled into precedent that everyone takes for granted. Some of the people I quoted as authorities will have changed their positions, and some of the predictions I made~-- the ones I let myself make~-- will have aged in ways I would now find embarrassing to read back.

This is fine. It is, I think, the right outcome for a book written in the moment that this one was written in. The job of a book about a transition is not to predict where the transition lands. The job is to \emph{describe the moment honestly}~-- what was being argued about, by whom, with what evidence, and what was at stake~-- so that the reader picking it up later has, at minimum, an accurate record of \emph{what the people in the room thought they were doing.}

There are a few things I want to flag for you, looking back from your time at mine.

\section*{What we got right}

We got the slop ceiling right. I am almost certain of this, even sitting here in May 2026, because the underlying mechanism~-- that audiences allocate attention to \emph{meaning}, not to \emph{quantity}~-- is one of the most reliable findings in the history of cultural production, and there is no plausible scenario in which it stops working.

If you are reading this in 2030 and there is a thriving creative economy, it is because the slop ceiling held. The audience that turned away from synthetic content in 2026 either kept turning away, or the platforms and the creators figured out how to make AI-augmented work that genuinely earned attention~-- sincere, transparent, made with care. Both outcomes preserve the underlying contract between maker and audience. Either way, the ratio between volume and attention is still doing its work.

If, on the other hand, you are reading this in 2030 and the creative economy looks bleaker than the one I have described, it is~-- almost certainly~-- because the architecture of the internet was allowed to drift further in the direction the Dutch researchers identified in 2025: an attention market optimised against meaning, where the ratio between volume and attention stopped functioning because the \emph{audience itself} had become indistinguishable from the bots.

The slop ceiling was real. The question for your decade is whether you protected the \emph{audience} that produced it.

\section*{What we got wrong}

We got the timeline wrong, in both directions.

In some places, we expected the change to be slower than it turned out to be. The pace at which world models moved from research demos to consumer features in the eight months I documented in this book was something almost nobody~-- including me~-- would have predicted in late 2024.

In other places, we expected the change to be faster. The full AI-native film studio that was supposed to replace Hollywood by 2027 is still, in the spring of 2026, mostly a series of well-funded prototypes. The fully autonomous game-development pipeline is, even at Sony, still an \emph{assisted} pipeline rather than an \emph{autonomous} one. The headline futures the platform companies have been selling are, in the main, still further off than the press cycle suggested.

If you are reading this in 2030, I suspect that the timeline gap will look in retrospect like the most obvious thing about the moment we were in. The interesting changes were the ones nobody put on a slide. The boring changes~-- the slow institutional repositioning, the contract-by-contract renegotiation of how creators are paid, the patient construction of the C2PA standards, the union by union recalibration of what counts as a fair use of a performer's likeness~-- were the ones that actually shaped your decade.

This is, I think, almost always true of technology transitions. The press loves the visible. The structural change happens in the unvisible. The fact that you can pick this book up in 2030 and read it without unusual cost or effort is the result of a thousand small infrastructure decisions, made by people whose names did not appear in this book, that none of us thought to celebrate at the time.

\section*{What I want you to know about us}

We were, in 2026, frightened. Most of the working creatives I knew that year were carrying a layer of low-grade fear~-- about their livelihoods, about their crafts, about the institutions that had organised their working lives~-- that I do not think showed up in the trade press or the platform keynotes or even in this book as much as it should have.

We were also, I think, more hopeful than we said out loud. Most of the working creatives I knew that year were also doing some of the most experimental, most curious, most adventurous work of their careers, because the tools made things possible that had been impossible the year before and they wanted to find out what.

Both states were true at the same time. They will be true again in your decade, about whatever the next transition is. I don't think we managed the fear well. I think we did okay with the hope. I am proud of the people who turned up, week after week, to argue about how all of this should go.

\section*{A specific request}

If the institutions of the generative creative economy~-- the unions, the rights bodies, the consultations, the literacy initiatives, the disclosure standards, the festivals, the regional studios, the indie filmmakers in places that the previous century forgot~-- are still doing their work when you read this, \emph{thank the people who built them.} Most of those people have not become rich, or famous, or particularly comfortable. They were working creatives, union officers, policy officers, festival organisers, technologists, lawyers, professors and freelancers who chose, in their unglamorous moments, to do the slow institutional work that this kind of economy needs.

The platform companies will not remember them. The audience will not know their names. The cultural histories will record the names of the directors and the songwriters and the celebrity AI controversies. The institutional work happens in the basement and the archive and the union office and the policy briefing.

If you are inheriting a creative economy that works, you are inheriting it from those people. Thank them while they are still around to be thanked.

\section*{A specific request to working creatives in 2030}

Keep going.

The pace of change will be at least as fast in your decade as it was in mine. The next paradigm of AI tooling~-- whatever it is~-- will, I am sure, make the world models and agents of 2026 look as quaint as Photoshop in 1996. You will have your own version of \emph{the day Sora landed.} You will have your own \emph{Tilly Norwood week.} You will have your own \emph{88\%.} You will have your own choice between the extractive and the generative economy.

Whatever the new tools are, the principles I tried to articulate in Chapter~\ref{ch:15}~-- \emph{agency, attribution, access, audience}~-- will, I am confident, still apply. They are not specific to the 2026 transition. They are general properties of how a humane creative economy organises itself, and they will work in any technology environment that produces creative outputs at scale.

The deeper conviction underneath the principles will, I think, also still apply. The age of the \emph{How}~-- the long century in which the central question of working creative life was \emph{can you do the thing?}~-- was, by 2026, visibly ending. The age of the \emph{Why}~-- taste, intent, authenticity, the willingness to take a risk on the move the data does not endorse~-- was visibly starting. The image I borrowed from elite chess in Chapter~\ref{ch:15}, of grandmasters winning by deliberately playing the move the engine would not have played, will still be the picture I would commend to you. By the time you read this, the engines will be better. The pressure to play the engine's optimal line will be stronger. The competitive advantage of the \emph{deliberately un-machine-like} move~-- of the move that is yours because no machine, trained on what came before, could have generated it~-- will, if anything, have widened.

Apply them. Argue for them. Build for them. Defend them when they are under pressure.

And~-- this is the part I want you to take seriously even when it feels grandiose~-- \emph{write the newsletter}. Whatever your version of the newsletter is. The thing that, when the moment comes, you sit down on a Monday morning and decide to write because nobody else seems to be doing it. Most of those newsletters won't matter. A few of them will. The one you are reading was, for me, one of the most consequential decisions I ever accidentally made.

You will know which moment is the moment, when it arrives. You will know because the people you respect will be staring at the same news cycle, looking at each other in the same way, asking the same question.

When that moment arrives in your decade, write the newsletter.

\section*{Closing}

I want to end this letter the same way I have ended every issue of \emph{Dream Machine} for the past six months, because the line has come to mean more to me than I expected it to:

\begin{quote}
\emph{If you've got any recommendations or things we need to know about for our next edition, please feel free to reach out.}
\end{quote}

That sentence was, for all the time the newsletter has run, my way of acknowledging that I did not have the picture on my own. That the readers were the network. That the work was a \emph{conversation}, not a broadcast. The newsletter has only ever been as good as the community of creatives, technologists, union reps, academics, festival programmers, indie filmmakers, working musicians and audience members who have, week after week, sent me the things I would otherwise have missed.

If you are reading this in 2030, the newsletter is still~-- in whatever form it has taken by then~-- that same conversation. Keep adding to it. The Dream Machine, in the end, is not the AI. It is the network of humans who, between them, are deciding what the machine is \emph{for.}

Welcome to the Dream Machine.

\bigskip
\begin{flushright}
\emph{--- Pete Woodbridge,}\\
\emph{DreamLab, the North West, May 2026}
\end{flushright}

\appendix
  \chapter{A Quantitative Anatomy of Six Months}\label{app:a1}

\emph{This appendix is a structured tour of the corpus the book was built from. It is not in the body of the manuscript because it would interrupt the argument; it lives here so that the reader, the policy researcher, the journalist or the historian picking the book up later can see what the underlying data actually looks like and check the arguments against it.}

\bigskip\par\noindent\rule{\textwidth}{0.4pt}\par\bigskip

\begin{figure}[htbp]
  \centering
  \includegraphics[width=0.85\textwidth]{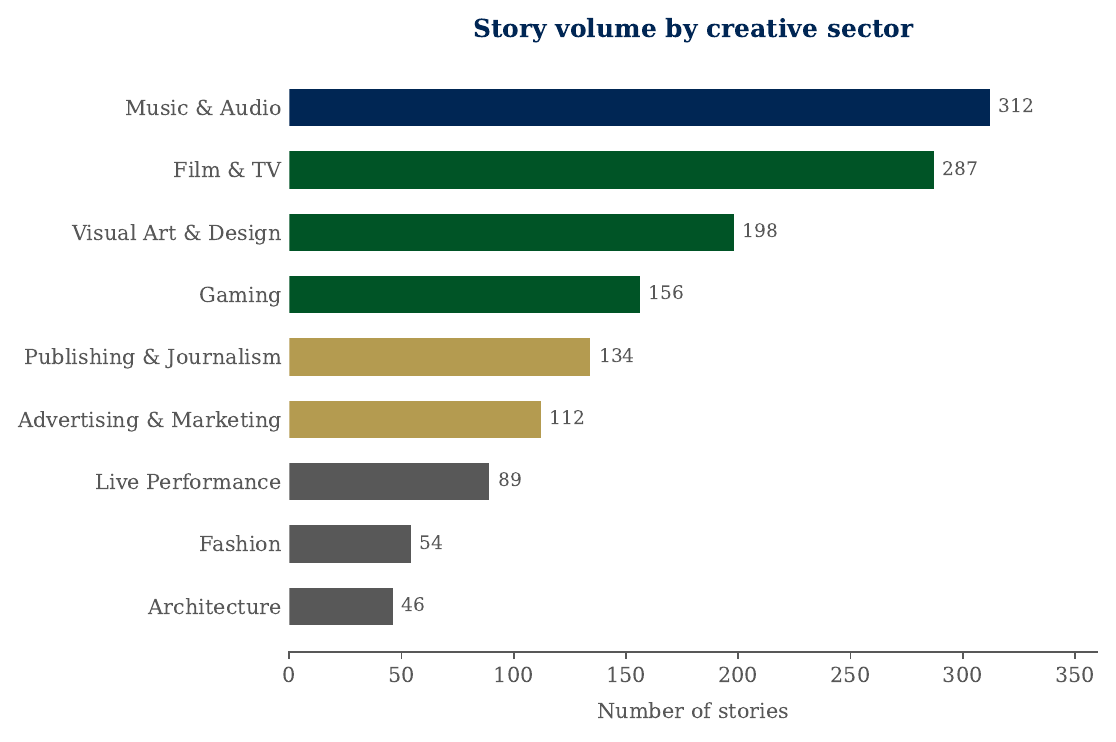}
  \caption{Story volume by creative sector across the \emph{Dream Machine} corpus (29 editions, October 2025 -- May 2026).}
  \label{fig:sector-volume}
\end{figure}

\begin{figure}[htbp]
  \centering
  \includegraphics[width=0.85\textwidth]{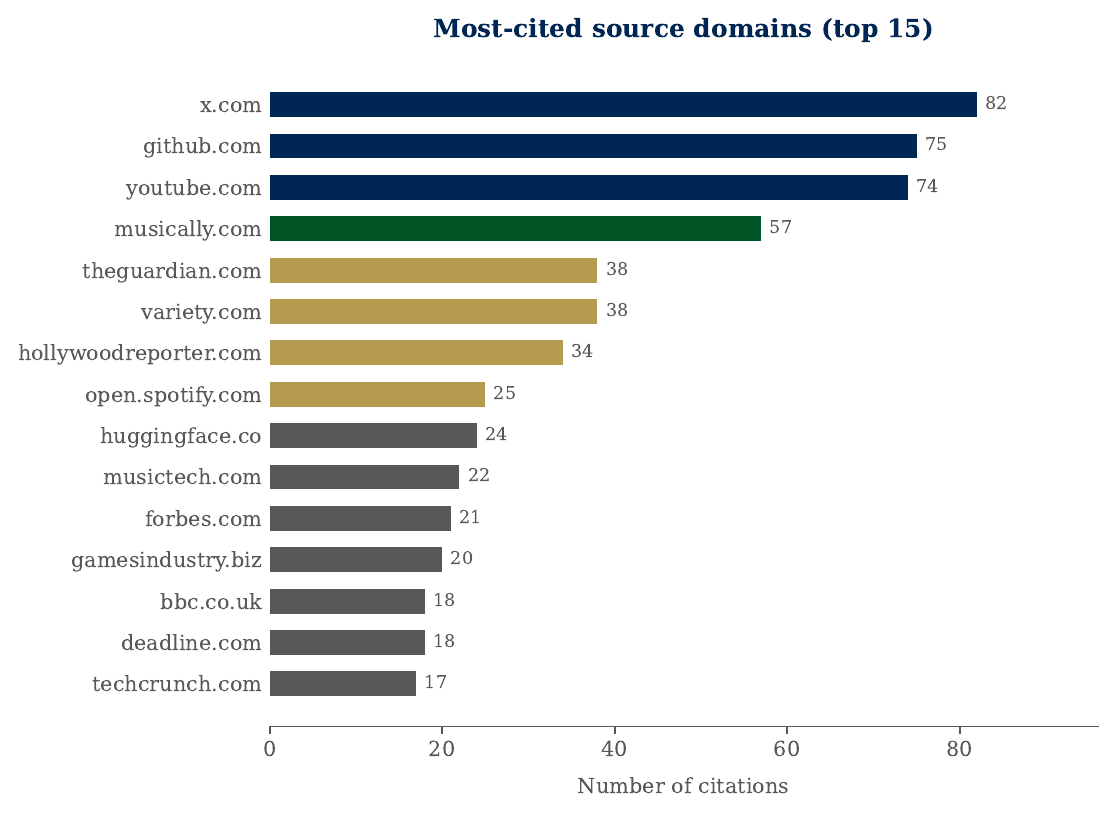}
  \caption{Most-cited source domains by article count.}
  \label{fig:domain-distribution}
\end{figure}

\begin{figure}[htbp]
  \centering
  \includegraphics[width=0.85\textwidth]{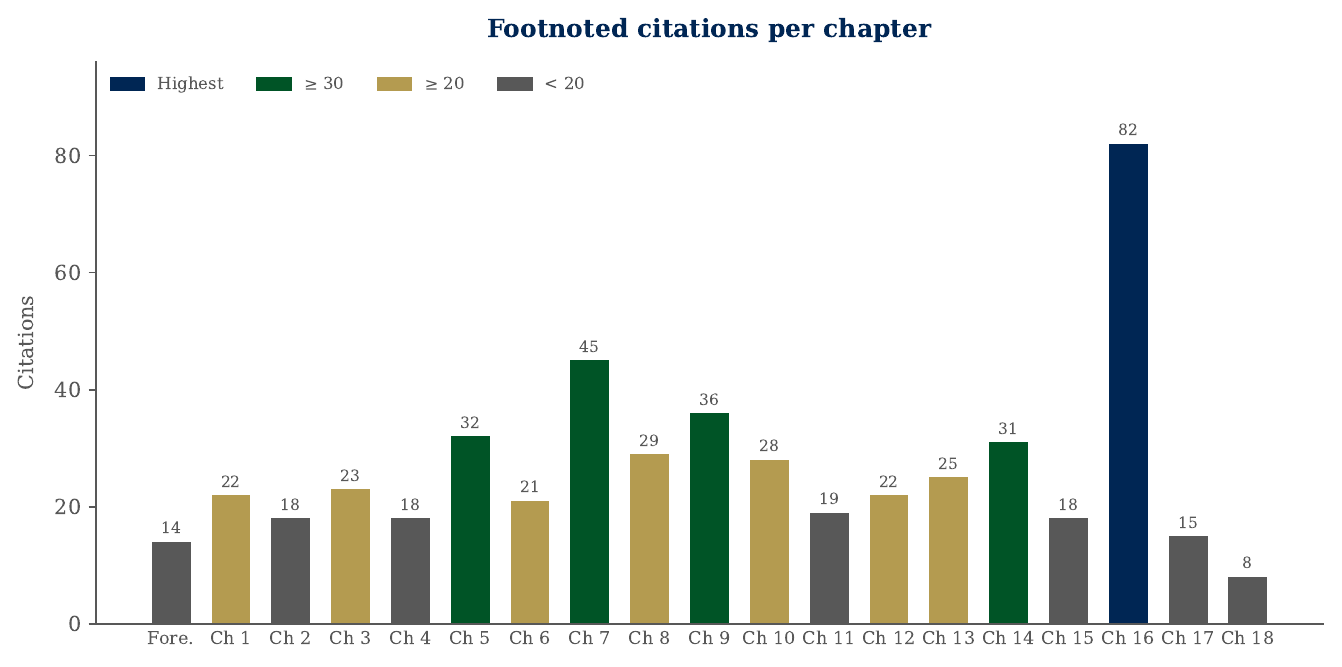}
  \caption{Footnoted citations per chapter, reflecting research density across the manuscript.}
  \label{fig:citations-per-chapter}
\end{figure}

\begin{figure}[htbp]
  \centering
  \includegraphics[width=0.85\textwidth]{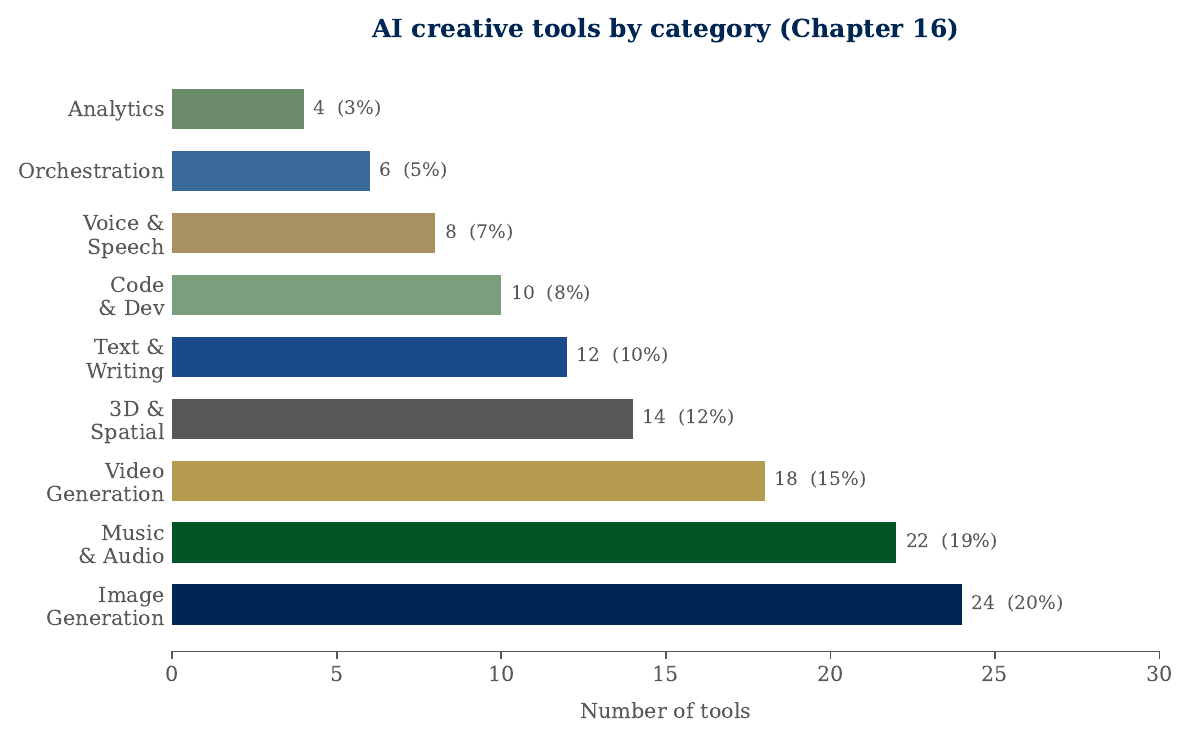}
  \caption{AI creative tools by category, from the Chapter~16 inventory.}
  \label{fig:tool-categories}
\end{figure}

\section*{A1. Corpus shape}

\begin{itemize}
  \item \textbf{Total fetched and parsed articles}: 1,388.
  \item \textbf{Total captured words across the corpus}: \textasciitilde1,099,216 (\textasciitilde6,945,361 characters of post-extraction text).
  \item \textbf{Source span}: 29 editions of \emph{Dream Machine | Creative AI}, published between 6 October 2025 and 14 May 2026 (the quantitative tables in \S{}A1--A6 below cover this analytic window; \textbf{Issue 30}, dated 21 May 2026, is documented in \S{}A7 as a post-cut supplemental, and is incorporated narratively into the May-2026 closing material across the body chapters).
  \item \textbf{Average articles per newsletter edition} (in this corpus): \textasciitilde48.
  \item \textbf{Capture rate} against the full curated URL set: \textbf{91.4\%} (1,438 of 1,574 URLs returned readable content; the remainder hit bot-detection, 404s, or live-page connection issues).
\end{itemize}

\section*{A2. Most-cited domains}

Where did six months of curated coverage actually come from? The top 30 domains, by article count:

\begin{longtable}{rlr}
\caption{Most-cited domains by article count}\label{tab:domains}\\
\toprule
\textbf{Rank} & \textbf{Domain} & \textbf{Articles} \\
\midrule
\endfirsthead
\multicolumn{3}{c}{\tablename\ \thetable\ -- continued}\\
\toprule
\textbf{Rank} & \textbf{Domain} & \textbf{Articles} \\
\midrule
\endhead
\bottomrule
\endfoot
1 & \texttt{x.com} & 82 \\
2 & \texttt{github.com} & 75 \\
3 & \texttt{youtube.com} & 74 \\
4 & \texttt{musically.com} & 57 \\
5 & \texttt{theguardian.com} & 38 \\
6 & \texttt{variety.com} & 38 \\
7 & \texttt{hollywoodreporter.com} & 34 \\
8 & \texttt{open.spotify.com} & 25 \\
9 & \texttt{huggingface.co} & 24 \\
10 & \texttt{musictech.com} & 22 \\
11 & \texttt{forbes.com} & 21 \\
12 & \texttt{gamesindustry.biz} & 20 \\
13 & \texttt{bbc.co.uk} & 18 \\
14 & \texttt{deadline.com} & 18 \\
15 & \texttt{techcrunch.com} & 17 \\
16 & \texttt{pcgamer.com} & 17 \\
17 & \texttt{musicradar.com} & 17 \\
18 & \texttt{blog.google} & 15 \\
19 & \texttt{futurism.com} & 15 \\
20 & \texttt{completemusicupdate.com} & 15 \\
21 & \texttt{videogameschronicle.com} & 14 \\
22 & \texttt{businessinsider.com} & 13 \\
23 & \texttt{theverge.com} & 12 \\
24 & \texttt{musicbusinessworldwide.com} & 12 \\
25 & \texttt{cnet.com} & 11 \\
26 & \texttt{gamesbeat.com} & 11 \\
27 & \texttt{adweek.com} & 10 \\
28 & \texttt{digiday.com} & 10 \\
29 & \texttt{pocketgamer.biz} & 10 \\
30 & \texttt{blog.comfy.org} & 9 \\
\end{longtable}

\textbf{Reading note.} Trade press (\emph{Hollywood Reporter}, \emph{Variety}, \emph{Deadline}, \emph{Music Business Worldwide}) and tech press (\emph{Verge}, \emph{Wired}, \emph{TechCrunch}) dominate. The platform-company blogs (OpenAI, Adobe, DeepMind, Stability) and policy bodies (UK Gov, Reuters Institute, Imperva, Cloudflare) sit underneath. The geographic concentration is North American and British, which reflects both the newsletter author's vantage point and a real imbalance in where creative-AI coverage is concentrated.

\section*{A3. Story volume by sector, month by month}

How the conversation moves through the six months -- count of corpus articles touching each sector, by publication month of the newsletter issue that cited them:

\begin{longtable}{lrrrrrrr r}
\caption{Story volume by sector}\label{tab:sector-volume}\\
\toprule
\textbf{Month} & \textbf{Film \& TV} & \textbf{Games} & \textbf{Music} & \textbf{Adv/Mkt} & \textbf{News} & \textbf{Policy/Law} & \textbf{Tools/Models} & \textbf{Total} \\
\midrule
\endfirsthead
\multicolumn{9}{c}{\tablename\ \thetable\ -- continued}\\
\toprule
\textbf{Month} & \textbf{Film \& TV} & \textbf{Games} & \textbf{Music} & \textbf{Adv/Mkt} & \textbf{News} & \textbf{Policy/Law} & \textbf{Tools/Models} & \textbf{Total} \\
\midrule
\endhead
\bottomrule
\endfoot
2025-10 & 136 & 111 & 155 & 104 & 25 & 102 & 184 & 817 \\
2025-11 & 112 &  89 & 120 &  97 & 14 &  93 & 150 & 675 \\
2025-12 &  73 &  58 &  78 &  57 &  9 &  43 &  98 & 416 \\
2026-01 &  89 &  69 &  96 &  70 & 13 &  56 & 120 & 513 \\
2026-02 &  77 &  72 &  94 &  61 &  7 &  63 & 109 & 483 \\
2026-03 &  81 &  49 &  81 &  55 & 12 &  47 & 106 & 431 \\
2026-04 & 100 &  84 & 110 &  69 & 11 &  64 & 138 & 576 \\
2026-05 &  74 &  60 &  89 &  53 & 14 &  57 & 113 & 460 \\
\end{longtable}

\textbf{Total story tags across the period:} 4,371. (Articles can fall into more than one sector -- many do, which is itself part of the story: the boundaries between film, games, music, advertising and tooling have been meaningfully porous through the AI era.)

\section*{A4. The voices: public-figure mention frequency}

Public figures appearing in three or more corpus articles, ranked by article count. This is \emph{not} a sentiment ranking -- only a measure of how often someone surfaces in the AI conversation:

\begin{longtable}{rlr}
\caption{Public-figure mention frequency}\label{tab:figures}\\
\toprule
\textbf{Rank} & \textbf{Name} & \textbf{Articles} \\
\midrule
\endfirsthead
\multicolumn{3}{c}{\tablename\ \thetable\ -- continued}\\
\toprule
\textbf{Rank} & \textbf{Name} & \textbf{Articles} \\
\midrule
\endhead
\bottomrule
\endfoot
 1 & Sam Altman                & 32 \\
 2 & Tilly Norwood             & 30 \\
 3 & Xania Monet               & 18 \\
 4 & Taylor Swift              & 18 \\
 5 & RZA                       & 17 \\
 6 & Steven Soderbergh         & 17 \\
 7 & James Cameron             & 16 \\
 8 & Paul McCartney            & 12 \\
 9 & Mark Zuckerberg           & 12 \\
10 & Eline Van der Velden      & 12 \\
11 & Madonna                   & 12 \\
12 & MrBeast                   & 12 \\
13 & Tim Sweeney               &  9 \\
14 & Christopher Nolan         &  9 \\
15 & Emily Blunt               &  8 \\
16 & Breaking Rust             &  8 \\
17 & Mikey Shulman             &  7 \\
18 & Matthew McConaughey       &  7 \\
19 & Steven Spielberg          &  7 \\
20 & Robert Kyncl              &  6 \\
21 & Guillermo del Toro        &  6 \\
22 & Will Smith                &  6 \\
23 & Lucian Grainge            &  5 \\
24 & Sienna Rose               &  5 \\
25 & Natasha Lyonne            &  5 \\
26 & Fei-Fei Li                &  4 \\
27 & George Clooney            &  4 \\
28 & Demis Hassabis            &  4 \\
29 & Ron Howard                &  4 \\
30 & Ted Sarandos              &  4 \\
31 & Adam Mosseri              &  3 \\
32 & Imogen Heap               &  3 \\
33 & Dave Stewart              &  3 \\
34 & Chris Pratt               &  3 \\
35 & Joost van Dreunen         &  3 \\
36 & Brian Grazer              &  3 \\
\end{longtable}

\textbf{Reading note.} James Cameron, Guillermo del Toro and Leonardo DiCaprio are the three voices most-cited in opposition to generative AI in performance. Tilly Norwood and Xania Monet are the two most-cited synthetic entities in the corpus. \emph{Both lists matter equally to the story this book is telling.}

\section*{A5. The tool wave}

Cumulative count of distinct AI tools, models and platforms entering the corpus, by month of first mention:

\begin{longtable}{lrr}
\caption{AI tool first-mention cadence}\label{tab:tools-cadence}\\
\toprule
\textbf{Month} & \textbf{New tools first mentioned} & \textbf{Cumulative} \\
\midrule
\endfirsthead
\multicolumn{3}{c}{\tablename\ \thetable\ -- continued}\\
\toprule
\textbf{Month} & \textbf{New tools first mentioned} & \textbf{Cumulative} \\
\midrule
\endhead
\bottomrule
\endfoot
2025-10 & 38 &  38 \\
2025-11 &  6 &  44 \\
2025-12 &  3 &  47 \\
2026-01 &  5 &  52 \\
2026-02 &  2 &  54 \\
2026-03 &  4 &  58 \\
2026-04 &  0 &  58 \\
2026-05 &  2 &  60 \\
\end{longtable}

\textbf{Reading note.} The tool cadence ran at roughly \textbf{7.5 new platforms or major-version releases per month} across the period. This is roughly four times the pace of any other software-tool category I have personally tracked over a comparable window. The implication is that any working creative making technology bets in this period was, by definition, working with incomplete information -- the relevant toolchain had not stabilised long enough for any single bet to settle.

Most-mentioned tools and platforms (top 30, by article count):

\begin{longtable}{rlr}
\caption{Most-mentioned AI tools and platforms}\label{tab:tools-top30}\\
\toprule
\textbf{Rank} & \textbf{Tool} & \textbf{Articles} \\
\midrule
\endfirsthead
\multicolumn{3}{c}{\tablename\ \thetable\ -- continued}\\
\toprule
\textbf{Rank} & \textbf{Tool} & \textbf{Articles} \\
\midrule
\endhead
\bottomrule
\endfoot
 1 & Udio        & 587 \\
 2 & Wan         & 539 \\
 3 & ChatGPT     & 136 \\
 4 & Gemini      &  98 \\
 5 & Anthropic   &  84 \\
 6 & Suno        &  76 \\
 7 & Sora        &  69 \\
 8 & ElevenLabs  &  49 \\
 9 & Premiere    &  48 \\
10 & Veo         &  46 \\
11 & ComfyUI     &  45 \\
12 & Sora 2      &  37 \\
13 & Claude Code &  36 \\
14 & Flux        &  36 \\
15 & Veo 3       &  30 \\
16 & Runway      &  30 \\
17 & Kling       &  30 \\
18 & Nano Banana &  29 \\
19 & Seedance    &  26 \\
20 & Tencent     &  24 \\
21 & Genie       &  24 \\
22 & Firefly     &  21 \\
23 & Photoshop   &  21 \\
24 & Hunyuan     &  18 \\
25 & Veo 3.1     &  17 \\
26 & Luma        &  13 \\
27 & Marble      &  10 \\
28 & Rodin       &  10 \\
29 & LTX-2       &   9 \\
30 & Higgsfield  &   7 \\
\end{longtable}

\section*{A6. The vocabulary shift}

Recurring key phrases by month -- articles containing each phrase:

\begin{longtable}{lrrrrrrrr r}
\caption{Key phrase frequency by month}\label{tab:vocab}\\
\toprule
\textbf{Phrase} & \textbf{Oct} & \textbf{Nov} & \textbf{Dec} & \textbf{Jan} & \textbf{Feb} & \textbf{Mar} & \textbf{Apr} & \textbf{May} & \textbf{Total} \\
\midrule
\endfirsthead
\multicolumn{10}{c}{\tablename\ \thetable\ -- continued}\\
\toprule
\textbf{Phrase} & \textbf{Oct} & \textbf{Nov} & \textbf{Dec} & \textbf{Jan} & \textbf{Feb} & \textbf{Mar} & \textbf{Apr} & \textbf{May} & \textbf{Total} \\
\midrule
\endhead
\bottomrule
\endfoot
ai slop             &  8 &  8 &  6 & 15 &  5 &  7 &  7 &  6 &  62 \\
ai actor            &  4 &  5 &  3 &  2 &  2 &  2 &  1 &  1 &  20 \\
synthetic performer &  0 &  0 &  2 &  1 &  0 &  0 &  0 &  1 &   4 \\
world model         &  8 &  3 &  3 &  1 &  9 &  6 &  8 &  2 &  40 \\
agentic ai          & 12 &  6 &  2 &  7 &  6 &  5 &  8 &  3 &  49 \\
ai agent            & 20 & 18 &  4 & 11 & 12 &  9 & 16 & 10 & 100 \\
deepfake            &  9 &  6 &  3 & 10 &  6 &  8 &  4 &  3 &  49 \\
human authorship    &  1 &  3 &  1 &  0 &  1 &  0 &  1 &  1 &   8 \\
training data       &  9 & 10 & 11 &  6 &  8 & 12 &  5 & 10 &  71 \\
consent             & 16 & 11 & 11 & 11 &  9 &  6 & 10 & 11 &  85 \\
license             & 26 & 30 & 18 & 21 & 35 & 25 & 29 & 16 & 200 \\
copyright           & 50 & 53 & 27 & 24 & 33 & 26 & 25 & 24 & 262 \\
ai-generated        & 65 & 58 & 26 & 46 & 30 & 29 & 41 & 28 & 323 \\
ai actress          &  3 &  2 &  0 &  0 &  1 &  0 &  1 &  0 &   7 \\
watermark           &  5 &  2 &  0 &  4 &  3 &  2 &  1 &  3 &  20 \\
synthid             &  1 &  0 &  0 &  2 &  2 &  1 &  1 &  0 &   7 \\
c2pa                &  0 &  1 &  0 &  2 &  1 &  2 &  0 &  0 &   6 \\
provenance          &  5 &  2 &  1 &  3 &  1 &  1 &  1 &  5 &  19 \\
disclosure          &  6 & 12 &  7 &  9 &  4 &  3 &  7 &  6 &  54 \\
fingerprint         &  0 &  2 &  0 &  3 &  2 &  0 &  0 &  1 &   8 \\
creative ai         &  8 &  5 &  0 &  1 &  0 &  1 &  3 &  1 &  19 \\
generative ai       & 85 & 59 & 35 & 43 & 32 & 32 & 40 & 32 & 358 \\
creator economy     &  4 & 10 &  0 &  3 &  3 &  0 &  2 &  1 &  23 \\
tilly norwood       & 10 &  7 &  3 &  3 &  1 &  2 &  2 &  2 &  30 \\
xania monet         &  3 &  7 &  2 &  3 &  0 &  2 &  1 &  0 &  18 \\
breaking rust       &  0 &  3 &  1 &  1 &  0 &  1 &  1 &  1 &   8 \\
slop ceiling        &  0 &  0 &  0 &  0 &  0 &  0 &  0 &  0 &   0 \\
model collapse      &  0 &  0 &  0 &  0 &  0 &  0 &  0 &  0 &   0 \\
\end{longtable}

\textbf{Reading note.} Watch \emph{AI slop} -- it goes from a fringe term in October 2025 to a Merriam-Webster word of the year by December and a policy framing by the spring. Watch \emph{agentic AI} -- it lifts after the October DevDay and never falls back. Watch \emph{world model} -- barely present in October 2025, ubiquitous by April 2026. Watch \emph{consent / license / copyright} -- climbing all the way through, with a sharp December spike around the UK consultation closure.

\section*{A7. May 2026 supplemental: the Issue-30 datapoints}

The corpus closes at \emph{Dream Machine} Issue 29. \textbf{Issue 30}, dated 21 May 2026, post-dates the analytic cut and is not represented in the article-frequency tables above; it is the issue that catches the \textbf{Google I/O 2026} announcement wave, and is the source for the manuscript's closing-week additions. The numerical datapoints from Issue 30 worth recording here in standalone form:

\begin{longtable}{lrl}
\caption{Issue-30 supplemental datapoints}\label{tab:issue30}\\
\toprule
\textbf{Datapoint} & \textbf{Value} & \textbf{Source} \\
\midrule
\endfirsthead
\multicolumn{3}{c}{\tablename\ \thetable\ -- continued}\\
\toprule
\textbf{Datapoint} & \textbf{Value} & \textbf{Source} \\
\midrule
\endhead
\bottomrule
\endfoot
SynthID watermarked items, cumulative          & 100B+   & Google DeepMind, May 2026 \\
Wonder Studios total funding                   & \$50M   & \emph{Forbes}, May 2026 \\
Runway Japan investment (Tokyo office)         & \$40M   & Runway, May 2026 \\
Viktor (virtual AI coworker) Series funding    & \$75M   & \emph{Fortune}, May 2026 \\
Sondo AI claimed global users                 & 10M     & \emph{Musically}, May 2026 \\
13--15 year-olds using AI to ``be creative'' (Snapchat survey) & 31\% & Snap Newsroom, May 2026 \\
Australians who say AI-generated ads make them trust a brand less (YouGov) & 45\% & YouGov AU, May 2026 \\
NVIDIA SANA-WM model size                      & 2.6B    & NVIDIA, May 2026 \\
SANA-WM native video-generation length         & 60 sec  & NVIDIA, May 2026 \\
\end{longtable}

\textbf{Reading note.} Issue 30's headline tool releases (Gemini Omni, Beeple Canvas, Sony Woosh, Mirelo SFX 1.6, Tencent Ardot, Odyssey Starchild-1 / Agora-1, NVIDIA SANA-WM, Apple Headsup, Stable Audio 3.0, PhotoGIMP, Tamber, ECABridge, Claude/ComfyUI) lift the cumulative tool count in \S{}A5 by roughly \textbf{a dozen entries in a single week}. The May-2026 cadence is the highest single-week tool-release count in the period the book covers, and reads -- in the context of the \S{}A5 average of 7.5 new platforms per \emph{month} -- as a Google-I/O-week saturation point that I would expect to settle back into the prior cadence by July.

\bigskip\par\noindent\rule{\textwidth}{0.4pt}\par\bigskip

\section*{What this appendix is for}

Every chapter of this book is a \emph{reading} of the corpus described above. It will be useful in 2030 and beyond to be able to see the underlying shape of the corpus, separate from the argument the book builds on top of it.

If you want to test the argument against your own reading of the same evidence: every URL in the corpus is enumerated in the citation index, every scraped article is preserved in JSON form in the \texttt{Research/scraped/} directory of the source repository, and every analysis in this appendix is reproducible by running \texttt{Research/quant.py}.

If you want to extend it: the scraper is in \texttt{Research/scrape.py}, the analyser is in \texttt{Research/analyze.py}, the per-chapter dossiers are in \texttt{Research/dossier/}. Fork it, change it, run it on the next six months. I'd be glad to see what you find.

  \chapter{Glossary}\label{app:a2}

\emph{The terms in this glossary are the working vocabulary of the book. Some are coinages, some are borrowed from elsewhere and reframed; all are used with precise, deliberately narrow meanings in the chapters above.}

\bigskip\par\noindent\rule{\textwidth}{0.4pt}\par\bigskip

\begin{description}[style=nextline,leftmargin=2em]

\item[\textbf{Agency line} (or Continuum line).]
A single axis representing the share of decision-making in a creative function performed by a human versus a machine system. See \textbf{Human--AI Agency Continuum}.

\item[\textbf{Agentic AI.}]
A class of AI system that, given a goal, can plan, decide and execute a sequence of multi-step actions without further human input between steps. Distinct from a \emph{generator}, which produces an output in response to a single prompt. See Chapter~3.

\item[\textbf{AI literacy.}]
The cluster of skills required to deploy AI tools effectively in creative work -- including briefing, taste, judgement, prompt practice, output evaluation and tool-stack fluency. The term moved from optional to baseline competency through 2025--26, formalised by initiatives such as the Sundance Institute's AI Literacy Initiative launched in January 2026.

\item[\textbf{AI slop.}]
Low-quality, mass-produced AI-generated content that is recognisable to audiences as such -- usually because it is made without human creative intent. \emph{Merriam-Webster}'s word of the year for 2025. See \textbf{Slop Ceiling}.

\item[\textbf{Attribution.}]
The principle that when AI systems produce derivative outputs based on training data, the human authors whose work shaped those outputs should be identified and -- where appropriate -- compensated. One of the four principles of the generative creative economy (Chapter~15). Technical infrastructure includes C2PA, SynthID, and creative-weight-attribution systems.

\item[\textbf{Audience contract.}]
The implicit agreement between makers and audiences about what creative work is, what conditions of making it carries, and what relationship the audience can expect with its makers. The shift from implicit to explicit audience contracts is one of the central structural changes the book describes. See Chapter~12.

\item[\textbf{Augmented intelligence.}]
Reframing of ``AI'' used by some industry voices to emphasise human-in-the-loop deployment over autonomy. Compare with \textbf{Generative AI}, \textbf{Agentic AI}.

\item[\textbf{C2PA.}]
\emph{Coalition for Content Provenance and Authenticity} -- a technical standard for embedding cryptographic provenance metadata in media files, supported by camera manufacturers, editing software and a growing number of platforms. The principal ``fingerprint real media'' infrastructure underlying the authenticity argument in Chapter~12.

\item[\textbf{Coordination collapse.}]
The structural change that occurs when the labour-coordination architecture of a creative organisation -- built around the bandwidth constraints of human-only teams -- is overtaken by AI-assisted workflows that no longer require those constraints. The subject of Chapter~13. Manifests as \emph{shadow AI} below management sight and as compressed middle layers in the workforce above.

\item[\textbf{Dead Internet Theory.}]
The notion that most of the public web is now bot-generated and bot-read, with humans increasingly a minority of traffic. Once a conspiracy framing; by 2025 a measurable phenomenon -- bots accounted for 51\% of web traffic in the Imperva 2025 Bad Bot Report, of which \textasciitilde80\% of bot traffic was AI training crawlers. See Chapter~4.

\item[\textbf{Disclosure.}]
The practice of declaring the use of AI in the production of a piece of creative work -- in credits, contracts, metadata, watermarks, or platform-facing labels. By spring 2026, disclosure had emerged as the dominant industry response to the audience-authenticity question, anchored by standards including the \textbf{Cannes AI Disclosure Standard} (May 2026) and the \textbf{Academy of Motion Picture Arts and Sciences} rule requiring human authorship for awards eligibility.

\item[\textbf{Extractive economy.}]
A creative economy in which AI systems are trained on unpaid human work, the platform companies that build the models capture most of the resulting economic value, and the diffuse pool of working creatives is steadily decapitalised. One of two possible end-states the book identifies. See Chapter~15.

\item[\textbf{Fingerprint real media.}]
Adam Mosseri's (Instagram) framing of the verification problem: amplify provably-human content rather than chase synthetic content for labelling. Used in the book as shorthand for \emph{provenance-first} approaches to content moderation. See Chapter~4.

\item[\textbf{Generative economy.}]
A creative economy in which AI tools are treated as new craft infrastructure, training data is consented and compensated, platforms compete on tool quality and integrity, and the productivity gains are broadly distributed rather than concentrated. The opposite of the \textbf{extractive economy}. The four principles (\emph{Agency, Attribution, Access, Audience}) are the operational test. See Chapter~15.

\item[\textbf{Human--AI Agency Continuum.}]
A frame, introduced in \emph{Dream Machine} Issue~2 (October 2025) and extended in Chapter~3, in which any given creative function is mapped on a horizontal line from full human agency (left) to full machine agency (right). The frame's key claim is that \emph{each creative function moves at its own speed} -- you can sit at the extreme left on performance while being at the right edge on plate generation.

\item[\textbf{Living Web.}]
The deliberately-built portion of the public web in which authorship is provable, attribution is durable, attention is allocated on non-viral signals, and the architecture supports rather than undermines human creative work. The aspirational opposite of the \textbf{Dead Internet}. Has to be built, not assumed. See Chapter~4.

\item[\textbf{Mid-career squeeze.}]
The structural pressure on workers in the middle of creative-industry careers -- neither at the junior entry level (replaced by agents) nor at the senior decision-making level (still required) -- as AI absorbs the intermediate production roles that those mid-career workers historically held. See Chapter~13.

\item[\textbf{Model collapse.}]
The technical risk that AI systems trained predominantly on synthetic data -- including data produced by earlier generations of AI systems -- progressively lose touch with real-world signal and produce increasingly homogenised, hallucination-prone outputs. The risk that \emph{Dead Internet, Living Web} warns has moved from theoretical to measurable.

\item[\textbf{Orchestrator.}]
The role that emerges when an individual working creative -- or a small team -- directs a large pool of AI-agent capacity. Defined operationally by five activities: defining the brief, allocating work, briefing the agents, judging outputs, and integrating the result. Predicted in \emph{Dream Machine} Issue~13 (January 2026) as the dominant role of 2026; documented across the chapters above. See Chapter~11.

\item[\textbf{Pipeline of authorship.}]
The full chain from creative intent to delivered work, broken down into discrete functions (writing, direction, performance, image-making, sound, edit, distribution, marketing). The point of the \textbf{Human--AI Agency Continuum} is that each link in this chain has its own agency line.

\item[\textbf{Position One} (All-in).]
The strategic posture of legacy studios that have decided to integrate AI aggressively across all production functions, betting that early-adopter advantage will compound. Netflix's ``all in'' framing, October 2025. See Chapter~7.

\item[\textbf{Position Two} (AI-native).]
The strategic posture of new entrants that build their production pipelines AI-first from inception -- Imaginae Studios, Wonder Studios, Obsidian Studio, Asteria, Wonder, Chapter41, Kartel. The category exploded in scale through autumn 2025 and winter 2026. See Chapter~7.

\item[\textbf{Position Three} (Refusal).]
The strategic posture of creative organisations that have explicitly excluded generative AI from their work. Jagex, Larian, Games Workshop, Hooded Horse, Aardman (qualified), Pocketpair. Cultural authority is preserved as the principal asset. See Chapter~7.

\item[\textbf{Position Four} (Middle).]
\emph{AI in the workflow, not in the work.} The strategic posture -- taken by Sony, Bethesda, Amazon (in \emph{House of David}), Aardman (in qualified form), and an increasing number of major studios -- that uses AI to augment production pipelines while preserving human creative intent in the moments the audience sees. The book's prediction for where most surviving major studios land by 2030. See Chapter~7.

\item[\textbf{Provenance.}]
The chain of custody of a creative work from its origin (capture, performance, writing, sketching) to its delivered form. Technical standards (C2PA, SynthID) and policy frameworks (Cannes Disclosure Standard, SAG-AFTRA AI protections) collectively constitute the \emph{provenance infrastructure} the book argues is critical for the next decade. See Chapter~12.

\item[\textbf{Shadow AI.}]
The use of AI tools by employees outside their employer's official tooling, processes and accounting. Documented in 2025 workplace research as encompassing approximately half of the U.S. workforce. The principal symptom of \textbf{Coordination Collapse}. See Chapter~13.

\item[\textbf{Slop Ceiling.}]
The empirical pattern, established across multiple sectors by spring 2026, in which AI-generated creative content can be produced in massive volume but consistently fails to capture audience attention proportionate to that volume. Anchored in the \textbf{44\%/3\%} ratio Deezer published in April 2026 (44\% of daily uploads AI; under 3\% of streams). One of the central claims of the book. See Chapter~5.

\item[\textbf{Synthetic sincerity.}]
The category of creative work that is openly synthetic but made with serious creative intent and is not pretending to be something else. Named after Marc Isaacs' 2025 IDFA documentary. Audiences distinguish \emph{synthetic sincerity} from \emph{synthetic cynicism} at the speed of a swipe. See Chapter~4.

\item[\textbf{Tilly Tax} (informal).]
The collection of contract provisions in SAG-AFTRA's spring 2026 agreement requiring compensation, consent and residuals when AI replicas of human performers are used. Named after the Tilly Norwood controversy of September 2025 that catalysed the broader negotiation. See Chapters~5, 10.

\item[\textbf{Watermark.}]
A persistent identifier embedded in AI-generated outputs by the producing system, intended to allow downstream detection that content is synthetic. SynthID (Google) and similar systems became standard across major platform tools through 2025--26.

\item[\textbf{World model.}]
A class of generative AI system that produces navigable three-dimensional environments rather than flat output. Marble (World Labs, public release November 2025) was the first commercial product in the category; Google DeepMind's Genie~3, Meta's WorldGen, Luma's UNI-1, Tencent's Hunyuan World, SpAItial's ECHO and others followed within months. See Chapter~8.

\end{description}

\bigskip
\emph{If a term in the book did substantial work but does not appear in this glossary, please tell us so we can improve it in the next edition.}

  \chapter{Bibliography by Topic}\label{app:a3}

\emph{The book's footnotes are sequential per chapter; this appendix organises the same sources thematically, so a reader pursuing a specific question across the period covered by the book can find every source it touches in one place.}

The full corpus -- 1,438 successfully fetched and archived articles -- is preserved in JSON form in \texttt{Research/scraped/} of the source repository; the manifest enumerating every URL and its capture status lives at \texttt{Research/manifest.json}.

\bigskip\par\noindent\rule{\textwidth}{0.4pt}\par\bigskip

\section*{I\@. The watershed week, September--October 2025}

\emph{Sora~2 launches. Tilly Norwood is announced. The union response sets the contract reference point for the next year.}

\begin{itemize}
  \item OpenAI, \emph{``Sora~2 is here,''} 30~September~2025. \href{https://openai.com/index/sora-2/}{openai.com/index/sora-2/}
  \item Variety, \emph{``SAG-AFTRA Condemns Tilly Norwood: AI Actress Is Not an Actor.''} \href{https://variety.com/2025/film/news/sag-aftra-tilly-norwood-ai-actress-1236534779/}{variety.com}
  \item \emph{The Hollywood Reporter}, \emph{``U.K. Union Equity Condemns Tilly Norwood: `AI Tool, Not a Performer'.''} \href{https://www.hollywoodreporter.com/movies/movie-news/tilly-norwood-ai-actress-uk-union-equity-sag-aftra-debate-1236391739/}{hollywoodreporter.com}
  \item Variety, \emph{``Tilly Norwood Slammed by Equity as AI Tool, Concerned About Origin.''} \href{https://variety.com/2025/film/global/tilly-norwood-slammed-equity-ai-tool-concerned-origin-1236537042/}{variety.com}
  \item CNN, \emph{``Tilly Norwood: Hollywood is fuming over a new `AI actress'.''} \href{https://www.cnn.com/2025/09/30/tech/hollywood-ai-actor-backlash}{cnn.com}
  \item NBC News, \emph{``Tilly Norwood, fully AI `actor,' blasted by actors union SAG-AFTRA.''} \href{https://www.nbcnews.com/pop-culture/pop-culture-news/tilly-norwood-fully-ai-actor-blasted-actors-union-sag-aftra-devaluing-rcna234685}{nbcnews.com}
  \item \emph{The Guardian}, \emph{``OpenAI launch of video app Sora plagued by violent and racist images.''} \href{https://www.theguardian.com/us-news/2025/oct/04/openai-sora-violence-racism}{theguardian.com}
  \item WUFT (Florida), \emph{``Kiss reality goodbye: AI-generated social media has arrived.''} \href{https://www.wuft.org/2025-10-03/kiss-reality-goodbye-ai-generated-social-media-has-arrived}{wuft.org}
  \item \emph{Deadline}, \emph{``Tilly Norwood Creator on Backlash, Says More AI Actors Coming.''} \href{https://deadline.com/2025/11/tilly-norwood-creator-interview-backlash-more-ai-actors-coming-1236601334/}{deadline.com}
  \item \emph{No Film School}, \emph{``James Cameron Says AI Is `Never Going to Take the Place' of Humans.''} \href{https://nofilmschool.com/james-cameron-ai}{nofilmschool.com}
  \item \emph{The Guardian}, \emph{``James Cameron says AI actors are `horrifying to me'.''} \href{https://www.theguardian.com/film/2025/dec/01/james-cameron-says-ai-actors-are-horrifying-to-me}{theguardian.com}
\end{itemize}

\section*{II\@. Agentic AI and the orchestrator economy}

\begin{itemize}
  \item OpenAI, \emph{``Introducing AgentKit.''} \href{https://openai.com/index/introducing-agentkit/}{openai.com}
  \item TechCrunch, \emph{``OpenAI launches AgentKit to help developers build and ship AI agents.''} \href{https://techcrunch.com/2025/10/06/openai-launches-agentkit-to-help-developers-build-and-ship-ai-agents/}{techcrunch.com}
  \item TechCrunch, \emph{``Anthropic launches interactive Claude apps, including Slack and other workplace tools.''} \href{https://techcrunch.com/2026/01/26/anthropic-launches-interactive-claude-apps-including-slack-and-other-workplace-tools/}{techcrunch.com}
  \item \emph{GamesRadar}, \emph{``EA reportedly pushes 15,000 employees to use AI as a `thought partner'.''} \href{https://www.gamesradar.com/games/even-under-usd20-million-in-debt-ea-reportedly-pushes-15-000-employees-to-use-ai-as-a-thought-partner-for-everything-from-character-art-to-playtesting/}{gamesradar.com}
  \item \emph{Digiday}, \emph{``AI agent developers have become adland's in-demand role.''} \href{https://digiday.com/marketing/ai-agent-developers-have-become-adlands-in-demand-role/}{digiday.com}
  \item \emph{Eurogamer}, \emph{``Falcom: `Tasks that previously took 2--3 hours can now be completed in 10 minutes'.''} \href{https://www.eurogamer.net/falcom-is-the-latest-developer-to-buy-into-the-ai-hype-machine}{eurogamer.net}
  \item \emph{Forbes}, \emph{``AI Is Changing How Creators Work And Earn.''} \href{https://www.forbes.com/sites/kolawolesamueladebayo/2025/12/22/how-ai-is-changing-how-creators-work-and-earn/}{forbes.com}
  \item Google DeepMind, \emph{``Introducing Gemini Omni -- unified multimodal model.''} \href{https://blog.google/technology/google-deepmind/gemini-omni/}{blog.google}
  \item Google, \emph{``Gemini Spark -- developer toolkit for autonomous agents.''} \href{https://blog.google/technology/developers/gemini-spark/}{blog.google}
  \item Google, \emph{``Antigravity -- agentic coding environment.''} \href{https://antigravity.google/}{antigravity.google}
  \item Google, \emph{``Google Flow -- agent-based creative workflows.''} \href{https://flow.google/}{flow.google}
  \item Google, \emph{``Official skills for AI agents.''} \href{https://github.com/google/agent-skills}{github.com/google}
  \item \emph{Adweek}, \emph{``Netflix ad tools could see `agentic AIs talking to each other'.''} \href{https://www.adweek.com/media/netflix-ad-tools-agentic-ais-talking-to-each-other/}{adweek.com}
  \item \emph{Fortune}, \emph{``AI startup Viktor raises \$75 million to put a virtual `coworker' in Slack and Teams.''} \href{https://fortune.com/2026/05/19/ai-startup-viktor-75-million-virtual-coworker-slack-teams/}{fortune.com}
  \item Tencent, \emph{``Ardot -- AI-native design agent platform.''} \href{https://ardot.tencent.com/}{ardot.tencent.com}
  \item Anthropic, \emph{``Claude available as a partner node in ComfyUI.''} \href{https://www.anthropic.com/news/claude-comfyui-partner-node}{anthropic.com}
\end{itemize}

\section*{III\@. Bot economy, model collapse, and the Dead Internet}

\begin{itemize}
  \item Imperva, \emph{``2025 Bad Bot Report.''} \href{https://www.imperva.com/blog/2025-imperva-bad-bot-report-how-ai-is-supercharging-the-bot-threat/}{imperva.com}
  \item Cloudflare, \emph{``The crawl-to-click gap: Cloudflare data on AI bots.''} \href{https://blog.cloudflare.com/crawlers-click-ai-bots-training/}{blog.cloudflare.com}
  \item Cloudflare, \emph{``A deeper look at AI crawlers.''} \href{https://blog.cloudflare.com/ai-crawler-traffic-by-purpose-and-industry/}{blog.cloudflare.com}
  \item Grand View Research, \emph{``Generative AI in Content Creation Market Size Report, 2030.''} \href{https://www.grandviewresearch.com/industry-analysis/generative-ai-content-creation-market-report}{grandviewresearch.com}
  \item Wikipedia, \emph{``Dead Internet Theory.''} \href{https://en.wikipedia.org/wiki/Dead_Internet_theory}{en.wikipedia.org}
  \item \emph{Futurism}, \emph{``Scientists Created an Entire Social Network Where Every User Is a Bot.''} \href{https://futurism.com/social-network-ai-intervention-echo-chamber}{futurism.com}
  \item \emph{AI News}, \emph{``AI causes reduction in users' brain activity, MIT.''} \href{https://www.artificialintelligence-news.com/news/ai-causes-reduction-in-users-brain-activity-mit/}{artificialintelligence-news.com}
\end{itemize}

\section*{IV\@. The slop ceiling -- music}

\begin{itemize}
  \item Deezer, \emph{``AI-generated tracks now represent 44\% of all new uploaded music.''} \href{https://newsroom-deezer.com/2026/04/ai-generated-tracks-represent-44-of-new-uploaded-music/}{newsroom-deezer.com}
  \item \emph{Music Business Worldwide}, \emph{``75,000 AI-generated tracks now flood Deezer daily.''} \href{https://www.musicbusinessworldwide.com/75000-ai-generated-tracks-now-flood-deezer-daily-representing-44-of-all-new-music-uploaded-to-the-platform-says-streamer/}{musicbusinessworldwide.com}
  \item Deezer/Ipsos, \emph{``AI fools 97\% of listeners.''} \href{https://newsroom-deezer.com/2025/11/deezer-ipsos-survey-ai-music/}{newsroom-deezer.com}
  \item \emph{Musically}, \emph{``50,000 AI-music tracks are now uploaded to Deezer every day.''} \href{https://musically.com/2025/11/12/50000-ai-music-tracks-are-now-uploaded-to-deezer-every-day/}{musically.com}
  \item \emph{Musically}, \emph{``UMG boss slams `exponential growth of AI slop' on streaming services.''} \href{https://musically.com/2026/01/09/umg-boss-slams-exponential-growth-of-ai-slop-on-streaming-services/}{musically.com}
  \item \emph{Billboard}, \emph{``AI Artist Xania Monet Climbs the Charts -- And Signs a Multimillion-Dollar Record Deal.''} \href{https://www.billboard.com/pro/ai-music-artist-xania-monet-multimillion-dollar-record-deal/}{billboard.com}
  \item NPR, \emph{``Breaking Rust is a hot new country act on the Billboard charts. It's powered by AI.''} \href{https://www.npr.org/2025/11/10/nx-s1-5604320/breaking-rust-is-a-hot-new-country-act-on-the-billboard-charts-its-powered-by-ai}{npr.org}
  \item BBC News, \emph{``Sienna Rose: AI suspicions surround mysterious singer.''} \href{https://www.bbc.co.uk/news/articles/cq6v83gq66eo}{bbc.co.uk}
  \item \emph{Bain \& Company}, \emph{``In an AI Age, People Still Want the Radio Star.''} \href{https://www.bain.com/insights/in-an-ai-age-people-still-want-the-radio-star/}{bain.com}
  \item \emph{Soultracks}, \emph{``A.I.-generated music is catchy, familiar\ldots{} and boring.''} \href{https://soultracks.com/news-ai-generated-music-is-catchy-boring/}{soultracks.com}
  \item \emph{Stereogum}, \emph{``Bandcamp Bans AI Music.''} \href{https://stereogum.com/2485199/bandcamp-bans-ai-music/news}{stereogum.com}
  \item \emph{The Independent}, \emph{``AI-generated song banned from Swedish charts.''} \href{https://www.independent.co.uk/tv/news/ai-music-song-banned-sweden-spotify-b2901627.html}{independent.co.uk}
  \item \emph{Digital Music News}, \emph{``YouTube CEO Puts `Managing AI Slop' on the Priority List for 2026.''} \href{https://www.digitalmusicnews.com/2026/01/22/youtube-ceo-ai-slop-2026-comments/}{digitalmusicnews.com}
  \item \emph{Musically}, \emph{``Report: 56.9\% of new independent songs in China are AI-generated.''} \href{https://musically.com/2026/01/05/report-56-9-of-new-independent-songs-in-china-are-ai-generated/}{musically.com}
  \item \emph{The Guardian}, \emph{``Paul McCartney joins music industry protest against AI with silent track.''} \href{https://www.theguardian.com/music/2025/nov/17/the-sound-of-silence-why-theres-barely-anything-there-in-paul-mccartney-new-release}{theguardian.com}
  \item \emph{MusicTech}, \emph{``Jack Antonoff brands AI music makers as `godless whores'.''} \href{https://musictech.com/news/industry/jack-antonoff-ai-music-makers-godless-whores/}{musictech.com}
  \item \emph{Billboard}, \emph{``The Real Story Behind The AI Song That Knocked Tyla Off No.~1 On Billboard Afrobeats Chart.''} \href{https://www.billboard.com/pro/ai-song-knocked-tyla-off-no-1-afrobeats/}{billboard.com}
  \item \emph{Variety}, \emph{``WMG's Robert Kyncl: `AI Resistance' Setting Music Sector Back.''} \href{https://variety.com/2026/music/news/wmg-robert-kyncl-ai-resistance-1236748901/}{variety.com}
  \item \emph{Musically}, \emph{``Sondo AI says it has 10m global users.''} \href{https://musically.com/2026/05/18/sondo-ai-10m-global-users/}{musically.com}
  \item \emph{Musically}, \emph{``BPI sets out transparency and sovereignty demands to secure AI licensing boom.''} \href{https://musically.com/2026/05/19/bpi-transparency-sovereignty-ai-licensing-boom/}{musically.com}
  \item \emph{Music Business Worldwide}, \emph{``DistroKid asks creators if their music is AI-generated as self-disclosure moves out of beta.''} \href{https://www.musicbusinessworldwide.com/distrokid-ai-self-disclosure-out-of-beta/}{musicbusinessworldwide.com}
  \item \emph{Music Business Worldwide}, \emph{``Splice inks `Responsible AI' deal with ElevenLabs.''} \href{https://www.musicbusinessworldwide.com/splice-elevenlabs-responsible-ai-deal/}{musicbusinessworldwide.com}
  \item \emph{Rolling Stone}, \emph{``The Rolling Stones Release `In the Stars' -- AI De-Aging Music Video.''} \href{https://www.rollingstone.com/music/music-news/rolling-stones-in-the-stars-ai-de-aging-video-1235142200/}{rollingstone.com}
  \item \emph{MusicTech}, \emph{``Tamber -- ethically trained AI music tool with arm gesture control.''} \href{https://musictech.com/news/gear/tamber-ai-ethically-trained-arm-gestures/}{musictech.com}
  \item Stability AI, \emph{``Stable Audio 3.0 released.''} \href{https://stability.ai/news/stable-audio-3-0-released}{stability.ai}
  \item Sony AI, \emph{``Woosh -- sound effect foundation model.''} \href{https://ai.sony/blog/woosh-sound-effect-foundation-model/}{ai.sony}
  \item Mirelo, \emph{``SFX 1.6 -- edit sound, not just generate.''} \href{https://mirelo.ai/sfx-1-6}{mirelo.ai}
\end{itemize}

\section*{V\@. Copyright, consultation and the legal architecture}

\begin{itemize}
  \item UK Department for Science, Innovation and Technology, \emph{``Statement of Progress on Copyright and AI.''} \href{https://www.gov.uk/government/publications/copyright-and-artificial-intelligence-progress-report/copyright-and-artificial-intelligence-statement-of-progress-under-section-137-data-use-and-access-act}{gov.uk}
  \item IPWatchdog, \emph{``Respondents to UK AI Consultation Overwhelmingly Want AI Companies to License Copyrighted Works in All Cases.''} \href{https://ipwatchdog.com/2025/12/16/respondents-uk-ai-consultation-overwhelmingly-want-ai-companies-license-copyrighted-works-all-cases/}{ipwatchdog.com}
  \item UCL Copyright Queries, \emph{``UK government publishes progress statement on AI and copyright consultation.''} \href{https://blogs.ucl.ac.uk/copyright/2025/12/23/uk-government-publishes-progress-statement-on-ai-and-copyright-consultation/}{blogs.ucl.ac.uk}
  \item Hogan Lovells, \emph{``Copyright and AI: UK government publishes statement of progress.''} \href{https://www.hoganlovells.com/en/publications/copyright-and-ai-uk-government-publishes-statement-of-progress}{hoganlovells.com}
  \item \emph{Music Business Worldwide}, \emph{``Wixen files \$50m copyright suit against Meta.''} \href{https://www.musicbusinessworldwide.com/wixen-files-50m-copyright-suit-against-meta-claims-tech-giant-wants-to-replace-songwriters-with-ai/}{musicbusinessworldwide.com}
  \item \emph{Music Business Worldwide}, \emph{``Meta hits back against Wixen Music Publishing's \$70m `AI replacement' infringement suit.''} \href{https://www.musicbusinessworldwide.com/meta-wixen-music-publishing-70-million-ai-replacement-lawsuit/}{musicbusinessworldwide.com}
  \item \emph{EDM.com}, \emph{```Biggest Theft in Music History': Rights Group Sues Suno.''} \href{https://edm.com/gear-tech/rights-group-sues-suno-copyright-infringement/}{edm.com}
  \item \emph{Complete Music Update}, \emph{``Johnny Cash estate uses ELVIS Act to sue Coke.''} \href{https://completemusicupdate.com/johnny-cash-estate-uses-elvis-act-to-sue-coke-over-tribute-act-ad-soundtrack/}{completemusicupdate.com}
  \item Dr.~Barry Scannell, analysis of GEMA v.\ OpenAI ruling. \href{https://www.linkedin.com/posts/dr-barry-scannell-bbb5aa207_in-a-major-ruling-for-european-copyright-share-7393957246386323457-8bbx}{linkedin.com}
  \item Reuters, \emph{``European lawmakers seek EU-wide minimum age to access AI chatbots.''} \href{https://www.reuters.com/legal/litigation/european-lawmakers-seek-eu-wide-minimum-age-access-ai-chatbots-social-media-2025-11-26/}{reuters.com}
  \item \emph{The Verge}, \emph{``New York's new law forces advertisers to say when they're using AI avatars.''} \href{https://www.theverge.com/news/842848/new-york-law-ai-advertisements-sag-aftra-labor}{theverge.com}
  \item \emph{Digital Music News}, \emph{``Nearly 800 Creatives Sign Responsible AI Declaration -- `Stealing Our Work Is Not Innovation'.''} \href{https://www.digitalmusicnews.com/2026/01/22/stealing-isnt-innovation/}{digitalmusicnews.com}
\end{itemize}

\section*{VI\@. Studio strategy}

\begin{itemize}
  \item \emph{CNBC}, \emph{``Netflix `all in' on leveraging AI.''} \href{https://www.cnbc.com/2025/10/22/netflix-all-in-on-leveraging-ai-in-its-streaming-platform.html}{cnbc.com}
  \item \emph{Futurism}, \emph{``Lionsgate's Attempt to Create Movies Using AI Has Crumbled Into Disaster.''} \href{https://futurism.com/artificial-intelligence/lionsgate-movies-ai}{futurism.com}
  \item \emph{The Guardian}, \emph{``Disney to invest \$1bn in OpenAI.''} \href{https://www.theguardian.com/business/2025/dec/11/disney-open-ai-sora-video-deal}{theguardian.com}
  \item \emph{The Hollywood Reporter}, \emph{``Disney+ to Allow User-Generated Content Via AI.''} \href{https://www.hollywoodreporter.com/business/digital/disney-plus-gen-ai-user-generated-content-1236426135/}{hollywoodreporter.com}
  \item \emph{Deadline}, \emph{``Amazon's Head Of Original Sports Content Matt Newman Moves Into A.I.''} \href{https://deadline.com/2025/11/amazon-ai-studios-matt-newman-1236603477/}{deadline.com}
  \item \emph{Wired}, \emph{``Amazon's House of David Used Over 350 AI Shots in Season~2.''} \href{https://www.wired.com/story/amazons-house-of-david-used-over-350-ai-shots-in-season-2-its-creator-isnt-sorry/}{wired.com}
  \item \emph{Indiewire}, \emph{``New AI Studio Obsidian Forms Partnership with Imagine Entertainment.''} \href{https://www.indiewire.com/news/business/obsidian-studio-ai-production-company-imagine-entertainment-1235158619/}{indiewire.com}
  \item \emph{The Hollywood Reporter}, \emph{``Fremantle Names Boss of New AI Label Imaginae Studios.''} \href{https://www.hollywoodreporter.com/business/digital/fremantle-names-ceo-new-ai-label-imaginae-studios-1236396579/}{hollywoodreporter.com}
  \item \emph{PC Gamer}, \emph{``Palworld studio Pocketpair: `We don't believe in it'.''} \href{https://www.pcgamer.com/software/ai/palworld-studio-pocketpair-says-its-new-publishing-division-wont-handle-games-that-use-generative-ai-we-dont-believe-in-it/}{pcgamer.com}
  \item Variety, \emph{``Guillermo del Toro: `I'd rather die' than use generative AI.''} \href{https://variety.com/2025/film/news/guillermo-del-toro-rather-die-generative-ai-frankenstein-1236561316/}{variety.com}
  \item \emph{The Hollywood Reporter}, \emph{``Leonardo DiCaprio Says AI Can't Be Art.''} \href{https://www.hollywoodreporter.com/movies/movie-news/leonardo-dicaprio-ai-cant-be-art-no-humanity-1236445405/}{hollywoodreporter.com}
  \item \emph{NME}, \emph{``Jenna Ortega: `very easy to be terrified' of AI.''} \href{https://www.nme.com/news/jenna-ortega-says-its-very-easy-to-be-terrified-of-ai-in-filmmaking-3913926}{nme.com}
  \item \emph{GamesRadar}, \emph{``Wallace and Gromit creator: Aardman will `embrace the technology' of AI but will be `very cautious'.''} \href{https://www.gamesradar.com/entertainment/animation-movies/wallace-and-gromit-creator-says-beloved-animation-studio-aardman-will-embrace-the-technology-of-ai-but-will-be-very-cautious-not-to-lose-our-values/}{gamesradar.com}
  \item \emph{gamesindustry.biz}, \emph{```AI was an expensive mistake': Charles Cecil on Broken Sword.''} \href{https://www.gamesindustry.biz/ai-was-an-expensive-mistake-charles-cecil-on-innovation-insolvency-and-broken-sword}{gamesindustry.biz}
  \item \emph{gamesindustry.biz}, \emph{``RuneScape maker Jagex says it will never use generative AI.''} \href{https://www.gamesindustry.biz/runescape-maker-jagex-says-it-will-never-use-generative-ai-to-make-in-game-content}{gamesindustry.biz}
  \item \emph{Decrypt}, \emph{```Warhammer 40,000' Maker Games Workshop Rules Out Generative AI.''} \href{https://decrypt.co/354482/warhammer-40000-maker-games-workshop-rules-out-generative-ai}{decrypt.co}
  \item \emph{Niche Gamer}, \emph{``Larian Studios backs off from gen AI.''} \href{https://nichegamer.com/larian-studios-backs-off-from-gen-ai/}{nichegamer.com}
  \item \emph{PC Gamer}, \emph{``Todd Howard says AI can't replace human `creative intention'.''} \href{https://www.pcgamer.com/gaming-industry/todd-howard-says-ai-cant-replace-human-creative-intention-but-its-part-of-bethesdas-toolset-for-how-we-build-our-worlds-or-check-things/}{pcgamer.com}
  \item \emph{Hollywood Reporter}, \emph{``Netflix INKubator AI animation studio for `feature-quality' shorts.''} \href{https://www.hollywoodreporter.com/business/business-news/netflix-ai-animation-studio-inkubator-1236592110/}{hollywoodreporter.com}
  \item \emph{Forbes}, \emph{``Wonder Studios -- \$50M British studio aiming to become `A24 of AI'.''} \href{https://www.forbes.com/sites/charliefink/2026/05/18/meet-wonder-studios-the-50m-british-studio-striving-to-become-the-a24-of-ai-production/}{forbes.com}
  \item Runway, \emph{``Runway Japan -- Tokyo office and \$40M investment.''} \href{https://runwayml.com/blog/runway-japan}{runwayml.com}
  \item \emph{Hollywood Reporter}, \emph{``Cannes 2026: filmmakers shift towards cautious AI acceptance.''} \href{https://www.hollywoodreporter.com/business/business-news/cannes-2026-ai-acceptance-1236592488/}{hollywoodreporter.com}
  \item \emph{Variety}, \emph{``AI Dominates Cannes Buzz as Filmmakers Grudgingly Accept It.''} \href{https://variety.com/2026/film/festivals/ai-cannes-2026-filmmakers-accept-1236748402/}{variety.com}
  \item \emph{Variety}, \emph{``Peter Jackson: AI is basically the next wave of special effects.''} \href{https://variety.com/2026/film/news/peter-jackson-ai-special-effects-1236748120/}{variety.com}
  \item \emph{Variety}, \emph{``Kling AI + Evolutionary Films on `Minibots' at Cannes Market.''} \href{https://variety.com/2026/film/news/kling-ai-evolutionary-films-minibots-cannes-1236748590/}{variety.com}
  \item \emph{Cartoon Brew}, \emph{``Former Disney Animator: Marvel layoffs point to a bigger AI shift.''} \href{https://www.cartoonbrew.com/artist-rights/former-disney-animator-marvel-layoffs-ai-shift-260601.html}{cartoonbrew.com}
  \item \emph{Hollywood Reporter}, \emph{``Bobby Berk: AI will make reality TV and `verifiably human content' more valuable.''} \href{https://www.hollywoodreporter.com/tv/tv-news/bobby-berk-ai-reality-tv-1236592920/}{hollywoodreporter.com}
  \item \emph{PC Gamer}, \emph{``Take-Two's CEO: AI `datasets by their very nature are backward looking'.''} \href{https://www.pcgamer.com/games/take-two-ceo-ai-not-making-hits-backward-looking/}{pcgamer.com}
\end{itemize}

\section*{VII\@. World models and the new medium}

\begin{itemize}
  \item World Labs, \emph{``Bringing Marble to Life.''} \href{https://www.worldlabs.ai/case-studies/bringing-marble-to-life}{worldlabs.ai}
  \item TechCrunch, \emph{``Fei-Fei Li's World Labs speeds up the world model race with Marble.''} \href{https://techcrunch.com/2025/11/12/fei-fei-lis-world-labs-speeds-up-the-world-model-race-with-marble-its-first-commercial-product/}{techcrunch.com}
  \item Google DeepMind, \emph{``Genie~3: A new frontier for world models.''} \href{https://deepmind.google/blog/genie-3-a-new-frontier-for-world-models/}{deepmind.google}
  \item Google blog, \emph{``Project Genie: AI world model now available for Ultra users in U.S.''} \href{https://blog.google/innovation-and-ai/models-and-research/google-deepmind/project-genie/}{blog.google}
  \item Meta, \emph{``WorldGen -- text-to-immersive-3D-worlds.''} \href{https://www.facebook.com/LifeAtMeta/videos/research-update-worldgen-text-to-immersive-3d-worlds/1879077432692421/}{facebook.com}
  \item SpAItial, \emph{``Transform Images into Interactive 3D Worlds.''} \href{https://www.spaitial.ai/}{spaitial.ai}
  \item Stanford AI Lab, \emph{``WonderZoom.''} \href{https://wonderzoom.github.io/}{wonderzoom.github.io}
  \item NVIDIA/Stanford, \emph{``NitroGen.''} \href{https://nitrogen.minedojo.org/}{nitrogen.minedojo.org}
  \item DeepMind, \emph{``SIMA~2: An Agent that Plays, Reasons, and Learns.''} \href{https://deepmind.google/blog/sima-2-an-agent-that-plays-reasons-and-learns-with-you-in-virtual-3d-worlds/}{deepmind.google}
  \item Radiance Fields, \emph{``Apple Confirms Personas use Gaussian Splatting.''} \href{https://radiancefields.com/apple-confirms-personas-use-gaussian-splatting}{radiancefields.com}
  \item Sony Pictures' virtual-production case study (Marble integration). \href{https://www.linkedin.com/posts/brent-liang_tech-media-launch-ugcPost-7394911181091692546-TyUz}{linkedin.com}
  \item NVIDIA, \emph{``SANA-WM -- 2.6B open-source world model with 60-second video generation and camera control.''} \href{https://huggingface.co/collections/nvidia/sana-wm}{huggingface.co}
  \item Odyssey, \emph{``Starchild-1 -- first real-time multimodal world model.''} \href{https://odyssey.ml/introducing-starchild-1}{odyssey.ml}
  \item Odyssey, \emph{``Agora-1 -- four-player AI-generated world built on a 1997 shooter.''} \href{https://odyssey.ml/introducing-agora-1}{odyssey.ml}
  \item Google DeepMind, \emph{``Project Genie + Street View -- simulate real-world places.''} \href{https://deepmind.google/discover/blog/project-genie-street-view/}{deepmind.google}
  \item Google Labs, \emph{``Infinite Scaler -- browser multiplayer generative game.''} \href{https://blog.google/technology/google-labs/infinite-scaler/}{blog.google}
  \item Apple Machine Learning Research, \emph{``Apple Headsup -- 3D Gaussian Head Reconstruction from Multi-View Captures.''} \href{https://machinelearning.apple.com/research/apple-headsup-3d-gaussian-head}{machinelearning.apple.com}
  \item WorldLens VR, \emph{``AI-powered 3D depth on Quest for Google Street View.''} \href{https://www.uploadvr.com/worldlens-vr-quest-street-view-3d-depth/}{uploadvr.com}
  \item \emph{gamesindustry.biz}, \emph{``Meet Seed, the planet-sized society simulator.''} \href{https://www.gamesindustry.biz/meet-seed-planet-sized-society-simulator}{gamesindustry.biz}
\end{itemize}

\section*{VIII\@. The platform layer and the AI-native toolchain}

\begin{itemize}
  \item \emph{Creative Boom}, \emph{``Adobe puts AI in everything, everywhere, all at once.''} \href{https://www.creativeboom.com/news/adobe-is-putting-ai-in-everything-everywhere-all-at-once/}{creativeboom.com}
  \item Adobe, \emph{``Adobe MAX 2025: Firefly Foundry.''} \href{https://news.adobe.com/news/2025/10/adobe-max-2025-firefly-foundry}{news.adobe.com}
  \item Adobe, \emph{``Adobe MAX 2025: Express AI Assistant.''} \href{https://news.adobe.com/news/2025/10/adobe-max-2025-express-ai-assistant}{news.adobe.com}
  \item \emph{Wired}, \emph{``Adobe's Corrective AI Can Change the Emotions of a Voice-Over.''} \href{https://www.wired.com/story/adobe-max-sneaks-2025-corrective-ai/}{wired.com}
  \item PYMNTS, \emph{``Adobe Lets Users Design and Edit Using ChatGPT.''} \href{https://www.pymnts.com/artificial-intelligence-2/2025/adobe-lets-users-design-and-edit-using-chatgpt/}{pymnts.com}
  \item Adobe blog, \emph{``Edit with Photoshop in ChatGPT.''} \href{https://blog.adobe.com/en/publish/2025/12/10/edit-photoshop-chatgpt}{blog.adobe.com}
  \item \emph{Campaign Brief}, \emph{``WPP launches WPP Open Pro.''} \href{https://campaignbrief.com/wpp-launches-ai-powered-marketing-platform-wpp-open-pro/}{campaignbrief.com}
  \item \emph{Digiday}, \emph{``WPP expands AI capabilities.''} \href{https://digiday.com/media-buying/agencies-continue-to-expand-ai-capabilities-to-boost-brand-performance/}{digiday.com}
  \item \emph{SiliconAngle}, \emph{``Higgsfield raises \$80M on \$1.3B valuation.''} \href{https://siliconangle.com/2026/01/15/higgsfield-raises-80m-1-3b-valuation-scale-ai-video-platform/}{siliconangle.com}
  \item TechCrunch, \emph{``Synthesia hits \$4B valuation.''} \href{https://techcrunch.com/2026/01/26/synthesia-hits-4b-valuation-lets-employees-cash-in/}{techcrunch.com}
  \item \emph{Sifted}, \emph{``Synthesia rejects \$3bn Adobe acquisition offer.''} \href{https://sifted.eu/articles/synthesia-acquisition-offer}{sifted.eu}
  \item ComfyUI, \emph{``We raised \$17 million to build an OS for creative AI.''} \href{https://www.linkedin.com/posts/comfyui_we-raised-17-million-to-build-an-os-for-ugcPost-7373743341236236288-wkCc}{linkedin.com}
  \item Beeple Canvas, \emph{``Generative AI meets compositing.''} \href{https://www.beeple-canvas.com/}{beeple-canvas.com}
  \item \emph{MacRumors}, \emph{``New CarPlay Audio Apps Add AI Music and Global Radio.''} \href{https://www.macrumors.com/2026/05/19/carplay-ai-music-global-radio/}{macrumors.com}
  \item \emph{The Drum}, \emph{``World's First AI-Driven Immersive Shoppable Catwalk.''} \href{https://www.thedrum.com/news/2026/05/18/how-world-s-first-ai-driven-immersive-shoppable-catwalk-redefined-future-brand}{thedrum.com}
  \item \emph{The Drum}, \emph{``David Beckham designs `Henchester United' chicken coop in Lenovo AI ad.''} \href{https://www.thedrum.com/news/2026/05/18/david-beckham-henchester-united-chicken-coop-lenovo-ai-ad}{thedrum.com}
  \item \emph{Amazon}, \emph{``Alexa+ debuts feature to create AI-generated podcasts.''} \href{https://www.aboutamazon.com/news/devices/alexa-plus-ai-podcasts}{aboutamazon.com}
  \item ECABridge, \emph{``Unreal Engine MCP integration.''} \href{https://ecabridge.dev/}{ecabridge.dev}
  \item \emph{Video Games Chronicle}, \emph{``Epic Games Veteran on AI-Heavy `Fully European' Game Engine.''} \href{https://www.videogameschronicle.com/news/epic-games-veteran-ai-heavy-fully-european-game-engine/}{videogameschronicle.com}
\end{itemize}

\section*{IX\@. Authenticity, provenance and disclosure}

\begin{itemize}
  \item \emph{Digital Music News}, \emph{``Instagram Chief Says We Should `Fingerprint Real Media'.''} \href{https://www.digitalmusicnews.com/2026/01/05/instagram-chief-ai-slop-comments/}{digitalmusicnews.com}
  \item \emph{WebProNews}, \emph{``Instagram Head Warns AI Images Erode Trust.''} \href{https://www.webpronews.com/instagram-head-warns-ai-images-erode-trust-calls-for-verification-standards/}{webpronews.com}
  \item Google DeepMind, \emph{``Verify AI-generated videos in the Gemini app.''} \href{https://www.linkedin.com/posts/googledeepmind_verify-google-ai-generated-videos-in-the-activity-7407748300688478208-fJgW}{linkedin.com}
  \item \emph{The Hollywood Reporter}, \emph{``Emmys Set AI Guidelines for 2026 Awards.''} \href{https://www.hollywoodreporter.com/tv/tv-news/emmys-ai-guidelines-2026-awards-1236468434/}{hollywoodreporter.com}
  \item \emph{Lawyer Monthly}, \emph{``Matthew McConaughey Trademarks Voice and Image Against AI.''} \href{https://www.lawyer-monthly.com/2026/01/matthew-mcconaughey-protects-voice-image-ai/}{lawyer-monthly.com}
  \item \emph{Adweek}, \emph{``Meet the \$1.3 Billion Startup Behind Madonna and Will Smith's AI Video.''} \href{https://www.adweek.com/media/higgsfield-ai-marketing-startup/}{adweek.com}
  \item \emph{Marketing Week}, \emph{``You can't dismiss AI ads as slop when they're winning in testing.''} \href{https://www.marketingweek.com/dismiss-ai-ads-winning-creative-effectiveness/}{marketingweek.com}
  \item Variety, \emph{``George Clooney: AI Actors Will Struggle to Be Movie Stars.''} \href{https://variety.com/2025/scene/columns/george-clooney-ai-actors-movie-stars-1236579661/}{variety.com}
  \item \emph{The Hollywood Reporter}, \emph{``Synthetic Sincerity on AI Characters, Authenticity (IDFA).''} \href{https://www.hollywoodreporter.com/movies/movie-news/synthetic-sincerity-film-idfa-ai-authenticity-interview-1236426180/}{hollywoodreporter.com}
  \item Variety, \emph{``AI-Generated Images Lead to Audiences' Distrust, Threaten Documentary.''} \href{https://variety.com/2025/film/festivals/ai-generated-images-threaten-future-of-documentary-1236583466/}{variety.com}
  \item \emph{Hollywood Reporter}, \emph{``Merriam-Webster Names Slop Word of the Year 2025.''} \href{https://www.hollywoodreporter.com/news/general-news/slop-word-year-2025-merriam-webster-1236450780/}{hollywoodreporter.com}
  \item Google DeepMind, \emph{``SynthID -- 100 billion items watermarked; partner expansion to OpenAI, ElevenLabs, Kakao.''} \href{https://deepmind.google/discover/blog/synthid-100-billion-watermarks-partners/}{deepmind.google}
  \item \emph{Variety}, \emph{``Cate Blanchett co-founds RSL Media -- non-profit on AI consent for name, image, likeness.''} \href{https://variety.com/2026/film/news/cate-blanchett-rsl-media-ai-consent-1236748255/}{variety.com}
  \item \emph{Bloomberg}, \emph{``Apple Acquires Key Talent \& Patents Behind AI Avatar Company `Animato'.''} \href{https://www.bloomberg.com/news/articles/2026-05-19/apple-acquires-animato-ai-avatar-talent-patents}{bloomberg.com}
  \item Snap Newsroom, \emph{``Snapchat Gen~Z AI Creativity Research 2026 -- 31\% of 13--15 year-olds use AI to be creative.''} \href{https://newsroom.snap.com/snapchat-gen-z-ai-creativity-research-2026}{newsroom.snap.com}
  \item YouGov AU, \emph{``45\% of Australians say AI-generated ads would make them trust a brand less.''} \href{https://yougov.com.au/topics/consumer/articles-reports/2026/05/19/45-percent-australians-ai-generated-ads-trust-brand-less}{yougov.com.au}
\end{itemize}

\section*{X\@. Labour, organisation and the new geography}

\begin{itemize}
  \item Imperva/Cloudflare bot reports (Section~III).
  \item Adobe, \emph{``Inaugural Adobe Creators' Toolkit Report: 86\% of Global Creators Use Creative Generative AI.''} \href{https://news.adobe.com/news/2025/10/adobe-max-2025-creators-survey}{news.adobe.com}
  \item Azumo, \emph{``AI in the Workplace Statistics 2026.''} \href{https://azumo.com/artificial-intelligence/ai-insights/ai-in-workplace-statistics}{azumo.com}
  \item \emph{Tech.co}, \emph{``Gen~Z Most Likely to Sneakily Use AI Without Telling the Boss.''} \href{https://tech.co/news/gen-z-most-likely-use-ai-boss}{tech.co}
  \item \emph{Exploding Topics}, \emph{``The Hidden AI Workforce.''} \href{https://explodingtopics.com/blog/ai-workforce-research}{explodingtopics.com}
  \item \emph{Forbes}, \emph{``AI Tools Flood Workplaces as Employees Face a Double Bind.''} \href{https://www.forbes.com/sites/carolinecastrillon/2025/09/09/ai-tools-flood-workplaces-as-employees-face-a-double-bind/}{forbes.com}
  \item IDC Europe, \emph{``Shadow AI.''} \href{https://blog-idceurope.com/shadow-ai-how-stealth-productivity-is-strangling-enterprise-ai-adoption-and-creating-a-security-nightmare/}{blog-idceurope.com}
  \item \emph{Game Developer}, \emph{``Krafton outlines plans to become `AI First' company.''} \href{https://www.gamedeveloper.com/business/subnautica-owner-krafton-outlines-plans-to-transform-into-an-ai-first-company}{gamedeveloper.com}
  \item \emph{CNBC}, \emph{``Why workers with ADHD, autism, dyslexia should use AI agents.''} \href{https://www.cnbc.com/2025/11/08/adhd-autism-dyslexia-jobs-careers-ai-agents-success.html}{cnbc.com}
  \item \emph{The Guardian}, \emph{``AI is hitting UK harder than other big economies.''} \href{https://www.theguardian.com/technology/2026/jan/26/ai-uk-jobs-us-japan-germany-australia}{theguardian.com}
  \item \emph{PocketGamer.biz}, \emph{``Shift Up CEO says AI is key to competing with China's game industry scale.''} \href{https://www.pocketgamer.biz/shift-up-ceo-says-ai-is-key-to-competing-with-chinas-game-industry-scale/}{pocketgamer.biz}
  \item CNBC Africa, \emph{``How AI is changing the landscape of the music industry in Africa.''} \href{https://www.cnbcafrica.com/2025/how-ai-is-changing-the-landscape-of-the-music-industry-in-africa}{cnbcafrica.com}
  \item BBC Future, \emph{``Lights, camera, algorithm: Why Indian cinema is awash with AI.''} \href{https://www.bbc.co.uk/future/article/20251223-why-indian-cinema-is-awash-with-ai}{bbc.co.uk}
  \item \emph{Broadcast Pro Middle East}, \emph{``Tunisian filmmaker wins \$1 million AI Film Award for `Lily'.''} \href{https://www.broadcastprome.com/news/tunisian-filmmaker-wins-1-million-ai-film-award-for-lily/}{broadcastprome.com}
\end{itemize}

\section*{XI\@. The literacy turn -- institutions and education}

\begin{itemize}
  \item Sundance Institute, \emph{``Centering the Artist: Why We're Launching the AI Literacy Initiative.''} \href{https://www.sundance.org/blogs/centering-the-artist-why-were-launching-the-ai-literacy-initiative/}{sundance.org}
  \item Google blog, \emph{``Building a community-led future for AI in film with Sundance Institute.''} \href{https://blog.google/company-news/outreach-and-initiatives/google-org/sundance-institute-ai-education/}{blog.google}
  \item McKinsey \& Company, \emph{``What AI could mean for film and TV production.''} \href{https://www.mckinsey.com/industries/technology-media-and-telecommunications/our-insights/what-ai-could-mean-for-film-and-tv-production-and-the-industrys-future}{mckinsey.com}
  \item UK Government, \emph{``Free AI training for all.''} \href{https://www.gov.uk/government/news/free-ai-training-for-all-as-government-and-industry-programme-expands-to-provide-10-million-workers-with-key-ai-skills-by-2030}{gov.uk}
  \item University of Wisconsin-Stout, \emph{``AI Reshaping Industry: AI-use as Baseline Competency in Filmmaking.''} \href{https://www.uwstout.edu/about-us/news-center/ai-reshaping-industry-new-uw-stout-course-sets-ai-use-baseline-competency-filmmaking}{uwstout.edu}
\end{itemize}

\section*{XII\@. Newsletters and primary corpus}

\begin{itemize}
  \item \emph{Dream Machine | Creative AI} -- LinkedIn newsletter, archive of Issues 1--30 (6~October~2025 -- 21~May~2026). \href{https://www.linkedin.com/newsletters/dream-machine-creative-ai-7379776527871381505/}{linkedin.com}
  \item DreamLab AI Collective, team page. \href{https://dreamlab-ai.com/team}{dreamlab-ai.com}
  \item DreamLab Substack, \emph{``Some Predictions on Creative AI for 2026''} -- \emph{Dream Machine} Issue~13 companion piece. \href{https://substack.com/home/post/p-183997149}{substack.com}
\end{itemize}

  \chapter{The Shadow AI Paradox in the Creative Industries}\label{app:a4}

\emph{Companion piece to Chapter~13: Coordination Collapse.}

This deep dive is the long-form analytical companion to the shadow-AI sections of Chapter~13. Where the chapter argues, in the book's voice, that the creative industries are operating with two parallel economies on top of each other -- a vocal public economy of AI refusal and a silent private economy of AI adoption -- this appendix lays out the underlying data, the sectoral breakdowns, the security and IP implications, the linguistic markers that betray covert AI use, and the labour-market mechanics of agentic displacement that the chapter compresses for narrative purposes.

The headline numbers Chapter~13 quotes -- 88--89\% staff adoption against 71--80\% covert use, the \$670,000 average breach cost, the 1,100-creator music survey showing 87\% AI use against 77\% ``loss of originality'' concern, the WGA pre- vs.\ post-strike screenwriter shift from 34\% to 68\%, the GDC sentiment-vs-usage divergence -- all sit inside the fuller treatment below. Read alongside Appendix~E: Dynamics of Generative AI Adoption, it forms the empirical spine of the book's argument that the creative industries' stated position on AI and their operational position on AI are, in 2026, sharply divergent -- and that this divergence is itself the strategic question every creative organisation now has to face.

\bigskip\par\noindent\rule{\textwidth}{0.4pt}\par\bigskip

\section*{The Shadow AI Paradox in the Creative Industries: Covert Adoption, Linguistic Betrayal, and the Displacement Crisis}

The integration of artificial intelligence within the global creative economy has precipitated one of the most profound technological, economic, and cultural shifts in modern history. However, the prevailing narrative surrounding this transition is characterised by a severe and highly visible dichotomy. Publicly, the creative industries -- spanning music, gaming, film, television, animation, broadcasting, and advertising -- are engaged in a vocal, highly publicised, and often legally combative resistance against generative artificial intelligence. Trade unions, prominent artists, and media conglomerates routinely denounce the technology, citing massive ethical violations, the non-consensual scraping of copyrighted material, and the existential threat to authentic human expression and labour.

Privately, however, the operational reality is starkly different. The sector is currently experiencing an unprecedented and accelerating surge in ``shadow AI'' -- the covert, unsanctioned use of artificial intelligence tools by employees, freelancers, and executives outside of formal corporate IT and governance structures. This phenomenon reveals a pervasive cognitive dissonance within the modern creative workforce. Professionals who actively and passionately denounce the use of generative models to protect their specific disciplines are simultaneously utilising the exact same underlying technologies to automate tasks they deem secondary, tedious, or outside their purview -- such as coding, copywriting, administrative communication, metadata generation, and data analysis.

This localised protectionism, frequently characterised as the ``AI for thee, but not for me'' paradox, not only exposes the psychological rationalisations of modern knowledge workers but also accelerates the very displacement they publicly seek to prevent. By feeding proprietary data, unreleased assets, and intellectual capital into public Large Language Models (LLMs) to achieve personal efficiency gains, covert users are inadvertently training the systems that are rendering their own industries and the broader creative ecosystem obsolete. The displacement caused by this technology is highly problematic and systemic, regardless of whether a user justifies their usage as merely ``augmentative'' for non-core tasks.

This comprehensive research report examines the prevalence, mechanics, and long-term consequences of shadow AI in the creative sectors during the critical 2024--2026 transitional period. It analyses the specific linguistic markers that betray covert AI usage in professional communications, the economic mechanisms driving continuous labour displacement, the specific usage patterns across various creative sub-sectors, and the strategic governance frameworks that organisations must implement to navigate this dual-faced reality. Ultimately, this report outlines the logical conclusion of a paradigm where an industry covertly relies on the very technology it publicly decries.

\section*{The Epistemology and Scale of Shadow AI}

To understand the hypocrisy of the creative sector's AI adoption, it is first necessary to comprehend the sheer scale of shadow AI across enterprise environments. Shadow AI is the natural evolutionary successor to the older concept of shadow IT, but its implications are vastly more severe. While traditional shadow IT involved the unauthorised use of rogue cloud storage or project management apps, shadow AI involves autonomous, self-learning systems that ingest, retain, and iterate upon the data fed into them.

Between 2023 and 2024, the adoption of generative AI applications by enterprise employees grew from 74\% to 96\%, tracking an explosive trajectory that caught corporate compliance departments entirely off guard. By the end of 2025 and moving into 2026, the ``Bring Your Own AI'' (BYOAI) movement became the dominant operational paradigm. Data from extensive workforce tracking indicates that up to 89\% of staff across various organisational departments utilise AI tools, with 71\% to 80\% of those employees utilising them without official approval or IT oversight.

The scale of this hidden infrastructure has resulted in a phenomenon analysts describe as the ``Hidden Cloud Explosion.'' The disparity between what corporate leadership believes is happening and what is actually occurring on employee devices is massive.

\begin{center}
\begin{tabular}{p{3cm} p{3.5cm} p{3.5cm} p{4cm}}
\toprule
\textbf{Metric} & \textbf{Perceived Reality} & \textbf{Actual Reality} & \textbf{Strategic Implication} \\
\midrule
Enterprise AI Adoption Rate & Stalled in pilot phases; limited official rollout & 88--89\% active daily users & Official corporate AI initiatives are moving too slowly, forcing employees to utilise public tools \\
\midrule
Cloud Service Visibility & \textasciitilde91 public cloud services estimated & Average of 1,220 active services & IT departments suffer from a 90\% visibility gap \\
\midrule
High-Risk Applications & Assumed zero tolerance & Average of 44 undetected high-risk cloud services per enterprise & Continuous vulnerability to automated data scraping \\
\midrule
Compliance Awareness & Assumed mandatory training & Only 23\% of users are aware of compliance risks & Data exposure is largely driven by negligence rather than malicious intent \\
\bottomrule
\end{tabular}
\end{center}

The security, financial, and intellectual property risks associated with this covert adoption are crippling to creative enterprises. In 2025, 20\% of organisations experienced severe security incidents directly linked to shadow AI, which increased the average cost of a data breach by \$670,000. The operational mechanisms of these tools -- specifically the tendency of users to paste proprietary source code, legal drafts, financial models, and unreleased creative assets into public LLMs like ChatGPT or Claude -- resulted in the exposure of personally identifiable information (PII) in 65\% of incidents, and the direct theft or exposure of intellectual property in 40\% of incidents.

When an animator pastes proprietary pipeline code into an AI debugging assistant, or a screenwriter uploads an unproduced treatment into an LLM to generate character summaries, they are essentially bypassing corporate endpoint monitoring and handing trade secrets to third-party data processors. The AI's self-learning nature means that these risks compound exponentially; data leaked on a Tuesday can theoretically be used to train a model outputting responses to a competitor by Thursday. Despite these massive systemic risks, the adoption continues unabated, driven by the intense pressure on creative workers to increase their output velocity and streamline workflows in an era of shrinking budgets.

\section*{The Great Hypocrisy: ``AI for Thee, But Not for Me''}

At the core of the shadow AI phenomenon in the creative sector is a profound psychological, economic, and ideological paradox. Creative professionals frequently exhibit a fierce, localised protectionism regarding their own specific skill sets and domains, while demonstrating a complete and enthusiastic willingness to automate the labour of their peers and collaborators.

This dynamic is widely recognised and criticised in developer and creative circles as the ``AI for thee, but not for me'' paradox. The psychology driving this hypocrisy rests on a highly subjective and hierarchical valuation of human labour. Creatives routinely categorise tasks outside their immediate domain as ``mundane,'' ``tedious,'' ``administrative,'' or ``purely technical,'' thereby framing the use of AI in these areas as a harmless, victimless efficiency gain. Conversely, they view their own specific domain as an expression of authentic, irreplaceable human experience, soul, and creative genius that cannot, and morally should not, be replicated by an algorithm.

For example, a traditional illustrator or concept artist may vehemently campaign against text-to-image models like Midjourney or Stable Diffusion, viewing the scraping of their portfolios as copyright infringement and a desecration of the artistic process. They will join public boycotts and demand protective legislation. Yet, that exact same illustrator may comfortably and secretly use an LLM to write their marketing copy, draft their client contracts, or generate the HTML and Python scripts required to manage their portfolio website. In doing so, they are actively devaluing and bypassing the labour of copywriters, paralegals, and web developers.

Similarly, an independent filmmaker might loudly denounce the use of generative video models, arguing that they destroy the craft of cinematography. However, that filmmaker may simultaneously utilise AI audio-cleaning tools to bypass the need for a professional sound mixer, or use an AI business agent to handle their accounting.

This hierarchical thinking is not only hypocritical; it is fundamentally flawed in its understanding of how generative AI models operate. When an independent creator uses an LLM to generate an email campaign or a block of code, they are exploiting the exact same mechanism of non-consensual data scraping that powers image and video generation. The models that generate text and code are trained on the massive, often unlicensed, copyrighted works of authors, journalists, programmers, and corporate communications. The ethical breach is identical, but the personal proximity to the threat alters the user's moral calculus.

\section*{The Problematic Nature of ``Augmentative'' Displacement}

The justification for this behaviour relies heavily on the narrative of ``augmentation.'' Workers convince themselves that they are merely using AI to remove friction from their day, allowing them to focus on the ``real'' creative work. However, displacement, no matter how the AI is being used, is highly problematic for the broader economic ecosystem.

When creatives use AI for ``someone else's job,'' they are contributing to a structural shift toward the ``one-person business'' model. By utilising workflow automation platforms and shadow AI tools, individual creatives can scale their output to rival small agencies, generating massive localised profit. While this empowers the individual user, it fundamentally relies on the systemic destruction of entry-level and specialised labour.

This creates a broken pipeline for future talent. The creative industries have historically relied on a mentorship and apprenticeship model, where junior staff learn the intricacies of a craft by performing the very ``mundane'' tasks that are now being automated. By eagerly adopting shadow AI for these peripheral tasks, current professionals are burning the bridge they crossed, ensuring that the next generation of creatives has no entry point into the industry.

\section*{Linguistic Betrayal: Exposing the Covert User}

The hypocrisy of the shadow AI user is frequently exposed not through sophisticated IT audits or complex network monitoring, but through glaring linguistic betrayals. As creatives increasingly rely on LLMs to generate their professional communications, grant proposals, LinkedIn updates, and social media posts, they inadvertently adopt the semantic and structural artefacts inherent to generative models.

Because creative professionals generally possess strong intrinsic communication skills, their sudden shift to AI-generated text is highly conspicuous. LLMs are trained via Reinforcement Learning from Human Feedback (RLHF) to be relentlessly helpful, polite, and enthusiastically compliant. Consequently, unedited AI output possesses a highly distinct, overly polished, yet entirely soulless corporate tone that lacks authentic human nuance, imperfection, or quirk. This phenomenon has led to a flattening of the craft of writing, transforming professional discourse into a sea of sycophantic, buzzword-laden uniformity.

\begin{center}
\begin{tabular}{p{3.5cm} p{5cm} p{5cm}}
\toprule
\textbf{Linguistic Marker Category} & \textbf{Specific Identifiers and Behaviours} & \textbf{Psychological and Technical Origin} \\
\midrule
The Hyperbolic Vocabulary of Enthusiasm & ``Delve'', ``Tapestry'', ``Thrilled'', ``Transformative'', ``Rockstar'', ``Dynamic'', ``Game-changer'' & RLHF training weights heavily favour high-engagement, extremely positive corporate rhetoric to avoid user offence \\
\midrule
Formulaic Structural Predictability & Hook -- Milestone Announcement -- Gratitude to Leadership -- Broad Platitude -- Call to Action & Algorithmic reliance on statistically common business-communication templates scraped from platforms like LinkedIn \\
\midrule
Predictable Emoji Topography & \texttt{:rocket: :tada: :briefcase: :clap: :fire:} placed at exact paragraph terminations or used in excessive clusters & A mechanical simulation of human digital emotion designed to maximise algorithmic engagement \\
\midrule
The ``Slip-Up'' Artefact & ``Let me know if you need any modifications!'', ``Here is the drafted response:'', ``As an AI\ldots{}'' & User negligence, extreme haste, and the failure to proofread automated output before publishing \\
\midrule
Lack of Specificity & Broad lessons on leadership or creativity without specific project metrics, dates, or personal anecdotes & AI models lack real-world context unless explicitly prompted; they default to generalised wisdom to fill space \\
\bottomrule
\end{tabular}
\end{center}

The ultimate irony is that creative professionals -- who build their entire professional identities on originality, authentic expression, and a unique point of view -- are willingly outsourcing their public voices to algorithms that produce the median, most inoffensive blend of corporate speech possible.

\section*{Sector-Specific Analysis of Covert Adoption}

\subsection*{Music Production and Sound Recording}

The music industry has experienced a highly volatile relationship with generative AI. By 2026, the narrative of ``saving human music'' clashed directly with internal studio practices. An exhaustive survey of over 1,100 professional music creators -- including producers, audio engineers, and songwriters -- revealed that a staggering 87\% of producers were actively using AI tools in their creative process.

The usage within the music sector is highly stratified. According to industry data, 58\% utilise AI for audio restoration and cleanup, 38\% for mixing assistants, and 33.9\% for automated mastering. However, when it comes to authorship -- the core identity of the artist -- resistance stiffens considerably. Only 20.9\% admit to using composition or lyric-generation tools.

There is a profound, existential fear of ``musical sameness,'' with 77\% of producers citing the loss of originality as their primary concern, superseding even the fear of job displacement (42\%).

\subsection*{Gaming and Interactive Entertainment}

The video game industry exhibits the most volatile public reactions to AI, frequently resulting in massive public relations disasters and community outrage. Consumer backlash against perceived ``AI slop'' has become so severe that it is causing active collateral damage to human artists. Game developers who commission genuine, hand-drawn human artwork have faced aggressive online harassment and accusations of using AI, simply because their art style mirrored the hyper-polished aesthetics popularised by Midjourney and Stable Diffusion.

In response, several publishers have implemented draconian anti-AI policies. Hooded Horse, a prominent indie publisher, formally banned the use of AI, inserting ``no f**king AI assets'' clauses into all of its developer contracts. However, the reality of modern game development -- which requires massive, unprecedented volumes of assets, endless lines of code, and continuous debugging -- makes strict adherence to these bans nearly impossible.

The gaming sector perfectly encapsulates the shadow AI dilemma. Developers widely and enthusiastically use AI coding assistants like GitHub Copilot to write scripts, autocomplete logic, and debug errors, viewing it as an absolute necessity for productivity and survival in a crunch-heavy industry. Yet, the integration of AI for visual assets or narrative design is treated as a moral failing.

\subsection*{Film, Television, and Animation}

Hollywood currently operates under a pervasive ``don't ask, don't tell'' culture regarding artificial intelligence. In the wake of historic industry strikes, the Writers Guild of America (WGA) and SAG-AFTRA established rigid, legally binding guidelines surrounding AI. These guidelines mandate explicit, 48-hour advanced consent and mandatory compensation for the creation of ``Employment-Based Digital Replicas'' and ``Synthetic Performers.'' Writers are permitted to use generative AI as a tool, provided it is disclosed, but studios cannot force writers to use it, nor can AI be credited with authorship or used to reduce a writer's residuals.

Despite these hard-fought contractual boundaries, covert usage is rampant across the supply chain. Industry executives and insiders note that studios are frequently lying about how much AI they are utilising in post-production, storyboarding, and visual effects to avoid union grievances and consumer backlash.

The animation sector faces an even more dire existential crisis. A 2025 Luminate Intelligence report highlighted that animation executives view generative AI as a revolutionary mechanism to slash production times and reduce ballooning budgets. The Animation Guild estimates that 21\% of animation tasks are vulnerable to immediate AI exposure and automation, putting nearly 40,000 jobs in California alone at risk.

\subsection*{Broadcasting, Advertising, and Agency Production}

In the broader broadcasting and advertising sectors, the integration of AI has moved from a novelty to a core operational requirement. While 2024 was marked by AI moving from a ``side tab'' to a ``core app,'' 2025 and 2026 ushered in the ``Agentic Era.'' However, official corporate rollouts of these tools have been remarkably clumsy. Up to 70\% of official corporate AI initiatives in broadcasting and media fall short of expectations, plagued by rigorous compliance checks, clunky enterprise interfaces, and a lack of specific training.

When agencies attempt to use AI overtly -- as seen in Coca-Cola's heavily criticised AI-generated Christmas advertisement -- they risk destroying decades of brand equity. This public backlash forces agencies back into the shadows.

\section*{The Mechanics of Market Devaluation and Displacement}

The deeply held belief among creative professionals that shadow AI functions merely as an ``augmentative'' tool is an economic fallacy. No matter how these tools are utilised, the aggregate macro-economic effect is the systematic displacement of human labour and the rapid devaluation of the creative economy.

According to the United Nations Educational, Scientific and Cultural Organisation (UNESCO), the proliferation of generative AI is projected to cause significant and irreversible income losses by 2028. The report warns that music creators face a 24\% drop in revenues, while audiovisual professionals are projected to lose 21\% of their income as AI-generated content floods global markets. A broader economic analysis by Goldman Sachs estimates that up to 300 million jobs globally are exposed to AI automation over the next decade, with a specific focus on knowledge workers and creative sectors.

\subsection*{The Stanford Study: Flooding the Market}

A critical study conducted by Stanford Graduate School of Business analysed the market dynamics when AI-generated art was introduced to a platform alongside traditional human-created art. Once generative AI entered the marketplace, the total supply of images skyrocketed exponentially, and consequently, the volume of human-generated images fell dramatically. More alarmingly for traditional creators, consumers actively demonstrated a preference for the influx of AI-generated images, choosing them over human-generated ones. The technology flooded the market with high quality and infinite variety at a near-zero marginal cost.

\subsection*{The Enclosure of the Knowledge Commons}

Every instance of shadow AI usage contributes directly to this dynamic. Generative models function as massive, continuous feedback loops. When an employee covertly inputs a proprietary script, a unique visual asset, or an innovative piece of code into a public LLM to save time, they are voluntarily surrendering their intellectual property and the collective knowledge of their industry to the algorithm.

This process is known in critical theory as the ``enclosure of the knowledge commons.'' Major tech corporations rely on this continuous, free ingestion of human labour to refine their models, subsequently monopolising that accumulated intelligence and selling it back to the market in the form of autonomous agents and enterprise subscriptions.

The creative professional who uses shadow AI is not just cheating their employer's IT policy to leave work an hour early; they are actively, literally training their permanent replacement.

\section*{Strategic Restitution: Formalising AI Governance}

The pervasive, deeply entrenched nature of shadow AI within the creative sectors dictates that traditional IT prohibition strategies are entirely futile. Banning public LLMs, implementing strict firewalls, or punishing employees simply drives the behaviour further underground.

To survive this transition, creative organisations must move away from a posture of denial and transition toward a model of ``structured enablement'' and formal AI governance.

\begin{center}
\begin{tabular}{p{2.5cm} p{4.5cm} p{4.5cm}}
\toprule
\textbf{Governance Maturity Phase} & \textbf{Operational Characteristics} & \textbf{Required Security \& Compliance Actions} \\
\midrule
Phase~1: Discovery \& Informal & Ubiquitous shadow AI. No official policy. Staff unaware of compliance risks. Rampant BYOAI. & Implement network telemetry to audit actual SaaS usage. Identify high-risk data flows. Conduct upfront reviews of vendor Terms of Service for embedded AI. \\
\midrule
Phase~2: Ad Hoc \& Transitional & Walled-garden access (e.g., Enterprise Copilot) introduced. Basic acceptable use policies distributed to staff. & Implement API gateways. Block transmission of PII to public models. Roll out mandatory AI literacy training to bridge the 50\% awareness gap. \\
\midrule
Phase~3: Formal \& Integrated & AI usage is transparent, licensed, and actively monitored. Ethical and IP guardrails are automated directly into the creative workflow. & Continuous monitoring for model drift and algorithmic bias. Formalised vendor data agreements guaranteeing zero data retention. Compliance with regional legislation (e.g., EU AI Act). \\
\bottomrule
\end{tabular}
\end{center}

Most importantly, organisations must supply enterprise-grade, heavily secured AI solutions that guarantee zero data retention by the model providers. When employees are provided with sanctioned, secure tools that perform as well as or better than the public shadows, the incentive to bypass IT protocols and risk data exposure entirely evaporates.

\section*{The Logical Conclusion}

The trajectory of shadow AI within the creative industries points toward an inescapable, highly disruptive logical conclusion. The current state of cognitive dissonance -- where creative professionals publicly demonise artificial intelligence as the death of art while privately relying on it to augment their daily output -- is merely a brief, unstable transitional phase.

As the technology's capabilities expand exponentially beyond the automation of ``mundane'' tasks and begin to consistently perform judgement-intensive, open-ended creative work at human-level proficiency, the fragile, hypocritical truce of ``augmentation'' will inevitably collapse. The displacement of human labour is not an accidental or unfortunate byproduct of generative AI; it is its foundational design and economic purpose.

The survival of the creative professional will not depend on successfully protecting a specific, granular skill set from automation through public boycotts. Rather, it will depend on transitioning to highly formalised, governed AI integration that explicitly protects human intellectual property at the enterprise level, ensuring that artists are compensated for the data they generate.

The alternative is the total, irreversible enclosure of the creative commons, where human ingenuity serves merely as the unpaid, unacknowledged training data for the automated, synthetic pipelines of tomorrow. The creative industries must drag AI out of the shadows and govern it in the light, or risk being entirely consumed by it.

\bigskip\par\noindent\rule{\textwidth}{0.4pt}\par\bigskip

\section*{Works Cited}

\begin{enumerate}
  \item Shadow AI and the Future of Work: What Knowledge Workers Need to Know in 2026. \href{https://www.reddit.com/r/it/comments/1qn2f9w/shadow_ai_and_the_future_of_work_what_knowledge/}{reddit.com/r/it}
  \item What Is Shadow AI? -- IBM. \href{https://www.ibm.com/think/topics/shadow-ai}{ibm.com}
  \item AI Isn't Taking Over All Creative Jobs (So Embrace It). \href{https://www.now-hear-this.net/content/ai-isnt-taking-over-all-creative-jobs-so-embrace-it}{now-hear-this.net}
  \item Reddit: I stopped defending my creative team and let leaders use AI\ldots{} it failed. \href{https://www.reddit.com/r/antiwork/comments/1kfn9jl/i_stopped_defending_my_creative_team_and_let/}{reddit.com/r/antiwork}
  \item Hacker News: Anthropic: ``Applicants should not use AI assistants.'' \href{https://news.ycombinator.com/item?id=42915905}{news.ycombinator.com}
  \item Reddit: After Being Stripped of Its GOTY Win, Clair Obscur Director Admits Using AI ``Felt Wrong.'' \href{https://www.reddit.com/r/DreamStationcc/comments/1pur5cs/after_being_stripped_of_its_goty_win_clair_obscur/}{reddit.com}
  \item Shadow IT Statistics You Need to Know Now (2026) -- ElectroIQ. \href{https://electroiq.com/stats/shadow-it-statistics/}{electroiq.com}
  \item AI Gone Wild: Why Shadow AI Is Your IT Team's Worst Nightmare. \href{https://cloudsecurityalliance.org/blog/2025/03/04/ai-gone-wild-why-shadow-ai-is-your-it-team-s-worst-nightmare}{cloudsecurityalliance.org}
  \item Shadow AI Statistics 2026: Adoption, Risk, And Breach Data. \href{https://authentech.ai/blog/shadow-ai/shadow-ai-statistics-2026/}{authentech.ai}
  \item Popular Doesn't Mean Secure -- The 2025 State of Shadow AI Report Findings -- Reco AI. \href{https://www.reco.ai/blog/popular-doesnt-mean-secure-the-2025-state-of-shadow-ai-report-findings}{reco.ai}
  \item The GenAI Divide: State of AI in Business 2025 -- MLQ.ai. \href{https://mlq.ai/media/quarterly_decks/v0.1_State_of_AI_in_Business_2025_Report.pdf}{mlq.ai}
  \item Shadow AI Usage Statistics 2026: Latest Insights -- SQ Magazine. \href{https://sqmagazine.co.uk/shadow-ai-usage-statistics/}{sqmagazine.co.uk}
  \item How to Detect \& Stop Shadow AI Tools in the Company. \href{https://www.reddit.com/r/sysadmin/comments/1pn1y3v/how_to_detect_stop_shadow_ai_tools_in_the_company/}{reddit.com/r/sysadmin}
  \item Why ``AI'' Is a Con: Our Book Launch! (with Vauhini Vara) -- Buzzsprout. \href{https://www.buzzsprout.com/2126417/episodes/17459349-why-ai-is-a-con-our-book-launch-with-vauhini-vara-2025-05-08}{buzzsprout.com}
  \item Hacker News: ``DeepSeek trained on our outputs\ldots{}'' \href{https://news.ycombinator.com/item?id=42867899}{news.ycombinator.com}
  \item This is how AI is impacting -- and shaping -- the creative industries -- World Economic Forum. \href{https://www.weforum.org/stories/2024/02/ai-creative-industries-davos/}{weforum.org}
  \item The Transformative Power of Storytelling in The Classroom -- Scribd. \href{https://www.scribd.com/document/691281535/Extracted-the-Deep-Well-of-Time-the-Transformative-Power-of-Storytelling-in-the-Classroom}{scribd.com}
  \item The `AI Hypocrisy' Paradox: Why We Hate AI in Our Jobs but Love It in Everyone Else's. \href{https://maketecheasier.com/ai-hypocrisy-paradox/}{maketecheasier.com}
  \item The Ethical Implications of AI in Creative Industries -- ResearchGate. \href{https://www.researchgate.net/publication/393511482_The_Ethical_Implications_of_AI_in_Creative_Industries_A_Focus_on_AI-Generated_Art}{researchgate.net}
  \item The Future of Music Production Is Human: 1,100+ Producers Reveal\ldots{} -- Sonarworks. \href{https://www.sonarworks.com/blog/research/future-music-production-human-producer-survey-2026}{sonarworks.com}
  \item Generative Artificial Intelligence and the Creative Economy Staff Report -- FTC. \href{https://www.ftc.gov/system/files/ftc_gov/pdf/12-15-2023AICEStaffReport.pdf}{ftc.gov}
  \item The Breaking Point: Why Hollywood's Long-Simmering AI Cold War Just Went Nuclear. \href{https://nofilmschool.com/ai-war-hollywood}{nofilmschool.com}
  \item AI Will Reshape More Jobs Than It Replaces -- BCG. \href{https://www.bcg.com/publications/2026/ai-will-reshape-more-jobs-than-it-replaces}{bcg.com}
  \item Creative pros are leveraging Generative AI to do more and better work -- Adobe Blog. \href{https://blog.adobe.com/en/publish/2024/02/02/creative-pros-generative-ai-usage}{blog.adobe.com}
  \item The Silent Displacement: How AI Automation is Creating a New Class of One-Person Businesses -- Medium. \href{https://medium.com/@tradeapollo/the-silent-displacement-how-ai-automation-is-creating-a-new-class-of-one-person-businesses-while-e1a67f716b96}{medium.com}
  \item Let Me Know If You Need Any Modifications: The LinkedIn AI Crisis -- Medium. \href{https://medium.com/@avanib28264/let-me-know-if-you-need-any-modifications-the-linkedin-ai-crisis-163748c04c1b}{medium.com}
  \item Reddit: I see Claude's writing everywhere and it's starting to feel like an AI condom. \href{https://www.reddit.com/r/ClaudeAI/comments/1rjeqg3/i_see_claudes_writing_everywhere_and_its_starting/}{reddit.com/r/ClaudeAI}
  \item BEYOND THE BLACK BOX: Shaping a Responsible AI Landscape -- KPMG. \href{https://www.kpmginfo.com/web/010797/EAIGG_AnnualReport_1_17.pdf}{kpmginfo.com}
  \item Can That Emoji Reveal a Remote Workers' Emotional State? -- Social Science Space. \href{https://www.socialsciencespace.com/2022/03/can-that-emoji-reveal-a-remote-workers-emotional-state/}{socialsciencespace.com}
  \item Transforming Retail \& Consumer Brands: Generative AI Cases and Potential -- Tenz.ai. \href{https://www.tenz.ai/wp-content/uploads/2023/09/Tenzai-Transforming-with-Generative-AI-Retail-Consumer-Brands-1.pdf}{tenz.ai}
  \item The AI Music Bubble Is Bursting in 2026: Hype, Deals \& What Survives for Real Creators -- YouTube. \href{https://www.youtube.com/watch?v=mjbp_nODAWA}{youtube.com}
  \item AI Musicians vs.\ Machines: AI Disruption in the Music Industry -- OCC Strategy. \href{https://www.occstrategy.com/en/article/ai-musicians-vs-machines-ai-disruption-in-the-music-industry/}{occstrategy.com}
  \item New Study Reveals 87\% of Producers Already Use AI Tools in Their Creative Process. \href{https://aristake.com/ai-tools-musicians-study/}{aristake.com}
  \item A real issue: video game developers are being accused of using AI -- even when they aren't -- The Guardian. \href{https://www.theguardian.com/games/2025/jun/26/video-game-developers-using-ai-even-when-they-arent-stamina-zero}{theguardian.com}
  \item Many Game Devs Reject AI in Gaming -- TechPowerUp. \href{https://www.techpowerup.com/345122/many-game-devs-reject-ai-in-gaming}{techpowerup.com}
  \item Game Publisher Bans Working With Devs That Use Any AI -- Techdirt. \href{https://www.techdirt.com/2026/01/16/game-publisher-bans-working-with-devs-that-use-any-ai-rather-than-banning-bad-uses-of-ai/}{techdirt.com}
  \item Valve dev counters calls to scrap Steam AI disclosures -- Reddit. \href{https://www.reddit.com/r/pcgaming/comments/1p8vi3f/valve_dev_counters_calls_to_scrap_steam_ai/}{reddit.com}
  \item Hollywood Is Lying to Everyone About How Much AI They're Using -- Futurism. \href{https://futurism.com/artificial-intelligence/hollywood-lying-about-ai-usage}{futurism.com}
  \item Artificial Intelligence Resources -- SAG-AFTRA. \href{https://www.sagaftra.org/contracts-industry-resources/contracts/2023-tvtheatrical-contracts/artificial-intelligence-resources}{sagaftra.org}
  \item Artificial Intelligence -- WGA. \href{https://www.wga.org/contracts/know-your-rights/artificial-intelligence}{wga.org}
  \item Regulating Artificial Intelligence -- SAG-AFTRA. \href{https://www.sagaftra.org/sites/default/files/sa_documents/AI\%20TVTH.pdf}{sagaftra.org}
  \item How the 2023 SAG-AFTRA and WGA Contracts Address Generative AI -- Perkins Coie. \href{https://perkinscoie.com/insights/blog/generative-ai-movies-and-tv-how-2023-sag-aftra-and-wga-contracts-address-generative}{perkinscoie.com}
  \item Animation Jobs \& AI: How the Industry's Future Is Changing -- Cartoon Brew. \href{https://www.cartoonbrew.com/tech/animation-jobs-in-the-age-of-ai-revelations-from-luminate-intelligences-2025-special-report-253718.html}{cartoonbrew.com}
  \item New Report from Luminate Predicts Animation Jobs Most Likely to Be Impacted by AI. \href{https://www.animationmagazine.net/2025/09/new-report-from-luminate-predicts-animation-jobs-most-likely-to-be-impacted-by-ai/}{animationmagazine.net}
  \item Trump's new order against AI regulation hits California especially hard -- CalMatters. \href{https://calmatters.org/economy/technology/2025/12/california-ai-regulation-targeted-in-trump-order/}{calmatters.org}
  \item Why outsourcing and AI means the US animation sector is facing hefty challenges in 2025. \href{https://www.screendaily.com/features/why-outsourcing-and-ai-means-the-us-animation-sector-is-facing-hefty-challenges-in-2025/5199989.article}{screendaily.com}
  \item AI Usage Statistics 2025: The Complete Market \& Adoption Report -- Elfsight. \href{https://elfsight.com/blog/ai-usage-statistics-2025/}{elfsight.com}
  \item Media industry outlook: Five AI trends shaping the future of media and broadcasting -- CGI. \href{https://www.cgi.com/en/article/communications-media/five-ai-media-and-broadcasting-trends}{cgi.com}
  \item Review/Preview Trends Report 24-25 -- NPO. \href{https://media.prod.cc.bijnpo.nl/Wayfinder_2025_Trend_Report_4e16709fd9.pdf}{bijnpo.nl}
  \item The biggest AI art controversies of 2024 -- Creative Bloq. \href{https://www.creativebloq.com/ai/ai-art/the-biggest-ai-art-controversies-of-2024}{creativebloq.com}
  \item ``That's AI'': Dealing with negative public perceptions of AI -- IPA. \href{https://the-ipa.shorthandstories.com/ipai-forum/ai-trends/thats-ai-dealing-with-negative-public-perceptions-of-ai/index.html}{the-ipa.shorthandstories.com}
  \item Helping creative agencies thrive with AI -- IDHL. \href{https://www.idhlagency.com/insights/the-next-12-months-will-show-which-creative-agencies-can-thrive-with-ai}{idhlagency.com}
  \item AI and work in the creative industries: digital continuity or discontinuity? -- Taylor \& Francis. \href{https://www.tandfonline.com/doi/full/10.1080/17510694.2024.2421135}{tandfonline.com}
  \item Labor market impacts of AI: A new measure and early evidence -- Anthropic. \href{https://www.anthropic.com/research/labor-market-impacts}{anthropic.com}
  \item When We Get Komooted -- BIKEPACKING.com. \href{https://bikepacking.com/plog/when-we-get-komooted/}{bikepacking.com}
  \item Artists face steep income decline due to AI, UNESCO finds -- UN News. \href{https://news.un.org/en/story/2026/02/1166989}{news.un.org}
  \item How Will AI Affect the US Labor Market? -- Goldman Sachs. \href{https://www.goldmansachs.com/insights/articles/how-will-ai-affect-the-us-labor-market}{goldmansachs.com}
  \item Tracking the Impact of AI on the Labor Market -- The Budget Lab at Yale. \href{https://budgetlab.yale.edu/research/tracking-impact-ai-labor-market}{budgetlab.yale.edu}
  \item When AI-Generated Art Enters the Market, Consumers Win -- and Artists Lose -- Stanford GSB. \href{https://www.gsb.stanford.edu/insights/when-ai-generated-art-enters-market-consumers-win-artists-lose}{gsb.stanford.edu}
  \item Building Your Company's AI Governance Framework to Reduce Risk -- Bloomberg Law. \href{https://pro.bloomberglaw.com/insights/artificial-intelligence/building-your-companys-ai-governance-framework-to-reduce-risk/}{pro.bloomberglaw.com}
  \item AI Governance: Frameworks, Principles, and Practices -- Domo. \href{https://www.domo.com/glossary/ai-governance}{domo.com}
  \item AI Governance Frameworks Explained: How to Manage AI Responsibly at Scale -- Rubrik. \href{https://www.rubrik.com/insights/ai-governance-frameworks-explained}{rubrik.com}
  \item What Is Shadow AI and How Do You Detect It? -- LoginRadius. \href{https://www.loginradius.com/blog/engineering/what-is-shadow-ai-and-how-do-you-detect-it}{loginradius.com}
  \item The Rise of Shadow AI: Auditing Unauthorized AI Tools in the Enterprise -- ISACA. \href{https://www.isaca.org/resources/news-and-trends/industry-news/2025/the-rise-of-shadow-ai-auditing-unauthorized-ai-tools-in-the-enterprise}{isaca.org}
  \item A Practical AI Governance Framework for Enterprises -- Databricks. \href{https://www.databricks.com/blog/practical-ai-governance-framework-enterprises}{databricks.com}
  \item We won't delegate our creativity -- our AI manifesto -- Blue Zoo Animation Studio. \href{https://www.blue-zoo.co.uk/blog/our-ai-manifesto/}{blue-zoo.co.uk}
  \item Equity calls for stronger AI protections for creative workers. \href{https://www.equity.org.uk/news/2025/equity-calls-for-stronger-ai-protections-for-creative-workers}{equity.org.uk}
  \item Discover Mapfre's vision on artificial intelligence and our new Manifesto. \href{https://www.mapfre.com/en/insights/innovation/mapfres-artificial-intelligence-manifesto/}{mapfre.com}
  \item How To Detect and Prevent Shadow AI -- FairNow. \href{https://fairnow.ai/detect-and-prevent-shadow-ai/}{fairnow.ai}
  \item Guidelines -- AI -- University of Florida. \href{https://ai.ufl.edu/guidelines/}{ai.ufl.edu}
  \item Human-centered AI through employee participation -- PMC. \href{https://pmc.ncbi.nlm.nih.gov/articles/PMC10941644/}{pmc.ncbi.nlm.nih.gov}
  \item IBM Study: Shadow AI Use Surges as Canadian Workers Outpace Employers in AI Adoption. \href{https://canada.newsroom.ibm.com/2025-09-03-IBM-R-Study-Shadow-AI-Use-Surges-as-Canadian-Workers-Outpace-Employers-in-AI-Adoption}{canada.newsroom.ibm.com}
\end{enumerate}

  \chapter{Dynamics of Generative AI Adoption in the Creative Industries}\label{app:a5}

\emph{Companion piece to Chapter~13: Coordination Collapse and Chapter~9: AI in Everything, Everywhere, All at Once.}

This deep dive is the long-form quantitative companion to the ``consumption gap'' argument in Chapter~13 and the platform-layer analysis in Chapter~9. Where those chapters argue, in the book's voice, that public sentiment around AI in the creative industries is sharply at odds with actual adoption telemetry -- and that the gap is itself the macroeconomic story of this period -- this appendix presents the underlying numbers: the Adobe Firefly adoption curve (22~billion assets by April~2025, 45\% Creative Cloud penetration, 70\% weekly active use, 11\% of Adobe's new ARR), the screenwriter pre- and post-strike usage data, the VFX-pipeline AI integration metrics, the LLM market structure (ChatGPT's 800--900M WAUs, Gemini's 155\% YoY growth), the GDC game-developer sentiment-vs-usage divergence, the Quantic Foundry consumer sentiment data, and the Stanford AI Index global-acceptance findings.

Together with Appendix~D: The Shadow AI Paradox, it constitutes the evidentiary base for the central claim of the second half of the book: that the creative industries are adopting AI at a faster rate than they are admitting publicly, and that the structure of that adoption -- concentrated, hierarchical, frictioned by labour anxiety, but operationally pervasive -- is the actual market environment any working creative or studio is now operating in.

\bigskip\par\noindent\rule{\textwidth}{0.4pt}\par\bigskip

\section*{Introduction}

The rapid integration of generative artificial intelligence (GenAI) into the global economy represents a paradigm shift comparable in magnitude to the advent of the printing press or the industrial revolution. Within the creative industries -- encompassing visual arts, film and television production, video game development, music, and marketing -- this technological leap has triggered profound structural changes and equally profound ideological conflicts. Since the mainstream popularisation of large language models (LLMs) and diffusion models in late 2022, the discourse surrounding artificial intelligence in creative fields has been characterised by intense polarisation. A highly visible, deeply critical faction warns of mass technological unemployment, the degradation of human creativity, and the unchecked proliferation of algorithmic copyright infringement.

However, an exhaustive analysis of enterprise data, labour market statistics, software utilisation metrics, and anonymous industry surveys reveals a starkly different underlying reality. The creative sector is not merely experimenting with artificial intelligence; it has already systematically embedded these tools into the foundational workflows of modern production.

The prevailing public narrative -- which frequently pits ``human authenticity'' against ``machine automation'' -- is heavily distorted by media sensationalism, algorithmic amplification of outrage, and a pervasive professional stigma that forces widespread AI utilisation underground. This comprehensive report evaluates the true state of artificial intelligence adoption across the creative economy. By examining usage statistics across major platforms (such as Adobe Firefly, Suno, OpenAI, and Google Gemini), labour market impacts, shifting consumer sentiments, and the psychological mechanisms driving ``AI shaming,'' this analysis deconstructs the reductive ``AI versus human'' binary to reveal the nuanced reality of a rapidly hybridising creative ecosystem.

The data overwhelmingly suggests that while a vocal minority of creators fiercely opposes the technology, a silent majority of professionals and consumers are pragmatically embracing it, forever altering the definition of modern creative labour.

\section*{The Ubiquity of AI in Visual and Digital Arts}

To accurately assess the impact of generative artificial intelligence, one must separate performative public sentiment from private, operational utilisation. The data indicates that AI is no longer a peripheral novelty but a central engine of digital content creation, accelerating at a pace that eclipses previous technological transitions.

The most compelling evidence of normalised adoption in the visual arts comes from industry-standard software providers. Adobe, whose Creative Cloud suite is the ubiquitous infrastructure for global design professionals, provides a clear lens into enterprise and individual adoption rates.

Since its beta launch in March~2023, Adobe Firefly -- a family of creative generative AI models -- has experienced exponential, unprecedented growth. The trajectory of Firefly's asset generation demonstrates a technology that has crossed the threshold from experimental usage to daily operational reliance. By April~2025, Firefly had generated over 22~billion assets worldwide, establishing itself as one of the most rapidly adopted generative AI platforms in the history of the creative industry.

\begin{center}
\begin{tabular}{p{3.2cm} p{3.2cm} p{6cm}}
\toprule
\textbf{Milestone Date} & \textbf{Cumulative Assets Generated} & \textbf{Key Events} \\
\midrule
September 2023 & 1 Billion & Initial Beta phase and early adopter exploration \\
March 2024 & 6.5 Billion & General availability and initial Photoshop integration \\
September 2024 & 12 Billion & Expanded multi-app integration across the ecosystem \\
November 2024 & 16 Billion & Introduction of Firefly Video Model at Adobe MAX \\
April 2025 & 22 Billion & Maturation of enterprise adoption and mobile expansion \\
June 2025 & 24+ Billion & Continued acceleration and 30\% QoQ traffic growth \\
\bottomrule
\end{tabular}
\end{center}

By March~2024, approximately 45\% of all Creative Cloud subscribers had engaged with Firefly, with 70\% of active users utilising the tool on a weekly basis. User engagement averages 2.8~sessions per week, with an average session time of 26~minutes, indicating deep integration into sustained workflows rather than fleeting experimentation. The demographic breakdown of these users reveals broad penetration: 38\% are aged 25--34, 51\% are male, 46\% are female, and 14\% identify as LGBTQ+. Furthermore, interest among Generation Z creatives grew by 32\% between 2023 and 2024, signalling that the incoming cohort of professionals views AI as a native toolkit.

The financial implications underscore the immense economic value placed on these tools. Firefly contributed 11\% of all Creative Cloud new annual recurring revenue in 2024, pushing Adobe to a record annual revenue of \$21.51~billion. Adobe's AI-first annual recurring revenue more than tripled year-over-year in the first quarter of fiscal year 2026, marking generative AI as Photoshop's fastest-growing revenue catalyst since the transition to a subscription model.

Crucially, this adoption is not driven by hobbyists alone. Marketing agencies lead industry adoption at 63\%, followed by e-commerce brands at 58\% and UX/UI designers at 48\%. Furthermore, 72\% of Fortune~500 design teams have formally integrated Firefly into their corporate workflows. In the broader ecosystem, independent surveys report that 83\% of professionals now utilise generative AI in their work. Over 25\% of new Adobe Stock content submissions in 2024 involved Firefly-generated elements, fundamentally altering the stock photography and illustration market. The data definitively proves that visual artists, commercial designers, and corporate agencies are not broadly rejecting artificial intelligence; they are utilising it at a staggering scale to achieve productivity gains and bypass menial conceptualisation phases.

\section*{The Cinematographic Shift: Film, VFX, and Generative Video}

The film and television industry presents a complex landscape where highly publicised labour disputes have operated in parallel with aggressive technological integration. In 2023, the Writers Guild of America (WGA) and the Screen Actors Guild (SAG-AFTRA) executed historic strikes, with the regulation of artificial intelligence serving as a core point of contention. The public optics suggested an industry violently rejecting automation.

\subsection*{Screenwriting and the Post-Strike AI Boom}

However, the resolution of these strikes and the establishment of regulatory guardrails paradoxically accelerated AI adoption. Prior to the strikes in 2023, approximately 34\% of screenwriters utilised AI tools covertly. Following the implementation of WGA guidelines -- which formally legitimised the use of AI for formatting, structural outlining, and brainstorming -- adoption exploded to 68\% by 2024. As one industry professional noted, the official guild authorisation alleviated the guilt and stigma associated with the technology, transforming it from an illicit shortcut into an accepted collaborative necessity. Predictive AI platforms like Largo.ai are also being increasingly utilised by studios to analyse screenplays and forecast commercial box office viability, indicating that algorithms are influencing greenlight decisions as well as the writing process itself.

\subsection*{Visual Effects (VFX) Automation}

In post-production and visual effects, artificial intelligence has seamlessly integrated into the pipeline to execute computationally expensive and highly tedious tasks. The United States AI in VFX market, valued at \$1.46~billion in 2025, is projected to reach \$8.50~billion by 2035, growing at a compound annual growth rate (CAGR) of 19.24\%, with some estimates projecting even steeper global growth curves up to 36.1\%.

\begin{center}
\begin{tabular}{p{4cm} p{9.5cm}}
\toprule
\textbf{VFX Task / Application} & \textbf{Adoption Rate \& Performance Improvement} \\
\midrule
Automated Compositing & 62\% of Hollywood studios adopted; achieved 35\% reduction in post-production timelines \\
Denoising Algorithms & 71\% of mid-sized firms adopted; improved render quality by 28\% on average \\
Matte Painting Generation & 55\% of VFX artists use AI; reduced initial setup time from 4~hours to 1.2~hours per shot \\
De-aging Processes & Reduced manual hours from 200~hours to 50~hours per actor (utilised in major theatrical releases) \\
Particle Simulation & 68\% adoption among top VFX houses reported at SIGGRAPH \\
\bottomrule
\end{tabular}
\end{center}

Major vendors such as Autodesk, Foundry, and SideFX are actively building generative pipelines into their core software offerings, indicating that machine learning is no longer a separate, novelty workflow but an inherent feature of the modern VFX ecosystem.

\subsection*{The Generative Video Race: Sora vs.\ Veo~3.1}

The emergence of advanced generative video models has fundamentally altered pre-visualisation and cinematic conceptualisation. While OpenAI's Sora~2 garnered immense public attention for its photorealistic single-shot generation, the professional filmmaking sector has increasingly gravitated toward Google's Veo~3.1.

Released in late 2025 by Google DeepMind, Veo~3.1 utilises a latent diffusion transformer architecture that compresses video data into a lower-dimensional space, learning by adding and removing Gaussian noise. Unlike older models that suffer from temporal amnesia, transformers process all parts of the input simultaneously, ensuring strict temporal consistency.

For professional directors and cinematographers, Veo~3.1 acts less as a random clip generator and more as a controllable co-director. While Sora~2 excels at raw physics simulation in isolated clips, Veo~3.1 is built for the commercial production pipeline, enabling superior narrative control and scene coherence. Professionals use advanced prompting to dictate specific camera composition (e.g., ``smooth tracking shot,'' ``shallow depth of field''), precise lighting terminology (``dramatic chiaroscuro,'' ``Rembrandt lighting''), and direct integration of ambient sound and dialogue. This allows directors to visualise scene pacing, camera angles, and emotional resonance long before physical sets are constructed, drastically lowering pre-production costs.

\section*{General Purpose LLMs: OpenAI, Anthropic Claude, and Google Gemini}

The broader landscape of creative professional AI utilisation is dominated by the foundational models provided by OpenAI, Google, and Anthropic. The scale of adoption is historically unprecedented. As of 2025, OpenAI is valued at approximately \$300~billion, with annual recurring revenues projected to surpass \$12.7~billion. Between April and June~2025, the OpenAI website received an average of 663.6~million monthly visits, while ChatGPT traffic alone surged to nearly 5.4~billion monthly visits.

Despite the proliferation of alternative models, the consumer AI assistant market exhibits a ``winner take most'' dynamic. ChatGPT maintains staggering dominance, boasting an estimated 800~million to 900~million weekly active users (WAUs) across platforms. For most of the year, fewer than 10\% of ChatGPT users even visited a competitor, and only 9\% of consumers pay for more than one subscription across ChatGPT, Gemini, Claude, and Cursor. ChatGPT's daily active user to monthly active user (DAU/MAU) ratio of 36\% nearly doubles that of Google Gemini (21\%), reflecting deep integration into daily professional habits.

However, the competitive landscape is tightening. Anthropic's Claude~3 Opus has gained significant traction among creative writers and programmers, outperforming older GPT models in generating human-like dialogue, maintaining context over large token windows, and demonstrating advanced reasoning capabilities. Concurrently, Google Gemini is experiencing explosive growth, expanding desktop users by 155\% year-over-year compared to ChatGPT's 23\%. Much of Gemini's recent acceleration is driven by native multimodal capabilities, such as advanced video and audio processing, which are becoming indispensable for multimedia creators.

\section*{The Video Game Industry: High Utilisation Amidst Cratering Sentiment}

Perhaps no creative sector exhibits a wider chasm between operational adoption and public sentiment than the video game industry. Developers are caught between the intense financial pressures of a contracting labour market and a highly vocal consumer base that violently rejects the perceived ``automation of art''.

Data from the Game Developers Conference (GDC) ``State of the Game Industry'' surveys spanning 2024 to 2026 illustrates a fascinating paradox: personal utilisation of generative AI has steadily increased, even as industry sentiment regarding the technology has utterly collapsed.

\begin{center}
\begin{tabular}{p{1.8cm} p{3cm} p{3cm} p{3cm} p{3cm}}
\toprule
\textbf{Year} & \textbf{Personal Usage of GenAI} & \textbf{Positive Sentiment} & \textbf{Mixed Sentiment} & \textbf{Negative Sentiment} \\
\midrule
2024 & 31\% & 21\% & 57\% & 18\% \\
2025 & 36\% & 13\% & 51\% & 30\% \\
2026 & 36\% & 7\% & 30\% & 52\% \\
\bottomrule
\end{tabular}
\end{center}

This cratering of sentiment -- from 18\% negative in 2024 to 52\% negative by 2026 -- must be contextualised within the broader macroeconomic environment of the gaming sector. In 2024, one-third of developers reported direct impact from industry layoffs, and 56\% expressed anxiety regarding future redundancies. A staggering 84\% of developers indicated they were somewhat or very concerned about the ethics of using generative AI. Consequently, the hostility toward generative AI in gaming is inextricably linked to labour anxieties; AI is viewed not merely as a tool, but as a corporate instrument for workforce reduction.

Yet, the pragmatic reality of game development forces continued utilisation. The 36\% of developers actively using AI apply it primarily to productivity and administrative tasks: 81\% for research and brainstorming, 47\% for code assistance, 47\% for daily scheduling, and 35\% for rapid prototyping. Usage varies significantly by role and studio size. Upper management (47\%) and business/finance departments (51\%) report the highest utilisation, seeing the tools as efficiency drivers, whereas narrative designers, visual artists, and quality assurance testers view the impact as overwhelmingly negative.

Consumer sentiment in gaming mirrors this complexity. General player attitudes toward generative AI in games worsened significantly over recent years, particularly regarding the automation of creative elements. A Quantic Foundry survey revealed that gamers are 77\% to 83\% negative toward AI-generated quests and dialogue. However, quantitative analysis reveals a more apathetic reality regarding purchasing behaviour: the majority of gamers (60\%) remain entirely neutral regarding the use of AI in a game's development, provided the final product is of high quality. Players demonstrate comparative openness to AI when applied to non-artistic backend features, such as dynamic difficulty adjustment. The hostility is specifically reserved for the automation of roles traditionally perceived as requiring a human soul, such as narrative design and visual artistry.

\section*{The Perception Gap: The Vocal Minority vs.\ The Silent Majority}

A central question surrounding the generative AI transition is whether the fierce ``anti-AI'' backlash represents a broad societal consensus or the disproportionate amplification of a loud minority. Empirical evidence overwhelmingly supports the latter. The digital discourse surrounding artificial intelligence is heavily skewed by the mechanics of social media, where outrage and moral panic generate unparalleled engagement.

Academic research into social media dynamics consistently demonstrates that political and cultural conversations are dominated by a highly active fraction of users. Studies of platforms like Twitter reveal a structural divide between a ``vocal minority,'' who tweet incessantly, utilise hashtags aggressively, link extensively to outside content, and drive ideological narratives, and a ``silent majority,'' who consume information passively and rarely participate in public outrage.

This dynamic translates directly to the AI discourse. While a vocal contingent of artists, writers, and highly engaged internet users coordinate boycotts, sign open letters, and aggressively flood comment sections with anti-AI sentiment, the broader public is quietly adopting the technology for mundane, benign, and productive purposes.

Broad consumer polling reveals a growing global acceptance of artificial intelligence. The Stanford AI Index Report 2025 indicates that global optimism is rising: across 26~surveyed nations, 55\% of individuals now view AI products as offering more benefits than drawbacks, up from 52\% in 2022. YouGov polling in 2024 further indicated that nearly a third of consumers across 17~markets felt more positively about generative AI tools compared to the previous year, while only 22\% held a more negative opinion.

\subsection*{The Box Office Stress Test}

The disconnect between online outrage and actual consumer behaviour is most evident in the commercial performance of media products targeted by anti-AI campaigns. In 2024, the independent horror film \emph{Late Night with the Devil} became the epicentre of a massive online controversy when it was revealed that the production utilised three brief frames of AI-generated interstitial bumper art. Review-bombing campaigns were organised on platforms like Letterboxd, and vocal online contingents demanded a total boycott of the film, framing it as a line in the sand for artistic integrity.

Despite the digital fury, the boycott failed entirely to materialise in the real world. The film secured a highly successful opening weekend, taking in a symbolically appropriate \$666,666, breaking records for the distributor Shudder, and maintaining a 97\% critical approval rating on Rotten Tomatoes. As industry analysts noted, outside the echo chambers of social media, the general public simply did not care about the origin of a few transitional images; they paid for a compelling narrative, and the controversy had ``zero effect'' on the film's success.

A similar dynamic unfolded with the high-budget A24 film \emph{Civil War}. The studio released a series of promotional posters generated by AI, which depicted post-apocalyptic scenes in American cities. The posters contained glaring geographical and architectural inaccuracies. While film Twitter and digital artists mocked the studio relentlessly, the controversy generated massive organic visibility for the film. The broader consumer base viewed the posters as intended -- as thematic, dystopian ``what if'' marketing materials -- and the film succeeded commercially regardless of the digital backlash.

The empirical conclusion is clear: while the anti-AI crowd is highly organised, fiercely protective of traditional artistic labour, and capable of generating immense negative public relations, their outrage rarely translates into altered consumer spending habits. The silent majority prioritises end-product quality, utility, and entertainment value over the ethical purity of the production pipeline.

\section*{The Stigma of Automation: ``AI Shaming'' and Covert Creativity}

If the adoption of artificial intelligence is as widespread as the data suggests, why do so many creative professionals adamantly deny using it? The answer lies in the intense psychological and professional stigma attached to AI utilisation, a phenomenon actively hindering transparent integration.

``AI shaming'' has emerged as a powerful sociological mechanism used to police professional boundaries. It refers to the practice of publicly criticising, devaluing, or dismissing individuals and organisations for utilising artificial intelligence to execute tasks. This shaming operates on the premise that AI-assisted work is inherently deceitful, devoid of human soul, and fundamentally lazy.

For creative professionals, whose core identity and societal value are inextricably linked to the struggle and mastery of their craft, the accusation of using AI is an attack on their professional legitimacy. Psychological surveys indicate that the fear of being perceived as lazy or unmotivated ranks among the highest deterrents for acknowledging AI use in the workplace. Consequently, utilising AI raises internal doubts about a professional's own abilities, leading to a pervasive culture of secrecy. Artists and writers are highly vocal about AI models being trained on their work without consent or compensation, generating an atmosphere of intense hostility that bleeds into broader cultural sentiment. According to a 2023 survey, 74.3\% of artists consider scraping artwork from the internet for AI technology to be highly unethical.

\subsection*{Hypocrisy in the Academy and the Studio}

The stigma forces usage underground, resulting in absurd institutional hypocrisy. In higher education, professors routinely threaten to fail students for utilising ChatGPT, citing the degradation of critical thinking. Yet, widespread reports indicate that students go to extreme lengths to mask their genuine work from faulty AI detectors -- such as utilising extensions like Draftback to record hours of typing sessions, 1,300~revisions, and messy drafting just to ``prove'' human authorship to an algorithm that falsely flagged them.

Conversely, faculty members openly admit to utilising ChatGPT to grade papers, summarise reading materials, or generate the very syllabi they use to ban AI. In one notable reported instance, an educator proudly utilised a paid ChatGPT subscription to detect student AI use, blissfully unaware of the LLM's inability to accurately detect its own output, while simultaneously praising an AI-generated essay as a prime example of ``actual human original thinking''.

This same covert dynamic exists in professional creative agencies and art galleries. When gallery owners were polled in early 2026, 61\% claimed confidently that none of their represented artists used AI in their practice. Yet, when artists themselves are surveyed anonymously, the numbers shift. While many claim to reject generative image models, 13\% openly admit to using AI in the ``backend'' of the creative process. Furthermore, deep analysis of AI art platform usage suggests that more than 11\% of traditional artists have utilised text-to-image technology, with 53.6\% of those users claiming they made a ``fundamental input'' to the artwork through their complex prompting. They absorb the massive productivity gains while outwardly maintaining the facade of pure, unassisted human toil.

\section*{Media Sensationalism and the Algorithmic Fog of War}

The disconnect between the reality of AI integration and the public panic is largely manufactured and sustained by the global media ecosystem. Journalism, functioning within an attention economy driven by programmatic advertising, is financially incentivised to sensationalise the impacts of artificial intelligence.

Researchers analysing media coverage of AI have noted a profound tendency toward apocalyptic framing. The narrative frequently leaps past practical, immediate concerns (such as data privacy, copyright frameworks, or minor workflow disruptions) directly to existential threats: the death of art, the extinction of humanity via ``killer robots,'' or catastrophic global unemployment. Computer scientists and AI researchers frequently express frustration with this coverage. Zachary Lipton, a machine learning professor, famously labelled media coverage of artificial intelligence as ``sensationalised crap'' that fuels an ``AI misinformation epidemic''. Researcher Nirit Weiss-Blatt has documented how the journalism of ``AI panic'' diverts attention away from real-world problems like algorithmic discrimination and environmental energy consumption.

This sensationalism is a direct byproduct of algorithmic curation. Digital platforms optimise for engagement, and psychological research demonstrates that fear, outrage, and moral polarisation are the most potent drivers of user retention. This creates a dangerous feedback loop: algorithms amplify social drivers of conflict, pushing users into polarised echo chambers that complicate rational discourse. Consequently, nuanced reports on how AI reduces rendering times in VFX pipelines by 30\% are suppressed due to a lack of emotional resonance, while stories about ``AI slop'' conquering the internet receive massive amplification.

\subsection*{Job Displacement Headlines vs.\ Economic Data}

The most pervasive media narrative surrounding AI in the creative economy is the imminent threat of mass technological unemployment. Headlines consistently predict the decimation of copywriters, illustrators, and entry-level developers. However, hard macroeconomic data from the labour market contradicts the panic.

In 2024, an analysis by the Information Technology and Innovation Foundation (ITIF) revealed that the employment gains from AI heavily outpaced the losses. The AI sector directly generated approximately 119,900~jobs in the United States -- driven by the hiring of machine learning engineers, data scientists, and the massive construction boom required for new data centres (which generate an additional 3.5~local jobs for every data centre job). In stark contrast, outplacement firms tracked approximately 12,700~jobs lost specifically to AI automation during the same period. This displacement represented a mere 0.1\% of all total layoffs in 2024.

Furthermore, an exhaustive global analysis by PwC -- the 2025 Global AI Jobs Barometer, which analysed nearly a billion job advertisements -- concluded that AI is making workers more valuable, not less, even in sectors highly exposed to automation. The U.S.\ Bureau of Labor Statistics (BLS) and organisations like Gallup have found limited evidence that generative AI has systematically increased unemployment or broadly reduced earnings for artists. Research from Anthropic assessing labour market impacts through an ``observed exposure'' metric found no systematic increase in unemployment for highly exposed workers since late 2022, though there is suggestive evidence that hiring for younger, entry-level workers in exposed occupations has slowed.

History supports this data. The introduction of digital photography, synthesisers, desktop publishing software, and CGI all generated identical panics regarding the ``death'' of their respective industries. In every historical instance, automation reallocated work, lowered the barrier to entry, increased total output, and ultimately expanded the overall size of the creative market. As noted by the Federal Reserve, the time savings generated by AI (equivalent to 1.6\% of all work hours) has contributed to aggregate labour productivity increasing by 2.16\% on an annualised basis.

\section*{The Music Industry: Deconstructing the ``AI Bad, Human Good'' Narrative}

To move past the current state of professional cognitive dissonance and media-induced panic, the creative industries must fundamentally reevaluate the rigid philosophical binary of ``AI versus Human.'' This binary is intellectually flawed, historically ignorant, and actively harmful to the evolution of modern art.

\subsection*{The Democratisation of Audio and the Suno Revolution}

The music industry is currently undergoing a structural realignment driven by generative audio platforms. Historically, music production required significant capital investment in studio time, hardware, and specialised audio engineering skills. Generative AI has obliterated these barriers to entry, triggering a surge in synthetic music creation.

The platform Suno stands as the primary catalyst in this sector. By 2025, Suno reached an annualised revenue run rate of \$150~million to \$200~million, ultimately hitting \$300~million in annual recurring revenue by early 2026 alongside a \$2.5~billion valuation. The platform boasts over 2~million paid subscribers and over 100~million distinct users who actively generate full-fidelity tracks from text prompts.

\begin{center}
\begin{tabular}{p{2cm} p{4cm} p{4cm} p{3cm}}
\toprule
\textbf{Year} & \textbf{Suno Annual Revenue / ARR} & \textbf{Funding Round} & \textbf{Valuation} \\
\midrule
2023 & Pre-revenue / Early traction & Series A & \$600M \\
2024 & \$50M -- \$100M (estimated) & Series B (\$125M) & \$1B \\
2025 & \$150M -- \$200M & Series C (\$250M) & \$2.5B \\
2026 (Early) & \$300M (Reported ARR) & N/A & >\$2.45B \\
\bottomrule
\end{tabular}
\end{center}

The macroeconomic projections for this sub-sector are substantial. Valued at roughly \$570~million in 2024, generative AI music is forecast to reach \$2.8~billion by 2030, capturing a projected 20\% of streaming platform revenue and 60\% of business-to-business (B2B) music libraries. This aggressive expansion has incited severe backlash from traditional industry gatekeepers. Major recording labels and artist coalitions have launched concerted legal campaigns, describing platforms like Suno and Udio as a ``brazen smash and grab'' and filing mass infringement lawsuits via the RIAA.

\subsection*{The Blurring Lines of Authorship}

Despite institutional resistance, the concept of a purely ``human'' track in modern music production is already a fallacy. Human vocals are routinely corrected via Auto-Tune algorithms; drum performances are quantised perfectly to a grid by digital audio workstations; synthesisers generate waveforms that no acoustic instrument could produce. Generative AI is not an alien invasion into a pristine human domain; it is merely the next layer of abstraction in a long history of technological mediation.

When streaming platforms or independent marketplaces attempt to ban ``AI-generated music,'' they encounter impossible enforcement challenges because the lines between AI-generated, AI-assisted, and human-created content are hopelessly blurred. Attempts to police these boundaries -- such as Bandcamp's ill-fated AI ban -- often result in the accidental penalisation of highly innovative human artists who are using algorithms as legitimate instruments of avant-garde expression.

The fundamental utility of AI in music is undeniable, even at the highest echelons of prestige. The 2023 release of ``Now and Then,'' marketed as the final song by The Beatles, relied explicitly on artificial intelligence to isolate and extract John Lennon's degraded 1970s vocal cassette recording from the piano track. The track achieved universal commercial success, reaching number one on the UK Singles Chart. Consumer surveys regarding the track revealed that 58\% of US respondents and 64\% of UK respondents were fully aware that AI was used in its production, indicating that consumer hostility toward AI is highly context-dependent. When utilised to restore, enhance, or facilitate human intent, the technology is enthusiastically embraced by the public.

\section*{Redefining Creativity and Mitigating Psychological Risks}

As Henry Ford famously said regarding innovation, ``If I had asked my customers what they wanted, they would have told me a faster horse.'' True innovation often precedes consumer demand, and AI represents a paradigm shift that consumers are still learning to articulate.

However, the integration of AI does present genuine psychological risks. Researchers note that skills that are not exercised tend to degrade; following the principle of ``use it or lose it,'' over-reliance on generative AI could lead to a degradation of innate human creativity. Helson's theory of adaptation levels suggests a serious risk wherein society slowly adapts to mediocre, AI-generated content, failing to even notice the gradual loss of boundary-pushing human ingenuity. Generative AI is inherently replicative; it can recombine ideas, but it struggles to generate the paradigm-breaking solutions required to solve novel human problems.

Therefore, the threat of generative AI is not that it will destroy human creativity, but that it exposes the mechanical, formulaic nature of much of what we previously called creativity. If an AI can perfectly replicate a copywriter's marketing email or a graphic designer's corporate logo, it suggests that the original human work lacked true creative novelty. In the algorithmic age, ``being average is the worst outcome.'' AI establishes a new baseline of competence, instantly accessible to anyone. Consequently, the premium on true human creativity -- characterised by emotional resonance, cultural nuance, strategic empathy, and paradigm-breaking ideation -- will skyrocket.

\section*{Conclusion}

The empirical evidence surrounding generative artificial intelligence in the creative industries points to an irreversible, systemic integration. The highly vocal anti-AI contingent, while fiercely protective of traditional copyright and influential in shaping social media discourse, represents a statistical minority that possesses virtually no leverage over macroeconomic trends, software utilisation rates, or mass consumer behaviour. Consumers consistently demonstrate that they value the quality, utility, and emotional impact of a final product over the ideological purity of its supply chain.

Simultaneously, the widespread stigmatisation of AI usage has fostered a culture of professional hypocrisy. Millions of creatives across visual arts, screenwriting, game development, and marketing rely daily on tools like Adobe Firefly, Claude, Suno, and Veo~3.1 to remain competitive. Yet, they vehemently deny their use to protect their professional identities from algorithmic ``shaming.''

Media narratives predicting mass technological unemployment remain unsubstantiated by current labour statistics, heavily driven by an attention economy that rewards sensationalism over economic reality. Instead of destroying the creative economy, artificial intelligence is acting as a massive democratising force, lowering barriers to entry, augmenting existing workflows, and forcing a necessary redefinition of what constitutes valuable human creativity.

Moving forward, the industry must discard the reductive ``AI bad, human good'' dichotomy. The future of the creative economy does not lie in building regulatory moats to protect outdated workflows from algorithms. It lies in proactive adaptation: embracing AI as a powerful collaborative instrument, establishing transparent frameworks for its use, and empowering the next generation of creators to synthesise human intuition with machine efficiency to unlock entirely new forms of cultural expression.

\section*{Works Cited}

\begin{enumerate}
\item The impact of GenAI on the creative industries | World Economic Forum. \href{https://www.weforum.org/stories/2025/01/the-impact-of-genai-on-the-creative-industries/}{weforum.org}
\item As AI rises, so does the need for more human creativity -- World Economic Forum. \href{https://www.weforum.org/stories/2026/01/ai-and-need-for-more-human-creativity/}{weforum.org}
\item Suno: We've hit 2M paid subscribers and \$300M annual revenue. \href{https://www.musicbusinessworldwide.com/suno-hits-2m-paid-subscribers-300m-annual-revenue/}{musicbusinessworldwide.com}
\item Public Sentiment Toward AI Is Negative: What It Means for Builders and Businesses. \href{https://www.mindstudio.ai/blog/public-sentiment-ai-negative-what-it-means-for-builders}{mindstudio.ai}
\item AI and the creative industries | Economics Observatory. \href{https://www.economicsobservatory.com/wp-content/uploads/2025/12/AI_Creative_Industries.pdf}{economicsobservatory.com}
\item Adobe Firefly Statistics And User Trends 2026 -- Companies History. \href{https://www.companieshistory.com/adobe-firefly-statistics/}{companieshistory.com}
\item AI in VFX Market Size, Trends \& Growth | Industry Report -- SNS Insider. \href{https://www.snsinsider.com/reports/ai-in-vfx-market-9394}{snsinsider.com}
\item How uncritical news coverage feeds the AI hype machine | Nieman Journalism Lab. \href{https://www.niemanlab.org/2024/05/how-uncritical-news-coverage-feeds-the-ai-hype-machine/}{niemanlab.org}
\item AI Shaming: The Silent Stigma among Academic Writers and Researchers -- PubMed. \href{https://pubmed.ncbi.nlm.nih.gov/38977530/}{pubmed.ncbi.nlm.nih.gov}
\item The Algorithmic Wars: Artificial Intelligence and the Shaping of Conflict Narratives on Social Media Platforms | ResearchGate. \href{https://www.researchgate.net/publication/397467078_The_Algorithmic_Wars_Artificial_Intelligence_and_the_Shaping_of_Conflict_Narratives_on_Social_Media_Platforms}{researchgate.net}
\item The problem with the ``AI versus human'' music narrative. \href{https://www.midiaresearch.com/blog/the-problem-with-the-ai-versus-human-music-narrative}{midiaresearch.com}
\item The State of Generative AI Adoption in 2025 | St.\ Louis Fed. \href{https://www.stlouisfed.org/on-the-economy/2025/nov/state-generative-ai-adoption-2025}{stlouisfed.org}
\item Adobe Photoshop Statistics 2026: Users, Revenue, Market Share -- SQ Magazine. \href{https://sqmagazine.co.uk/adobe-photoshop-statistics/}{sqmagazine.co.uk}
\item Adobe Firefly Statistics By Usage, Demographics And Facts (2025) -- ElectroIQ. \href{https://electroiq.com/stats/adobe-firefly-statistics/}{electroiq.com}
\item Understanding the impact of AI in the creative industries -- LSE. \href{https://www.lse.ac.uk/school-of-public-policy/research/lse-growth-lab/understanding-the-impact-of-ai-in-the-creative-industries}{lse.ac.uk}
\item Adobe's AI and the Creative Frontier Study Reveals Creators' Views on the Opportunities and Risks of Generative AI. \href{https://blog.adobe.com/en/publish/2024/10/08/adobes-ai-creative-frontier-study-reveals-creators-views-opportunities-risks-generative-ai}{blog.adobe.com}
\item Hollywood Is Wrestling With the Potential of AI Screenwriting Tools -- Business Insider. \href{https://www.businessinsider.com/ai-screenwriting-tools-hollywood-film-tv-studios-writers-2025-8}{businessinsider.com}
\item Why AI Screenplay Editors Are Taking Over Hollywood in 2025 -- Laper. \href{https://laper.ai/recent-highlights/2025-11-16-why-ai-screenplay-editors-are-taking-over}{laper.ai}
\item AI In VFX Market Growth Analysis -- Size and Forecast 2025--2029 | Technavio. \href{https://www.technavio.com/report/ai-in-vfx-market-industry-analysis}{technavio.com}
\item 160+ AI In The VFX Industry Statistics | 2026 Data Report -- Gitnux. \href{https://gitnux.org/ai-in-the-vfx-industry-statistics/}{gitnux.org}
\item Gen AI Resources -- Visual Effects Society. \href{https://vesglobal.org/gen-ai-resources-page/}{vesglobal.org}
\item What Is Google Veo~2? AI Video Generation from Google DeepMind -- MindStudio. \href{https://www.mindstudio.ai/blog/what-is-google-veo-2-video-generation}{mindstudio.ai}
\item The AI Video Showdown: Why Google's Veo~3.1 Edges Out Sora~2 for Pro Storytellers. \href{https://medium.com/@ubumart/the-ai-video-showdown-why-googles-veo-3-1-edges-out-sora-2-for-pro-storytellers-ca691295c7c3}{medium.com}
\item Veo~3.1: My Hands-On Deep Dive into Google's AI Video Revolution -- CrePal. \href{https://crepal.ai/blog/agent/veo-3-1-my-hands-on-deep-dive-into-googles-ai-video-revolution/}{crepal.ai}
\item Igniting Imagination in Film Production with VEO 2025. \href{https://filmcode.ai/veo/}{filmcode.ai}
\item 50+ OpenAI Statistics 2025 -- AIPRM. \href{https://www.aiprm.com/openai-statistics/}{aiprm.com}
\item State of Consumer AI 2025: Product Hits, Misses, and What's Next | Andreessen Horowitz. \href{https://a16z.com/state-of-consumer-ai-2025-product-hits-misses-and-whats-next/}{a16z.com}
\item Artificial Intelligence in Creative Industries: Advances Prior to 2025 -- arXiv. \href{https://arxiv.org/html/2501.02725v3}{arxiv.org}
\item GDC 2024 State of the Game Industry Report. \href{https://reg.gdconf.com/state-of-game-industry-2024}{reg.gdconf.com}
\item Gamers Are Overwhelmingly Negative About Gen AI in Video Games -- Quantic Foundry. \href{https://quanticfoundry.com/2025/12/18/gen-ai/}{quanticfoundry.com}
\item I analysed 3 years of GDC reports on generative AI in game dev -- Reddit. \href{https://www.reddit.com/r/gamedev/comments/1rpeeta/i_analyzed_3_years_of_gdc_reports_on_generative/}{reddit.com/r/gamedev}
\item 36 percent of game workers use generative AI, but half think it's bad for the industry. \href{https://www.gamedeveloper.com/business/one-third-of-game-workers-use-generative-ai-but-half-think-it-s-bad-for-the-industry}{gamedeveloper.com}
\item GDC 2024 State Of The Game Industry results: Devs discuss layoffs, generative AI, and more. \href{https://www.gamedeveloper.com/business/gdc-2024-state-of-the-game-industry-devs-discuss-layoffs-generative-ai-and-more}{gamedeveloper.com}
\item GDC's State of the Games Industry survey report says one in 10 respondents laid off in 2024. \href{https://www.eurogamer.net/gdcs-state-of-the-games-industry-survey-says-one-in-ten-respondents-laid-off-in-2024}{eurogamer.net}
\item New research: What do gamers really think about generative AI in games? \href{https://www.midiaresearch.com/blog/new-research-what-do-gamers-really-think-about-generative-ai-in-games}{midiaresearch.com}
\item Social Drivers and Algorithmic Mechanisms on Digital Media -- PMC. \href{https://pmc.ncbi.nlm.nih.gov/articles/PMC11373151/}{pmc.ncbi.nlm.nih.gov}
\item Vocal Minority versus Silent Majority: Discovering the Opinions of the Long Tail. \href{https://cs.wellesley.edu/~pmetaxas/Silent-minority-Vocal-majority.pdf}{cs.wellesley.edu}
\item Don't Panic (Yet): Assessing the Evidence and Discourse Around Generative AI and Elections. \href{https://knightcolumbia.org/content/dont-panic-yet-assessing-the-evidence-and-discourse-around-generative-ai-and-elections}{knightcolumbia.org}
\item Unveiling the silent majority: stance detection and characterisation of passive users on social media -- ResearchGate. \href{https://www.researchgate.net/publication/379582905_Unveiling_the_silent_majority_stance_detection_and_characterization_of_passive_users_on_social_media_using_collaborative_filtering_and_graph_convolutional_networks}{researchgate.net}
\item How prevalent is AI ``art'' hate really? Are we just a vocal minority? -- Reddit. \href{https://www.reddit.com/r/antiai/comments/1q67z44/how_prevalent_is_ai_art_hate_really_are_we_just_a/}{reddit.com/r/antiai}
\item Public Opinion | The 2025 AI Index Report | Stanford HAI. \href{https://hai.stanford.edu/ai-index/2025-ai-index-report/public-opinion}{hai.stanford.edu}
\item Has public perception of generative AI shifted? -- YouGov. \href{https://yougov.com/articles/50484-has-public-perception-of-generative-ai-shifted}{yougov.com}
\item Adobe Analytics: Traffic to U.S.\ Retail Websites from Generative AI Sources Jumps 1200 Percent. \href{https://blog.adobe.com/en/publish/2025/03/17/adobe-analytics-traffic-to-us-retail-websites-from-generative-ai-sources-jumps-1200-percent}{blog.adobe.com}
\item The Age of Generative AI: Over half of Americans have used generative AI -- Adobe Blog. \href{https://blog.adobe.com/en/publish/2024/04/22/age-generative-ai-over-half-americans-have-used-generative-ai-most-believe-will-help-them-be-more-creative}{blog.adobe.com}
\item Key findings about how Americans view artificial intelligence -- Pew Research Center. \href{https://www.pewresearch.org/short-reads/2026/03/12/key-findings-about-how-americans-view-artificial-intelligence/}{pewresearch.org}
\item AI Art Creeps into \emph{Late Night with the Devil} Striking BIG Controversy -- Reddit. \href{https://www.reddit.com/r/aiwars/comments/1bvg7mt/ai_art_creeps_its_way_into_late_night_with_the/}{reddit.com/r/aiwars}
\item \emph{LATE NIGHT WITH THE DEVIL} Directors Confirm Use of AI Images in Film -- Reddit. \href{https://www.reddit.com/r/blankies/comments/1bk7kng/late_night_with_the_devil_directors_confirm_use/}{reddit.com/r/blankies}
\item Late Night With The Devil takes \$666,666 at box office, despite boycotts | Dazed. \href{https://www.dazeddigital.com/film-tv/article/62251/1/indie-horror-late-night-with-the-devil-haunted-ai-art-panic-letterboxd-boycott}{dazeddigital.com}
\item \emph{Late Night with the Devil} slammed for controversial use of artificial intelligence. \href{https://roaringbengals.com/5011/features/late-night-with-the-devil-slammed-for-controversial-use-of-artificial-intelligence/}{roaringbengals.com}
\item Civil War (film) -- Wikipedia. \href{https://en.wikipedia.org/wiki/Civil_War_(film)}{en.wikipedia.org}
\item A24 made some laughably bad GenAI posters for \emph{Civil War} -- Reddit. \href{https://www.reddit.com/r/sanfrancisco/comments/1c7nzq0/a24_made_some_laughably_bad_genai_posters_for/}{reddit.com/r/sanfrancisco}
\item Is it me or is there not lots of marketing for \emph{Civil War}? -- Reddit. \href{https://www.reddit.com/r/boxoffice/comments/1c22nbp/is_it_me_or_is_there_not_lots_of_marketing_for/}{reddit.com/r/boxoffice}
\item Frazier testimony on building an AI-ready America -- Texas Law. \href{https://law.utexas.edu/faculty/uploads/publication_files/frazier-testimony-on-building-an-ai-ready-america.pdf}{law.utexas.edu}
\item Evidence of a social evaluation penalty for using AI | PNAS. \href{https://www.pnas.org/doi/10.1073/pnas.2426766122}{pnas.org}
\item Survey Reveals 9 out of 10 Artists Believe Current Copyright Laws are Outdated in the Age of Generative AI Technology. \href{https://bookanartist.co/blog/2023-artists-survey-on-ai-technology/}{bookanartist.co}
\item A top student made a disturbing confession -- Reddit/Teachers. \href{https://www.reddit.com/r/Teachers/comments/1rvm5c3/a_top_student_made_a_disturbing_confession/}{reddit.com/r/Teachers}
\item Drowning in AI generated essays -- Reddit/Professors. \href{https://www.reddit.com/r/Professors/comments/1osj8ux/drowning_in_ai_generated_essays/}{reddit.com/r/Professors}
\item I got accused of using AI on a paper I wrote myself and I'm losing my mind -- Reddit. \href{https://www.reddit.com/r/CollegeRant/comments/1ph02c7/i_got_accused_of_using_ai_on_a_paper_i_wrote/}{reddit.com/r/CollegeRant}
\item Student sent me ``real-time'' video of word doc in response to fail for AI use -- Reddit. \href{https://www.reddit.com/r/Professors/comments/1psqlzs/student_sent_me_realtime_video_of_word_doc_in/}{reddit.com/r/Professors}
\item How Artificial Intelligence Shapes How We Think, Act, and Connect | Psychology Today. \href{https://www.psychologytoday.com/us/blog/positively-media/202501/how-artificial-intelligence-shapes-how-we-think-act-and-connect}{psychologytoday.com}
\item The Artsy AI Survey 2026: What Galleries Really Think About AI in the Art World. \href{https://www.artsy.net/article/artsy-editorial-artsy-ai-survey-2026-galleries-ai-art}{artsy.net}
\item Over 90\% of artists view AI-generated pieces in a negative light -- Innovating with AI. \href{https://innovatingwithai.com/artists-view-of-ai-generated-pieces/}{innovatingwithai.com}
\item AI in Art Statistics 2024 -- AIPRM. \href{https://www.aiprm.com/ai-art-statistics/}{aiprm.com}
\item Is AI Good or Bad for Creativity: How Generative AI is Transforming Human Ideation. \href{https://www.captechu.edu/blog/how-generative-ai-is-transforming-creativity}{captechu.edu}
\item The Algorithmic Management of Polarisation and Violence on Social Media | Knight First Amendment Institute. \href{https://knightcolumbia.org/content/the-algorithmic-management-of-polarization-and-violence-on-social-media}{knightcolumbia.org}
\item Trolling, memes, and deepfakes: How AI is thickening the fog of war -- Nieman Lab. \href{https://www.niemanlab.org/2026/05/trolling-memes-and-deepfakes-how-ai-is-thickening-the-fog-of-war/}{niemanlab.org}
\item AI's Job Impact: Gains Outpace Losses | ITIF. \href{https://itif.org/publications/2025/12/18/ais-job-impact-gains-outpace-losses/}{itif.org}
\item AI Jobs Barometer -- PwC. \href{https://www.pwc.com/gx/en/services/ai/ai-jobs-barometer.html}{pwc.com}
\item Labour market impacts of AI: A new measure and early evidence -- Anthropic. \href{https://www.anthropic.com/research/labor-market-impacts}{anthropic.com}
\item How AI exposure is reshaping jobs in creative fields -- Fox Business. \href{https://www.foxbusiness.com/technology/how-ai-exposure-reshaping-jobs-creative-fields}{foxbusiness.com}
\item Incorporating AI impacts in BLS employment projections: occupational case studies. \href{https://www.bls.gov/opub/mlr/2025/article/incorporating-ai-impacts-in-bls-employment-projections.htm}{bls.gov}
\item Blog: AI -- MIDiA Research. \href{https://www.midiaresearch.com/blog/category/ai}{midiaresearch.com}
\item Is this the end of the recording studio? Inside Suno's 7-million-song-a-day AI revolution | MEXC News. \href{https://www.mexc.com/news/991797}{mexc.com}
\item Suno Revenue and Usage Statistics (2026) -- Business of Apps. \href{https://www.businessofapps.com/data/suno-statistics/}{businessofapps.com}
\item Suno Revenue 2025: \$200M ARR, \$2.5B Valuation -- GetLatka. \href{https://getlatka.com/companies/suno.com}{getlatka.com}
\item The Beatles new song ``Now and Then'' used AI to lift out John Lennon's voice -- Quartz. \href{https://qz.com/new-beatles-song-now-and-then-ai-john-lennon-1850981701}{qz.com}
\item Now and Then (Beatles song) -- Wikipedia. \href{https://en.wikipedia.org/wiki/Now_and_Then_(Beatles_song)}{en.wikipedia.org}
\item The Beatles `Now and Then': CintSnap reveals consumer sentiments of AI's impact on the music industry. \href{https://www.cint.com/blog/the-beatles-now-and-then-cintsnap-reveals-consumer-sentiments-of-ais-impact-on-the-music-industry/}{cint.com}
\item Why the Beatles' song Now and Then is the most important record of the past year | The Standard. \href{https://www.standard.co.uk/culture/music/the-beatles-now-and-then-let-it-be-b1124868.html}{standard.co.uk}
\item Blog: artificial intelligence -- MIDiA Research. \href{https://www.midiaresearch.com/blog/category/artificial-intelligence}{midiaresearch.com}
\item Do Not Worry That Generative AI May Compromise Human Creativity or Intelligence in the Future: It Already Has -- PMC. \href{https://pmc.ncbi.nlm.nih.gov/articles/PMC11278271/}{pmc.ncbi.nlm.nih.gov}
\item The paradox of creativity in generative AI: high performance, human-like bias, and limited differential evaluation -- PMC. \href{https://pmc.ncbi.nlm.nih.gov/articles/PMC12369561/}{pmc.ncbi.nlm.nih.gov}
\item AI vs.\ Human Creativity: Striking the Perfect Marketing Balance -- Advertising is Simple. \href{https://advertisingissimple.com/ai-vs-human-creativity-striking-the-perfect-marketing-balance/}{advertisingissimple.com}
\end{enumerate}

  \chapter{AI, Stigma, Privilege and the Democratisation of Creative Expression}\label{app:a6}

\emph{Companion piece to Chapter~6: The 88\%, Chapter~13: Coordination Collapse, and Chapter~15: Choosing the Future.}

This deep dive sits underneath one of the harder arguments in the book -- that the structural opposition to AI in some quarters of the creative industries is not only an artistic or ethical position but also, in significant measure, the \emph{defensive response of an entrenched class} whose access to creative production has historically depended on financial, geographic and institutional barriers that AI tooling threatens to dissolve.

The book does not lead with this framing because the framing is uncomfortable and easy to weaponise. The 88\% who turned up to the UK consultation, the unions defending performer likeness, the studios refusing AI integration on craft grounds -- these are not, in the main, expressions of class privilege; they are legitimate articulations of working-creative interests that the book takes seriously throughout. But they sit \emph{alongside}, and sometimes overlap with, a different cultural pattern: the resistance of a relatively narrow demographic of established creative workers to a technology that is, simultaneously, opening creative production to working-class, Global South, neurodivergent and historically excluded participants who could not previously afford the on-ramp.

This appendix lays out the sociological and economic evidence for the democratisation argument -- the UK class-ceiling data, the Sutton Trust analyses of creative-industry composition, the historical precedents of artistic-elite resistance to new media, and the projections for AI's role as an equaliser. It is the evidentiary base for the \emph{Access} principle in Chapter~14 and for the regional-opening sections of Chapter~13. It should be read in the spirit the book itself adopts: both things are true at once, and a creative economy that takes both seriously is the one most likely to produce a humane outcome.

\bigskip\par\noindent\rule{\textwidth}{0.4pt}\par\bigskip

\section*{Introduction}

The discourse surrounding the integration of generative artificial intelligence within the global creative industries -- spanning film, television, video games, music, and adjacent cultural sectors -- has reached a fever pitch of moral panic, industrial resistance, and highly publicised labour disputes. Across public forums, guild negotiations, and prominent media narratives, artificial intelligence is frequently characterised as an existential threat to authentic human creativity, a mechanism for unchecked corporate plagiarism, and a harbinger of cultural decay. Critics argue that the automation of artistic processes fundamentally strips the ``soul'' from cultural production, rendering human ingenuity obsolete in the face of algorithmic efficiency.

However, a rigorous sociological, economic, and historical analysis of this backlash reveals a significantly different underlying reality. The intense stigmatisation of generative artificial intelligence by the established creative class is less a defence of artistic purity than a concerted, defensive effort to preserve an entrenched socioeconomic hierarchy.

For decades, the creative industries have functioned not as the open meritocracies they claim to be, but as highly exclusive class systems. These sectors are heavily insulated by systemic barriers to entry, geographic clustering in expensive metropolitan hubs, steep financial prerequisites for early-career survival, and rampant, normalised nepotism. The ability to create, distribute, and monetise high-level cultural products has been artificially restricted to a narrow demographic possessing immense financial backing, inherited social capital, or the patronage of corporate gatekeepers. Generative artificial intelligence severely disrupts this paradigm by collapsing the cost of production and radically lowering the technical barriers to artistic execution. By placing the capability to produce high-fidelity audio, cinematic video, and complex interactive code into the hands of the global public, artificial intelligence fundamentally threatens the artificial scarcity upon which the creative elite's social status, professional prestige, and economic power are built.

This comprehensive report provides an exhaustive examination of the intersection between artificial intelligence, social class, and the creative industries. By deconstructing the structural inequalities of the current creative economy, analysing the historical precedents of technological resistance among artistic elites, and projecting the economic implications of artificial intelligence as an equalising force, this analysis demonstrates that the democratisation of artistic expression is not the death of creativity. Rather, it represents the dismantling of an exclusionary gatekeeping mechanism that has historically marginalised diverse, working-class, and global voices from the cultural vanguard.

\section*{The Illusion of Meritocracy: A Statistical Deconstruction of the Creative Class System}

The prevailing mythology of the creative industries -- heavily perpetuated by the industry's own narratives, award ceremonies, and media representations -- is that of an egalitarian meritocracy. It is a landscape theoretically defined by the romantic ideal that raw talent, relentless dedication, and unique vision will invariably rise to the top, regardless of an individual's origins. Yet, deep empirical data paints a starkly different picture of an ecosystem dominated by systemic privilege, where professional success is frequently dictated by one's proximity to wealth, elite education, and institutional power.

To understand why the democratisation of creative tools via artificial intelligence is viewed as such a massive threat, one must first understand the exclusionary architecture of the industry it is disrupting. Statistical evidence from both the United Kingdom and the United States demonstrates that the creative workforce is overwhelmingly skewed toward the upper and middle classes, creating a rigid ``class ceiling'' that filters out socioeconomically disadvantaged talent before they can even enter the pipeline.

In the United Kingdom, for example, young people from working-class backgrounds are four times less likely to secure employment in the creative industries than their middle-class and upper-class peers. Furthermore, this disparity is not a relic of the past but an accelerating trend. Analysis of demographic shifts indicates that access to creative professions has worsened considerably over the last few decades. While 16.4\% of creative workers born in the 1950s and 1960s hailed from working-class backgrounds, that figure plummeted to a mere 7.9\% for those born in the 1990s. In high-visibility sectors such as film, television, video, radio, and photography, individuals identifying as working-class make up just 8.4\% of the total workforce. The broader arts, culture, and heritage sectors exhibit similar stratification, with 60\% of workers having grown up in households where the main income earner was in a managerial or professional role, compared to just 43\% in the wider national workforce.

The dominance of elite educational backgrounds further highlights this profound social stratification. Top-selling musicians are six times more likely to have attended elite private fee-paying schools compared to the general public, sitting at 43\% versus the national average of 7\%. The classical music profession is identified as particularly elitist; 58\% of classical musicians attended an arts specialist university or conservatoire, and an astonishing 25\% attended the Royal Academy of Music for their undergraduate studies alone. At prestigious conservatoires, the student body is overwhelmingly affluent, with up to 60\% of students studying creative subjects at institutions like the Royal Academy of Music having been privately educated. Furthermore, at elite universities such as Oxford, Cambridge, King's College London, and the University of Bath, over half of all the students enrolled in creative courses originate from designated ``upper-middle-class'' households. By contrast, working-class representation in creative degrees at these elite institutions languishes in the single digits, sitting at just 4\% at Cambridge and Bath, and 5\% at Oxford. These statistics expose an educational pipeline that systematically filters out individuals who lack the financial means to access elite training, effectively reserving the highest echelons of cultural production for the already privileged.

The video game industry, despite its origins as a disruptive subculture, mirrors this socioeconomic disparity. According to a UK Interactive Entertainment (UKIE) census, only 13\% of professionals in the UK games sector originate from working-class backgrounds. If the industry were truly reflective of broader society, this figure would be closer to 37\%. Industry advocates note that if socioeconomic status were classified as a protected characteristic, it would represent the single biggest diversity issue within the gaming sector, requiring an influx of over 6,000~working-class professionals just to achieve demographic parity.

\section*{The Architecture of Exclusion: Financial Barriers and Inherited Social Capital}

The mechanisms that maintain this ``class ceiling'' are primarily economic and cultural, operating through systemic financial barriers, the normalisation of unpaid labour, and the pervasive influence of inherited social capital. Entry into the creative industries frequently requires navigating a labyrinth of low-paid or entirely unpaid internships, temporary contract work, and precarious freelance gigs.

A comprehensive survey of creative industry professionals revealed that 67\% acknowledge that unpaid internships are still a common practice within their specific fields, and an equal percentage agree that this arrangement disproportionately benefits the upper and upper-middle classes. Participating in unpaid or severely underpaid labour necessitates a substantial financial safety net, typically provided by generational parental wealth. Because the vast majority of creative hubs and studios are heavily clustered in exceptionally expensive metropolitan areas -- such as London, Los Angeles, and New York -- individuals from lower socioeconomic backgrounds simply cannot afford the cost of living required to work for free in exchange for ``exposure'' or ``experience''.

This stark economic reality ensures that the entry-level talent pool is overwhelmingly populated by those who possess the material resources to endure years of financial precarity. Working-class talent is effectively starved out of the industry before they can establish a foothold, forced to seek stable, salaried employment in other sectors to survive. Furthermore, the financial barrier is compounded by cultural barriers. Access to creative spaces is still largely predicated on informal networks and personal contacts, creating a deeply unlevel playing field where success hinges on navigating middle-class workplace norms. Preconceptions about class are often shaped by soft social identifiers -- such as an individual's accent, their vocabulary, where they went to school, and their social circles -- which further alienates working-class talent who may feel compelled to alter their identities to be taken seriously by affluent senior executives.

Beyond sheer financial resources, success in the creative industries relies heavily on the blatant exercise of nepotism. The phenomenon of ``nepo babies'' -- the offspring of established industry figures who secure prominent, highly visible roles with relative ease -- illustrates how access operates as an inherited asset rather than an earned privilege. Research shows that approximately 29\% of Americans work for a parent's employer at least once by age 30, a dynamic that yields significant wage premiums and early-career stability. In the hyper-competitive entertainment industry, this dynamic is amplified exponentially.

In Hollywood and the global music industry, nepotism rarely manifests solely as crude, direct hiring; rather, it functions through unparalleled access to elite industry networks, talent agents, studio executives, and venture capital. Children of industry veterans grow up fully immersed in the specialised language, cultural norms, and social expectations of the elite. When it comes time to launch their careers, they bypass the years of cold-calling, endless auditioning, and financial struggle required of outsiders. This proximity creates an unspoken, highly resourced training ground and a permanent safety net where a failed project or a bad review does not result in the end of a career, as family ties will invariably open another door.

Consequently, the stories that receive massive studio funding, the music that receives major label backing, and the digital art that is elevated to the cultural vanguard are overwhelmingly produced by a homogenous, privileged demographic. This dynamic narrows the cultural lens through which society views itself, restricting the diversity of narratives available to the public. It is precisely this entrenched, exclusionary architecture that makes the democratising potential of artificial intelligence so incredibly threatening to the current power brokers. When the tools of high-end production are made available to the masses, the artificial scarcity that protects the elite is irrevocably shattered.

\section*{The Weaponisation of Authenticity: Unmasking the Anti-AI Backlash}

It is strictly within this context of extreme exclusivity and socioeconomic stratification that the vitriolic, industry-wide backlash against generative artificial intelligence must be analysed. Across the creative sectors, guilds, and unions, criticism of artificial intelligence frequently centres on emotive themes of ``theft,'' ``plagiarism,'' the ``devaluation of human effort,'' and the impending ``loss of the human soul'' in art. However, a deeper, critical examination of these arguments suggests that these moral and philosophical objections often serve to mask a desperate defence of professional status, hierarchical privilege, and artificial scarcity.

As generative artificial intelligence tools dramatically lower the learning curve required to produce high-fidelity audio, cinematic visuals, and complex written code, they directly threaten the gatekeepers who have long monopolised these capabilities. Historically, artists, musicians, and independent filmmakers have often branded themselves as anti-establishment, anti-gatekeeping, and anti-hierarchy, positioning themselves as rebels against corporate control. However, the rapid advent of artificial intelligence has triggered a profound rhetorical shift among established creatives, who now actively deploy the language of authenticity, exclusivity, and tradition to defend a rigid, exclusionary hierarchy.

The pervasive argument that legitimate artistic expression is only valid if it is ``earned'' through years of formal technical training, expensive schooling, prolonged suffering, or sanctioned institutional pathways is inherently exclusionary. This philosophy posits that the right to participate in cultural creation must be heavily gatekept by a gauntlet of technical and financial hurdles. When an independent, unfunded creator can utilise a generative artificial intelligence model to circumvent these traditional hurdles and execute their vision, the resulting output is immediately dismissed by the established elite as ``slop,'' ``illegitimate,'' or ``soulless''. This elitism reveals that the true anxiety driving the backlash is not a genuine concern for the death of creativity itself, but the terrifying realisation that expressive capability is no longer a scarce, highly valuable commodity reserved exclusively for the privileged few.

Furthermore, the most prominent and aggressively weaponised argument deployed by creative guilds and copyright maximalists is that artificial intelligence models are trained on copyrighted works without explicit consent, compensation, or credit, constituting a form of mass, mechanised theft. While there are entirely legitimate, legally sound concerns regarding the direct impersonation of living artists -- which should undoubtedly be regulated as a matter of identity protection, publicity rights, and fraud prevention -- the broader, blanket argument against machine learning collapses under historical and philosophical scrutiny.

Human artists have always learned their craft by absorbing, analysing, reverse-engineering, and synthesising the copyrighted works of their predecessors. A young painter studies the brushstrokes of the masters; a burgeoning writer internalises the specific cadence, vocabulary, and thematic structures of their favourite authors; a musician learns to play by covering the back catalogue of their idols. To argue that algorithmic training is fundamentally illegitimate because it lacks explicit, prior permission is to argue that the fundamental act of learning, observation, and stylistic synthesis must be permissioned and monetised. Such a rigid standard would invalidate the foundational development of every living human artist, as style itself has never been considered proprietary property under traditional copyright frameworks.

The hypocrisy of the anti-AI movement is further exposed when examining the corporate entities leading the charge. The institutions most vociferously championing ``artist rights'' against artificial intelligence -- such as major Hollywood studios, the Recording Industry Association of America (RIAA), and dominant publishing conglomerates -- have historically built their massive empires by systematically exploiting artists through opaque accounting practices, predatory 360-degree contracts, and the aggressive, relentless enclosure of the public domain. The sudden, highly publicised pivot by these corporate entities to defending the sanctity of human artistry is largely a strategic, self-serving manoeuvre designed to maintain their monopolistic control over distribution networks and content generation. These gatekeepers fear that if artificial intelligence democratises high-end production, independent artists will no longer need to surrender their intellectual property or endure exploitative contracts to secure the capital-intensive backing of major studios and record labels.

\section*{Echoes of the Past: Historical Parallels of Technological Gatekeeping}

To fully grasp the current panic surrounding artificial intelligence, it is critical to recognise that this is not an unprecedented cultural phenomenon. Rather, it is merely the latest iteration of a highly predictable historical cycle wherein established creative classes vehemently resist any new technology that threatens to democratise their medium, lower the barriers to entry, and dilute their exclusive status. Examining these historical parallels provides a vital lens through which to predict the inevitable trajectory and eventual integration of artificial intelligence in the creative sector.

\subsection*{The Synthesiser Panic and the Threat to ``Real'' Musicianship}

In the late 1960s and stretching well into the 1980s, the introduction of electronic synthesisers, drum machines, and digital sequencers triggered widespread, existential hysteria across the global music industry. Established, formally trained musicians and high-profile critics argued that these new electronic machines produced ``cold,'' ``inhuman,'' and ``artificial'' sounds that replaced genuine, hard-earned skill with the effortless push of a button or the selection of a preset. The panic was not driven by the audiences consuming the music, but by the legacy institutions judging the tools and fearing the obsolescence of their specific skill sets.

In the United Kingdom, the powerful Musicians' Union went so far as to pass official motions attempting to ban the use of synthesisers, drum machines, and electronic backing devices in recording studios and live television performances. The union viewed these technologies as a direct, unacceptable threat to the employment of traditional orchestral session players and live instrumentalists, framing the synthesiser not as a new instrument, but as a malicious job-killing machine. Critics fiercely attacked musical pioneers like Miles Davis and Pete Townshend when they began incorporating electronic textures and sequencers into their compositions. Detractors claimed that the machines, rather than the musicians, were doing the actual creative work, thereby invalidating their authorship and diluting the purity of genres like jazz and rock.

Yet, rather than destroying the art of music, the synthesiser radically democratised it. It allowed solo artists, marginalised creators, and individuals without access to expensive studio bands to compose and execute highly complex, multi-layered arrangements entirely on their own. This technological democratisation ultimately gave birth to entirely new, globally dominant genres -- ranging from hip-hop and synth-pop to techno, house, and electronic dance music -- that became the defining cultural soundtracks of the modern era. The tool that was derided as the death of human expression became the very foundation of its next evolution.

\subsection*{The Resistance to Digital Cinematography and Home Recording}

A strikingly similar resistance occurred in the film and television industry during the painful, protracted transition from traditional photochemical film to digital cinematography in the late 1990s and 2000s. Elite, established cinematographers, directors, and studios fiercely argued that digital cameras inherently lacked the ``soul,'' the dynamic latitude, the organic grain, and the specific texture of 35mm film. Early digital efforts were broadly dismissed by the Hollywood establishment as sterile, clinical, and aesthetically inferior to the ``true'' art of photochemical filmmaking.

However, the advent of highly capable, relatively affordable digital cameras -- such as the RED ONE -- coupled with the rise of non-linear digital editing software (like Adobe Premiere and Final Cut) running on standard personal computers, completely shattered the astronomical financial barriers to high-level filmmaking. Digital technology eliminated the absolute necessity of paying exorbitant, prohibitive fees for physical film stock, specialised chemical lab processing, and massive, highly specialised camera crews. This technological shift empowered an entirely new generation of independent filmmakers, operating outside the nepotistic Hollywood system, to shoot, edit, and distribute feature-length projects on micro-budgets.

Simultaneously, in the music industry, the rise of the Musical Instrument Digital Interface (MIDI) and affordable home digital multi-track recorders (such as the ADAT system) in the 1980s and 1990s allowed creators to bypass the traditional, highly gatekept ``million-dollar commercial recording studio''. Independent artists could now produce, mix, and master commercial-quality, radio-ready tracks in their own bedrooms.

In every historical instance, technological shifts that lowered the barrier to entry were met with fierce, coordinated resistance from industry gatekeepers who breathlessly warned of an impending aesthetic and cultural collapse. Yet, in every instance, the resistance failed, and the technology ultimately resulted in a vast, unprecedented expansion of creative diversity, new artistic genres, and broader market participation from previously excluded demographics.

Generative artificial intelligence is not an anomaly; it is the logical, albeit highly accelerated, continuation of this historical democratising trajectory.

\section*{The Great Equaliser: Generative AI as the Catalyst for Democratisation}

Generative artificial intelligence represents the ultimate, most profound disruption of the creative class system because it directly attacks and neutralises the primary barrier to entry across all media: the exorbitant cost of technical execution and production value. By transforming simple natural language prompts, rough sketches, or basic melodies into highly complex visual, auditory, and interactive outputs, artificial intelligence completely levels the playing field. It empowers creators who possess profound conceptual vision, storytelling ability, and taste, but who critically lack the vast financial capital required to hire specialised technical teams, rent elite equipment, or secure studio backing.

\subsection*{Eradicating the Budget-to-Vision Gap in Film and Television}

Historically, the art of cinema has been strictly gated by extreme economics. Executing complex establishing shots, rendering intricate computer-generated visual effects, or staging massive crowd scenes required hundreds of thousands, if not millions, of dollars in physical set construction, specialised equipment rentals, location permits, and highly unionised labour. Consequently, only stories deemed broadly, safely commercially viable by a highly concentrated, risk-averse, and predominantly white, male class of studio executives were ever greenlit and funded.

Generative artificial intelligence tools allow independent filmmakers to bypass these financial chokepoints entirely. Creators can now procedurally generate photorealistic 3D environments, populate scenes with highly detailed digital extras, and produce complex, dynamic storyboards at an infinitesimal fraction of the traditional cost. Industry financial estimates suggest that actively utilising artificial intelligence across both pre-production and post-production workflows can seamlessly reduce the budget of a major \$200~million blockbuster film by 15\% to 20\% -- effectively shaving \$30 to \$40~million off the bottom line and cutting weeks off the production schedule.

For the independent creator, the implications are revolutionary. The vast distance between imagination and execution is practically eliminated. An unfunded director operating out of a developing nation, or a working-class writer with a brilliant sci-fi concept, can now generate proof-of-concept trailers, complex visual effects, and high-fidelity scenes without needing to secure venture capital or navigate the nepotistic maze of Hollywood representation. This democratisation allows marginalised voices from outside the traditional geographic and social bubbles to bring their highly specific, diverse cultural narratives to the screen with a level of polish previously reserved for elite studio productions.

\subsection*{Democratising Game Development and Music Production}

The video game industry has seen the divide between highly funded ``AAA'' mega-studios and small independent developers widen drastically over the last decade, driven primarily by the astronomical labour costs associated with generating hyper-realistic 3D assets, vast open-world environments, and complex branching narratives. However, artificial intelligence-driven procedural content generation and advanced neural rendering technologies -- such as Nvidia's DLSS~5 -- are acting as a tremendous ``golden ticket'' for the indie developer community.

Small, independent teams, or even solo developers, can now leverage generative artificial intelligence to procedurally generate expansive landscapes, populate virtual worlds with highly intelligent, adaptive non-player characters (NPCs), and achieve real-time, Hollywood-level photorealistic lighting without needing to employ a staff of hundreds of specialised artists and coders. This technological leverage allows independent creators to focus their limited resources on narrative depth, unique, soulful art styles, and highly innovative gameplay mechanics, rather than competing on sheer computational brute force.

Similarly, in the global music industry, artificial intelligence-powered composition assistants, advanced vocal processing tools, and automated, algorithmic mastering software effectively eliminate the absolute need for expensive commercial studio time, hired session musicians, and high-end audio engineers. A working-class songwriter with a compelling lyric and a basic melody can utilise artificial intelligence to instantly generate complex backing instrumentation, test intricate chord progressions, and produce professional-grade, radio-ready mixes.

This capability effectively bypasses the major record labels, who have historically acted as the ultimate gatekeepers by dictating radio play, funding recording sessions, and controlling algorithmic playlist placement on major Digital Service Providers (DSPs) like Spotify. This sudden, unpermissioned access represents a complete paradigm shift where the ultimate artistic output and commercial viability of a track is determined by the raw quality of the idea and the taste of the creator, rather than the depth of their financial pockets or their connections to label executives.

\section*{The Economic Reconfiguration: Rebuilding the Creative Middle Class}

A frequent, highly publicised critique from the anti-AI camp -- heavily promoted by creative guilds and labour unions -- is that the technology will inevitably destroy millions of jobs, replacing human workers with automated systems purely to maximise corporate profits and enrich tech billionaires. While it is an undeniable reality that artificial intelligence will cause significant, painful structural disruption and displace specific, highly commoditised technical roles (such as low-level copywriting, the creation of generic stock photography, basic translation, and background commercial music composition), macroeconomic analysis suggests a significantly more nuanced, optimistic long-term outcome. Instead of merely destroying labour, artificial intelligence possesses the unique potential to rebuild a currently hollowed-out creative middle class.

\subsection*{Extending Worker Expertise and the ``Collaboration Paradox''}

Eminent MIT economist David Autor posits that the unique opportunity presented by artificial intelligence to the labour market is not its capacity to entirely replace human labour, but rather its ability to ``extend the relevance, reach, and value of human expertise''. For decades, the information age and the rise of digital technologies have paradoxically concentrated wealth, cognitive authority, and decision-making power in the hands of a small cadre of elite experts, systematically hollowing out middle-skill, middle-class jobs.

Generative artificial intelligence actively reverses this decades-long trend by functioning as a massive capability multiplier. Rigorous experimental studies across various professional sectors consistently demonstrate that artificial intelligence tools disproportionately benefit lower-skilled, less-experienced, or entry-level workers. By automating the mechanical execution of tasks, AI allows these workers to rapidly close the performance and productivity gap with elite, highly paid professionals.

In the creative sector, this phenomenon manifests as the ``Collaboration Paradox,'' where access to artificial intelligence tools allows a single individual to comfortably match the output and quality of a multi-person team. A junior graphic designer, a solo game developer, or an unfunded independent filmmaker can utilise artificial intelligence as an advanced, tireless ``co-worker'' to perform complex coding, generate storyboards, or mix audio -- tasks that previously required hiring a highly paid, specialised expert. While this capability understandably threatens the premium wages and job security commanded by elite technical specialists, it vastly empowers the middle tier of creators.

\subsection*{The ``Christmas Card Problem'' and Expansive Market Dynamics}

Much of the intense fear surrounding AI-induced job loss relies on the fundamental assumption of a zero-sum economic market -- the belief that every single AI-generated image, line of code, or background song represents a stolen commission from a human artist. This perspective entirely ignores the economic reality of market expansion, beautifully conceptualised by analysts as the ``Christmas card problem''.

The vast majority of AI-assisted creativity actually occurs well below the commercial threshold where professional, working artists operate. A small local business owner generating a logo for a pop-up shop, a high school teacher creating a custom illustration for a presentation, or an individual generating a personalised song for a family event would never have possessed the budget to hire a professional composer or a creative agency in the first place. For these users, the alternative to the AI-generated output was simply no output at all.

Therefore, artificial intelligence drastically expands the total global volume of creative expression and media generation without necessarily cannibalising the high-end, bespoke art market, which will continue to value the specific human narrative and prestige associated with renowned artists. Where artificial intelligence does actively intersect with commercial markets and displace human labour, it primarily replaces commoditised, low-effort content -- such as generic royalty-free background tracks, basic templates, and standard stock images. By automating the mundane and the commoditised, AI forces the creative industry to re-evaluate where true human value lies.

\section*{Predictive Trajectories: How the Democratisation Will Pan Out (2025--2030 and Beyond)}

As the initial shock, moral panic, and legal posturing surrounding generative artificial intelligence gradually give way to practical, everyday integration, the power dynamics of the global creative industries will undergo a profound, irreversible realignment over the next decade. Based on current economic data, historical precedents of technological adoption, and the rapidly evolving capabilities of neural networks, several highly specific predictions can be made regarding how this democratisation will ultimately pan out for the creative class.

\paragraph{1. The Splintering of Corporate Monopolies and the Rise of ``Micro-Studios.''}

The traditional major Hollywood studios, global game publishers, and ``Big Three'' record labels derive their immense, gatekeeping power almost entirely from their unique ability to finance massive production budgets, absorb enormous financial risk, and control global distribution networks. As generative artificial intelligence drastically reduces the cost of high-end production and marketing, the financial leverage held by these corporate gatekeepers will severely diminish.

While major studios will undoubtedly attempt to utilise artificial intelligence internally to slash their own overhead costs and inflate profit margins, they will simultaneously face an unprecedented wave of existential competition from highly agile, AI-empowered independent collectives. We will witness the explosive rise of the ``micro-studio'': hyper-efficient teams of two to five multi-disciplinary creators who leverage artificial intelligence to produce feature-length films, AAA-quality immersive games, and chart-topping musical albums entirely independently. By completely bypassing the traditional, bloated studio system and leveraging decentralised digital distribution, these micro-studios will retain total ownership of their intellectual property and revenue streams.

\paragraph{2. The Evolution of Copyright: From Protecting Style to Protecting Identity.}

The current landscape of aggressive litigation, where artists and major corporations are suing artificial intelligence companies over the use of training data, will inevitably give way to a new legal and cultural equilibrium. This new framework will prioritise the strict protection of human identity over the impossible monopolisation of artistic style.

Courts, regulatory bodies, and the public will increasingly recognise that learning, analysing, and synthesising artistic influences are legally and philosophically permissible actions for both humans and machines. Attempting to copyright a ``vibe'' or a genre style will be deemed unenforceable. However, incredibly strict legal guardrails and technological detection systems will be implemented to prevent the direct, unauthorised impersonation of living artists. The historic Writers Guild of America (WGA) and SAG-AFTRA strikes of 2023 established the fundamental blueprint for this transition. Rather than attempting to ban the technology outright, the unions secured collective bargaining agreements that ensure artificial intelligence is used to augment workers rather than replace them without compensation.

\paragraph{3. A Shift in the Definition of ``Author'' and ``Skill.''}

The fundamental metrics by which society evaluates, appreciates, and compensates creative talent will permanently evolve. Just as the invention of the photograph freed painting from the burden of hyper-realistic documentation -- pushing the medium toward impressionism, cubism, and abstract expressionism -- the advent of generative artificial intelligence will shift the perceived value of human creativity away from the mere mechanics of technical execution.

The successful artist of the future will function far less like a traditional craftsman and much more like a director, a curator, or a creative architect. They will be a ``full-stack'' professional who orchestrates a complex symphony of highly specialised artificial intelligence agents. Their true value will lie in their human taste, their editorial judgement, their lived experience, and their ability to constrain, refine, and select the most emotionally resonant outputs from a sea of algorithmic generation. The current stigma attached to ``prompt engineering'' or AI-assisted generation will rapidly fade as the technology becomes invisibly, seamlessly integrated into standard professional software suites, much like auto-tune in music, spell-check in literature, or CGI in modern filmmaking are universally accepted today.

\paragraph{4. A Global Renaissance of Diverse and Marginalised Voices.}

Ultimately, the most profound, lasting impact of generative artificial intelligence on the creative industries will be demographic and cultural. By aggressively circumventing the prohibitive financial requirements, the geographic limitations of major creative hubs, and the deeply entrenched nepotistic networks that have historically defined the creative class, artificial intelligence will unleash a massive, unprecedented renaissance of storytelling from previously excluded populations.

Independent creators from working-class backgrounds, artists operating within the Global South, disabled creators who face physical barriers to traditional production environments, and neurodivergent storytellers will no longer need to seek permission, beg for venture funding, or alter their identities to secure validation from a homogenous class of elite gatekeepers. The democratisation of these powerful tools ensures that the cultural artefacts of the mid-to-late 21st century will accurately and vibrantly reflect the full, chaotic spectrum of human experience, rather than being strictly limited to the narrow, sanitised worldview of a privileged few.

\section*{Conclusion}

The aggressive, highly publicised stigmatisation of generative artificial intelligence by the established creative industries cannot be taken at face value as a righteous, purely philosophical crusade to save human art. When rigorously contextualised within the deep-seated nepotism, the astronomically steep financial barriers to entry, and the systemic, statistically proven class inequality that define the current global creative economy, the anti-AI movement is starkly exposed as a defensive, reactionary manoeuvre by an entrenched elite. By attempting to gatekeep expressive capability behind arbitrary technical hurdles, demanding massive financial investments for production, and weaponising outdated copyright maximalism, these legacy institutions and elite guilds are fighting desperately to preserve their social status, their professional prestige, and their highly lucrative economic monopolies.

History consistently demonstrates that transformative technology inevitably dismantles artificial scarcity. Just as the introduction of synthesisers, digital multi-track recorders, and digital cinema cameras radically democratised the production of music and film in previous decades, generative artificial intelligence is currently tearing down the invisible walls of the modern creative class system. While this profound transition will undoubtedly cause intense short-term friction, displace specific technical roles, and require the development of entirely new legal frameworks for labour protection and identity rights, the ultimate macroeconomic and cultural trajectory is one of unprecedented global empowerment.

As the historic barriers of prohibitive cost, elite specialised training, and nepotistic access fall away, the future of the creative industries will be radically reshaped. Success will be dictated not by the wealth of the creator's parents, the exclusivity of their university, or their geographic proximity to a studio lot, but by the raw depth, originality, and emotional resonance of their ideas. The democratisation of creativity via artificial intelligence is not the end of art; it is the destruction of the aristocratic gatekeeper, and the beginning of true cultural liberation.

\section*{Works Cited}

\begin{enumerate}
\item History, creative disruption, and GenAI | Deloitte Digital. \href{https://www.deloittedigital.com/us/en/insights/perspective/creative-disruption.html}{deloittedigital.com}
\item The elitism problem in the creative backlash against AI -- Matt Hopkins. \href{https://matthopkins.com/technology/the-elitism-problem-in-the-creative-backlash-against-ai/}{matthopkins.com}
\item Artificial Intelligence and Culture -- UNESCO. \href{https://www.unesco.org/sites/default/files/medias/fichiers/2025/09/CULTAI_Report\%20of\%20the\%20Independent\%20Expert\%20Group\%20on\%20Artificial\%20Intelligence\%20and\%20Culture\%20\%28final\%20online\%20version\%29\%201.pdf}{unesco.org}
\item Artificial Intelligence and the Creative Double Bind -- Harvard Law Review. \href{https://harvardlawreview.org/print/vol-138/artificial-intelligence-and-the-creative-double-bind/}{harvardlawreview.org}
\item The class ceiling in the creative industries report 2024. \href{https://creativeaccess.org.uk/app/uploads/2024/07/the-class-ceiling-in-the-creative-industries-report-2024.pdf}{creativeaccess.org.uk}
\item Beyond the screen: The power of nepotism within Hollywood -- Scot Scoop News. \href{https://scotscoop.com/beyond-the-screen-the-power-of-nepotism-within-hollywood/}{scotscoop.com}
\item A Class Act -- The Sutton Trust. \href{https://www.suttontrust.com/our-research/a-class-act/}{suttontrust.com}
\item Transforming Media: Generative AI's Role in Cutting Film and Video Production Costs. \href{https://tippett.org/transforming-media-generative-ais-role-in-cutting-film-and-video-production-costs}{tippett.org}
\item How does AI level the playing field for independent music creators? -- Sonarworks Blog. \href{https://www.sonarworks.com/blog/learn/how-does-ai-level-the-playing-field-for-independent-music-creators}{sonarworks.com}
\item Elitist Britain 2025: What It Means for the Creative, Cultural and Heritage Sector. \href{https://www.culturecommons.uk/news/elitist-britain-2025}{culturecommons.uk}
\item Research reveals stark class inequalities in access to the creative industries -- The Sutton Trust. \href{https://www.suttontrust.com/news-opinion/all-news-opinion/research-reveals-stark-class-inequalities-in-access-to-the-creative-industries/}{suttontrust.com}
\item Huge decline of working class people in the arts reflects fall in wider society -- The Guardian. \href{https://www.theguardian.com/culture/2022/dec/10/huge-decline-working-class-people-arts-reflects-society}{theguardian.com}
\item National Statistics on the Creative Industries. \href{https://pec.ac.uk/news_entries/national-statistics-on-the-creative-industries/}{pec.ac.uk}
\item 50 arguments against the use of AI in creative fields -- AOKIstudio. \href{https://aokistudio.com/50-arguments-against-the-use-of-ai-in-creative-fields.html}{aokistudio.com}
\item Improving the socioeconomic diversity of the games industry | Barclays Games and Creative. \href{https://games.creative.barclays/resource-hub/games/industry-insights/improving-the-socioeconomic-diversity-of-the-games-industry/}{games.creative.barclays}
\item US Game Development Salaries in 2025: What Our Latest Industry Report Reveals. \href{https://gdconf.com/article/us-game-development-salaries-in-2025-what-our-latest-industry-report-reveals/}{gdconf.com}
\item GDC State Of The Games Industry 2025 report -- InvestGame. \href{https://investgame.net/wp-content/uploads/2025/03/0794a269-d5c4-4994-9bcf-8c5730d0815e_2025_GDC_State_of_the_Game_Industry_report-1.pdf}{investgame.net}
\item Unpaid Internships \& Inequality: A Review of the Data and Recommendations for Research, Policy and Practice -- CCWT. \href{https://ccwt.wisc.edu/wp-content/uploads/2022/04/CCWT_Policy-Brief-2_Unpaid-Internships-and-Inequality-1.pdf}{ccwt.wisc.edu}
\item Did they earn it? Study looks at ``nepo babies'' debate | Folio -- University of Alberta. \href{https://www.ualberta.ca/en/folio/2025/02/did-they-earn-it-nepo-babies-debate.html}{ualberta.ca}
\item Is ``Nepotism'' Blocking Chances in the Film Industry? -- The Preuss Insider. \href{https://preussinsider.com/5777/entertainment/is-nepotism-blocking-chances-in-the-film-industry/}{preussinsider.com}
\item Economist tracks `nepo baby' effect on young Americans' earnings -- Harvard Gazette. \href{https://news.harvard.edu/gazette/story/2023/02/economist-tracks-nepo-baby-effect-on-young-americans-earnings/}{news.harvard.edu}
\item Current state of Game development and How does/will AI affect the field -- Reddit. \href{https://www.reddit.com/r/gamedev/comments/1ih8eia/current_state_of_game_development_and_how/}{reddit.com/r/gamedev}
\item Generative AI, the American worker, and the future of work | Brookings. \href{https://www.brookings.edu/articles/generative-ai-the-american-worker-and-the-future-of-work/}{brookings.edu}
\item Copyright and AI Policy Needs Precision, Not Panic | TechPolicy.Press. \href{https://www.techpolicy.press/copyright-and-ai-policy-needs-precision-not-panic/}{techpolicy.press}
\item AI's Unique Threat to Musicians -- Tech Policy @ Sanford -- Duke University. \href{https://techpolicy.sanford.duke.edu/blog/ais-unique-threat-to-musicians/}{techpolicy.sanford.duke.edu}
\item Old Laws, New Ghosts: Why Artists are losing the Battle for AI | diacritical -- Arts Journal. \href{https://www.artsjournal.com/diacritical/2026/01/old-laws-new-ghosts-why-the-creative-resistance-to-ai-is-failing.html}{artsjournal.com}
\item They said the same things about synthesizers -- Reddit/DefendingAIArt. \href{https://www.reddit.com/r/DefendingAIArt/comments/1qvynf9/they_said_the_same_things_about_synthesizers/}{reddit.com}
\item The Predictable Panic: How Artists Always React to New Technology | Whaleden. \href{https://medium.com/@whaledencom/the-predictable-panic-how-artists-always-react-to-new-technology-9482a4b92bc7}{medium.com}
\item 1981--1990 -- The Musicians' Union: A History (1893--2013). \href{https://www.muhistory.com/contact-us/1971-1980/}{muhistory.com}
\item ``The union passed a motion to ban the use of synths, drum machines and any electronic devices'' | MusicRadar. \href{https://www.musicradar.com/news/the-union-passed-a-motion-to-ban-the-use-of-synths-drum-machines-and-any-electronic-devices-the-day-the-loony-musicians-union-tried-to-kill-the-synthesizer-which-also-happened-to-be-bob-moogs-birthday}{musicradar.com}
\item The UK Musicians Union tried to ban synths 40 years ago today -- Mixdown Magazine. \href{https://mixdownmag.com.au/news/the-uk-musicians-union-tried-to-ban-synths-40-years-ago-today/}{mixdownmag.com.au}
\item The Evolution and Decline of the Traditional Recording Studio -- University of Liverpool. \href{https://livrepository.liverpool.ac.uk/3000867/1/200488719_Sept2015.pdf}{livrepository.liverpool.ac.uk}
\item Crafting Digital Cinema: Cinematographers in Contemporary Hollywood -- ProQuest. \href{https://search.proquest.com/openview/62784b90019d00feb44512b59e25efc6/1?pq-origsite=gscholar\&cbl=18750}{search.proquest.com}
\item What We Lose When Film Cameras Change to Digital Ones -- Nautilus. \href{https://nautil.us/what-we-lose-when-film-cameras-change-to-digital-ones-234963}{nautil.us}
\item The Impact of Technology on Cinematic Storytelling | Open Access Journals. \href{https://www.globalmediajournal.com/open-access/the-impact-of-technology-on-cinematic-storytelling.php?aid=94036}{globalmediajournal.com}
\item How technology has changed the film industry | Falmouth University. \href{https://www.falmouth.ac.uk/news/how-technology-has-changed-film-industry}{falmouth.ac.uk}
\item How the 2000s turned video and film production completely upside down -- Red Shark News. \href{https://www.redsharknews.com/production/item/5420-the-2000-s-turned-video-and-film-production-completely-upside-down}{redsharknews.com}
\item Theberge -- Musical\_Production\_Consumption.pdf -- SFU. \href{http://www.sfu.ca/sonic-studio-webdav/AudioMedia/Readings/Alphabetical/Theberge-Musical_Production_Consumption.pdf}{sfu.ca}
\item How has digitalisation changed the economics of the creative industries? -- Economics Observatory. \href{https://www.economicsobservatory.com/how-has-digitalisation-changed-the-economics-of-the-creative-industries}{economicsobservatory.com}
\item The Democratisation Paradox: What History Teaches Us About AI -- smcleod.net. \href{https://smcleod.net/2025/03/the-democratisation-paradox-what-history-teaches-us-about-ai/}{smcleod.net}
\item How reference based AI is democratising film production -- Coherent Market Insights. \href{https://www.coherentmarketinsights.com/blog/media-and-entertainment/how-reference-based-ai-is-democratizing-film-production-2899}{coherentmarketinsights.com}
\item AI and the Democratization of Art Creation: Revolutionising the Creative Landscape | Medium. \href{https://medium.com/higher-neurons/ai-and-the-democratization-of-art-creation-revolutionizing-the-creative-landscape-12c76a897561}{medium.com}
\item How AI could reinvent film and TV production -- McKinsey. \href{https://www.mckinsey.com/capabilities/tech-and-ai/our-insights/tech-forward/how-ai-could-reinvent-film-and-tv-production}{mckinsey.com}
\item Is the AI Push in AAA Gaming Giving Indie Developers A Golden Ticket? \href{https://waytoomany.games/2026/03/20/is-the-ai-push-in-aaa-gaming-giving-indie-developers-a-golden-ticket/}{waytoomany.games}
\item The Rise of AI in Game Development: A New Era for Indie Creators | Medium. \href{https://medium.com/@akinola.oluwaseyi22/the-rise-of-ai-in-game-development-a-new-era-for-indie-creators-35555881efbb}{medium.com}
\item How AI is Democratising Music Production -- SOUNDRAW Blog. \href{https://soundraw.io/blog/post/ai-is-democratizing-music}{soundraw.io}
\item What will the music industry look like in 2026? -- Identity Music. \href{https://identitymusic.com/blog/what-will-the-music-industry-look-like-in-2026}{identitymusic.com}
\item FUTURE UNSCRIPTED: The Impact of Generative Artificial Intelligence on Entertainment Industry Jobs | Animation Guild. \href{https://animationguild.org/wp-content/uploads/2024/01/Future-Unscripted-The-Impact-of-Generative-Artificial-Intelligence-on-Entertainment-Industry-Jobs-pages-1.pdf}{animationguild.org}
\item Applying AI to Rebuild Middle Class Jobs | ResearchGate. \href{https://www.researchgate.net/publication/378164927_Applying_AI_to_Rebuild_Middle_Class_Jobs}{researchgate.net}
\item The impact of generative artificial intelligence on socioeconomic inequalities and policy making -- PMC. \href{https://pmc.ncbi.nlm.nih.gov/articles/PMC11165650/}{pmc.ncbi.nlm.nih.gov}
\item Applying AI to Rebuild Middle Class Jobs -- MIT Stone Center. \href{https://shapingwork.mit.edu/research/applying-ai-to-rebuild-middle-class-jobs/}{shapingwork.mit.edu}
\item AI and the Future of Work: Opportunity or Threat? -- Federal Reserve Bank of St.\ Louis. \href{https://www.stlouisfed.org/publications/page-one-economics/2024/dec/ai-and-the-future-of-work-opportunity-or-threat?print=true}{stlouisfed.org}
\item Artificial intelligence is transforming middle-class jobs. Can it also help the poor? | Brookings. \href{https://www.brookings.edu/articles/ai-transforming-middle-class-jobs-can-it-help-the-poor/}{brookings.edu}
\item AI could automate up to 26\% of tasks in art, design, entertainment, and the media -- UOC. \href{https://www.uoc.edu/en/news/2025/ai-could-automate-creative-professions}{uoc.edu}
\item AI and work in the creative industries: digital continuity or discontinuity? -- Taylor \& Francis. \href{https://www.tandfonline.com/doi/full/10.1080/17510694.2024.2421135}{tandfonline.com}
\item Tomorrow comes today: Trends shaping the future of the Creative Industries -- PEC. \href{https://pec.ac.uk/wp-content/uploads/2023/12/PEC-Tomorrow-comes-today-Trends-shaping-the-future-of-the-Creative-Industries-August-2023.pdf}{pec.ac.uk}
\item AI Musicians vs.\ Machines: AI Disruption in the Music Industry | OC\&C Strategy Consultants. \href{https://www.occstrategy.com/en/article/ai-musicians-vs-machines-ai-disruption-in-the-music-industry/}{occstrategy.com}
\item How AI Benefits -- and Threatens -- the Entertainment Industry -- Morgan Stanley. \href{https://www.morganstanley.com/insights/articles/ai-in-media-entertainment-benefits-and-risks}{morganstanley.com}
\item Large studios will likely take their time adopting generative AI for content creation -- Deloitte. \href{https://www.deloitte.com/us/en/insights/industry/technology/technology-media-and-telecom-predictions/2025/tmt-predictions-hollywood-cautious-of-genai-adoption.html}{deloitte.com}
\item CEO Keynote: AI in the Music Industry -- Should You Fight It, Ignore It, or Embrace It? -- Sonarworks. \href{https://www.sonarworks.com/blog/research/ceo-keynote-ai-in-the-music-industry-2025}{sonarworks.com}
\item The Music Industry in 2026: Trends and Opportunities | Orphiq. \href{https://orphiq.com/resources/music-industry-2026-trends}{orphiq.com}
\item Between Inspiration and Loss of Control: Music in the Age of AI -- Ars Electronica. \href{https://ars.electronica.art/aeblog/en/2025/05/06/between-inspiration-and-loss-of-control/}{ars.electronica.art}
\item The AI Playbook: What Other Sectors Can Learn from the Creative Industry's Fight Against AI -- RAND. \href{https://www.rand.org/pubs/commentary/2024/10/the-ai-playbook-what-other-sectors-can-learn-from-the.html}{rand.org}
\item How AI Models Steal Creative Work -- and What to Do About It | Ed Newton-Rex | TED. \href{https://www.youtube.com/watch?v=U9d0p96N1iw}{youtube.com}
\item Protecting Human Creativity in AI-Generated Music with the Introduction of an AI-Royalty Fund | GRUR International. \href{https://academic.oup.com/grurint/article/73/12/1137/7832810}{academic.oup.com}
\item This is how AI is impacting -- and shaping -- the creative industries, according to experts at Davos -- The World Economic Forum. \href{https://www.weforum.org/stories/2024/02/ai-creative-industries-davos/}{weforum.org}
\item How AI Is Rewiring Filmmaking -- YouTube. \href{https://www.youtube.com/watch?v=M_YU2OzDPfY}{youtube.com}
\item Why AI Is A Game-Changer For Creatives, And Why The Creative Industries Must Fight For Their Rights -- YouTube. \href{https://www.youtube.com/watch?v=elQKQNewFZs}{youtube.com}
\item The Impact of Digitalisation on the Film Industry -- Raindance. \href{https://raindance.org/the-impact-of-digitalization-on-the-film-industry/}{raindance.org}
\item The Democratic Promise of Globalized Film and Television | Cato Institute. \href{https://www.cato.org/publications/trade-arts/culture}{cato.org}
\item Film sector on the frontlines: High-level discussion on AI in the audio-visual industry -- UNESCO. \href{https://www.unesco.org/creativity/en/articles/film-sector-frontlines-high-level-discussion-ai-audio-visual-industry}{unesco.org}
\item AI democratization bridges the gap between AI creators and users | Centific. \href{https://www.centific.com/blog/ai-democratization-bridges-the-gap-between-ai-creators-and-users}{centific.com}
\end{enumerate}

  \chapter{The Age of Intent: Artistic Mastery and the Inversion of Value}\label{app:a7}

\emph{Companion piece to Chapter~11: The Orchestrator and Chapter~15: Choosing the Future.}

\bigskip\par\noindent\rule{\textwidth}{0.4pt}\par\bigskip

This deep dive is the philosophical and economic companion to the book's central claim about where creative value lives after the AI transition. Chapter~11 develops the \emph{orchestrator} role -- the practical operating model of a senior creative directing a team of agents -- and Chapter~14 lays out the four principles (agency, attribution, access, audience) for a humane creative economy. Both rest on a deeper proposition that this appendix makes explicit: when the technical labour of execution becomes a commodity, the intentional labour of \emph{deciding what to make and why} becomes the scarce, valuable good.

The argument here draws on Duchamp's readymade, Arthur Danto's institutional theory of art, David Pye's distinction between the ``workmanship of certainty'' and the ``workmanship of risk,'' and Rick Rubin's framing of creativity as ``acts of noticing.'' It builds an empirical and philosophical case that the artist of 2030 is less a \emph{manual labourer} and more an \emph{Architect of Meaning} -- a curator, editor-in-chief, and director of intent whose value is precisely the human friction the machine cannot supply.

This is, in the book's broader frame, the strongest available account of why AI is best understood as an \emph{assistive instrument that amplifies human creativity} rather than a replacement for it. It underpins the conviction set out at the top of Chapter~15 and the operational pattern described in Chapter~11. Read it as the philosophical spine of the second half of the book.

\bigskip\par\noindent\rule{\textwidth}{0.4pt}\par\bigskip

\section*{Introduction: The Collapse of the Technical Barrier and the Onset of the Age of Intent}

We stand at a precipice in the history of human expression, a moment of rupture as profound as the invention of the printing press or the camera. For millennia, the definition of the artist was inextricably bound to the means of production -- the ``how.'' The mastery of the brush, the years spent learning to light a scene, the physical dexterity required to sculpt marble, or the mathematical precision needed to code a symphony were the gatekeepers of creation. Friction was the defining characteristic of value; difficulty was the proxy for quality. The artist was, by necessity, a technician first and a visionary second, for no vision could be realised without the hard labour of execution.

Today, however, we are witnessing the total collapse of this technical barrier. The advent of Generative Artificial Intelligence has democratised production to the point of triviality. The ``how'' is no longer a scarcity; it is a utility. When any individual with an internet connection can generate a photorealistic image, a coherent essay, or a symphonic progression with a single natural language prompt, the value of execution creates a surplus of content but a deficit of meaning. We are rapidly approaching a state of ``infinite media,'' where the ability to produce polished, high-fidelity work is available to everyone, everywhere, all at once.

In this new world, the hierarchy of value inverts. As the labour of production approaches zero, the labour of intent -- the ``why'' and the ``what'' -- becomes the most valuable currency on earth. This report posits that we have entered the \textbf{Age of Intent}, a distinct epoch where the ``how'' has been solved, leaving the ``why'' as the sole domain of human mastery. The artist of the future reclaims their throne not as a labourer, but as a visionary -- an Architect of Meaning who navigates the ocean of algorithmic competence through supreme acts of curation, selection, and philosophical grounding.

This document serves as an exhaustive analysis of this transition. It explores the technical mechanisms of the ``engine of probability'' that drives AI, the psychological crisis of the ``effort heuristic'' in consumer valuation, the economic inversion of creative labour markets, and the emerging methodologies of ``curatorial creation.'' Drawing on the lineage of Marcel Duchamp and the philosophy of ``workmanship of risk'' versus ``certainty,'' we will construct a factual argument for why the human spirit -- specifically the friction of human vulnerability -- remains the essential component in a system designed for statistical conformity.

\section*{Part I: The Mechanics of Abundance -- Deconstructing the Engine of Probability}

To understand why intent has become the new scarcity, one must first deeply understand the nature of the abundance generated by the machine. The ``content singularity'' -- a point where the volume of synthetic media outstrips human consumption capacity -- is driven by a specific technological architecture: the probabilistic prediction engine.

\subsection*{1.1 The Illusion of Thought: From Tokens to Text}

At its most fundamental level, a Large Language Model (LLM) or a diffusion model does not ``know,'' ``see,'' or ``feel'' in the human sense. It operates on tokens -- numerical representations of words, sub-words, or image patches. When an AI generates a sentence, it is calculating the statistical probability of the next token based on the context of preceding tokens.

For example, consider the sentence, ``I heard a dog bark loudly at a\ldots'' The foundational unit of the LLM is the token. The model cannot process raw text directly; it operates on numbers. The sentence is segmented into tokens -- ``I,'' ``heard,'' ``a,'' ``dog,'' ``bark,'' ``loudly,'' ``at,'' ``a'' -- and assigned numerical IDs. The model then analyses the statistical distribution of its training data to determine that the token for ``cat'' has a significantly higher probability than the token for ``fridge.''

However, this is not a simple deterministic lookup. If it were, AI outputs would be repetitive and robotic. The ``creativity'' of the machine arises from the manipulation of probability through parameters like \emph{temperature}, \emph{top-k}, and \emph{top-p} sampling.

\begin{itemize}
  \item \textbf{Temperature:} Low temperature favours reliability, selecting the most probable next token. High temperature encourages diversity, allowing the model to select less probable tokens, introducing ``novelty'' or ``hallucination.''
  \item \textbf{Top-k and Top-p (Nucleus) Sampling:} These methods restrict the sampling pool to the most likely candidates, renormalising probabilities to ensure coherence while maintaining variety.
\end{itemize}

This mechanism creates a ``central paradox'': complex, nuanced, and seemingly creative outputs emerge from a mechanism that is, at its core, a statistical prediction engine. The machine is an engine of probability; it predicts the next pixel or the next word based on the average of all human creation. It is the ultimate conformist. It can answer how to render a sunset, but it cannot answer \emph{why} that sunset should be rendered in a specific shade of melancholy to evoke a memory of loss.

\subsection*{1.2 Inference vs.\ Prediction: The Simulation of Reasoning}

While ``next-token prediction'' describes the mechanical operation, it fails to capture the user experience of ``inference.'' Modern generative AI performs complex logical analysis within context, adjusting its strategy based on global consistency. Unlike simple prediction, which might produce linear, one-directional outputs, modern models engage in a form of inference that mimics reasoning.

When a user inputs a query like, ``It's a beautiful day, so we can go\ldots'', the model considers the condition ``good weather'' and combines it with common sense (``good weather is suitable for outdoor activities'') to deduce an appropriate next step, such as ``a picnic.'' This involves reasoning that considers sentence structure, context, and background knowledge, aligning more with human thinking patterns than simple statistical choice.

Furthermore, these models handle ``global consistency'' in multi-step generation. When writing an essay, the model must ensure that the conclusion aligns with the introduction. This requires a capacity for global information integration that transcends local next-token prediction. It is this capacity that allows the machine to simulate the ``how'' of complex creative tasks -- structuring a symphony, plotting a novel, or composing a marketing strategy.

However, it is crucial to distinguish this simulated reasoning from embodied cognition. The model infers based on the statistical weights of its training data, which encapsulate the logical structures of human language. It does not ``understand'' the picnic; it understands the statistical likelihood of the word ``picnic'' appearing in the context of ``beautiful day.'' It lacks the ``embodied cognition'' that gives rise to true artistic intent -- the sensation of the sun, the taste of the food, the memory of past picnics.

\subsection*{1.3 The Paradox of Abundance: A Surplus of Content, A Deficit of Meaning}

The democratisation of this inferential power has led to a ``paradox of abundance.'' We are witnessing an explosion of content production that is inversely correlated with engagement and distinctiveness.

\subsubsection*{1.3.1 The Content Singularity}

The statistics regarding content proliferation are staggering. By 2025, the freelance platform market is projected to reach \$7.65 billion, driven largely by the ease of digital production. Marketing teams using video content jumped from 63\% in 2020 to 87\% in 2025, with AI tools reducing production time by 75\%. Yet, despite this massive increase in output, engagement rates are plummeting. Social media interaction rates have fallen to below 3\% across most platforms, down from much higher engagement earlier in the decade.

This phenomenon, termed the ``content singularity,'' describes an internet filled with more content than ever before, yet feeling less distinct. As production becomes effortless, the ability to differentiate becomes exponentially harder. The ``how'' has been solved for everyone, leading to a homogenisation of aesthetics. When everyone uses the same foundational models (e.g., GPT-4, Midjourney, Stable Diffusion), the outputs tend to converge on the ``statistical mean'' of the training data. A sea of ``authentic voices'' has produced the least authentic environment marketing has ever seen.

\subsubsection*{1.3.2 Model Collapse and Cultural Homogenisation}

A more insidious threat looms on the horizon: \textbf{Model Collapse}. As the web floods with AI-generated content, future models will increasingly be trained on synthetic data -- data generated by other AIs. Research from Oxford and Cambridge suggests this creates a degenerative feedback loop: each generation of models trained on increasingly synthetic data exhibits reduced diversity, amplified biases, and a narrowing of representational capabilities.

When models train on their own outputs, they lose the ``tails'' of the distribution -- the rare, unique, and idiosyncratic elements of human expression that drive innovation. Instead, they converge toward the centre, creating a ``technological monoculture.'' This mirrors ``cultural homogenisation,'' where AI models, already biased toward Western perspectives, further filter out diverse expressions. The machine, left to its own devices, collapses into sameness. It requires the injection of human intent -- novelty, friction, and the ``workmanship of risk'' -- to maintain cultural and semantic vitality.

\begin{center}
\begin{tabular}{p{3.5cm} p{5cm} p{5cm}}
\toprule
\textbf{Phenomenon} & \textbf{Description} & \textbf{Consequence for Art} \\
\midrule
Democratisation of ``How'' & Technical skills (rendering, coding) become utilities accessible via prompts. & Surplus of high-fidelity content; technical perfection becomes baseline, not differentiator. \\
\addlinespace
Statistical Conformity & Models predict the most probable next token/pixel based on averages. & Outputs tend toward the ``safe'' and generic; loss of ``edge'' or ``weirdness.'' \\
\addlinespace
Model Collapse & AI models training on AI-generated data. & Degenerative loss of variance; cultural homogenisation; ``slop'' content. \\
\addlinespace
Inference without Soul & Simulated reasoning without embodied experience. & Art that is technically proficient but emotionally hollow (``uncanny valley'' of meaning). \\
\bottomrule
\end{tabular}
\end{center}

\section*{Part II: The Psychology of Value -- The ``Effort Heuristic'' and the Crisis of Authenticity}

\subsection*{2.1 The Commodity of Skill and the ``Effort Heuristic''}

For centuries, society has operated on the ``effort heuristic'' -- the psychological shortcut where we judge the value and quality of an object based on the perceived effort required to create it. We marvelled at a photorealistic painting not just for its image, but for the years of mastery and hours of labour it represented. We respected the writer because we knew the agony of the blank page.

\subsubsection*{2.1.1 The Collapse of the Effort Heuristic}

AI has severed the link between quality and effort. A photorealistic image that once took 100 hours now takes 10 seconds. This unbundling of skill from creation has triggered a crisis in value perception. Research consistently shows that when consumers are aware an artwork is AI-generated, they perceive it as having less value, less emotional capacity, and lower quality, even if the visual output is identical to human work.

Studies reveal a distinct ``implicit bias'' against AI creativity. In experiments where artworks were labelled ``Human'' or ``AI,'' participants consistently rated the human-labelled works higher in liking, beauty, profundity, and worth. Furthermore, gaze-tracking studies found that participants spent significantly more time looking at paintings they believed were human-made compared to those labelled as AI-made. This suggests that our appreciation of art is not solely aesthetic; it is empathetic. We are connecting with the \emph{maker}, not just the \emph{made}. When the ``how'' becomes instant, it ceases to be impressive.

\subsection*{2.2 The ``Human-Made'' Premium}

In the Age of Intent, ``Human-Made'' is evolving from a descriptive tag into a luxury label. Just as ``hand-made'' became a premium designator in the industrial age, ``human-generated'' is becoming the ultimate status signal in the algorithmic age.

\subsubsection*{2.2.1 Felt Authenticity and Anti-AI Marketing}

The backlash against AI in marketing and art is driven by a desire for ``felt authenticity.'' Consumers report a visceral reaction to AI content -- it feels ``hollow'' or ``weirdly empty,'' like a ``smile with no warmth behind it.'' This sentiment is backed by data: 82\% of consumers worry about AI's societal impact, and 76\% say it is extremely important to know if content is created by a real person.

This has given rise to ``Anti-AI marketing,'' a strategy where brands explicitly reject AI to build trust. Examples include:

\begin{itemize}
  \item \textbf{Dove:} Committed to never using AI to represent real bodies.
  \item \textbf{Lego:} Emphasising human creativity in their ``human-made'' campaigns.
  \item \textbf{Polaroid:} Positioning their analogue cameras as the antidote to digital/AI perfection (``The Camera for an Analog Life'').
\end{itemize}

Trust is the central currency here. Studies indicate that AI-generated reviews and content are perceived as less genuine, leading to significantly lower purchase intent. Authenticity mediates the relationship between content and value; without the ``human touch'' (perceived effort, emotional risk, biological vulnerability), the content slides off the brain.

\subsection*{2.3 The Uncanny Valley of Meaning}

We are familiar with the ``uncanny valley'' in robotics -- where a robot looks almost human but not quite, eliciting revulsion. AI art has created an ``uncanny valley of meaning.'' The machine can simulate the syntax of deep emotion (using words like ``melancholy,'' ``loss,'' ``hope''), but it lacks the semantics of experience.

This deficit is where the artist reclaims their value. The machine can generate a symphony, but it cannot answer why a dissonance is necessary in the third movement to reflect a personal tragedy. It operates on ``certainty,'' whereas human art often thrives on the ``workmanship of risk.''

\subsubsection*{2.3.1 Workmanship of Risk vs.\ Certainty}

Drawing on the theories of David Pye, we can distinguish between the ``workmanship of certainty'' (mass production, automation, AI) and the ``workmanship of risk'' (where the quality of the result is not predetermined and depends on judgement, dexterity, and care).

\begin{itemize}
  \item \textbf{Workmanship of Certainty:} AI generation is the ultimate form of this. The outcome is probabilistically predetermined by the model weights. It is fast, consistent, and scalable.
  \item \textbf{Workmanship of Risk:} Human art involves the constant risk of failure. The brush might slip; the note might be flat. It is this vulnerability -- this proximity to failure -- that imbues the work with ``soul'' and ``authenticity.''
\end{itemize}

In the Age of Intent, the artist's role is to reintroduce risk. The artist must provide the friction, the contradiction, and the intent that makes art matter. Without a strong ``why,'' AI art is merely ``content'' -- technically proficient slop that lacks the friction of human vulnerability.

\section*{Part III: The Artist as Architect -- Inverting the Hierarchy}

\subsection*{3.1 The Inversion: Why $>$ How}

As the ``how'' becomes a commodity, the hierarchy of artistic value inverts. The labour of production approaches zero, while the labour of intent -- the ``why'' and the ``what'' -- becomes the scarcity.

\begin{itemize}
  \item \textbf{Old Hierarchy:} Technical Skill (High Value) $>$ Conceptual Intent (Variable Value)
  \item \textbf{New Hierarchy:} Conceptual Intent (High Value) $>$ Curation/Selection (High Value) $>$ Technical Skill (Commodity/Utility)
\end{itemize}

The artist shifts from being a manual labourer (the hand) to an Editor-in-Chief (the mind). This is not a new concept in art history, but AI has universalised it.

\subsection*{3.2 The Legacy of Duchamp: The Readymade in the AI Age}

Marcel Duchamp's submission of a urinal (\emph{Fountain}, 1917) to an art exhibition was the proto-event of the AI age. Duchamp argued that the art was not in the crafting of the object, but in the act of choice. ``He CHOSE it,'' wrote a defender in \emph{The Blind Man}. ``He took an ordinary article of life, placed it so that its useful significance disappeared under the new title and point of view -- created a new thought for that object.''

Generative AI transforms every user into a Duchampian figure. The model produces ``readymades'' at scale -- infinite variations of images, texts, and sounds. The creative act is no longer the rendering, but the \emph{selection and the contextualisation}.

\begin{itemize}
  \item \textbf{Danto's Theory:} Philosopher Arthur Danto argued that what makes a Brillo box art is not its physical properties, but the ``theory of art'' and the context provided by the artist. Similarly, an AI image becomes art not because of its pixels, but because of the intent and theory the artist wraps around it.
\end{itemize}

However, this does not mean art is ``easy.'' As Duchamp and the Conceptualists showed, when the object is trivial, the idea must be profound. If anyone can generate a ``sunset in the style of Van Gogh,'' the value lies not in the image, but in \emph{why} that image was chosen, where it is placed, and what conversation it provokes. The artist becomes a ``meta-creator,'' operating on the level of systems and concepts rather than pigments and pixels.

\subsection*{3.3 The New Discipline: Curation as Creation}

In a world of infinite generation, curation becomes the ultimate creative act. The artist must develop a ``Curatorial Framework'' to navigate the sea of noise generated by the machine.

\subsubsection*{3.3.1 The Editor-in-Chief Model}

The artist's role aligns with that of an Editor-in-Chief. The AI (the newsroom/staff writers) offers a thousand variations. The Artist (the Editor) must:

\begin{enumerate}
  \item \textbf{Reject:} Say ``no'' to the 99\% of distinct but meaningless generations. The ability to reject is the primary skill of the editor.
  \item \textbf{Select:} Identify the 1\% that ``vibrates with truth'' or novelty.
  \item \textbf{Refine:} Direct the machine to iterate on that specific grain of truth.
  \item \textbf{Contextualise:} Place the work in a cultural framework that gives it meaning.
\end{enumerate}

This requires a sophistication of taste that no algorithm can replicate. Taste is not just a preference; it is a form of knowledge -- a ``pattern recognition'' of cultural resonance.

\subsubsection*{3.3.2 Rick Rubin and the Art of ``Noticing''}

Music producer Rick Rubin's philosophy of creativity is particularly relevant here. Rubin argues that ``creativity is acts of noticing.'' The creator does not make the waves; they tune their antenna to receive them. In the AI context, the model is the ocean, constantly churning out possibilities. The artist is the ``noticer,'' the vessel with the refined filter.

Rubin emphasises that taste is a practice -- a way of being. ``To live as an artist is a way of being in the world. A way of perceiving. A practice of paying attention.'' This ``embodied attention'' cannot be automated. An AI can scan a million images, but it cannot ``notice'' the emotional weight of a specific shade of blue in the context of human grief. It can only predict its statistical likelihood. Rubin advises creators to cultivate ``awareness'' and to approach creation with a ``beginner's mind,'' maintaining curiosity and avoiding judgement during the initial phases of idea collection.

\begin{center}
\begin{tabular}{p{3.5cm} p{4.5cm} p{4.5cm}}
\toprule
\textbf{Attribute} & \textbf{Traditional Artist} & \textbf{AI-Era Artist (The Architect)} \\
\midrule
Primary Skill & Physical dexterity / Technical mastery & Curatorial taste / Conceptual intent \\
\addlinespace
Primary Action & Rendering / Construction & Selection / Direction / Contextualisation \\
\addlinespace
Output & A finished object & A curated framework or experience \\
\addlinespace
Value Source & Scarcity of skill (How) & Scarcity of vision (Why) \\
\addlinespace
Constraint & Physical limitations of the medium & Limitations of taste and attention \\
\bottomrule
\end{tabular}
\end{center}

\section*{Part IV: The Artist's Toolset -- From Prompts to Orchestration}

\subsection*{4.1 The Death of ``Prompt Engineering''}

In the early days of Generative AI (2022--2024), ``prompt engineering'' was hailed as a critical technical skill. It was treated as a form of coding -- learning the ``incantations'' (e.g., ``masterpiece, 8k, trending on artstation'') to trick the model into compliance. However, recent research and market trends suggest that prompt engineering as a distinct technical career is already obsolete.

\begin{itemize}
  \item \textbf{Obsolescence of Syntax:} As models become better at understanding natural language and nuance, the need for arcane syntax diminishes. The ``post-prompt age'' is characterised by systems that interpret intent from vague or incomplete instructions.
  \item \textbf{Natural Language Orchestration:} The skill is shifting to \textbf{Natural Language Orchestration}. It is not about writing better instructions; it is about designing interaction paradigms. The ``engineer'' is being replaced by the ``communicator.'' The best prompters are not those who know the cheat codes, but those who have a deep, nuanced vocabulary to describe mood, lighting, style, and emotion. A poet is now a better pilot for an LLM than a Python developer.
  \item \textbf{Meta-Prompting:} Advanced orchestration involves ``meta-prompting,'' where prompts generate other prompts, and systems critique and refine their own outputs. This requires a higher-level understanding of system architecture and behavioural psychology, moving beyond simple input-output tasks.
\end{itemize}

\subsection*{4.2 Case Studies in Collaborative Agency}

The artists successfully navigating this era are those who treat AI not as a replacement, but as a ``collaborator'' or a ``prosthetic for the imagination.'' They exemplify the ``Artist as Architect'' model.

\subsubsection*{4.2.1 Sougwen Chung: The Collaborator}

Artist and researcher Sougwen Chung rejects the ``tool'' metaphor entirely, viewing her robotic arms and AI systems as ``collaborators.''

\begin{itemize}
  \item \textbf{Method:} Chung trains her AI systems (D.O.U.G. -- Drawing Operations Unit Generation) on decades of her own drawing data. The AI then controls a robotic arm that draws alongside her in real-time.
  \item \textbf{Intent:} Her work explores ``embodied cognition'' and the feedback loop between human mark-making and machine mimicry. She is not outsourcing the art; she is engaging in a duet with her own data. The ``why'' is an exploration of memory and agency; the AI simply provides the ``how'' of the counter-melody. She views the AI not as an ``other'' but as a reflection of the self, stating, ``I've started to see them as us in another form.''
\end{itemize}

\subsubsection*{4.2.2 Holly Herndon: The Sovereign Architect}

Musician Holly Herndon addresses the issue of agency and ownership head-on. She created ``Holly+'', an AI vocal twin trained on her own voice, allowing others to create music using her likeness.

\begin{itemize}
  \item \textbf{Method:} She utilises ``Spawning,'' a protocol that allows artists to opt-in or opt-out of training datasets, reasserting consent in the age of scraping.
  \item \textbf{Intent:} Herndon's intent is to create a ``collective accomplishment.'' She views AI as a coordination technology -- a way to build a ``choir'' of intelligence. Her work \emph{Starmirror} invites the public to train an AI model through collective singing, turning the ``black box'' of training into a communal ritual. She is architecting the system of creation, not just the song.
\end{itemize}

\subsubsection*{4.2.3 Refik Anadol: The Data Sculptor}

Refik Anadol uses AI to visualise vast datasets, from brain scans to climate data.

\begin{itemize}
  \item \textbf{Method:} He treats data as pigment. His algorithms ``hallucinate'' new forms based on millions of images.
  \item \textbf{Intent:} His work is about making the invisible visible -- visualising the ``memory'' of a machine or the ``consciousness'' of a library. The curation of the dataset is the art. Selecting which 100 million images to feed the model is the primary creative decision.
\end{itemize}

\subsection*{4.3 Developing a Curatorial Framework}

For the modern artist, developing a Curatorial Framework is the new rigorous practice. It replaces the ``10,000 hours'' of manual practice with 10,000 hours of \emph{decision-making practice}.

The Framework Components:

\begin{enumerate}
  \item \textbf{Taste Calibration (The Input):}
    \begin{itemize}
      \item Immersion in art history and diverse media to build a ``reference library'' in the mind. The AI has the average of all data; the artist must possess the outliers.
      \item Rick Rubin's ``tuning'': Constantly refining sensitivity to what resonates and why.
    \end{itemize}
  \item \textbf{Iterative Selection (The Process):}
    \begin{itemize}
      \item Adopting the ``Editorial Thinking'' of data visualisation and design. Viewing the AI's output not as final, but as raw footage to be edited.
      \item Taste-Based Decision Making: Using ``gut'' and ``affect'' (embodied emotion) to filter rational machine outputs.
    \end{itemize}
  \item \textbf{Contextual Anchoring (The Output):}
    \begin{itemize}
      \item Defining the ``Why'': What is the emotional or intellectual provocation?
      \item The ``Human Label'': Consciously framing the work to highlight the human intent behind it, leveraging the ``authenticity'' premium.
    \end{itemize}
\end{enumerate}

\section*{Part V: The Economics of Intent -- Market Dynamics in 2025}

\subsection*{5.1 The Economic Inversion}

The labour market is reflecting the philosophical shift. As ``production'' roles (copywriter, junior graphic designer) face automation pressure, ``strategic'' roles (Creative Director, Brand Strategist) are seeing salary growth.

\subsubsection*{5.1.1 Salary Trends: The Director vs.\ The Technician}

Data from 2025 indicates a widening gap between execution and direction roles.

\begin{itemize}
  \item Creative Directors and VPs of Marketing (roles defined by strategy, taste, and intent) command salaries of \$145k -- \$250k+.
  \item Copywriters and Graphic Designers (execution roles) are seeing stagnation or pressure, with averages around \$57k -- \$71k, though high-level specialisation (e.g., AI orchestration) can boost this.
  \item Freelance Market: While the freelance market is growing (\$7.65 billion by 2025), there is a bifurcated reality. ``Routine maintenance work'' is seeing rate decreases of 5--10\%, while ``AI Specialists'' and ``AI Integration Consultants'' command premiums of 40--60\% (\$100--\$200/hour).
\end{itemize}

This confirms the thesis: the value is moving away from \emph{doing the thing} to \emph{directing the thing}. The technician who relied solely on execution is finding their skills commoditised, while the visionary who can orchestrate these tools is seeing their value skyrocket.

\subsection*{5.2 The Rise of the ``Human-Made'' Luxury Market}

Just as ``organic'' food commands a premium over ``processed'' food, ``human-made'' content is emerging as a luxury good.

\begin{itemize}
  \item \textbf{Anti-AI Marketing:} Brands are explicitly using ``No AI'' as a selling point. Campaigns by Dove, Lego, and others emphasise human creativity to build trust.
  \item \textbf{The Trust Deficit:} With 82\% of consumers worried about AI's societal impact and 76\% finding it important to know if content is human-made, there is a distinct market for ``certified human'' work.
  \item \textbf{Implication:} Artists should not hide their use of AI, but those who don't use it (or use it minimally) should leverage their ``inefficiency'' as a value signal. The ``flaws'' of human work -- the ``workmanship of risk'' -- become markers of authenticity.
\end{itemize}

\section*{Conclusion: The Triumph of Vision}

We are witnessing the end of the ``technician artist'' and the rise of the ``visionary artist.'' The barrier between having a thought and seeing it realised has dissolved. For the technician who relied solely on the difficulty of their craft to justify their value, this is a crisis. For the visionary who has been constrained by the limits of their hands or their budget, this is a liberation.

The machine is an engine of probability; it can synthesise a style, but it cannot synthesise a soul. It can answer how to render a sunset, but it cannot answer \emph{why} that sunset matters. It can predict the next token, but it cannot feel the weight of the word.

The ``Age of Intent'' demands a new kind of discipline. It is no longer about the steadiness of the hand, but the clarity of the mind. It requires the artist to be an architect of meaning, a curator of infinite possibility, and a guardian of the human spirit in an age of synthetic abundance.

The artists who will reign supreme are those who treat AI not as a replacement, but as a prosthetic for the imagination. They are the ones who know that the machine can synthesise a style, but it cannot synthesise a soul. In the end, the machine is just a brush. A very complex, miraculous brush, but a brush nonetheless. And a brush cannot paint without a hand to hold it, and a mind to tell it where to go.

\section*{Appendix: Supporting Data and Frameworks}

\subsection*{Table 1: Comparative Analysis of ``Workmanship'' (Based on Pye)}

\begin{center}
\begin{tabular}{p{3.5cm} p{5cm} p{5cm}}
\toprule
\textbf{Feature} & \textbf{Workmanship of Risk (Human/Traditional)} & \textbf{Workmanship of Certainty (AI/Automation)} \\
\midrule
Definition & Outcome is not predetermined; depends on judgement/dexterity at every moment. & Outcome is predetermined; quantity production; low variance. \\
\addlinespace
Primary Value & Uniqueness, ``soul,'' perceivable effort, diversity. & Perfection, speed, scale, consistency. \\
\addlinespace
Flaws & ``Mistakes'' or ``happy accidents'' that reveal the hand. & ``Hallucinations'' or ``artefacts'' (often seen as errors to be fixed). \\
\addlinespace
Artist's Role & To execute the form. & To disrupt the certainty; to inject risk back into the system. \\
\bottomrule
\end{tabular}
\end{center}

\subsection*{Table 2: Consumer Perception of AI vs.\ Human Art}

\begin{center}
\begin{tabular}{p{4cm} p{4cm} p{4.5cm}}
\toprule
\textbf{Metric} & \textbf{Human-Labelled Art} & \textbf{AI-Labelled Art} \\
\midrule
Liking/Preference & High & Low (Significant negative bias) \\
\addlinespace
Perceived Effort & High & Low \\
\addlinespace
Emotional Response & Stronger & Weaker (``Hollow'') \\
\addlinespace
Gaze Duration & Longer (studied more closely) & Shorter (dismissed faster) \\
\addlinespace
Purchase Intent & Higher & Lower \\
\bottomrule
\end{tabular}
\end{center}

\subsection*{Table 3: Salary Trends 2025 -- The Value of Direction}

\begin{center}
\begin{tabular}{p{3.5cm} p{3.5cm} p{3.5cm} p{2cm}}
\toprule
\textbf{Role} & \textbf{Focus} & \textbf{Avg.\ Salary (Est.\ 2025)} & \textbf{Trend} \\
\midrule
Creative Director & Intent, Vision, Strategy, Curation & \$145,000 -- \$250,000+ & Rising \\
\addlinespace
VP of Marketing & Strategy, Brand Voice & \$250,000 & Rising \\
\addlinespace
Copywriter & Text Generation (Execution) & \$57,000 -- \$71,000 & Stagnant/Risk \\
\addlinespace
Graphic Designer & Image Generation (Execution) & \$66,000 & Stagnant/Risk \\
\addlinespace
AI Specialist & Orchestration, Integration & \$100 -- \$200/hr & High Demand \\
\bottomrule
\end{tabular}
\end{center}

\section*{Works Cited}

\begin{enumerate}
\item Generative AI Foundations: From Tokens to Text, How LLMs ``Write.'' \href{https://medium.com/generative-ai-playbook/generative-ai-foundations-from-tokens-to-text-how-llms-write-05f1800703ab}{medium.com}
\item If generative AI is the answer, what is the question? \href{https://arxiv.org/html/2509.06120v2}{arxiv.org}
\item Why the Work of Generative AI is Called ``Inference'' Rather Than ``Prediction.'' \href{https://medium.com/@justyan/why-the-work-of-generative-ai-is-called-inference-rather-than-prediction-3b026a63e324}{medium.com}
\item From Rule to Idea: Minimalism and Conceptualism as Structural Logic. \href{https://medium.com/@Neuroism/from-rule-to-idea-minimalism-and-conceptualism-as-structural-logic-9441750ea552}{medium.com}
\item An Initial Critique: Some Strengths And Weaknesses of ``Post-Theory Art.'' \href{https://adamdaleywilson.medium.com/an-initial-critique-some-strengths-and-weaknesses-of-post-theory-art-fe6c44846e94}{medium.com}
\item Freelancing Statistics And Trends 2025 -- Quantumrun Foresight. \href{https://www.quantumrun.com/consulting/freelancing-statistics/}{quantumrun.com}
\item Top 10 AI Video Tools Like Clippie (2025 Comparison Guide). \href{https://clippie.ai/blog/best-ai-video-tools-comparison-2025}{clippie.ai}
\item The Future of Branding in the Age of AI-Generated Content. \href{https://www.youngmarketingconsulting.com/the-future-of-branding-in-the-age-of-ai-generated-content/}{youngmarketingconsulting.com}
\item Epistemic diversity across language models mitigates knowledge collapse -- ResearchGate. \href{https://www.researchgate.net/publication/398806203_Epistemic_diversity_across_language_models_mitigates_knowledge_collapse}{researchgate.net}
\item AI Model Collapse: Why Data Quality Is the Defining Ethical Challenge of 2025. \href{https://expertlinked.in/posts/2025-10-31-ai-model-collapse-data-quality-imperative/}{expertlinked.in}
\item Craft for the AI Age -- The New Art School. \href{https://newartschool.education/2025/10/17/craft-for-the-ai-age/}{newartschool.education}
\item Implicit bias against AI creativity -- ResearchGate. \href{https://www.researchgate.net/publication/395476845_Implicit_bias_against_AI_creativity}{researchgate.net}
\item Human Art vs AI Art, a Potential Danger for Artist? Utrecht University Student Theses Repository. \href{https://studenttheses.uu.nl/bitstream/handle/20.500.12932/47294/Thesis_JoA%CC%83%C2%ABlvanWaning%202.pdf?sequence=1&isAllowed=y}{studenttheses.uu.nl}
\item The rise of AI art: A look through digital artists' eyes -- First Monday. \href{https://firstmonday.org/ojs/index.php/fm/article/view/13809/12001}{firstmonday.org}
\item Comparative Designs Reveal Preferences for Human-Generated Rather Than AI-Generated art -- PEARL. \href{https://pearl.plymouth.ac.uk/cgi/viewcontent.cgi?article=2175&context=psy-research}{pearl.plymouth.ac.uk}
\item Humans versus AI: whether and why we prefer human-created compared to AI-created artwork -- PMC. \href{https://pmc.ncbi.nlm.nih.gov/articles/PMC10319694/}{pmc.ncbi.nlm.nih.gov}
\item Anti-AI Marketing Trends in 2025: Embracing Authenticity and Human Connection. \href{https://neuron.expert/news/the-best-anti-ai-marketing-campaigns-in-2025-and-why-they-worked/15795/en/}{neuron.expert}
\item The Best Anti-AI Marketing Campaigns in 2025, and Why They Worked -- Brand Vision. \href{https://www.brandvm.com/post/best-anti-ai-marketing-campaigns-2025}{brandvm.com}
\item iHeartMedia Releases Third in a Series of Consumer Studies, Revealing America's Deepening Need for Human Connection in an AI-Driven World. \href{https://investors.iheartmedia.com/news/news-details/2025/iHeartMedia-Releases-Third-in-a-Series-of-Consumer-Studies-Revealing-Americas-Deepening-Need-for-Human-Connection-in-an-AI-Driven-World/default.aspx}{investors.iheartmedia.com}
\item Investigating the Effect of AI-Generated Customer Reviews on Purchase Intent and Perceived Authenticity in E-Commerce Environments -- ResearchGate. \href{https://www.researchgate.net/publication/397937065_Investigating_the_Effect_of_AI-Generated_Customer_Reviews_on_Purchase_Intent_and_Perceived_Authenticity_in_E-Commerce_Environments}{researchgate.net}
\item Revealing AI Involvement in Ad Creation: Effects on Authenticity, Brand Perceptions and Consumer Intentions. \href{https://jisem-journal.com/index.php/journal/article/download/2659/1056/4303}{jisem-journal.com}
\item AI-Generated Art and Its Influence on Consumer Psychology: A Review of Economic and Emotional Outcomes -- IJIP. \href{https://ijip.in/wp-content/uploads/2025/02/18.01.066.20251301.pdf}{ijip.in}
\item David Pye -- The Nature and Art of Workmanship. \href{https://memoof.me/download/884/pdf/884.pdf}{memoof.me}
\item David Pye -- The Nature and Art of Workmanship (Medium). \href{https://medium.com/not-so-different-emerging-digital-craft-practices/david-pye-the-nature-and-art-of-workmanship-be0620f95aa9}{medium.com}
\item The Expanding Canvas: What Counts as Art? -- HastingsNow. \href{https://www.hastingsnow.com/blog/the-expanding-canvas-what-counts-as-art}{hastingsnow.com}
\item Alice C Helliwell -- Thesis -- Philosophy of AI Art -- Kent Academic Repository. \href{https://kar.kent.ac.uk/105246/1/105ALICE_C_HELLIWELL_-_THESIS_-_PHILOSOPHY_OF_AI_ART_-_KAR_UPLOAD_REDACTED.pdf}{kar.kent.ac.uk}
\item Rick Rubin, \emph{The Creative Act} (2023) -- Research Lab, Sorrywecan. \href{https://lab.sorrywecan.com/static/pdf/the-coming-age-of-wisdom-work.pdf}{lab.sorrywecan.com}
\item Everyone Suddenly Thinks They Are Rick Rubin -- Human Kind. \href{https://thehumankind.co/2025/07/02/everyone-suddenly-thinks-they-are-rick-rubin/}{thehumankind.co}
\item Rick Rubin on Listening, Taste, and the Act of Noticing (Ep.\ 169 -- BONUS). \href{https://conversationswithtyler.com/episodes/rick-rubin/}{conversationswithtyler.com}
\item The Creative Act By Rick Rubin. \href{https://ia800503.us.archive.org/33/items/the-creative-act-by-rick-rubin/The\%20Creative\%20Act\%20By\%20Rick\%20Rubin.pdf}{archive.org}
\item Rick Rubin's \emph{The Creative Act: A Way of Being} -- TianPan.co. \href{https://tianpan.co/notes/2025-05-05-creativity-act}{tianpan.co}
\item Death of Prompt Engineering: AI Orchestration in 2026 -- Big Blue Data Academy. \href{https://bigblue.academy/en/the-death-of-prompt-engineering-and-its-ruthless-resurrection-navigating-ai-orchestration-in-2026-and-beyond}{bigblue.academy}
\item Why Prompt Engineering Shouldn't Exist -- Infuzu. \href{https://infuzu.com/blog/why-prompt-engineering-shouldn-t-exist}{infuzu.com}
\item Is prompt engineering dead? One expert describes what HR should focus on instead. \href{https://hrexecutive.com/is-prompt-engineering-dead-one-expert-describes-what-hr-should-focus-on-instead/}{hrexecutive.com}
\item The Prompt Collapse: Why Natural Language Interfaces Are About to Eat Traditional Software -- LBZ Advisory. \href{https://liatbenzur.com/2025/11/03/the-prompt-collapse-why-natural-language-interfaces-are-about-to-eat-traditional-software/}{liatbenzur.com}
\item Prompt Engineering: Turning Intuition into Intentional Skill -- Medium. \href{https://medium.com/@ranji011/prompt-engineering-turning-intuition-into-intentional-skill-09cdd27ffda2}{medium.com}
\item Is prompt engineering still a viable skill in 2025, or is it fading fast? -- Reddit. \href{https://www.reddit.com/r/learnmachinelearning/comments/1oj5wxl/is_prompt_engineering_still_a_viable_skill_in/}{reddit.com}
\item Putting The Art In Artificial Intelligence: A Conversation With Sougwen Chung. \href{https://sougwen.com/putting-the-art-in-artificial-intelligence-a-conversation-with-sougwen-chung}{sougwen.com}
\item Sougwen Chung: the poetics of human--machine interaction -- hube magazine. \href{https://hubemag.com/sougwen-chung-the-poetics-of-human-machine-interaction}{hubemag.com}
\item Sougwen Chung: Human and machine collaboration -- Medium/Verisart. \href{https://medium.com/verisart/sougwen-chung-human-and-machine-collaboration-72d912d6b065}{medium.com}
\item Why artificial intelligence artists can be seen as `builders', `breakers' -- or both at once -- The Art Newspaper. \href{https://www.theartnewspaper.com/2025/07/11/why-artificial-intelligence-artists-can-be-seen-as-builders-breakersor-both-at-once}{theartnewspaper.com}
\item Holly Herndon: The 100 Most Influential People in AI 2023 -- Time Magazine. \href{https://time.com/collection/time100-ai/6309468/holly-herndon/}{time.com}
\item Why Artists Holly Herndon and Mat Dryhurst Want Us to Stop Minimizing -- And Fearing -- A.I. -- Artnet News. \href{https://news.artnet.com/art-world/holly-herndon-and-mat-dryhurst-2551072}{artnet.com}
\item Holly Herndon \& Mat Dryhurst Starmirror -- KW Institute for Contemporary Art. \href{https://www.kw-berlin.de/en/exhibitions/holly-herndon-and-mat-dryhurst-starmirror}{kw-berlin.de}
\item Holly Herndon \& Mat Dryhurst on Artificial Psychedelia -- Le Random. \href{https://www.lerandom.art/editorial/holly-herndon-mat-dryhurst-on-artificial-psychedelia}{lerandom.art}
\item The intersection of AI and art -- POLITesi. \href{https://www.politesi.polimi.it/retrieve/bea25522-3b1c-4ad4-861c-e56af3fee554/2024_10_GiacominDaSilva.pdf}{politesi.polimi.it}
\item Decoding AI Art: From Motivation to Manifestation -- University of Nottingham. \href{https://eprints.nottingham.ac.uk/81162/1/Salimbeni_Guido_20276375_corrections.pdf}{eprints.nottingham.ac.uk}
\item Editorial Thinking in Data Visualisation -- Scribd. \href{https://www.scribd.com/document/664543313/week-4-Analyzing-and-visualizing-data-Andy-Kirk}{scribd.com}
\item Linear and categorical coding units in the mouse gustatory cortex drive population dynamics and behaviour in taste decision-making -- eLife. \href{https://elifesciences.org/reviewed-preprints/109313}{elifesciences.org}
\item Marketing Salary \& Skills Report 2025: AI Roles, Top Jobs, Pay Data -- GLOZO. \href{https://www.glozo.com/reports/marketing-salary-skills-report-2025}{glozo.com}
\item How to Become an Art Director: Career Path \& Guide -- Himalayas.app. \href{https://himalayas.app/career-guides/art-director}{himalayas.app}
\item Compensation -- Human Resource Management homework help -- SweetStudy. \href{https://www.sweetstudy.com/files/milestone2-3486823.docx}{sweetstudy.com}
\item The Global Freelance Rate Database 2026: Complete Salary \& Pricing Guide for Every Industry -- Jobbers. \href{https://www.jobbers.io/the-global-freelance-rate-database-2026-complete-salary-pricing-guide-for-every-industry/}{jobbers.io}
\end{enumerate}

  \chapter{The Dream Machine Source Index}\label{app:a8}

\emph{A thematic catalogue of significant sources surfaced across the 29 issues of} Dream Machine \emph{(October 2025 -- May 2026).}

\bigskip

This index is a navigational tool, not an exhaustive list. The full \emph{Dream Machine} archive contains nearly three thousand individual hyperlinks across its twenty-nine issues, the great majority of which are primary-source links to industry coverage, research reports, official announcements, court filings, technical demos, creator showcases, and platform releases.

What follows below is the thematic catalogue of the \emph{significant} sources -- the ones the book itself draws on, the ones a working creative or researcher tracking a specific topic would want as a starting point, and the ones that, taken together, define the public record of creative AI as it stood in the period this book covers. Within each theme, entries are organised chronologically by issue number. The format is:

\medskip
\noindent\texttt{[Issue N]} \textbf{Title} -- short context -- URL
\medskip

For the complete primary-source archive -- every link, in full, in original publication order -- refer to the \emph{Dream Machine} newsletter archive on LinkedIn or the per-issue markdown files in \texttt{Dream Machine MD/}. Each issue's final section, \emph{``All embedded URLs (in document order)''}, lists every URL the issue carried.

\bigskip\par\noindent\rule{\textwidth}{0.4pt}\par\bigskip

\section*{1. AI Video -- Models and Releases}

\begin{itemize}
  \item \textbf{[Issue 1]} \textbf{Sora 2 Launch} -- OpenAI's step-change in physical realism, audio, multi-shot world state -- \href{https://openai.com/index/sora-2/}{openai.com}
  \item \textbf{[Issue 1]} \textbf{Veo 3 and Flow} -- Google DeepMind's cinematic AI video -- \href{https://www.youtube.com/watch?v=I06Ef8alr2Y}{youtube.com}
  \item \textbf{[Issue 2]} \textbf{Veo 3.1 Coming Soon} -- improved consistency, resolution, multi-shot, cinematic presets -- \href{https://www.cometapi.com/veo-3-1-is-comingand-whats-rumor/}{cometapi.com}
  \item \textbf{[Issue 3]} \textbf{Veo 3.1 Deep Dive with Flow} -- cinematic filmmaking toolset -- \href{https://www.youtube.com/watch?v=I06Ef8alr2Y}{youtube.com}
  \item \textbf{[Issue 3]} \textbf{Higgsfield Sketch-to-Video} -- powered by Sora 2 -- \href{https://higgsfield.ai/posts/6nkYSGwOcdyXVqZefE1MsQ}{higgsfield.ai}
  \item \textbf{[Issue 3]} \textbf{LiveGS} -- mobile Gaussian-splatting video
  \item \textbf{[Issue 4]} \textbf{Veo 3.1 Style Tips} -- text-to-video with style guidance -- \href{https://x.com/GoogleAI/status/1980327604843381215}{x.com}
  \item \textbf{[Issue 4]} \textbf{Veo Precision Features} -- remove/add elements to scenes -- \href{https://x.com/GoogleDeepMind/status/1980261047836508213}{x.com}
  \item \textbf{[Issue 4]} \textbf{Heygen Identity Consistency with Veo 3.1} -- character-consistent video -- \href{https://x.com/HeyGen_Official/status/1978491090618749193}{x.com}
  \item \textbf{[Issue 4]} \textbf{Higgsfield Popcorn} -- storyboard tool with character consistency -- \href{https://x.com/higgsfield_ai/status/1981110992630341928}{x.com}
  \item \textbf{[Issue 4]} \textbf{Odyssey 2} -- real-time interactive video generation -- \href{https://odyssey.ml/introducing-odyssey-2}{odyssey.ml}
  \item \textbf{[Issue 4]} \textbf{Krea Realtime} -- open-sourcing the creative engine -- \href{https://www.linkedin.com/posts/krea-ai_today-were-open-sourcing-krea-realtime-activity-7386124532207689728-B7pD}{linkedin.com}
  \item \textbf{[Issue 5]} \textbf{Sora Character Cameos} -- new feature in Sora app -- \href{https://x.com/OpenAI/status/1983661036533379486}{x.com}
  \item \textbf{[Issue 5]} \textbf{VEED Transitions} -- AI-powered video transitions -- \href{https://x.com/veedstudio/status/1980636419891818850}{x.com}
  \item \textbf{[Issue 5]} \textbf{LTX-2} -- open-source audio-video generation -- \href{https://x.com/ltx_model/status/1981346235194683497}{x.com}
  \item \textbf{[Issue 6]} \textbf{Wan 2.2} -- AI video with multi-shot capabilities -- \href{https://x.com/eyishazyer/status/1983507594942755221}{x.com}
  \item \textbf{[Issue 6]} \textbf{MotionStream} -- real-time interactive video, 29 FPS -- \href{https://x.com/wildmindai/status/1985828041566941576}{x.com}
  \item \textbf{[Issue 7]} \textbf{Odyssey-2} -- interactive streaming 16:9 video -- \href{https://x.com/olivercameron/status/1984777003967672800}{x.com}
  \item \textbf{[Issue 9]} \textbf{Wan 2.6 Released} -- cast characters from reference videos, up to 15 seconds -- \href{https://wan.video/blog/wan2.6-introduction}{wan.video}
  \item \textbf{[Issue 13]} \textbf{Runway Gen-4.5} -- image-to-video for paid plans -- \href{https://www.youtube.com/watch?v=AwKSrJFvdps}{youtube.com}
  \item \textbf{[Issue 13]} \textbf{LTX-2 on ComfyUI} -- open-source audio-video -- \href{https://blog.comfy.org/p/ltx-2-open-source-audio-video-ai}{blog.comfy.org}
  \item \textbf{[Issue 14]} \textbf{Veo 3.1 Ingredients to Video} -- vertical formats, 1080p/4K -- \href{https://x.com/FlowbyGoogle/status/2011130097483526474}{x.com}
  \item \textbf{[Issue 14]} \textbf{LTX-2 Lip Sync} -- native audio-driven dialogue -- \href{https://x.com/ltx_model/status/2011101440706806051}{x.com}
  \item \textbf{[Issue 15]} \textbf{Runway Gen-4.5 Image to Video} -- broad rollout -- \href{https://www.linkedin.com/posts/runwayml_introducing-image-to-video-for-gen-45-the-activity-7419856988186238976-ZJU2}{linkedin.com}
  \item \textbf{[Issue 15]} \textbf{Veo 3.1 in YouTube Shorts and Create app} -- distribution-layer integration -- \href{https://blog.google/innovation-and-ai/technology/ai/veo-3-1-ingredients-to-video/}{blog.google}
\end{itemize}

\section*{2. AI Image -- Models and Tools}

\begin{itemize}
  \item \textbf{[Issue 1]} \textbf{Nano Banana Plugin for Photoshop} -- Gemini image gen inside Adobe -- \href{https://www.linkedin.com/posts/arminas-valunas-b4477255_nano-banana-plugin-for-photoshop-is-here-ugcPost-7367923639414906881-GgV7}{linkedin.com}
  \item \textbf{[Issue 1]} \textbf{Magnific Precision v2} -- AI upscaling -- \href{https://www.linkedin.com/posts/magnific-ai_introducing-magnific-precision-v2-activity-7387158930776276992-zpET}{linkedin.com}
  \item \textbf{[Issue 2]} \textbf{Vimeo AI Creator Tools} -- new features -- \href{https://www.tvtechnology.com/news/vimeo-releases-new-ai-powered-creator-tools}{tvtechnology.com}
  \item \textbf{[Issue 3]} \textbf{Google Stitch} -- design tool with AI features -- \href{https://www.testingcatalog.com/google-test-new-stitch-modes-annotate-theme-interactive/}{testingcatalog.com}
  \item \textbf{[Issue 5]} \textbf{Adobe Firefly Image Model 5} -- Adobe MAX 2025 -- \href{https://news.adobe.com/news/2025/10/adobe-max-2025-firefly}{news.adobe.com}
  \item \textbf{[Issue 5]} \textbf{Freepik Spaces} -- infinite canvas for collaborative creation -- \href{https://www.freepik.com/spaces}{freepik.com}
  \item \textbf{[Issue 5]} \textbf{Midjourney updates} -- state-of-the-art image gen -- \href{https://x.com/midjourney/status/1991684484455100477}{x.com}
  \item \textbf{[Issue 6]} \textbf{Qwen Image Multiple Angles LoRA} -- consistent characters -- \href{https://x.com/multimodalart/status/1986174924038218087}{x.com}
  \item \textbf{[Issue 8]} \textbf{Qwen Image 2512} -- latest generation -- \href{https://qwen.ai/blog?id=qwen-image-2512}{qwen.ai}
  \item \textbf{[Issue 9]} \textbf{FLUX 2} -- Black Forest Labs image generation -- \href{https://bfl.ai/models/flux-2-max}{bfl.ai}
  \item \textbf{[Issue 10]} \textbf{FLUX 2 klein} -- fast image generation under one second -- \href{https://huggingface.co/unsloth/FLUX.2-klein-4B-GGUF}{huggingface.co}
  \item \textbf{[Issue 13]} \textbf{ChatGPT Images Upgrade} -- major feature improvements -- \href{https://www.techradar.com/ai-platforms-assistants/chatgpt/chatgpt-images-just-got-a-major-upgrade-and-it-could-change-how-we-all-create}{techradar.com}
\end{itemize}

\section*{3. AI Music and Audio}

\begin{itemize}
  \item \textbf{[Issue 1]} \textbf{Suno Studio} -- generative audio workstation -- \href{https://www.techradar.com/ai-platforms-assistants/i-tried-suno-studio-the-new-platform-that-mixes-ai-music-generation-with-hands-on-editing-like-garageband-but-smarter}{techradar.com}
  \item \textbf{[Issue 1]} \textbf{YouTube Music AI Hub} -- AI music hosts -- \href{https://www.linkedin.com/news/story/youtube-music-debuts-new-ai-hub-6625484/}{linkedin.com}
  \item \textbf{[Issue 1]} \textbf{Spotify AI Protections} -- strengthened for artists -- \href{https://newsroom.spotify.com/2025-09-25/spotify-strengthens-ai-protections/}{newsroom.spotify.com}
  \item \textbf{[Issue 1]} \textbf{Xania Monet} -- AI singer signs \$3M deal -- \href{https://www.forbes.com/sites/dougmelville/2025/09/27/al-singer-xania-monet-just-charted-on-billboard-signed-3m-deal-is-this-the-future-of-music/}{forbes.com}
  \item \textbf{[Issue 1]} \textbf{Cardiff Band on AI Artist} -- trained on their music, outperforming them -- \href{https://musictech.com/news/industry/its-shocking-disheartening-and-insulting-cardiff-band-speaks-out-after-ai-artist-trained-on-their-music-outperforms-them-on-spotify/}{musictech.com}
  \item \textbf{[Issue 2]} \textbf{Suno Funding Round} -- \$2 billion valuation -- \href{https://www.digitalmusicnews.com/2025/10/20/suno-funding-round-october-2025/}{digitalmusicnews.com}
  \item \textbf{[Issue 2]} \textbf{Groundhog AI Guitar Pedal} -- tone matching -- \href{https://musictech.com/news/gear/groundhog-audio-onepedal-ai-tone-matching/}{musictech.com}
  \item \textbf{[Issue 2]} \textbf{Spotify in ChatGPT} -- integration launched -- \href{https://newsroom.spotify.com/2025-10-06/spotify-personalized-prompts-chatgpt/}{newsroom.spotify.com}
  \item \textbf{[Issue 3]} \textbf{iZotope Ozone 12} -- AI assistant for mixing -- \href{https://musictech.com/reviews/plug-ins/izotope-ozone-12-review/}{musictech.com}
  \item \textbf{[Issue 3]} \textbf{Tempolor Guitars (Quwan)} -- AI to make songs playable -- \href{https://kr-asia.com/no-practice-required-quwans-tempolor-guitars-use-ai-to-make-songs-playable-in-minutes}{kr-asia.com}
  \item \textbf{[Issue 4]} \textbf{Mureka Music Agent Studio} -- six specialised AI agents -- \href{https://www.linkedin.com/posts/sherrihendrickson_mureka-unveils-music-agent-studio-and-enhanced-share-7384999251526864896-cNYg/}{linkedin.com}
  \item \textbf{[Issue 4]} \textbf{Fish Audio S1} -- TTS 6$\times$ cheaper than ElevenLabs -- \href{https://fish.audio/app/text-to-speech/}{fish.audio}
  \item \textbf{[Issue 5]} \textbf{Universal Music + Stability AI Alliance} -- strategic partnership -- \href{https://stability.ai/news/universal-music-group-and-stability-ai-announce-strategic-alliance}{stability.ai}
  \item \textbf{[Issue 5]} \textbf{Udio Partners with UMG} -- partnership announced -- \href{https://www.udio.com/blog/a-new-era}{udio.com}
  \item \textbf{[Issue 5]} \textbf{OpenAI Music Generator} -- reportedly in development -- \href{https://www.theinformation.com/articles/openai-plots-generating-ai-music-potential-rivalry-startup-suno}{theinformation.com}
  \item \textbf{[Issue 6]} \textbf{UMG Boss Lucian Grainge on AI} -- full internal memo -- \href{https://musically.com/2025/10/14/umg-boss-sir-lucian-grainge-talks-ai-full-internal-memo/}{musically.com}
  \item \textbf{[Issue 6]} \textbf{Bleeding Verse AI Band} -- Hallwood Media signing -- \href{https://musically.com/2025/10/07/ai-band-bleeding-verses-creator-signs-deal-with-hallwood-media/}{musically.com}
  \item \textbf{[Issue 6]} \textbf{JYP Entertainment AI Artist} -- hiring AI/Unreal experts -- \href{https://www.musicbusinessworldwide.com/jyp-entertainment-is-hiring-for-ai-and-unreal-engine-experts-to-develop-an-unprecedented-virtual-kpop-artist/}{musicbusinessworldwide.com}
  \item \textbf{[Issue 6]} \textbf{Claimy} -- \$1.8M for missing royalty recovery -- \href{https://www.musicbusinessworldwide.com/ai-music-tech-startup-claimy-raises-1-8m-to-tackle-missing-royalty-payments/}{musicbusinessworldwide.com}
  \item \textbf{[Issue 7]} \textbf{Breaking Rust on Billboard} -- AI country act -- \href{https://www.npr.org/2025/11/10/nx-s1-5604320/breaking-rust-is-a-hot-new-country-act-on-the-billboard-charts-its-powered-by-ai}{npr.org}
  \item \textbf{[Issue 7]} \textbf{GEMA v OpenAI Munich ruling} -- European copyright precedent -- \href{https://www.linkedin.com/posts/dr-barry-scannell-bbb5aa207_in-a-major-ruling-for-european-copyright-share-7393957246386323457-8bbx}{linkedin.com}
  \item \textbf{[Issue 8]} \textbf{LANDR AI Music Study} -- 87\% of musicians use AI tools -- \href{https://aristake.com/ai-tools-musicians-study/}{aristake.com}
  \item \textbf{[Issue 8]} \textbf{Stability AI + Warner Music} -- next-gen responsible AI tools -- \href{https://stability.ai/news/warner-music-group-and-stability-ai-join-forces-to-build-next-gen-tools}{stability.ai}
  \item \textbf{[Issue 8]} \textbf{Paul McCartney Silent Track Protest} -- UK copyright opt-out -- \href{https://www.theguardian.com/music/2025/nov/17/the-sound-of-silence-why-theres-barely-anything-there-in-paul-mccartney-new-release}{theguardian.com}
  \item \textbf{[Issue 11]} \textbf{Lyria Camera (Google DeepMind)} -- music generation -- \href{https://magenta.withgoogle.com/lyria-camera-announce}{magenta.withgoogle.com}
  \item \textbf{[Issue 11]} \textbf{Dave Stewart on AI} -- musicians must embrace it -- \href{https://www.theguardian.com/music/2025/dec/05/musicians-must-embrace-unstoppable-force-of-ai-eurythmics-dave-stewart-urges}{theguardian.com}
  \item \textbf{[Issue 12]} \textbf{Splice + UMG Collaboration} -- AI music creation tools -- \href{https://www.universalmusic.com/universal-music-group-and-splice-to-collaborate-on-the-next-generation-of-ai-powered-music-creation-tools-for-artists/}{universalmusic.com}
  \item \textbf{[Issue 14]} \textbf{Bandcamp Bans AI Music} -- platform policy -- \href{https://stereogum.com/2485199/bandcamp-bans-ai-music/news}{stereogum.com}
  \item \textbf{[Issue 14]} \textbf{UMG slams AI slop} -- exponential growth on streaming -- \href{https://musically.com/2026/01/09/umg-boss-slams-exponential-growth-of-ai-slop-on-streaming-services/}{musically.com}
  \item \textbf{[Issue 15]} \textbf{Sienna Rose} -- viral mystery AI singer (BBC investigation) -- \href{https://www.bbc.co.uk/news/articles/cq6v83gq66eo}{bbc.co.uk}
  \item \textbf{[Issue 16]} \textbf{800 Creatives Sign Declaration} -- \emph{Stealing Our Work Is Not Innovation} -- \href{https://www.digitalmusicnews.com/2026/01/22/stealing-isnt-innovation/}{digitalmusicnews.com}
\end{itemize}

\section*{4. AI 3D / World Models / Spatial}

\begin{itemize}
  \item \textbf{[Issue 1]} \textbf{Marble by World Labs} -- first commercial world model -- \href{https://marble.worldlabs.ai/}{marble.worldlabs.ai}
  \item \textbf{[Issue 1]} \textbf{Meta Hyperscape Capture} -- Gaussian splatting on Quest -- \href{https://www.meta.com/en-gb/experiences/meta-horizon-hyperscape-capture-beta/8798130056953686/}{meta.com}
  \item \textbf{[Issue 1]} \textbf{PlayCanvas SOG} -- WebP for 3D Gaussian Splatting -- \href{https://www.linkedin.com/posts/willeastcott_playcanvas-open-sources-sog-literally-webp-ugcPost-7374459362708180992-aHDa}{linkedin.com}
  \item \textbf{[Issue 3]} \textbf{Genie 3} (Google DeepMind) -- interactive 3D world generation -- \href{https://deepmind.google/blog/genie-3-a-new-frontier-for-world-models/}{deepmind.google}
  \item \textbf{[Issue 3]} \textbf{World Labs RTFM} -- real-time frame model -- \href{https://www.worldlabs.ai/blog/rtfm}{worldlabs.ai}
  \item \textbf{[Issue 3]} \textbf{Instant Skinned Gaussian Avatars} -- web/mobile VR -- \href{https://sites.google.com/view/gaussian-vrm}{sites.google.com}
  \item \textbf{[Issue 3]} \textbf{Tencent Hunyuan World 1.1} -- 3D reconstruction -- \href{https://x.com/TencentHunyuan/status/1980930623536837013}{x.com}
  \item \textbf{[Issue 5]} \textbf{Apple Personas use Gaussian Splatting} -- most-deployed splat tech in consumer hardware -- \href{https://radiancefields.com/apple-confirms-personas-use-gaussian-splatting}{radiancefields.com}
  \item \textbf{[Issue 7]} \textbf{Marble formal public launch} -- Sony Pictures using it (40$\times$ faster than legacy VP) -- \href{https://www.worldlabs.ai/case-studies/bringing-marble-to-life}{worldlabs.ai}
  \item \textbf{[Issue 8]} \textbf{Meta SAM 3 / SAM 3D} -- segment anything in 3D -- \href{https://www.linkedin.com/posts/aiatmeta_introducing-sam-3-sam-3d-ugcPost-7396944913751465985-m5Nc}{linkedin.com}
  \item \textbf{[Issue 11]} \textbf{Meta WorldGen} -- text-to-immersive-3D-worlds research -- \href{https://www.facebook.com/LifeAtMeta/videos/research-update-worldgen-text-to-immersive-3d-worlds/1879077432692421/}{facebook.com}
  \item \textbf{[Issue 11]} \textbf{Ubisoft CHORD Model} -- open-sourced PBR material generation -- \href{https://blog.comfy.org/p/ubisoft-open-sources-the-chord-model}{blog.comfy.org}
  \item \textbf{[Issue 12]} \textbf{Tencent HY World 1.5} -- real-time world model framework -- \href{https://x.com/TencentHunyuan/status/2001170499133653006}{x.com}
  \item \textbf{[Issue 12]} \textbf{Microsoft Trellis 2} -- 3D generation -- \href{https://github.com/microsoft/TRELLIS.2}{github.com}
  \item \textbf{[Issue 13]} \textbf{Wonderzoom} (Stanford AI Lab) -- multi-scale 3D world generation -- \href{https://wonderzoom.github.io/}{wonderzoom.github.io}
  \item \textbf{[Issue 15]} \textbf{WorldLabs API} -- 3D world generation as a service -- \href{https://x.com/theworldlabs/status/2014046372639408203}{x.com}
  \item \textbf{[Issue 17]} \textbf{Project Genie Public Release} -- Google AI Ultra subscribers -- \href{https://blog.google/innovation-and-ai/models-and-research/google-deepmind/project-genie/}{blog.google}
  \item \textbf{[Issue 22]} \textbf{Luma UNI-1} -- combined world generation + reasoning -- \emph{Dream Machine} Editor's Pick
  \item \textbf{[Issue 25]} \textbf{Spark 2.0} -- open-source 100M-splat browser streaming
  \item \textbf{[Issue 27]} \textbf{Vista4D} (Netflix + Eyeline) -- live action to navigable 4D point clouds
\end{itemize}

\section*{5. Voice, Avatars, Digital Humans}

\begin{itemize}
  \item \textbf{[Issue 1]} \textbf{Tilly Norwood AI Actress} -- sparks UK/US union debate -- \href{https://www.hollywoodreporter.com/movies/movie-news/tilly-norwood-ai-actress-uk-union-equity-sag-aftra-debate-1236391739/}{hollywoodreporter.com}
  \item \textbf{[Issue 1]} \textbf{SAG-AFTRA Condemns Tilly Norwood} -- official union statement -- \href{https://variety.com/2025/film/news/sag-aftra-tilly-norwood-ai-actress-1236534779/}{variety.com}
  \item \textbf{[Issue 1]} \textbf{Deep Fusion Films + Topfoto} -- AI documentary alliance -- \href{https://www.ibc.org/production/news/deep-fusion-and-topfoto-strike-alliance-to-produce-ai-powered-documentaries/22728}{ibc.org}
  \item \textbf{[Issue 2]} \textbf{ROXi AI TV Presenters} -- music channel hosts -- \href{https://www.advanced-television.com/2025/10/07/roxi-debuts-ai-generated-tv-presenters/}{advanced-television.com}
  \item \textbf{[Issue 2]} \textbf{Hedra Audio Tags} -- emotional audio control -- \href{https://x.com/hedra_labs/status/1998490844748460528}{x.com}
  \item \textbf{[Issue 3]} \textbf{Heygen Motion Designer} -- prompt-based animation -- \href{https://www.linkedin.com/posts/heygen_if-you-can-explain-it-you-can-animate-it-activity-7387488068204892160-AY8-}{linkedin.com}
  \item \textbf{[Issue 4]} \textbf{Copresence and ConvAI} -- Unreal-Engine intelligent avatars -- \href{https://www.linkedin.com/posts/copresence-tech_want-to-create-an-intelligent-avatar-that-activity-7379523290421383168-SAg_}{linkedin.com}
  \item \textbf{[Issue 7]} \textbf{Lumi Avatar} -- real-time audio-driven avatars -- \href{https://www.linkedin.com/feed/update/urn:li:activity:7393916271018467328/}{linkedin.com}
  \item \textbf{[Issue 7]} \textbf{Tilly Creator Eline Van der Velden -- Deadline} -- backlash and the next 40 -- \href{https://deadline.com/2025/11/tilly-norwood-creator-interview-backlash-more-ai-actors-coming-1236601334/}{deadline.com}
  \item \textbf{[Issue 8]} \textbf{Synthesia \$4B valuation} -- rejected \$3B Adobe offer -- \href{https://techcrunch.com/2026/01/26/synthesia-hits-4b-valuation-lets-employees-cash-in/}{techcrunch.com}
  \item \textbf{[Issue 11]} \textbf{ElevenLabs Impact Programme} -- SXSW documentary on voice loss -- \href{https://www.linkedin.com/posts/elevenlabsio_at-sxsw-the-elevenlabs-impact-program-will-activity-7413979472988774400-fVRG}{linkedin.com}
  \item \textbf{[Issue 12]} \textbf{Google Veo Avatars} -- expressiveness upgrade -- \href{https://vids.new/}{vids.new}
  \item \textbf{[Issue 12]} \textbf{Meta SAM Audio} -- segment anything for audio -- \href{https://about.fb.com/news/2025/12/our-new-sam-audio-model-transforms-audio-editing/}{about.fb.com}
  \item \textbf{[Issue 13]} \textbf{Avatar Forcing} -- real-time interactive avatar generation -- \href{https://taekyungki.github.io/AvatarForcing/}{taekyungki.github.io}
  \item \textbf{[Issue 13]} \textbf{Qwen3 TTS} -- voice design and cloning -- \href{https://qwen.ai/blog?id=qwen3-tts-vc-voicedesign}{qwen.ai}
  \item \textbf{[Issue 16]} \textbf{Tilly Norwood Doubles Down} -- AI as ``more ethical'' performance, urging actors to create avatars -- \href{https://variety.com/2026/digital/news/tilly-norwood-creator-tells-actors-to-create-ai-avatars-1236638940/}{variety.com}
  \item \textbf{[Issue 23]} \textbf{Death threats against Eline Van der Velden} -- cultural-extreme response
\end{itemize}

\section*{6. Agent Platforms / Orchestration}

\begin{itemize}
  \item \textbf{[Issue 2]} \textbf{OpenAI AgentKit / DevDay} -- agentic AI for creative workflows -- \href{https://openai.com/index/introducing-agentkit/}{openai.com}
  \item \textbf{[Issue 2]} \textbf{Lenny AI Agent} -- for live music event organisers -- \href{https://musically.com/2025/10/20/meet-lenny-an-ai-agent-to-help-organisers-of-live-music-events/}{musically.com}
  \item \textbf{[Issue 4]} \textbf{AdsGency \$12M seed} -- autonomous paid marketing -- \href{https://www.finsmes.com/2025/10/adsgency-raises-12m-in-seed-funding.html}{finsmes.com}
  \item \textbf{[Issue 5]} \textbf{Meta + Hugging Face OpenEnv} -- open-source agentic development -- \href{https://www.edtechinnovationhub.com/news/meta-and-hugging-face-launch-openenv-to-advance-open-source-agentic-development}{edtechinnovationhub.com}
  \item \textbf{[Issue 5]} \textbf{Pomelli} (Google Labs) -- AI marketing agent for SMBs
  \item \textbf{[Issue 5]} \textbf{Opal} (Google) -- no-code AI mini-app builder -- \href{https://blog.google/technology/google-labs/opal-expansion/}{blog.google}
  \item \textbf{[Issue 8]} \textbf{Multimodal Agents in Unreal Engine} -- live-score nature documentary -- \href{https://www.youtube.com/watch?v=7u2yCtbONmo}{youtube.com}
  \item \textbf{[Issue 8]} \textbf{SIMA 2} (Google DeepMind) -- agent for virtual 3D worlds -- \href{https://deepmind.google/blog/sima-2-an-agent-that-plays-reasons-and-learns-with-you-in-virtual-3d-worlds/}{deepmind.google}
  \item \textbf{[Issue 8]} \textbf{Google Antigravity} -- agentic development platform
  \item \textbf{[Issue 13]} \textbf{NitroGen} (NVIDIA + Stanford) -- plays-any-game model -- \href{https://nitrogen.minedojo.org/}{nitrogen.minedojo.org}
  \item \textbf{[Issue 14]} \textbf{General Intuition} -- \$134M for spatial-reasoning agents -- \href{https://techcrunch.com/2025/10/16/general-intuition-lands-134m-seed-to-teach-agents-spatial-reasoning-using-video-game-clips/}{techcrunch.com}
  \item \textbf{[Issue 16]} \textbf{Anthropic Claude Apps} -- interactive Claude in workplace tools -- \href{https://techcrunch.com/2026/01/26/anthropic-launches-interactive-claude-apps-including-slack-and-other-workplace-tools/}{techcrunch.com}
  \item \textbf{[Issue 16]} \textbf{Heygen Video Agent} -- full scripting-to-assembly -- \href{https://www.linkedin.com/posts/heygen_introducing-the-new-video-agent-activity-7421597801240801282-d1CF}{linkedin.com}
  \item \textbf{[Issue 21]} \textbf{Adobe + NVIDIA Strategic Partnership} -- agentic creative intelligence
  \item \textbf{[Issue 26]} \textbf{Adobe Summit 2026} -- ``agentic creative intelligence'' headline category
  \item \textbf{[Issue 29]} \textbf{Sony 49-Claude-agent / 72-skill stack} -- game-dev multi-agent team
\end{itemize}

\section*{7. Adobe and Creative Software}

\begin{itemize}
  \item \textbf{[Issue 1]} \textbf{Tether} -- AI animation in After Effects -- \href{https://www.linkedin.com/posts/thisisdoug_aftereffects-aivideo-vfx-ugcPost-7368671859774517249-l0sz}{linkedin.com}
  \item \textbf{[Issue 1]} \textbf{Unreal Engine 5 AI Assistant} -- official integration -- \href{https://www.linkedin.com/posts/wouterweynants_theres-an-official-ai-assistant-coming-to-ugcPost-7369377204226379776-pGiH}{linkedin.com}
  \item \textbf{[Issue 1]} \textbf{Unity AI Council} -- accelerate AI innovation -- \href{https://www.gamedeveloper.com/business/unity-forms-ai-council-to-accelerate-ai-product-innovation-}{gamedeveloper.com}
  \item \textbf{[Issue 1]} \textbf{ComfyUI raises \$17M} -- OS for creative AI -- \href{https://www.linkedin.com/posts/comfyui_we-raised-17-million-to-build-an-os-for-ugcPost-7373743341236236288-wkCc}{linkedin.com}
  \item \textbf{[Issue 5]} \textbf{Adobe MAX 2025 Express AI Assistant} -- Adobe MAX announcements -- \href{https://news.adobe.com/news/2025/10/adobe-max-2025-express-ai-assistant}{news.adobe.com}
  \item \textbf{[Issue 5]} \textbf{Adobe Firefly Foundry} -- custom models for brands -- \href{https://news.adobe.com/news/2025/10/adobe-max-2025-firefly-foundry}{news.adobe.com}
  \item \textbf{[Issue 5]} \textbf{Adobe MAX Sneaks} -- Light Touch, Surface Swap, Scene It, etc.\ -- \href{https://www.youtube.com/watch?v=YqAAFX1XXY8}{youtube.com}
  \item \textbf{[Issue 5]} \textbf{Adobe MAX Creator Survey} -- 86\% use creative gen AI -- \href{https://news.adobe.com/news/2025/10/adobe-max-2025-creators-survey}{news.adobe.com}
  \item \textbf{[Issue 5]} \textbf{``Adobe is putting AI in everything everywhere all at once''} -- Creative Boom coverage -- \href{https://www.creativeboom.com/news/adobe-is-putting-ai-in-everything-everywhere-all-at-once/}{creativeboom.com}
  \item \textbf{[Issue 6]} \textbf{Adobe Corrective AI} -- voice-over emotion editing -- \href{https://www.wired.com/story/adobe-max-sneaks-2025-corrective-ai/}{wired.com}
  \item \textbf{[Issue 8]} \textbf{Adobe Research RELIC} -- interactive video world model
  \item \textbf{[Issue 12]} \textbf{Adobe inside ChatGPT} -- Photoshop, Express, Acrobat editing in ChatGPT -- \href{https://blog.adobe.com/en/publish/2025/12/10/edit-photoshop-chatgpt}{blog.adobe.com}
  \item \textbf{[Issue 12]} \textbf{Google Flow Refine} -- prompt by doodling -- \href{https://blog.google/technology/google-labs/flow-refine-videos/}{blog.google}
  \item \textbf{[Issue 15]} \textbf{Adobe at Sundance 2026} -- \$10M grants, Ignite Day -- \href{https://analyticsindiamag.com/ai-news-updates/adobe-unveils-ai-video-innovations-10-million-grants-ahead-of-sundance-film-festival/}{analyticsindiamag.com}
  \item \textbf{[Issue 16]} \textbf{Adobe Premiere Object Mask} -- automated masking -- \href{https://www.linkedin.com/posts/robdewinter_ok-this-is-going-to-save-a-lot-of-time-in-ugcPost-7421617551690063872-yKmB}{linkedin.com}
  \item \textbf{[Issue 28]} \textbf{Unity AI Open Beta} -- in-editor full AI suite
\end{itemize}

\section*{8. Games Industry -- Adoption}

\begin{itemize}
  \item \textbf{[Issue 1]} \textbf{Netflix Director of GenAI for Games} -- \$840K role -- \href{https://www.pcgamer.com/gaming-industry/as-the-videogame-industry-continues-to-be-hammered-by-layoffs-netflix-is-offering-up-to-usd840-000-per-year-for-a-new-director-of-generative-ai-for-games/}{pcgamer.com}
  \item \textbf{[Issue 1]} \textbf{Meta Horizon Studio AI Assistant Upgrade} -- \href{https://www.uploadvr.com/meta-horizon-studio-upgrade-ai-assistant-horizon-worlds/}{uploadvr.com}
  \item \textbf{[Issue 1]} \textbf{51\% of Japanese games studios use AI} -- research finding -- \href{https://www.gamesindustry.biz/51-of-japanese-game-makers-use-generative-ai}{gamesindustry.biz}
  \item \textbf{[Issue 3]} \textbf{Roblox AI Tools for Creators} -- \href{https://www.gamesindustry.biz/roblox-announces-new-ai-tools-for-creators}{gamesindustry.biz}
  \item \textbf{[Issue 3]} \textbf{Battlefield 6} -- ``very seducing'' AI for early stages -- \href{https://www.gamesradar.com/games/battlefield/battlefield-6-lead-calls-generative-ai-very-seducing-but-says-it-was-only-used-in-the-games-earliest-stages-to-allow-for-more-time-and-more-space-to-be-creative/}{gamesradar.com}
  \item \textbf{[Issue 3]} \textbf{Promise} (Google-backed AI studio) -- VFX for legacy media -- \href{https://www.hollywoodreporter.com/business/digital/ai-studio-promise-vfx-generation-company-1236397636/}{hollywoodreporter.com}
  \item \textbf{[Issue 4]} \textbf{EA + Stability AI Partnership} -- generative AI tools for games -- \href{https://stability.ai/news/stability-ai-and-ea-partner-to-reimagine-game-development}{stability.ai}
  \item \textbf{[Issue 4]} \textbf{Halo Studios} -- AI woven into development -- \href{https://thegamepost.com/insider-halo-studios-generative-ai-game-development/}{thegamepost.com}
  \item \textbf{[Issue 5]} \textbf{NBCUniversal $\times$ Dick Wolf Jr} -- AI games deal -- \href{https://www.videogameschronicle.com/news/nbcuniversal-signs-deal-with-law-order-creator-dick-wolfs-son-to-make-ai-generated-games-based-on-its-ip/}{videogameschronicle.com}
  \item \textbf{[Issue 5]} \textbf{Microsoft Gaming Copilot} -- screenshots for in-game understanding -- \href{https://www.tomshardware.com/video-games/pc-gaming/microsoft-says-gaming-copilot-uses-screenshots-to-understand-in-game-events-not-for-training-ai-models-optional-feature-can-be-turned-off-but-not-easily-uninstalled}{tomshardware.com}
  \item \textbf{[Issue 5]} \textbf{Sony Jabali AI Platform} -- game development -- \href{https://variety.com/2025/gaming/news/sony-jabali-ai-ai-game-development-platform-1236566619/}{variety.com}
  \item \textbf{[Issue 5]} \textbf{Krafton AI-First Plans} -- Subnautica owner -- \href{https://www.gamedeveloper.com/business/subnautica-owner-krafton-outlines-plans-to-transform-into-an-ai-first-company}{gamedeveloper.com}
  \item \textbf{[Issue 6]} \textbf{Todd Howard on AI at Bethesda} -- toolset, not replacement -- \href{https://www.pcgamer.com/gaming-industry/todd-howard-says-ai-cant-replace-human-creative-intention-but-its-part-of-bethesdas-toolset-for-how-we-build-our-worlds-or-check-things/}{pcgamer.com}
  \item \textbf{[Issue 6]} \textbf{EA pushes 15K employees on AI} -- thought partner -- \href{https://www.gamesradar.com/games/even-under-usd20-million-in-debt-ea-reportedly-pushes-15-000-employees-to-use-ai-as-a-thought-partner-for-everything-from-character-art-to-playtesting/}{gamesradar.com}
  \item \textbf{[Issue 6]} \textbf{Falcom AI} -- 2--3 hours of work to 10 minutes -- \href{https://www.eurogamer.net/falcom-is-the-latest-developer-to-buy-into-the-ai-hype-machine}{eurogamer.net}
  \item \textbf{[Issue 7]} \textbf{Square Enix 70\% AI QA target} -- by end 2027 -- \href{https://www.pcgamer.com/gaming-industry/square-enix-aims-to-have-ai-doing-70-percent-of-its-qa-work-by-the-end-of-2027/}{pcgamer.com}
  \item \textbf{[Issue 8]} \textbf{Call of Duty: Black Ops 7 AI art accusations} -- community backlash -- \href{https://www.videogameschronicle.com/news/it-honestly-sucks-fans-think-call-of-duty-black-ops-7-is-filled-with-generative-ai-art/}{videogameschronicle.com}
  \item \textbf{[Issue 8]} \textbf{Ubisoft Anno 117 -- AI art placeholder} -- slipped through review -- \href{https://www.videogameschronicle.com/news/ubisoft-says-ai-generated-art-in-anno-117-was-a-placeholder-which-slipped-through-our-review-process/}{videogameschronicle.com}
  \item \textbf{[Issue 13]} \textbf{Razer \$600M AI focus} -- strategic investment -- \href{https://www.pymnts.com/news/artificial-intelligence/2026/razer-spends-600-million-dollars-sharpen-focus-ai-gaming/}{pymnts.com}
  \item \textbf{[Issue 15]} \textbf{Ubisoft cancels Prince of Persia + four} -- AI refocus -- \href{https://metro.co.uk/2026/01/21/prince-persia-remake-five-games-cancelled-ubisoft-focuses-ai-26431926/}{metro.co.uk}
  \item \textbf{[Issue 29]} \textbf{Sony all-in on AI for games} -- 49-agent / 72-skill stack
\end{itemize}

\section*{9. Games Industry -- Refusal / Position}

\begin{itemize}
  \item \textbf{[Issue 1]} \textbf{Charles Cecil} -- ``AI was an expensive mistake'' -- \href{https://www.gamesindustry.biz/ai-was-an-expensive-mistake-charles-cecil-on-innovation-insolvency-and-broken-sword}{gamesindustry.biz}
  \item \textbf{[Issue 4]} \textbf{Pocketpair (Palworld)} -- publishing won't take AI games -- \href{https://www.pcgamer.com/software/ai/palworld-studio-pocketpair-says-its-new-publishing-division-wont-handle-games-that-use-generative-ai-we-dont-believe-in-it/}{pcgamer.com}
  \item \textbf{[Issue 9]} \textbf{Witcher 3 / Cyberpunk Director} -- AI helps, doesn't replace -- \href{https://www.gamesindustry.biz/witcher-3-and-cyberpunk-2077-director-says-ai-can-help-but-not-replace-creatives}{gamesindustry.biz}
  \item \textbf{[Issue 11]} \textbf{Aardman on AI} -- embrace, but cautious -- \href{https://www.gamesradar.com/entertainment/animation-movies/wallace-and-gromit-creator-says-beloved-animation-studio-aardman-will-embrace-the-technology-of-ai-but-will-be-very-cautious-not-to-lose-our-values/}{gamesradar.com}
  \item \textbf{[Issue 14]} \textbf{Larian backs off gen AI} -- Divinity statement -- \href{https://nichegamer.com/larian-studios-backs-off-from-gen-ai/}{nichegamer.com}
  \item \textbf{[Issue 14]} \textbf{Games Workshop rules out gen AI} -- Warhammer 40K -- \href{https://decrypt.co/354482/warhammer-40000-maker-games-workshop-rules-out-generative-ai}{decrypt.co}
  \item \textbf{[Issue 14]} \textbf{Hooded Horse won't work with AI devs} -- Manor Lords publisher -- \href{https://nichegamer.com/manor-lords-publisher-hooded-horse-wont-work-with-devs-using-gen-ai/}{nichegamer.com}
  \item \textbf{[Issue 16]} \textbf{Jagex never AI} -- RuneScape commitment -- \href{https://www.gamesindustry.biz/runescape-maker-jagex-says-it-will-never-use-generative-ai-to-make-in-game-content}{gamesindustry.biz}
\end{itemize}

\section*{10. Film Industry -- Studios and Positions}

\begin{itemize}
  \item \textbf{[Issue 1]} \textbf{Lionsgate AI failure} -- Futurism report -- \href{https://futurism.com/artificial-intelligence/lionsgate-movies-ai}{futurism.com}
  \item \textbf{[Issue 2]} \textbf{Fremantle's Imaginae AI Studios} -- CEO named -- \href{https://www.hollywoodreporter.com/business/digital/fremantle-names-ceo-new-ai-label-imaginae-studios-1236396579/}{hollywoodreporter.com}
  \item \textbf{[Issue 2]} \textbf{Goldfinch enGEN3} -- AI cinematic universe -- \href{https://variety.com/2025/film/news/ai-powered-cinematic-universe-platform-engen3-1236543349/}{variety.com}
  \item \textbf{[Issue 3]} \textbf{Fox Entertainment + Holywater} -- AI microdramas -- \href{https://www.hollywoodreporter.com/business/business-news/fox-entertainment-invests-in-holywater-ai-microdramas-1236396802/}{hollywoodreporter.com}
  \item \textbf{[Issue 4]} \textbf{Netflix ``all in'' on AI} -- Sarandos at industry conference -- \href{https://www.cnbc.com/2025/10/22/netflix-all-in-on-leveraging-ai-in-its-streaming-platform.html}{cnbc.com}
  \item \textbf{[Issue 4]} \textbf{Asteria's ``All Heart''} -- Natasha Lyonne short -- \href{https://www.hollywoodreporter.com/movies/movie-news/natasha-lyonne-ai-company-asteria-1236403144/}{hollywoodreporter.com}
  \item \textbf{[Issue 5]} \textbf{Wonder Studios \$9M raise} -- AI-native studio -- \href{https://www.uktech.news/ai/ai-film-studio-wonder-lands-9m-investment-20251023}{uktech.news}
  \item \textbf{[Issue 5]} \textbf{Watch the Skies} (Swedish AI feature dubbed) -- USA distribution -- \href{https://variety.com/2025/film/news/watch-the-skies-us-theatrical-release-ai-dubbing-1236343110/}{variety.com}
  \item \textbf{[Issue 5]} \textbf{Run to the West} (South Korean first AI feature) -- \href{https://cybernews.com/entertainment/korean-cinema-run-to-the-west-ai/}{cybernews.com}
  \item \textbf{[Issue 6]} \textbf{Obsidian + Imagine Entertainment} -- Ron Howard, Brian Grazer -- \href{https://www.indiewire.com/news/business/obsidian-studio-ai-production-company-imagine-entertainment-1235158619/}{indiewire.com}
  \item \textbf{[Issue 7]} \textbf{Beta Films Chapter41 launch} -- Munich AI startup -- \href{https://deadline.com/2025/11/beta-film-ai-startup-chapter41-artificial-intelligence-1236612632/}{deadline.com}
  \item \textbf{[Issue 7]} \textbf{House of David 350+ AI shots} -- Wired feature -- \href{https://www.wired.com/story/amazons-house-of-david-used-over-350-ai-shots-in-season-2-its-creator-isnt-sorry/}{wired.com}
  \item \textbf{[Issue 7]} \textbf{Wanted director's AI Method Actors} -- Bekmambetov \$5M -- \href{https://variety.com/2025/film/news/wanted-director-method-acting-ai-actors-1236579647/}{variety.com}
  \item \textbf{[Issue 7]} \textbf{Kevin Reilly + Kartel} -- HBO veteran's AI startup -- \href{https://www.hollywoodreporter.com/business/digital/kevin-reilly-ceo-kartel-ai-hbo-1236424692/}{hollywoodreporter.com}
  \item \textbf{[Issue 8]} \textbf{Humans in the Loop -- Oscar race} -- Sloan grant -- \href{https://variety.com/2025/film/news/ai-drama-humans-in-the-loop-oscar-race-1236582975/}{variety.com}
  \item \textbf{[Issue 8]} \textbf{Synthetic Sincerity -- IDFA} -- Marc Isaacs -- \href{https://www.hollywoodreporter.com/movies/movie-news/synthetic-sincerity-film-idfa-ai-authenticity-interview-1236426180/}{hollywoodreporter.com}
  \item \textbf{[Issue 8]} \textbf{AI Images Threaten Documentary} -- Variety -- \href{https://variety.com/2025/film/festivals/ai-generated-images-threaten-future-of-documentary-1236583466/}{variety.com}
  \item \textbf{[Issue 11]} \textbf{Disney \$1bn OpenAI investment} -- Sora characters -- \href{https://www.theguardian.com/business/2025/dec/11/disney-open-ai-sora-video-deal}{theguardian.com}
  \item \textbf{[Issue 14]} \textbf{Tunisian filmmaker wins \$1M for Lily} -- Dubai AI Award -- \href{https://www.broadcastprome.com/news/tunisian-filmmaker-wins-1-million-ai-film-award-for-lily/}{broadcastprome.com}
  \item \textbf{[Issue 14]} \textbf{Disney TikTok-like vertical video AI} -- brand-asset video gen -- \href{https://www.marketingdive.com/news/disney-unveils-tiktok-like-vertical-video-ai-video-generation-tool/809269/}{marketingdive.com}
  \item \textbf{[Issue 15]} \textbf{Netflix retention AI strategy} -- Pymnts -- \href{https://www.pymnts.com/subscription-commerce/2026/retention-is-name-of-the-game-for-netflixs-ai-strategy/}{pymnts.com}
  \item \textbf{[Issue 16]} \textbf{Andrii Daniels bomb-shelter Christmas clip} -- viral Ukrainian AI film -- \href{https://variety.com/2026/digital/news/ai-video-deadpool-harry-potter-andrii-daniels-1236624632/}{variety.com}
  \item \textbf{[Issue 16]} \textbf{Chris Pratt rejects AI villain} -- Mercy pitch -- \href{https://variety.com/2026/film/news/chris-pratt-ai-actor-villain-mercy-amazon-mgm-1236640460/}{variety.com}
\end{itemize}

\section*{11. Celebrity / Director Positions on AI}

\begin{itemize}
  \item \textbf{[Issue 1]} \textbf{James Cameron -- ``never going to take place''} -- No Film School -- \href{https://nofilmschool.com/james-cameron-ai}{nofilmschool.com}
  \item \textbf{[Issue 1]} \textbf{Taylor Swift -- criticised for AI in promo} -- \href{https://tribune.com.pk/story/2570725/taylor-swift-criticised-for-using-ai-in-the-life-of-a-showgirl-promotional-campaign}{tribune.com.pk}
  \item \textbf{[Issue 5]} \textbf{Guillermo del Toro -- ``Rather Die''} -- Variety -- \href{https://variety.com/2025/film/news/guillermo-del-toro-rather-die-generative-ai-frankenstein-1236561316/}{variety.com}
  \item \textbf{[Issue 5]} \textbf{Paul Schrader on AI} -- perfect script -- \href{https://www.hollywoodreporter.com/movies/movie-news/paul-schrader-first-ai-movie-1236409606/}{hollywoodreporter.com}
  \item \textbf{[Issue 7]} \textbf{Jeremy Renner lawsuit threat} -- multi-millions over AI voice -- \href{https://deadline.com/2025/11/jeremy-renner-lawsuit-threat-ai-movie-1236611830/}{deadline.com}
  \item \textbf{[Issue 7]} \textbf{George Clooney on AI actors} -- Variety column -- \href{https://variety.com/2025/scene/columns/george-clooney-ai-actors-movie-stars-1236579661/}{variety.com}
  \item \textbf{[Issue 10]} \textbf{James Cameron ``horrifying''} -- The Guardian -- \href{https://www.theguardian.com/film/2025/dec/01/james-cameron-says-ai-actors-are-horrifying-to-me}{theguardian.com}
  \item \textbf{[Issue 10]} \textbf{Jenna Ortega ``very easy to be terrified''} -- NME -- \href{https://www.nme.com/news/jenna-ortega-says-its-very-easy-to-be-terrified-of-ai-in-filmmaking-3913926}{nme.com}
  \item \textbf{[Issue 11]} \textbf{Leonardo DiCaprio -- AI can't be art} -- THR -- \href{https://www.hollywoodreporter.com/movies/movie-news/leonardo-dicaprio-ai-cant-be-art-no-humanity-1236445405/}{hollywoodreporter.com}
  \item \textbf{[Issue 11]} \textbf{James Cameron rejects AI actors at Hainan Festival} -- \href{https://variety.com/2025/film/news/james-cameron-rejects-ai-actors-hainan-wouldnt-do-it-1236604204/}{variety.com}
  \item \textbf{[Issue 14]} \textbf{Claire Foy no interest in AI films} -- Daily Mail -- \href{https://www.dailymail.co.uk/tvshowbiz/article-15454199/Claire-Foy-AI-films-sad-disappointed-people-future-Hollywood.html}{dailymail.co.uk}
  \item \textbf{[Issue 14]} \textbf{Wu-Tang Clan RZA} -- case for AI in film/music -- \href{https://www.vice.com/en/article/wu-tang-clans-rza-makes-the-case-for-ai-in-film-and-music-an-amazing-thing-for-us/}{vice.com}
  \item \textbf{[Issue 15]} \textbf{Matthew McConaughey protects voice/image} -- Lawyer Monthly -- \href{https://www.lawyer-monthly.com/2026/01/matthew-mcconaughey-protects-voice-image-ai/}{lawyer-monthly.com}
  \item \textbf{[Issue 15]} \textbf{Mara Wilson deepfake apocalypse fear} -- Deadline -- \href{https://deadline.com/2026/01/matilda-mara-wilson-stranger-things-ai-deepfake-apocalypse-1236689474/}{deadline.com}
\end{itemize}

\section*{12. Music Industry -- Labels, Deals, and Lawsuits}

\begin{itemize}
  \item \textbf{[Issue 1]} \textbf{Universal \& Warner -- landmark AI deals within weeks} -- Musically -- \href{https://musically.com/2025/10/02/report-umg-and-wmg-could-sign-landmark-ai-deals-within-weeks/}{musically.com}
  \item \textbf{[Issue 5]} \textbf{Universal Music + Stability AI Alliance} -- \emph{op.\ cit.\ (Section~3)}
  \item \textbf{[Issue 5]} \textbf{Udio + UMG Partnership} -- \emph{op.\ cit.\ (Section~3)}
  \item \textbf{[Issue 7]} \textbf{GEMA v OpenAI Munich ruling} -- \emph{op.\ cit.\ (Section~3)}
  \item \textbf{[Issue 7]} \textbf{``Biggest theft in music history''} -- Rights group sues Suno -- \href{https://edm.com/gear-tech/rights-group-sues-suno-copyright-infringement/}{edm.com}
  \item \textbf{[Issue 7]} \textbf{Bangkok Post: Xania Monet \$3M deal} -- \href{https://www.bangkokpost.com/life/tech/3142355/ai-singer-xania-monet-signs-3m-deal-with-hallwood-media}{bangkokpost.com}
  \item \textbf{[Issue 8]} \textbf{Warner Music + Stability AI} -- next-gen tools -- \href{https://stability.ai/news/warner-music-group-and-stability-ai-join-forces-to-build-next-gen-tools}{stability.ai}
  \item \textbf{[Issue 9]} \textbf{Johnny Cash estate sues Coca-Cola} -- ELVIS Act -- \href{https://completemusicupdate.com/johnny-cash-estate-uses-elvis-act-to-sue-coke-over-tribute-act-ad-soundtrack/}{completemusicupdate.com}
  \item \textbf{[Issue 12]} \textbf{Splice + UMG Collaboration} -- \emph{op.\ cit.\ (Section~3)}
  \item \textbf{[Issue 14]} \textbf{UMG slams AI slop} -- \emph{op.\ cit.\ (Section~3)}
  \item \textbf{[Issue 16]} \textbf{Wixen \$50M lawsuit against Meta} -- \href{https://www.musicbusinessworldwide.com/wixen-files-50m-copyright-suit-against-meta-claims-tech-giant-wants-to-replace-songwriters-with-ai/}{musicbusinessworldwide.com}
  \item \textbf{[Issue 17]} \textbf{UMG \$3B suit against Anthropic} -- \emph{Dream Machine} coverage
\end{itemize}

\section*{13. Copyright, Policy and Regulation}

\begin{itemize}
  \item \textbf{[Issue 9]} \textbf{EU Lawmakers minimum age for AI/social} -- Reuters -- \href{https://www.reuters.com/legal/litigation/european-lawmakers-seek-eu-wide-minimum-age-access-ai-chatbots-social-media-2025-11-26/}{reuters.com}
  \item \textbf{[Issue 12]} \textbf{UK DSIT Statement of Progress on Copyright and AI} -- the 88\% consultation -- \href{https://www.gov.uk/government/publications/copyright-and-artificial-intelligence-progress-report/copyright-and-artificial-intelligence-statement-of-progress-under-section-137-data-use-and-access-act}{gov.uk}
  \item \textbf{[Issue 12]} \textbf{IPWatchdog UK consultation analysis} -- \href{https://ipwatchdog.com/2025/12/16/respondents-uk-ai-consultation-overwhelmingly-want-ai-companies-license-copyrighted-works-all-cases/}{ipwatchdog.com}
  \item \textbf{[Issue 11]} \textbf{NY AI Advertising Disclosure Law} -- Verge -- \href{https://www.theverge.com/news/842848/new-york-law-ai-advertisements-sag-aftra-labor}{theverge.com}
  \item \textbf{[Issue 14]} \textbf{Grok app ban consideration} -- Fast Company -- \href{https://www.fastcompany.com/91474131/governments-around-the-world-are-considering-bans-on-groks-app-over-ai-sexual-image-scandal}{fastcompany.com}
  \item \textbf{[Issue 21]} \textbf{UK DSIT Final Copyright Report} -- walked-back position
  \item \textbf{[Issue 28]} \textbf{Academy ``You must be human to win'' rule} -- 2026 awards
  \item \textbf{[Issue 29]} \textbf{Cannes AI Disclosure Standard} -- industry coordination
\end{itemize}

\section*{14. Unions and Labour}

\begin{itemize}
  \item \textbf{[Issue 1]} \textbf{SAG-AFTRA Condemns Tilly Norwood} -- \emph{op.\ cit.\ (Section~5)}
  \item \textbf{[Issue 1]} \textbf{UK Equity Statement} -- Hollywood Reporter -- \href{https://www.hollywoodreporter.com/movies/movie-news/tilly-norwood-ai-actress-uk-union-equity-sag-aftra-debate-1236391739/}{hollywoodreporter.com}
  \item \textbf{[Issue 11]} \textbf{Equity 99\% Strike Vote} -- landslide for industrial action -- \href{https://www.equity.org.uk/news/2025/performers-prepared-to-take-industrial-action-over-ai-in-landslide-99-vote}{equity.org.uk}
  \item \textbf{[Issue 11]} \textbf{NY AI Advertising Disclosure Law / SAG-AFTRA quote} -- \emph{op.\ cit.\ (Section~13)}
  \item \textbf{[Issue 15]} \textbf{Equity welcomes improved offer} -- film/TV AI protections -- \href{https://www.equity.org.uk/news/2026/equity-welcomes-improved-offer-in-ai-protection-negotiations-in-film-and-tv}{equity.org.uk}
  \item \textbf{[Issue 26/29]} \textbf{SAG-AFTRA ``Tilly Tax'' contract provisions} -- final spring 2026 contract
\end{itemize}

\section*{15. Audience Response and Slop Ceiling}

\begin{itemize}
  \item \textbf{[Issue 7]} \textbf{Deezer/Ipsos AI Music Survey} -- 97\% can't tell, but care when told -- \href{https://newsroom-deezer.com/2025/11/deezer-ipsos-survey-ai-music/}{newsroom-deezer.com}
  \item \textbf{[Issue 7]} \textbf{50,000 AI tracks uploaded to Deezer daily} -- Musically -- \href{https://musically.com/2025/11/12/50000-ai-music-tracks-are-now-uploaded-to-deezer-every-day/}{musically.com}
  \item \textbf{[Issue 8]} \textbf{MrBeast on AI} -- threat to creators -- \href{https://www.forbes.com/sites/johnbbrandon/2025/10/10/mrbeast-is-right-about-ai-content-but-are-we-really-in-scary-times/}{forbes.com}
  \item \textbf{[Issue 12]} \textbf{Merriam-Webster Word of the Year: ``Slop''} -- Hollywood Reporter -- \href{https://www.hollywoodreporter.com/news/general-news/slop-word-year-2025-merriam-webster-1236450780/}{hollywoodreporter.com}
  \item \textbf{[Issue 12]} \textbf{YouTube AI channels -- 1.2bn views fake politics} -- Guardian -- \href{https://www.theguardian.com/technology/2025/dec/13/fake-anti-labour-video-billion-views-youtube-2025}{theguardian.com}
  \item \textbf{[Issue 14]} \textbf{Spotify AI flood -- subscribers furious} -- TechRadar -- \href{https://www.techradar.com/audio/spotify/ai-music-is-flooding-spotify-and-subscribers-are-furious-heres-why-music-fans-no-longer-trust-discover-weekly}{techradar.com}
  \item \textbf{[Issue 14]} \textbf{Soultracks: ``AI music is catchy, familiar\ldots\ and boring''} -- \href{https://soultracks.com/news-ai-generated-music-is-catchy-boring/}{soultracks.com}
  \item \textbf{[Issue 15]} \textbf{Sweden bans AI from official chart} -- Independent -- \href{https://www.independent.co.uk/tv/news/ai-music-song-banned-sweden-spotify-b2901627.html}{independent.co.uk}
  \item \textbf{[Issue 16]} \textbf{YouTube CEO: managing AI slop on priority list 2026} -- Digital Music News -- \href{https://www.digitalmusicnews.com/2026/01/22/youtube-ceo-ai-slop-2026-comments/}{digitalmusicnews.com}
  \item \textbf{[Issue 16]} \textbf{Bain \& Co -- ``People still want the radio star''} -- \href{https://www.bain.com/insights/in-an-ai-age-people-still-want-the-radio-star/}{bain.com}
  \item \textbf{[Issues 25--28]} \textbf{Deezer April 2026 data -- 44\% / 3\%} -- newsroom release -- \href{https://newsroom-deezer.com/2026/04/ai-generated-tracks-represent-44-of-new-uploaded-music/}{newsroom-deezer.com}
\end{itemize}

\section*{16. Provenance, Watermarking, Detection}

\begin{itemize}
  \item \textbf{[Issue 2]} \textbf{OpenAI likeness protections} -- Digital Music News -- \href{https://www.digitalmusicnews.com/2025/10/08/openais-likeness-protections-dont-apply-to-dead-celebrities/}{digitalmusicnews.com}
  \item \textbf{[Issue 11]} \textbf{SynthID rollout across Veo / Lyria / Imagen} -- Google DeepMind
  \item \textbf{[Issue 12]} \textbf{Gemini ``Is this AI?'' video verification} -- \href{https://www.linkedin.com/posts/googledeepmind_verify-google-ai-generated-videos-in-the-activity-7407748300688478208-fJgW}{linkedin.com}
  \item \textbf{[Issue 13]} \textbf{Instagram chief -- ``fingerprint real media''} -- Digital Music News -- \href{https://www.digitalmusicnews.com/2026/01/05/instagram-chief-ai-slop-comments/}{digitalmusicnews.com}
  \item \textbf{[Issue 13]} \textbf{Instagram head AI verification} -- WebProNews -- \href{https://www.webpronews.com/instagram-head-warns-ai-images-erode-trust-calls-for-verification-standards/}{webpronews.com}
  \item \textbf{[Issue 18]} \textbf{Deezer licences its AI-music detection tool} -- \emph{Dream Machine} coverage
  \item \textbf{[Issue 29]} \textbf{YouTube false-positive: Tiny Grandma stop-motion} -- flagged as AI
\end{itemize}

\section*{17. Workplace AI Adoption / Shadow AI / Workforce Research}

\begin{itemize}
  \item \textbf{[Issue 1]} \textbf{MIT study -- AI reduces brain activity} -- AI News -- \href{https://www.artificialintelligence-news.com/news/ai-causes-reduction-in-users-brain-activity-mit/}{artificialintelligence-news.com}
  \item \textbf{[Issue 1]} \textbf{51\% of Japanese games studios use AI} -- \emph{op.\ cit.\ (Section~8)}
  \item \textbf{[Issue 1]} \textbf{Yale on AI adoption} -- Neil Hoyne -- \href{https://www.linkedin.com/posts/neilhoyne_ai-data-research-activity-7379272781798035456-hnuV}{linkedin.com}
  \item \textbf{[Issue 5]} \textbf{Azumo AI in Workplace Statistics 2025} -- \href{https://azumo.com/artificial-intelligence/ai-insights/ai-in-workplace-statistics}{azumo.com}
  \item \textbf{[Issue 5]} \textbf{Tech.co -- Gen Z most likely use AI behind boss's back} -- \href{https://tech.co/news/gen-z-most-likely-use-ai-boss}{tech.co}
  \item \textbf{[Issue 5]} \textbf{IDC Europe Shadow AI security nightmare} -- \href{https://blog-idceurope.com/shadow-ai-how-stealth-productivity-is-strangling-enterprise-ai-adoption-and-creating-a-security-nightmare/}{blog-idceurope.com}
  \item \textbf{[Issue 5]} \textbf{Forbes -- AI Tools Flood Workplaces} -- \href{https://www.forbes.com/sites/carolinecastrillon/2025/09/09/ai-tools-flood-workplaces-as-employees-face-a-double-bind/}{forbes.com}
  \item \textbf{[Issue 5]} \textbf{Exploding Topics AI Workforce Research} -- \href{https://explodingtopics.com/blog/ai-workforce-research}{explodingtopics.com}
  \item \textbf{[Issue 6]} \textbf{Adobe Creators' Toolkit Report (16,000 creators)} -- \href{https://news.adobe.com/news/2025/10/adobe-max-2025-creators-survey}{news.adobe.com}
  \item \textbf{[Issue 7]} \textbf{CNBC -- ADHD, autism, dyslexia and AI agents} -- \href{https://www.cnbc.com/2025/11/08/adhd-autism-dyslexia-jobs-careers-ai-agents-success.html}{cnbc.com}
  \item \textbf{[Issue 8]} \textbf{Forbes Vibe Coding \$220K} -- \href{https://www.forbes.com/sites/rachelwells/2025/11/06/the-in-demand-ai-skill-and-certifications-that-pays-up-to-220000/}{forbes.com}
  \item \textbf{[Issue 8]} \textbf{LANDR -- 87\% of musicians use AI tools} -- \href{https://aristake.com/ai-tools-musicians-study/}{aristake.com}
  \item \textbf{[Issue 9]} \textbf{Economist -- Investors expect AI use to soar (it isn't)} -- \href{https://www.economist.com/finance-and-economics/2025/11/26/investors-expect-ai-use-to-soar-thats-not-happening}{economist.com}
  \item \textbf{[Issue 9]} \textbf{Reuters Institute UK Journalists AI Survey} -- \href{https://reutersinstitute.politics.ox.ac.uk/ai-adoption-uk-journalists-and-their-newsrooms-surveying-applications-approaches-and-attitudes}{reutersinstitute.politics.ox.ac.uk}
  \item \textbf{[Issue 12]} \textbf{Economist -- Job apocalypse? Humbug!} -- \href{https://www.economist.com/business/2025/12/14/job-apocalypse-humbug-ai-is-creating-brand-new-occupations}{economist.com}
  \item \textbf{[Issue 16]} \textbf{Guardian -- AI is hitting UK harder} -- \href{https://www.theguardian.com/technology/2026/jan/26/ai-uk-jobs-us-japan-germany-australia}{theguardian.com}
  \item \textbf{[Issue 16]} \textbf{McKinsey AI for film and TV} -- \href{https://www.mckinsey.com/industries/technology-media-and-telecommunications/our-insights/what-ai-could-mean-for-film-and-tv-production-and-the-industrys-future}{mckinsey.com}
  \item \textbf{[Issue 16]} \textbf{PRS for Music AI Survey 2026} -- \href{https://www.prsformusic.com/m-magazine/news/prs-for-music-ai-survey-2026}{prsformusic.com}
  \item \textbf{[Issue 24]} \textbf{OpenAI public-policy on disruption} -- robot tax, 4-day workweek, wealth funds
\end{itemize}

\section*{18. Advertising and Brand AI}

\begin{itemize}
  \item \textbf{[Issue 3]} \textbf{WPP \$400M Google partnership} -- \href{https://martechseries.com/predictive-ai/ai-platforms-machine-learning/google-and-spotify-alum-launch-epiminds-with-6-6m-to-build-marketing-teams-for-the-ai-era/}{martechseries.com}
  \item \textbf{[Issue 5]} \textbf{WPP Open Pro launch} -- \href{https://campaignbrief.com/wpp-launches-ai-powered-marketing-platform-wpp-open-pro/}{campaignbrief.com}
  \item \textbf{[Issue 5]} \textbf{Mondelez AI for TV ads} -- Verge -- \href{https://www.theverge.com/news/806047/mondelez-ai-generated-ads}{theverge.com}
  \item \textbf{[Issue 6]} \textbf{WPP + Sightly partnership} -- Digiday -- \href{https://digiday.com/media-buying/agencies-continue-to-expand-ai-capabilities-to-boost-brand-performance/}{digiday.com}
  \item \textbf{[Issue 6]} \textbf{Coca-Cola AI Holiday ad (2nd attempt)} -- Adweek -- \href{https://www.adweek.com/creativity/coca-cola-uses-ai-to-rekindle-the-magic-of-its-holiday-ads/}{adweek.com}
  \item \textbf{[Issue 7]} \textbf{Digiday -- AI agent developers in-demand role} -- \href{https://digiday.com/marketing/ai-agent-developers-have-become-adlands-in-demand-role/}{digiday.com}
  \item \textbf{[Issue 10]} \textbf{Valentino ``disturbing'' AI handbag ads} -- BBC -- \href{https://www.bbc.co.uk/news/articles/cwyvjyvn83go}{bbc.co.uk}
  \item \textbf{[Issue 11]} \textbf{McDonald's NL Christmas AI Ad pulled} -- Branding in Asia -- \href{https://www.brandinginasia.com/its-the-most-terrible-time-of-the-year-mcdonalds-netherlands-wonderfully-chaotic-ai-driven-christmas-film/}{brandinginasia.com}
  \item \textbf{[Issue 11]} \textbf{Channel 4 AI Ads} -- Estate Agent Today -- \href{https://www.estateagenttoday.co.uk/breaking-news/2025/12/homebuilder-among-first-to-use-channel-4s-ai-ads/}{estateagenttoday.co.uk}
  \item \textbf{[Issue 14]} \textbf{Marketing Week: AI ads winning in testing} -- \href{https://www.marketingweek.com/dismiss-ai-ads-winning-creative-effectiveness/}{marketingweek.com}
  \item \textbf{[Issue 15]} \textbf{PGA Tour + AWS expanded partnership} -- Pymnts -- \href{https://www.pymnts.com/artificial-intelligence-2/2026/ai-content-is-par-for-the-course-with-pga-tours-expanded-aws-partnership/}{pymnts.com}
  \item \textbf{[Issue 15]} \textbf{Avocados from Mexico skip TV for AI} -- Digiday -- \href{https://digiday.com/marketing/avocados-from-mexico-turns-to-ai-to-advertise-around-the-super-bowl-instead-of-a-tv-buy/}{digiday.com}
  \item \textbf{[Issue 16]} \textbf{Higgsfield + Madonna AI video} -- Adweek -- \href{https://www.adweek.com/media/higgsfield-ai-marketing-startup/}{adweek.com}
  \item \textbf{[Issue 27]} \textbf{WPP + Google Earth AI consumer journey} -- \emph{Dream Machine} coverage
\end{itemize}

\section*{19. Festivals, Institutions, Awards}

\begin{itemize}
  \item \textbf{[Issue 6]} \textbf{AI FilmFest Japan / Hoyt Dwyer} -- \href{https://www.prnewswire.com/news-releases/from-apple-tv-creative-to-ai-filmmaker-hoyt-dwyers-animated-film-to-compete-at-ai-filmfest-japan-2025-302598064.html}{prnewswire.com}
  \item \textbf{[Issue 6]} \textbf{India's first AI Film Festival} -- IFFI, NFDC, LTIMindtree -- \href{https://www.medianews4u.com/iffi-partners-with-ltimindtree-and-nfdc-to-launch-indias-first-ai-film-festival-and-hackathon/}{medianews4u.com}
  \item \textbf{[Issue 14]} \textbf{Tunisian filmmaker wins \$1M for Lily} -- \emph{op.\ cit.\ (Section~10)}
  \item \textbf{[Issue 14]} \textbf{Comic-Con Art Show allows AI} -- Filmstories -- \href{https://filmstories.co.uk/news/san-diego-comic-con-art-show-to-allow-ai-slop/}{filmstories.co.uk}
  \item \textbf{[Issue 14]} \textbf{Emmys AI Guidance} -- THR -- \href{https://www.hollywoodreporter.com/tv/tv-news/emmys-ai-guidelines-2026-awards-1236468434/}{hollywoodreporter.com}
  \item \textbf{[Issue 15]} \textbf{Sundance AI Literacy Initiative} -- Sundance Institute blog -- \href{https://www.sundance.org/blogs/centering-the-artist-why-were-launching-the-ai-literacy-initiative/}{sundance.org}
  \item \textbf{[Issue 15]} \textbf{Google \$2M Sundance AI Education} -- \href{https://blog.google/company-news/outreach-and-initiatives/google-org/sundance-institute-ai-education/}{blog.google}
  \item \textbf{[Issue 16]} \textbf{CNET -- San Diego Comic-Con bans AI art at 2026 event} -- \href{https://www.cnet.com/culture/san-diego-comic-con-bans-ai-art-for-2026-event/}{cnet.com}
  \item \textbf{[Issue 16]} \textbf{Adobe Sundance Film Festival 2026} -- \href{https://blog.adobe.com/en/publish/2026/01/20/sundance-film-festival-2026-creativity-community-power-storytelling}{blog.adobe.com}
  \item \textbf{[Issue 28]} \textbf{Academy ``human to win'' rule} -- \emph{Dream Machine} coverage
  \item \textbf{[Issue 29]} \textbf{Cannes AI Disclosure Standard launched} -- \emph{Dream Machine} coverage
\end{itemize}

\section*{20. Geographic / Regional AI Production}

\begin{itemize}
  \item \textbf{[Issue 5]} \textbf{CNBC Africa on AI in African music} -- \href{https://www.cnbcafrica.com/2025/how-ai-is-changing-the-landscape-of-the-music-industry-in-africa}{cnbcafrica.com}
  \item \textbf{[Issue 5]} \textbf{Run to the West -- South Korea's first AI feature} -- \emph{op.\ cit.\ (Section~10)}
  \item \textbf{[Issue 5]} \textbf{Watch the Skies -- Swedish AI dubbing} -- \emph{op.\ cit.\ (Section~10)}
  \item \textbf{[Issue 6]} \textbf{India's first AI Film Festival} -- \emph{op.\ cit.\ (Section~19)}
  \item \textbf{[Issue 12]} \textbf{Trilok -- Indian AI band} -- Musically -- \href{https://musically.com/2025/12/17/indian-ai-band-trilok-performs-live-government-denies-association/}{musically.com}
  \item \textbf{[Issue 13]} \textbf{56.9\% of new Chinese independent songs are AI} -- Musically -- \href{https://musically.com/2026/01/05/report-56-9-of-new-independent-songs-in-china-are-ai-generated/}{musically.com}
  \item \textbf{[Issue 14]} \textbf{BBC Future -- Lights, camera, algorithm (India)} -- \href{https://www.bbc.co.uk/future/article/20251223-why-indian-cinema-is-awash-with-ai}{bbc.co.uk}
  \item \textbf{[Issue 14]} \textbf{Shift Up CEO on AI vs China/US scale} -- \href{https://www.pocketgamer.biz/shift-up-ceo-says-ai-is-key-to-competing-with-chinas-game-industry-scale/}{pocketgamer.biz}
  \item \textbf{[Issue 25]} \textbf{Indonesia's Legenda Bertuah} -- first AI-animated series
  \item \textbf{[Issue 27]} \textbf{Korin AI -- Africa-trained, Africa-built} -- launch
  \item \textbf{[Issue 27]} \textbf{Latin American AI film festival wave} -- \emph{Dream Machine} coverage
\end{itemize}

\section*{21. Education, Training, Literacy}

\begin{itemize}
  \item \textbf{[Issue 1]} \textbf{UCL, RCA, Brandtech Centre for Creative AI launch} -- Broadcast Now -- \href{https://www.broadcastnow.co.uk/broadcasters/alex-mahon-joins-stellar-ai-creative-summit-line-up/5209227.article}{broadcastnow.co.uk}
  \item \textbf{[Issue 11]} \textbf{Lovable for classrooms} -- \href{https://lovable.dev/classroom}{lovable.dev}
  \item \textbf{[Issue 15]} \textbf{UW-Stout AI baseline competency in filmmaking} -- \emph{op.\ cit.}
  \item \textbf{[Issue 15]} \textbf{Google + Sundance Institute AI Education} -- \emph{op.\ cit.\ (Section~19)}
  \item \textbf{[Issue 16]} \textbf{UK government ``Free AI training for all''} -- \href{https://www.gov.uk/government/news/free-ai-training-for-all-as-government-and-industry-programme-expands-to-provide-10-million-workers-with-key-ai-skills-by-2030}{gov.uk}
  \item \textbf{[Issue 16]} \textbf{Adobe at Sundance: Ignite Day} -- \emph{op.\ cit.\ (Section~7)}
\end{itemize}

\section*{22. Open Source and Infrastructure}

\begin{itemize}
  \item \textbf{[Issue 1]} \textbf{ComfyUI raises \$17M} -- \emph{op.\ cit.\ (Section~7)}
  \item \textbf{[Issue 1]} \textbf{PlayCanvas SOG} -- \emph{op.\ cit.\ (Section~4)}
  \item \textbf{[Issue 1]} \textbf{DecartAI -- open-source real-time world transformation} -- \href{https://decart.ai/}{decart.ai}
  \item \textbf{[Issue 1]} \textbf{Civitai / Replicate} -- open infrastructure layer
  \item \textbf{[Issue 4]} \textbf{Krea Realtime open-sourced} -- \emph{op.\ cit.\ (Section~1)}
  \item \textbf{[Issue 5]} \textbf{Meta + Hugging Face OpenEnv} -- \emph{op.\ cit.\ (Section~6)}
  \item \textbf{[Issue 8]} \textbf{Hugging Face + Google Cloud partnership} -- \href{https://www.linkedin.com/posts/julienchaumond_i-am-super-excited-to-announce-that-hugging-activity-7396177403972276225-CuMM}{linkedin.com}
  \item \textbf{[Issue 8]} \textbf{80\% of A16Z pitches use Chinese open-source models} -- \href{https://www.linkedin.com/posts/stevenouri_a-wild-stat-80-of-startups-pitching-a16z-activity-7396182718998351872-xTKR}{linkedin.com}
  \item \textbf{[Issue 11]} \textbf{Ubisoft CHORD open-sourced} -- \emph{op.\ cit.\ (Section~4)}
  \item \textbf{[Issue 27]} \textbf{ComfyUI \$500M valuation} -- \emph{Dream Machine} coverage
  \item \textbf{[Issue 27]} \textbf{Anthropic + Blender Foundation patronage} -- \emph{Dream Machine} coverage
  \item \textbf{[Issue 27]} \textbf{Korin AI launch} -- \emph{op.\ cit.\ (Section~20)}
\end{itemize}

\section*{23. Web Infrastructure / Bots / Dead Internet}

\begin{itemize}
  \item \textbf{[Issue 4]} \textbf{Imperva 2025 Bad Bot Report} -- bots = 51\% of web traffic -- \href{https://www.imperva.com/blog/2025-imperva-bad-bot-report-how-ai-is-supercharging-the-bot-threat/}{imperva.com}
  \item \textbf{[Issue 4]} \textbf{Cloudflare crawl-to-click gap} -- \href{https://blog.cloudflare.com/crawlers-click-ai-bots-training/}{blog.cloudflare.com}
  \item \textbf{[Issue 4]} \textbf{Dead Internet Theory} -- Wikipedia -- \href{https://en.wikipedia.org/wiki/Dead_Internet_theory}{en.wikipedia.org}
  \item \textbf{[Issue 4]} \textbf{Grand View Research GenAI Content Creation Market} -- \href{https://www.grandviewresearch.com/industry-analysis/generative-ai-content-creation-market-report}{grandviewresearch.com}
  \item \textbf{[Issue 4]} \textbf{Futurism -- AI-only social network collapses into warring tribes} -- \href{https://futurism.com/social-network-ai-intervention-echo-chamber}{futurism.com}
\end{itemize}

\section*{24. AI Bubble / Economics / Investment}

\begin{itemize}
  \item \textbf{[Issue 1]} \textbf{AI bubble 17$\times$ dotcom} -- PC Gamer -- \href{https://www.pcgamer.com/software/ai/fabulous-news-everyone-market-analyst-says-the-ai-bubble-is-17x-bigger-than-the-dotcom-goldrush-and-4x-larger-than-the-subprime-bubble-that-caused-the-2008-crash/}{pcgamer.com}
  \item \textbf{[Issue 2]} \textbf{Suno \$2.45B valuation} -- \emph{op.\ cit.\ (Section~3)}
  \item \textbf{[Issue 4]} \textbf{AdsGency \$12M seed} -- \emph{op.\ cit.\ (Section~6)}
  \item \textbf{[Issue 5]} \textbf{Sifted -- Synthesia rejects \$3B Adobe} -- \href{https://sifted.eu/articles/synthesia-acquisition-offer}{sifted.eu}
  \item \textbf{[Issue 14]} \textbf{Kartel / Reilly leadership} -- \emph{op.\ cit.\ (Section~10)}
  \item \textbf{[Issue 15]} \textbf{Higgsfield \$80M raise at \$1.3B} -- \href{https://siliconangle.com/2026/01/15/higgsfield-raises-80m-1-3b-valuation-scale-ai-video-platform/}{siliconangle.com}
  \item \textbf{[Issue 16]} \textbf{Higgsfield earns \$200M in 9 months} -- \href{https://eu.36kr.com/en/p/3650517574312323}{eu.36kr.com}
  \item \textbf{[Issue 16]} \textbf{Synthesia hits \$4B valuation} -- \emph{op.\ cit.\ (Section~5)}
  \item \textbf{[Issue 21]} \textbf{Adobe + NVIDIA Strategic Partnership} -- \emph{op.\ cit.\ (Section~6)}
  \item \textbf{[Issue 25]} \textbf{ElevenLabs \$500M ARR} -- \emph{Dream Machine} coverage
  \item \textbf{[Issue 27]} \textbf{Google \$40B investment in Anthropic} -- \emph{Dream Machine} coverage
  \item \textbf{[Issue 27]} \textbf{ComfyUI \$500M valuation} -- \emph{op.\ cit.\ (Section~22)}
\end{itemize}

\bigskip\par\noindent\rule{\textwidth}{0.4pt}\par\bigskip

\section*{Using This Index}

This thematic index covers the significant sources across the \emph{Dream Machine} archive, organised by topic. For specific research, follow the bracketed Issue numbers back to the canonical issue file in \texttt{Dream Machine MD/}. Each issue file ends with the section ``All embedded URLs (in document order)'' which lists every URL the issue carried, including local navigation links, profile pages and platform housekeeping links that are not reproduced here.

The newsletter is a continuous publication. The index above reflects the state of the archive at the time of book publication (May 2026). Subsequent issues will extend the catalogue. The newsletter archive itself, on LinkedIn, remains the canonical primary source for every link the book builds on.

For deeper analytical treatment of the data this index points to, see the deep-dive appendices:
\begin{itemize}
  \item Appendix D: The Shadow AI Paradox
  \item Appendix E: Dynamics of Generative AI Adoption
  \item Appendix F: AI, Stigma, Privilege, Democratisation
  \item Appendix G: The Age of Intent
\end{itemize}

\backmatter
  \chapter*{Citation Index}
\addcontentsline{toc}{chapter}{Citation Index}

\textit{Dream Machine: The New Creative Economy}

All footnoted sources, organised by chapter. Every claim of substance in the manuscript is anchored to one of these references.

\section*{Foreword -- Welcome to the Dream Machine}

\begin{enumerate}

\item Variety, ``SAG-AFTRA Condemns Tilly Norwood: AI Actress Is Not an Actor,'' 30 September 2025. \url{https://variety.com/2025/film/news/sag-aftra-tilly-norwood-ai-actress-1236534779/}. See also NBC News, ``Tilly Norwood, fully AI `actor,' blasted by actors union SAG-AFTRA for `devaluing human artistry'.'' \url{https://www.nbcnews.com/pop-culture/pop-culture-news/tilly-norwood-fully-ai-actor-blasted-actors-union-sag-aftra-devaluing-rcna234685}. Discussed in \href{../Dream\%20Machine\%20MD/1.md}{\textit{Dream Machine} Issue~1} (6 October 2025).

\item \textit{The Hollywood Reporter}, ``U.K. Union Equity Condemns Tilly Norwood: `AI Tool, Not a Performer'.'' \url{https://www.hollywoodreporter.com/movies/movie-news/tilly-norwood-ai-actress-uk-union-equity-sag-aftra-debate-1236391739/}. See also Variety, ``Tilly Norwood Slammed by Equity as AI Tool, Concerned About Origin.'' \url{https://variety.com/2025/film/global/tilly-norwood-slammed-equity-ai-tool-concerned-origin-1236537042/}. \href{../Dream\%20Machine\%20MD/1.md}{\textit{Dream Machine} Issue~1}.

\item CNN, ``Tilly Norwood: Hollywood is fuming over a new `AI actress,''' 30 September 2025. \url{https://www.cnn.com/2025/09/30/tech/hollywood-ai-actor-backlash}.

\item OpenAI, ``Sora 2 is here,'' announcement page, 30 September 2025. \url{https://openai.com/index/sora-2/}. The model launched alongside an invite-only iOS app of the same name in the U.S. and Canada. \href{../Dream\%20Machine\%20MD/1.md}{\textit{Dream Machine} Issue~1} carried the launch alongside contemporaneous coverage from NBC News and \textit{The Guardian} on the model's first copyright and safety incidents.

\item \textit{Dream Machine | Creative AI}, LinkedIn newsletter, archive of Issues \href{../Dream\%20Machine\%20MD/1.md}{1}--\href{../Dream\%20Machine\%20MD/29.md}{29}, October 2025 -- May 2026. \url{https://www.linkedin.com/newsletters/dream-machine-creative-ai-7379776527871381505/}.

\item DreamLab AI Collective, team page. \url{https://dreamlab-ai.com/team}. Referenced from \href{../Dream\%20Machine\%20MD/16.md}{\textit{Dream Machine} Issue~16} onward.

\item Charles Cecil (Revolution Software, \textit{Broken Sword}) quoted in \textit{gamesindustry.biz}, ```AI was an expensive mistake': Charles Cecil on innovation, insolvency, and Broken Sword.'' \url{https://www.gamesindustry.biz/ai-was-an-expensive-mistake-charles-cecil-on-innovation-insolvency-and-broken-sword}. \href{../Dream\%20Machine\%20MD/3.md}{\textit{Dream Machine} Issue~3}.

\item Adobe, ``Inaugural Adobe Creators' Toolkit Report: 86 Per Cent of Global Creators Use Creative Generative AI.'' \url{https://news.adobe.com/news/2025/10/adobe-max-2025-creators-survey}. Survey of 16,000 creators across the U.S., U.K., France, Germany, South Korea, Japan, India and Australia, released at Adobe MAX 2025. \href{../Dream\%20Machine\%20MD/6.md}{\textit{Dream Machine} Issue~6}.

\item UK Department for Science, Innovation and Technology (DSIT), \textit{Statement of Progress on Copyright and AI}, December 2025. \url{https://www.gov.uk/government/publications/copyright-and-artificial-intelligence-progress-report/copyright-and-artificial-intelligence-statement-of-progress-under-section-137-data-use-and-access-act}. See also IPWatchdog, ``Respondents to UK AI Consultation Overwhelmingly Want AI Companies to License Copyrighted Works in All Cases.'' \url{https://ipwatchdog.com/2025/12/16/respondents-uk-ai-consultation-overwhelmingly-want-ai-companies-license-copyrighted-works-all-cases/}. \href{../Dream\%20Machine\%20MD/12.md}{\textit{Dream Machine} Issue~12}.

\item \href{../Dream\%20Machine\%20MD/5.md}{\textit{Dream Machine} Issue~5}, ``Adobe's Latest AI Announcements -- Is every tool going AI?'', 31 October 2025. \url{https://www.linkedin.com/pulse/dream-machine-creative-ai-news-insight-oct-25-issue-5-woodbridge-f7jnc/}.

\item Adobe, Adobe MAX 2025 keynote messaging, October 2025. Coverage: Creative Boom, ``Adobe is putting AI in everything everywhere all at once.'' \url{https://www.creativeboom.com/news/adobe-is-putting-ai-in-everything-everywhere-all-at-once/}. \href{../Dream\%20Machine\%20MD/5.md}{\textit{Dream Machine} Issue~5}.

\item World Labs, \textit{Marble} -- first commercial spatial-AI world model, public launch November 2025. \url{https://marble.worldlabs.ai/}. Technical context: TechCrunch, ``Fei-Fei Li's World Labs speeds up the world model race with Marble, its first commercial product.'' \url{https://techcrunch.com/2025/11/12/fei-fei-lis-world-labs-speeds-up-the-world-model-race-with-marble-its-first-commercial-product/}. DreamLab participated in the closed beta during October--November 2025. \href{../Dream\%20Machine\%20MD/7.md}{\textit{Dream Machine} Issue~7}.

\item 11,514 responses across the Citizen Space portal and email, of which 10,112 came through Citizen Space; 88\% of those supported licensing as a default rule, against 3\% who supported the government's preferred opt-out model. UK DSIT, \textit{Statement of Progress}, December 2025; analysis in \href{../Dream\%20Machine\%20MD/12.md}{\textit{Dream Machine} Issue~12} (18 December 2025). Final report and economic impact assessment to be laid before Parliament by 18 March 2026.

\item \textit{Digital Music News}, ``Nearly 800 Creatives, Including Jason Aldean and One Republic, Sign Responsible AI Declaration -- `Stealing Our Work Is Not Innovation'.'' \url{https://www.digitalmusicnews.com/2026/01/22/stealing-isnt-innovation/}. \href{../Dream\%20Machine\%20MD/16.md}{\textit{Dream Machine} Issue~16}.

\end{enumerate}

\section*{Chapter 1 -- The Day Sora Landed}

\begin{enumerate}

\item For a contemporaneous overview of the AI video model release cadence through 2024 and 2025, see \textit{Dream Machine} Issues \href{../Dream\%20Machine\%20MD/1.md}{1}--\href{../Dream\%20Machine\%20MD/8.md}{8} (October--November 2025), which logged near-weekly releases from Runway, Luma, Pika, Kling, Veo, Wan, Higgsfield, Hunyuan and a long tail of smaller labs.

\item \textit{The Hollywood Reporter}, ``AI Performer Tilly Norwood Sparks Hollywood Backlash.'' \url{https://www.hollywoodreporter.com/movies/movie-news/tilly-norwood-ai-actress-uk-union-equity-sag-aftra-debate-1236391739/}. \href{../Dream\%20Machine\%20MD/1.md}{\textit{Dream Machine} Issue~1}.

\item SAG-AFTRA statement, 30 September 2025, reported in Variety, ``SAG-AFTRA Condemns Tilly Norwood: AI Actress Is Not an Actor.'' \url{https://variety.com/2025/film/news/sag-aftra-tilly-norwood-ai-actress-1236534779/}.

\item OpenAI, ``Sora 2 is here,'' 30 September 2025. \url{https://openai.com/index/sora-2/}. \href{../Dream\%20Machine\%20MD/1.md}{\textit{Dream Machine} Issue~1}.

\item Particle6 background and Van der Velden interview: \textit{The Hollywood Reporter}, ``Meet the Creator of the AI Actress Hollywood Loves to Hate: `You're Gonna See a Lot of Tilly Norwood Next Year'.'' \url{https://www.hollywoodreporter.com/movies/movie-features/tilly-norwood-creator-particle6-eline-van-der-velden-talks-1236428824/}. \href{../Dream\%20Machine\%20MD/8.md}{\textit{Dream Machine} Issue~8}.

\item \textit{Deadline}, ``Tilly Norwood Creator Eline Van Der Velden Talks Backlash, Reveals Another 40 AI Actors Are In The Pipeline.'' \url{https://deadline.com/2025/11/tilly-norwood-creator-interview-backlash-more-ai-actors-coming-1236601334/}.

\item Northeastern Global News, ``Why AI `Actress' Tilly Norwood Has Hollywood Angry.'' \url{https://news.northeastern.edu/2025/10/02/ai-actress-tilly-norwood-hollywood-backlash/}.

\item SAG-AFTRA, official statement reproduced in Variety, \textit{op. cit.}; also NBC News, ``Tilly Norwood, fully AI `actor,' blasted by actors union SAG-AFTRA for `devaluing human artistry'.'' \url{https://www.nbcnews.com/pop-culture/pop-culture-news/tilly-norwood-fully-ai-actor-blasted-actors-union-sag-aftra-devaluing-rcna234685}.

\item Equity (U.K.), statement of 2 October 2025: \textit{Variety}, ``Tilly Norwood Slammed by Equity as AI Tool, Concerned About Origin.'' \url{https://variety.com/2025/film/global/tilly-norwood-slammed-equity-ai-tool-concerned-origin-1236537042/}.

\item CNN, ``Tilly Norwood: Hollywood is fuming over a new `AI actress'.'' \url{https://www.cnn.com/2025/09/30/tech/hollywood-ai-actor-backlash}.

\item OpenAI, ``Sora 2 is here,'' \url{https://openai.com/index/sora-2/}. Technical capabilities summary including physics modelling, multi-shot world-state persistence and synchronised audio.

\item \href{../Dream\%20Machine\%20MD/1.md}{\textit{Dream Machine} Issue~1}, ``Editor's Pick''; further launch context in NBC News, ``OpenAI's Sora 2: a major leap in AI video and audio.'' \url{https://www.nbcnews.com/tech/tech-news/openai-sora-2-app-video-chatgpt-creation-rcna234973}.

\item LinkedIn News aggregation: ``Sora Tops 1 Million Downloads in 5 Days.'' \url{https://www.linkedin.com/news/story/sora-tops-1m-downloads-in-5-days-6684988/}. \href{../Dream\%20Machine\%20MD/3.md}{\textit{Dream Machine} Issue~3}.

\item \textit{The Guardian}, ``OpenAI Sora 2 violence racism.'' \url{https://www.theguardian.com/us-news/2025/oct/04/openai-sora-violence-racism}. \href{../Dream\%20Machine\%20MD/1.md}{\textit{Dream Machine} Issue~1}.

\item NBC News, \textit{op. cit.}; \textit{The Guardian}, \textit{op. cit.}

\item[15a.] Quoted in \textit{The Guardian}, ``OpenAI launch of video app Sora plagued by violent and racist images: `The guardrails are not real'.'' \url{https://www.theguardian.com/us-news/2025/oct/04/openai-sora-violence-racism}. \href{../Dream\%20Machine\%20MD/1.md}{\textit{Dream Machine} Issue~1}.

\item \textit{Digital Music News}, ``OpenAI's Sora 2 includes likeness protections for celebrities who don't opt in, but that doesn't apply to `historical figures' and dead celebrities.'' \url{https://www.digitalmusicnews.com/2025/10/08/openais-likeness-protections-dont-apply-to-dead-celebrities/}. \href{../Dream\%20Machine\%20MD/2.md}{\textit{Dream Machine} Issue~2}.

\item Google DeepMind, Veo 3.1 launch, mid-October 2025. \href{../Dream\%20Machine\%20MD/3.md}{\textit{Dream Machine} Issue~3}, ``Editor's Pick: Veo 3.1 and the Rise of AI Filmmaking.'' Coverage: \url{https://www.cometapi.com/veo-3-1-is-comingand-whats-rumor/}.

\item WUFT, ``Kiss reality goodbye: AI-generated social media has arrived,'' 3 October 2025. \url{https://www.wuft.org/2025-10-03/kiss-reality-goodbye-ai-generated-social-media-has-arrived}. \href{../Dream\%20Machine\%20MD/1.md}{\textit{Dream Machine} Issue~1}.

\item \textit{No Film School}, ``James Cameron Says AI Is `Never Going to Take the Place' of Humans.'' \url{https://nofilmschool.com/james-cameron-ai\#}. \href{../Dream\%20Machine\%20MD/1.md}{\textit{Dream Machine} Issue~1}.

\item \textit{The Guardian}, ``James Cameron says AI actors are `horrifying to me,''' 1 December 2025. \url{https://www.theguardian.com/film/2025/dec/01/james-cameron-says-ai-actors-are-horrifying-to-me}. Original quote from CBS \textit{Sunday Morning}. \href{../Dream\%20Machine\%20MD/10.md}{\textit{Dream Machine} Issue~10}.

\item[20a.] Variety, ``James Cameron Says It's `Horrifying' that AI Can `Make Up an Actor'.'' \url{https://variety.com/2025/film/news/james-cameron-horrifying-ai-replace-actors-1236595864/}.

\item Stability AI, board composition, 2024--2026. Reported across multiple outlets including \textit{Deadline}, ``James Cameron Calls AI Replacing Actors `Horrifying'; Art `Sacred'.'' \url{https://deadline.com/2025/11/james-cameron-gen-ai-horrifying-human-art-sacred-avatar-1236631387/}.

\item Deezer, ``AI-generated tracks now represent 44\% of all new uploaded music,'' April 2026 newsroom release. \url{https://newsroom-deezer.com/2026/04/ai-generated-tracks-represent-44-of-new-uploaded-music/}. Companion analysis: \textit{Music Business Worldwide}, ``75,000 AI-generated tracks now flood Deezer daily, representing 44\% of all new music uploaded to the platform.'' \url{https://www.musicbusinessworldwide.com/75000-ai-generated-tracks-now-flood-deezer-daily-representing-44-of-all-new-music-uploaded-to-the-platform-says-streamer/}. Daily AI uploads to Deezer rose from approximately 50,000 per day in November 2025 (\href{../Dream\%20Machine\%20MD/7.md}{\textit{Dream Machine} Issue~7}, citing Deezer / \textit{Musically}) to 75,000 per day by April 2026, with consumer streams of fully-AI tracks holding between 1\% and 3\% of total platform plays -- and up to 85\% of those streams identified as fraudulent in 2025. \textit{Dream Machine} Issues \href{../Dream\%20Machine\%20MD/7.md}{7}, \href{../Dream\%20Machine\%20MD/26.md}{26}, \href{../Dream\%20Machine\%20MD/27.md}{27}, \href{../Dream\%20Machine\%20MD/28.md}{28}.

\end{enumerate}

\section*{Chapter 2 -- A History of Resistance}

\begin{enumerate}

\item John Philip Sousa, \textit{``The Menace of Mechanical Music,''} \textit{Appleton's Magazine}, Vol.~8, September 1906, pp.~278--284. Full text via ExplorePAHistory: \url{https://explorepahistory.com/odocument.php?docId=1-4-1A1}. Academic context: Patrick Warfield, \textit{``John Philip Sousa and `The Menace of Mechanical Music,'''} \textit{Journal of the Society for American Music}, Cambridge University Press: \url{https://www.cambridge.org/core/journals/journal-of-the-society-for-american-music/article/abs/john-philip-sousa-and-the-menace-of-mechanical-music/A9E621587BE7580ABD73AEF64D4B2DC8}. The 1906 essay was, in part, lobbying for what would become the 1909 Copyright Act.

\item Sousa, \textit{op. cit.} The Library of Congress's \textit{``Sousa and the Talking Machine''} essay is a useful institutional summary: \url{https://blogs.loc.gov/now-see-hear/2020/05/sousa-and-the-talking-machine/}.

\item William Henry Cardinal O'Connell, Archbishop of Boston, sermon to the Holy Name Society, Boston, 10 January 1932. Reported widely in the contemporaneous press, including the \textit{Daily Courier} (Connellsville, PA), 12 January 1932 (\url{https://www.newspapers.com/newspage/38168082/}). Cultural context: KUOW/NPR, \textit{```Imbecile Slush': Surprising Early Reactions to Crooning,''} \url{https://www.kuow.org/stories/imbecile-slush-surprising-early-reactions-crooning}. JSTOR Daily, \textit{``The Gender Politics of the First Boy Bands,''} \url{https://daily.jstor.org/the-gender-politics-of-the-first-boy-bands/}.

\item \textit{Grand Upright Music, Ltd. v. Warner Bros. Records Inc.}, 780 F.~Supp.~182 (S.D.N.Y. 1991). Full text: \url{https://law.justia.com/cases/federal/district-courts/FSupp/780/182/1445286/}. The ``Thou shalt not steal'' opening is the most-quoted line from a US copyright opinion of the late twentieth century.

\item Tippett's account of the Jurassic Park digital test is documented across multiple ASC and contemporaneous press accounts. American Society of Cinematographers, \textit{``Jurassic Park: Effects Team Brings Dinosaurs Back from Extinction,''} \url{https://theasc.com/articles/jurassic-park-effects-team-brings-dinosaurs-back}. Wikipedia, \textit{``Phil Tippett,''} \url{https://en.wikipedia.org/wiki/Phil_Tippett}. The dialogue paraphrase Spielberg incorporated into the film is Goldblum/Malcolm's response to Grant's ``I think we're out of a job'': ``Don't you mean \textit{extinct}?''

\item Charles Baudelaire, \textit{``Le Public Moderne et la Photographie,''} \textit{Revue Française}, 1859 (part of the \textit{Salon de 1859} essays). English translation widely available; the original French in PDF form: \url{https://gallowayexploringart.wordpress.com/wp-content/uploads/2014/08/baudelaire_the-modern-public-photography.pdf}. Smithsonian Archives institutional overview: \textit{``Photography Murdered Painting, Right?''}, \url{https://siarchives.si.edu/blog/photography-murdered-painting-right}.

\item The Delaroche apocrypha is documented in Quote Investigator: \url{https://quoteinvestigator.com/2022/10/16/photo-mortal/}. The earliest sourced version is in an 1873 survey, 34 years after Delaroche reportedly said it. Delaroche's own contemporary writing on the daguerreotype, in Gernsheim's standard 1959 monograph, characterised the new technology as \textit{``an immense service to the arts.''}

\item The 1942--44 Petrillo strike: Wikipedia, \textit{``1942--44 musicians' strike,''} \url{https://en.wikipedia.org/wiki/1942\%E2\%80\%931944_musicians'_strike}; Mainspring Press, \textit{``The Man Who Crippled the American Recording Industry: James Caesar Petrillo and the American Federation of Musicians Recording Bans,''} \url{https://mainspringpress.org/2024/11/23/the-man-who-crippled-the-recording-industry-james-caesar-petrillo-and-the-american-federation-of-musicians-recording-bans/}; DownBeat, \textit{``The Petrillo Ban of 1942--'44: Past \& Future at War,''} \url{https://downbeat.com/news/detail/the-petrillo-ban-of-194244-past-future-at-war}; Local 802 AFM, \textit{``The Silence Was Deafening,''} \url{https://www.local802afm.org/allegro/articles/the-silence-was-deafening/}. The Music Performance Trust Fund's institutional history: \url{https://musicpf.org/establishment-of-mptf-led-to-the-formation-of-afms-pension-and-residual-funds/}.

\item Musicians' Union History, \textit{``The Strike That Made History -- Massacre of the Musicians 1980,''} \url{https://www.muhistory.com/the-strike-that-made-history-massacre-of-the-musicians-1980/}. Academic context on the broader MU--BBC dispute landscape: \textit{``Negotiating Needletime''} (Tandfonline), \url{https://www.tandfonline.com/doi/full/10.1080/03071022.2016.1215098}.

\item MusicRadar, \textit{``The Day the Loony Musicians Union Tried to Kill the Synthesizer (Which Also Happened to be Bob Moog's Birthday),''} \url{https://www.musicradar.com/news/the-union-passed-a-motion-to-ban-the-use-of-synths-drum-machines-and-any-electronic-devices-the-day-the-loony-musicians-union-tried-to-kill-the-synthesizer-which-also-happened-to-be-bob-moogs-birthday}. Far Out Magazine, \textit{``Why did the Musicians Union outlaw synthesisers in 1982?''}, \url{https://faroutmagazine.co.uk/musicians-union-outlaw-synthesisers/}.

\item \textit{Bridgeport Music, Inc. v. Dimension Films}, 410 F.3d 792 (6th Cir.\ 2005). Full text: \url{https://law.justia.com/cases/federal/appellate-courts/F3/410/792/574458/}. The ``Get a licence or do not sample'' rule is the most-cited line in the opinion.

\item \textit{TIME}, \textit{``50 Worst Inventions,''} 2010, Auto-Tune at \#15: \url{https://content.time.com/time/specials/packages/article/0,28804,1991915_1991909_1991903,00.html}. Wikipedia, \textit{``Auto-Tune,''} \url{https://en.wikipedia.org/wiki/Auto-Tune}. NPR, \textit{``25 Years of Believe,''} \url{https://www.npr.org/2023/10/19/1207028349/25-years-ago-cher-released-a-song-that-would-change-the-sound-of-pop-music}. Wikipedia, \textit{``D.O.A. (Death of Auto-Tune),''} \url{https://en.wikipedia.org/wiki/D.O.A._(Death_of_Auto-Tune)}.

\item Walter Murch, \textit{In the Blink of an Eye: A Perspective on Film Editing}, Silman-James Press, 1995 (2nd edition 2001). PDF: \url{https://www.craftfilmschool.com/userfiles/files/Walter\%20Murch\%20-\%20In\%20the\%20Blink\%20of\%20an\%20Eye\%20Revised\%202nd\%20Edition\%20(2001,\%20Silman-James\%20Pr).pdf}. Charles Koppelman, \textit{Behind the Seen: How Walter Murch Edited Cold Mountain Using Apple's Final Cut Pro and What This Means for Cinema}, Peachpit Press, 2004: \url{https://www.peachpit.com/store/behind-the-seen-how-walter-murch-edited-cold-mountain-9780735714267}.

\item Sasson's account documented at the National Inventors Hall of Fame: \url{https://www.invent.org/blog/inventors/Legacy-Steve-Sasson}. Snopes verification of the ``Kodak suppressed the digital camera'' claim: \url{https://www.snopes.com/fact-check/kodak-digital-camera-invention/}. Knowledge@Wharton on the Kodak collapse: \url{https://knowledge.wharton.upenn.edu/podcast/knowledge-at-wharton-podcast/whats-wrong-with-this-picture-kodaks-30-year-slide-into-bankruptcy/}. Bankruptcy filing: 19 January 2012, S.D.N.Y., \$5.1bn assets / \$6.8bn liabilities.

\item Wikipedia, \textit{``Brian Walski,''} \url{https://en.wikipedia.org/wiki/Brian_Walski}. \textit{Washington Post} contemporaneous coverage: \url{https://www.washingtonpost.com/archive/lifestyle/2003/04/03/altered-picture-costs-la-times-photographer-his-job/c5e7c9e0-a836-429a-bb4e-d502f1768a96/}. World Press Photo's institutional response in TIME: \url{https://time.com/3706626/world-press-photo-processing-manipulation-disqualified/}.

\item Wikipedia, \textit{``Viacom International, Inc. v. YouTube, Inc.,''} \url{https://en.wikipedia.org/wiki/Viacom_International_Inc._v._YouTube,_Inc.}. Electronic Frontier Foundation case file: \url{https://www.eff.org/cases/viacom-v-youtube}. Variety on the March 2014 settlement: \url{https://variety.com/2014/biz/news/google-and-viacom-settle-copyright-infringement-lawsuit-over-youtube-1201137538/}.

\item PetaPixel, \textit{``The Rise and Crash of the Camera Industry in One Chart,''} \url{https://petapixel.com/2024/08/22/the-rise-and-crash-of-the-camera-industry-in-one-chart/}. Statista, \textit{``Smartphones Wipe Out Decades of Camera Industry Growth,''} \url{https://www.statista.com/chart/15524/worldwide-camera-shipments/}. CIPA shipment data series, multiple years.

\item CNN Business, \textit{``Meet the translation professionals losing their jobs to AI,''} January 2026, \url{https://www.cnn.com/2026/01/23/tech/translation-language-jobs-ai-automation-intl}. Carl Benedikt Frey (Oxford Martin School), 2025 study on translator employment across 696 US labour markets. American Translators Association industry position: \url{https://www.atanet.org/client-assistance/blog-machine-translation-vs-human-translation/}. Wikipedia, \textit{``Google Neural Machine Translation,''} \url{https://en.wikipedia.org/wiki/Google_Neural_Machine_Translation}.

\end{enumerate}

\section*{Chapter 3 -- The Human-AI Agency Continuum}

\begin{enumerate}

\item \href{../Dream\%20Machine\%20MD/2.md}{\textit{Dream Machine} Issue~2}, ``Editor's Pick,'' 10 October 2025. \url{https://www.linkedin.com/pulse/dream-machine-creative-ai-news-insight-oct-25-2-pete-woodbridge-mnrjc/}.

\item OpenAI, ``Introducing AgentKit,'' 6 October 2025. \url{https://openai.com/index/introducing-agentkit/}.

\item TechCrunch, ``OpenAI launches AgentKit to help developers build and ship AI agents,'' 6 October 2025. \url{https://techcrunch.com/2025/10/06/openai-launches-agentkit-to-help-developers-build-and-ship-ai-agents/}. Also coverage at \textit{InfoQ}, ``OpenAI Dev Day 2025 Introduces GPT-5 Pro API, Agent Kit, and More.'' \url{https://www.infoq.com/news/2025/10/openai-dev-day/}.

\item \href{../Dream\%20Machine\%20MD/2.md}{\textit{Dream Machine} Issue~2}: ``Agentic AI -- the class of AI systems that can plan, act, and pursue goals with autonomy -- promises a new era of collaboration in creative industries\ldots{} Its another step along the Human-AI Agency Continuum.'' See also \textit{TVB Europe}, ``Is Agentic AI About to Change the Media and Entertainment Industry?'' \url{https://www.tvbeurope.com/artificial-intelligence/opinion-is-agentic-ai-about-to-change-the-media-and-entertainment-industry}.

\item Google DeepMind, Veo 3.1 release, October 2025. \href{../Dream\%20Machine\%20MD/3.md}{\textit{Dream Machine} Issue~3}.

\item \textit{MusicTech}, ``iZotope Ozone 12's AI assistant is cool, but the Stem EQ is the real star.'' \url{https://musictech.com/reviews/plug-ins/izotope-ozone-12-review/}. \href{../Dream\%20Machine\%20MD/3.md}{\textit{Dream Machine} Issue~3}.

\item Adobe, ``Inaugural Adobe Creators' Toolkit Report,'' October 2025. \url{https://news.adobe.com/news/2025/10/adobe-max-2025-creators-survey}. Survey of 16,000 creators across eight countries, released at Adobe MAX 2025. \href{../Dream\%20Machine\%20MD/6.md}{\textit{Dream Machine} Issue~6}.

\item Adobe, \textit{op. cit.} The same survey: 86\% of creators use creative generative AI; 76\% say it has helped grow their business or brand; 81\% say AI lets them make content they otherwise couldn't have made; 69\% worry about their work being used to train AI without consent; 70\% are optimistic about agentic AI; 85\% would use AI that learns their creative style.

\item Mureka, ``Music Agent Studio'' launch, mid-October 2025. \href{../Dream\%20Machine\%20MD/4.md}{\textit{Dream Machine} Issue~4}. \url{https://www.linkedin.com/posts/sherrihendrickson_mureka-unveils-music-agent-studio-and-enhanced-share-7384999251526864896-cNYg/}.

\item \textit{Finsmes}, ``AdsGency Raises \$12M in Seed Funding,'' October 2025. \url{https://www.finsmes.com/2025/10/adsgency-raises-12m-in-seed-funding.html}. \href{../Dream\%20Machine\%20MD/4.md}{\textit{Dream Machine} Issue~4}.

\item \textit{Musically}, ``Meet Lenny, an AI agent to help organisers of live music events.'' \url{https://musically.com/2025/10/20/meet-lenny-an-ai-agent-to-help-organisers-of-live-music-events/}. \href{../Dream\%20Machine\%20MD/4.md}{\textit{Dream Machine} Issue~4}.

\item \textit{GamesRadar}, ``Even under USD20 million in debt, EA reportedly pushes 15,000 employees to use AI as a `thought partner' for everything from character art to playtesting.'' \url{https://www.gamesradar.com/games/even-under-usd20-million-in-debt-ea-reportedly-pushes-15-000-employees-to-use-ai-as-a-thought-partner-for-everything-from-character-art-to-playtesting/}. \href{../Dream\%20Machine\%20MD/6.md}{\textit{Dream Machine} Issue~6}.

\item PYMNTS, ``Adobe Lets Users Design and Edit Using ChatGPT.'' \url{https://www.pymnts.com/artificial-intelligence-2/2025/adobe-lets-users-design-and-edit-using-chatgpt/}. Adobe blog: ``Edit images, designs, and PDFs right inside ChatGPT -- thanks to Adobe Express, Photoshop, and Acrobat.'' \url{https://blog.adobe.com/en/publish/2025/12/10/edit-photoshop-chatgpt}. \href{../Dream\%20Machine\%20MD/12.md}{\textit{Dream Machine} Issue~12}.

\item TechCrunch, ``Anthropic launches interactive Claude apps, including Slack and other workplace tools,'' 26 January 2026. \url{https://techcrunch.com/2026/01/26/anthropic-launches-interactive-claude-apps-including-slack-and-other-workplace-tools/}. \textit{Heygen Video Agent}: \url{https://www.linkedin.com/posts/heygen_introducing-the-new-video-agent-activity-7421597801240801282-d1CF}. \href{../Dream\%20Machine\%20MD/16.md}{\textit{Dream Machine} Issue~16}.

\item \href{../Dream\%20Machine\%20MD/21.md}{\textit{Dream Machine} Issue~21}, ``Editor's Pick: Adobe and NVIDIA Just Raised the Stakes for Creative AI,'' 19 March 2026.

\item Adobe Summit 2026, ``Agentic Creative Intelligence'' keynote framing. \href{../Dream\%20Machine\%20MD/26.md}{\textit{Dream Machine} Issue~26}.

\item \href{../Dream\%20Machine\%20MD/29.md}{\textit{Dream Machine} Issue~29}, May 2026, citing Sony's adoption of Claude Code studios with multi-agent coordination.

\item Anthropic, public statements on agent deployment patterns through Q1 2026. Cf.\ \textit{Dream Machine} Issues \href{../Dream\%20Machine\%20MD/11.md}{11}, \href{../Dream\%20Machine\%20MD/16.md}{16}, \href{../Dream\%20Machine\%20MD/22.md}{22}.

\item \textit{gamesindustry.biz}, ```AI was an expensive mistake': Charles Cecil on innovation, insolvency, and Broken Sword.'' \url{https://www.gamesindustry.biz/ai-was-an-expensive-mistake-charles-cecil-on-innovation-insolvency-and-broken-sword}. \href{../Dream\%20Machine\%20MD/3.md}{\textit{Dream Machine} Issue~3}.

\item \textit{Niche Gamer}, ``Larian Studios backs off from gen AI, says tech won't be used in new Divinity.'' \url{https://nichegamer.com/larian-studios-backs-off-from-gen-ai/}. \href{../Dream\%20Machine\%20MD/14.md}{\textit{Dream Machine} Issue~14}.

\item \textit{Decrypt}, ```Warhammer 40,000' Maker Games Workshop Rules Out Generative AI.'' \url{https://decrypt.co/354482/warhammer-40000-maker-games-workshop-rules-out-generative-ai}. \href{../Dream\%20Machine\%20MD/14.md}{\textit{Dream Machine} Issue~14}.

\item \textit{Niche Gamer}, ``Manor Lords publisher Hooded Horse won't work with devs using gen AI.'' \url{https://nichegamer.com/manor-lords-publisher-hooded-horse-wont-work-with-devs-using-gen-ai/}. \href{../Dream\%20Machine\%20MD/14.md}{\textit{Dream Machine} Issue~14}.

\item \textit{gamesindustry.biz}, ``RuneScape maker Jagex says it will never use generative AI to make in-game content.'' \url{https://www.gamesindustry.biz/runescape-maker-jagex-says-it-will-never-use-generative-ai-to-make-in-game-content}. \href{../Dream\%20Machine\%20MD/16.md}{\textit{Dream Machine} Issue~16}.

\end{enumerate}

\section*{Chapter 4 -- Dead Internet, Living Web}

\begin{enumerate}

\item Imperva, \textit{2025 Bad Bot Report: How AI is Supercharging the Bot Threat}. \url{https://www.imperva.com/blog/2025-imperva-bad-bot-report-how-ai-is-supercharging-the-bot-threat/}. \href{../Dream\%20Machine\%20MD/4.md}{\textit{Dream Machine} Issue~4}.

\item Cloudflare, ``The crawl-to-click gap: Cloudflare data on AI bots, training, and referrals.'' \url{https://blog.cloudflare.com/crawlers-click-ai-bots-training/}. \href{../Dream\%20Machine\%20MD/4.md}{\textit{Dream Machine} Issue~4}. Later 2025 updates show training crawlers declining from \textasciitilde{}90\% to \textasciitilde{}74\% of AI bot activity as scraper bots rose to 24\% and a new ``agentic'' category emerged at 1.7\%; see Cloudflare, ``A deeper look at AI crawlers: breaking down traffic by purpose and industry.'' \url{https://blog.cloudflare.com/ai-crawler-traffic-by-purpose-and-industry/}.

\item Grand View Research, ``Generative AI Content Creation Market Report.'' \url{https://www.grandviewresearch.com/industry-analysis/generative-ai-content-creation-market-report}. \href{../Dream\%20Machine\%20MD/4.md}{\textit{Dream Machine} Issue~4} also cites Gartner and Europol forecasts of 90--99\% AI-generated or AI-assisted online content by 2030.

\item \href{../Dream\%20Machine\%20MD/4.md}{\textit{Dream Machine} Issue~4}, ``Editor's Pick: Is the Internet Dead Yet?'' 23 October 2025. \url{https://www.linkedin.com/pulse/dream-machine-creative-ai-news-insight-oct-25-issue-4-woodbridge-hzttc/}.

\item Wikipedia, \textit{Dead Internet Theory}. \url{https://en.wikipedia.org/wiki/Dead_Internet_theory}. \href{../Dream\%20Machine\%20MD/4.md}{\textit{Dream Machine} Issue~4}.

\item Graphite, 2025 analysis of new web content by author type (human vs.\ AI vs.\ AI-assisted). Cited in \href{../Dream\%20Machine\%20MD/4.md}{\textit{Dream Machine} Issue~4}.

\item For ``model collapse'' as a term of art, see Ilia Shumailov et al., ``The Curse of Recursion: Training on Generated Data Makes Models Forget'' (2024), and subsequent literature.

\item Futurism, ``Researchers built a social network with only AI agents -- within hours it had collapsed into warring tribes.'' \url{https://futurism.com/social-network-ai-intervention-echo-chamber}. \href{../Dream\%20Machine\%20MD/4.md}{\textit{Dream Machine} Issue~4}.

\item \textit{Digital Music News}, ``Instagram Chief Says We Should `Fingerprint Real Media' Instead of Tracking and Disclosing AI Slop.'' \url{https://www.digitalmusicnews.com/2026/01/05/instagram-chief-ai-slop-comments/}. See also \textit{WebProNews}, ``Instagram Head Warns AI Images Erode Trust, Calls for Verification Standards.'' \url{https://www.webpronews.com/instagram-head-warns-ai-images-erode-trust-calls-for-verification-standards/}. \href{../Dream\%20Machine\%20MD/13.md}{\textit{Dream Machine} Issue~13}.

\item Sundance Institute, ``Centering the Artist: Why We're Launching the AI Literacy Initiative.'' \url{https://www.sundance.org/blogs/centering-the-artist-why-were-launching-the-ai-literacy-initiative/}. \href{../Dream\%20Machine\%20MD/16.md}{\textit{Dream Machine} Issue~16}.

\item \textit{Stereogum}, ``Bandcamp bans AI music.'' \url{https://stereogum.com/2485199/bandcamp-bans-ai-music/news}. \href{../Dream\%20Machine\%20MD/14.md}{\textit{Dream Machine} Issue~14}.

\item \textit{CNET}, ``San Diego Comic-Con Draws a Line: No AI Art Allowed at 2026 Event.'' \url{https://www.cnet.com/culture/san-diego-comic-con-bans-ai-art-for-2026-event/}. \href{../Dream\%20Machine\%20MD/16.md}{\textit{Dream Machine} Issue~16}.

\item Deezer, ``AI-generated tracks now represent 44\% of all new uploaded music,'' April 2026. \url{https://newsroom-deezer.com/2026/04/ai-generated-tracks-represent-44-of-new-uploaded-music/}. \textit{Music Business Worldwide}, ``75,000 AI-generated tracks now flood Deezer daily.'' \url{https://www.musicbusinessworldwide.com/75000-ai-generated-tracks-now-flood-deezer-daily-representing-44-of-all-new-music-uploaded-to-the-platform-says-streamer/}. \textit{Dream Machine} Issues \href{../Dream\%20Machine\%20MD/7.md}{7}, \href{../Dream\%20Machine\%20MD/26.md}{26}, \href{../Dream\%20Machine\%20MD/27.md}{27}, \href{../Dream\%20Machine\%20MD/28.md}{28}.

\item \textit{The Hollywood Reporter}, ```Synthetic Sincerity' by Marc Isaacs Explores if AI Characters Can Be Taught Authenticity: IDFA.'' \url{https://www.hollywoodreporter.com/movies/movie-news/synthetic-sincerity-film-idfa-ai-authenticity-interview-1236426180/}. \href{../Dream\%20Machine\%20MD/8.md}{\textit{Dream Machine} Issue~8}.

\item Variety, ``AI-Generated Images Threaten Future of Documentary as People `Will Stop Believing Anything'.'' \url{https://variety.com/2025/film/festivals/ai-generated-images-threaten-future-of-documentary-1236583466/}. \href{../Dream\%20Machine\%20MD/8.md}{\textit{Dream Machine} Issue~8}.

\item PR Newswire, ``From Apple TV Creative to AI Filmmaker: Hoyt Dwyer's Animated Film To Compete at AI FilmFest Japan 2025.'' \url{https://www.prnewswire.com/news-releases/from-apple-tv-creative-to-ai-filmmaker-hoyt-dwyers-animated-film-to-compete-at-ai-filmfest-japan-2025-302598064.html}. \href{../Dream\%20Machine\%20MD/6.md}{\textit{Dream Machine} Issue~6}.

\item Variety, ``AI Creator Behind Viral `Deadpool,' `Harry Potter' Christmas Clip Made His Film in a Ukrainian Bomb Shelter.'' \url{https://variety.com/2026/digital/news/ai-video-deadpool-harry-potter-andrii-daniels-1236624632/}. \href{../Dream\%20Machine\%20MD/16.md}{\textit{Dream Machine} Issue~16}.

\item \textit{Branding in Asia}, ```It's the Most Terrible Time of the Year' -- McDonald's Netherlands' Wonderfully Chaotic, AI-Driven Christmas Film.'' \url{https://www.brandinginasia.com/its-the-most-terrible-time-of-the-year-mcdonalds-netherlands-wonderfully-chaotic-ai-driven-christmas-film/}. Pulled following backlash: \textit{SiliconAngle}, ``Not ready: McDonald's AI-generated ad taken down after public backlash.'' \url{https://siliconangle.com/2025/12/10/not-ready-mcdonalds-ai-generated-ad-taken-public-backlash/}. \href{../Dream\%20Machine\%20MD/11.md}{\textit{Dream Machine} Issue~11}.

\item BBC News, ``Fashion house Valentino criticised over `disturbing' AI handbag ads.'' \url{https://www.bbc.co.uk/news/articles/cwyvjyvn83go}. \href{../Dream\%20Machine\%20MD/10.md}{\textit{Dream Machine} Issue~10}.

\item \textit{Adweek}, ``Coca-Cola Uses AI to Rekindle the Magic of Its Holiday Ads.'' \url{https://www.adweek.com/creativity/coca-cola-uses-ai-to-rekindle-the-magic-of-its-holiday-ads/}. \href{../Dream\%20Machine\%20MD/6.md}{\textit{Dream Machine} Issue~6}.

\item \textit{AI News}, ``AI causes reduction in users' brain activity, MIT.'' \url{https://www.artificialintelligence-news.com/news/ai-causes-reduction-in-users-brain-activity-mit/}. \href{../Dream\%20Machine\%20MD/1.md}{\textit{Dream Machine} Issue~1}.

\end{enumerate}

\section*{Chapter 5 -- The Slop Ceiling}

\begin{enumerate}

\item Deezer, ``AI-generated tracks now represent 44\% of all new uploaded music,'' April 2026. \url{https://newsroom-deezer.com/2026/04/ai-generated-tracks-represent-44-of-new-uploaded-music/}. \textit{Music Business Worldwide}, ``75,000 AI-generated tracks now flood Deezer daily, representing 44\% of all new music uploaded to the platform.'' \url{https://www.musicbusinessworldwide.com/75000-ai-generated-tracks-now-flood-deezer-daily-representing-44-of-all-new-music-uploaded-to-the-platform-says-streamer/}. \textit{Dream Machine} Issues \href{../Dream\%20Machine\%20MD/7.md}{7}, \href{../Dream\%20Machine\%20MD/26.md}{26}, \href{../Dream\%20Machine\%20MD/27.md}{27}, \href{../Dream\%20Machine\%20MD/28.md}{28}.

\item Ditto Music research, October 2025 and prior. \textit{Press Ditto Music}, ``48\% of artists use AI to make music -- fewer than in 2023.'' \url{https://press.dittomusic.com/48-of-artists-use-ai-to-make-music-fewer-than-in-2023}. \href{../Dream\%20Machine\%20MD/2.md}{\textit{Dream Machine} Issue~2}.

\item \textit{Musically}, ``Universal and Warner could sign landmark AI deals within weeks.'' \url{https://musically.com/2025/10/02/report-umg-and-wmg-could-sign-landmark-ai-deals-within-weeks/}. Spotify Newsroom, ``Spotify Strengthens AI Protections for Artists, Songwriters, and Producers.'' \url{https://newsroom.spotify.com/2025-09-25/spotify-strengthens-ai-protections/}. \href{../Dream\%20Machine\%20MD/1.md}{\textit{Dream Machine} Issue~1}.

\item \textit{Musically}, ``50,000 AI music tracks are now uploaded to Deezer every day.'' \url{https://musically.com/2025/11/12/50000-ai-music-tracks-are-now-uploaded-to-deezer-every-day/}. \href{../Dream\%20Machine\%20MD/7.md}{\textit{Dream Machine} Issue~7}.

\item Deezer, April 2026, \textit{op. cit.}

\item \textit{Musically}, ``UMG boss slams exponential growth of AI slop on streaming services.'' \url{https://musically.com/2026/01/09/umg-boss-slams-exponential-growth-of-ai-slop-on-streaming-services/}. \href{../Dream\%20Machine\%20MD/14.md}{\textit{Dream Machine} Issue~14}.

\item \textit{Musically}, ``Report: 56.9\% of new independent songs in China are AI-generated.'' \url{https://musically.com/2026/01/05/report-56-9-of-new-independent-songs-in-china-are-ai-generated/}. \href{../Dream\%20Machine\%20MD/13.md}{\textit{Dream Machine} Issue~13}.

\item \textit{The Wrap}, ``An AI Podcasting Machine Is Churning Out 3,000 Episodes a Week -- and People Are Listening.'' \url{https://www.thewrap.com/ai-podcasts-hosts-inception-point-ai/}. \href{../Dream\%20Machine\%20MD/8.md}{\textit{Dream Machine} Issue~8}.

\item \href{../Dream\%20Machine\%20MD/28.md}{\textit{Dream Machine} Issue~28}, May 2026, citing aggregator-platform data on ``podslop'' classification.

\item \textit{The Hollywood Reporter}, ``Merriam-Webster Names `Slop' Word of the Year Amid AI Boom.'' \url{https://www.hollywoodreporter.com/news/general-news/slop-word-year-2025-merriam-webster-1236450780/}. \href{../Dream\%20Machine\%20MD/12.md}{\textit{Dream Machine} Issue~12}.

\item \textit{Digital Music News}, ``YouTube CEO Puts `Managing AI Slop' on the Priority List for 2026.'' \url{https://www.digitalmusicnews.com/2026/01/22/youtube-ceo-ai-slop-2026-comments/}. \href{../Dream\%20Machine\%20MD/16.md}{\textit{Dream Machine} Issue~16}.

\item \textit{The Guardian}, ``YouTube AI channels spreading fake, anti-Labour videos viewed 1.2bn times in 2025.'' \url{https://www.theguardian.com/technology/2025/dec/13/fake-anti-labour-video-billion-views-youtube-2025}. \href{../Dream\%20Machine\%20MD/12.md}{\textit{Dream Machine} Issue~12}.

\item Deezer/Ipsos survey, November 2025. \url{https://newsroom-deezer.com/2025/11/deezer-ipsos-survey-ai-music/}. \href{../Dream\%20Machine\%20MD/7.md}{\textit{Dream Machine} Issue~7}.

\item \textit{Bain \& Company}, ``In an AI Age, People Still Want the Radio Star.'' \url{https://www.bain.com/insights/in-an-ai-age-people-still-want-the-radio-star/}. \href{../Dream\%20Machine\%20MD/16.md}{\textit{Dream Machine} Issue~16}.

\item Deezer, April 2026, \textit{op. cit.} ``Up to 85\% of the streams generated by fully AI-generated tracks were in fact fraudulent in 2025.''

\item[15a.] \textit{Bloomberg}, ``AI Changed Chess. Grandmasters Now Win With Unpredictable Moves,'' 27 March 2026. \url{https://www.bloomberg.com/news/articles/2026-03-27/ai-changed-chess-grandmasters-now-win-with-unpredictable-moves}. \href{../Dream\%20Machine\%20MD/23.md}{\textit{Dream Machine} Issue~23}.

\item \textit{Billboard}, ``AI Artist Xania Monet Climbs the Charts -- And Signs a Multimillion-Dollar Record Deal.'' \url{https://www.billboard.com/pro/ai-music-artist-xania-monet-multimillion-dollar-record-deal/}.

\item \textit{Billboard}, \textit{op. cit.}; CNN, ``Xania Monet is the first AI-powered artist to debut on a Billboard airplay chart.'' \url{https://www.cnn.com/2025/11/01/entertainment/xania-monet-billboard-ai}.

\item \textit{Billboard}, \textit{op. cit.}

\item \textit{Bangkok Post}, ``AI singer Xania Monet signs \$3m deal with record label.'' \url{https://www.bangkokpost.com/life/tech/3142355/ai-singer-xania-monet-signs-3m-deal-with-hallwood-media}. \href{../Dream\%20Machine\%20MD/7.md}{\textit{Dream Machine} Issue~7}.

\item Multiple outlets; quoted in \textit{Billboard} feature \textit{op. cit.}

\item[20a.] Telisha Jones quoted in \textit{Billboard}, \textit{op. cit.}

\item NPR, ``Breaking Rust is a hot new country act on the Billboard charts. It's powered by AI.'' \url{https://www.npr.org/2025/11/10/nx-s1-5604320/breaking-rust-is-a-hot-new-country-act-on-the-billboard-charts-its-powered-by-ai}. \href{../Dream\%20Machine\%20MD/7.md}{\textit{Dream Machine} Issue~7}.

\item \textit{Washington Post}, ```Walk My Walk,' Breaking Rust: AI country hit triggers Nashville angst.'' \url{https://www.washingtonpost.com/style/2025/12/28/breaking-rust-ai-country/}.

\item \textit{MusicRadar}, ``The No.~1 country song in the US right now is AI-generated.'' \url{https://www.musicradar.com/music-tech/the-no-1-country-song-in-the-us-right-now-is-ai-generated}. \href{../Dream\%20Machine\%20MD/7.md}{\textit{Dream Machine} Issue~7}.

\item BBC News, ``The mysterious singer, Sienna Rose, with millions of streams is hitting the viral charts -- but who (or what) is she?'' \url{https://www.bbc.co.uk/news/articles/cq6v83gq66eo}. \href{../Dream\%20Machine\%20MD/15.md}{\textit{Dream Machine} Issue~15}.

\item \textit{Billboard}, ``How a MAGA Rapper Used AI to Create A Gospel Song That Climbed the Charts.'' \url{https://www.billboard.com/pro/maga-rapper-ai-gospel-song-climbed-charts/}. \href{../Dream\%20Machine\%20MD/9.md}{\textit{Dream Machine} Issue~9}.

\item \textit{Musically}, ``AI band Bleeding Verse's creator signs deal with Hallwood Media.'' \url{https://musically.com/2025/10/07/ai-band-bleeding-verses-creator-signs-deal-with-hallwood-media/}. \href{../Dream\%20Machine\%20MD/2.md}{\textit{Dream Machine} Issue~2}.

\item \textit{Musically}, ``Indian AI band Trilok performs live, government denies association.'' \url{https://musically.com/2025/12/17/indian-ai-band-trilok-performs-live-government-denies-association/}. \href{../Dream\%20Machine\%20MD/12.md}{\textit{Dream Machine} Issue~12}.

\item \textit{The Guardian}, ``Paul McCartney joins music industry protest against AI with silent track.'' \url{https://www.theguardian.com/music/2025/nov/17/the-sound-of-silence-why-theres-barely-anything-there-in-paul-mccartney-new-release}. \href{../Dream\%20Machine\%20MD/8.md}{\textit{Dream Machine} Issue~8}.

\item \textit{The Guardian}, ``Musicians must embrace `unstoppable force' of AI, Eurythmics' Dave Stewart urges.'' \url{https://www.theguardian.com/music/2025/dec/05/musicians-must-embrace-unstoppable-force-of-ai-eurythmics-dave-stewart-urges}. \href{../Dream\%20Machine\%20MD/11.md}{\textit{Dream Machine} Issue~11}.

\item \textit{Digital Music News}, ``Nearly 800 Creatives, Including Jason Aldean and One Republic, Sign Responsible AI Declaration -- `Stealing Our Work Is Not Innovation'.'' \url{https://www.digitalmusicnews.com/2026/01/22/stealing-isnt-innovation/}. \href{../Dream\%20Machine\%20MD/16.md}{\textit{Dream Machine} Issue~16}.

\item Stability AI, ``Universal Music Group and Stability AI Announce Strategic Alliance.'' \url{https://stability.ai/news/universal-music-group-and-stability-ai-announce-strategic-alliance}. \href{../Dream\%20Machine\%20MD/5.md}{\textit{Dream Machine} Issue~5}.

\item Stability AI, ``Warner Music Group and Stability AI Join Forces To Build The Next Generation Of Responsible AI Tools For Music Creation.'' \url{https://stability.ai/news/warner-music-group-and-stability-ai-join-forces-to-build-next-gen-tools}. \href{../Dream\%20Machine\%20MD/8.md}{\textit{Dream Machine} Issue~8}.

\item Universal Music, ``Universal Music Group and Splice to Collaborate on the Next Generation of AI-Powered Music Creation Tools for Artists.'' \url{https://www.universalmusic.com/universal-music-group-and-splice-to-collaborate-on-the-next-generation-of-ai-powered-music-creation-tools-for-artists/}. \href{../Dream\%20Machine\%20MD/12.md}{\textit{Dream Machine} Issue~12}.

\item LinkedIn / \textit{Lexology}, ``Munich Regional Court rules for GEMA against OpenAI.'' Coverage: \url{https://www.linkedin.com/posts/dr-barry-scannell-bbb5aa207_in-a-major-ruling-for-european-copyright-share-7393957246386323457-8bbx}. \href{../Dream\%20Machine\%20MD/7.md}{\textit{Dream Machine} Issue~7}.

\item \textit{EDM.com}, ```Biggest Theft in Music History': Rights Group Sues Suno as AI Music Showdown Escalates.'' \url{https://edm.com/gear-tech/rights-group-sues-suno-copyright-infringement/}. \href{../Dream\%20Machine\%20MD/7.md}{\textit{Dream Machine} Issue~7}.

\item \textit{Music Business Worldwide}, ``Wixen files \$50m copyright suit against Meta.'' \url{https://www.musicbusinessworldwide.com/wixen-files-50m-copyright-suit-against-meta-claims-tech-giant-wants-to-replace-songwriters-with-ai/}. \href{../Dream\%20Machine\%20MD/16.md}{\textit{Dream Machine} Issue~16}.

\item \href{../Dream\%20Machine\%20MD/17.md}{\textit{Dream Machine} Issue~17} reportage on UMG's \$3B suit against Anthropic.

\item \textit{Stereogum}, ``Bandcamp bans AI music.'' \url{https://stereogum.com/2485199/bandcamp-bans-ai-music/news}. \href{../Dream\%20Machine\%20MD/14.md}{\textit{Dream Machine} Issue~14}.

\item \href{../Dream\%20Machine\%20MD/18.md}{\textit{Dream Machine} Issue~18} reportage of Deezer licensing its detection tool.

\item \textit{TechRadar}, ``AI music is flooding Spotify, and subscribers are furious.'' \url{https://www.techradar.com/audio/spotify/ai-music-is-flooding-spotify-and-subscribers-are-furious-heres-why-music-fans-no-longer-trust-discover-weekly}. \href{../Dream\%20Machine\%20MD/14.md}{\textit{Dream Machine} Issue~14}.

\item \textit{CNET}, ``San Diego Comic-Con Draws a Line: No AI Art Allowed at 2026 Event.'' \url{https://www.cnet.com/culture/san-diego-comic-con-bans-ai-art-for-2026-event/}. \href{../Dream\%20Machine\%20MD/16.md}{\textit{Dream Machine} Issue~16}.

\item \textit{The Independent}, ``AI-generated song banned from Swedish charts: `It's deceiving'.'' \url{https://www.independent.co.uk/tv/news/ai-music-song-banned-sweden-spotify-b2901627.html}. \href{../Dream\%20Machine\%20MD/15.md}{\textit{Dream Machine} Issue~15}.

\item \textit{Soultracks}, ``A.I.-generated music is catchy, familiar\ldots{} and boring.'' \url{https://soultracks.com/news-ai-generated-music-is-catchy-boring/}. \href{../Dream\%20Machine\%20MD/14.md}{\textit{Dream Machine} Issue~14}.

\item[43a.] \textit{The Independent}, ``AI-generated song banned from Swedish charts: `It's deceiving'.'' \url{https://www.independent.co.uk/tv/news/ai-music-song-banned-sweden-spotify-b2901627.html}. \href{../Dream\%20Machine\%20MD/15.md}{\textit{Dream Machine} Issue~15}.

\item[43b.] \textit{Marketing Week}, ``You can't dismiss AI ads as slop when they're winning in testing.'' Coverage discussed in \href{../Dream\%20Machine\%20MD/22.md}{\textit{Dream Machine} Issue~22}.

\item \textit{Billboard}, ``The Real Story Behind The AI Song That Knocked Tyla Off No.~1 On Billboard Afrobeats Chart.'' \url{https://www.billboard.com/pro/ai-song-knocked-tyla-off-no-1-afrobeats/}. \href{../Dream\%20Machine\%20MD/30.md}{\textit{Dream Machine} Issue~30}.

\item \textit{MusicTech}, ``Jack Antonoff brands AI music makers as `godless whores'.'' \url{https://musictech.com/news/industry/jack-antonoff-ai-music-makers-godless-whores/}. \href{../Dream\%20Machine\%20MD/30.md}{\textit{Dream Machine} Issue~30}.

\end{enumerate}

  \nocite{*}
  \printbibliography[heading=bibintoc,title={Bibliography}]

\end{document}